\documentclass[12pt]{report}
\newif\iffigs\figstrue
\usepackage{latexsym,amssymb} 
\iffigs  
  \input{epsf} 
\else
\fi
\font\tenmsbm=msbm10 scaled 1200
\font\sevenmsbm=msbm9
\newfam\msbmfam  
\textfont\msbmfam=\tenmsbm 
\scriptfont\msbmfam=\sevenmsbm

\newtheorem{definizione}{Definition}[section]
\newcommand{\bdefi}{\begin{definizione}}
\newcommand{\edefi}{\end{definizione}}
\makeatletter
\@addtoreset{equation}{section}
\makeatother

\newcommand{\eqn}[1]{(\ref{#1})}
\newcommand{\ft}[2]{{\textstyle\frac{#1}{#2}}}
\newsavebox{\uuunit}
\sbox{\uuunit}
    {\setlength{\unitlength}{0.825em}
     \begin{picture}(0.6,0.7)
        \thinlines
        \put(0,0){\line(1,0){0.5}}
        \put(0.15,0){\line(0,1){0.7}}
        \put(0.35,0){\line(0,1){0.8}}
       \multiput(0.3,0.8)(-0.04,-0.02){10}{\rule{0.5pt}{0.5pt}}
     \end {picture}}
\newcommand {\unity}{\mathord{\!\usebox{\uuunit}}}

\def\veck{\stackrel{\rightarrow}{k}}

\def\IP{\relax{\rm I\kern-.18em P}}

\font\tenmsbm=msbm10 scaled 1200
\font\sevenmsbm=msbm9
\newfam\msbmfam
\textfont\msbmfam=\tenmsbm
\scriptfont\msbmfam=\sevenmsbm

\def\dslash{\hbox{\ooalign{$\displaystyle\partial$\cr$/$}}}
\def\Aslash{\hbox{\ooalign{$\displaystyle A$\cr$\hspace{.03in}/$}}}
\def\Aplusslash{\hbox{\ooalign{$\displaystyle A^+$\cr$\hspace{.03in}/$}}}
\def\Aminusslash{\hbox{\ooalign{$\displaystyle A^-$\cr$\hspace{.03in}/$}}}

\def\Zminusslash{\hbox{\ooalign{$\displaystyle Z^-$\cr$\hspace{.03in}/$}}}

\def\nslash{\nabla\!\!\!\!/}
\def\inbar{\vrule height1.5ex width.4pt depth0pt}
\def\IC{\relax\,\hbox{$\inbar\kern-.3em{\rm C}$}}
\def\IG{\relax\,\hbox{$\inbar\kern-.3em{\rm G}$}}
\def\IB{\relax{\rm I\kern-.18em B}}
\def\ID{\relax{\rm I\kern-.18em D}}
\def\IL{\relax{\rm I\kern-.18em L}}
\def\IF{\relax{\rm I\kern-.18em F}}
\def\IH{\relax{\rm I\kern-.18em H}}
\def\II{\relax{\rm I\kern-.17em I}}
\def\IN{\relax{\rm I\kern-.18em N}}
\def\IM{\relax{\rm I\kern-.18em M}}
\def\IP{\relax{\rm I\kern-.18em P}}
\def\IQ{\relax\,\hbox{$\inbar\kern-.3em{\rm Q}$}}
\def\bfzero{\relax\,\hbox{$\inbar\kern-.3em{\rm 0}$}}
\def\IK{\relax{\rm I\kern-.18em K}}
\def\IG{\relax\,\hbox{$\inbar\kern-.3em{\rm G}$}}
 \font\cmss=cmss10 \font\cmsss=cmss10 at 7pt
\def\IR{\relax{\rm I\kern-.18em R}}
\def\IGam{\relax{{\rm I}\kern-.18em \Gamma}}
\def\ZZ{\relax\ifmmode\mathchoice
{\hbox{\cmss Z\kern-.4em Z}}{\hbox{\cmss Z\kern-.4em Z}}
{\lower.9pt\hbox{\cmsss Z\kern-.4em Z}}
{\lower1.2pt\hbox{\cmsss Z\kern-.4em Z}}\else{\cmss Z\kern-.4em
Z}\fi}
\def\bfone{\relax{\rm 1\kern-.35em 1}}

\def\LL{{\cal L}}
\def\a{\alpha}
\def\b{\beta}
\def\d{\delta}
\def\l{\lambda}

\def\c{\chi}
\def\g{\gamma}

\def\s{\sigma}
\def\e{\epsilon}
\def\r{\rho}
\def\o{\omega}
\font\cmss=cmss10 \font\cmsss=cmss10 at 7pt

\def\inbar{\vrule height1.5ex width.4pt depth0pt}
\def\IC{\relax\,\hbox{$\inbar\kern-.3em{\rm C}$}}
\def\bfzero{\relax\,\hbox{$\inbar\kern-.3em{\rm 0}$}}
\def\bfone{\relax{\rm 1\kern-.35em 1}}

\def\ii{{\rm i}}
\def\diag{{\rm diag}}

\def\cA{{\cal A}} \def\cB{{\cal B}}
\def\cC{{\cal C}} \def\cD{{\cal D}}
 \def\cG{{\cal G}}
\def\cH{{\cal H}} 
 
\def\cL{{\cal L}} \def\cM{{\cal M}}
\def\cN{{\cal N}} \def\cO{{\cal O}}
 
\def\cR{{\cal R}} 
 
\def\cY{{\cal Y}}
\def\tilde{\widetilde}

\def\del{\partial }

\def\IE{\relax{{\rm I\kern-.18em E}}}

\def\IGam{\relax{{\rm I}\kern-.18em \Gamma}}

\def\ii{{\rm i}}
\newsavebox{\bobox}
\sbox{\bobox}
    {\setlength{\unitlength}{0.825em}
     \begin{picture}(0.6,0.7)
        \thinlines
        \put(0,0){\line(1,0){0.7}}
        \put(0,0){\line(0,1){0.7}}
        \put(0.7,0){\line(0,1){0.7}}
        \put(0,0.7){\line(1,0){0.7}}
       \multiput(0,0)(0.05,0.05){14}{\rule{0.4pt}{0.4pt}}
       \multiput(0,0.65)(0.05,-0.05){13}{\rule{0.4pt}{0.4pt}}
     \end {picture}}
\newcommand {\xbox}{\mathord{\!\usebox{\bobox}}\,}
\def\bet{\begin{tabular}}
\def\eet{\end{tabular}}

\def\a{\alpha}
\def\b{\beta}
\def\l{\lambda}

\def\c{\chi}
\def\g{\gamma}

\def\s{\sigma}
\def\e{\epsilon}
\def\ve{\varepsilon}

\def\be{\bar{\epsilon}}

\def\lll{<\!\!<}
\def\ll{\left}
\def\rr{\right}
\def\nn{\nonumber}
\def\be{\begin{equation}}
\def\ee{\end{equation}}
\def\beq{\begin{eqnarray}}
\def\eeq{\end{eqnarray}}
\begin{document}
%%%%%%%%%%%%%%%%%%%% title page %%%%%%%%%%%%%%%%%%%%%%%%%%%%%%%%
\begin{titlepage}
\begin{flushright}
ULB-TH/00-02\\
hep-th/0002116\\
Febraury 1999\\
\end{flushright}
\vskip 2cm
\begin{center}
{\LARGE 
Harmonic Analysis and Superconformal\\ 
Gauge Theories in Three Dimensions\\ 
from $AdS/CFT$ Correspondence
}\\
\vfill%\vskip 15mm%27.mm
{\large  Leonardo Gualtieri   } \\
\vfill%\vskip 7mm%1cm
{\small
Physique Th\'eorique et Math\'ematique,\\
Universit\'e Libre de Bruxelles,\\
Campus Plaine C.P. 231, B-1050 Bruxelles, Belgium
}
\end{center}
\vfill
\begin{center}
{\bf Abstract}
\end{center}
{\small 
In this thesis I review various aspects of the $AdS_4/CFT_3$
correspondence, where $AdS_4$ supergravity arises from compactification 
of $M$--theory on a coset space $G/H$ and preserves $\cN<8$ supersymmetries.
One focal point of my review is that the
complete spectrum of such $\cN$--extended supergravity can be
determined by means of harmonic analysis on the homogeneous space $G/H$. 
This spectrum can be matched with the candidate conformal theory on the boundary,
in this way providing very non--trivial checks of the $AdS/CFT$ correspondence.
Furthermore, this spectrum can be useful to study the representation
theory of $\cN$--extended supersymmetry on $AdS_4$, namely representation theory 
for the superalgebra of $Osp(\cN\vert 4)$.
I review $Osp(\cN\vert 4)$ representation theory, and
derive the translation vocabulary between states of $AdS_4$ supergravity and conformal
superfields on the boundary, by means of the double interpretation
of $Osp(\cN\vert 4)$ unitary irreducible representations. 
In the cases of $\cN=2,3$, using results from harmonic analysis I 
give the complete structure of all supermultiplets. 
Harmonic analysis as a method to determine spectra of supergravity 
compactifications is explained.
Calculations are explicitly performed in the case  
$G/H=M^{111}=(SU(3)\times SU(2)\times U(1))/(SU(2)\times U(1)\times 
U(1))$, preserving $\cN=2$ supersymmetries. For this manifold, and also for 
the case $G/H=Q^{111}=(SU(2)\times SU(2)\times 
SU(2))/(U(1)\times U(1))$, I describe the construction of a
candidate dual superconformal theory on the boundary. This construction is
based on geometrical insight provided by the properties of the
metric cone $\cC(G/H)$ transverse to the $M2$--brane wordvolume.
}
\vspace{2mm} \vfill \hrule width 3.cm
\end{titlepage}
%%%%%%%%%%%%%%%%%%%%%%%% end titlepage %%%%%%%%%%%%%%%%%%%%%%%%%%
\tableofcontents
\newpage
$~$
\newpage
\chapter*{Introduction}
\addcontentsline{toc}{chapter}{Introduction}
\par
The $AdS/CFT$ correspondence is a recent interesting development in the
context of string theory and $M$-theory. It is based on the conjecture,
proposed by J. Maldacena \cite{Maldyads} at the end of '97 and further developed by E. Witten
\cite{Wittenads} and S. Gubser, I. Klebanov, A. Polyakov \cite{GKPads}, 
that string$/M$ theory on some backgrounds of the form
\be
AdS_d\times X_{11-d}
\ee
or
\be
AdS_d\times X_{10-d},
\ee
where $X$ is a compact space and $AdS$ is the anti--de Sitter space \cite{defAdS},
is equivalent to a {\it superconformal quantum field theory} (SCFT) on the boundary
of the anti--de Sitter space. Actually, this boundary
coincides with compactified Minkowski space in $d-1$ dimensions.
The superconformal theory can also be viewed as a theory defined on 
a stack of $N$ coincident
branes, $D$--branes in ten dimensions or $M$--branes in eleven dimensions, 
with gauge group $U\ll(N\rr)^k$ or $SU\ll(N\rr)^k$.
For example \cite{Maldyads}, the string theory on $AdS_5\times S^5$ is equivalent to the 
$\cN=4~SYM$ theory with gauge group $U\ll(N\rr)$ and living on the four dimensional boundary of $AdS_5$.
\par
A striking point of this correspondence is that in the limit $g_{YM}^2N\longrightarrow\infty$, $g_{YM}^2
\longrightarrow 0$ the duality is between {\it classical supergravity} and a strongly coupled
SCFT, since the string/$M$-theory corrections are of order $1/N$.
So it is possible to determine physical observables of the quantum theory 
on the boundary by classical supergravity calculations on the bulk.  
\par
The starting point of this conjecture has been the observation that
the isometry group of string$/M$ theory on $AdS_d$ is $SO\ll(2,d-1\rr)$,
and this is also the conformal group in $d-1$ dimensions. So, these two theories have the same symmetry.
The conjecture, in its more complete formulation, is that  there is a
one to one correspondence between supergravity fields $\Phi$
on the bulk and conformal
primary operators $\cO$ on the boundary, and the generating functional of the
correlators of the boundary theory can be written, in the appropriate limit, in terms of the classical
supergravity action.
\par
For every off shell field configuration on the boundary of $AdS$ $\Phi_0$
there is a unique on shell field configuration $\Phi$ that is regular on the bulk and which
has $\Phi_0$ as boundary value. 
Then the field $\Phi$ on the bulk depends on its value $\Phi_0$ 
on the boundary, and the generating functional is \cite{Wittenads}
\be
\label{Wittform}
Z\ll(\Phi_0\rr)\equiv\ll<e^{\int_{AdS_{d-1}}{\Phi_0\cO}}\rr>
\stackrel{AdS/CFT}{=}
e^{-S\ll(\Phi\ll(\Phi_0\rr)\rr)},
\ee
where, in the appropriate limit, $S\ll(\Phi_0\rr)$ is the classical supergravity action on the bulk. 
\par
We can view the correspondence as between the states of supergravity
compactified on the bulk, which are
the states created by the on--shell fields $\Phi$, and 
conformal operators on the
boundary. In particular,
the BPS states of compactified supergravity correspond to 
short primary superconformal
operators of the boundary theory, which are protected against quantum corrections.
\par
Another feature of this correspondence is that the set of the
Kaluza Klein states on the bulk must completely match with the set of 
conformal operators on the boundary; 
the energies of these Kaluza Klein states correspond to the conformal weights 
of the operators on the boundary;
in particular, in the case of BPS states,
this correspondence can be checked without calculations at the quantum level,
since the conformal weights are protected against quantum corrections.
This is true also for the other quantum numbers.
\par
The most studied cases of $AdS/CFT$ correspondence are those with $D=10,~d=5$,
where the bulk theory is  the compactification of string theory or, at low energy,
ten dimensional supergravity, on
\be
AdS_5\times X_5. 
\ee
This case has become very popular since the bulk theory is string theory which is well known, and the boundary theory has
four dimensions. In the present thesis I rather consider the case $D=11,~d=4$. Then
the bulk theory is the compactification of $M-$theory, or, at low energy, eleven dimensional supergravity, on
\be
AdS_4\times X_7.
\ee
There are various reasons of this choice.
\begin{itemize}
\item Since a formulation of the quantum theory for the fundamental degrees of freedom of 
$M$--theory is lacking, it is of utmost interest to explore the properties of all its vacua.
\item The study of $AdS_4/CFT_3$ correspondence is useful to understand the three dimensional 
conformal field theories, which are not well known. For example, a classification of the 
central charges as known for four dimensional CFTs \cite{anselmi} is lacking for three 
dimensional CFTs. Furthermore, three dimensional conformal field theories are intrinsically 
interesting, being related to statistical mechanics.
\item In the eighties the Kaluza Klein compactifications of eleven dimensional supergravity on 
$AdS_4\times X_7$ background was studied in order to find a realistic theory as a supergravity
compactification. In particular, the manifold $X_7=M^{111}$ (that I will introduce in the 
following) was studied \cite{kkwitten},\cite{cdfm111},\cite{spec321}, having this manifold 
isometry $SU\ll(3\rr)\times SU\ll(2\rr) \times U\ll(1\rr)$ as the standard model group. That 
way resulted to be wrong, because such compactifications yield theories with unphysical 
cosmological constants and no chiral fermions. Today, in the completely new context of 
$AdS/CFT$ correspondence, the anti--de Sitter space resulting by these compactifications is 
no more a flaw of the theory, but an asset, and all those results can be utilized in the 
new perspective.
\end{itemize}
\par
In general, there should be $AdS/CFT$ correspondence if $AdS_4\times X_7$ is a classical 
supergravity solution. This restricts the possible choices of the compact manifold $X_7$.
At the moment, there are three kinds of $AdS_4\times X_7$ correspondences that have 
been studied:
\begin{itemize}
\item $X_7=S^7$, that yields maximal $\cN=8$ supergravity 
\cite{Maldyads}, \cite{Ahar}, \cite{correspN8}.
\item $X_7=S^7/\Gamma$ with $\Gamma$ a discrete group, namely, an orbifold of $S^7$;
these cases yield consistent truncations of $\cN=8$ supergravity \cite{Ahar}, \cite{corresporbifoldN8}.
\item $X_7=G/H$ coset manifold that is also an Einstein space; these cases yield
$\cN<8$ supergravities which are not truncations of the $\cN=8$ theory, but completely
new ones \cite{tatar} (see \cite{witkleb} for the $AdS_5\times X_5$ case).
\end{itemize}
The same considerations hold true for the correspondence with 
$AdS_5\times X_5$ or other choices of $D,d$.
\par
Up to now, several checks have been found of the $AdS_4/CFT_3$ correspondence. The most complete of them
refer to the case $X_7=S^7$, or to the cases of $S^7$ orbifolds. In most of these cases, the correspondence
of the BPS Kaluza Klein states of supergravity with the BPS superfield operators of the SCFT has been
checked. Furthermore, in those cases where the correlators of the SCFT are known in the strong coupling limit,
the formula (\ref{Wittform}) comes out true. This holds also for other choices of $D,d$.
\par
Nevertheless, these are not the strongest possible checks of this type, particularly 
for the check of the spectrum.
In the maximal supersymmetric case $X_7=S^7$, the energies of the Kaluza Klein states and the conformal weights of 
the superconformal operators depend only on their $R$--symmetry representations. 
Being the superisometry $Osp\ll(8\vert 4\rr)$ the same, it is not really surprising that the energies and the conformal 
weights actually coincide.
And the truncations of these theories do not contain any really new information.
On the contrary, for supergravities on $X_7=G/H\neq S^7$, 
these energies and conformal weights depend also on the so called
{\it flavour group} representations, and this dependence is ruled not only by the $Osp\ll(\cN\vert 4\rr)$ representation
theory, but also - on the bulk side - by the geometry of the compactification on $X_7$. Furthermore, the theories with $\cN<8$ are
less constrained than the maximal supersymmetric one. Then, a check of the spectrum in a case with $X_7=G/H\neq S^7$ 
is more significant than in the maximally supersymmetric case.
\par
This thesis is mainly based on the work done during my last Ph.D. year in the collaborations
\cite{noi1},\cite{noi2},\cite{noi3},\cite{noi4}, in order 
to discuss non--trivial checks of $AdS/CFT$ correspondence in the cases
\be
AdS_4\times \ll(G\over H\rr)_7
\ee
preserving $\cN<8$ supersymmetries. 
It is also my aim to make a systematic review of the algebraic and geometric foundations of
this correspondence.
\par
We have studied in detail the case of the manifold
\be
\label{X7=M}
X_7=M^{111}={SU\ll(3\rr)\times SU\ll(2\rr)\times U\ll(1\rr)\over SU\ll(2\rr)\times U\ll(1\rr)\times U\ll(1\rr)}
\ee
preserving $\cN=2$ supersymmetries. Furthermore, we have studied, with less detail, the cases of the manifold
\be
X_7=Q^{111}={SU\ll(2\rr)\times SU\ll(2\rr)\times SU\ll(2\rr)\over U\ll(1\rr)\times U\ll(1\rr)}
\ee
preserving $\cN=2$ supersymmetries, and of the manifold
\be
X_7=N^{010}={SU\ll(3\rr)\over U\ll(1\rr)}
\ee
preserving $\cN=3$ supersymmetries. 
\par
In \cite{noi3}, we have constructed superconformal theories candidate to be 
dual to supergravity on $AdS_4\times M^{111}$ and to supergravity on
$AdS_4\times Q^{111}$
\footnote{The construction of the SCFT dual to supergravity on $AdS_4\times 
N^{010}$ is in preparation \cite{N010Q111sp}.}. Matching these theories with
the supergravity on the bulk prevoiously derived \cite{noi1},
we found new non--trivial checks of $AdS/CFT$ correspondence.
Furthermore, in order to reach these results, 
other results were obtained as a byproduct \cite{noi1},\cite{noi2},\cite{noi3},\cite{noi4}.
\begin{itemize} 
\item We built a case of $AdS/CFT$ correspondence following all the path, from the development of
the supergravity theory on the bulk to the development of the candidate superconformal field
theory on the boundary. This gave us a deeper understanding of the mechanism of $AdS/CFT$ correspondence,
expecially on the relations between the conformal superfields on the boundary and the Kaluza Klein 
spectrum on the bulk \cite{noi2}.
\item We used the techniques of harmonic analysis in order to find the
complete spectrum of supergravity on $AdS_4\times M^{111}$ \cite{noi1}.
These techniques had been developed in the eighties \cite{salamstrat}, \cite{bosonicmassformula}, \cite{spectfer}, \cite{univer}
but this is the first time the complete
spectrum of an intricate case as (\ref{X7=M}) has been worked out; now we know more about how to handle such problems. 
Furthermore, this spectrum has value as a supergravity result, even out of the $AdS/CFT$ correspondence context.
\item Up to know, the structure of several $\cN=2$ and $\cN=3$ $AdS_4$ supermultiplets was not known. 
The spectra of supergravities we found give us the lacking information on the general representation theory of $\cN=2$ and $\cN=3$
supersymmetry on $AdS_4$, and the complete structure of all $\cN=2$
\cite{noi1} and $\cN=3$ \cite{noi4} supermultiplets\footnote{With the exception 
of the $\cN=3$ short supermultiplets with $J_0=1/2$ and $J_0=3/2$.}, completing the results of
\cite{multanna}, \cite{frenico}.
\end{itemize}
\par
An analogous program has been carried out by \cite{witkleb}, \cite{T11spectrum}, \cite{sergiotorino}
in the case of $AdS_5\times X_5$ with $X_5=T^{11}={SU\ll(2\rr)\times SU\ll(2\rr)\over U\ll(1\rr)}$.
In this case the conformal theory has been found in \cite{witkleb} as a deformation of an orbifold theory with larger supersymmetry.
The Kaluza Klein spectrum of the corresponding supergravity has been worked out in \cite{T11spectrum}, and a comparison with
superfields on the boundary theory \cite{sergiotorino} gave another non trivial check of the $AdS/CFT$ correspondence.
Furthermore, the same program has recently been carried out in \cite{stiefel} for an
other $AdS_4\times X_7$ case, the one of the Stiefel manifold $X_7=SO(5)/SO(3)$, finding similar results.
\par
\vskip .5cm
In this thesis I present our results in a systematical and didactic form, sometimes going into
details more than the papers \cite{noi1}, \cite{noi2}, \cite{noi3},
\cite{noi4}. Furthermore, some technical details are new and unpublished.
\par
\section*{Contents of the thesis}
\par
The present thesis is organized as follows.
\par
In chapter $1$ I review some basic concepts of the $AdS/CFT$ correspondence, in order to explain the background and
the motivation of the subsequent work. Then, I consider explicitly the case I am interested in, which is
\be
AdS_4\times \ll(G\over H\rr)_7.
\ee
\par
In chapter $2$ I study the representation theory of $Osp\ll(\cN\vert 4\rr)$, that is, of supersymmetric theories on $AdS_4$.
With the help of Lie algebra techniques I show the double interpretation of the $Osp\ll(\cN\vert 4\rr)$ unitary irreducible representations,
as states of bulk supergravity and as superfields of a boundary superconformal theory. In the $\cN=2$ case I explicitly
explain how to know, given a state on the bulk, which is the corresponding conformal operator on the boundary \cite{noi2}.
Furthermore I give the complete structure of all the $Osp\ll(\cN\vert 4\rr)$ unitary irreducible representations, namely, the
supermultiplets of $AdS_4$ supersymmetry, in the cases $\cN=2$ \cite{noi1} and $\cN=3$ \cite{noi4}. 
I explain how to find this structure by the matching of results found with the Freedmann Nicolai method of norms \cite{frenico} 
with results given by Kaluza Klein spectra.
\par
In chapter $3$ I explicitly derive the complete spectrum of supergravity compactified on
\be
AdS_4\times M^{111}
\ee
using harmonic analysis \cite{noi1}. This is a very powerful method, which allow to solve a differential equation problem by purely
algebraic calculations. A detailed description of all the mathematical tools used and of our derivation is given.
\par
In chapter $4$ I build candidate SCFT's dual by $AdS/CFT$ correspondence to supergravity compactified on $AdS_4\times M^{111}$ and
to supergravity compactified on $AdS_4\times Q^{111}$ \cite{noi3}. 
Unfortunately, while when the compact manifold is the seven sphere or an orbifold of the seven sphere there is a
straightforward way to build the conformal theory, by relating it to a ten dimensional string theory with $D$--branes, 
this seems not to be possible when the compact manifold is a coset manifold as $M^{111}$ or $Q^{111}$.
One have to use geometrical intuition to argue the fundamental field content and the gauge group of the theory.  
However, I show that having these theories a toric description, there are strong arguments to argue them, and the
results fit surprisingly with the bulk theory.
\newpage
$~$
\newpage
\par
\chapter{$AdS/CFT$ Correspondence and $G/H~M-$branes}
\par
\section{The $AdS/CFT$ Correspondence}
In this section I review some basic concepts of the $AdS/CFT$ correspondence,
in order to explain the background and the motivation of the subsequent work.
Several excellent and complete reviews on this wide field of research are available in 
the literature \cite{Maldyrev}, \cite{sevrev}. 
\par
\subsection{The Maldacena Conjecture for $AdS_5\times S^5$}
\par
Let us consider $IIB$ string theory on flat ten dimensional Minkowski space,
with $N$ coincident $D3$--branes.
The perturbative excitations of this theory are the closed strings, which are the excitations
of Minkowski empty space, and the open strings, which can end only on the $D$--branes,
and are the excitations of the $D$--branes themselves. 
\par
Let us consider the low energy limit of the system, namely, 
take into account only the energies lower than the string
scale $1/\sqrt{\alpha'}$ 
\be
\label{lel}
E\sqrt{\alpha'}\lll 1
\ee
keeping all the dimensionless parameters ($g_s,N$) fixed.
Then only the massless string states can be excited.
The effective action of massless modes, 
obtained by integrating out the massive fields, has the form
\be
S=S_{\rm bulk}+S_{\rm brane}+S_{\rm int}.
\ee
\begin{itemize}
\item $S_{\rm bulk}$ is the action of ten dimensional supergravity; in the low energy
limit
\footnote{In the actual calculations, the simplest way to perform the low energy limit
(\ref{lel}) is to send $\a'\longrightarrow 0$; however we must remind that
if one wants to be rigorous, only dimensionless quantities can be sent to
zero; $\a'\longrightarrow 0$ is a shorthand notation for $E\sqrt{\a'}\lll 0$.}
it becomes the action of free ten dimensional supergravity in Minkowski space.
\item $S_{\rm brane}$ is defined on the $3+1$--dimensional brane worldvolume, and
in the low energy limit becomes the action of $\cN=4$ super Yang Mills (SYM) theory 
with gauge group $U\ll(N\rr)$, and 
\be
g_{YM}^2=4\pi g_s.
\ee
Notice that this is a superconformal field theory (SCFT).
\item $S_{\rm int}$ describes the interaction between the brane and the bulk, and in the
low energy limit disappears.
\end{itemize}
Then in the low energy limit there are two decoupled systems, free $IIB$ supergravity on
Minkowski space and $\cN=4$ SYM theory with gauge group $U\ll(N\rr)$.
\par
But string theory can be also viewed from the so--called macroscopic point of view.
The absorption of closed strings by 
the $D$--branes can be seen also as the interaction of the string modes with a
non--trivial supergravity background, due to a massive and charged source localized at
the position of the $D$--branes. In other words, the $D$--branes behave as
massive and charged objects, sources of the supergravity fields. 
The $IIB$ supergravity background is the following $p$--brane solution:
\beq
ds^2&=&f^{-1/2}\ll(-dt^2+dx_1^2+dx_2^2+dx_3^2\rr)+f^{1/2}\ll(dr^2+r^2d\Omega_5^2\rr)\nn\\
F_5&=&\ll(1+*\rr)dtdx_1dx_2dx_3df^{-1}\nn\\
f&=&1+{R^4\over r^4},~~~~R^4\equiv 4\pi g_sN{\alpha'}^2,
\label{pbrane10}
\eeq
where both the mass and the five--form charge (per unit volume) 
are proportional to the number of the branes $N$.
Furthermore this solution is a BPS solution, namely, it preserves half of the $IIB$ supersymmetry.
\par
This is a black brane solution, with an horizon at $r=0$.
The energy $E_p$ of an object as measured by an observer at a constant position $r$ and the energy
$E$ measured by an observer at infinity are related by the redshift factor
\be
E=f^{-1/4}E_p,
\ee
then as an object is brought near the horizon, it appears with lower energy to the observer 
at infinity.
\par
We want to consider only the {\it low energy} $IIB$ string theory excitations on this background,
where I mean low energy from the point of view of an observer at infinity. There are two kinds
of low energy excitations: the massless particles propagating in the bulk region (that is, $r/R\gg 1$)
with large wavelength, and any kind of excitation if it is close enough to the horizon.
These two kinds of excitations are decoupled. The bulk massless particles cannot excite the near
horizon region, because the cross section $\sigma\sim R^8\omega^3$ 
is small in the low energy limit (corresponding to big particle wavelengths); we can
reformulate this phenomenom saying that the particles cannot be absorbed in this limit because
their wavelengths are bigger than the typical gravitational size of the brane. On the other hand,
the near horizon excitations have to climb an high potential barrier to escape from the asymptotic region.
\par
In the near horizon region, defined by $r\lll R$, $f\sim R^4/r^4$, so the metric becomes
\be
\label{nearhorr10}
ds^2={r^2\over R^2}\ll(-dt^2+dx_1^2+dx_2^2+dx_3^2\rr)+R^2{dr^2\over r^2}+R^2d\Omega_5^2\,,
\ee
that is the metric of
\be
AdS_5\times S^5
\ee
with $R=\ll(4\pi g_sN\alpha'^2\rr)^{1/4}$ curvature radius of $AdS_5$ and of $S^5$. From this
point of view, $N$ is the flux of the five--form through $S^5$.
\par
To be more precise about the near horizon limit, a string excitation has $E_p\sim{1\over\sqrt{\alpha'}}$;
an observer at infinity sees the energy
\be
E=f^{-1/4}E_p\sim{r\over\sqrt{\alpha'}}E_p\sim{r\over\alpha'}\,.
\ee
In the low energy limit $E\sqrt{\a'}\lll 1$, then,
\be
{r\over\sqrt{\a'}}\lll 1\,.
\label{roversqrt}
\ee
This means that any excitation of string theory does survive if it is enough
close to the horizon to satisfy (\ref{roversqrt}).
We can express this by a coordinate redefinition:
\be
U\equiv {r\over\sqrt{4\pi g_sN}\alpha'}={r\over R^2}\,.
\ee
In terms of $U$, the energies of the near horizon string excitations are
finite, and the metric (\ref{nearhorr10}) becomes
\beq
ds^2&=&\sqrt{4\pi g_sN}\alpha'\ll[U^2\ll(-dt^2+dx_1^2+dx_2^2+dx_3^2\rr)+
{dU^2\over U^2}+d\Omega_5^2\rr]\nn\\
&=&R^2\ll[U^2\ll(-dt^2+dx_1^2+dx_2^2+dx_3^2\rr)+
{dU^2\over U^2}+d\Omega_5^2\rr]
\label{nearhorU10}
\,.
\eeq
We have derived this metric as a near horizon geometry, that is, the
(\ref{nearhorU10}) is the metric in the region $U\lll 1/R$. 
Here $R$ is only a constant factor in front of
the metric. We can rescale the coordinates, so that the region $U\lll 1/R$
describes an entire $AdS_5$ space. In other words, we blow up the near horizon 
(or ''throat'') region of the (\ref{pbrane10}) geometry into 
the entire $AdS_5\times S^5$ space. The supergravity excitations of the throat coincide
with the excitations of $AdS_5\times S^5$ supergravity.
\par
The low energy theory, then, consists on these two decoupled parts, the free $IIB$ supergravity on
Minkowski space and, near the horizon, the $AdS_5\times S^5$ $IIB$ string theory 
(with all the excitations).
\par
From both the points of view, then, in the low energy limit there are two decoupled systems,
one of which is the free empty $IIB$ supergravity. This suggests that the second systems appearing
in both description may be dual, namely, mathematically equivalent.
\par
This is the {\bf Maldacena conjecture}: the $\cN=4$ $U\ll(N\rr)$ SYM quantum field
theory in $3+1$ dimensions is dual to $IIB$ string theory on $AdS_5\times S^5$ background.
\par
In the above reasoning we have kept fixed the two dimensionless parameters of the
theory, $g_s$ and $N$, or, equivalently, $g_s$ and $X\equiv 4\pi Ng_s$ (which is
the t'Hooft coupling). Let us consider various limits of these parameters.
\begin{itemize}
\item When, as in the above reasoning, 
\be
X,~g_s~~\hbox{finite}\,,
\ee
we have the
correspondence between \begin{it} $IIB$ string theory on $AdS_5\times S^5$ and 
the $\cN=4$ four dimensional SYM theory with gauge group $U(N)$, $N$ finite and t'Hooft coupling
finite.\end{it} 
\item When 
\beq
X&&\hbox{finite}\nn\\
g_s&\longrightarrow& 0\nn\\ 
N&\longrightarrow&\infty\,,
\eeq
the above reasoning does not change, because $g_s,~N$ appear only in the
combination $X$; in particular, $\sqrt{\a'}\sim R$ remains true.
\par
The correspondence is between classical \begin{it} $IIB$ string theory (that is, with only
tree diagrams, because $g_s\longrightarrow 0$) on $AdS_5\times S^5$ and the $\cN=4$
four dimensional SYM theory with gauge group $U(N)$, 
$N\longrightarrow\infty,~X$ finite. \end{it}
\item When 
\beq
X&\longrightarrow&\infty\,,\nn\\
g_s&\longrightarrow& 0\nn\\ 
N&\longrightarrow&\infty\,,\label{bigtHooftlimit}
\eeq
we have
\be
R=X^{1/4}\sqrt{\a'}\gg\sqrt{\a'}\,.
\ee
On the bulk, the classical supergravity excitations decouple 
from the other string excitations. 
In fact, ten dimensional supergravity on the $AdS_5\times S^5$ background is a theory whose
dynamical fields are the fluctuations around this background. These fields can be expanded
in $S^5$ harmonics, yielding a tower of five dimensional supergravity Kaluza Klein fields, 
whose masses are of order $m\sim 1/R$, and whose energies are of order
$E^{KK}_p\sim 1/R$. The string excitations, instead, have energies $E_p^s\sim 1/\sqrt{\a'}$.
The energies as seen by an observer at infinity are then respectively
\beq
E^{KK}&\sim&{r\over R^2}\nn\\
E^s&\sim&{r\over R\sqrt{\a'}}\gg E^{KK}\,.
\eeq
So the supergravity (Kaluza Klein) excitations have finite energy in terms of the
coordinate $U=r/R^2$, while the string excitations decouple.
In other words, the $IIB$ string theory on $AdS_5\times S^5$
becomes classical supergravity on that background, because
the  string length is much smaller than the characteristic length of the space, $R$.
\par
The correspondence is between \begin{it} classical supergravity on 
$AdS_5\times S^5$ and  the $\cN=4$
four dimensional SYM theory with gauge group $U(N)$, 
$N\longrightarrow\infty,~X\longrightarrow\infty$.\end{it} 
\end{itemize}
In the following I will consider mainly the last limit (\ref{bigtHooftlimit}),
that is the one which has received most confirmations, and then is the most
firmly founded version of the correspondence.
Notice that in this limit the brane theory is a {\it strongly coupled} theory, being $X$ large.
The $AdS/CFT$ correspondence in the limit (\ref{bigtHooftlimit}), 
then, relates a weakly coupled theory with a strongly coupled one, 
in different dimensions. 
\par
The argument of the decoupling limit sketched above is not the unique 
motivation of the Maldacena conjecture. There are a lot of previous results, observations, 
open problems, that can be better understood in the context of this conjecture.
\begin{itemize}
\item The idea that string theories can describe gauge theories dates back to the origin of string 
theory. In particular, as t'Hooft showed \cite{tHooftstring} that the large $N$ limit of $SU(N)$ gauge theory is formally
similar to perturbative string theory, long efforts have  been done to find an exact gauge field/string
duality. In this context, it has also been suggested \cite{firstcorresp} that four dimensional $SU(N)$ Yang Mills theory
could be dual to a five dimensional string theory.
\item A great advance in non--perturbative string theory has been the discovery \cite{D=pbrane} that a
system of several $D$--branes in string theory can be described as a black $p$--brane solution 
of suypergravity. In particular, this yielded the first microscopic explanation of the Bekenstein Hawking 
entropy: A. Strominger and C. Vafa \cite{VafaStro}  considered $IIB$ string theory compactified on a five 
dimensional compact manifold,  and a system of intersecting $D$--branes wrapped around the 
compact manifold; they worked out the entropy in both the description, in the 'microscopic' one
by counting the $D$--branes states, in the 'macroscopic' one by applying the Bekenstein Hawking
formula, and the two results coincide. 
\par
But in the case of $N$ $D$--branes on the
non--compact space, the results \cite{klebentropy} was similar but different:
\be
S_{\rm Bekenstein~Hawking}={\pi^2\over 2}N^2V_3T^3~~~~~~~
S_{D{\rm -branes}}={2\pi^2\over 3}N^2V_3T^3\,.
\ee
This result is meaningful in the context of $AdS/CFT$ correspondence. In fact, the Bekenstein Hawking
calculation applies in the supergravity limit, that as I said is the strong coupling limit of the gauge theory
on the brane. Conversely, the $D$--brane calculation is perturbative, then gives the entropy
in the weak coupling limit of the gauge theory. The two results, then, differ by the renormalization
group flow of a smooth function, that yields the factor $2\over 3$.
\item It has been derived \cite{crosssec} the absorption cross--section of massless bulk excitations from the 
system of coincident $D$--branes, in two ways. First, with the $D$--branes description,
looking at the process of closed strings that become open strings on the branes. Second,
with the supergravity description (\ref{pbrane10}); as I said, there is a potential barrier
separating the bulk from the near horizon geometry, so waves incident from $r\gg R$
penetrate into the near horizon geometry with a certain cross--section.
\par
These two cross sections coincide:
\be
\sigma={g_s^2{\alpha'}^4\omega^3N^2\over 32\pi}.
\ee
The meaning of this result is clear in the context of $AdS/CFT$ correspondence. In the 
$D$--brane description, a particle incident from the asymptotic infinity is converted
into an excitation of the stack of $D$--branes, namely, into an excitation of the gauge
theory on the world volume. In the supergravity description, a particle incident from the
asymptotic region tunnels into the $r\lll R$ region and produces an excitation of the
near horizon geometry. These two descriptions of the absorption process give the same
cross--sections because the excitations of $AdS_5\times S^5$ supergravity correspond
to the excitations of the $\cN=4$ SYM theory.
\end{itemize}
\par
But the key to the Maldacena conjecture has been a crucial observation on the symmetry
groups. The isometry group of $AdS_5$ space is $SO(4,2)$, but this is also the conformal
group in four dimensions. Furthermore, the isometry of the compact space $S^5$ is 
$SO(6)=SU(4)$, and this is also the $R$--symmetry (namely, the automorphism group of the
superalgebra) of $\cN=4$ SYM theory. More generally, the isometry of the supergravity 
background $AdS_5\times S^5$ is the supergroup $SU(2,2\vert 4)$, that is also the
superconformal symmetry of the $\cN=4$ SYM theory. The fact that these theories have the same 
symmetry is the first hint that they could be equivalent, although they live in different dimensions.
\par
\subsection{The Maldacena Conjecture for $AdS_4\times S^7$}
\par
The Maldacena conjecture can be extended to other cases. 
Let us consider $M$--theory on flat eleven dimensional Minkowski space, with $N$ coincident
$M2$--branes.
In this case, instead of the string length $\sqrt{\alpha'}$ there is the Planck length $l_p$, 
and there is no parameter analogous to the string coupling $g_s$; the only
dimensionless parameter is $N$. Let us consider the low energy limit,
\be
\label{lel11}
El_p\lll 1\,,
\ee
with $N$ fixed.
\par
It is not known which are the perturbative excitations of this theory, because the
quantum $M$--theory has not been found yet, however it is known that the low energy excitations
on the bulk are described by eleven dimensional supergravity, and that it is possible to define superconformal
field theories on the worldvolume of the $M2$--branes.
\par
From the macroscopic point of view, the $M2$--branes
behave as massive and charged objects, sources of a supergravity background of the form
\beq
ds^2&=&f^{-2/3}\ll(-dt^2+dx_1^2+dx_2^2\rr)+f^{1/3}\ll(dr^2+r^2d\Omega_7^2\rr)\nn\\
F_5&=&dtdx_1dx_2df^{-1}\\
f&=&1+{R^6\over r^6},~~~~R^6\equiv 32\pi^2Nl_p^6.
\label{pbrane11}
\eeq
In the near horizon region $r\lll R$, $f\simeq R^6/r^6$, so the metric becomes
\be
\label{nearhor11}
ds^2={r^4\over R^4}\ll(-dt^2+dx_1^2+dx_2^2\rr)+R^2{dr^2\over r^2}+R^2d\Omega_7^2\,,
\ee
that is the metric of
\be
AdS_4\times S^7
\ee
with 
\be
\label{Rlp}
{1\over 2}R={\ll(32\pi^2N\rr)^{1/6}\over 2}l_p
\ee
curvature radius of $AdS_4$ and $R$ curvature radius of $S^7$.
\par
Similarly to the case of $AdS_5\times S^5$, an $M$--theory excitation has
$E_p\sim{1\over l_p}$, and a near horizon excitation, as seen from
infinity, has
\be
E=f^{-1/3}E_p\sim{r^2\over l_p^2}E_p\sim{r^2\over l_p^3}\,.
\ee
The limit (\ref{lel11}) is satisfied if
\be
{r\over l_p}\lll 1\,.
\ee
The coordinate redefinition is
\be
U\equiv {2r^2\over\sqrt{32\pi^2N}l_p^3}={2r^2\over R^3}\,,
\ee
in terms of $U$ the energies of the $M$--theory excitations are finite, and 
the metric (\ref{nearhor11}) becomes
\be
\label{nearhorU11}
ds^2={\ll(32\pi^2N\rr)^{1/3}l_p^2\over 4}
\ll[U^2\ll(-dt^2+dx_1^2+dx_2^2\rr)+{dU^2\over U^2}+4d\Omega_7^2\rr]\,.
\ee
The near horizon region can be rescaled to the entire anti--de Sitter space.
\par
The low energy theory, then, consists on these two 
decoupled systems, the free empty eleven dimensional supergravity and a system that
\begin{itemize}
\item[-] from the macroscopic point of view is $M$--theory on $AdS_4\times S^7$,
\item[-] from the microscopic point
of view is the quantum theory on the $M2$--brane worldvolume. 
\end{itemize}
The Maldacena conjecture states that these two systems are equivalent. 
\par
We have taken $N$ finite. If, instead, we take 
\be
\label{sugralimit11}
N\longrightarrow\infty\,,
\ee
we have the supergravity limit. In fact, 
\be
R\sim N^{1/6}l_p\gg l_p\,,
\ee
and the higher energy $M$-theory excitations decouple from the supergravity
(Kaluza Klein) excitations. Only the latter remain finite in terms of the
coordinate $U$.
\par
In this limit, the correspondence is between
classical eleven dimensional supergravity on $AdS_4\times S^7$ and a
superconformal quantum field theory on the $M2$--brane.
We will mainly consider this limit, that is the most firmly stated.
\par
The theory on the brane has the same symmetry of the bulk theory, and this allows us to single it 
out. The isometry supergroup of the superalgebra is $Osp(8\vert 4)$, that is the supergroup whose
bosonic subgroup is $Sp(4,\IR)\times SO(8)=SO(3,2)\times SO(8)$. It has $\cN=8$ supersymmetry.
But $SO(3,2)$ is also the conformal group in three dimensions, and $Osp(8\vert 4)$ is the
superconformal supergroup of the $\cN=8$ SCFT in three dimensions with gauge group $U(N)$, 
that is the infrared limit of the $\cN=8$ SYM theory.
This is the theory on the brane, dual to the bulk supergravity in the limit $N\longrightarrow\infty$.
\par
Differently from the ten dimensional $AdS_5\times S^5$ case, now
the theory on the brane is conformal only at the infrared fixed point $g_{YM}=0$;
when $g_{YM}\neq 0$ we have a gauge theory not equivalent to any bulk theory; all the 
forms of the correspondence occur when $g_{YM}\longrightarrow 0$, and differ only in
the $N$ range.
\par
All I said about the $M2$--brane is valid, with small differences, also for the $M5$--brane; in this case
the near horizon geometry is $AdS_7\times S^4$. Furthermore, the Maldacena conjecture is valid
also for $D3$--branes, $M2$--branes and $M5$--branes corresponding to less symmetric 
supergravity configurations, giving less supersymmetric theories; I will 
examine these cases afterwards,
in section \ref{seclessN}, and the rest of this thesis concerns them.
\par
\subsection{The realization of the correspondence} 
\par 
After the formulation of the Maldacena conjecture, E. Witten \cite{Wittenads} and, independently, 
S. Gubser, I. Klebanov and A. Polyakov \cite{GKPads} proposed a precise formulation of the 
$AdS/CFT$ correspondence, telling in what sense the bulk and brane theories should be 
identified, and giving a method to calculating correlation functions of the quantum 
theory on the brane by classical supergravity (or string) calculations on the bulk. 
I will follow the line of Witten's reasoning. 
\par 
Let us consider the correspondence between $IIB$ string theory on $AdS_5\times S^5$ and 
$\cN=4$ SCFT on $3+1$ dimensions. First of all, we can note that the conformal theory is 
not defined on $3+1$ Minkowski space $\cM_4$,  
but on its compactified version $\tilde{\cM}_4$, that is $\cM_4$ with some 
''points at infinity'' added: without these points  
Minkowski is not a representation space of the conformal group $SO(4,2)$. 
The compactified Minkowski space $\tilde{\cM}_4$ coincides with the 
boundary of the $AdS_5$ space 
\be 
\tilde{\cM}_4\equiv \partial AdS_5. 
\ee 
On the other hand, we can consider string theory (or supergravity) on $AdS_5\times S^5$ 
from the Kaluza Klein point of view, as a five dimensional supergravity theory on $AdS_5$ 
with compact internal space $S^5$. 
\par 
We can rephrase the Maldacena conjecture as the correspondence between a supergravity 
theory (or string or $M$ theory) on an $AdS$ space  times a compact space 
and a superconformal quantum field theory on the {\bf boundary} of the $AdS$ space itself. 
It becomes a bulk/boundary correspondence. 
\par 
Then the equivalence between a theory on $AdS_5$ and a theory on $\partial  
AdS_5$ can be seen as a realization of the so--called {\it holographic principle} 
\cite{holography}, which states that in a quantum gravity theory all physics within 
some volume can be described in terms of some theory on the boundary with less than one 
degree of freedom per Planck area.  
\par 
On the other hand, the $AdS$ space is very peculiar. It has a time--like boundary at spatial 
infinity; consequently, it is not possible to define the Cauchy problem, that is, to 
determine all the dynamics giving the field values on a Cauchy hypersurface, because the 
fields depend on their boundary values. On the contrary, if we give the boundary values of 
the fields and impose that they are regular on the bulk, there is an unique solution of the 
field equations. In this sense, the $AdS$ space is intrinsically holographic. 
\par 
We can now attempt to make more precise the Maldacena conjecture, relating the 
field theory on the boundary with supergravity (or string theory) on the bulk. The simplest 
recipe, that combines all the ingredients we have, is the following. Let us consider a 
field $\Phi$ on $AdS_5$. Its equation of motion $\Box\Phi=0$, as I said,  
has an unique solution on the bulk with any given boundary values. Let $\Phi_0$ be the  
restriction of $\Phi$ to the boundary $\partial AdS_5$. We will assume that in the  
correspondence between $AdS_5$ and conformal field theory on the boundary, $\Phi_0$ 
couples to a conformal field $\cal O$, singlet under the gauge group, via a coupling  
\be 
\label{PhiO} 
\int_{\tilde{\cM_4}}{\Phi_0{\cal O}}. 
\ee 
In other words, we consider that the boundary values of string theory fields (in 
particular, supergravity fields) act as sources of gauge invariant operators in the 
field theory. From a $D$--brane perspective, we think of closed string states on the 
bulk as sourcing gauge singlet operators on the brane which originate as composite 
operators built form open strings. 
Then $\Phi_0$ is the current source of the 
quantum field $\cal O$ excitations, and we assume the generating functional of 
the correlation functions $\ll<{\cal O}\ll(x_1\rr){\cal O}\ll(x_2\rr)\dots{\cal O} 
\ll(x_n\rr)\rr>$,  
\be 
Z\ll(\Phi_0\rr)\equiv\ll<\exp\ll(\int_{\tilde{M}_4}{\Phi_0{\cal O}} 
\rr)\rr>_{CFT}, 
\ee 
to be 
\be 
\label{conjectureZ} 
Z\ll(\Phi_0\rr)=\exp\ll(-S\ll(\Phi\ll(\Phi_0\rr)\rr)\rr). 
\ee 
Here $S\ll(\Phi\rr)$ is the action of classical supergravity on the bulk in the 
limit $g_sN\longrightarrow\infty,~g_s\longrightarrow 0$, with classical string  
corrections if $g_sN$ finite, $g_s\longrightarrow 0$ and string loop corrections 
if $g_s$, $N$ finite.  
However, I will consider in the following the classical supergravity case. 
\par
\begin{figure}[ht]
\begin{center}
\leavevmode
\hbox{%
\epsfxsize=11cm
\epsfbox{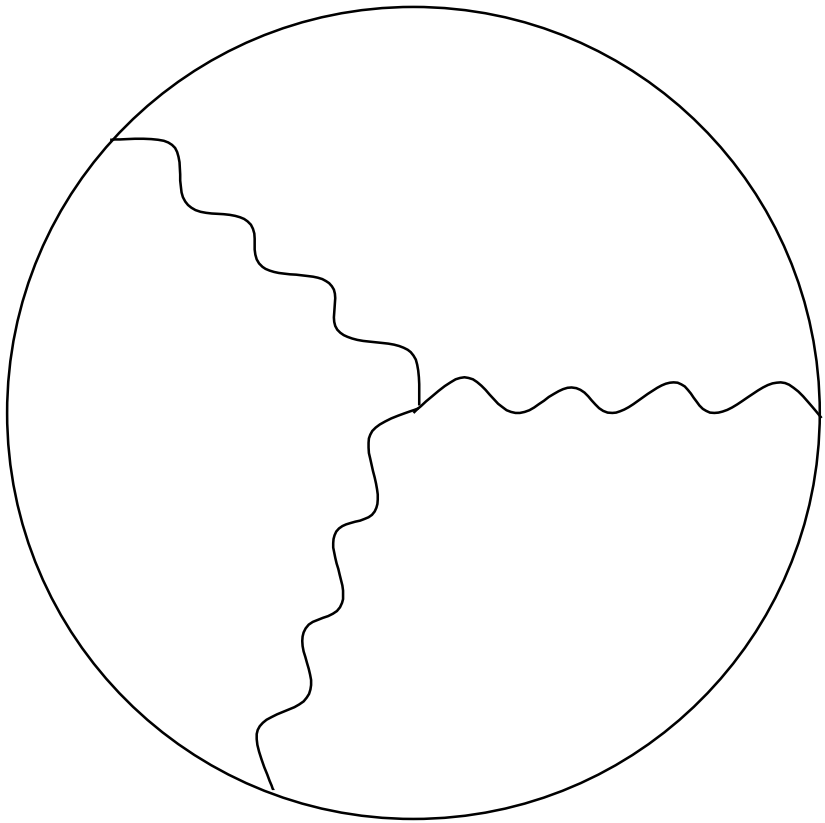}}
\end{center}
\caption{Three point function on the boundary theory via $AdS/CFT$ correspondence}
\label{wittenfigure}
\end{figure}

 
In this picture the interaction between two points of the boundary quantum theory 
is mediated by the bulk. An excitation on the boundary interacts with the bulk,  
propagating via the equation of motion $\Box\Phi=0$, and in the same way it  
propagates from the bulk to another point on the boundary, as in Fig.\ref{wittenfigure}.
%%%%%%%%%%%%%%%%%%%%%%%%%%%%%%%%%%%%%%%%%
%%%%%%%%%%%%%%%%%%%%%%%%%%%%%%%%%%%%%%%%%%%%%%%%%%%%%%%%%%%%%%%%%
To visualize better the system, and to do simpler calculation, it is useful 
to consider euclidean signature; so the 
$AdS_5$ space can be seen as the open unit ball $B_{5}$, 
with metric 
\be 
\label{metricball} 
ds^2={4\sum_{a=0}^4dy_a^2\over\ll(1-\ll|y\rr|^2\rr)^2} 
\ee 
and its boundary as the sphere $S^4$ 
\be 
\sum_{a=0}^4y_a^2=1. 
\ee 
\par 
Let us consider the simplest case, the massless scalar field $\phi$. 
Its boundary value $\phi_0$ is conformally invariant, so, by conformal invariance 
of the action, $\cal O$ has conformal dimension $d-1=4$. The equation of motion 
of $\phi$ is the Laplace equation, which can be solved with the Green function 
method. Doing the calculations in euclidean signature, we find 
\footnote{In these coordinates one regards the euclidean $AdS_5$ not as the open unit 
ball, but as an infinite open half-space, with boundary $S^5$.} 
\be 
\phi\ll(x_0,x_i\rr)=c\int d{\bf x'}{x_0^4\over\ll(x_0^2+\ll|{\bf x}-{\bf x'}\rr|^2 
\rr)^4}\phi_0\ll(x_i'\rr) 
\ee 
(where $c$ is a constant depending on the normalization)  
and substituting this expression in the action one finds
\be 
I\ll(\phi\rr)=2c\int d{\bf x}d{\bf x'}{\phi_0\ll({\bf x}\rr)\phi_0\ll({\bf x'}\rr) 
\over\ll|{\bf x}-{\bf x'}\rr|^8}. 
\ee 
So the two point function of the operator ${\cal O}$ with conformal dimension $4$ is 
proportional to $\ll|{\bf x}-{\bf x'}\rr|^{-8}$, and the other are zero, 
and this is what was expected in conformal field theory. 
\par 
The same can be done for all the fields of supergravity, massless and massive.  
In this case the correlators are well known, and result to coincide with the ones  
derived with this recipe. Furthermore, one finds a relation between the masses of the 
fields $\Phi$ and the conformal weights of the corresponding operators $\cal O$. 
\par 
To understand this, we have to define more precisely the extension of bulk fields to 
the boundary, first of all the metric. 
The metric on the open ball 
$B_5$ (\ref{metricball}) does not extend over $\bar{B}_5$, because it becomes
singular on the boundary. To get a metric which extends over $\bar{B}_5$ we 
have to replace (\ref{metricball}) with a metric of the form 
\be 
d\tilde{s}^2=f^2ds^2 
\ee 
with $f$ having a zero on the boundary, for example $f=1-\ll|y\rr|^2$. 
Then $d\tilde{s}^2$ restricts to a metric on the boundary $S^4$.  
As there is no natural choice of $f$, this metric is only well--defined  
up to conformal transformations: one could replace $f$ by 
\be 
f\longrightarrow e^wf 
\ee 
with $w$ any real function on $\bar{B}_5$, and this would induce the conformal  
transformation 
\be 
d\tilde{s}^2\longrightarrow e^{2w}d\tilde{s}^2 
\ee 
in the metric on $S^5$. Then the metric on $AdS_5$ does not define a metric 
on its boundary, but only a conformal structure (namely, an equivalence class 
of metrics). Notice that the metric on the boundary has conformal weight $-2$, 
the contravariant metric has conformal weight $2$, and  
the corresponding operator $\cal O$ (with covariant indices)  
has by (\ref{PhiO}) conformal weight $4-2=2$. 
\par  
Now let us consider the massive scalar fields. Differently from the massless  
fields, they diverge on the boundary. Their asymptotic behaviour is 
$\phi\sim e^{\lambda_+z}$ where $z$ is a coordinate that goes to infinity  
on the boundary, and $\lambda_+$ is the positive root of the equation 
\footnote{with the normalization of \cite{castdauriafre}, differing from 
the normalization of \cite{Wittenads} and \cite{Maldyrev} by a factor $16$} 
\be 
m^2=16\ll(\lambda+1\rr)\ll(\lambda+3\rr). 
\ee 
Then we have to take solutions of the field equation with asymptotic behaviour 
\be 
\phi\sim f^{-\lambda_+}\phi_0 
\ee 
where $f$ is a function with a zero at the boundary, and $\phi_0$ is a function on 
the boundary. So, like the metric, even $\phi_0$ is not univocally defined, it  
depends on the choice of the function $f$ (that we can assume to be the same 
function defining the metric on the boundary). $\phi_0$, then, is a conformal field, 
which under conformal transformations becomes
\be 
\phi_0\longrightarrow e^{w\lambda_+}\phi_0 
\ee 
and has then conformal weight $-\lambda_+$. Consequently, the corresponding  
operator $\cal O$ of (\ref{PhiO}) has conformal weight $\Delta\equiv 4+\lambda_+$. 
The relation between the mass of the bulk field $\phi$ and the conformal weight 
of the corresponding boundary operator is 
\be 
m^2=16\ll(\Delta-1\rr)\ll(\Delta-3\rr). 
\ee 
But representation theory of $AdS_5$ space tells us that the energy of an $AdS_5$ 
field is related to its mass by 
\be 
m^2=16\ll(E-1\rr)\ll(E-3\rr), 
\ee 
then {\it the energy of the bulk field coincides with the conformal weight of the 
corresponding boundary operator} 
\be 
E=\Delta. 
\ee 
In the next chapter I will give a deeper explanation of this result. It refers to all 
the string theory fields $\Phi$, and to all the theories dual by Maldacena conjecture. 
We can then give the more complete formulation of 
the {\bf $AdS/CFT$ correspondence}: 
\begin{it}
\begin{itemize} 
\item[ ] Every string theory on a background of the form  
\be 
AdS_d\times X_{10-d} 
\ee 
or $M$--theory on a background of the form 
\be 
AdS_d\times X_{11-d} 
\ee 
(were $d$ and the compact space $X_{D-d}$ are 
such that $AdS_d\times X_{D-d}$ is a supergravity solution) is 
equivalent to a superconformal quantum field theory on the boundary on the $AdS_d$ space. 
There is a one-to-one correspondence between the on--shell 
fields on the bulk theory and the off-shell conformal operators 
(which are gauge singlets) on the boundary theory; they have the same 
quantum numbers, and the energy of each bulk field is equal to the conformal weight of 
the corresponding boundary operator. In the limit $g_sN\longrightarrow\infty$,  
$g_s\longrightarrow 0$ for string theory and $N\longrightarrow\infty$
for $M$--theory,  the bulk theory reduces to classical supergravity. 
The generating functional of the boundary theory 
is given by the expression (\ref{conjectureZ}) in terms of the bulk theory. 
\end{itemize} 
\end{it}
\par 
\subsection{Comparison with ''experiment''} 
\label{seclessN}
\par 
The first check of the $AdS/CFT$ correspondence has been done in \cite{Wittenads} and 
\cite{S5correlators}, 
where it has been shown the duality, in the limit 
\be 
g_sN\longrightarrow\infty,~~g_s\longrightarrow 0 
\ee 
for the bulk theory and 
\be 
\label{bigthooft} 
g_{YM}\longrightarrow 0,~~g_{YM}N\longrightarrow\infty 
\ee for the boundary theory, between 
$AdS_5\times S^5$ supergravity and $\cN=4$ $U(N)$ SYM theory on $\tilde{M}_4$. 
\par 
The Kaluza Klein spectrum of $AdS_5\times S^5$ supergravity has been worked out long ago  
\cite{spectrumS5}. There are only the so--called ''short'' multiplets (see the next chapter) 
of five dimensional supergravity, which are protected by supersymmetry against 
quantum and stringy corrections 
\footnote{In fact, their masses (which in our normalization are 
expressed in units of $R$, $m=m_{\rm dimensional}R$, are all $m_{dimensional}\sim 1/R$, then $E$ and 
$\Delta$ do not depend on $R$. On the contrary, the stringy excitations have 
$m\sim\ll(g_{YM}N\rr)^{1/4}$, and decouple in the limit (\ref{bigthooft}).}. 
The conformal fields that correspond to these excitations  
are similarly in ''small'' representations, with dimensions protected against quantum 
corrections.  
\par 
The $\cN=4$ $U(N)$ SYM (that is a superconformal theory) is well known \cite{conformalS5} 
in the weak coupling limit. But we need information about its strong coupling limit 
(\ref{bigthooft}) to compare with the bulk supergravity. Fortunately, some information is  
protected against quantum corrections, and then does not change as the coupling varies.  
First of all, there are operators in ''small'' representations of the superconformal 
group. In the case of $AdS_5\times S^5$, all the operators dual 
to supergravity are protected, and can then be compared. Furthermore, some correlation 
functions are also protected against quantum corrections and do not depend on the coupling; 
they can then be compared with the expression predicted by $AdS/CFT$ correspondence  
(\ref{conjectureZ}). Both these tests have been worked out, the first in \cite{Wittenads}, 
the second in \cite{S5correlators}, and were successful. 
\par 
Other checks has been done in several other cases of $AdS/CFT$ correspondences 
(see the references in \cite{Maldyrev}): 
every time the spectrum at strong coupling is known, it corresponds 
to the Kaluza Klein supergravity spectrum, and every time some 
correlators at strong coupling are known, they coincide with the ones given by the 
(\ref{conjectureZ}). 
\par 
\section{$G/H$ $M$--branes} 
\subsection{More on the $AdS_4\times X_7$ case} 
\par 
The duality between string theory on $AdS_5\times S^5$ and $\cN=4$ SYM theory 
can be generalized, as I said, to dualities between string theory on $AdS_5\times X_5$ 
backgrounds, where $X_5$ is a compact space, and other conformal gauge theories, 
provided $AdS_5\times X_5$ to be a supergravity solution  
\footnote{One assumes that it 
is possible to define string theory around these backgrounds, even if it has not 
been done yet; however, the most part of the calculations are performed in the 
supergravity limit.}. 
In the same way, the duality between $M$--theory on $AdS_4\times S^7$ and the  
infrared limit of the $\cN=8$ SYM theory on $\tilde{\cM}_3$ can be generalized 
to dualities between $M$--theory on  
\be 
AdS_4\times X_7 
\label{ads4M7} 
\ee 
backgrounds, where $X_7$ is a compact space such that (\ref{ads4M7}) is a  
supergravity solution, and other SCFTs.
\par 
In the literature of $AdS/CFT$ correspondence, the most studied case is  
the correspondence between string theory on $AdS_5\times X_5$ and four  
dimensional SCFT. The main reasons are that string theory allows to build 
the gauge theories on $D$--branes (while the theories on $M$--branes are to 
be guessed, or derived relating them to theories on $D$--branes), and that the 
strong coupling of four dimensional SYM theories is an obvious field of interest. 
However, even the case of correspondence between $M$--theory on $AdS_4\times S^7$ 
is interesting, on one hand because three dimensional conformal theories are also 
interesting by themselves, on the other hand because it would be interesting to know  
something about $M$--theory, that today is rather mysterious; this could be the way  
to find the conformal theory intrinsic to $M$--branes. Furthermore, there 
are some results derived in the eighties on supergravity on $AdS_4\times X_7$,  
that can be simply utilized in this new context.  
In the following, throughout all the thesis, I will consider only the  
correspondence between $M$--theory on (\ref{ads4M7}) and three dimensional 
superconformal field theories. 
\par 
The $AdS_4\times S^7$ case has maximal supersymmetry:  
the theories of the correspondence have $32$ supersymmetry charges, corresponding 
to $\cN=8$ four dimensional supergravity on the bulk and $\cN=8$ three dimensional 
SCFT on the boundary. On the contrary, the other $AdS_4\times X_7$ cases are  
{\it less supersymmetric cases}. 
\par 
Let $G$ be the isometry of the $X_7$ space. Being  
\be 
AdS_4\equiv{SO\ll(3,2\rr)\over SO\ll(3,1\rr)}\,, 
\ee
the isometry of $AdS_4\times X_7$ is
\be
SO\ll(3,2\rr)\times G\,.
\ee 
As I will explain in chapter $3$, if $X_7$ admits $\cN$ Killing spinors, namely, 
there are $\cN$ solutions of the equation 
\be 
\label{killingeq} 
{\cal D}_{\a}\eta\ll(y\rr)=c\tau_{\a}\eta\ll(y\rr)
\ee
($c$ is a constant depending on the normalization), then $G$ has the form 
\be
G=SO\ll(\cN\rr)\times G'\,.
\ee
Furthermore there is a supergravity solution with background $AdS_4\times X_7$, 
called Freund Rubin solution 
(I will describe this solution afterwards),  
which is an $\cN$ extended supergravity. Its isometry supergroup is
\be
\label{ospgp} 
Osp\ll(\cN\vert 4\rr)\times G'.
\ee 
I remind that $Osp\ll(\cN\vert 4\rr)$ is the isometry supergroup of $AdS_4$ supergravity 
(see \cite{castdauriafre}). It is the supergroup made up by its bosonic subgroup  
$SO(3,2)\times SO(\cN)$ and by the supercharges $Q$: 
\be 
Osp\ll(\cN\vert 4\rr)=\ll(\begin{array}{c|c} 
SO\ll(3,2\rr) & Q \\ 
\hline \bar{Q} & SO\ll(\cN\rr) \\ 
\end{array}\rr). 
\ee 
Notice that the $SO\ll(\cN\rr)$ part of $G$ has become the
$R-$symmetry of the supergravity: it has been absorbed by the supergroup.
The remaining isometry, $G'$, is an additional internal local
symmetry of the four dimensional theory. 
\par 
The bosonic part of (\ref{ospgp}) is the remnant of the $AdS$ isometry in
eleven dimensions, and is gauged by the fields resulting by the decomposition
of the eleven dimensional massless graviton: the four dimensional
massless graviton, and four dimensional 
massless vectors in the adjoint representation of $G$. The supersymmetries are gauged
by $\cN$ massless gravitinos. 
The fields of the four dimensional supergravity are organized in unitary 
irreducible representations (UIRs) (with spin not bigger than two) 
of $Osp\ll(\cN\vert 4\rr)\times G'$,  
which are the supermultiplets
organized in $G'$ representations.
\par 
On the other side, the corresponding operators on the boundary are 
organized in the same UIRs of the same supergroup 
$Osp\ll(\cN\vert 4\rr)\times G'$,
that has also the interpretation of the superconformal group in three
dimensions times $G'$. The energies of the four dimensional 
fields correspond to the conformal weights of the three dimensional
operators. I remind, however, that the bulk fields are {\it on--shell}, 
the boundary operators are {\it off-shell}; notice that the degrees of  
freedom of an on--shell field 
on $AdS_4$ and the degrees of freedom of an off--shell field on 
$\tilde{\cal M}_3$ coincide. On the other hand, also the $R$--symmetry  
(namely, the automorphism symmetry of the superalgebra) of four 
dimensional anti--de Sitter superspace and of three dimensional 
Poincar\'e superspace coincide, being $SO(\cN)$. Furthermore, 
it is worth noting that the Majorana spinors in three dimensions 
are half the Majorana spinors in four dimensions, so if we look at the  
supergroup $Osp\ll(\cN\vert 4\rr)$ as the bulk superisometry it has  
$\cN$ fermionic generators, but if we look at it as the 
boundary superconformal group it has $2\cN$ fermionic generators: $\cN$ are  
the supersymmetry charges of three dimensional $\cN$ extended SCFT,  
the other $\cN$ are the special conformal supercharges. 
\par
A key point of $AdS/CFT$ correspondence is that the superisometry of 
the two dual theories, in this case (\ref{ospgp}), is a local symmetry of 
the bulk theory, and a global symmetry of the boundary theory; in fact,  
on the bulk there is a supergravity theory, on the boundary a supersymmetric 
theory, whose only local symmetry is the gauge group. Then, a part of 
the superconformal symmetry $Osp\ll(\cN\vert 4\rr)$, the brane theory 
has a local symmetry, the gauge group that we call {\it colour}, and a global 
symmetry, the $G'$ group that we call {\it flavour}, in analogy with  
ordinary QCD. 
\par 
Let us consider the simplest case, the one with maximal supersymmetry,
\be
AdS_4\times S^7.
\ee
The seven sphere preserves $\cN=8$ supersymmetry, and
\be
G=SO\ll(8\rr),
\ee
then the flavour group $G'$ group is not present. The
symmetry group of the $d=4$ supergravity is then
\be
Osp\ll(8\vert 4\rr).
\ee
This theory is dual to a three dimensional $\cN=8$ SCFT, IR fixed
point of an $\cN=8$ SYM theory, defined on the worldvolume of $N$  
$M2-$branes.
The spectrum of this compactification has been determined, and 
the energies have been checked to be consistent with what we know
on conformal weights of the primary conformal operators of
the boundary theory \cite{correspN8}. This is a check of the $AdS/CFT$ correspondence,
but not the strongest possible check of this kind. In fact, the UIRs of  
$Osp\ll(8\vert 4\rr)$ with spin not bigger than two 
are very constrained. As it happens for the case of $AdS_5\times S^5$, 
there are only shortened representations, and 
the energy values of shortened representations  
(as I will explain in the next chapter)
are univocally determined by the $R$--symmetry representations.  
And there is no flavour group.
Then the spectrum of $AdS_4\times S^7$ supergravity can be deduced
by an algebraic study of the UIRs of $Osp\ll(8\vert 4\rr)$, it is
not necessary to consider really the compactification of the
supergravity; and $Osp\ll(8\vert 4\rr)$ is also the symmetry of
the boundary theory. Furthermore, the maximally symmetric supergravity 
is a theory more constrained than less supersymmetric cases.
\par
Much more intriguing should be to check the $AdS/CFT$ correspondence
in lower supersymmetry cases, when the spectrum of the compactified
supergravity is given not only by the $Osp\ll(\cN\vert 4\rr)$ algebra, but
also by the geometry of the compactification.  
There are two kinds of known $AdS_4/CFT_3$ correspondences with $X_7\neq S^7$: 
\begin{itemize} 
\item The orbifolds $S^7/\Gamma$, where $\Gamma$ is 
a discrete subgroup of $SO(8)$ \cite{corresporbifoldN8}. 
Such manifolds have the local geometry of $S^7$, and the corresponding supergravities  
are truncations of $\cN=8$ supergravity.  
\item Compact coset spaces  
\be 
\label{G/H7} 
X_7=\ll(G\over H\rr)_7 
\ee 
that are also Einstein spaces. They are not locally equivalent to $S^7$, and 
the corresponding supergravities are not truncations of $\cN=8$ supergravity. 
\end{itemize} 
The latter case is the more interesting, because it is not related with 
the $S^7$ case. It is the case of the 
so called {\bf $G\over H$ $M-$branes}, and is the one I have been studying.
\par
\subsection{Supergravity on $AdS_4\times\ll(G\over H\rr)_7$} 
\par 
If we put $N$ $M$--branes on flat eleven dimensional Minkowski space, with $N$ big, 
we get a system that, from the macroscopical point of view, and in the supergravity 
limit, looks like a $p$--brane solution of eleven dimensional supergravity  
(\ref{pbrane11}), whose near horizon limit is $AdS_4\times\ll(G/H\rr)_7$,  
and that asymptotically tends to flat space.  
It has been shown \cite{g/hpape},\cite{noi3} that for every  
supergravity solution of the form $AdS_4\times\ll(G/H\rr)_7$,  
there is a brane solution of supergravity with the same symmetries of the former  
solution, whose near horizon limit is $AdS_4\times\ll(G/H\rr)_7$,  
and whose asymptotic limit is a Ricci flat - but not flat - space,  
$\cC({G/H})$, namely the {\it cone} on $G/H$
\be 
\label{cone} 
ds_{\cC\ll({G\over H}\rr)}^2=dr^2+r^2ds_{{G\over H}}^2
\ee 
times the three dimensional Minkowski space.
Notice that when ${G/H}={SO\ll(8\rr)/SO\ll(7\rr)}=S^7$  
the cone is the flat euclidean space, and $r=0$ is a coordinate  
singularity, but in the other cases the singularity $r=0$ is physical. 
\par 
Then, if we put $N$ $M$--branes not on ${\cal M}_{11}$ but on ${\cal M}_3 
\times\cC\ll(G/H\rr)$, we get on the branes a SCFT equivalent, by $AdS/CFT$  
correspondence, to the supergravity solution given by the near horizon 
geometry blown up to all the space. This supergravity solution has been  
found in the eighties, it is called {\it Freund Rubin solution} \cite{freundrubin}: 
\be 
\label{FreundRubin} 
\begin{array}{ccc} 
\begin{array}{ccc} 
g_{\mu\nu}\ll(x,y\rr)&=&g^0_{\mu\nu}\ll(x\rr)\\ 
g_{\alpha\beta}\ll(x,y\rr)&=&g^0_{\alpha\beta}\ll(y\rr)\\ 
g_{\mu\alpha}&=&0\\ 
\end{array}&& 
\begin{array}{ccc} 
F_{\mu\nu\rho\sigma}&=&e\sqrt{g^0}\varepsilon_{\mu\nu\rho\sigma}\\ 
{\rm other~}F&=&0\\ 
\psi_{\mu}=\psi_{\alpha}&=&0\\ 
\end{array}\\ 
\end{array} 
\ee 
where $x^{\mu}$, $\mu=0,\dots,3$ are the coordinates of $AdS_4$ space, 
$y^{\alpha}$, $\alpha=1,\dots,7$ are the coordinates of the internal $G/H$ 
space, $g^0_{\mu\nu}$ is the $AdS_4$ metric, $g^0_{\alpha\beta}$ is the 
$G/H$ metric. 
\par
The parameter $e$ here introduced is related to the anti--de Sitter Radius by
\be
R_{AdS}={1\over 4e}.
\ee
As I said, when one performs explicit calculations, usually \cite{Maldyrev},
\cite{frenico}, \cite{gunawar}, \cite{popelast}, \cite{dewitads} measures 
dimensionful physical quantities in terms of the scale length which, in our case,
is the anti--de Sitter radius; that is, setting  $R_{AdS}=1$. However, here I
follow the conventions of \cite{castdauriafre}, \cite{multanna}, where 
$e=1$ and then $R_{AdS}=1/4$. This is the reason for the discrepancy by a factor $4$ in
the mass normalizations of these papers. Notice that this does not mean that the parameter $e$ is dimensionful;
as I will explain in section \ref{branenormalizations}, we get rid of dimensionful quantities by putting to one
\be
\kappa^2=8\pi G_{11}\sim l_p^9;
\ee
reinstalling $\kappa$, the relation between $e$ and anti--de Sitter radius is
\footnote{These formulas and conventions was derived in the context of eleven dimensional
supergravity and $M$--theory, before the $AdS/CFT$ correspondence was proposed.
$R_{AdS}\over l_p$ is a free parameter in the context of eleven dimensional supergravity.
However, in the context of $AdS/CFT$ correspondence, such a quantity is
related to $N$. Then I don't give an explicit expression of $\kappa$:
different conventions ($e=1$, $e=1/4$) correspond to different values of
$\kappa^2/l_p^9$.}
\be
R_{AdS}={\kappa^{2/9}\over 4e}.
\ee
\par 
For every seven dimensional compact coset space that is also an Einstein 
space, the (\ref{FreundRubin}) is a classical solution of eleven dimensional 
supergravity, and then it is possible a Kaluza Klein dimensional reduction to 
four dimensional supergravity. If $G/H$ admits $\cN$ Killing spinors, the four  
dimensional theory is an $\cN$--extended supergravity (see chapter $3$). The coset manifolds giving 
a supersymmetric Freund Rubin compactification have been completely classified 
in the eighties \cite{castromwar}, \cite{castdauriafre}: 
\be 
\begin{array}{|c|c|c|} 
\hline 
{\rm space} & \cN & G' \\ 
\hline 
S^7={SO\ll(8\rr)\over SO\ll(7\rr)} & 8 & \\ 
\hline 
S^7_{\rm squashed}={SO\ll(5\rr)\times SO\ll(3\rr)\times SO\ll(2\rr)\over  
SO\ll(3\rr)\times SO\ll(3\rr)\times SO\ll(2\rr)} 
 & 1 & SO\ll(5\rr)\times SO\ll(3\rr)\\ 
\hline 
N^{0p0}={SU\ll(3\rr)\times SU\ll(2\rr)\over SU\ll(2\rr)\times U\ll(1\rr)} 
& 3 & SU\ll(3\rr) \\ 
\hline 
M^{ppr}={SU\ll(3\rr)\times SU\ll(2\rr)\times U\ll(1\rr)\over SU\ll(2\rr) 
\times U\ll(1\rr)\times U\ll(1\rr)} & 2 & SU\ll(3\rr)\times SU\ll(2\rr) \\ 
\hline 
Q^{ppp}={SU\ll(2\rr)\times SU\ll(2\rr)\times SU\ll(2\rr)\times U\ll(1\rr)\over  
U\ll(1\rr)\times U\ll(1\rr)\times U\ll(1\rr)} & 2 & SU\ll(2\rr)\times  
SU\ll(2\rr)\times SU\ll(2\rr) \\ 
\hline 
V_{5,2}={SO\ll(5\rr)\times U\ll(1\rr)\over SO\ll(3\rr)\times U\ll(1\rr)} & 2 & SO\ll(5\rr) \\ 
\hline 
\end{array} 
\ee 
Most of these spaces are described in chapter 3, where the mass spectra of 
supergravity on $AdS_4\times \ll(G/H\rr)_7$ in the cases $\ll(G/H\rr)_7=M^{111}$ ($\cN=2$),  
$\ll(G/H\rr)_7=N^{010}$ ($\cN=3$)
\footnote{and partially of $\ll(G/H\rr)_7=Q^{111}$ ($\cN=2$)}
are given and, for $M^{111}$, explicitly worked out. The case
$\ll(G/H\rr)_7=V_{5,2}$ has been recently studied in \cite{stiefel}.
\par
\chapter{Representation theory of $Osp\left({\cal N}\vert 4\right)$}
\par
A field on four dimensional anti--de Sitter space
\be
AdS_4\equiv{SO\ll(3,2\rr)\over SO\ll(3,1\rr)}
\ee
is an unitary irreducible representation (UIR) of the isometry group \hbox{$SO\ll(3,2\rr)$.}  
Notice that such representations cannot be finite dimensional, being  $SO\ll(3,2\rr)$
non compact, and then have to be fields.
\par
Supergravity on $AdS_4$ is defined on the $\cN$ extended anti--de Sitter superspace, which,
in the coset space formulation, is
\be
AdS_{4\vert\cN}\equiv{Osp\ll(\cN\vert 4\rr)\over SO\ll(3,1\rr)\times SO\ll(\cN\rr)}.
\ee
It has $4$ bosonic coordinates labelling the points on $AdS_4$ and $4\cN$ fermionic coordinates
transforming as Majorana spinors under $SO\ll(1,3\rr)$ and as vectors under $SO\ll(\cN\rr)$.
Its isometry supergroup is $Osp\ll(\cN\vert 4\rr)$, so the superfields are UIRs of such a
supergroup. In other words, an UIR of $Osp\ll(\cN\vert 4\rr)$ is a {\it supermultiplet} of
$AdS_4$ fields.
\par 
The $Osp\ll(\cN\vert 4\rr)$ supergroup is described in the next section. Here I
stress that its bosonic subalgebra is 
\be
SO\ll(3,2\rr)\times SO\ll(\cN\rr),
\ee
namely, the anti--de Sitter isometry times the so--called {\it $R$--symmetry}. The $R$--symmetry 
is the external automorphism algebra of the supersymmetry charges. In anti--de Sitter supersymmetry 
it belongs to the irreducible part
of the supersymmetry algebra itself, while in Poincar\'e supersymmetry it does not.
\par
In general, the $R$--symmetry depends on the kind of supersymmetry (Poincar\'e or anti--de Sitter)
and on the dimensionality of the theory:
\be
\begin{array}{c}
\begin{array}{|c|c|c|c|}
\hline
                & d=3   & d=4   & d=5   \\
\hline
        \hbox{Poincar\'e}       & SO\ll(\cN\rr) & SU\ll(\cN\rr)   & Usp\ll(\cN\rr) \\
\hline
                 AdS    & SO\ll(\cN_L\rr)\times SO\ll(\cN_R\rr) & SO\ll(\cN\rr) 
& SU\ll(\cN/2\rr) \\ 
\hline
\end{array}
\\ \\
R\hbox{-symmetry}\\
\\
\end{array}
\ee
\par
The same supergroup $Osp\ll(\cN\vert 4\rr)$ has also another interpretation: it is the conformal 
supergroup of a three dimensional Poincar\'e theory with $\cN$ extended supersymmetry.
In this context, $SO\ll(3,2\rr)$ is the conformal
group in three dimensions. The fourth coordinate translation is interpreted as
conformal scaling, and the Lorentz rotations involving this coordinate are interpreted
as conformal boosts. Half the fermionic generators are the supersymmetry charges,
the other half are the special conformal supercharges. The $R$--symmetry 
of three dimensional Poincar\'e theories, like those of four dimensional $AdS$ theories,
is $SO\ll(\cN\rr)$. Then, the UIRs of $Osp\ll(\cN\vert 4\rr)$ can also be organized as supermultiplets 
of three dimensional conformal fields, namely, as three dimensional conformal superfields.
\par
In order to make the comparison between compactified supergravity on the bulk and
superconformal field theory on the boundary explicit, we need a general vocabulary
between these two descriptions of $Osp\ll(\cN\vert 4\rr)$. 
We need to know, given a state on the bulk, which should be the 
corresponding conformal operator on the boundary, in order to
check whether it is actually present. This is the main aim of the present chapter. 
\par
In section $1$ the $osp\ll(\cN\vert 4\rr)$ superalgebra is defined with
its basic properties and the conventions are established. Furthermore, its
compact and non compact five--grading structures, fundamental in 
understanding the algebraic basis of $AdS_4/CFT_3$ correspondence, 
are described. In section $2$ the theory of $SO\ll(3,2\rr)$ UIRs is
briefly sketched. In section $3$ we extend our analysis to the UIRs of 
$Osp\ll(\cN\vert 4\rr)$, stressing both the interpretations of $Osp\ll(\cN\vert 4\rr)$ 
as isometry of four dimensional anti--de Sitter superspace and as the superconformal
group in three dimensions. In section $4$ the method
of explicit construction of $Osp\ll(\cN\vert 4\rr)$ UIRs as supergravity supermultiplets 
is given, and these latter are explicitly retrieved in the cases of 
$\cN=1,~\cN=2,$ and $\cN=3$ supersymmetry. In section $5$ 
the superspace on the bulk and on the boundary of $AdS_4$ is constructed, and, in
the case $\cN=2$, the short superfields are found to correspond with the $Osp\ll(2\vert 4\rr)$
UIRs derived in the precedent section. Part of the content of the present chapter refers to
results obtained within the collaborations \cite{noi1},\cite{noi2}.
\par
\section{The $osp({\cal N}\vert 4)$ superalgebra: definition, properties and
notations}
\par
The non compact superalgebra $osp({\cal N} \vert 4)$ relevant to the
$AdS_4/CFT_3$ correspondence is a real section of the complex orthosymplectic
superalgebra $osp({\cal N} \vert 4, \IC)$ that admits the Lie algebra
\begin{equation}
g_{even} = sp(4,\IR) \oplus so({\cal N}, \IR)
\label{geven}
\end{equation}
as even subalgebra. Alternatively, due to the isomorphism
$sp(4,\IR)\equiv usp(2,2)$ we can take a different real section of
$osp({\cal N} \vert 4, \IC)$ such that the even subalgebra is:
\begin{equation}
  g_{even} = usp(2,2) \oplus so({\cal N}, \IR)\,.
\label{gevenp}
\end{equation}
Here we rely on the second formulation (\ref{gevenp})
which is more convenient to discuss unitary irreducible
representations. The two formulations are related by a unitary
transformation that, in spinor language, corresponds to a different
choice of the gamma matrix representation. Formulation (\ref{geven}) is
obtained in a Majorana representation where all the gamma matrices are real
(or purely imaginary), while formulation (\ref{gevenp}) is related to
a Dirac representation.
\par
Our choice for the gamma matrices in a Dirac representation is the
following one\footnote{we adopt as explicit representation of the $SO(3)~\tau$
matrices a permutation of the canonical Pauli matrices $\s^a$:
$\tau^1=\s^3$, $\tau^2=\s^1$ and $\tau^3=\s^2$.}:
\begin{equation}
\Gamma^0=\left(\begin{array}{cc}
\unity&0\\
0&-\unity
\end{array}\right)\,,\qquad
\Gamma^{1,2,3}=\left(\begin{array}{cc}
0&\tau^{1,2,3}\\
-\tau^{1,2,3}&0
\end{array}\right)\,,
\qquad C_{[4]}=i\Gamma^0\Gamma^3\,,
\label{dirgamma}
\end{equation}
having denoted by $C_{[4]}$ the charge conjugation matrix in $4$--dimensions
$C_{[4]}\, \Gamma^\mu
\, C_{[4]}^{-1} = - ( \Gamma^\mu)^T$.
\par
Then the $Osp({\cal N}\vert 4)$ superalgebra is defined as the set of
graded $(4+{\cal N})\times(4+{\cal N})$ matrices $\mu$ that satisfy the
following two conditions:
\begin{equation}
\begin{array}{rccclcc}
\mu^T & \left( \matrix {C_{[4]} & 0 \cr 0 & \bfone_{{\cal N}\times {\cal N}}
\cr}
\right)& +&\left( \matrix {C_{[4]} & 0 \cr 0 & \bfone_{{\cal N}\times {\cal N}}
\cr}
\right)& \mu & = & 0  \\
\null&\null&\null&\null&\null&\null&\null\\
\mu^\dagger & \left( \matrix {\Gamma^0 & 0 \cr 0 & -\bfone_{{\cal N}\times {\cal
N}} \cr}
\right)& +&\left( \matrix {\Gamma^0 & 0 \cr 0 & -\bfone_{{\cal N}\times {\cal
N}} \cr}
\right)& \mu & = & 0 \\
\end{array}
\label{duecondo}
\end{equation}
the first condition defining the complex orthosymplectic algebra, the
second condition defining the real section with even subalgebra as in
eq.(\ref{gevenp}). Eq.s (\ref{duecondo}) are solved by setting:
\begin{equation}
\mu   =  \left( \matrix {\varepsilon^{AB} \, \frac{1}{4} \,
\left[\IGam_A \, , \, \IGam_B \right ] & \epsilon^i \cr
{\bar \epsilon}^i & \mbox{i}\, \varepsilon_{ij}\, t^{ij} \cr }\right)\,.   \nonumber\\
\label{mumatri}
\end{equation}
In eq.(\ref{mumatri}) $\varepsilon_{ij}=-\varepsilon_{ji}$
is an arbitrary real antisymmetric ${\cal N} \times {\cal N}$ tensor,
 $t^{ij} = -t^{ji}$ is the antisymmetric ${\cal N} \times {\cal N}$
 matrix:
\begin{equation}
 ( t^{ij})_{\ell m} = \mbox{i}\left( \delta ^i_\ell \delta ^j_m - \delta
 ^i_m \delta ^j_\ell \right)
\label{tgene}
\end{equation}
namely a standard generator of the $SO({\cal N})$ Lie algebra,
\begin{equation}
  \IGam_A=\cases{\begin{array}{cl}
  \mbox{i} \, \Gamma_5 \Gamma_\mu & A=\mu=0,1,2,3 \\
\Gamma_5\equiv\mbox{i}\Gamma^0\Gamma^1\Gamma^2\Gamma^3 & A=4 \\
\end{array} }
\label{bigamma}
\end{equation}
denotes a realization of the $SO(2,3)$ Clifford algebra:
\begin{eqnarray}
  \left \{ \IGam_A \, ,\, \IGam_B \right\} &=& 2
  \eta_{AB}\nonumber\\
\eta_{AB}&=&{\rm diag}(+,-,-,-,+)
\label{so23cli}
\end{eqnarray}
and
\begin{equation}
  \epsilon^i = C_{[4]} \left( {\bar \epsilon}^i\right)^T
  \quad (i=1,\dots\,{\cal N})
\label{qgene}
\end{equation}
are ${\cal N}$ anticommuting Majorana spinors.
\par
The index conventions we have so far introduced can be
summarized as follows. Capital indices $A,B=0,1,\ldots,4$ denote
$SO(2,3)$ vectors. The   latin indices of type
$i,j,k=1,\ldots,{\cal N}$ are $SO({\cal N})$ vector indices.
The indices $a,b,c,\ldots=1,2,3$ are used to denote spatial directions
of $AdS_4$: $\eta_{ab}={\rm diag}(-,-,-)$, while the indices of type
$m,n,p,\ldots=0,1,2$ are space-time indices for the Minkowskian
boundary $\partial \left( AdS_4\right) $: $\eta_{mn}={\rm diag}(+,-,-)$.
\par
To write the $osp({\cal N} \vert 4)$ algebra in abstract form it
suffices to read the graded matrix (\ref{mumatri}) as a linear
combination of generators:
\begin{equation}
  \mu \equiv -\mbox{i}\varepsilon^{AB}\, M_{AB}
  +\mbox{i}\varepsilon_{ij}\,T^{ij}
  +{\bar \epsilon}_i \, Q^i
\label{idegene}
\end{equation}
where $Q^i = C_{[4]} \left(\overline Q^i\right)^T$ are also Majorana spinor
operators.
Then the superalgebra reads as follows:
\begin{eqnarray}
\left[ M_{AB} \, , \, M_{CD} \right]  & = & \mbox{i} \, \left(\eta_{AD} M_{BC}
+ \eta_{BC} M_{AD} -\eta_{AC} M_{BD} -\eta_{BD} M_{AC} \right)  \nonumber\\
\left [T^{ij} \, , \, T^{kl}\right] &=&
-\mbox{i}\,(\delta^{jk}\,T^{il}-\delta^{ik}\,T^{jl}-
\delta^{jl}\,T^{ik}+\delta^{il}\,T^{jk})\,  \nonumber \\
\left[M_{AB} \, , \, Q^i \right] & = & -\mbox{i} \frac{1}{4} \,
\left[\IGam_A \, , \, \IGam_B \right ] \, Q^i \nonumber\\
\left[T^{ij}\, , \,  Q^k\right] &=&
-\mbox{i}\, (\delta^{jk}\, Q^i - \delta^{ik}\,   Q^j ) \nonumber\\
\left \{Q^{\alpha i}, \overline Q_{\b}^{\,j}\right \} & = & \mbox{i}
\delta^{ij}\frac{1}{4}\,\left[\IGam^A \, , \, \IGam^B \right ]{}^{\a}_{\
\b}M_{AB}
+\mbox{i}\delta^{\a}_{\,\b}\,T^{ij}\,.
\label{algbr}
\end{eqnarray}
The form (\ref{algbr}) of the $osp({\cal N}\vert 4)$ superalgebra
coincides with that given in papers \cite{frenico}, \cite{multanna}.
\par
In the gamma matrix basis (\ref{dirgamma})
the Majorana supersymmetry charges have the following form:
\begin{eqnarray}
Q^i =
\left(\matrix{a_\a^i\cr\varepsilon_{\alpha\beta}\bar a^{\b i}}\right)\,,
\qquad \bar a^{\alpha i} \equiv \left( a_\alpha^i \right)^\dagger \,,
\end{eqnarray}
where $a_{\a}^i$ are two-component $SL(2,\IC)$ spinors: $\a,\b,\ldots = 1,2$.
We do not use dotted and undotted indices to
denote conjugate $SL(2,\IC)$ representations; we rather use different symbols $a,\,
\bar{a}$.
Raising and lowering  is performed by means of the
$\varepsilon$-symbol:
\begin{equation}
\psi_\a = \varepsilon_{\a\b} \psi^\b \,, \qquad
\psi^\a = \varepsilon^{\a\b} \psi_\b\,,
\end{equation}
where $\varepsilon_{12}=\varepsilon^{21}=1$, so that
$\varepsilon_{\a\g}\varepsilon^{\g\b}=\delta_\a^\b$.
Unwritten indices are contracted with the low index at the left
of the high index.
\par
\subsection{Compact and non compact five gradings of the $osp({\cal N}|4)$
superalgebra}
\par
As it is extensively explained in \cite{gunaydinminiczagerman}, a non-compact
group $G$  admits unitary irreducible representations of the lowest
weight type if it has a $G^0$ with respect to whose Lie algebra $g^0$ there
exists a {\it three grading} of the Lie algebra $g$ of $G$.
In the case of a non--compact superalgebra the lowest weight
UIRs can be constructed if the three grading is   generalized to
a {\it five grading} where the even (odd) elements are integer
(half-integer) graded:
\begin{eqnarray}
g = g^{-1} \oplus g^{-\ft12} \oplus g^0 \oplus g^{+\ft12} \oplus g^{+1}\,,\\
\nonumber\\
\left[g^k,g^l\right]\subset g^{k+l}\qquad g^{k+l}=0\ {\rm for}\ |k+l|>1\,.
\end{eqnarray}
For the supergroup
$Osp({\cal N}|4)$ this grading can be made in two ways, choosing
as grade zero subalgebra either the maximal compact subalgebra
\begin{eqnarray}
g^0 \equiv so(3) \oplus so(2) \oplus so({\cal N}) \subset osp({\cal N} \vert 4)
\label{so3so2}
\end{eqnarray}
or the non-compact subalgebra
\begin{eqnarray}
{\tilde g}^0 \equiv
so(1,2) \oplus so(1,1) \oplus so({\cal N}) \subset osp({\cal N} \vert 4)
\label{so12so11}
\end{eqnarray}
which also exists, has the same complex extension and is also
maximal.
\par
The existence of the double five--grading is the algebraic core of
the $AdS_4/CFT_3$ correspondence. Decomposing a UIR  of
$Osp({\cal N} \vert 4)$ into representations of $g^0$ exibits its
interpretation as a supermultiplet of {\it particles states} on the bulk of
$AdS_4$, while decomposing it into representations of ${\tilde g}^0$
makes explicit its interpretation as a supermultiplet of {\it conformal
primary fields} on the boundary $\partial (AdS_4)$.
\par
In both cases the grading is determined by the generator $X$ of the abelian
factor $SO(2)$ or $SO(1,1)$:
\begin{equation}
[X,g^k]=k\,g^k\,.
\end{equation}
In the compact case (see \cite{frenico}) the $SO(2)$ generator
$X$ is given by $M_{04}$.
It is interpreted as the energy generator of the four-dimensional $AdS$
theory.
It was used in \cite{multanna} and \cite{noi1} for the construction
of the $Osp(2 \vert 4)$ representations, yielding the long
multiplets of \cite{multanna} and the short and ultra-short
multiplets of \cite{noi1}.
I repeat such decompositions here.
\par
\begin{figure}[ht]
\begin{center}
\leavevmode
\hbox{%
\epsfxsize=11cm
\epsfbox{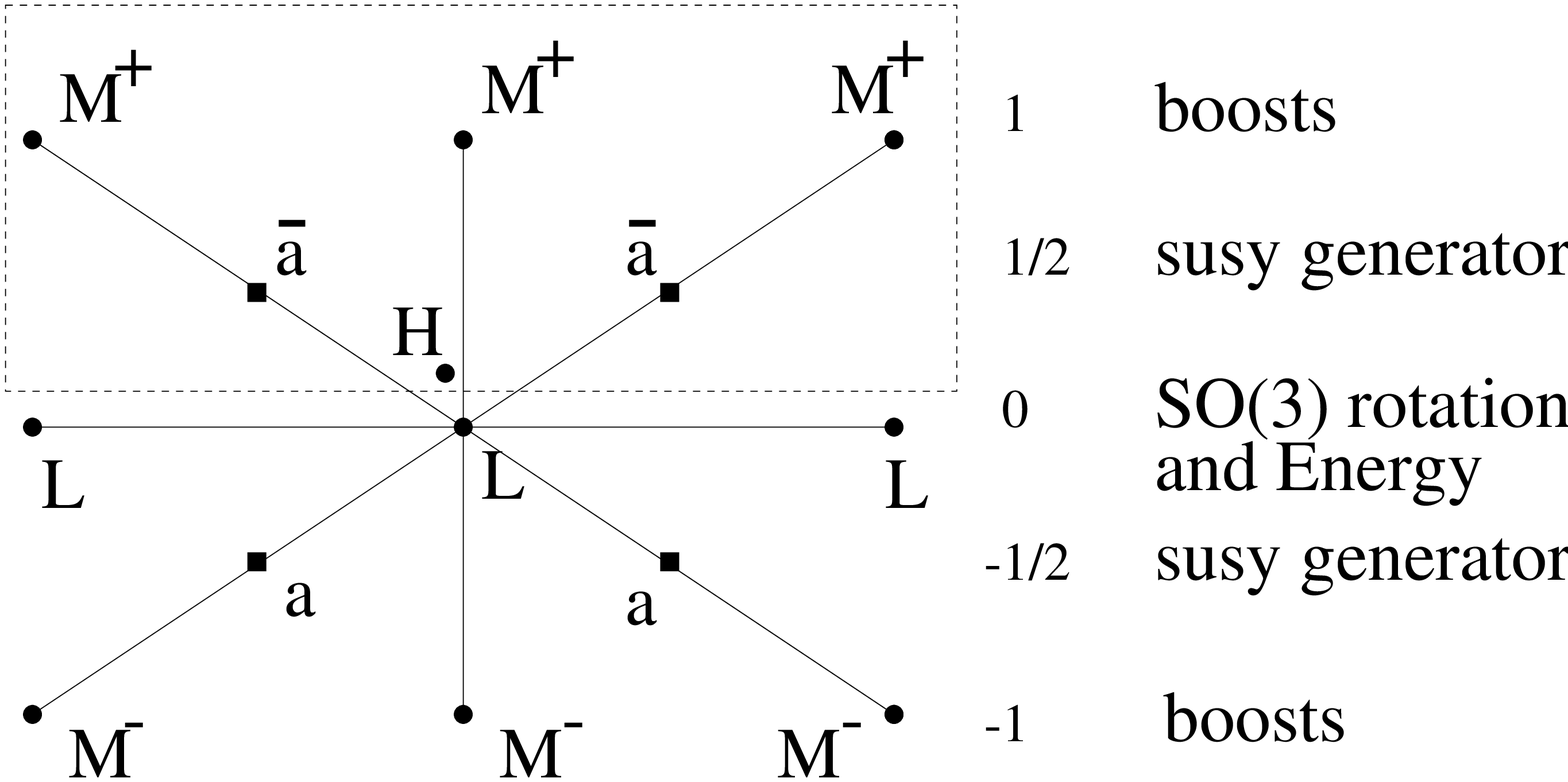}}
\caption{{\small Schematic representation of the root diagram
of $Osp({\cal N}|4)$ in the $SO(2) \times SO(3)$ basis.
\label{pistac}
The grading with respect to the energy $H$ is given on the right. }}
\end{center}
\end{figure}

We call $H$ the energy generator of $SO(2)$, $L_a$ the rotations
of $SO(3)$:
\begin{eqnarray}
H &=& M_{04} \,, \nonumber \\
L_a &=& \ft12 \varepsilon_{abc} \, M_{bc} \,,
\end{eqnarray}
and $M_a^\pm$ the boosts:
\begin{eqnarray}
M_a^+ &=& - M_{a4} + i M_{0a} \,, \nonumber \\
M_a^- &=&  M_{a4} + i M_{0a} \,.
\end{eqnarray}
The supersymmetry generators are $a^i_\alpha$ and $\bar a^{\alpha i}$.
Rewriting the $osp({\cal N} \vert 4)$ superalgebra (\ref{algbr})
in this basis we obtain:
\begin{eqnarray}
{}[H, M_a^+] &=& M_a^+ \,,\nonumber \\
{}[H, M_a^-] &=& -M_a^- \,,\nonumber \\
{}[L_a, L_b] &=& i \, \varepsilon_{abc} L_c
\,, \nonumber \\
{}[M^+_a, M^-_b] &=& 2 \, \delta_{ab}\, H + 2 i \,
\varepsilon_{abc} \, L_c \,,
\nonumber \\
{}[L_a, M^+_b ] &=& i \, \varepsilon_{abc} \, M^+_c \,,
\nonumber \\
{}[L_a, M^-_b ] &=& i \, \varepsilon_{abc} \, M^-_c \,,
\nonumber \\
{}[T^{ij}, T^{kl}] &=&
-i\,(\delta^{jk}\,T^{il}-\delta^{ik}\,T^{jl}-
\delta^{jl}\,T^{ik}+\delta^{il}\,T^{jk})
\,, \nonumber \\
{}[T^{ij}, \bar a^{\alpha k}] &=&
-i\, (\delta^{jk}\, \bar a^{\alpha i} - \delta^{ik}\, \bar a^{\alpha j} )
\,, \nonumber \\
{}[T^{ij},  a_\alpha^k] &=&
-i\, (\delta^{jk}\, a_\alpha^i - \delta^{ik}\,   a_\alpha^i )
\,,\nonumber \\
{}[H, a_\alpha^i] &=& -\ft12 \, a_\alpha^i \,, \nonumber \\
{}[H, \bar a^{\alpha i}] &=& \ft12 \, \bar a^{\alpha i}
\,, \nonumber \\
{}[M_a^+, a_\alpha^i] &=& (\tau_a)_{\alpha\beta}\, \bar a^{\beta i}
\,, \nonumber \\
{}[M_a^-, \bar a^{\alpha i}] &=&
- (\tau_a)^{\alpha\beta} \,  a_\beta^i \,,\nonumber \\
{}[L_a, a_\alpha^i]&=& \ft12 \, (\tau_a)_{\alpha}{}^\beta \,  a^i_\beta
\,, \nonumber \\
{}[L_a, \bar a^{\alpha i}] &=& -\ft12 \, (\tau_a)^\alpha{}_\beta
\,  \bar a^{\beta i} \,, \nonumber \\
\{a_\alpha^i, a_\beta^j \} &=& \delta^{ij} \, (\tau^k)_{\alpha\beta}\,
M_k^- \,, \nonumber \\
\{\bar a^{\alpha i}, \bar a^{\beta j} \} &=&
\delta^{ij} (\tau^k)^{\alpha\beta} \, M_k^+ \,,
\nonumber \\
\{ a_\alpha^i, \bar a^{\beta j} \}
&=& \delta^{ij}\, \delta_\alpha{}^\beta \, H
+ \delta^{ij} \, (\tau^k)_\alpha{}^\beta \, L_k
+i \, \delta_\alpha{}^\beta \, T^{ij} \,.
\label{ospH}
\end{eqnarray}
The five--grading structure of the algebra (\ref{ospH}) is shown in
fig. \ref{pistac} .
\par
\begin{figure}[ht]
\begin{center}
\leavevmode
\hbox{%
\epsfxsize=11cm
\epsfbox{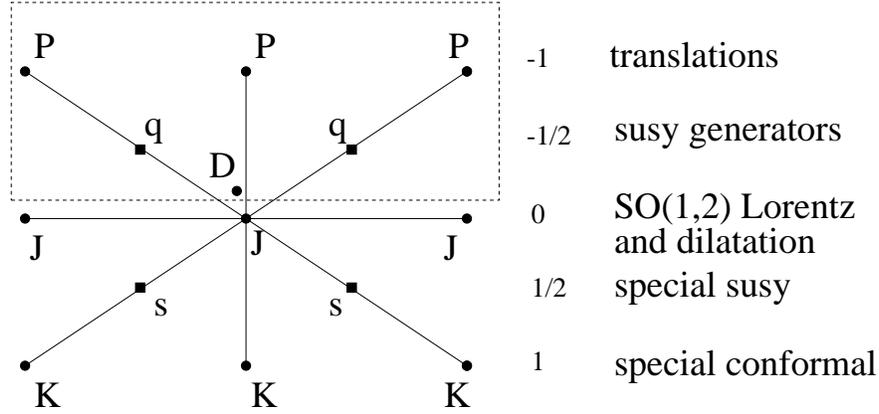}}
\caption{{\small Schematic  representation of the root diagram
of $Osp({\cal N}|4)$ in the $SO(1,1)\times SO(1,2)$ basis.
The grading with respect to the dilatation $D$ is given on the right.\label{pirillo}}}
\end{center}
\end{figure}

In the superconformal field theory context we are interested in the
action of the $Osp({\cal N} \vert 4)$
generators on superfields living on the minkowskian boundary  $\partial(AdS_4)$.
To be precise the boundary   is a compactification of $d=3$
Minkowski space and admits a conformal family of metrics
$ g_{mn} = \phi(z) \eta_{mn}$ conformally
equivalent to the the flat Minkowski metric
\begin{equation}
\eta_{mn} = (+,-,-) \,, \qquad m,n,p,q = 0,1,2 \,.
\label{minkio3}
\end{equation}
Precisely because we are interested in conformal field theories the
choice of representative metric inside the conformal
family is immaterial and the flat one (\ref{minkio3}) is certainly the
most convenient.
The requested action  of the superalgebra generators
is obtained upon starting from  the non--compact
grading with respect to (\ref{so12so11}). To this effect we define
the {\it dilatation} $SO(1,1)$ generator $D$ and the {\it Lorentz} $SO(1,2)$
generators $J_m$ as follows:
\begin{equation}
D \equiv i\, M_{34} \,, \qquad J^m = \ft{i}{2} \, \varepsilon^{mpq} M_{pq}\,.
\label{dilalor}
\end{equation}
%%%%%%%%%%%%%%%%%%%%%%%%%%%%%%%%%%%%%%%%%
%%%%%%%%%%%%%%%%%%%%%%%%%%%%%%%%%%%%%%%%%%%%%%%%%%%%%%%%%%%%%%%%%
In addition we define the
the $d=3$ {\it translation generators} $P_m$  and {\it special conformal
boosts} $K_m$ as follows:
\begin{eqnarray}
P_m = M_{m4} - M_{3m} \,, \nonumber \\
K_m = M_{m4} + M_{3m} \,.
\label{Pkdefi}
\end{eqnarray}
Finally we define the generators of $d=3$ {\it ordinary} and {\it special
conformal supersymmetries}, respectively given by:
\begin{eqnarray}
q^{\alpha i} = \ft{1}{\sqrt{2}}\left(a_\alpha^i + \bar a^{\alpha i}\right)
\,, \nonumber \\
s_\alpha^i = \ft{1}{\sqrt{2}}\left(-a_\alpha^i + \bar a^{\alpha i}\right)
\,.
\label{qsdefi}
\end{eqnarray}
%%%%%%%%%%%%%%%%%%%%%%%%%%%%%%%%%%%%%%%%%%%%%%%%%%%%%%%%%%%%%%%%
%%%%%%%%%%%%%%%%%%%%%%%%%%%%%%%%%%%%%%%%%%%%%%%%%%%%%%%%%%%%%%%%
The $SO({\cal N})$ generators are left unmodified as above.
In this new basis the $osp({\cal N}\vert 4)$-algebra (\ref{algbr}) reads
as follows
\begin{eqnarray}
{}[D, P_m] &=& -P_m \,, \nonumber \\
{}[D, K_m] &=& K_m \,, \nonumber \\
% {}[D, J_m] &=& 0 \,,  \nonumber \\
{}[J_m, J_n] &=& \varepsilon_{mnp} \, J^p \,, \nonumber \\
{}[K_m, P_n] &=& 2 \, \eta_{mn}\, D - 2 \, \varepsilon_{mnp} \, J^p \,,
\nonumber \\
{}[J_m, P_n] &=&  \varepsilon_{mnp} \, P^p \,, \nonumber \\
{}[J_m, K_n] &=&  \varepsilon_{mnp} \, K^p \,, \nonumber \\
{}[T^{ij}, T^{kl}] &=&
-i\,(\delta^{jk}\,T^{il}-\delta^{ik}\,T^{jl}-
\delta^{jl}\,T^{ik}+\delta^{il}\,T^{jk})
\,, \nonumber \\
{}[T^{ij}, q^{\alpha k}] &=&
-i\, (\delta^{jk}\, q^{\alpha i} - \delta^{ik}\, q^{\alpha j} )
\,, \nonumber \\
{}[T^{ij}, s_\alpha^k] &=&
-i\, (\delta^{jk}\, s_\alpha^i - \delta^{ik}\,  s_\alpha^i )
\,,\nonumber \\
{}[D, q^{\alpha i}] &=& -\ft12 \, q^{\alpha i} \,,\nonumber \\
{}[D, s_\alpha^i] &=& \ft12 \, s_\alpha^i \,,\nonumber \\
{}[K_m, q^{\alpha i} ] &=&
- i\, (\gamma_m)^{\alpha \beta}\, s_\beta^i \,, \nonumber
\\
{}[P_m, s_\alpha^i] &=&
- i\, (\gamma_m)_{\alpha \beta}\, q^{\beta i} \,, \nonumber \\
{}[J_m, q^{\alpha i} ] &=&
- \ft{i}{2} \, (\gamma_m)^\alpha{}_\beta q^{\beta i} \,,
\nonumber \\
{}[J_m, s_\alpha^i] &=& \ft{i}{2} \,  (\gamma_m)_\alpha{}^\beta s_\beta^i \,,
\nonumber \\
\{q^{\alpha i}, q^{\beta j} \} &=& - i\, \delta^{ij}\,
(\gamma^m)^{\alpha\beta} P_m \,, \nonumber \\
\{s_\alpha^i, s_\beta^j \} &=&
i\, \delta^{ij} \, (\gamma^m)_{\alpha\beta} K_m \,,
\nonumber \\
\{q^{\alpha i}, s_\beta^j \} &=& \delta^{ij} \delta^\alpha{}_\beta \, D
- i\, \delta^{ij} (\gamma^m)^\alpha{}_\beta J_m
+ i \delta^\alpha{}_\beta T^{ij}
\,,
\label{ospD}
\end{eqnarray}
and the five grading structure of eq.s (\ref{ospD}) is displayed in
fig.\ref{pirillo}.
In both cases of fig.\ref{pistac} and fig.\ref{pirillo} if one takes
the subset of generators of positive grading plus the abelian grading
generator $X=\cases{H\cr D\cr}$ one obtains a {\it solvable superalgebra} of dimension
$4+2{\cal N}$.
\par
\section{UIRs of $SO\ll(3,2\rr)$}
\par
In order to construct the UIRs of $Osp\ll(\cN\vert 4\rr)$, the first step is to build the $SO\ll(3,2\rr)$ UIRs.
To do it, we use the method of induced representations, using the compact grading structure; then, we
are building the fields in $AdS_4$ space. We consider then the graded decomposition
of $SO\ll(3,2\rr)$ with respect to its $SO\ll(2\rr)$ generator $H$,
\be
so\ll(3,2\rr)= g_-\oplus g_0\oplus g_+.
\ee
$g_0$ is the Lie algebra of the compact subgroup
\be
\underbrace{SO\ll(3\rr)}_{\rm spin}\times\underbrace{SO\ll(2\rr)}_{\rm energy}\subset SO\ll(3,2\rr)
\ee
and commutes with the energy, while $ g_{\pm}$ are raising and lowering generators with respect to $H$.
\par
In practice, as we have seen, we define
\beq
H&=&M_{04}\subset g_0\nn\\
L_{a}&=&{1\over 2}\ve_{abc}M_{bc}\subset  g_0\nn\\
M^{\pm}_a&=&\ii M_{0a}\mp M_{a4}\subset g_{\pm}.
\eeq
We have
\beq
\ll[H,L_a\rr]&=&0\nn\\
 \ll[H,M^{\pm}_a\rr]&=&\pm M^{\pm}_a.
\eeq
Notice that $\ll(H\rr)^+=H,~\ll(L_a\rr)^+=L_a,~\ll(M_a^{\pm}\rr)^+=-M_a^{\mp}$.
\par
Furthermore, it is useful to organize the $M^{\pm}_a$ generators in the following way:
\beq
M^{\pm}_{\ll(+\rr)}&=&{1\over\sqrt{2}}\ll(M^{\pm}_1+\ii M^{\pm}_2\rr)\nn\\
M^{\pm}_{\ll(-\rr)}&=&{1\over\sqrt{2}}\ll(M^{\pm}_1-\ii M^{\pm}_2\rr)\nn\\
M^{\pm}_3,&&
\eeq
so that they have a definite action also on the spin:
\beq
\ll[L_3,M^{\pm}_{\ll(+\rr)}\rr]&=&M^{\pm}_{\ll(+\rr)}\nn\\
\ll[L_3,M^{\pm}_{\ll(-\rr)}\rr]&=&-M^{\pm}_{\ll(-\rr)}\nn\\
\ll[L_3,M^{\pm}_3\rr]&=&0.
\eeq
We are interested into representations with energy bounded from below; then, we consider
the UIRs of the compact subgroup $SO\ll(3\rr)\times SO\ll(2\rr)$ annihilated by the
energy lowering generators $M^-_a$. We call them the ground states of the representation.
Applying the raising generators $M^+_a$ (more precisely, the generators of the enveloping 
algebra of $SO\ll(3,2\rr)$ built by its $ g_+$ subspace) on the states of an
$SO\ll(3\rr)\times SO\ll(2\rr)$ UIR, we get an $SO\ll(3,2\rr)$ UIR, and in this way one finds all 
the UIRs of $SO\ll(3,2\rr)$.
\par
A representation of $SO\ll(3\rr)\times SO\ll(2\rr)$ is defined by the labels $\ll(E,s\rr)$,
where the eigenvalue of $H$ is $E$ and the eigenvalue of $L^2$ is $s\ll(s+1\rr)$.
Its states are labeled by the $L_3$ eigenvalue $m=-s,\dots,s$. 
Then the values of $E,s$ (the energy and spin of the ground states) define the
generic UIR of $SO\ll(3,2\rr)$.
\par
We denote a generic state with the quantum numbers of $H,~L^2,~L_3,~\ll(\bar{E},\bar{s},\bar{m}\rr)$,
and with the quantum numbers $\ll(E,s\rr)$ that single out the $SO\ll(3,2\rr)$ UIR to which it belongs 
(that is, the $H,L^2$ quantum numbers of the ground states of that representation):
\be
\ll|\ll(E,s\rr)\bar{E},\bar{s},\bar{m}\rr>.
\ee
We denote an UIR of $SO\ll(3,2\rr)$ with ground states having $E,s$ by
\be
D\ll(E,s\rr).
\ee
\par
\subsection{Unitarity bounds, massless representations and singletons}
\par
A representation $D\ll(E,s\rr)$ is well defined only if the Hilbert space does not contain
negative norm states; if it contains null norm states, the physical Hilbert space
is the quotient space between the complete space and the space of the null
norm states.
Evaluating the norms of the excited states one finds \cite{nicolai} that
the necessary and sufficient conditions for the absence of negative norm states are:
\be
\begin{array}{ll}
E\ge s+1 & {\rm if~}s\ge 1 \\
E\ge s+{1\over 2} &{\rm if~}s=0,{1\over 2}\\
\end{array}\,.
\ee
For special values of $E$ one finds that some of the states obtained
applying the raising operators $M_a^+$ on the ground states have vanishing norms.
This means that they are decoupled from the representation, form another UIR of
$SO\ll(3,2\rr)$, and our UIR is {\bf shortened}. It happens when:
\be
\label{shortenedirreps}
\begin{array}{c}
\begin{array}{ll}
E= s+1 & s\ge 1 \\
E= s+{1\over 2} &s=0,{1\over 2}\\
\end{array}\\ \\
\hbox{Short $AdS_4$ UIRs .}\\ \\
\end{array}
\ee
For these values of $E,s$ the equation of motion acquires gauge invariance; the states
decoupled because of the shortening are the gauge degrees of freedom, which can be
removed by gauge fixing.
\par
The shortened representations partially coincide with the {\it massless representations}.
It is not obvious how to define the mass in anti--de Sitter theories, since the quadratic Casimir
operator ${\cal C}=M_{AB}M^{AB}$ is different from the usual mass $P_aP^a$, which is
not a conserved quantity. The usual way to define a mass for $AdS_4$ UIRs 
\cite{defmassless}, \cite{nicolai} refers to the concept of
masslessness, inherited by analogy from Poincar\'e theories. In Poincar\'e space, the massless
field equations have enhanced symmetry, from $ISO\ll(3,1\rr)$ to conformal symmetry
$SO\ll(4,2\rr)$. Furthermore, the massless Poincar\'e representations are UIRs of the
conformal group, irreducible under $ISO\ll(3,1\rr)\subset SO\ll(4,2\rr)$.
This phenomenom occurs also in $AdS_4$ space, and when it occurs
we name the corresponding $AdS$ representation massless. 
Another reason for this choice is that
these $AdS_4$ representations become, by Inon\"u Wigner contraction, 
the massless Poincar\'e representations; indeed, the corresponding mass generator goes to zero in
this limit.
With this definition, the massless $AdS_4$ representations are the following:
\be
\label{masslessirreps}
\begin{array}{c}
\begin{array}{ll}
E= s+1 & s\ge {1\over 2} \\
E= 1,2 &s=0\\
\end{array}\\ \\
\hbox{Massless $AdS_4$ UIRs .}\\ \\
\end{array}
\ee
We can see that the $D\ll(s+1,s\rr)$ with $s\ge 1$ are both shortened and massless representations.
For $s=1/2,0$, the massless representations are not shortened: they do not have gauge invariance,
but their equations of motion are conformal and their contractions are Poincar\'e massless representations.
There is only a little subtlety: the $D\ll(s+1,s\rr)~s\ge 1/2$ are UIRs of $SO\ll(4,2\rr)$, but $D\ll(1,0\rr)$ and 
$D\ll(2,0\rr)$ are not separately $SO\ll(4,2\rr)$ UIRs: only their direct sum $D\ll(1,0\rr)\oplus D\ll(2,0\rr)$
is an $SO\ll(4,2\rr)$ UIR. The Inon\"u Wigner contractions of $D\ll(s+1,s\rr)~s\ge 1/2$ and 
$D\ll(1,0\rr)\oplus D\ll(2,0\rr)$ are the Poincar\'e massless representations.
\par
The shortened representations with $s=0,1/2$ are not massless representations. They have very 
peculiar properties: they do not describe a sufficient number of 
degrees of freedom to admit a field realization
on $AdS_4$; once the gauge degrees of freedom are removed, the only remaining degrees of 
freedom live on the boundary $\del AdS_4$, and not on the bulk of $AdS_4$ itself. These representations, found by Dirac 
\cite{diracsingletons}, are called {\bf singletons}:
\be
\label{singletonirreps}
\begin{array}{c}
\begin{array}{ll}
E= 0 & s= {1\over 2} \\
E= {1\over 2} &s=1\\
\end{array}\\ \\
\hbox{Singleton $AdS_4$ UIRs .}\\ \\
\end{array}
\ee
They cannot live on the bulk of $AdS_4$, but only on the boundary. Furthermore, the
tensor products of the singletons yield all the massless $AdS_4$ representations.
\par
Now that we have defined when a representation is massless, we define the squared mass 
as the additional constant term in the quadratic field equations,
\be
\xbox_s^{\rm massless}\Phi=m_{\ll(s\rr)}^2\Phi\,.
\ee
The mass squared is linear in the quadratic Casimir $m^2_{\ll(s\rr)}=\b_{\ll(s\rr)}
\ll({\cal C}_2+\a_{\ll(s\rr)}\rr)$; the overall normalization $\b_{\ll(s\rr)}$ is arbitrary. 
We take the normalization of \cite{castdauriafre},\cite{noi1},\cite{noi3}, that gives
\footnote{In \cite{Maldyrev}, \cite{frenico}, \cite{gunawar}, \cite{dewitads} the 
mass normalization differs by a factor $4$, for the reason explained in chapter $1$.}
\be
\begin{array}{|c|rl|}
\hline
&&\\
s=0 & m^2_{\ll(0\rr)}&=16\ll(E_{\ll(0\rr)}-2\rr)\ll(E_{\ll(0\rr)}-1\rr)\\
&&\\
\hline
&&\\
s=1/2 & \ll|m_{\ll(1/2\rr)}\rr|&=4E_{\ll(1/2\rr)}-6\\
&&\\
\hline
&&\\
s=1 & m^2_{\ll(1\rr)}&=16\ll(E_{\ll(1\rr)}-2\rr)\ll(E_{\ll(1\rr)}-1\rr)\\
&&\\
\hline
&&\\
s=3/2 & \ll|m_{\ll(3/2\rr)}+4\rr|&=4E_{\ll(3/2\rr)}-6\\
&&\\
\hline
\end{array}
\label{massenergyAdS4}
\ee
Notice that when $s=0$, for each energy value ${1\over 2}<E<{5\over 2}$ 
there are two mass square values; they correspond to the same form of the field equation.
Furthermore, when $1<E<2$ the mass square is negative, $-4<m^2<0$; however, it has been shown
\cite{adsstability} that in anti--de Sitter space, due to the presence of the boundary, 
the stability bound is, in our normalizations, $m^2>-4$ and not $m^2>0$.
\par
\section{UIRs of $Osp\ll(\cN\vert 4\rr)$ viewed in the compact and non
compact five--grading bases}
\par
We start by briefly
recalling the procedure of \cite{frenico}, \cite{heidernreich}
to construct UIRs of $Osp({\cal N}\vert 4)$ in the compact grading
(\ref{so3so2}) (these procedure will be discussed more extensively in the next section). 
Then, in a parallel way to
what was done in \cite{gunaydinminiczagerman} for the case
of the $SU(2,2\vert 4)$ superalgebra we show that also
for $Osp({\cal N}\vert 4)$ in each UIR carrier space  there exists an
unitary rotation
that maps  eigenstates of $H, L^2, L_3$
into eigenstates of $D,J^2,J_2$. By means of such a rotation the
decomposition of the UIR into $SO(2)\times SO(3)$ representations is
mapped into an analogous decomposition into $SO(1,1) \times SO(1,2)$
representations.
While $SO(2)\times SO(3)$ representations describe the {\it on--shell}
degrees of freedom of a {\it bulk particle} with an energy
$E_0$ and a spin $s_0$,  irreducible  representations of $SO(1,1) \times
SO(1,2)$ describe the {\it off-shell} degrees of freedom  of a
{\it boundary field} with scaling weight $D$ and Lorentz character $J$.
Relying on this  we show how to
construct the on-shell four-dimensional
superfield multiplets that generate the states of these representations
and the off-shell three-dimensional superfield multiplets that build
the conformal field theory on the boundary.
\par
Lowest weight representations of $Osp({\cal N}\vert 4)$  are
constructed starting from the basis (\ref{ospH}) and choosing a
{\it Clifford vacuum state} such that
\begin{eqnarray}
M_i^- \vert (E_0, s_0, \Lambda_0) \rangle &=& 0 \,,
\nonumber \\
a^i_\alpha \vert (E_0, s_0, \Lambda_0) \rangle &=& 0 \,,
\label{energyreps}
\end{eqnarray}
where $E_0$ denotes the eigenvalue of the energy operator
$M_{04}$ while $s_0$ and $\Lambda_0$ are the labels of an irreducible $SO(3)$ and
$SO({\cal N})$ representation, respectively\footnote{In this context we call it state even if it is
a collection of states.}.
In particular we have:
\begin{eqnarray}
M_{04}\, \vert (E_0, s_0, \Lambda_0)\rangle
& = & E_0 \, \vert (E_0, s_0, \Lambda_0) \rangle
\nonumber \\
L_a \, L_a \, \vert (E_0, s_0, \Lambda_0) \rangle & = & s_0(s_0+1) \,
\vert (E_0, s_0, \Lambda_0) \rangle  \nonumber\\
L_3  \vert (E_0, s_0, \Lambda_0) \rangle & =& s_0 \,\vert (E_0, s_0, \Lambda_0) \rangle\,.
\label{eigval}
\end{eqnarray}
The states filling up the UIR
are then built by applying the operators $M^-$ and the anti-symmetrized
products of the operators $\bar a^i_\alpha$:
\begin{eqnarray}
\left( M_1^+ \right)^{n_1} \left( M_2^+ \right)^{n_2}
\left( M_3^+ \right)^{n_3} [ \bar a^{i_1}_{\alpha_1}
\dots \bar a^{i_p}_{\alpha_p}]
\vert (E_0, s_0, \Lambda_0) \rangle\,.
\label{so3so2states}
\end{eqnarray}
The antisymmetrization of the fermionic operators is due to the fact that 
\be
\{\bar a^{\alpha i}, \bar a^{\beta j} \} =
\delta^{ij} (\tau^k)^{\alpha\beta} \, M_k^+ 
\ee
so the symmetrized fermionic generators yield excited states of the same $AdS_4$ fields,
not new $AdS_4$ fields.
\par
Lowest weight representations are similarly constructed
with respect to five--grading (\ref{ospD}).
One starts from a vacuum state that is annihilated by the conformal boosts and
by
the special conformal supersymmetries
\begin{eqnarray}
K_m \, \vert (D_0, j_0, \Lambda_0) \rangle &=& 0 \,,
\nonumber \\
s^i_\alpha \, \vert (D_0, j_0, \Lambda_0) \rangle &=& 0 \,,
\label{primstate}
\end{eqnarray}
and that is an eigenstate of the dilatation operator $D$ and an
irreducible $SO(1,2)$ representation of spin $j_0$:
\begin{eqnarray}
  D \, \vert (D_0, j_0, \Lambda_0) \rangle &=& D_0 \, \vert (D_0, j_0, \Lambda_0) \rangle
  \nonumber\\
  J_m \, J_n \, \eta^{mn} \,\vert (D_0, j_0, \Lambda_0) \rangle &=&
 j_0 (j_0 +1) \,\vert (D_0, j_0, \Lambda_0) \rangle\nonumber\\
  J_2 \,\vert (D_0, j_0, \Lambda_0) \rangle & = & j_0 \vert (D_0, j_0, \Lambda_0) \rangle\,.
\label{Djvac}
\end{eqnarray}
%%%%%FIN QUI 2%%%%%%%
As for the $SO({\cal N})$ representation the new vacuum is the same
as before.
The states filling the UIR are now constructed by applying to the vacuum
the operators $P_m$ and the
anti-symmetrized products of $q^{\alpha i}$,
\begin{eqnarray}
\left( P_0\right)^{p_0} \left( P_1 \right)^{p_1}
\left( P_2 \right)^{p_2} [ q^{\alpha_1 i_1} \dots q^{\alpha_q i_q}]
\vert (D_0, j_0, \Lambda_0) \rangle\,.
\label{so12so11states}
\end{eqnarray}
\par
In the language of  conformal field theories the vacuum state
satisfying eq.(\ref{primstate}) is named a {\it primary state}
(corresponding to the value at $z^m=0$ of a primary conformal field).
The states (\ref{so12so11states}) are called the {\it descendants}.
\par
\begin{figure}[ht]
\begin{center}
\leavevmode
\hbox{%
\epsfxsize=12cm
\epsfbox{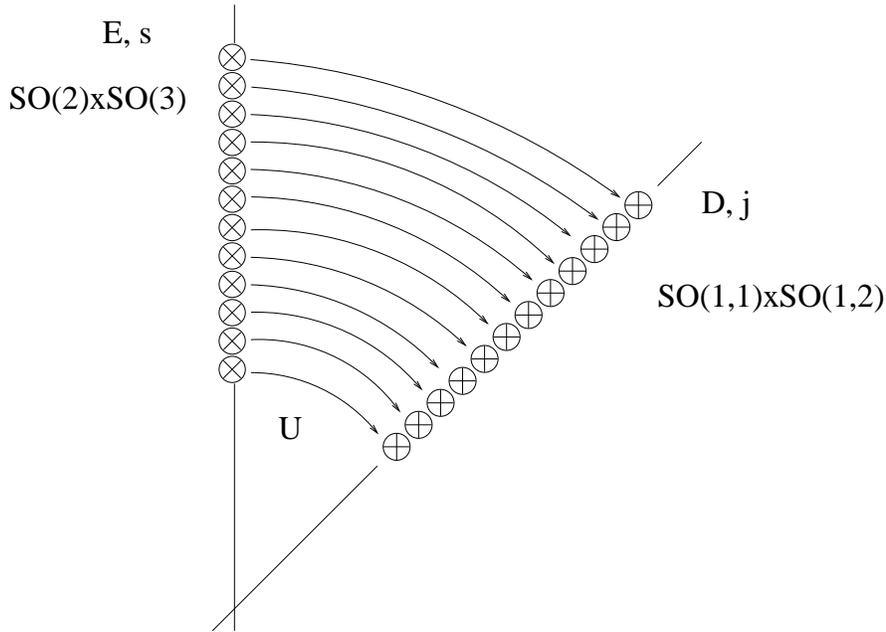}}
\caption{{\small The operator $U=\exp\{\ll(i\pi/\sqrt 2\rr)(H-D)\}$ rotates
the Hilbert space of the physical states.
It takes states labeled by the Casimirs ($E,\,s$) of
the $SO(2)\times SO(3)\subset Osp({\cal N}|4)$ into states
labeled by the Casimirs ($D,\,j$) of $SO(1,1)\times SO(1,2)$.\label{rota}}}
\end{center}
\end{figure}

The rotation between the $SO(3)\times SO(2)$ basis
and the $SO(1,2)\times SO(1,1)$ basis is performed by the
operator:
\begin{eqnarray}
U\equiv  \exp \left[{\ft{i}{\sqrt{2}}\pi(H-D)} \right] \,,
\label{rotationmatrix}
\end{eqnarray}
which has the following properties
\begin{eqnarray}
D U &=& - U H \,, \nonumber \\
J_0 U &=& i \, U L_3 \,, \nonumber \\
J_1 U &=& U L_1 \,, \nonumber \\
J_2 U &=& U L_2 \,,
\label{L0}
\end{eqnarray}
with respect to the grade $0$ generators. Furthermore, with respect
to the non vanishing grade generators we have:
\begin{eqnarray}
K_0 U &=& -i \, U M_3^- \,, \nonumber \\
K_1 U &=& - U M_1^- \,, \nonumber \\
K_2 U &=& - U M_2^- \,, \nonumber \\
P_0 U &=& i\,U M_3^+ \,, \nonumber \\
P_1 U &=& U M_1^+ \,, \nonumber \\
P_2 U &=& U M_2^+ \,, \nonumber \\
q^{\alpha i} U &=& -i\, U \bar a^{\alpha i} \nonumber \\
s_\alpha^i U &=& i \, U a_\alpha^i \,.
\label{L+L-}
\end{eqnarray}
As one  immediately sees from (\ref{L+L-}), U interchanges the
compact five--grading structure of the superalgebra with its non
compact one. In particular
the $SO(3)\times SO(2)$-vacuum with energy $E_0$
is mapped into an $SO(1,2)\times SO(1,1)$ primary state and
one obtains all the descendants (\ref{so12so11states}) by acting
with $U$ on the particle states (\ref{so3so2states}). Furthermore
from (\ref{L0}) we read the conformal weight and the Lorentz
group representation of the primary state
$U \vert (E_0, s_0, \Lambda_0) \rangle$. Indeed its
 eigenvalue with respect to the dilatation generator $D$ is:
\begin{equation}
D_0 = - E_0 \,,
\end{equation}
and  we find the following relation between  the Casimir operators
 of $SO(1,2)$ and $SO(3)$,
\begin{equation}
J^2 U = U L^2 \,, \qquad J^2 \equiv -J_0^2 + J_1^2 + J_2^2 \,,
\end{equation}
which implies that
\begin{equation}
j_0 = s_0 \,.
\end{equation}
Hence under the action of $U$ a particle state of energy $E_0$ and
spin $s_0$ of the bulk is mapped into a {\it primary conformal field}
of conformal weight $-E_0$ and Lorentz spin $s_0$ on the boundary.
This discussion is visualized in fig.\ref{rota}.
\label{rotation}
%%%%%%%%%%%%%%%%%%%%%%%%%%%%%%%%%%%%%%%%%%%%%%%%%%%%%%%%%%%%%%%%
%
%                    F I G U R E
%
%%%%%%%%%%%%%%%%%%%%%%%%%%%%%%%%%%%%%%%%%%%%%%%%%%%%%%%%%%%%%%%%
\par
As in $SO\ll(3,2\rr)$ representation theory, even the $Osp\ll(\cN\vert 4\rr)$ UIRs
have to satisfy unitarity bounds, because all the states (\ref{so3so2states}) or
(\ref{so12so11states}) must have nonnegative norms. 
When some of these bounds are saturated, some norms vanish, and
the corresponding representations are shortened. These representations are BPS states,
namely they are protected against quantum corrections; indeed the number of the states does
not change with renormalization, so the shortening condition, which is
a condition on the quantum numbers $E_0,s_0,\Lambda_0$ or $D_0,j_0,\Lambda_0$,
must remain satisfied; then, being $\Lambda_0$ a non renormalized quantity, also
$E_0$ and $D_0$ are not renormalized.
\par
\section{The explicit construction of $Osp\ll(\cN\vert 4\rr)$ UIRs}
\label{constructsupermultiplets}
\par
Now I show how the $Osp\ll(\cN\vert 4\rr)$ UIRs are explicitly worked out, from
the compact grading viewpoint, namely, as supermultiplets of $\cN$ extended
$AdS_4$ fields.
I remind that we are interested only on supermultiplets not containing fields with 
spin greater than two.
\par
\subsection{Finding all the states}
\par
As I said, to find the field structure of a supermultiplet, one does not have to consider
excited states of the fields: it suffices to restrict one's attention to their ground states, which are
$SO\ll(3\rr)\times SO\ll(2\rr)\times SO\ll(\cN\rr)$ UIRs annihilated by the 
$M_i^-$ operators. Their spin and $R$--symmetry labels
$E,s,\Lambda$ identify the corresponding $AdS_4$ field, namely an $SO\ll(3,2\rr)$ UIR.
A supermultiplet is identified by its lowest energy $AdS_4$ field, whose ground states
are an $SO\ll(3\rr)\times SO\ll(2\rr)\times SO\ll(\cN\rr)$ UIR 
$\vert \ll(E_0, s_0, \Lambda_0\rr) \rangle$ satisfying
\begin{eqnarray}
M_i^- \vert (E_0, s_0, \Lambda_0) \rangle &
 =& 0 \,,
\nonumber \\
a^i_\alpha \vert (E_0, s_0, \Lambda_0) \rangle &=& 0 \,.
\end{eqnarray}
We name this state the vacuum, namely, the ground state of the lowest energy 
$SO\ll(3,2\rr)$ representation $D\ll(E_0,s_0\rr)$ contained in the supermultiplet.
The  other fields $D\ll(E,s\rr)$ of the supermultiplet are
found by applying the antisymmetrized products of fermionic raising generators on the
vacuum. We denote the entire supermultiplet, namely, the $Osp\ll(\cN\vert 4\rr)$ UIR, 
as
\be
SD\ll(E_0,s_0,\Lambda_0\vert\cN\rr).
\ee
The first step is then to find, given $\cN$,  all the operators of the form 
\be
^pK^{i_1\dots i_p}_{\alpha_1\dots\alpha_p}=[ \bar a^{i_1}_{\alpha_1}
\dots \bar a^{i_p}_{\alpha_p}].
\label{operators}
\ee
Then, for each vacuum state, applying on it the operators (\ref{operators}) 
we find all the possible fields of the 
corresponding supermultiplet. Each operator $^pK$ has given spin and $R$--symmetry
labels, which have to be composed with that of the vacuum.
\par
Since the $\bar{a}$ generators in the (\ref{operators}) are antisymmetrized, 
if two $\bar{a}$'s are symmetric in the $SO\ll(\cN\rr)$ $R$--symmetry indices, they have to be
antisymmetric in the $SU\ll(2\rr)$ spin indices. Notice that we consider $SU\ll(2\rr)$ instead of 
its locally isomorphic group $SO\ll(3\rr)$ as spin group, in order to make the calculation simpler: 
all the spin representations can be expressed by means of $SU\ll(2\rr)$ Young tableaux.
A useful trick to find, given $\cN$, all the possible $^pK^{i_1\dots i_p}_{\alpha_1\dots\alpha_p}$,
is to consider temporarily the $\bar{a}_{\alpha}^i$ as a representation of
\be
\label{toobig}
SU\ll(2\rr)\times SU\ll(\cN\rr)\supset SU\ll(2\rr)\times SO\ll(\cN\rr).
\ee
Actually, they are a representation of $SU\ll(2\rr)\times SU\ll(\cN\rr)$ 
only as a vector space, not as an algebra,
because the superalgebra (\ref{ospH}) is not $SU\ll(\cN\rr)$--invariant. However,
we can find the $\tilde{^pK}$ operators irreducible under $SU\ll(2\rr)\times SU\ll(\cN\rr)$ and then branch 
them in $SU\ll(2\rr)\times SO\ll(\cN\rr)$ representations obtaining in this way 
the $^pK$ operators. The reason for this procedure is that, given an 
operator $\tilde{^pK}^{i_1\dots i_p}_{\alpha_1\dots\alpha_p}$, if the representation of its 
$SU\ll(\cN\rr)$ indices is described by a given Young tableau, the representation if its 
$SU\ll(2\rr)$ indices has to be described by the transposed Young tableau, in order to have a 
representation antisymmetric in the exchange $i_a\leftrightarrow i_b,~\alpha_a\leftrightarrow
\alpha_b$. Then if we write all the allowed
$SU\ll(2\rr)$ Young diagrams whose transposed are allowed $SU\ll(\cN\rr)$ Young diagrams,
we find all the allowed operators $\tilde{^pK}$, and decomposing them in 
$SU\ll(2\rr)\times SO\ll(\cN\rr)$ irreducible representations we find all the $^pK$'s.
\par
The states created by operators $^pK$ (containing $p$ generators $\bar{a}$) are denoted as
the $B_p$ sector of the representation. The maximum possible value of $p$ 
is $p=2\cN$, corresponding to the $SU(2)$ Young tableau with two rows and $\cN$
columns.
\par
\subsection{Finding unitarity bounds and norms}
\par
The determination of all the possible states is the simplest part of our work.
The most cumbersome calculation is the derivation of the unitarity bounds and the
shortenings of the representations. This should be done by calculating the norms of the
states we have found, by means of the algebra (\ref{ospH}) which allows us to express the norms in terms of 
$E_0,s_0,\Lambda_0$. But this calculation is affordable only up to the $B_2$ sector, while
when there are three or more $\bar{a}$ factors the calculation is too long.
This because an operator $\bar{a}$ on a ground state of an $AdS_4$ field in general gives not only
the ground state of another field of a multiplet, but also excited states of less energetic fields
of the multiplet. We have then to use other information in order to find
the structure of short multiplets:
\begin{itemize}
\item The $Osp\ll(\cN\vert 4\rr)$ UIRs are also UIRs of $Osp\ll(\cN'\vert 4\rr)\subset 
Osp\ll(\cN\vert 4\rr)$ with $\cN'<\cN$, so the 
$\cN$--supermultiplets must be decomposable in $\cN'$ supermultiplets, and the vanishing 
of some states implies the vanishing of other states.
\item By Inon\"u Wigner contraction, the $Osp\ll(\cN\vert 4\rr)$ UIRs become massless
representations (eventually reducible) of the Poincar\'e superalgebra.Then, all the $\cN$ extended anti--de Sitter supermultiplets must be decomposable in 
$\cN$ extended massless Poincar\'e supermultiplets; the only exception is the supersingleton 
representation.
\item The massless UIRs of  $Osp\ll(\cN\vert 4\rr)$ are well known, being the fields of
exact four dimensional supergravity; they coincide with Poincar\'e massless supermultiplets. 
This is understandable: the exact
supergravity is a field theory whose field content cannot depend on a particular vacuum,
Poincar\'e space or anti--de Sitter space; only Kaluza Klein supergravity, which is a linearized 
theory, obtained as a truncation of exact massless eleven dimensional supergravity expanded around a
given background, is reminiscent of that background.
\item Explicit calculations of Kaluza Klein spectrum of specific $\cN$  extended supergravities,
as the ones in the next chapter, allow to fill the gaps in the knowledge on the supermultiplet
structure.
\item As I will show afterwards, the superfield formalism allows another formulation of short
representations, which can be useful to derive them. This has not been done for all values of 
$\cN$, but in the known cases is a good check of the multiplet structure.
\label{otherways}
\end{itemize}
\par
\subsection{The structure of $\cN=1$ supermultiplets}
\par
This case has been completely worked out in \cite{heidernreich}. There is no
$R$--symmetry group. The maximum number of fermionic generators allowed
is $p=2\cN=2$. There are:
\begin{itemize}
\item the $B_0$ sector, created by the identity $\unity$, 
that is the lowest lying representation $D\ll(E_0,s_0\rr)$;
\item the $B_1$ sector, created by
\be
^1K_{\a}=\bar{a}_{\a},
\ee
having spin $1/2$.
\item the $B_2$ sector, created by
\be
^2K=\ve^{\a\b}\bar{a}_{\a}\bar{a}_{\b},
\ee
having spin $0$.
\end{itemize}
There are two cases:
\begin{enumerate}
\item $s_0=0$\\  The lowest lying field is $D\ll(E_0,0\rr)$, the spin $1/2$ 
operator of $B_1$ 
gives $D\ll(E_0\!+\!1/2,\!1/2\rr)$.  The spin $0$ operator of $B_2$ gives 
$D\ll(E_0+1,0\rr)$. The entire 
supermultiplet is then
\be
D\ll(E_0,0\rr)\oplus D\ll(E_0+1/2,1/2\rr) \oplus D\ll(E_0+1,0\rr).
\ee
\item ${1\over 2}\ge s_0\ge {3\over 2}$\\
The spin $1/2$ operator of $B_1$ gives 
$D\ll(E_0+1/2,s_0+1/2\rr)\oplus D\ll(E_0+1/2,s_0-1/2\rr)$. The spin $0$ operator of $B_2$ 
gives $D\ll(E_0+1,s_0\rr)$. The entire supermultiplet is then
\be
D\ll(E_0,s_0\rr)\oplus D\ll(E_0+1/2,s_0+1/2\rr) \oplus D\ll(E_0+1/2,s_0-1/2\rr) \oplus D\ll(E_0+1,s_0\rr).
\ee
\end{enumerate}
We have now to carry on the calculation of the norms, using the (\ref{ospH}) algebra.
We denote the vacuum, with norm $1$, by
\be
\vert\Omega\rangle\equiv\vert\ll(E_0,s_0\rr)~E_0,s_0,m\rangle.
\ee
\par
\vskip .5cm
\centerline{\large\fbox{$s=0$ case}}
\par
\begin{itemize}
\item $B_1$ sector\\
The operator $\bar{a}_1$ gives
\be
\bar{a}_1\vert\Omega\rangle=R_{1/2}\vert\ll(E_0+1/2,1/2\rr)~E_0+1/2,1/2,1/2\rangle\,.
\ee
Using the algebra (\ref{ospH}), we find
\be
\langle\Omega\vert a_1\bar{a}_1\vert\Omega\rangle=E_0=\ll|R_{1/2}\rr|^2,
\ee
that yields the condition $E_0\ge 0$. The condition arising from $\bar{a}_2$ is identical.
\item $B_2$ sector\\
The operator $\ve^{\a\b}\bar{a}_{\a}\bar{a}_{\b}$ gives
\beq
\ll[\bar{a}_1,\bar{a}_2\rr]\vert\Omega\rangle&=&R_0\vert\ll(E_0+1,0\rr)E_0+1,0,m\rangle+\nn\\
&&+\b M_3^+\vert\ll(E_0,0\rr)E_0,0,0\rangle
\eeq
Let us determine $\b$. 
\beq
M^-_3\ll[\bar{a}_1,\bar{a}_2\rr]\vert\Omega\rangle&=&\ll[M^-_3\ll[\bar{a}_1,\bar{a}_2\rr]\rr]\vert\Omega\rangle=0=\nn\\
&=&-2E_0\b\vert\ll(E_0,0\rr)E_0,0,0\rangle,
\eeq
then $\b=0$. Now we can find $\ll|R_0\rr|^2$:
\be
\langle\Omega\vert\ll[a_2,a_1\rr]\ll[\bar{a}_1,\bar{a}_2\rr]\vert\Omega\rangle=
4E_0^2-2E_0=\ll|R_0\rr|^2.
\ee
The unitarity condition arising from this sector is then
\be
\label{E0ge12}
E_0\ge{1\over 2},
\ee
stronger than the condition arising from the $B_1$ sector.
This is then the only unitarity condition of the representation. 
When it is saturated, $R_0=0$ and $D\ll(E_0+1,0\rr)$ decouples.
Furthermore, in this case ($E_0=1/2$) the representations $D\ll(E_0,0\rr)$ and
$D\ll(E_0+1/2,1/2\rr)$ are the singletons.
\end{itemize}
\par
\vskip .5cm
\centerline{\large\fbox{$s>0$ case}}
\par
\begin{itemize}
\item $B_1$ sector\\
The operator $\bar{a}_1$ gives
\beq
\bar{a}_1\vert\Omega\rangle&=&R_{1/2}\sqrt{s_0+m+1\over 2s_0+1}
\vert\ll(E_0+1/2,s_0+1/2\rr)~E_0+1/2,s_0+1/2,m+1/2\rangle+\nn\\
&&-R_{-1/2}\sqrt{s_0-m\over 2s_0+1}
\vert\ll(E_0+1/2,s_0-1/2\rr)~E_0+1/2,s_0-1/2,m-1/2\rangle.\nn\\
&&
\eeq
To find the values of the constants $R_{1/2},R_{-1/2}$, we take
specific values of $m$ and determine the norms using the algebra (\ref{ospH}). \\
Taking $m=s$, 
\be
\langle\Omega\vert a_1\bar{a}_1\vert\Omega\rangle=E_0+s_0=\ll|R_{1/2}\rr|^2,
\ee
that yields the condition $E_0+s_0\ge 0$.\\
Taking $m=-1/2$,
\be
\langle\Omega\vert a_1\bar{a}_1\vert\Omega\rangle=E_0-{1/2}={1\over 2}
\ll(\ll|R_{1/2}\rr|^2+\ll|R_{-1/2}\rr|^2\rr),
\ee
that yields the condition $E_0-s_0-1\ge 0$. \\
The unitarity condition arising from 
$B_1$ is then
\be
E_0\ge s_0+1.
\label{N1bound}
\ee
The condition arising from $\bar{a}_2$ is identical.
This condition coincides with the unitarity bound of $AdS_4$ fields for $s_0>1$, and is stronger for $s_0=1/2$.
When it is saturated, $R_{-1/2}=0$ and the $D\ll(E_0+1/2,s_0-1/2\rr)$ field decouples. 
Furthermore, $E_0=s_0+1$ is the masslessness condition for $AdS_4$ fields with $s_0\ge 1/2$,
so when it is saturated the fields $D\ll(E_0,s_0\rr)$ and $D\ll(E_0+1/2,s_0+1/2\rr)$ 
in (\ref{longN1multiplets}) are massless.
\item $B_2$ sector\\
The operator $\ve^{\a\b}\bar{a}_{\a}\bar{a}_{\b}$ gives
\beq
\ll[\bar{a}_1,\bar{a}_2\rr]\vert\Omega\rangle&=&R_0\vert\ll(E_0+1,s_0\rr)E_0+1,s_0,m\rangle+\nn\\
&&+\hbox{boosted elements of}~D\ll(E_0,s_0\rr).
\eeq
In order to calculate $\ll|R_0\rr|^2$ we take $m=s_0$. Then
\beq
\ll[\bar{a}_1,\bar{a}_2\rr]\vert\Omega\rangle&=&R_0\vert\ll(E_0+1,s_0\rr)E_0+1,s_0,s_0\rangle+\nn\\
&&+\a M_{\ll(+\rr)}^+\vert\ll(E_0,s_0\rr)E_0,s_0,s_0-1\rangle+\nn\\
&&+\b M_3^+\vert\ll(E_0,s_0\rr)E_0,s_0,s_0\rangle\,.
\eeq
Let us determine $\a$ and $\b$. 
\beq
M^-_{\ll(-\rr)}\ll[\bar{a}_1,\bar{a}_2\rr]\vert\Omega\rangle&=&\ll[M^-_{\ll(-\rr)}\ll[\bar{a}_1,\bar{a}_2\rr]\rr]\vert\Omega\rangle=\nn\\
&=&\sqrt{2}J_-\vert\Omega\rangle=2\sqrt{s_0}\vert\ll(E_0,s_0\rr)E_0,s_0,s_0-1\rangle=\nn\\
&=&\ll(-2\ll(E_0+s_0-1\rr)\a+2\sqrt{s_0}\b\rr)\vert\ll(E_0,s_0\rr)E_0,s_0,s_0-1\rangle\nn\\
&&\\
M^-_3\ll[\bar{a}_1,\bar{a}_2\rr]\vert\Omega\rangle&=&\ll[M^-_3\ll[\bar{a}_1,\bar{a}_2\rr]\rr]\vert\Omega\rangle=\nn\\
&=&2s_0\vert\ll(E_0,s_0\rr)E_0,s_0,s_0\rangle=\nn\\
&=&\ll(2\sqrt{s_0}\a-2E_0\b\rr)\vert\ll(E_0,s_0\rr)E_0,s_0,s_0\rangle,
\eeq
then
\beq
2\a\sqrt{s_0}-2E_0\b&=&2s_0\nn\\
-\a\ll(E_0+s_0-1\rr)+\b\sqrt{s_0}&=&\sqrt{s_0}
\eeq
which gives
\beq
\a&=&-{\sqrt{s_0}\over E_0-1}\nn\\
\b&=&-{s_0\over E_0-1}.
\eeq
Now we can find $\ll|R_0\rr|^2$:
\be
\begin{array}{l}
\langle\Omega\vert\ll[a_2,a_1\rr]\ll[\bar{a}_1,\bar{a}_2\rr]\vert\Omega\rangle=\\
=4E_0^2-2E_0-4s_0\ll(s_0+1\rr)=\ll|R_0\rr|^2+2\a^2\ll(E_0+s_0-1\rr)+2\b^2E_0-4\a\b\sqrt{s_0}=\\
=\ll|R_0\rr|^2+2s{E_0+s_0-1\over\ll(E_0-1\rr)^2}+{2s_0^2E_0\over\ll(E_0-1\rr)^2}-{4s^2\over
\ll(E_0-1\rr)^2},\\
\end{array}
\ee
that with some elementary but tedious manipulation gives
\be
\ll|R_0\rr|^2={2\over E_0-1}\ll(2E_0-1\rr)\ll(E_0+s_0\rr)\ll(E_0-s_0-1\rr).
\ee
The condition (\ref{N1bound}) guarantees $\ll|R_0\rr|^2\ge 0$, and then is the
only unitarity condition of the representation. When it is saturated, $R_0=0$ and $D\ll(E_0+1,s_0\rr)$
decouples.
\end{itemize}
In conclusion, the complete list of $Osp\ll(1\vert 4\rr)$ UIRs is:
\begin{enumerate}
\item $E_0>s_0+1,~{1\over 2}<s_0<{3\over 2}$: 
massive vector ($s_0=1/2$), gravitino ($s_0=1$) and graviton ($s_0=3/2$) multiplets
\beq
SD\ll(E_0,s_0\vert 1\rr)&=&D\ll(E_0,s_0\rr)\oplus D\ll(E_0+1/2,s_0+1/2\rr) \oplus \nn\\
&&\oplus D\ll(E_0+1/2,s_0-1/2\rr) \oplus D\ll(E_0+1,s_0\rr).
\label{longN1multiplets}
\eeq
The fields of these multiplets are all massive.
\item $E_0=s_0+1,~{1\over 2}<s_0<{3\over 2}$: 
massless vector ($s_0=1/2$), gravitino ($s_0=1$) and graviton ($s_0=3/2$) multiplets
\beq
SD\ll(s_0+1,s_0\vert 1\rr)&=&D\ll(s_0+1,s_0\rr)\oplus D\ll(s_0+3/2,s_0+1/2\rr) 
\label{masslessN1multiplets}
\eeq
The fields of these multiplets are all massless. Then, as we have seen, they tend 
with Inon\"u Wigner contraction to Poincar\'e massless fields. Actually, the entire 
multiplet tends to the corresponding massless multiplet of $\cN=1$ Poincar\'e supersymmetry,
and has then the same structure.
\item $s_0=0,~E_0>{1\over 2}$: Wess Zumino multiplet
\be
SD\ll(E_0,0\vert 1\rr)=D\ll(E_0,0\rr)\oplus D\ll(E_0+1/2,1/2\rr) \oplus D\ll(E_0+1,0\rr).
\label{wesszuminomultiplet}
\ee
When $E_0=1$, all the fields of this multiplet are massless. When $E_0=2$, some
of the fields of this multiplet are massless. In the other cases, they are all massive.
\item $s_0=0,~E_0={1\over 2}$: supersingleton representation
\be
SD\ll(1/2,0\vert 1\rr)=D\ll(1/2,0\rr)\oplus D\ll(1,1/2\rr).
\label{supersingletonN1}
\ee
The $SO\ll(3,2\rr)$ UIRs of this $Osp\ll(1\vert 4\rr)$ UIR are all singletons.
Then, this representation does not have a realization as fields on
the bulk, only on the boundary.
\end{enumerate}
\par
\subsection{The structure of $\cN=2$ supermultiplets}
\par
This case has been first studied in \cite{multanna}, where the list of the $~^pK$ operators has been
derived and some norms have been worked out, yielding the unitarity bounds, the shortening conditions
and the structure of some multiplets. Then, in \cite{noi1}, the complete spectrum on a particular $\cN=2$ 
supergravity compactification has been worked out (see chapter $3$), and as a byproduct the remaining information
on $\cN=2$ UIRs has been found, namely, the absence of further unitarity bounds and shortening conditions,
and the structure of all the multiplets.
\par
The maximum number of fermionic generators allowed is $p=2\cN=4$.
The $R$--symmetry group is $SO\ll(2\rr)$, locally isomorphic to $U\ll(1\rr)$;
the $U\ll(1\rr)$ UIRs are labeled by a rational number $y$ usually called {\it hypercharge},
eigenvalue of the $U\ll(1\rr)$ generator $Y$.
In the fermionic generators $\bar{a}_{\a}^i$ the index $i=1,2$ runs in the vector 
representation on $SO\ll(2\rr)$; the well--suited fermionic generators 
for the $U\ll(1\rr)$ form of the $R$--symmetry are
\beq
a^{\pm}_{\a}&=&{1\over\sqrt{2}}\ll(a^1_{\a}\pm \ii a^2_{\a}\rr)\nn\\
{\overline{a}}^{\pm}_{\a}&=&{1\over\sqrt{2}}\ll({\overline{a}}^1_{\a}\pm \ii{\overline{a}}^2_{\a}\rr),
\eeq
satisfying
\beq 
\ll(a_{\a}^{\pm}\rr)^+&=&\bar{a}_{\a}^{\mp}\nn\\
\ll[H,\bar{a}_{\a}^{\pm}\rr]&=&{1\over 2}\bar{a}_{\a}^{\pm}\nn\\
\ll[Y,\bar{a}_{\a}^{\pm}\rr]&=&\pm\bar{a}_{\a}^{\pm}.
\eeq
Then, they are raising and lowering generators of hypercharge with weight $1$.
\par
In order to find all the operators $^pK$ we use the method previously described. 
We denote the operators by the representations of their indices; write all the 
representations allowed, and determine their $SO\ll(2\rr)\times U\ll(1\rr)$ labels. I remind
that an $SO\ll(2\rr)$ UIR whose Young diagram (which is one row) has $n>0$ boxes,
coincide to an $U\ll(1\rr)$ UIR with $y=n$ plus its conjugate representation, having
$y=-n$; the $SO\ll(2\rr)$ singlet coincide to a real $U\ll(1\rr)$ UIR with $y=0$.
\par
The complete list of $^pK$ operators is
\be
\begin{array}{|c|c|c|c|}
\hline
        & SU\ll(2\rr)\times SU\ll(2\rr) & ~~~SU\ll(2\rr)\times SO\ll(2\rr)
~\ll(^pK~{\rm UIR}\rr)& \ll(s,y\rr)~{\rm ~of}~~^pK \\
\hline
B_0     & \ll(1,1\rr)           & \ll(1,1\rr)                   & \ll(0,0\rr)           \\
\hline
B_1     & \ll(\Box,\Box\rr)     & \ll(\Box,\Box\rr)             & \ll({1\over 2},\pm 1\rr)      \\
\hline
B_2     & \ll(\Box\!\Box,1\rr)  & \ll(\Box\!\Box,1\rr)          & \ll(1,0\rr)           \\
        & \ll(1,\Box\!\Box\rr)  & \ll(1,\Box\!\Box\rr)\oplus\ll(1,1\rr) & \ll(0,\pm 2\rr)\oplus\ll(0,0\rr)      \\
\hline
B_3     & \ll(\Box,\Box\rr)     & \ll(\Box,\Box\rr)             & \ll({1\over 2},\pm 1\rr)      \\
\hline
B_4     & \ll(1,1\rr)           & \ll(1,1\rr)                   & \ll(0,0\rr)           \\
\hline
\end{array}
\label{pKN2}
\ee
For example, the operator $\ll(\Box,\Box\rr)$ in $B_3$ is $~^3K_{\a}\equiv\ve^{\b\g}\bar{a}^{\pm}_{\a}
\bar{a}^+_{\b}\bar{a}^-_{\g}$. 
\par
We can derive the complete list of states tensorizing the representations (\ref{pKN2}) with
the quantum numbers of the possible vacua. In practice, the hypercharges adds up trivially, 
while the spins follow the usual rules for angular momentum composition. This means that
the number of states depends on the value of $s_0$, and the multiplets with spin not bigger than two have
\be
0\le s_0\le 1.
\ee
\par
Let us carry on the calculation of the norms. I do it only for the sectors
 $B_1$ and, partially, $B_2$: the other norm
calculations are too long, and we search the missing information on unitarity bounds
and shortening in other directions. We denote the vacuum, having norm $1$, by
\be
\vert\Omega\rangle\equiv\vert\ll(E_0,s_0,y_0\rr)~E_0,s_0,m,y_0\rangle.
\ee
\par
\vskip .5cm
\centerline{\large\fbox{$s=0$ case}}
\par
\begin{itemize}
\item $B_1$ sector\\
The operators $\bar{a}_1^{\pm}$ give
\be
\bar{a}_1^{\pm}\vert\Omega\rangle=R_{1/2}^{\pm}
\vert\ll(E_0+1/2,1/2,y_0\pm 1\rr)~E_0+1/2,1/2,1/2,y_0\pm 1\rangle.
\ee
We have 
\be
\langle\Omega\vert a_1^{\mp}\bar{a}_1^{\pm}\vert\Omega\rangle=E_0\mp y_0=\ll|R_{1/2}^{\pm}\rr|^2,
\ee
that yields the condition $E_0\mp y_0\ge 0$.\\
The unitarity condition arising from 
$B_1$ is then
\be
E_0\ge\ll|y_0\rr|
\label{E0gey0}
\ee
The condition arising from $\bar{a}_2^{\pm}$ is identical.
If it is strictly satisfied, the $B_1$ sector yields
\be
SD\ll(E_0+1/2,1/2,y_0+1\vert 2\rr)\oplus SD\ll(E_0+1/2,1/2,y_0-1\vert 2\rr).
\ee
When the (\ref{E0gey0}) is saturated, 
\begin{itemize}
\item if $y_0>0$\\
$R_{1/2}^+=0$ and the $SD\ll(E_0+1/2,1/2,y_0+1\vert 2\rr)$ field decouples; 
\item if $y_0<0$\\
$R_{1/2}^-=0$ and the $SD\ll(E_0+1/2,1/2,y_0-1\vert 2\rr)$ field decouples. 
\end{itemize}
We will see that the case $y_0=0$ is excluded.
\item $B_2$ sector\\
The operators $^2K$ are
\beq
\ve^{\a\b}\bar{a}_{\a}^+\bar{a}_{\b}^+=&\ll[\bar{a}_1^+,\bar{a}_2^+\rr]~~&{\rm with}~s=0,y=2\nn\\
\ve^{\a\b}\bar{a}_{\a}^-\bar{a}_{\b}^-=&\ll[\bar{a}_1^-,\bar{a}_2^-\rr]~~&{\rm with}~s=0,y=-2\nn\\
\ve^{\a\b}\bar{a}_{\a}^+\bar{a}_{\b}^-=&\ll[\bar{a}_1^+,\bar{a}_2^-\rr]~~&{\rm with}~s=0,y=0\nn\\
\bar{a}_{\ll(\a\rr.}^+\bar{a}_{\ll.\b\rr)}^-&&{\rm with}~s=1,y=0.
\eeq
The operator $\ll[\bar{a}_1^+,\bar{a}_2^+\rr]$ gives
\beq
\ll[\bar{a}_1^+,\bar{a}_2^+\rr]\vert\Omega\rangle&=&R_0^+\vert\ll(E_0+1,0,y_0+2\rr)E_0+1,s_0,m,y_0+2\rangle+\nn\\
&&+\beta M_3^+\vert\ll(E_0,0,y_0+2\rr)E_0,0,0,y_0+2\rangle.
\eeq
Let us determine $\beta$.
\beq
M^-_3\ll[\bar{a}_1^+,\bar{a}_2^+\rr]\vert\Omega\rangle&=&\ll[M^-_3\ll[\bar{a}_1^+,\bar{a}_2^+\rr]\rr]\vert\Omega\rangle=0=\nn\\
&=&-2E_0\b\vert\ll(E_0,0,y_0+2\rr)E_0,0,0,y_0+2\rangle,
\eeq
then $\b=0$. Now we can find $\ll|R_0^+\rr|^2$:
\be
\langle\Omega\vert\ll[a_2^-,a_1^-\rr]\ll[\bar{a}_1^+,\bar{a}_2^+\rr]\vert\Omega\rangle=
4\ll(E_0-y_0\rr)\ll(E_0-y_0-1\rr)=\ll|R_0^+\rr|^2.
\label{Rpge0}
\ee
The operator $\ll[\bar{a}_1^-,\bar{a}_2^-\rr]$ is the complex conjugate of $\ll[\bar{a}_1^+,\bar{a}_2^+\rr]$,
and as we have seen the conjugation of a $SO\ll(2\rr)\times U\ll(1\rr)$ changes the sign of the hypercharge.
Then,
\be
\ll[\bar{a}_1^-,\bar{a}_2^-\rr]\vert\Omega\rangle=R_0^-\vert\ll(E_0+1,0,y_0-2\rr)E_0+1,0,0,y_0-2\rangle
\ee
and
\be
\ll|R_0^-\rr|^2=4\ll(E_0+y_0\rr)\ll(E_0+y_0-1\rr).
\label{Rmge0}
\ee
This yields a unitarity condition stronger than the (\ref{E0gey0}).
It is:
\beq
E_0&\ge&\ll|y_0\rr|+1\nn\\
{\rm or}&&\nn\\
E_0&=&\ll|y_0\rr|\ge{1\over 2}.
\label{N2s0bound}
\eeq
In fact, if $E_0\neq\ll|y_0\rr|$ the (\ref{Rpge0}) and (\ref{Rmge0}) are both satisfied only if $E_0\ge\ll|y_0\rr|+1$,
while if $E_0=y_0>0$ the (\ref{Rpge0}) is zero, and the (\ref{Rmge0}) gives the bound $2y_0-1>0$; the
same thing happens if $E_0=-y_0>0$. Notice that when $\ll|y_0\rr|<E_0<\ll|y_0\rr|+1$ the unitarity bound
is not satisfied; the set of allowed $E_0,y_0$ values is not connected.
\par
When the (\ref{N2s0bound}) is strictly satisfied, the operators 
$\ll[\bar{a}_1^+,\bar{a}_2^+\rr]$, $\ll[\bar{a}_1^-,\bar{a}_2^-\rr]$ yield
\be
D\ll(E_0+1,0,y_0+2\rr)\oplus\ll(E_0+1,0,y_0-2\rr).
\ee
When $E_0=\ll|y_0\rr|+1$,
\begin{itemize}
\item if $y_0>0$\\
$R_0^+=0$ and the $D\ll(E_0+1,s_0,y_0+2\rr)$ field decouples;
\item if $y_0<0$\\
$R_0^-=0$ and the $D\ll(E_0+1,s_0,y_0-2\rr)$ field decouples;
\item if $y_0=0$\\
$R_0^+=R_0^-=0$ and the fields $D\ll(E_0+1,s_0,y_0+2\rr)$,
$D\ll(E_0+1,s_0,y_0-2\rr)$ decouple. 
\end{itemize}
When $E_0=\ll|y_0\rr|$,
\begin{itemize}
\item if $y_0>1/2$\\
$R_0^+=0$ and the $D\ll(E_0+1,s_0,y_0+2\rr)$ field decouples;
\item if $y_0<-1/2$\\
$R_0^-=0$ and the $D\ll(E_0+1,s_0,y_0-2\rr)$ field decouples; 
\item if $\ll|y_0\rr|=1/2$\\
$R_0^+=R_0^-=0$ and the fields $D\ll(E_0+1,s_0,y_0+2\rr)$,
$D\ll(E_0+1,s_0,y_0-2\rr)$ decouple. 
\end{itemize}
\end{itemize}
\par
I do not perform the calculation of the norms for the operators $\ve^{\a\b}\bar{a}_{\a}^+\bar{a}_{\b}^-$,
$\bar{a}_{\ll(\a\rr.}^+\bar{a}_{\ll.\b\rr)}^-$ of the $B_2$ sector and for the operators in the $B_3,B_4$ sectors.
\vskip .5cm
\centerline{\large\fbox{$s>0$ case}}
\par
\begin{itemize}
\item $B1$ sector\\
The operators $\bar{a}_1^{\pm}$ give
\[
\bar{a}_1^{\pm}\vert\Omega\rangle\!=\!R_{1/2}^{\pm}\sqrt{s_0\!+\!m\!+\!1
\over 2s_0\!+\!1}
\vert\ll(E_0\!+\!1/2,s_0\!+\!1/2,y_0\!\pm \!1\rr)\,E_0\!+\!1/2,s_0\!+\!1/2,
m\!+\!1/2,y_0\!\pm\!1\rangle\!+
\]
\[R_{-1/2}^{\pm}\sqrt{s_0\!-\!m\over 2s_0\!+\!1}
\vert\ll(E_0\!+\!1/2,s_0\!-\!1/2,y_0\!\pm \!1\rr)\,E_0\!+\!1/2,s_0\!-\!1/2,
m\!-\!1/2,y_0\!\pm\!1\rangle.\]
\be
~
\ee
To find the values of the constants $R_{1/2},R_{-1/2}$, we take
specific values of $m$ and determine the norms using the algebra (\ref{ospH}). \\
Taking $m=s$, 
\be
\langle\Omega\vert a_1^{\mp}\bar{a}_1^{\pm}\vert\Omega\rangle=E_0+s_0\mp y_0=\ll|R_{1/2}^{\pm}\rr|^2,
\ee
that yields the condition $E_0+s_0\mp y_0\ge 0$.\\
Taking $m=-1/2$,
\be
\langle\Omega\vert a_1^{\mp}\bar{a}_1^{\pm}\vert\Omega\rangle=E_0-{1/2}\mp y_0={1\over 2}
\ll(\ll|R_{1/2}\rr|^2+\ll|R_{-1/2}\rr|^2\rr),
\ee
that yields the condition $E_0-s_0-1\mp y_0\ge 0$. \\
The unitarity condition arising from 
$B_1$ is then
\be
E_0\ge s_0+\ll|y_0\rr|+1.
\label{N2bound}
\ee
The condition arising from $\bar{a}_2^{\pm}$ is identical.
This condition is coincident or stronger than the unitarity bound of $AdS_4$ fields for $s_0>1/2$.
If it is strictly satisfied, the $B_1$ sector yields
\beq
D\ll(E_0+1/2,s_0+1/2,y_0+1\rr)\oplus D\ll(E_0+1/2,s_0+1/2,y_0-1\rr)\oplus\nn\\
D\ll(E_0+1/2,s_0-1/2,y_0+1\rr)\oplus D\ll(E_0+1/2,s_0-1/2,y_0-1\rr).
\eeq
When the (\ref{N2bound}) is saturated, 
\begin{itemize}
\item if $y_0>0$\\
$R_{-1/2}^+=0$ and the $D\ll(E_0+1/2,s_0-1/2,y_0+1\rr)$ field decouples; 
\item if $y_0<0$\\
$R_{-1/2}^-=0$ and the $D\ll(E_0+1/2,s_0-1/2,y_0-1\rr)$ field decouples;
\item if $y_0=0$\\
$R_{-1/2}^+=R_{-1/2}^-=0$ and the fields $D\ll(E_0\!+\!1/2,s_0\!-\!1/2,y_0\!+\!1\rr)$,\\
$D\ll(E_0\!+\!1/2,s_0\!-\!1/2,y_0\!-\!1\rr)$ decouple;
furthermore, in this case the fields \\$D\ll(E_0,s_0,\pm\! 1\rr)$ and $D\ll(E_0\!+\!1/2,s_0\!+\!1/2,\pm\! 1\rr)$ 
are massless.
\end{itemize}
\item $B2$ sector\\
The operators $^2K$ are
\beq
\ve^{\a\b}\bar{a}_{\a}^+\bar{a}_{\b}^+=&\ll[\bar{a}_1^+,\bar{a}_2^+\rr]~~&{\rm with}~s=0,y=2\nn\\
\ve^{\a\b}\bar{a}_{\a}^-\bar{a}_{\b}^-=&\ll[\bar{a}_1^-,\bar{a}_2^-\rr]~~&{\rm with}~s=0,y=-2\nn\\
\ve^{\a\b}\bar{a}_{\a}^+\bar{a}_{\b}^-=&\ll[\bar{a}_1^+,\bar{a}_2^-\rr]~~&{\rm with}~s=0,y=0\nn\\
\bar{a}_{\ll(\a\rr.}^+\bar{a}_{\ll.\b\rr)}^-&&{\rm with}~s=1,y=0.
\eeq
The operator $\ll[\bar{a}_1^+,\bar{a}_2^+\rr]$ gives
\beq
\ll[\bar{a}_1^+,\bar{a}_2^+\rr]\vert\Omega\rangle&=&R_0^+\vert\ll(E_0+1,s_0,y_0+2\rr)E_0+1,s_0,m,y_0+2\rangle+\nn\\
&&+\hbox{boosted elements of}~D\ll(E_0,s_0,y_0+2\rr).
\eeq
In order to calculate $\ll|R_0^+\rr|^2$ we take $m=s_0$. Then
\beq
\ll[\bar{a}_1^+,\bar{a}_2^+\rr]\vert\Omega\rangle&=&R_0^+\vert\ll(E_0+1,s_0,y_0+2\rr)E_0+1,s_0,s_0,y_0+2\rangle+\nn\\
&&+\a^+ M_{\ll(+\rr)}^+\vert\ll(E_0,s_0,y_0+2\rr)E_0,s_0,s_0-1,y_0+2\rangle+\nn\\
&&+\b^+ M_3^+\vert\ll(E_0,s_0,y_0+2\rr)E_0,s_0,s_0,y_0+2\rangle\,.
\eeq
Let us determine $\a^+$ and $\b^+$. By means of the algebra (\ref{ospH}) we find
\be
\begin{array}{l}
M^-_{\ll(-\rr)}\ll[\bar{a}_1^+,\bar{a}_2^+\rr]\vert\Omega\rangle=
\ll[M^-_{\ll(-\rr)}\ll[\bar{a}_1^+,\bar{a}_2^+\rr]\rr]\vert\Omega\rangle=0=\\
=\ll(-2\ll(E_0+s_0-1\rr)\a^++2\sqrt{s_0}\b^+\rr)\vert\ll(E_0,s_0,y_0+2\rr)E_0,s_0,s_0-1,y_0+2\rangle\\
\end{array}
\ee
\be
\begin{array}{l}
M^-_3\ll[\bar{a}_1^+,\bar{a}_2^+\rr]\vert\Omega\rangle=
\ll[M^-_3\ll[\bar{a}_1^+,\bar{a}_2^+\rr]\rr]\vert\Omega\rangle=0=\\
=\ll(2\sqrt{s_0}\a^+-2E_0\b^+\rr)\vert\ll(E_0,s_0,y_0+2\rr)E_0,s_0,s_0,y_0+2\rangle,\\
\end{array}
\ee
then
\be
\a^+=\b^+=0.
\ee
Now we can find $\ll|R_0^+\rr|^2$:
\beq
\ll|R_0^+\rr|^2&=&\langle\Omega\vert\ll[a_2^-,a_1^-\rr]\ll[\bar{a}_1^+,\bar{a}_2^+\rr]\vert\Omega\rangle=\ll(\ll(E_0\!+\!s_0\!-\!1\!-\!y_0\rr)\ll(E_0\!-\!s_0
\!-\!y_0\rr)\!-\!2s_0\rr)\!+\nn\\
&&-\!\ll(\!-\!\ll(E_0\!-\!s_0\!-\!y_0\rr)
\ll(E_0\!+\!s_0\!-\!y_0\!-\!1\rr)\!+\!2s_0\rr)\!+\!\nn\\
&\!-\!&\ll(\!-\!\ll(E_0\!+\!s_0\!-\!y_0\rr)
\ll(E_0\!-\!s_0\!-\!y_0\!-\!1\rr)\rr)\!+\!\ll(\ll(E_0\!+\!s_0\!-\!y_0\rr)
\ll(E_0\!-\!s_0\!-\!y_0\!-\!1\rr)\rr)=\nn\\
&=&4\ll(E_0\!-\!y_0\!+\!s_0\rr)\ll(E_0\!-\!y_0\!-\!s_0\!-\!1\rr)\,.
\eeq
\par
The operator $\ll[\bar{a}_1^-,\bar{a}_2^-\rr]$ is the complex conjugate of $\ll[\bar{a}_1^+,\bar{a}_2^+\rr]$,
and as we have seen the conjugation of a $U\ll(1\rr)$ representation changes the sign of the hypercharge.
Then,
\be
\ll[\bar{a}_1^-,\bar{a}_2^-\rr]\vert\Omega\rangle=R_0^-\vert\ll(E_0+1,s_0,y_0-2\rr)E_0+1,s_0,m,y_0-2\rangle
\ee
and
\be
\ll|R_0^-\rr|^2=4\ll(E_0+y_0+s_0\rr)\ll(E_0+y_0-s_0-1\rr).
\ee
The condition (\ref{N2bound}) 
\be
E_0\ge s_0+\ll|y_0\rr|+1
\ee
guarantees $\ll|R_0^+\rr|^2\ge 0$ and  $\ll|R_0^-\rr|^2\ge 0$. 
If it is strictly satisfied, the operators $\ll[\bar{a}_1^+,\bar{a}_2^+\rr]$, $\ll[\bar{a}_1^-,\bar{a}_2^-\rr]$ yield
\be
D\ll(E_0+1,s_0,y_0+2\rr)\oplus\ll(E_0+1,s_0,y_0-2\rr).
\ee
When it is saturated, 
\begin{itemize}
\item if $y_0>0$\\
$R_0^+=0$ and the $D\ll(E_0+1,s_0,y_0+2\rr)$ field decouples;
\item if $y_0<0$\\
$R_0^-=0$ and the $D\ll(E_0+1,s_0,y_0-2\rr)$ field decouples;
\item if $y_0=0$\\
$R_0^+=R_0^-=0$ and the fields $D\ll(E_0+1,s_0,y_0+2\rr)$,
$D\ll(E_0+1,s_0,y_0-2\rr)$ decouple. 
\end{itemize}
\end{itemize}
We do not perform the calculation of the norms for the operators $\ve^{\a\b}\bar{a}_{\a}^+\bar{a}_{\b}^-$,
$\bar{a}_{\ll(\a\rr.}^+\bar{a}_{\ll.\b\rr)}^-$ in the $B_2$ sector and for the operators in the $B_3,B_4$ sectors.
\par
At this point we do not know if there are other unitarity bounds and shortening conditions in addition to
the (\ref{N2s0bound}) for $s_0=0$ and (\ref{N2bound}) for $s_0>0$. Furthermore, we do not know which
other fields decouple in the shortened representation just found, namely, the complete structure of the
short $\cN=2$ multiplets. We know from the literature on supergravity the complete structure of the massless
supermultiplets, but not of the massive ones.
A possible way to get this information is by deriving the norms of the remainings
states, but it would be a very lengthy calculation. I prefer to utilize the tricks listed in page \ref{otherways}.
\par
First of all, we can use the results of harmonic analysis on the $M^{111}$ manifold (and, in part, $Q^{111}$), 
described in next chapter, which gives the complete mass spectrum on the corresponding
Kaluza Klein supergravity solutions (which have, in both cases, $\cN=2$ supersymmetry).
We have found the masses, energies and hypercharges (and flavour quantum numbers, which however 
are not relevant in the present context) of almost all the particles of these supergravity; this is
enough to organize them in supermultiplets, and by means of the part just derived on supermultiplet
structure and of the decomposition under $\cN=1$ supermultiplets we can complete the spectrum;
as a byproduct, we find the complete structure of the multiplets appearing in these supergravities,
related with their energies and hypercharges. This confirms that the only unitarity bounds in
$\cN=2$ supersymmetry are 
\beq
E_0\ge s_0+\ll|y_0\rr|+1&&s_0\ge 0\nn\\
{\rm or}&&\nn\\
E_0=\ll|y_0\rr|\ge {1\over 2}&&s_0=0,
\label{N2bounds}
\eeq
and the only shortening conditions are the ones we found, corresponding to the saturations of these bounds. 
This procedure is explained in the next chapter, here I report the results. 
\par
I give the list of the $Osp\ll(2\vert 4\rr)$ UIRs with maximal spin not greater than two.
Their explicit structures are showed in tables \ref{N2longgraviton},$\dots$,\ref{N2supersingleton}. 
I remind that since we are 
considering the $R$--symmetry $SO\ll(2\rr)$ in the complex form $U\ll(1\rr)$, when the hypercharge is 
different from zero, the supermultiplet is complex; there are then two supermultiplets, conjugate each
other with opposite values of $y_0$; in this case, I display in the tables the one with $y_0>0$.
When, on the contrary, $y_0=0$, the supermultiplet is real, and then there is only one of them.
\par
\begin{enumerate}
\item $E_0>s_0+\ll|y_0\rr|+1,~0<s_0<1$: {\it long multiplets}; long graviton multiplet ($s_0=1$),
long gravitino multiplet ($s_0=1/2$), long vector multiplet ($s_0=0$). They have the structures displayed in 
tables \ref{N2longgraviton},\ref{N2longgravitino},\ref{N2longvector}. The fields of these multiplets are all massive. 
Their decomposition under $\cN=2\longrightarrow\cN=1$ is, if $y_0\neq 0$,
\beq
SD\ll(E_0,1,y_0\vert 2\rr)&\longrightarrow&
SD\ll(E_0+1/2,3/2\vert 1\rr)\oplus SD\ll(E_0+1,1\vert 1\rr)\oplus\nn\\
&&SD\ll(E_0+1/2,1/2\vert 1\rr)\oplus \oplus SD\ll(E_0,1\vert 1\rr)\oplus \nn\\
&&\oplus SD\ll(E_0+1/2,3/2\vert 1\rr)\oplus SD\ll(E_0+1,1\vert 1\rr)\oplus \nn\\
&&SD\ll(E_0+1/2,1/2\vert 1\rr) \oplus SD\ll(E_0,1\vert 1\rr)\label{N21longgraviton}\\
&&\nn\\
SD\ll(E_0,1/2,y_0\vert 2\rr)&\longrightarrow&
SD\ll(E_0+1/2,1\vert 1\rr)\oplus SD\ll(E_0+1,1/2\vert 1\rr)\oplus \nn\\
&&SD\ll(E_0,1/2\vert 1\rr) \oplus SD\ll(E_0+1/2,0\vert 1\rr)\oplus \nn\\
&&\oplus SD\ll(E_0+1/2,1\vert 1\rr)\oplus SD\ll(E_0+1,1/2\vert 1\rr)\oplus \nn\\
&&SD\ll(E_0,1/2\vert 1\rr)\oplus \oplus SD\ll(E_0+1/2,0\vert 1\rr)\label{N21longgravitino}\\
&&\nn\\
SD\ll(E_0,0,y_0\vert 2\rr)&\longrightarrow&
SD\ll(E_0+1/2,1/2\vert 1\rr)\oplus SD\ll(E_0+1,0\vert 1\rr)\oplus \nn\\
&&SD\ll(E_0,0\vert 1\rr)\oplus \nn\\
&&\oplus SD\ll(E_0+1/2,1/2\vert 1\rr)\oplus SD\ll(E_0+1,0\vert 1\rr)\oplus \nn\\
&&SD\ll(E_0,0\vert 1\rr)\label{N21longvector}
\eeq
while if $y_0=0$ it is
\beq
SD\ll(E_0,1,0 \vert 2\rr)&\longrightarrow&
SD\ll(E_0+1/2,3/2\vert 1\rr)\oplus SD\ll(E_0+1,1\vert 1\rr)\oplus\nn\\
&&SD\ll(E_0+1/2,1/2\vert 1\rr)\oplus \oplus SD\ll(E_0,1\vert 1\rr)\label{N21longgravitonb}\\
&&\nn\\
SD\ll(E_0,1/2,0 \vert 2\rr)&\longrightarrow& 
SD\ll(E_0+1/2,1\vert 1\rr)\oplus SD\ll(E_0+1,1/2\vert 1\rr)\oplus \nn\\
&&SD\ll(E_0,1/2\vert 1\rr) \oplus SD\ll(E_0+1/2,0\vert 1\rr)\label{N21longgravitinob}\\
&&\nn\\
SD\ll(E_0,0,0 \vert 2\rr)&\longrightarrow&
SD\ll(E_0+1/2,1/2\vert 1\rr)\oplus SD\ll(E_0+1,0\vert 1\rr)\oplus \nn\\
&&SD\ll(E_0,0\vert 1\rr)\label{N21longvectorb}.
\eeq
\item $E_0=s_0+\ll|y_0\rr|+1,~\ll|y_0\rr|>0,~0<s_0<1$: {\it short multiplets}; short  graviton multiplet ($s_0=1$),
short  gravitino multiplet ($s_0=1/2$), short  vector multiplet ($s_0=0$). They have the structures displayed in 
tables \ref{N2shortgraviton},\ref{N2shortgravitino},\ref{N2shortvector}.
The fields of these multiplets are all massive.
Their decomposition under $\cN=2\longrightarrow\cN=1$ is
\beq
SD\ll(|y_0|+2,1,y_0\vert 2\rr)&\longrightarrow&
SD\ll(|y_0|+5/2,3/2\vert 1\rr)\oplus SD\ll(|y_0|+2,1\vert 1\rr)\oplus\nn\\
&&\oplus SD\ll(|y_0|+5/2,3/2\vert 1\rr)\oplus SD\ll(|y_0|+2,1\vert 1\rr)\nn\\
&&\label{N21shortgraviton}\\
SD\ll(|y_0|+3/2,1/2,y_0\vert 2\rr)&\longrightarrow&
SD\ll(|y_0|+2,1\vert 1\rr)\oplus SD\ll(|y_0|+3/2,1/2\vert 1\rr)\oplus \nn\\
&&\oplus SD\ll(|y_0|+2,1\vert 1\rr)\oplus SD\ll(|y_0|+3/2,1/2\vert 1\rr)\nn\\
&&\label{N21shortgravitino}\\
SD\ll(|y_0|+1,0,y_0\vert 2\rr)&\longrightarrow&
SD\ll(|y_0|+3/2,1/2\vert 1\rr)\oplus SD\ll(|y_0|+1,0\vert 1\rr)\oplus \nn\\
&&\oplus SD\ll(|y_0|+3/2,1/2\vert 1\rr)\oplus SD\ll(|y_0|+1,0\vert 1\rr)\,.\nn\\
&&\label{N21shortvector}
\eeq
\item $E_0=s_0+1,~y_0=0,~0<s_0<1$: {\it massless multiplets}; massless graviton multiplet ($s_0=1$),
massless gravitino multiplet ($s_0=1/2$), massless vector multiplet ($s_0=0$). They have the structures displayed in 
tables \ref{N2masslessgraviton},\ref{N2masslessgravitino},\ref{N2masslessvector}. 
The fields of these multiplets are all massless.
Their decomposition under $\cN=2\longrightarrow\cN=1$ is
\beq
SD\ll(2,1,0\vert 2\rr)&\longrightarrow& SD\ll(5/2,3/2\vert 1\rr) \oplus SD\ll(2,1\vert 1\rr)\label{N21masslessgraviton}\\
SD\ll(3/2,1/2,0\vert 2\rr)&\longrightarrow& SD\ll(2,1\vert 1\rr) \oplus SD\ll(3/2,1/2\vert 1\rr)\label{N21masslessgravitino}\\
SD\ll(1,0,0\vert 2\rr)&\longrightarrow& SD\ll(3/2,1/2\vert 1\rr) \oplus SD\ll(1,0\vert 1\rr).\label{N21masslessvector}
\eeq
\item $E_0=\ll|y_0\rr|>1/2,~s_0=0$: {\it hypermultiplet}. It has the structure displayed in table \ref{N2hyper}.
For $E_0\neq 1,2$ the fields of this multiplet are all massive. For $E_0=2$, some of them are massless, 
some other massive, and for $E_0=1$ all the fields of the multiplet are massless. 
However, this multiplet is always complex, because $y_0\neq 0$.
Its decomposition under $\cN=2\longrightarrow \cN=1$ is
\be
SD\ll(|y_0|,0,y_0\vert 2\rr)\longrightarrow SD\ll(|y_0|,0\vert 1\rr)\oplus SD\ll(|y_0|,0\vert 1\rr).\label{N21hyper}
\ee
Notice that this multiplet arises from a particular shortening (the one with $E_0=\ll|y_0\rr|$, different from
the one with $E_0=\ll|y_0\rr|+1$) of the vector multiplet, under which also the maximal spin state,
the vector, decouple. So the multiplet does not contain spins greater than $1/2$; this phenomenom is not
possible in long multiplets with $\cN=2$ supersymmetry, due to the existence of an operator in
the enveloping algebra having spin one (see (\ref{pKN2}) ). For historical reasons, multiplets with 
spin not greater than $1/2$ are called hypermultiplets.
\item $E_0=\ll|y_0\rr|=1/2$: {\it supersingleton representation}. It has the structure displayed in
table \ref{N2supersingleton}. The $SO\ll(3,2\rr)$ UIRs of this $Osp\ll(2\vert 4\rr)$ UIR are all singletons;
then, this representation is not a multiplet of supergravity fields, it does not have a realization
as fields on the bulk, but only on the boundary.
Its decomposition under $\cN=2\longrightarrow \cN=1$ is
\be
SD\ll(1/2,0,1/2\vert 2\rr)\longrightarrow SD\ll(1/2,0\vert 1\rr)\oplus SD\ll(1/2,0\vert 1\rr).\label{N21supersingleton}
\ee
\end{enumerate}
\par
To be precise, in principle there could be other unitarity bounds or shortening phenomena arising from
the norms not evaluated, but this seems very unlikely because the Kaluza Klein analysis of two spectra 
does not show anything of that. 
\par
\subsection{The structure of $\cN=3$ supermultiplets}
\par
This case has been first studied in \cite{frenico}, where some norms have been worked out giving
the unitarity bounds and some shortening conditions, and the structure of the short vector multiplet
has been worked out. Then, in \cite{noi4}, the list of the $~^pK$ operators and the 
$\cN=3\longrightarrow\cN=2$ decompositions have been derived, 
relying on the results of \cite{piet}, and the complete spectrum of a particular $\cN=3$ supergravity 
compactifiation has been worked out (see chapter $3$); as a byproduct, the lacking information
on $\cN=3$ UIRs has been found, namely, the absence of further unitarity bounds, the remaining
shortening conditions, and the structure of all the multiplets (with the exception of the ones with 
$J_0=3/2,1/2$, not appearing in the spectrum of our compactification).
\par
The maximum number of fermionic generators allowed is $p=2\cN=8$.
The $R$--symmetry group is $SO\ll(3\rr)$, locally isomorphic to $SU\ll(2\rr)$,
which is the form we consider. An $R$--symmetry UIR is  labeled by
its $SU\ll(2\rr)_R$ spin, which we call {\it isospin} $J$, and its states are
labeled by the third isospin component $M\in\ll[-J,J\rr]$.
In the fermionic generators $\bar{a}_{\a}^i$ the index $i=1,\dots,3$ runs in the vector 
representation of $SO\ll(3\rr)$; the well--suited fermionic generators 
for the $SU\ll(2\rr)$ form of the $R$--symmetry are
\beq
a^{\pm}_{\a}&=&{1\over\sqrt{2}}\ll(a^1_{\a}\pm a^2_{\a}\rr)\nn\\
a^3_{\a}&&
\eeq
satisfying
\beq 
\ll(a_{\a}^{\pm}\rr)^+&=&\bar{a}_{\a}^{\mp}\nn\\
\ll[H,\bar{a}_{\a}^{\pm,3}\rr]&=&{1\over 2}\bar{a}_{\a}^{\pm,3}\nn\\
\ll[M,\bar{a}_{\a}^{\pm}\rr]&=&\pm\bar{a}_{\a}^{\pm}\nn\\
\ll[M,\bar{a}_{\a}^3\rr]&=&0.
\eeq
\par
Let us find all the operators $^pK$. 
As usual, we denote the operators by the representations of their indices; write all the 
representations allowed, and determine their $SO\ll(2\rr)\times SU\ll(2\rr)$ labels. I remind
that for $SU\ll(3\rr)$ representations $\stackrel{\textstyle \Box}{\Box}~\simeq\Box$. Furthermore
I remind that, under the isomorphism $SO\ll(3\rr)\simeq SU\ll(2\rr)$, the simplest UIRs transforms as
\be
\begin{array}{ccc}
SO\ll(3\rr)     & SU\ll(2\rr)           &       \\
1       & 1             & J=0   \\
\Box    & \Box\!\Box            & J=1   \\
\Box\!\Box      & \Box\!\Box\!\Box\!\Box        & J=2  \,. \\
\end{array}
\ee
For example, $\bar{a}_{\a}^i$ are in the $\bf{3}$ of $SO\ll(3\rr)\simeq SU\ll(2\rr)$, that is the $\Box$ of $SO\ll(3\rr)$
and the $\Box\!\Box$ of $SU\ll(2\rr)$.
\par
The complete list of $^pK$ operators is
\be
\begin{array}{|c|c|c|c|}
\hline
        & SU\ll(2\rr)\times SU\ll(3\rr) & ~~~SU\ll(2\rr)\times SU\ll(2\rr)
~\ll(^pK~{\rm UIR}\rr)& \ll(s,J\rr)~{\rm ~of}~~^pK \\
\hline
B_0     & \ll(1,1\rr)           & \ll(1,1\rr)                   & \ll(0,0\rr)           \\
\hline
B_1     & \ll(\Box,\Box\rr)     & \ll(\Box,\Box\!\Box\rr)               & \ll({1\over 2},1\rr)  \\
\hline
B_2     & \ll(\Box\!\Box,\Box\rr)       & \ll(\Box\!\Box,\Box\!\Box\rr) & \ll(1,1\rr)           \\
        & \ll(1,\Box\!\Box\rr)  & \ll(1,\Box\!\Box\!\Box\!\Box\rr)\oplus\ll(1,1\rr)     
                                                & \ll(0,2\rr)\oplus\ll(0,0\rr)  \\
\hline
&&&\\
B_3     & \ll(\Box,\stackrel{\textstyle \Box\!\Box}{\Box~~}\rr)& \ll(\Box,\Box\!\Box\!\Box\!\Box\rr)\oplus
                                     \ll(\Box,\Box\!\Box\rr)& \ll({1\over 2},2\rr)\oplus\ll({1\over 2},1\rr)\\
        & \ll(\Box\!\Box\!\Box,1\rr)    & \ll(\Box\!\Box\!\Box,1\rr)            & \ll({3\over 2},0\rr)  \\
\hline
B_4     & \ll(1,\Box\!\Box\rr)  & \ll(1,\Box\!\Box\!\Box\!\Box\rr)\oplus\ll(1,1\rr)& \ll(0,2\rr)\oplus\ll(0,0\rr)\\
        & \ll(\Box\!\Box,\Box\rr)       & \ll(\Box\!\Box,\Box\!\Box\rr) & \ll(1,1\rr)           \\      
\hline
B_5     & \ll(\Box,\Box\rr)     & \ll(\Box,\Box\!\Box\rr)               & \ll({1\over 2},1\rr)  \\
\hline
B_6     & \ll(1,1\rr)           & \ll(1,1\rr)                   & \ll(0,0\rr)           \\
\hline
\end{array}
\label{pKN3}
\ee
\par
We can derive the complete list of states tensorizing the representations (\ref{pKN3}) with
the quantum numbers of the possible vacua. Both the spins and the isospins
follow the usual rules of angular momentum composition. This means that
the number of states depends on the value of $s_0$, and the multiplets with spin not bigger than two have
\be
0\le s_0\le {1\over 2}.
\ee
\par
The norms of the states created by the sector $B_1$ have been derived in \cite{frenico}. I give here their results,
without repeating their proof. The highest weight operator $\bar{a}_1^+$ yields
\beq
\bar{a}_1^+\vert\ll(E_0,s_0,J_0\rr)E_0,s_0,m,J_0,M\rangle
=\sum_{\mu\nu}R_{\mu\nu}\langle s_0+\mu,m+{1\over 2}\vert s_0,m,{1\over 2},{1\over 2}\rangle\cdot\nn\\
\cdot\langle J_0\!+\!\nu,M\!+\!1\vert J_0,M,1,1\rangle\ll|\ll(E_0\!+\!
{1\over 2},s_0\!+\!\mu,J_0\!+\!\nu\rr)
E_0\!+\!{1\over 2},s_0\!+\!\mu,m\!+\!{1\over 2},M\!+1\!\rr>.\nn
\eeq
\be
~\label{Rmunudef}
\ee
By giving appropriate values to $m,M$, one finds the expression of the $\ll|R_{\mu\nu}\rr|$:
\beq
\ll|R_{{1\over 2},1}\rr|^2&=& E_0+s_0-J_0               ,\nn\\
\ll|R_{-{1\over 2},1}\rr|^2&=&   E_0-s_0-J_0-1  ,\nn\\
\ll|R_{{1\over 2},0}\rr|^2&=&   E_0+s_0+1               ,\nn\\
\ll|R_{-{1\over 2},0}\rr|^2&=&   E_0-s_0                ,\nn\\
\ll|R_{{1\over 2},-1}\rr|^2&=&   E_0+s_0+J_0+1  ,\nn\\
\ll|R_{-{1\over 2},-1}\rr|^2&=& E_0-s_0+J_0.
\eeq
I remind that when $s_0=0$ the Clebsch Gordan coefficients multiplying $R_{-{1\over 2},\nu}$ in
the expansion (\ref{Rmunudef}) vanish.
\par
These norms yield the following unitarity bounds:
\beq
E_0\ge s_0+J_0+1&&s_0>0\nn\\
E_0\ge J_0&&s_0=0.
\eeq
From the norm evaluation of some operators in other sectors done in \cite{frenico} another unitarity 
bound arises: when $s_0=0$, we can have $E_0=J_0$ or $E_0>J_0+1$, but not $J_0<E_0<J_0+1$;
as in the $\cN=2$ case, there is a ''disconnected'' unitarity condition. Furthermore, as in the $\cN=2$ case,
in the corresponding short multiplet also the maximal spin field decouple, yielding a vector multiplet
instead of a gravitino multiplet. The structure of this supermultiplet has been completely
determined in \cite{frenico}.
\par
Using the results of the harmonic analysis on the $N^{010}$ manifold given in \cite{piet}, we have found in
\cite{noi4} the complete spectrum (as described in the next chapter) of the corresponding Kaluza Klein
solution, which has $\cN=3$ supersymmetry. We have found the masses, energies and isospins of almost
all the particles of this supergravity; this is enough to organize them in supermultiplets, and by means of
the results of \cite{frenico} and of the decomposition under $\cN=2$ supermultiplets we can complete the 
spectrum; as a byproduct, we found the complete structure of all the multiplets appearing in this supergravity,
related with their energies and isospins. This confirms that the only unitarity bounds in $\cN=3$ 
supersymmetry are 
\beq
E_0&\ge s_0+J_0+1&~~s_0\ge 0\nn\\
{\rm or}&&\nn\\
E_0&= J_0&~~s_0=0.
\label{N3bounds}
\eeq
We have short representations when 
\begin{equation} 
E_0=J_0+s_0+1~~{\rm or}~E_0=J_0,~s_0=0. 
\end{equation} 
I stress that there is another  shortening mechanism of 
a completely different origin. The creation operators that act on the 
vacuum have isospin $0\le J_0\le 2$. If the isospin of the vacuum 
is $J_0\ge 2$, the creation operators give rise to states with isospin 
in the range 
$J_0-J\le J_{\rm composite}\le J_0+J$. Yet, in the case where $0\le J_0<2$, 
some of these states cannot appear. This mechanism is not related 
to an unitarity bound, and these representations are not $BPS$ states of 
supergravity, nor primary conformal operators on the boundary. Then we 
call {\it long} the representations with $E_0>s_0+J_0+1$, even if $0\le J_0<2$. 
In this context, the massless representations are the short ones with $J_0=0$ 
for the case of the massless graviton and gravitino multiplets and with $J_0=1$ for 
the case of the massless vector multiplets; the supersingleton representation
is the short vector multiplet with $J_0=1/2$, $SD\ll(1/2,0,1/2\vert 3\rr)$.
Unfortunately, only states with integer isospins appear in the $N^{010}$ spectrum,
then we do not have enough information to know the complete structure of
multiplets with $J_0=3/2$ and $J_0=1/2$, with the exception of the supersingleton 
which we worked out a part. However, the $\cN=2$ decomposition we found is
true for all the values of $J_0$.
\par
The complete list of the $Osp\left(3\vert 4\right)$ UIRs with $s_{max}\le 2$ 
is given below: 
\begin{itemize} 
\item {\em long graviton multiplet} $SD\left(E_0,1/2,J_0\vert 3\right)$ where $E_0>J_0+{3/2}$, see 
table \ref{N3longgraviton}; 
\item {\em long gravitino multiplet} $SD\left(E_0,0,J_0\vert 3\right)$ where $E_0>J_0+1$, see 
table \ref{N3longgravitino}; 
\item {\em short graviton multiplet} $SD\left(J_0+{3/ 2},1/2,J_0\vert 3\right)$, see 
table \ref{N3shortgraviton}; 
\item {\em short gravitino multiplet} $SD\left(J_0+1,0,J_0\vert 3\right)$, see 
table \ref{N3shortgravitino}; 
\item {\em short vector multiplet} $SD\left(J_0,0,J_0\vert 3\right)$, $J_0\ge 1$, see 
table \ref{N3shortvector};
\item {\em supersingleton representation} $SD\ll(1/2,0,1/2\vert 3\right)$, see table
\ref{N3shortvector}.
\end{itemize} 
Note that there are no long vector multiplets, and no hypermultiplets at all. 
\par
The ${\cal N}=3\longrightarrow{\cal N}=2$ decompositions of 
the above multiplets are listed below. I remind that an $Osp\ll(2\vert 4\rr)$ UIR is
denoted by $SD\ll(E_0,s_0,y_0\vert 2\rr)$, and if $y_0\neq 0$ this is a complex representation,
the conjugate one having opposite hypercharge. So, in the following list 
when there is a 
complex representation we write $SD\ll(E_0,s_0,y_0\vert 2\rr)\oplus SD\ll(E_0,s_0,-y_0\vert 2\rr)$.
Then, for example, the $\cN=3$ supersingleton representation with this convention decomposes
$SD\ll(1/2,0,1/2\vert 3\rr)\longrightarrow SD\ll(1/2,0,1/2\vert 2\rr)\oplus SD\ll(1/2,0,-1/2\vert 2\rr)$,
but actually it coincide with the $\cN=2$ supersingleton representation.
\begin{eqnarray} 
SD\left(E_0,1/2,J_0\vert 3\right)&\!\longrightarrow&\!\bigoplus\limits_{y=-J_0}^{J_0}SD 
\left(E_0+1/2,1,y\vert 2\right) 
\oplus\bigoplus\limits_{y=-J_0}^{J_0}SD\left(E_0,1/2,y\vert 2\right) 
\nonumber\\ 
&&\!\oplus\!\bigoplus\limits_{y=-J_0}^{J_0}\!SD\left(E_0+\!1\!,1/2,y\vert 2
\right)\!\oplus \!\bigoplus\limits_{y=-J_0}^{J_0}\!SD\left(E_0\!+\!1/2,0,y
\vert 2\right) 
\nonumber\\ 
&&{\rm where~}E_0>J_0+3/2
\nonumber\\
SD\left(J_0+3/2,1/2,J_0\vert 3\right)&\!\longrightarrow&\bigoplus 
\limits_{y=-J_0}^{J_0}SD\left(J_0+2,1,y\vert 2\right) 
\oplus\bigoplus\limits_{y=-J_0}^{J_0}SD\left(J_0+3/2,1/2,y\vert 2\right) 
\nonumber\\
SD\left(E_0,0,J_0\vert 3\right)&\!\longrightarrow&\bigoplus 
\limits_{y=-J_0}^{J_0}SD\left(E_0+1/2,1/2,y\vert 2\right) 
\oplus\bigoplus\limits_{y=-J_0}^{J_0}SD\left(E_0+1,0,y\vert 2\right)
\nonumber\\ 
&&\oplus\bigoplus\limits_{y=-J_0}^{J_0}SD\left(E_0,0,y\vert 2\right)~~~~~ 
{\rm where~}E_0>J_0+1
\nonumber\\
SD\left(J_0+1,0,J_0\vert 3\right)&\!\longrightarrow&\bigoplus 
\limits_{y=-J_0}^{J_0}SD\left(J_0+3/2,1/2,y\vert 2\right) 
\oplus\bigoplus\limits_{y=-J_0}^{J_0}SD\left(J_0+1,0,y\vert 2\right) 
\nonumber\\
SD\left(J_0,0,J_0\vert 3\right)&\!\longrightarrow&\bigoplus 
\limits_{y=-J_0}^{J_0}SD\left(J_0,0,y\vert 2\right)\,.
\label{decomposition}
\end{eqnarray} 
Notice that while $Osp\ll(3\vert 4\rr)\supset Osp\ll(2\vert 4\rr)$,  
$Osp\ll(3\vert 4\rr)\supset\!\!\!\!\!\!\!/~~Osp\ll(1\vert 4\rr)\times SO\ll(3\rr)$. It
is then impossible in general to decompose the $\cN=3$ UIRs in 
$\cN=1$ UIRs with definite isospin.
\par
\section{The $AdS_4$ and $\partial AdS_4$ superspaces}
\par
\label{superspaces}
As I said, the anti--de Sitter superspace is the following supercoset:
\begin{equation}
  AdS_{4\vert{\cal N} } \equiv \frac{Osp({\cal N} \vert 4)}{SO(1,3) \times
SO({\cal N})}
\label{ads4N}
\end{equation}
and has $4$ bosonic coordinates labelling the points in $AdS_4$ and $4
\times {\cal N}$ fermionic coordinates $\Theta^{\alpha i}$ that
transform as Majorana spinors under $SO(1,3)$ and as vectors under
$SO({\cal N})$. There are many possible coordinate choices for
parametrizing such a manifold, but as far as the bosonic submanifold is
concerned it was shown in \cite{g/hpape} that a particularly useful
parametrization is the solvable one where the $AdS_4$ coset is
regarded as a {\it non--compact solvable group manifold}:
\begin{equation}
  AdS_4 \equiv \frac{SO(2,3)}{SO(1,3)} = \exp \left [ Solv_{adS}
  \right]\,.
\label{solvads}
\end{equation}
The solvable algebra $Solv_{adS}$  is spanned by the unique
non--compact Cartan generator $D$ belonging to the coset and
by three abelian operators $P_m$ ($m=0,1,2$) generating the
translation subalgebra in $d=1+2$ dimensions. The solvable
coordinates are
\begin{equation}
   \begin{array}{rclcrcl}
     \rho & \leftrightarrow  & D & ; & z^m & \leftrightarrow & P_m \
   \end{array}
\label{solvcord}
\end{equation}
and in such coordinates the $AdS_4$ metric takes the form
\footnote{which is the form appearing in (\ref{nearhorU11}), where $\rho$ is called $U$}
\be
\rho^2\ll(-dz_0^2+dz_1^2+dz_2^2\rr)+{1\over\rho^2}d\rho^2.
\label{adsmet}
\ee
Hence $\rho$ is interpreted as measuring the distance
from the brane--stack and $z^m$ are interpreted as cartesian
coordinates on the brane boundary $\partial (AdS_4)$. A possible
question is: can such a solvable parametrization of $AdS_4$ be extended 
to a supersolvable parametrization of anti--de Sitter superspace as defined in
(\ref{ads4N})? In practice that means to single out a solvable
superalgebra with $4$ bosonic and $4 \times {\cal N}$ fermionic generators.
As shown in \cite{torinos7}, this turns out to be impossible, yet there is a supersolvable
algebra $Ssolv_{adS} $ with $4$ bosonic and $2 \times {\cal N}$ fermionic
generators
whose exponential defines the {\it solvable anti--de Sitter
superspace}:
\begin{equation}
  AdS^{(Solv)}_{4\vert 2{\cal N}}\equiv \exp\left[ Ssolv_{adS}\right]\,.
\label{solvsup}
\end{equation}
The supermanifold (\ref{solvsup}) is also a supercoset of the same
supergroup $Osp({\cal N} \vert 4)$ but with respect to a different
subgroup:
\begin{equation}
AdS^{(Solv)}_{4\vert 2{\cal N}} = \frac{Osp(4\vert {\cal N})}
{CSO(1,2\vert{\cal N})}
\label{supcos2}
\end{equation}
where $CSO(1,2\vert {\cal N})\subset Osp({\cal N}\vert 4)$ is generated by 
an algebra containing $3 + 3+ \ft{{\cal N}({\cal N}-1)}{2}$ bosonic generators
and $2\times {\cal N}$ fermionic ones. This algebra is the semidirect
product:
\begin{equation}
\begin{array}{ccc}
 cso(1,2\vert {\cal N}) & = & {\underbrace {iso(1,2\vert {\cal N})  \oplus  
so({\cal N})}}\\
 \null & \null & \mbox{semidirect}
 \end{array}
\label{CSOdefi}
\end{equation}
of  the ${\cal N}$--extended {\it superPoincar\'e} algebra
in three dimensions $iso(1,2\vert {\cal N})$ with $so({\cal N})$. 
It should be clearly distinguished from the central extension of the
Poincar\'e superalgebra, $Z[iso(1,2\vert {\cal N})]$, which
has the same number of generators but different commutation
relations.
Indeed there are three essential differences that it is worth to
recall at this point:
\begin{enumerate}
  \item In $Z\left [ISO(1,2\vert {\cal N})\right ]$
  the ${{\cal N}({\cal N}-1)}/{2}$ internal generators
  $Z^{ij}$ are abelian, while in $CSO(1,2\vert {\cal N})$ the
  corresponding $T^{ij}$ are non abelian and generate $SO({\cal N})$.
  \item In $Z\left [ISO(1,2\vert {\cal N})\right ]$ the supercharges $q^{\alpha i}$
  commute with $Z^{ij}$ (these are in fact central charges), while in
  $CSO(1,2\vert {\cal N})$ they transform as vectors under $T^{ij}$.
  \item In $Z\left [ISO(1,2\vert {\cal N})\right ]$ the anticommutator of two supercharges
  yields, besides the translation generators $P_m$, also the central charges
  $Z^{ij}$, while in $CSO(1,2\vert {\cal N})$ this is not true.
\end{enumerate}
\par
In both cases of fig.\ref{pistac} and fig.\ref{pirillo} if one takes
the subset of generators of positive grading plus the abelian grading
generator $X=\cases{E\cr D\cr}$ one obtains a {\it solvable superalgebra} of dimension
$4+2{\cal N}$. It is however only in the non compact case of
fig.\ref{pirillo} that the bosonic subalgebra of the solvable
superalgebra generates anti--de Sitter space $AdS_4$ as a solvable group
manifold.
\par
The  structure of $ISO(1,2\vert {\cal N}) \subset Osp({\cal N} \vert 4)$ can be easily
seen in picture \ref{pirillo}, displaying the root diagram of
the superconformal interpretation of the $Osp({\cal N}\vert
4)$. $CSO(1,2\vert{\cal N})$ is spanned by the generators out of the square, which
have null or negative grading, namely the conformal
boosts $K_m$, the Lorentz generators $J_m$ and the special conformal
supersymmetries $s^i_\alpha$. Notice that the generators in the square define
a solvable subalgebra at sight, because
in a root diagram the commutator of two generators, if not zero, corresponds to the vector
sum of the vectors corresponding to the two generators. 
 The solvable superalgebra $Ssolv_{adS}$ mentioned
in eq. (\ref{solvsup}) is the vector span of the following
generators:
\begin{equation}
  Ssolv_{adS} \equiv \mbox{span} \left\{ P_m, D, q^{\alpha i} \right\}.
\label{Span}
\end{equation}
\par
Being a coset, the solvable $AdS$--superspace
$AdS^{(Solv)}_{4\vert 2{\cal N}}$ supports a non linear representation of the full
$Osp({\cal
N}\vert 4 )$ superalgebra. As shown in \cite{torinos7}, we can regard
$AdS^{(Solv)}_{4\vert 2{\cal N}}$ as ordinary anti--de Sitter superspace $AdS_{4\vert
{\cal N}}$
where $2\times {\cal N}$ fermionic coordinates have being eliminated by fixing
$\kappa$--supersymmetry.
\par
\par
The strategy to construct the boundary superfields is the following.
First we construct  the supermultiplets on the bulk by acting on the
abstract states spanning the UIR with the coset
representative of the solvable superspace $AdS^{(Solv)}_{4\vert 2{\cal N}}$
and then we reach the boundary by performing the limit $\rho \to 0$
(see fig. \ref{boufil}).
\iffigs
\begin{figure}
\caption{Boundary superfields are obtained as limiting values of superfields on the bulk
\label{boufil}}
\begin{center}
\epsfxsize = 10cm
\epsffile{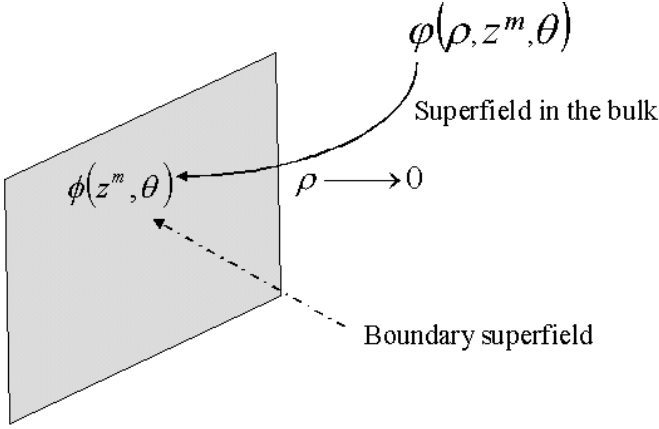}
\vskip  0.2cm
\hskip 2cm
\unitlength=1.1mm
\end{center}
\end{figure}
\fi
\par
Then, we restrict us to the case $\cN=2$. 
According to our previous discussion each of the $\cN=2$ shortened multiplets,
namely, the short multiplets, the hypermultiplet and the massless multiplets, 
correspond to a primary superfield on the boundary.
We determine such superfields with the above described method.
Short supermultiplets correspond to {\it constrained superfields}.
The shortening conditions  relating masses and hypercharges are
retrieved here as the necessary condition to maintain the constraints
after a superconformal transformation.
\par
\subsection{$AdS_4$ and $\partial AdS_4$ as cosets and their Killing vectors}
\par
\label{supercoset}
We have previously studied $Osp({\cal N}\vert 4)$ and its
representations in two different bases. The form (\ref{ospH}) of the
superalgebra
is  that used  to construct the $Osp(2\vert4)$ and $Osp(3\vert 4)$ supermultiplets
in section \ref{constructsupermultiplets}.
I showed, in section \ref{rotation}, how to translate these results
in terms of the form (\ref{ospD}) of the $Osp({\cal N}\vert 4)$ algebra
in order to allow a comparison with the three-dimensional
CFT on the boundary.
Now we introduce the description of the anti--de
Sitter superspace and of  its boundary in terms of supersolvable Lie algebra
parametrization as in eq.s (\ref{solvsup}), (\ref{supcos2}).
It turns out that such a description is the most appropriate
for a comparative study between $AdS_4$ and its boundary. We calculate
the Killing vectors of these two coset spaces
since they are needed to determine the
superfield multiplets living on both $AdS_4$ and $\partial AdS_4$.
\par
So we write both the bulk and the boundary superspaces as
supercosets\footnote{For an
extensive explanation about supercosets I refer the reader to
\cite{castdauriafre}. In the context of $D=11$ and $D=10$ compactifications
see also \cite{renatpiet}.},
\begin{eqnarray}
\frac{G}{H}  \,.
\label{GoverH}
\end{eqnarray}
Applying supergroup elements  $g \in Osp({\cal N}\vert 4)$ to the coset
representatives
$L(y)$ these latter transform as follows:
\begin{eqnarray}
g \, L(y) = L(y^\prime) h(g,y) \,,
\label{defcoset}
\end{eqnarray}
where  $h(y)$ is some element of $H \subset Osp({\cal N}\vert 4)$, named
the compensator  that, generically depends both on $g$ and on the coset point
$y\in G/H$.
For our purposes it is useful to consider the infinitesimal
form of (\ref{defcoset}), i.e. for infinitesimal $g$ we can write (see chapter $3$):
\begin{eqnarray}
g&=& 1 + \epsilon^A T_A \,, \nonumber \\
h &=& 1 - \epsilon^A W^H_A(y) T_H \,, \nonumber \\
y^{\mu\prime}  \, & = & y^\mu+\epsilon^A k^\mu_A(y)
\end{eqnarray}
and we obtain:
\begin{eqnarray}
T_A L(y) &=& k_A L(y) - L(y) T_H W^H_A(y) \,
,\label{infinitesimalcosettransfo}\\
k_A & \equiv & k^\mu_A(y) \, \frac{\partial}{\partial y^\mu}\,.
\label{kildefi}
\end{eqnarray}
The shifts in the superspace coordinates $y$ determined by the supergroup elements
(see eq.(\ref{defcoset})) define the Killing vector fields
(\ref{kildefi}) of the coset manifold.\footnote{The Killing vectors satisfy the
algebra with structure functions with opposite sign, see \cite{castdauriafre}.}
\par
Let us now consider the solvable anti--de Sitter superspace
defined in eq.s (\ref{solvsup}), (\ref{supcos2}). It describes a
$\kappa$--gauge fixed supersymmetric extension of the bulk  $AdS_4$.
As explained by eq.(\ref{supcos2}) it is a supercoset (\ref{GoverH})
where $G\!=\!Osp({\cal N}\vert 4)$ and $H\!=\!CSO(1,2\vert{\cal N}) 
\times SO({\cal N})$.
Using the non--compact basis (\ref{ospD}),
the subgroup $H$ is given by,
\begin{eqnarray}
H^{AdS} =CSO(1,2\vert{\cal N}) \,\equiv\,
 \mbox{span}\, \left\{ \, J^m, K_m, s_\alpha^{i}, T^{ij} \,\right\}  \,.
\label{HAdS}
\end{eqnarray}
A coset representative can be written as follows\footnote{We use the notation
$x\cdot y \equiv x^m y_m$ and $\theta^i q^i \equiv \theta^i_\alpha q^{\alpha
i}$. }:
\begin{eqnarray}
L^{AdS}(y) = \exp \left [{\rho D + i \, x \cdot P + \theta^i q^i} \right ]\,,
\qquad
y=(\rho, x,\theta) \,.
\label{AdScosetrepresentative}
\end{eqnarray}
In $AdS_{4\vert 2{\cal N}}$
 $s$-supersymmetry and  $K$-symmetry have a non linear realization
 since the corresponding generators are not part of the solvable superalgebra
$Ssolv_{adS}$  that is exponentiated (see eq.(\ref{Span})).
\par
The form of the Killing vectors simplifies considerably if we
rewrite the coset representative as a product of exponentials
\begin{eqnarray}
L(y) = \exp \left[ i \, z \cdot P\right]\, \cdot \, \exp \left[ \xi^i q^i\right]
 \, \cdot \,  \exp \left[ {\rho D} \right].
\label{simplecostrepresentative}
\end{eqnarray}
This amounts to the following coordinate change:
\begin{eqnarray}
z&=& \left( 1-\ft12 \rho + \ft16 \rho^2 + {\cal O}(\rho^3) \right)\,
x \,, \nonumber \\
\xi^i&=& \left(1-\ft14 \rho + \ft{1}{24} \rho^2 + {\cal O}(\rho^3)\right)\,
\theta^i \,.
\label{changecos}
\end{eqnarray}
This is the parametrization that was used in
\cite{torinos7} to get the $Osp(8\vert 4)$-singleton action
from the supermembrane. For this choice of coordinates the anti--de Sitter metric
takes the standard form (\ref{adsmet}).
The Killing vectors are
\begin{eqnarray}
\veck[ P_m ]&=& -i\, \partial_m \,, \nonumber \\
\veck [q^{\alpha i}] &=&
\frac{\partial}{\partial \xi_\alpha^i}
-\frac{1}{2}\left(\gamma^m\xi^i\right)^\alpha
\partial_m \,,
\nonumber \\
\veck [J^m] &=& \varepsilon^{mpq} z_p \partial_q
-\frac{i}{2} \left(\xi^i\gamma^m \right)_\alpha \frac{\partial}{\partial
\xi_\alpha^i}
\,, \nonumber \\
\veck [D] &=& \frac{\partial}{\partial \rho}
- z \cdot \partial
-\frac{1}{2}\xi_\alpha^i
\frac{\partial}{\partial \xi_\alpha^i}
\,,
\nonumber \\
\veck [s^{\alpha i}] &=&
-\xi^{\alpha i} \frac{\partial}{\partial \rho}
+\frac{1}{2}  \xi^{\alpha i}\, z \cdot \partial
+ \frac{i}{2} \varepsilon^{pqm}\, z_p(\gamma_q \xi^i)^\alpha \partial_m+
\nonumber \\
& &
-\frac{1}{8} (\xi^j \xi^j) (\gamma^m \xi^i)^\alpha \partial_m
- z^m (\gamma_m)^\alpha{}_\beta \frac{\partial}{\partial \xi^i_\beta}
-\frac{1}{4} (\xi^j\xi^j) \frac{\partial}{\partial \xi^i_\alpha}+
\nonumber \\
& &
+\frac{1}{2} \xi^{\alpha i} \xi^{\beta j}
\frac{\partial}{\partial\xi^j_\beta}
-\frac{1}{2}(\gamma^m \xi^i)^\alpha \xi^j_\beta \gamma_m
\frac{\partial}{\partial\xi^j_\beta}
\,,
\label{simplekilling}
\end{eqnarray}
and for the compensators we find:
\begin{eqnarray}
W[P]&=& 0 \,, \nonumber \\
W[q^{\alpha i}] &=& 0 \,, \nonumber \\
W[J^m] &=& - J^m \,, \nonumber \\
W[D] &=& 0 \,, \nonumber \\
W[s^{\alpha i}] &=&  - s^{\alpha i}
  + i\, \left(\gamma^m \theta^i \right)^\alpha \,J_m
  - i  \theta^{\alpha j} \, T^{ij} \,.
\label{compensatorsAdSrzxi}
\end{eqnarray}
For a detailed
derivation of these Killing vectors and compensators I refer
the reader to  \cite{noi2}.
\par
The boundary superspace $\partial(AdS_{4\vert 2{\cal N}})$ is formed by the
points on the
supercoset with $\rho=0$:
\begin{eqnarray}
L^{CFT}(y) = \exp \left[\mbox{ i}\, x \cdot P + \theta^i q^i\right]\,.
\label{boundarycosetrepresentative}
\end{eqnarray}
In order to see how the supergroup acts on fields that
live on this boundary we  use the fact that
this submanifold is by itself a supercoset. Indeed
instead of $H^{AdS} \subset Osp({\cal N} \vert 4)$ as given in  (\ref{HAdS}), we
can choose the larger subalgebra
\begin{equation}
H^{CFT} = \mbox{span} \, \left \{ D, J^m, K_m, s_\alpha^{i}, T^{ij} \right \}
\,,
\label{HCFT}
\end{equation}
and consider the new supercoset $G/H^{CFT}$.
By definition also on this smaller space we have a non linear realization of the
full orthosymplectic superalgebra.
For the Killing vectors we find (see \cite{noi2}):
\begin{eqnarray}
\veck [P_m] &=& -i\, \partial_m \,, \nonumber \\
\veck [q^{\alpha i}] &=&
\frac{\partial}{\partial \theta_\alpha^i}
-\frac{1}{2}\left(\gamma^m\theta^i\right)^\alpha
\partial_m \,,
\nonumber \\
\veck [J^m] &=& \varepsilon^{mpq} x_p \partial_q
-\frac{i}{2} \left(\theta^i\gamma^m \right)_\alpha \frac{\partial}{\partial
\theta_\alpha^i}
\,, \nonumber \\
\veck[D] &=&
- x \cdot \partial
-\frac{1}{2}\theta_\alpha^i
\frac{\partial}{\partial \theta_\alpha^i} \,,
\nonumber \\
\veck [s^{\alpha i}] &=&
\frac{1}{2} \theta^{\alpha i}\, x \cdot \partial
+ \frac{i}{2} \varepsilon^{pqm}\, x_p(\gamma_q \theta^i)^\alpha \partial_m
-\frac{1}{8} (\theta^j \theta^j) (\gamma^m \theta^i)^\alpha \partial_m+
\nonumber \\
& &
- x^m (\gamma_m)^\alpha{}_\beta \frac{\partial}{\partial \theta^i_\beta}
-\frac{1}{4} (\theta^j\theta^j) \frac{\partial}{\partial \theta^i_\alpha}
+\frac{1}{2} \theta^{\alpha i} \theta^{\beta j}
\frac{\partial}{\partial\theta^j_\beta}
-\frac{1}{2}(\gamma^m \theta^i)^\alpha \theta^j_\beta \gamma_m
\frac{\partial}{\partial\theta^j_\beta}
\,,  \nonumber\\
\label{bounkil}
\end{eqnarray}
and for the compensators we have:
\begin{eqnarray}
W[P_m] &=& 0 \,, \nonumber \\
W[q^{\alpha i}] &=& 0 \,, \nonumber \\
W[J^m] &=& -J^m \,, \nonumber \\
W[D] &=& D \,, \nonumber\\
W[s^{\alpha i}] &=&  \theta^{\alpha i}\, D -
     s^{\alpha i} + i\, \left( \gamma^m \theta^i \right)^\alpha\,J_m
     - i \theta^j T^{ij} \,.
\label{compensatorsdAdS}
\end{eqnarray}
If we compare the Killing vectors on the boundary (\ref{bounkil})  with those on
the
bulk (\ref{simplekilling}) we see that they are very similar. The
only formal difference is the suppression of the $\frac{\partial}{\partial\rho
}$ terms. The conceptual difference, however, is relevant. On the
boundary the transformations generated by (\ref{bounkil}) are the
{\it standard superconformal transformations} in three--dimensional
(compactified) Minkowski space.
On the bulk the transformations generated by (\ref{simplekilling})
are {\it superisometries} of anti--de Sitter superspace. They might be
written in completely different but equivalent forms if we used other
coordinate frames. The form they have is due to the use of the {\it
solvable coordinate frame} $(\rho, z, \xi)$ which is the most appropriate to study the
restriction of bulk supermultiplets to the boundary.
For more details on this point I refer the reader to 
\cite{noi2}.
\par
\subsection{Conformal $Osp(2 \vert 4)$ superfields: general discussion}
\par
\label{superfields}
Let us restrict our attention to ${\cal N}\!\!=\!\!2$.
In this case the $SO(2)$ group has just one generator that we name
the hypercharge:
\begin{equation}
Y \equiv T^{12} \,.
\end{equation}
Since it is convenient to work with eigenstates of the
hypercharge operator, we reorganize the two Grassmann spinor coordinates of
superspace in complex combinations:
\begin{eqnarray}
\theta^\pm_\alpha = \frac{1}{\sqrt{2}}(\theta^1_\alpha
\pm i \theta^2_\alpha) \,,
\qquad Y \, \theta^\pm_\alpha = \pm \theta^\pm_\alpha~.
\label{complexthet}
\end{eqnarray}
In this new notations the Killing vectors generating
$q$--supersymmetries on the boundary (see eq.(\ref{bounkil})) take the
form:
\begin{equation}
{\vec k}\left[q^{\alpha i}\right]  \quad \longrightarrow \quad q^{\alpha \pm}=
\frac{\partial}{\partial \theta^\mp_\alpha}
- \ft{1}{2} \,  (\gamma^m)^\alpha{}_\beta \theta^{\beta \pm} \partial_m \,.
\label{qpm}
\end{equation}
A generic superfield is a function $\Phi(x,\theta)$ of the bosonic
coordinates $x$ and of all the $\theta .s$. Expanding such a field in
power series of the $\theta .s$ we obtain a multiplet of $x$--space
fields that, under the action of the Killing vector (\ref{qpm}), form a
representation of Poincar\'e supersymmetry. Such a representation can
be shortened by imposing on the superfield $\Phi(x,\theta)$
constraints that are invariant with respect to the action of the
Killing vectors (\ref{qpm}). This is possible because of the
existence of the so called superderivatives, namely of fermionic vector
fields that commute with the supersymmetry Killing vectors. In our
notations the superderivatives are defined as follows:
\begin{eqnarray}
{\cal D}^{\alpha \pm} =
\frac{\partial}{\partial \theta^\mp_\alpha}
+\ft{1}{2} \,  (\gamma^m)^\alpha{}_\beta \theta^{\beta \pm} \partial_m \,,
\label{supdervcf}
\end{eqnarray}
and satisfy the required property
\begin{equation}
\begin{array}{rclcr}
\{{\cal D}^{\alpha \pm}, q^{\beta \pm}\}&=&
\{{\cal D}^{\alpha \pm}, q^{\beta \mp}\}&=&0 \,.
\end{array}
\label{commDq}
\end{equation}
As explained in \cite{castdauriafre} the existence of
superderivatives is the manifestation at the fermionic level of a
general property of coset manifolds. For $G/H$ the true isometry
algebra is not $G$, rather it is $G \times (N(H)_G/H)$ (minus the explicit
$U(1)$'s) where $N(H)_G$
denotes the normalizer of the stability subalgebra $H$ (see chapter $3$). The
additional isometries are generated by {\it right--invariant} rather
than {\it left--invariant} vector fields that as such commute with
the {\it left--invariant} ones. If we agree that the Killing vectors
are left--invariant vector fields then the superderivatives are
right--invariant ones and generate the additional superisometries of
Poincar\'e superspace. Shortened representations of Poincar\'e
supersymmetry are superfields with a prescribed behaviour under the
additional superisometries: for instance they may be invariant under
such transformations. We can formulate these shortening conditions by
writing constraints such as
\begin{equation}
  {\cal D}^{\alpha +}\Phi(x,\theta)=0 \, .
\label{typconstr}
\end{equation}
The key point in our  discussion is that a constraint of type
(\ref{typconstr}) is guaranteed from eq.s (\ref{commDq}) to be
invariant with respect to the superPoincar\'e algebra, yet it is not
a priori guaranteed that it is invariant under the action of the full
superconformal algebra (\ref{bounkil}). Investigating the additional
conditions that make a constraint such as (\ref{typconstr})
superconformal invariant is the main goal of the present section.
This is the main tool that allows a transcription of the
Kaluza--Klein results for supermultiplets into a superconformal
language.
\par
To develop such a programme it is useful to perform a further
coordinate change that is quite traditional in superspace literature \cite{wessbagger}.
Given the coordinates  $x$ on the boundary (or the coordinates $z$ for
the bulk) we set:
\begin{eqnarray}
y^m=x^m + \ft{1}{2} \, \theta^+ \gamma^m \theta^- \,.
\end{eqnarray}
Then the superderivatives become
\begin{eqnarray}
{\cal D}^{\alpha +} &=&  \frac{\partial}{\partial \theta^-_\alpha}
\,, \nonumber \\
{\cal D}^{\alpha -} &=& \frac{\partial}{\partial \theta^+_\alpha}
+ (\gamma^m)^\alpha{}_\beta\theta^{\beta -}
\partial_m \,.
\end{eqnarray}
It is our aim to describe superfield multiplets both on the bulk
and on the boundary. It is clear that one can do the
same redefinitions for the Killing vector of $q$-supersymmetry (\ref{qpm})
and that  one can introduce superderivatives also for the theory on the
bulk. 
\par
So let us finally turn to superfields.
 We begin by  focusing on boundary
superfields since their treatment is
slightly easier than  the treatment of bulk superfields.
\par
\begin{it}
A {\bf primary superfield} is defined as follows
(see \cite{gunaydinminiczagerman}, \cite{macksalam}):
\begin{eqnarray}
\Phi^{\partial AdS}(x, \theta) =\exp \left [\mbox{i}\, x\cdot P + \theta^i
q^i\right ]
 \Phi(0)\,,
\label{3Dsuperfield}
\end{eqnarray}
where $\Phi(0)$ is a primary field (see eq.(\ref{primstate}))
\footnote{For an operator $\Phi$, the action of algebra operators is actually the
adjoint action $\ll[\cO,\Phi\rr]$; however, as a shorthand notation we call the states with the names of
the corresponding fields, so we write $\cO\Phi$.}
\begin{eqnarray}
s_\alpha^i \Phi(0) &=& 0 \,, \nonumber \\
K_m \Phi(0) &=& 0 \,,
\end{eqnarray}
of  scaling weight $D_0$, hypercharge
$y_0$ and eigenvalue $j$ for the ``third-component'' operator $J_2$
\begin{equation}
\begin{array}{ccccc}
  D \, \Phi(0) = D_0 \, \Phi(0)&;& Y \, \Phi(0) = y_0 \, \Phi(0) &;&
J_2\, \Phi(0) = j \, \Phi(0)\,.
\end{array}
\label{primlab}
\end{equation}
\end{it}
From the above definition one sees that the primary superfield
$\Phi^{\partial AdS}(x, \theta)$ is
actually obtained by acting with the coset representative
(\ref{boundarycosetrepresentative}) on the
$SO(1,2)\times SO(1,1)$-primary field.
Hence we know how it transforms under the infinitesimal
transformations of the group $Osp(2\vert 4)$. Indeed one simply uses
(\ref{infinitesimalcosettransfo}) to obtain the result.
For example under dilatation we have:
\begin{eqnarray}
D\, \Phi^{\partial AdS}(x,\theta) = \left(-x\cdot\partial - \ft12 \theta^i
\frac{\partial}{\partial \theta^i} + D_0   \right) \Phi(x,\theta),
\end{eqnarray}
where the term $D_0$ comes from the compensator in
(\ref{compensatorsdAdS}).
Of particular interest is the transformation under
special supersymmetry since it imposes the constraints
for shortening,
\begin{equation}
s^\pm \Phi^{\partial AdS}(x,\theta) = \veck [s^\pm] \Phi(x, \theta)
                      + e^{i\, x\cdot P + \theta^i q^i}
                       \left( -\theta^\pm D
                              - i\, \gamma^m \theta^\pm \,J_m
                              +s^\pm
                              \pm \theta^\pm Y \right) \Phi(0)
                      \,. \label{spm}
\end{equation}
For completeness we give the form
of $s^\pm$ in the $y$-basis where it gets a relatively concise form,
\begin{eqnarray}
\veck [s^{\alpha-}] &=& - \left( y\cdot \gamma\right)^\alpha{}_\beta
\frac{\partial}{\partial \theta_\beta^+}
+ \frac{1}{2} \left( \theta^- \theta^- \right)
\frac{\partial}{\partial \theta^-_\alpha}
\nonumber \\
\veck [s^{\alpha+}] &=& \theta^{\alpha+} y\cdot\partial
+ i\, \varepsilon^{pqm} y_p \left( \gamma_p \theta^+ \right)^\alpha\partial_m
+ \frac{1}{2} \left(\theta^+ \theta^+\right)
\frac{\partial}{\partial \theta^+_\alpha}+
\nonumber \\
& & + \theta^+\gamma^m \theta^- \left(\gamma_m\right)^\alpha{}_\beta
\frac{\partial}{\partial \theta_\beta^-}
\,.
\label{finalkillings}
\end{eqnarray}
\par
Let us now turn to a direct discussion of multiplet shortening and
consider the superconformal invariance of Poincar\'e constraints
constructed with the superderivatives ${\cal D}^{\alpha \pm}$. The
simplest example is provided by the {\it chiral supermultiplet}.
By definition this is a scalar superfield $\Phi_{chiral}(y,\theta)$
obeying the constraint (\ref{typconstr}) which is solved by boosting
only along $q^-$ and not along $q^+$:
\begin{eqnarray}
\Phi_{chiral}(y,\theta)=e^{i\, y \cdot P + \theta^+ q^-} \Phi(0).
\end{eqnarray}
Hence we have
\begin{eqnarray}
\Phi_{chiral}(\rho, y, \theta) = X(\rho, y) + \theta^+ \lambda(\rho, y) +
\theta^+
\theta^+ H(\rho, y)
\end{eqnarray}
on the bulk or
\begin{eqnarray}
\Phi_{chiral}(y, \theta) = X(y) + \theta^+ \lambda(y) + \theta^+
\theta^+ H(y)
\label{chiralmultiplet3D}
\end{eqnarray}
on the boundary. The field components of the chiral multiplet are:
\begin{eqnarray}
X = e^{i\, y\cdot P} \, \Phi(0)\,,\qquad
\lambda = i \, e^{i\, y\cdot P} \,  q^- \Phi(0) \,, \qquad
H = -\ft14 e^{i\, y\cdot P} \, q^- q^- \phi(0)
\,.
\label{chiralcomponents}
\end{eqnarray}
For completeness,
we write the superfield $\Phi$ also in the
$x$-basis\footnote{where $\Box=\partial^m \partial_m\,$.},
\begin{eqnarray}
\Phi(x) &=& X(x)
+\theta^+ \lambda(x) + (\theta^+\theta^+) H(x)
+\ft{1}{2} \theta^+ \gamma^m \theta^- \partial_m X(x)+
\nonumber \\
& &
+\ft{1}{4}  (\theta^+\theta^+) \theta^- \dslash \lambda(x)
+\ft{1}{16} (\theta^+ \theta^+) (\theta^- \theta^-)
\Box X(x)=
\nonumber \\
&=& \exp\left(\ft{1}{2} \theta^+ \gamma^m\theta^-
\partial_m \right) \Phi(y) \,.
\label{formchiral}
\end{eqnarray}
\par
Because of (\ref{commDq}), we are guaranteed that under $q$--supersymmetry
the chiral superfield
$\Phi_{chiral}$ transforms into a chiral superfield. We should  verify
that this is true also for $s$--supersymmetry.
To say it simply we just have to check that   $s^- \Phi_{chiral}$ does not
depend
on $\theta^-$. This is not generically true, but it becomes true
if certain  extra conditions on the quantum numbers of the primary
state are satisfied. Such conditions are the same one obtains as
multiplet shortening conditions when constructing the UIRs of the
superalgebra.
\par
In the specific instance of the chiral multiplet,
looking at (\ref{spm}) and (\ref{finalkillings})
we see that  in $s^-\Phi_{chiral}$ the terms  depending on $\theta^-$
are the following ones:
\begin{eqnarray}
s^- \Phi \Big\vert_{\theta^-} = - \left( D_0+ y_0 \right)  \theta^- \Phi =0 \,,
\end{eqnarray}
they cancel if
\begin{equation}
D_0 = - y_0 \,.
\label{Disy}
\end{equation}
Eq.(\ref{Disy}) is easily recognized as the unitarity condition for
the existence of $Osp(2\vert 4)$ hypermultiplets (see section \ref{constructsupermultiplets}).
The algebra (\ref{ospD}) ensures that the chiral multiplet
also transforms into a chiral multiplet under $K_m$. Moreover
we know that the action of the compensators of $K_m$ on
the chiral multiplet is zero. Furthermore,
the compensators of the generators $P_m, q^i, J_m$
on the chiral multiplet are zero and from (\ref{infinitesimalcosettransfo})
we conclude that their generators act on the chiral multiplet
as the Killing vectors.
\par
Notice that the linear part of the
$s$-supersymmetry transformation on the chiral multiplet
has the same form of the $q$-supersymmetry but with the
parameter taken to be $\epsilon_q = -i\, y \cdot \g \epsilon_s$.
As already stated the non-linear form of  $s$-supersymmetry
is the consequence of its gauge fixing which we have implicitly imposed
from the start by choosing the supersolvable Lie algebra
parametrization of superspace and by  taking the coset representatives
as in (\ref{AdScosetrepresentative}) and (\ref{boundarycosetrepresentative}).
In addition to the chiral multiplet there exists also
the complex conjugate {\it antichiral multiplet}
${\bar \Phi}_{chiral}=\Phi_{antichiral}$
with opposite hypercharge and the relation $D_0=y_0$.
\par
\subsection{Matching the Kaluza Klein results for $Osp(2 \vert 4)$
supermultiplets with boundary conformal superfields}
\par
\label{vocabulary}
It is now our purpose to reformulate the ${\cal N}=2$ multiplets in terms of
superfields living on the boundary of the $AdS_4$ space--time manifold.
This is the key step to convert information coming from classical
harmonic analysis on the compact manifold $X_7$ into predictions on
the spectrum of conformal primary operators present in the
three--dimensional gauge theory of the M2--brane.
\par
Interpreted as superfields on the boundary the  {\it long multiplets}
correspond to {\it unconstrained superfields} and their discussion is quite
straightforward.
We are mostly interested in short multiplets that correspond to
composite operators of the microscopic gauge theory with protected
scaling dimensions. In superfield language, as we have shown in the
previous section, {\it short multiplets} are constrained superfields.
\par
Just as on the boundary, also on the bulk, we obtain such constraints
by means of the bulk superderivatives. In order to show how
this works we begin  by discussing the {\it chiral superfield  on the
bulk} and then show how it is obtained from the hypermultiplet.
\par
\subsubsection{Chiral superfields are the Hypermultiplets: the basic example}
The treatment for the bulk chiral field is completely analogous
to that of chiral superfield on the boundary.
\par
Generically bulk superfields are given by:
\begin{eqnarray}
\Phi^{AdS}(\rho, x, \theta) = \exp \left[\rho D + \mbox{i}\, x\cdot P +
 \theta^i q^i\right ]
\, \Phi(0)\,.
\label{4Dsuperfield}
\end{eqnarray}
Using the parametrization (\ref{changecos}) we can rewrite
(\ref{4Dsuperfield}) in the following way:
\begin{eqnarray}
\Phi^{AdS}(\rho, z, \xi) = \exp \left[\mbox{i}\, z \cdot P + \xi^i q^i\right ]
                 \, \cdot \, \exp \left[\rho D_0\right ]\, \Phi(0)\,.
\label{simpleAdSsuperfield}
\end{eqnarray}
Then the  generator $D$ acts on this field as follows:
\begin{eqnarray}
D \, \Phi^{AdS}(\rho, z, \xi) =
\left(- z \cdot \partial
      - \ft12
        \xi^i \frac{\partial}{\partial \xi^i}
        + D_0
\right) \Phi^{AdS}(\rho, z, \xi) \,.
\end{eqnarray}
Just as for boundary  chiral superfields, also on the bulk  we find that
the constraint (\ref{typconstr}) is
invariant under   the $s$-supersymmetry rule (\ref{simplekilling}) if
and only if:
\begin{equation}
  D_0= - y_0\,.
\label{dugaly}
\end{equation}
Furthermore, looking at (\ref{simpleAdSsuperfield})
one sees that for the bulk superfields $D_0=0$ is forbidden.
This constraint on the scaling dimension
together with the relation $E_0=-D_0$, coincides with the
constraint:
\begin{equation}
  E_0=\vert y_0 \vert
\label{Eugaly}
\end{equation}
defining the $Osp(2\vert 4)$ hypermultiplet UIR of $Osp(2\vert 4)$.
The transformation of the bulk chiral superfield under
$s, P_m, q^i, J_m$ is simply given by the bulk
Killing vectors. In particular the form of the $s$-supersymmetry Killing vector
coincides with that given in
(\ref{finalkillings}) for the boundary.
\par
As we saw a chiral superfield on the bulk
describes an $Osp(2\vert 4)$
hypermultiplet.
\par
Applying the rotation matrix $U$  of eq. (\ref{rotationmatrix})
to the states in table \ref{N2hyper} 
 we  indeed find the field components (\ref{chiralcomponents}) of
the chiral supermultiplet
\footnote{I remind that the fields on the bulk
are on--shell, the fields on the boundary are off--shell; then, for
example, the spinor in table \ref{N2hyper} has the same number of degrees of
freedom of the spinor in (\ref{chiralmultiplet3D}), that is two, because 
the first is four dimensional on--shell, the second is three dimensional 
off--shell.}.
\par
Having clarified how to obtain the four-dimensional
chiral superfield from the $Osp(2\vert 4)$ hypermultiplet
we can now obtain the other shortened $Osp(2\vert 4)$
superfields from the supermultiplets found in section $2.4.4$.
\par
\subsubsection{Superfield description of the short vector multiplet}
\par
Let us start with the short massive vector multiplet. 
The constraint for shortening is 
\begin{equation}
  E_0=\vert y_0 \vert + 1
\label{eisyplusone}
\end{equation}
and the particle states
of the multiplet are given in table \ref{N2shortvector}.
Applying the rotation matrix $U$ to the states in 
table  \ref{N2shortvector} we find the following states:
\begin{eqnarray}
S=\vert {\rm vac} \rangle \,, \qquad
\lambda^{\pm}_L = i \, q^\pm \vert {\rm vac} \rangle \,, \qquad
\pi^{--} = -\ft14 \, q^- q^- \vert {\rm vac} \rangle \,, \qquad
{\rm etc} \dots
\label{rotatedstates}
\end{eqnarray}
where we used the same notation for the rotated
as for the original states
and up to an irrelevant factor $\ft14$.
We  follow the same procedure   also for the other short and
massless multiplets. Namely in the superfield transcription
of our multiplets we  use the same
names for the superspace field components as
 for the particle fields appearing  in the
$SO(3)\times SO(2)$ basis. Moreover when convenient we rescale some field
components without mentioning it explicitly.
The list of states appearing in  (\ref{rotatedstates})
are the components of a superfield
\begin{eqnarray}
\Phi_{vector} &=& S + \theta^- \lambda_L^+ + \theta^+ \lambda_L^-
                + \theta^+ \theta^- \pi^0
                + \theta^+ \theta^+ \pi^{--}
                + \theta^+ \Aslash \theta^-
                + \theta^+ \theta^+ \, \theta^- \lambda_T^- \,,
\nonumber \\ & &
\label{vecsupfil}
\end{eqnarray}
which is the explicit solution of the following constraint
\begin{equation}
{\cal D}^+ {\cal D}^+ \Phi_{vector} = 0 \,.
\end{equation}
imposed on a superfield of the form (\ref{4Dsuperfield})
with hypercharge $y_0$.
\par
In superspace literature a superfield of type
(\ref{vecsupfil}) is named a linear superfield.
If we consider the variation of a linear superfield with respect to
 $s^-$, such variation contains, a priori, a term of the form
\begin{equation}
s^- \Phi_{vector} \Big\vert_{\theta^-\theta^-}=
\ft12 \left(D_0 + y_0+1\right)(\theta^-\theta^-) \lambda_L^+ \,,
\end{equation}
which has to cancel if $\Phi_{vector}$ is to transform into
a linear multiplet  under $s^-$. Hence the following condition has to
be imposed
\begin{equation}
D_0 = - y_0 - 1 \,.
\label{dismymone}
\end{equation}
which is identical with the bound for the vector multiplet
shortening $E_0=y_0+1$.
\par
\subsubsection{Superfield description of the short gravitino
multiplet}
\par
Let us consider the short gravitino multiplets.
The particle state content of these multiplets is given in table \ref{N2shortgravitino}.
Applying the rotation matrix $U$ (\ref{rotationmatrix})
to these states,
and identifying the particle states with the corresponding rotated
field states as we have done in the previous cases,
we find  the following spinorial
superfield
\begin{eqnarray}
\Phi_{gravitino} &=& \lambda_L + \Aplusslash \theta^- + \Aminusslash \theta^+
                     + \phi^- \theta^+
                     + 3 \, (\theta^+ \theta^-) \lambda_T^{+-}
                     - (\theta^+ \gamma^m \theta^-) \gamma_m \lambda_T^{+-}+
\nonumber \\ & &
                     + (\theta^+ \theta^+) \lambda_T^{--}
      + (\theta^+ \gamma^m \theta^-) \chi_m^{(+)}
      + (\theta^+ \theta^+) \Zminusslash  \theta^- \,,
\end{eqnarray}
where the vector--spinor field $\chi^m$ is expressed in terms of
the spin-$\ft32$ field with symmetrized spinor indices in the following way
\begin{equation}
\chi^{(+)m \alpha} =
\left( \gamma^m \right)_{\beta\gamma}\, \chi^{(+)(\alpha\beta\gamma)}
\label{chispin32}
\end{equation}
and where, as usual, $\Aplusslash = \gamma^m \, A_m^+$.
\par
The superfield $\Phi_{gravitino}$
is linear in  the sense that it does not depend
on the monomial $\theta^-\theta^-$, but to be precise it is
a  spinorial superfield (\ref{4Dsuperfield})
with hypercharge $y_0$
that fulfills the stronger constraint
\begin{equation}
{\cal D}^+_\alpha \Phi_{gravitino}^\alpha = 0 \,.
\label{lineargravitino}
\end{equation}
\par
The generic  linear spinor superfield
contains, in its expansion, also terms  of the form
$\varphi^+ \theta^-$ and $(\theta^+\theta^+) \varphi^-\theta^-$,
where $\varphi^+$ and $\varphi^-$ are
 scalar fields and a term $(\theta^+ \g^m \theta^-) \chi_m$
where the spinor-vector $\chi_m$ is not an irreducible $\ft{3}{2}$
representation since it cannot be written as in (\ref{chispin32}).
 \par
Explicitly we have:
\begin{eqnarray}
\Phi^\alpha_{linear} &=& \lambda_L + \Aplusslash \theta^- + \Aminusslash \theta^+
                     + \phi^- \theta^+ + \varphi^+ \theta^-
                     + 3 \, (\theta^+ \theta^-) \lambda_T^{+-}
                     + (\theta^+ \theta^+) \lambda_T^{--}+
\nonumber \\ & &
      + (\theta^+ \gamma^m \theta^-) \chi_m
      + (\theta^+ \theta^+) \Zminusslash  \theta^- + (\theta^+\theta^+) \varphi^-\theta^-\,.
\end{eqnarray}
The  field component $\chi^{\alpha m}$ in a generic unconstrained
spinor superfield can be decomposed in
a spin-$\ft12$ component and a spin-$\ft32$ component according to,
\begin{eqnarray}
\begin{array}{|c|c|}
\hline
\,\,\,   & \,\,\,  \cr
\hline
\end{array}
\times
\begin{array}{|c|}
\hline
\,\,\,  \cr
\hline
\end{array}
=
\begin{array}{c}
\begin{array}{|c|c|}
\hline
\,\,\,   & \,\,\,  \cr
\hline
\end{array}
\cr
\begin{array}{|c|}
\,\,\, \cr
\hline
\end{array}
\begin{array}{c}
\,\,\, \cr
\end{array}
\end{array}
+
\begin{array}{|c|c|c|}
\hline
\,\,\,   & \,\,\,  & \,\,\, \cr
\hline
\end{array}
\label{gravitinodecomposition}
\end{eqnarray}
where $m={ \begin{array}{|c|c|}
\hline
  &   \cr
\hline
\end{array}}\, $.
Then the constraint (\ref{lineargravitino}) eliminates
the scalars $\varphi^\pm$ and eliminates
the ${  \begin{array}{c}
\begin{array}{|c|c|}
\hline
   &   \cr
\hline
\end{array}
\cr
\begin{array}{|c|}
 \cr
\hline
\end{array}
\begin{array}{c}
 \cr
\end{array}
\end{array}}$-component of $\chi$ in terms of $\lambda_T^{+-}$.
From
\begin{equation}
s_\beta^-\, \Phi_{gravitino}^\alpha\Big\vert_{\theta^-\theta^-}
= \ft12 \left( -D_0 - y_0 - \ft32 \right)(\theta^- \theta^-)
     (\Aplusslash)_\beta{}^\alpha
\end{equation}
we conclude that the constraint (\ref{lineargravitino}) is superconformal
invariant if and only if
\begin{equation}
D_0 = - y_0 - \ft32 \,.
\end{equation}
\par
Once again we have retrieved the shortening condition already known
in the $SO(3) \times SO(2)$ basis: $E_0 =   \vert y_0 \vert + \ft32$.
\par
\subsubsection{Superfield description of the short graviton
multiplet}
\par
Applying the rotation $U$ (\ref{rotationmatrix}) to the states of table \ref{N2shortgraviton},
and identifying the particle states with the corresponding boundary fields, as
we have done so far, we derive the short graviton superfield:
\begin{eqnarray}
\Phi^m_{graviton} &=&
A^m + \theta^+ \gamma^m \lambda_T^-
+  \theta^- \chi^{(+)+m} + \theta^+ \chi^{(+)-m}+
\nonumber \\
& &
+ (\theta^+ \theta^-) \, Z^{+-m}
+ \ft{i}{2}\, \varepsilon^{mnp} \, ( \theta^+ \gamma_n \theta^-) \,  Z_p^{+-}+
+ (\theta^+ \theta^+) \, Z^{--m}
\nonumber \\
& &
+ ( \theta^+ \gamma_n \theta^-) \, h^{mn} +
(\theta^+ \theta^+) \, \theta^- \chi^{(-)-m}
\,,
\end{eqnarray}
where
\begin{eqnarray}
\chi^{(+)\pm m \alpha} &=&
\left( \g^m \right)_{\beta\gamma}\chi^{(+)\pm(\alpha\beta\gamma)} \,,
\nonumber \\
\chi^{(-)- m \alpha} &=&
\left( \g^m \right)_{\beta\gamma}\chi^{(-)-(\alpha\beta\gamma)} \,,
\nonumber \\
h^m{}_m &=& 0 \,.
\label{gravitonchispin32}
\end{eqnarray}
This superfield  satisfies the following constraint,
\begin{eqnarray}
{\cal D}^+_\alpha \Phi_{graviton}^{\alpha \beta}=0 \,,
\end{eqnarray}
where we have defined:
\begin{eqnarray}
\Phi^{\alpha\beta}= \left(\gamma_m\right)^{\alpha\beta} \, \Phi^m\,.
\end{eqnarray}
Furthermore we check that $s^- \Phi^m_{graviton}$ is still a short graviton
superfield if and only if:
\begin{equation}
D_0 = - y_0 - 2 \,.
\end{equation}
corresponding to the known unitarity bound:
\begin{equation}
  E_0 = \vert y_0 \vert +2\,.
\label{graunibou}
\end{equation}
\par
\subsubsection{Superfield description of the massless vector
multiplet}
\par
Considering now massless multiplets we focus on the massless vector
multiplet, described in table \ref{N2masslessvector}.
Applying the rotation $U$ (\ref{rotationmatrix}) we get,
\begin{eqnarray}
V= S + \theta^+ \lambda_L^- + \theta^- \lambda_L^+
+ (\theta^+ \theta^-) \, \pi +   \theta^+ \Aslash \theta^- \,.
\label{Vfixed}
\end{eqnarray}
This multiplet can be obtained by a real superfield
\begin{eqnarray}
V &=& S + \theta^+ \lambda_L^- + \theta^- \lambda_L^+
+ (\theta^+ \theta^-) \, \pi +   \theta^+ \Aslash \theta^-+
\nonumber \\ & &
+ (\theta^+\theta^+) \, M^{--} + (\theta^-\theta^-) \, M^{++}+
\nonumber \\
& &
+ (\theta^+ \theta^+) \, \theta^- \mu^-
+ (\theta^- \theta^-) \, \theta^+ \mu^++
\nonumber \\ & &
+ (\theta^+\theta^+) (\theta^-\theta^-) \, F \,, \nonumber\\
V^\dagger &=& V
\end{eqnarray}
that transforms as follows under a gauge transformation,
\begin{eqnarray}
V \rightarrow V + \Lambda + \Lambda^\dagger \,,
\end{eqnarray}
where $\Lambda$ is a chiral superfield of the form
(\ref{formchiral}). In components
this reads,
\begin{eqnarray}
S &\rightarrow& S + X + X^* \,, \nonumber\\
\lambda_L^- &\rightarrow& \lambda_L^- + \lambda \,, \nonumber \\
\pi &\rightarrow& \pi \,, \nonumber \\
A_m &\rightarrow& A_m + \ft{1}{2}\, \partial_m\left(X-X^*\right) \,,
\nonumber \\
M^{--} &\rightarrow& M^{--} + H \,, \nonumber \\
\mu^{-} &\rightarrow& \mu^- +\ft{1}{4} \, \dslash\lambda \,, \nonumber \\
F &\rightarrow& F +\ft{1}{16} \, \Box  X \,,
\end{eqnarray}
which may be used to gauge fix the real multiplet in the
following way,
\begin{equation}
M^{--}= M^{++}=\mu^-=\mu^+= F =0\,,
\label{gaugefixingvector}
\end{equation}
to obtain (\ref{Vfixed}).
For the scaling weight $D_0$ of the massless vector multiplet
we find $-1$. Indeed this follows from the fact that $\Lambda$
is a chiral superfield with $y_0=0, D_0=0$. Which is also in
agreement with $E_0=1$.
\par
\subsubsection{Superfield description of the massless graviton
multiplet}
\par
The massless graviton multiplet is composed of the bulk
particle states listed in table \ref{N2masslessgraviton}, from which, 
with the usual procedure we obtain
\begin{eqnarray}
g_m = A_m + \theta^+ \chi^{(+)-}_m
    + \theta^- \chi^{(+)+}_m +  \theta^+ \gamma^n \theta^-\, h_{mn}\,.
\end{eqnarray}
Similarly as for the vector multiplet we may write this
multiplet as a gauge fixed multiplet with local gauge symmetries
that include local coordinate transformations, local supersymmetry
and local $SO(2)$, in other words full supergravity. However this is not the
goal of this chapter where we prepare to interprete the bulk gauge fields as composite states
in the boundary conformal field theory.
\par
This completes the treatment of the short $Osp(2\vert 4)$
boundary superfields. We have found that
all of them are linear superfields with the extra constraint
that they have to transform into superfields of the
same type under $s$-supersymmetry. Such constraint is identical to
the shortening conditions found by representation theory of $Osp\ll(2\vert 4\rr)$.
%%%%%%%%%%%%%%%%%%%%%%%%%%%%%%%%%%%%%%%%%%%%%%%%%%%%%%%%
\begin{table}[htbp] 
  \begin{center} 
  \begin{footnotesize}
    \begin{tabular}{|c|c|c|} 
      \hline 
      \multicolumn{3}{|c|}{} \cr 
      \multicolumn{3}{|c|}{$SD(E_0>y_0+2,1,y_0\ge 0\vert 2)$} \cr 
      \multicolumn{3}{|c|}{} \cr 
      \hline 
      spin & energy & hypercharge  \cr 
      \hline 
      \hline 
&&\cr
$2$      & $E_0+1    $  & $y_0$    \cr
&&\cr
      \hline 
&&\cr
$3/2 $  & $E_0+3/2 $  & $y_0-1$   \cr
$3/2 $  & $E_0+3/2 $  & $y_0+1$     \cr
$3/2 $  & $E_0+1/2 $  & $y_0-1$     \cr
$3/2 $  & $E_0+1/2 $  & $y_0+1$    \cr
&&\cr
      \hline 
&&\cr
$1$      & $E_0+2    $  & $y_0$      \cr
$1$      & $E_0+1    $  & $y_0-2$   \cr
$1$      & $E_0+1    $  & $y_0+2$    \cr
$1$      & $E_0+1    $  & $y_0$      \cr
$1$      & $E_0+1    $  & $y_0$      \cr
$1$      & $E_0      $  & $y_0$    \cr
&&\cr
      \hline 
&&\cr
$1/2 $  & $E_0+3/2 $  & $y_0-1$     \cr
$1/2 $  & $E_0+3/2 $  & $y_0+1$    \cr
$1/2 $  & $E_0+1/2 $  & $y_0-1$    \cr
$1/2 $  & $E_0+1/2 $  & $y_0+1$     \cr
&&\cr
      \hline 
&&\cr
$0$      & $E_0+1    $  & $y_0$      \cr
&&\cr
      \hline 
    \end{tabular} 
\qquad 
\caption{${\cal N}=2$ long graviton multiplet with $y_0\ge 0$}
\label{N2longgraviton}
 \end{footnotesize}
  \end{center} 
\end{table} 
%%%%%%%%%%%%%%%%%%%%%%%%%%%%%%%%%%%%%%%%%%%%%%%%%% 
\begin{table}[htbp] 
  \begin{center} 
  \begin{footnotesize}
    \begin{tabular}{|c|c|c|} 
      \hline 
      \multicolumn{3}{|c|}{} \cr 
      \multicolumn{3}{|c|}{$SD(E_0>y_0+3/2,1/2,y_0\ge 0\vert 2)$} \cr 
      \multicolumn{3}{|c|}{} \cr 
      \hline 
      spin & energy & hypercharge  \cr 
      \hline 
      \hline 
&&\cr
$3/2 $  & $E_0+1$      & $y_0$    \cr
&&\cr
\hline
&&\cr
$1$      & $E_0+3/2 $  & $y_0-1$ \cr
$1$      & $E_0+3/2 $  & $y_0+1$  \cr
$1$      & $E_0+1/2 $  & $y_0-1$ \cr
$1$      & $E_0+1/2 $  & $y_0+1$  \cr
&&\cr
\hline
&&\cr
$1/2 $  & $E_0+2$      & $y_0$    \cr
$1/2 $  & $E_0+1$      & $y_0-2$  \cr
$1/2 $  & $E_0+1$      & $y_0$    \cr
$1/2 $  & $E_0+1$      & $y_0+2$  \cr
$1/2 $  & $E_0+1$      & $y_0$    \cr
$1/2 $  & $E_0$        & $y_0$    \cr
&&\cr
\hline
&&\cr
$0$      & $E_0+3/2 $  & $y_0-1$ \cr
$0$      & $E_0+3/2 $  & $y_0+1$  \cr
$0$      & $E_0+1/2 $  & $y_0-1$ \cr
$0$      & $E_0+1/2 $  & $y_0+1$  \cr
&&\cr
      \hline 
    \end{tabular} 
\qquad 
\caption{${\cal N}=2$ long gravitino multiplet with $y_0\ge 0$}
\label{N2longgravitino}
 \end{footnotesize}
  \end{center} 
\end{table} 
%%%%%%%%%%%%%%%%%%%%%%%%%%%%%%%%%%%%%%%%%%%%%%%%%% 
\begin{table}[htbp] 
  \begin{center} 
  \begin{footnotesize}
    \begin{tabular}{|c|c|c|} 
      \hline 
      \multicolumn{3}{|c|}{} \cr 
      \multicolumn{3}{|c|}{$SD(E_0>y_0+3/2,0,y_0\ge 0\vert 2)$} \cr 
      \multicolumn{3}{|c|}{} \cr 
      \hline 
      spin & energy & hypercharge  \cr 
      \hline 
      \hline 
&&\cr
$1$      & $E_0+1$      & $y_0$    \cr
&&\cr
\hline
&&\cr
$1/2 $  & $E_0+3/2 $  & $y_0-1$   \cr
$1/2 $  & $E_0+3/2 $  & $y_0+1$    \cr
$1/2 $  & $E_0+1/2 $  & $y_0-1$  \cr
$1/2 $  & $E_0+1/2 $  & $y_0+1$   \cr
&&\cr
\hline
&&\cr
$0$      & $E_0+2$      & $y_0$     \cr
$0$      & $E_0+1$      & $y_0-2$   \cr
$0$      & $E_0+1$      & $y_0+2$    \cr
$0$      & $E_0+1$      & $y_0$      \cr
$0$      & $E_0$        & $y_0$     \cr
&&\cr
      \hline 
    \end{tabular} 
\qquad 
\caption{${\cal N}=2$ long vector multiplet with $y_0\ge 0$}
\label{N2longvector}
 \end{footnotesize}
  \end{center} 
\end{table} 
%%%%%%%%%%%%%%%%%%%%%%%%%%%%%%%%%%%%%%%%%%%%%%%%%% 
\begin{table}[htbp] 
  \begin{center} 
  \begin{footnotesize}
    \begin{tabular}{|c|c|c|} 
      \hline 
      \multicolumn{3}{|c|}{} \cr 
      \multicolumn{3}{|c|}{$SD(y_0+2,1,y_0>0\vert 2)$} \cr 
      \multicolumn{3}{|c|}{} \cr 
      \hline 
      spin & energy & hypercharge  \cr 
      \hline 
      \hline 
&&\cr
$2$      & $y_0+3$        & $y_0$         \cr
&&\cr
\hline
&&\cr
$3/2 $  & $y_0+7/2 $    & $y_0-1$      \cr
$3/2 $  & $y_0+5/2 $    & $y_0+1$            \cr
$3/2 $  & $y_0+5/2 $    & $y_0-1$             \cr
&&\cr
\hline
&&\cr
$1$      & $y_0+3    $    & $y_0-2$        \cr
$1$      & $y_0+3    $    & $y_0$   \cr
$1$      & $y_0+2      $  & $y_0$           \cr
&&\cr
\hline
&&\cr
$1/2 $  & $y_0+5/2 $  & $y_0-1$    \cr
&&\cr
      \hline 
    \end{tabular} 
\qquad 
\caption{${\cal N}=2$ short graviton multiplet with $y_0>0$}
\label{N2shortgraviton}
 \end{footnotesize}
  \end{center} 
\end{table} 
%%%%%%%%%%%%%%%%%%%%%%%%%%%%%%%%%%%%%%
\begin{table}[htbp] 
  \begin{center} 
  \begin{footnotesize}
    \begin{tabular}{|c|c|c|} 
      \hline 
      \multicolumn{3}{|c|}{} \cr 
      \multicolumn{3}{|c|}{$SD(y_0+3/2,1/2,y_0>0\vert 2)$} \cr 
      \multicolumn{3}{|c|}{} \cr 
      \hline 
      spin & energy & hypercharge  \cr 
      \hline 
      \hline 
&&\cr
$3/2 $  & $y_0+5/2 $    & $y_0$       \cr
&&\cr
\hline
&&\cr
$1$      & $y_0+3$        & $y_0-1$     \cr
$1$      & $y_0+2$        & $y_0+1$       \cr
$1$      & $y_0+2$        & $y_0-1$     \cr
&&\cr
\hline
&&\cr
$1/2 $   & $y_0+5/2 $   & $y_0$       \cr
$1/2 $  & $y_0+5/2 $    & $y_0-2$        \cr
$1/2 $  & $y_0+3/2 $    & $y_0$            \cr
&&\cr
\hline
&&\cr
$0$       &  $y_0+3$   & $y_0\pm 1$        \cr
&&\cr
      \hline 
    \end{tabular} 
\qquad 
\caption{${\cal N}=2$ short gravitino multiplet with $y_0>0$}
\label{N2shortgravitino}
 \end{footnotesize}
  \end{center} 
\end{table} 
%%%%%%%%%%%%%%%%%%%%%%%%%%%%%%%%%%%%%%
\begin{table}[htbp] 
  \begin{center} 
  \begin{footnotesize}
    \begin{tabular}{|c|c|c|} 
      \hline 
      \multicolumn{3}{|c|}{} \cr 
      \multicolumn{3}{|c|}{$SD(y_0+1,0,y_0>0\vert 2)$} \cr 
      \multicolumn{3}{|c|}{} \cr 
      \hline 
      spin & energy & hypercharge  \cr 
      \hline 
      \hline 
&&\cr
$1$      & $y_0+2$      & $y_0$      \cr
&&\cr
\hline
&&\cr
$1/2 $  & $y_0+5/2 $  & $y_0- 1$   \cr
$1/2 $  & $y_0+3/2 $  & $y_0+1$       \cr
$1/2 $  & $y_0+3/2 $  & $y_0-1$    \cr
&&\cr
\hline
&&\cr
$0$      & $y_0+2$      & $y_0-2$   \cr
$0$      & $y_0+2$      & $y_0$           \cr
$0$      & $y_0+1$      & $y_0$       \cr
&&\cr
      \hline 
    \end{tabular} 
\qquad 
\caption{${\cal N}=2$ short vector multiplet with $y_0>0$}
\label{N2shortvector}
 \end{footnotesize}
  \end{center} 
\end{table} 
%%%%%%%%%%%%%%%%%%%%%%%%%%%%%%%%%%%%%%
\begin{table}[htbp] 
  \begin{center} 
  \begin{footnotesize}
    \begin{tabular}{|c|c|c|} 
      \hline 
      \multicolumn{3}{|c|}{} \cr 
      \multicolumn{3}{|c|}{$SD(y_0,0,y_0>1/2\vert 2)$} \cr 
      \multicolumn{3}{|c|}{} \cr 
      \hline 
      spin & energy & hypercharge  \cr 
      \hline 
      \hline 
&&\cr
$1/2 $  & $y_0+1/2 $  & $y_0-1$    \\
&&\cr
\hline
&&\cr
$0$      & $y_0+1$      & $y_0-2$    \\
$0$      & $y_0$        & $y_0$           \\
&&\cr
      \hline 
    \end{tabular} 
\qquad 
\caption{${\cal N}=2$ hypermultiplet with $y_0>1/2$}
\label{N2hyper}
 \end{footnotesize}
  \end{center} 
\end{table} 
%%%%%%%%%%%%%%%%%%%%%%%%%%%%%%%%%%%%%%
\begin{table}[htbp] 
  \begin{center} 
  \begin{footnotesize}
    \begin{tabular}{|c|c|c|} 
      \hline 
      \multicolumn{3}{|c|}{} \cr 
      \multicolumn{3}{|c|}{$SD(2,1,0\vert 2)$} \cr 
      \multicolumn{3}{|c|}{} \cr 
      \hline 
      spin & energy & hypercharge  \cr 
      \hline 
      \hline 
&&\cr
$2$      & $3$        & $0$    \cr
&&\cr
\hline
&&\cr
$3/2 $  & $5/2 $    & $-1$   \cr
$3/2 $  & $5/2 $    & $+1$  \cr
&&\cr
\hline
&&\cr
$1$      & $2$        & $0$       \cr
&&\cr
      \hline 
    \end{tabular} 
\qquad 
\caption{${\cal N}=2$ massless graviton multiplet}
\label{N2masslessgraviton}
 \end{footnotesize}
  \end{center} 
\end{table} 
%%%%%%%%%%%%%%%%%%%%%%%%%%%%%%%%%%%%%%
\begin{table}[htbp] 
  \begin{center} 
  \begin{footnotesize}
    \begin{tabular}{|c|c|c|} 
      \hline 
      \multicolumn{3}{|c|}{} \cr 
      \multicolumn{3}{|c|}{$SD(3/2,1/2,0\vert 2)$} \cr 
      \multicolumn{3}{|c|}{} \cr 
      \hline 
      spin & energy & hypercharge  \cr 
      \hline 
      \hline 
&&\cr
$3/2 $      & $5/2$        & $0$    \cr
&&\cr
\hline
&&\cr
$1$  & $2$    & $-1$   \cr
$1$  & $2$    & $+1$  \cr
&&\cr
\hline
&&\cr
$1/2 $      & $3/2$        & $0$       \cr
&&\cr
      \hline 
    \end{tabular} 
\qquad 
\caption{${\cal N}=2$ massless gravitino multiplet}
\label{N2masslessgravitino}
 \end{footnotesize}
  \end{center} 
\end{table} 
%%%%%%%%%%%%%%%%%%%%%%%%%%%%%%%%%%%%%%
\begin{table}[htbp] 
  \begin{center} 
  \begin{footnotesize}
    \begin{tabular}{|c|c|c|} 
      \hline 
      \multicolumn{3}{|c|}{} \cr 
      \multicolumn{3}{|c|}{$SD(1,0,0\vert 2)$} \cr 
      \multicolumn{3}{|c|}{} \cr 
      \hline 
      spin & energy & hypercharge  \cr 
      \hline 
      \hline 
&&\cr
$1$      & $2$      & $0$             \cr
&&\cr
\hline
&&\cr
$1/2 $  & $3/2 $  & $-1$    \cr
$1/2 $  & $3/2 $  & $+1$    \cr
&&\cr
\hline
&&\cr
$0$      & $2$      & $0$     \cr
$0$      & $1$        & $0$     \cr
&&\cr
      \hline 
    \end{tabular} 
\qquad 
\caption{${\cal N}=2$ massless vector multiplet}
\label{N2masslessvector}
 \end{footnotesize}
  \end{center} 
\end{table} 
%%%%%%%%%%%%%%%%%%%%%%%%%%%%%%%
\begin{table}[htbp] 
  \begin{center} 
  \begin{footnotesize}
    \begin{tabular}{|c|c|c|} 
      \hline 
      \multicolumn{3}{|c|}{} \cr 
      \multicolumn{3}{|c|}{$SD(1/2,0,1/2\vert 2)$} \cr 
      \multicolumn{3}{|c|}{} \cr 
      \hline 
      spin & energy & hypercharge  \cr 
      \hline 
      \hline
&&\cr 
$1/2 $  & $1$  & $-1/2 $    \cr
&&\cr
\hline
&&\cr
$0$      & $1/2 $        & $1/2 $           \cr
&&\cr
      \hline 
    \end{tabular} 
\qquad 
\caption{${\cal N}=2$ supersingleton representation}
\label{N2supersingleton}
 \end{footnotesize}
  \end{center} 
\end{table} 
%%%%%%%%%%%%%%%%%%%%%%%%%%%%%%%%%%%%%%
%%%%%%%%%%%%%%%%%%%%%%%%%%%%%%%%%%%%%%%%%%%%%%%%%% 
% tables for the long graviton multiplets             % 
%%%%%%%%%%%%%%%%%%%%%%%%%%%%%%%%%%%%%%%%%%%%%%%%%% 
\begin{table}[htbp] 
  \begin{center} 
\begin{tiny}
%%%%%%%%%%%%%%%%%%%%%%%%%%%%%%%%%%%%%%%% 
    \begin{tabular}{|c|c|c|} 
      \hline 
      \multicolumn{3}{|c|}{} \cr 
      \multicolumn{3}{|c|}{$SD(E_0>J_0+3/2,1,J_0\ge 2\vert 3)$} \cr 
      \multicolumn{3}{|c|}{} \cr 
      \hline 
      spin & energy & isospin  \cr 
      \hline 
      \hline 
       &&\cr 
      $2$ & $E_0+\ft32$ & $J_0$ \cr 
       &&\cr 
      \hline 
       &&\cr 
        & $E_0+2$ & $\cases{J_0+1 \cr J_0 \cr J_0-1}$ 
      \cr 
      $\ft32$ & & \cr 
         & $E_0+1$ & $\cases{J_0+1 \cr J_0 \cr J_0-1}$ 
      \cr 
      &&\cr 
      \hline 
       &&\cr 
          & $E_0+\ft52$ & $\cases{J_0+1 \cr J_0 \cr J_0-1}$ 
    \cr 
       &&\cr 
      $1$ & $E_0+\ft32$ & 
      $\cases{J_0+2\cr J_0+1\cr J_0+1\cr J_0\cr J_0\cr J_0\cr J_0-1\cr J_0-1\cr J_0-2}$ 
    \cr 
       &&\cr 
          & $E_0+\ft12$ & 
      $\cases{J_0+1\cr J_0\cr J_0-1}$ \cr 
       &&\cr 
      \hline 
       &&\cr 
         & $E_0+3$ & $\cases{J_0}$ \cr 
       &&\cr 
         & $E_0+2$ & 
      $\cases{J_0+2\cr J_0+1\cr J_0+1\cr J_0\cr J_0\cr J_0\cr J_0-1\cr J_0-1\cr J_0-2}$ 
    \cr 
      $\ft12$ &&\cr 
       &&\cr 
         & $E_0+1$ & 
      $\cases{J_0+2\cr J_0+1\cr J_0+1\cr J_0\cr J_0\cr J_0\cr J_0-1\cr J_0-1\cr J_0-2}$ 
    \cr 
       &&\cr 
         & $E_0$ & $\cases{J_0}$ \cr 
      \hline 
       &&\cr 
          & $E_0+\ft52$ & 
      $\cases{J_0+1\cr J_0\cr J_0-1}$ \cr 
       &&\cr 
      $0$ & $E_0+\ft32$ & 
      $\cases{J_0+2\cr J_0+1\cr J_0+1\cr J_0 \cr J_0 \cr J_0-1\cr J_0-1\cr J_0-2}$ 
    \cr 
       &&\cr 
          & $E_0+\ft12$ & 
      $\cases{J_0+1\cr J_0\cr J_0-1}$\cr 
       &&\cr 
      \hline 
    \end{tabular} 
\qquad 
%%%%%%%%%%%%%%%%%%%%%%%%%%%%%%%%%%%%%%%% 
    \begin{tabular}{|c|c|c|} 
      \hline 
      \multicolumn{3}{|c|}{} \cr 
      \multicolumn{3}{|c|}{$SD(E_0>5/2,1,1\vert 3)$} \cr 
      \multicolumn{3}{|c|}{} \cr 
      \hline 
      spin & energy & isospin  \cr 
      \hline 
      \hline 
       &&\cr 
      $2$ & $E_0+\ft32$ & $1$ \cr 
       &&\cr 
      \hline 
       &&\cr 
        & $E_0+2$ & $\cases{2 \cr 1 \cr 0}$ 
      \cr 
      $\ft32$ & & \cr 
         & $E_0+1$ & $\cases{2 \cr 1 \cr 0}$ 
      \cr 
       &&\cr 
      \hline 
       &&\cr 
          & $E_0+\ft52$ & $\cases{2 \cr 1 \cr 0}$ 
    \cr 
       &&\cr 
      $1$ & $E_0+\ft32$ & 
      $\cases{3\cr 2\cr 2\cr 1\cr 1\cr 1 \cr 0}$ 
    \cr 
       &&\cr 
          & $E_0+\ft12$ & 
      $\cases{2\cr 1 \cr 0}$ \cr 
       &&\cr 
      \hline 
       &&\cr 
         & $E_0+3$ & $\cases{1}$ \cr 
       &&\cr 
         & $E_0+2$ & 
       $\cases{3\cr 2\cr 2\cr 1\cr 1\cr 1 \cr 0}$ 
    \cr 
      $\ft12$ &&\cr 
         & $E_0+1$ & 
       $\cases{3\cr 2\cr 2\cr 1\cr 1\cr 1 \cr 0}$ 
    \cr 
       &&\cr 
         & $E_0$ & $\cases{1}$ \cr 
       &&\cr 
      \hline 
       &&\cr 
          & $E_0+\ft52$ & 
      $\cases{2\cr 1 \cr 0}$ \cr 
       &&\cr 
      $0$ & $E_0+\ft32$ & 
       $\cases{3\cr 2\cr 2\cr 1\cr 1 \cr 0}$ 
    \cr 
       &&\cr 
          & $E_0+\ft12$ & 
      $\cases{2\cr 1\cr 0}$\cr 
       &&\cr 
      \hline 
    \end{tabular} 
\qquad 
%%%%%%%%%%%%%%%%%%%%%%%%%%%%%%%%%%%%%%%% 
    \begin{tabular}{|c|c|c|} 
      \hline 
      \multicolumn{3}{|c|}{} \cr 
      \multicolumn{3}{|c|}{$SD(E_0>3/2,1,0\vert 3)$} \cr 
      \multicolumn{3}{|c|}{} \cr 
      \hline 
      spin & energy & isospin  \cr 
      \hline 
      \hline 
       &&\cr 
      $2$ & $E_0+\ft32$ & $0$ \cr 
       &&\cr 
      \hline 
       &&\cr 
        & $E_0+2$ & $\cases{1}$ 
      \cr 
      $\ft32$ & & \cr 
         & $E_0+1$ & $\cases{1}$ 
      \cr 
       &&\cr 
      \hline 
       &&\cr 
          & $E_0+\ft52$ & $\cases{1}$ 
    \cr 
       &&\cr 
      $1$ & $E_0+\ft32$ & 
      $\cases{2\cr 1\cr 0}$ 
    \cr 
       &&\cr 
          & $E_0+\ft12$ & 
      $\cases{1}$ \cr 
       &&\cr 
      \hline 
       &&\cr 
         & $E_0+3$ & $\cases{0}$ \cr 
       &&\cr 
         & $E_0+2$ & 
       $\cases{2\cr 1 \cr 0}$ 
    \cr 
      $\ft12$ &&\cr 
         & $E_0+1$ & 
       $\cases{2\cr 1 \cr 0}$ 
    \cr 
       &&\cr 
         & $E_0$ & $\cases{0}$ \cr 
       &&\cr 
      \hline 
       &&\cr 
          & $E_0+\ft52$ & 
      $\cases{1}$ \cr 
       &&\cr 
      $0$ & $E_0+\ft32$ & 
       $\cases{2\cr 1}$ 
    \cr 
       &&\cr 
          & $E_0+\ft12$ & 
      $\cases{1}$\cr 
       &&\cr 
      \hline 
    \end{tabular} 
\qquad 
%%%%%%%%%%%%%%%%%%%%%%%%%%%%%%%%%%%%%%%% 
    \caption{The long ${\cal N}=3$ graviton multiplet: $E_0 > J_0 + \ft32$. 
         From left to right: $J_0\geq 2, J_0=1, J_0=0$.} 
    \label{N3longgraviton} 
    \end{tiny}
  \end{center} 
\end{table} 
%%%%%%%%%%%%%%%%%%%%%%%%%%%%%%%%%%%%%%%%%%%%%%%%%% 
% tables for the long gravitino multiplets       % 
%%%%%%%%%%%%%%%%%%%%%%%%%%%%%%%%%%%%%%%%%%%%%%%%%% 
\begin{table}[htbp] 
  \begin{center} 
  \begin{footnotesize}
    \begin{tabular}{|c|c|c|} 
      \hline 
      \multicolumn{3}{|c|}{} \cr 
      \multicolumn{3}{|c|}{$SD(E_0>J_0+1,1/2,J_0\ge 2\vert 3)$} \cr 
      \multicolumn{3}{|c|}{} \cr 
      \hline 
      spin & energy & isospin  \cr 
      \hline 
      \hline 
       &&\cr 
      $\ft32$ & $E_0+\ft32$ & $J_0$ \cr 
       &&\cr 
      \hline 
       &&\cr 
          & $E_0+2$ & $\cases{J_0+1\cr J_0\cr J_0-1}$ \cr 
      $1$ &&\cr 
          & $E_0+1$ & $\cases{J_0+1\cr J_0\cr J_0-1}$ \cr 
       &&\cr 
      \hline 
       &&\cr 
      $\ft12$ & $E_0+\ft52$ & $\cases{J_0+1\cr J_0\cr J_0-1}$ \cr 
       &&\cr 
      $\ft12$ & $E_0+\ft32$ & 
      $\cases{J_0+2\cr J_0+1\cr J_0+1\cr J_0\cr J_0 \cr J_0-1\cr J_0-1\cr J_0-2}$ \cr 
       &&\cr 
      $\ft12$ & $E_0+\ft12$ & $\cases{J_0+1\cr J_0\cr J_0-1}$ \cr 
       &&\cr 
      \hline 
       &&\cr 
         & $E_0+3$ & $\cases{J_0}$ \cr 
       &&\cr 
         & $E_0+2$ & $\cases{J_0+2\cr J_0+1\cr J_0\cr J_0\cr J_0-1\cr J_0-2}$ \cr 
      $0$ &&\cr 
         & $E_0+1$ & $\cases{J_0+2\cr J_0+1\cr J_0\cr J_0\cr J_0-1\cr J_0-2}$ \cr 
       &&\cr 
         & $E_0$ & $\cases{J_0}$ \cr 
       &&\cr 
      \hline 
    \end{tabular} 
\qquad 
%%%%%%%%%%%%%%%%%%%%%%%%%%%%%%%%%%%%%%%% 
% long gravitino J_0=1                   % 
%%%%%%%%%%%%%%%%%%%%%%%%%%%%%%%%%%%%%%%% 
    \begin{tabular}{|c|c|c|} 
      \hline 
      \multicolumn{3}{|c|}{} \cr 
      \multicolumn{3}{|c|}{$SD(E_0>2,0,1\vert 3)$} \cr 
      \multicolumn{3}{|c|}{} \cr 
      \hline 
      spin & energy & isospin  \cr 
      \hline 
      \hline 
       &&\cr 
      $\ft32$ & $E_0+\ft32$ & $1$ \cr 
       &&\cr 
      \hline 
       &&\cr 
          & $E_0+2$ & $\cases{2\cr 1\cr 0}$ \cr 
      $1$ &&\cr 
          & $E_0+1$ & $\cases{2\cr 1\cr 0}$ \cr 
       &&\cr 
      \hline 
       &&\cr 
      $\ft12$ & $E_0+\ft52$ & $\cases{2\cr 1\cr 0}$ \cr 
       &&\cr 
      $\ft12$ & $E_0+\ft32$ & 
      $\cases{3\cr 2\cr 2\cr 1\cr 1 \cr 0}$ \cr 
       &&\cr 
      $\ft12$ & $E_0+\ft12$ & $\cases{2\cr 1\cr 0}$ \cr 
       &&\cr 
      \hline 
       &&\cr 
         & $E_0+3$ & $\cases{1}$ \cr 
       &&\cr 
         & $E_0+2$ & $\cases{3\cr 2\cr 1\cr 1}$ \cr 
      $0$ &&\cr 
         & $E_0+1$ & $\cases{3\cr 2\cr 1\cr 1}$ \cr 
       &&\cr 
         & $E_0$ & $\cases{1}$ \cr 
       &&\cr 
      \hline 
    \end{tabular} 
\qquad 
%%%%%%%%%%%%%%%%%%%%%%%%%%%%%%%%%%%%%%%% 
% long gravitino J_0=0                   % 
%%%%%%%%%%%%%%%%%%%%%%%%%%%%%%%%%%%%%%%% 
    \begin{tabular}{|c|c|c|} 
      \hline 
      \multicolumn{3}{|c|}{} \cr 
      \multicolumn{3}{|c|}{$SD(E_0>1,0,0\vert 3)$} \cr 
      \multicolumn{3}{|c|}{} \cr 
      \hline 
      spin & energy & isospin  \cr 
      \hline 
      \hline 
       &&\cr 
      $\ft32$ & $E_0+\ft32$ & $0$ \cr 
       &&\cr 
      \hline 
       &&\cr 
          & $E_0+2$ & $\cases{1}$ \cr 
      $1$ &&\cr 
          & $E_0+1$ & $\cases{1}$ \cr 
       &&\cr 
      \hline 
       &&\cr 
      $\ft12$ & $E_0+\ft52$ & $\cases{1}$ \cr 
       &&\cr 
      $\ft12$ & $E_0+\ft32$ & 
      $\cases{2\cr 1}$ \cr 
       &&\cr 
      $\ft12$ & $E_0+\ft12$ & $\cases{1}$ \cr 
       &&\cr 
      \hline 
       &&\cr 
         & $E_0+3$ & $\cases{0}$ \cr 
       &&\cr 
         & $E_0+2$ & $\cases{2\cr 0}$ \cr 
      $0$ &&\cr 
         & $E_0+1$ & $\cases{2\cr 0}$ \cr 
       &&\cr 
         & $E_0$ & $\cases{0}$ \cr 
       &&\cr 
      \hline 
    \end{tabular} 
    \caption{The long ${\cal N}=3$ gravitino multiplet: $E_0 > J_0 + 1$. 
         From left to right: $J_0\geq 2, J_0=1, J_0=0$.} 
    \label{N3longgravitino} 
    \end{footnotesize}
  \end{center} 
\end{table} 
%%%%%%%%%%%%%%%%%%%%%%%%%%%%%%%%%%%%%%%% 
% tables for the short graviton multiplets 
%%%%%%%%%%%%%%%%%%%%%%%%%%%%%%%%%%%%%%%% 
\begin{table}[htbp] 
  \begin{center}  
  \begin{footnotesize}
%%%%%%%%%%%%%%%%%%%%%%%%%%%%%%%%%%%%%%%% 
    \begin{tabular}{|c|c|c|} 
      \hline 
      \multicolumn{3}{|c|}{} \cr 
      \multicolumn{3}{|c|}{$SD(J_0+3/2,1/2,J_0\ge 2\vert 3)$} \cr 
      \multicolumn{3}{|c|}{} \cr 
      \hline 
      spin & energy & isospin  \cr 
      \hline 
      \hline 
       &&\cr 
      $2$ & $J_0+3$ & $J_0$ \cr 
       &&\cr 
      \hline 
       &&\cr 
        & $J_0+\ft72$ & $\cases{ J_0 \cr J_0-1}$ 
      \cr 
      $\ft32$ & & \cr 
         & $J_0+\ft52$ & $\cases{J_0+1 \cr J_0 \cr J_0-1}$ 
      \cr 
       &&\cr 
      \hline 
       &&\cr 
          & $J_0+4$ & $\cases{J_0-1}$ 
    \cr 
       &&\cr 
      $1$ & $J_0+3$ & 
      $\cases{J_0+1\cr J_0\cr J_0\cr J_0-1\cr J_0-1\cr J_0-2}$ 
    \cr 
       &&\cr 
          & $J_0+2$ & 
      $\cases{J_0+1\cr J_0\cr J_0-1}$ \cr 
       &&\cr 
      \hline 
       &&\cr 
         & $J_0+\ft72$ & 
      $\cases{ J_0\cr J_0-1\cr J_0-2}$ 
    \cr 
      $\ft12$ &&\cr 
         & $J_0+\ft52$ & 
      $\cases{ J_0+1\cr  J_0\cr J_0\cr J_0-1\cr J_0-1\cr J_0-2}$ 
    \cr 
       &&\cr 
         & $J_0+\ft32$ & $\cases{J_0}$ \cr 
       &&\cr 
      \hline 
       &&\cr 
      $0$ & $J_0+3$ & 
      $\cases{J_0\cr J_0-1\cr J_0-2}$ 
    \cr 
       &&\cr 
          & $J_0+2$ & 
      $\cases{ J_0\cr J_0-1}$\cr 
       &&\cr 
      \hline 
    \end{tabular} 
\qquad 
%%%%%%%%%%%%%%%%%%%%%%%%%%%%%%%%%%%%%%%% 
    \begin{tabular}{|c|c|c|} 
      \hline 
      \multicolumn{3}{|c|}{} \cr 
      \multicolumn{3}{|c|}{$SD(5/2,1/2,1\vert 3)$} \cr 
      \multicolumn{3}{|c|}{} \cr 
      \hline 
      spin & energy & isospin  \cr 
      \hline 
      \hline 
       &&\cr 
      $2$ & $4$ & $1$ \cr 
       &&\cr 
      \hline 
       &&\cr 
        & $\ft92$ & $\cases{1\cr 0}$ 
      \cr 
      $\ft32$ & & \cr 
         & $\ft72$ & $\cases{2\cr 1\cr 0}$ 
      \cr 
       &&\cr 
      \hline 
       &&\cr 
          & $5$ & $\cases{0}$ 
    \cr 
       &&\cr 
      $1$ & $4$ & 
      $\cases{2\cr 1\cr 1\cr 0}$ 
    \cr 
       &&\cr 
          & $3$ & 
      $\cases{2\cr 1\cr 0}$ \cr 
       &&\cr 
      \hline 
       &&\cr 
         & $\ft92$ & 
      $\cases{1}$ 
    \cr 
      $\ft12$ &&\cr 
         & $\ft72$ & 
      $\cases{2\cr 1\cr 1\cr 0}$ 
    \cr 
       &&\cr 
         & $\ft52$ & $\cases{1}$ \cr 
       &&\cr 
      \hline 
       &&\cr 
      $0$ & $4$ & 
      $\cases{1}$ 
    \cr 
       &&\cr 
          & $3$ & 
      $\cases{1 \cr 0}$\cr 
       &&\cr 
      \hline 
    \end{tabular} 
\qquad 
%%%%%%%%%%%%%%%%%%%%%%%%%%%%%%%%%%%%%%%% 
    \begin{tabular}{|c|c|c|} 
      \hline 
      \multicolumn{3}{|c|}{} \cr 
      \multicolumn{3}{|c|}{$SD(3/2,1/2,0\vert 3)$} \cr 
      \multicolumn{3}{|c|}{} \cr 
      \hline 
      spin & energy & isospin  \cr 
      \hline 
      \hline 
       &&\cr 
      $2$ & $3$ & $0$ \cr 
       &&\cr 
      \hline 
       &&\cr 
      $\ft32$ 
         & $\ft52$ & $\cases{1}$ 
      \cr 
       &&\cr 
      \hline 
       &&\cr 
      $1$  & $2$ & 
      $\cases{1}$ \cr 
       &&\cr 
      \hline 
       &&\cr 
      $\ft12$  & $\ft32$ & $\cases{0}$ \cr 
       &&\cr 
      \hline 
    \end{tabular} 
\qquad 
%%%%%%%%%%%%%%%%%%%%%%%%%%%%%%%%%%%%%%%% 
    \caption{The short ${\cal N}=3$ graviton multiplet: $E_0 = J_0 + \ft32$. 
         From left to right: $J_0\geq 2, J_0=1,$ and $J_0=0$ 
         (that is massless).} 
    \label{N3shortgraviton} 
    \end{footnotesize}
  \end{center} 
\end{table} 
%%%%%%%%%%%%%%%%%%%%%%%%%%%%%%%%%%%%%%%% 
% tables for the short gravitino multiplets                      % 
%%%%%%%%%%%%%%%%%%%%%%%%%%%%%%%%%%%%%%%% 
\begin{table}[htbp] 
  \begin{center} 
  \begin{footnotesize}
    \begin{tabular}{|c|c|c|} 
      \hline 
      \multicolumn{3}{|c|}{} \cr 
      \multicolumn{3}{|c|}{$SD(J_0+1,0,J_0\ge 2\vert 3)$} \cr 
      \multicolumn{3}{|c|}{} \cr 
      \hline 
      spin & energy & isospin  \cr 
      \hline 
      \hline 
       &&\cr 
      $\ft32$ & $J_0+\ft52$ & $J_0$ \cr 
       &&\cr 
      \hline 
       &&\cr 
          & $J_0+3$ & $\cases{J_0\cr J_0-1}$ \cr 
      $1$ &&\cr 
          & $J_0+2$ & $\cases{J_0+1\cr J_0\cr J_0-1}$ \cr 
       &&\cr 
      \hline 
       &&\cr 
      $\ft12$ & $J_0+\ft72$ & $\cases{J_0-1}$ \cr 
       &&\cr 
      $\ft12$ & $J_0+\ft52$ & 
      $\cases{J_0+1\cr J_0\cr J_0 \cr J_0-1\cr J_0-1\cr J_0-2}$ \cr 
       &&\cr 
      $\ft12$ & $J_0+\ft32$ & $\cases{J_0+1\cr J_0\cr J_0-1}$ \cr 
       &&\cr 
      \hline 
       &&\cr 
         & $J_0+3$ & $\cases{J_0\cr J_0-1\cr J_0-2}$ \cr 
      $0$ &&\cr 
         & $J_0+2$ & $\cases{J_0+1\cr J_0\cr J_0\cr J_0-1\cr J_0-2}$ \cr 
       &&\cr 
         & $J_0+1$ & $\cases{J_0}$ \cr 
       &&\cr 
      \hline 
    \end{tabular} 
\qquad 
%%%%%%%%%%%%%%%%%%%%%%%%%%%%%%%%%%%%%%%% 
% short gravitino J_0=1                  % 
%%%%%%%%%%%%%%%%%%%%%%%%%%%%%%%%%%%%%%%% 
    \begin{tabular}{|c|c|c|} 
      \hline 
      \multicolumn{3}{|c|}{} \cr 
      \multicolumn{3}{|c|}{$SD(2,0,1\vert 3)$} \cr 
      \multicolumn{3}{|c|}{} \cr 
      \hline 
      spin & energy & isospin  \cr 
      \hline 
      \hline 
       &&\cr 
      $\ft32$ & $\ft72$ & $1$ \cr 
       &&\cr 
      \hline 
       &&\cr 
          & $4$ & $\cases{1\cr 0}$ \cr 
      $1$ &&\cr 
          & $3$ & $\cases{2\cr 1\cr 0}$ \cr 
       &&\cr 
      \hline 
       &&\cr 
      $\ft12$ & $\ft92$ & $\cases{0}$ \cr 
       &&\cr 
      $\ft12$ & $\ft72$ & 
      $\cases{2\cr 1\cr 1 \cr 0}$ \cr 
       &&\cr 
      $\ft12$ & $\ft52$ & $\cases{2\cr 1\cr 0}$ \cr 
       &&\cr 
      \hline 
       &&\cr 
         & $4$ & $\cases{1}$ \cr 
      $0$ &&\cr 
         & $3$ & $\cases{2\cr 1\cr 1}$ \cr 
       &&\cr 
         & $2$ & $\cases{1}$ \cr 
       &&\cr 
      \hline 
    \end{tabular} 
\qquad 
%%%%%%%%%%%%%%%%%%%%%%%%%%%%%%%%%%%%%%%% 
% short gravitino J_0=0                  % 
%%%%%%%%%%%%%%%%%%%%%%%%%%%%%%%%%%%%%%%% 
    \begin{tabular}{|c|c|c|} 
      \hline 
      \multicolumn{3}{|c|}{} \cr 
      \multicolumn{3}{|c|}{$SD(1,0,0\vert 3)$} \cr 
      \multicolumn{3}{|c|}{} \cr 
      \hline 
      spin & energy & isospin  \cr 
      \hline 
      \hline 
       &&\cr 
      $\ft32$ & $\ft52$ & $0$ \cr 
       &&\cr 
      \hline 
       &&\cr 
      $1$ & $2$ & $\cases{1}$ \cr 
       &&\cr 
      \hline 
       &&\cr 
      $\ft12$ & $\ft32$ & $\cases{1}$ \cr 
       &&\cr 
      \hline 
       &&\cr 
         & $2$ & $\cases{0}$ \cr 
    $0$ &  & \cr 
         & $1$ & $\cases{0}$ \cr 
       &&\cr 
      \hline 
    \end{tabular} 
    \caption{The short ${\cal N}=3$ gravitino multiplet: $E_0 = J_0 + \ft32$. 
    From left to right, $J_0\geq 2, J_0=1,$ and $J_0=0$ (that is massless).} 
    \label{N3shortgravitino} 
    \end{footnotesize}
  \end{center} 
\end{table} 
%%%%%%%%%%%%%%%%%%%%%%%%%%%%%%%%%%%%%%%% 
% table for the vector multiplets           % 
%%%%%%%%%%%%%%%%%%%%%%%%%%%%%%%%%%%%%%%% 
\begin{table}[htbp] 
  \begin{center} 
  \begin{footnotesize}
%%%%%%%%%%%%%%%%%%%%%%%%%%%%%%%%%%%%%%%% 
% massive vector multiplet             % 
%%%%%%%%%%%%%%%%%%%%%%%%%%%%%%%%%%%%%%%% 
    \begin{tabular}{|c|c|c|} 
      \hline 
      \multicolumn{3}{|c|}{} \cr 
      \multicolumn{3}{|c|}{$SD(J_0,0,J_0\ge 2\vert 3)$} \cr 
      \multicolumn{3}{|c|}{} \cr 
      \hline 
      spin & energy & isospin  \cr 
      \hline 
      \hline 
       &&\cr 
      $1$  & $J_0+1$ & $\cases{J_0-1}$ \cr 
       &&\cr 
      \hline 
       &&\cr 
      $\ft12$ & $J_0+\ft32$ & 
      $\cases{J_0-1\cr J_0-2}$ \cr 
       &&\cr 
      $\ft12$ & $J_0+\ft12$ & $\cases{J_0\cr J_0-1}$ \cr 
       &&\cr 
      \hline 
       &&\cr 
         & $J_0+2$ & $\cases{J_0-2}$ \cr 
      $0$ &&\cr 
         & $J_0+1$ & $\cases{J_0\cr J_0-1\cr J_0-2}$ \cr 
       &&\cr 
         & $J_0$ & $\cases{J_0}$ \cr 
       &&\cr 
      \hline 
    \end{tabular} 
\qquad 
%%%%%%%%%%%%%%%%%%%%%%%%%%%%%%%%%%%%%%%% 
% massless vector multiplet            % 
%%%%%%%%%%%%%%%%%%%%%%%%%%%%%%%%%%%%%%%% 
    \begin{tabular}{|c|c|c|} 
      \hline 
      \multicolumn{3}{|c|}{} \cr 
      \multicolumn{3}{|c|}{$SD(1,0,1\vert 3)$} \cr 
      \multicolumn{3}{|c|}{} \cr 
      \hline 
      spin & energy & isospin  \cr 
      \hline 
      \hline 
       &&\cr 
      $1$  & $2$ & $\cases{0}$ \cr 
       &&\cr 
      \hline 
       &&\cr 
      $\ft12$ & $\ft32$ & $\cases{1\cr 0}$ \cr 
       &&\cr 
      \hline 
       &&\cr 
      $0$   & $2$ & $\cases{1}$ \cr 
       &&\cr 
         & $1$ & $\cases{1}$ \cr 
       &&\cr 
      \hline 
    \end{tabular} 
\qquad 
%%%%%%%%%%%%%%%%%%%%%%%%%%%%%%%%%%%%%%%% 
% supersingleton representation            % 
%%%%%%%%%%%%%%%%%%%%%%%%%%%%%%%%%%%%%%%% 
    \begin{tabular}{|c|c|c|} 
      \hline 
      \multicolumn{3}{|c|}{} \cr 
      \multicolumn{3}{|c|}{$SD(1/2,0,1/2\vert 3)$} \cr 
      \multicolumn{3}{|c|}{} \cr 
      \hline 
      spin & energy & isospin  \cr 
      \hline 
      \hline 
       &&\cr 
      $\ft12$ & $1$ & $\cases{\ft12}$ \cr 
       &&\cr 
      \hline 
       &&\cr 
      $0$   & $\ft12$ & $\cases{\ft12}$ \cr 
       &&\cr 
      \hline 
    \end{tabular} 
    \caption{The ${\cal N}=3$ vector multiplets: $E_0 = J_0$. The massive 
    vector multiplet with $J_0\geq 2$, the massless vector multiplet with $J_0=1$ 
    and the supersingleton representation with $J_0=1/2$.} 
    \label{N3shortvector} 
    \end{footnotesize}
  \end{center} 
\end{table} 
\par
\chapter{The complete spectra of the $AdS_4\times\left(G\over H\right)_7$ 
solutions from harmonic analysis}
\par
In this chapter I consider the Freund Rubin solutions of
eleven dimensional supergravity compactified on backgrounds
\be
AdS_4\times X_7
\ee
in the three cases
\beq
X_7&=M^{111}&~~~~ \ll(\cN=2\rr)\nn\\
X_7&=Q^{111}&~~~~ \ll(\cN=2\rr)\nn\\
X_7&=N^{010}&~~~~ \ll(\cN=3\rr).
\label{threespaces}
\eeq
The complete mass spectra of the corresponding four dimensional supergravities
are determined,
by means of harmonic analysis. All the particles found fit into supermultiplets, and such
supermultiplets are organized into UIRs of the flavour group $G'$ (\ref{ospgp}).
I stress that harmonic analysis enable us to solve this problem by means of group
theory and differential geometry, without solving differential equations.
\par
In section $1$ I describe the Freund Rubin compactifications with
$X_7=G/H$ and discuss their symmetries.
In section $2$ I
define and describe the $M^{111}$ manifold and, in less detail, the $Q^{111}$ and $N^{010}$ manifolds.
However, a further description of $M^{111}$ and $Q^{111}$ is given in the next chapter. In 
section $3$ I review the theory of harmonic analysis on coset spaces, and how it can be
applied to derive the mass spectra of Freund Rubin supergravities. In section $4$ I describe the explicit derivation of 
the complete mass spectrum of $AdS_4\times M^{111}$ supergravity. In section $5$ I give the complete mass spectrum 
of $AdS_4\times N^{010}$ supergravity, without describing its derivation by harmonic analysis; furthermore, I give a part 
of the mass spectrum of $AdS_4\times Q^{111}$ supergravity, which has been found long ago without the help of harmonic analysis.
Part of the content of the present chapter refers to results obtained within the collaborations
\cite{noi1}, \cite{noi4}. 
\par
\section{Supergravity on $AdS_4\times G/H_7$}
\par
\subsection{A summary of coset space differential geometry}
\par
Here I sketch very briefly the basic ideas of differential geometry on coset spaces,
with some results that will be used afterwards. For the proof of these results and
for a complete discussion of these topics, see \cite{castdauriafre},
\cite{vannieuw}, \cite{newcast}.
\par
\subsubsection{Definitions}
\par
Let us consider a coset manifold $G/H$, whose dimension is $n={\rm dim}G-{\rm dim}H$.
It can be parametrized by the $n$ coordinates $y^{\a}$, on which the coset
representatives $L\ll(y\rr)$ do depend. Under left multiplication by $g\in G$ we have:
\be
gL\ll(y\rr)=L\ll(y'\rr)h\ll(y\rr).
\ee
The Lie algebra $\IG$ of the group $G$ admits the following orthogonal split:
\beq
&\IG=\IH\oplus\IK,\nn\\
&T_i\in\IH,~T_a\in\IK,~T_{\Lambda}\in\IG
\eeq
where $\IH$ contains the generators of $H$ and $\IK$ the remaining $n$ generators.
Then we can express the elements of $G$ as $g=e^{y^aT_a}e^{x^iT_i}$, and the
coset representatives as $L\ll(y\rr)=e^{y^aT_a}$. 
\par
We will consider {\it reductive} coset manifolds, namely, such that
\be
\label{reductve}
\ll[\IH,\IK\rr]\subset\IK.
\ee
Furthermore, we will consider semisimple coset manifolds.
\par
Since $G/H$ is reductive, the $n$ generators $T_a\in\IK$ are in a representation of $H$,
realized by means of the structure constants 
\be
\label{HinSOn}
C^{~~b}_{ia}=-\ll(T^H_i\rr)_a^{~b}=\ll(T^H_i\rr)_{~a}^b
\ee
(since by Jacobi identity $C^{~~c}_{ia}C^{~~b}_{jc}={1\over 2}C^{~~k}_{ij}C^{~~b}_{ak}$).
Being $G$ semisimple we have $C^{~~b}_{ia}=C_{iab}=C_{i\ll[ab\rr]}$ (in a basis in which the 
Killing metric is the Kronecker delta), so the $T^H$ are 
also $SO\ll(n\rr)$ generators in the fundamental representation. Hence,
\be
H\subset SO\ll(n\rr)
\ee
and this embedding is realized by the generators (\ref{HinSOn}). Notice that in the cases studied in this thesis we set $n=7$.
\par
\subsubsection{Killing vectors}
\par
The  transformation law $gL\ll(y\rr)=L\ll(y'\rr)h\ll(y\rr)$ for infinitesimal $g$ becomes
\be
T_{\Lambda}L\ll(y\rr)=K_{\Lambda}\ll(y\rr)L\ll(y\rr)-L\ll(y\rr)T_iW_{\Lambda}^i\ll(y\rr)
\label{killingonL}
\ee
where 
\beq
g&=&1+\epsilon^{\Lambda}T_{\Lambda}\nn\\
h&=&1-\epsilon^{\Lambda}W^i_{\Lambda}\ll(y\rr)T_i\nn\\
{y'}^a&=&y^a+\epsilon^{\Lambda}K_{\Lambda}^a\ll(y\rr).
\eeq
The $y$--dependent matrices $W^i_{\Lambda}\ll(y\rr)$ are called {\it $H$--compensators}, and the $y$--dependent
differential operators
\be
K_{\Lambda}\ll(y\rr)\equiv K_{\Lambda}^a\ll(y\rr){\partial\over\partial y^a}
\ee
are called {\it Killing vectors on} $G/H$. We have
\beq
\ll[T_{\Lambda},T_{\Sigma}\rr]L\ll(y\rr)=C_{\Lambda\Sigma}^{~~\Delta}T_{\Delta}L\ll(y\rr),\nn\\
\ll[K_{\Lambda},K_{\Sigma}\rr]=-C_{\Lambda\Sigma}^{~~\Delta}K_{\Delta}.
\eeq
\par
\subsubsection{Vielbein, $H$--connection, $H$ Lie derivative}
\par
The one--form 
\be
\Omega\ll(y\rr)=L^{-1}\ll(y\rr)dL\ll(y\rr)
\ee
is $\IG$--valued, and can be expanded in a generator basis as follows:
\be
\Omega\ll(y\rr)=\cB^a\ll(y\rr)T_a+\Omega^i\ll(y\rr)T_i
\ee
where $\cB^a\ll(y\rr)=\cB_{\a}^{~a}\ll(y\rr)dy^{\a}$ is a vielbein on $G/H$ and
$\Omega^i\ll(y\rr)=\Omega_{\a}^{~i}\ll(y\rr)dy^{\a}$ is called the {\it $H$--connection}.
Under left multiplication of an infinitesimal $g\in G$ this vielbein transforms as
\beq
\cB^a\ll(y+\delta y\rr)&=&\cB^a\ll(y\rr)-\epsilon^{\Lambda}W_{\Lambda}^i\ll(y\rr)C^{~~a}_{ib}\cB^b\ll(y\rr)\nn\\
\delta y^a&=&\epsilon^{\Lambda}K_{\Lambda}^a\ll(y\rr).
\label{leftinvariantvielbein}
\eeq
A vielbein transforming as above, namely, $G$--invariant modulo an $H$--compensator,
is named a {\it $G$--left invariant vielbein}.
Notice that the left action of $G$ on $\cB^a\ll(y\rr)$ is an $H$ transformation 
in the fundamental representation of $SO\ll(n\rr)$. We can also define the metric on $G/H$,
\be
\label{metricGH}
g_{\a\b}\ll(y\rr)=\gamma_{ab}\cB_{\a}^a\ll(y\rr)\cB_{\b}^b\ll(y\rr)
\ee
(where $\gamma_{ab}$ is the Killing metric of $G$ restricted to $G/H$); it can be shown that this metric is
left $G$--invariant, so $G$ is an isometry of this metric; furthermore, this metric
is insensitive to the choice of the coset parametrization.
\par
The $H$--connection defines a parallel transport on the coset manifold, and then an $H$--covariant derivative
\be
\cD^H=d+\Omega^iT_i.
\ee
It can be written in terms of the embedding (\ref{HinSOn}) $H\subset SO\ll(n\rr)$:
\be
\cD^H=d+\Omega^i\ll(T_i\rr)^{ab}t^{SO\ll(n\rr)}_{ab}
\ee
where $t^{SO\ll(n\rr)}_{ab}$ are the $SO\ll(n\rr)$ generators. For example, for the vector representation
they are $\ll(t^{SO\ll(n\rr)}_{ab}\rr)^{cd}=-\ii\delta^{cd}_{ab}$, while for the spinor representation
they are given by the two-indices gamma matrices.
\par
Let us determine the action of the $H$--covariant derivative on the inverse of the coset representative. Being
\be
LdL^{-1}=-dLL^{-1},
\ee
we have
\be
\Omega L^{-1}=L^{-1}dLL^{-1}=-dL^{-1}=\ll(\Omega^iT_i+\cB^aT_a\rr)L^{-1},
\ee
then
\be
\label{Hcovaction}
\cD^HL^{-1}=\ll(d+\Omega^iT_i\rr)L^{-1}=-\cB^aT_aL^{-1}.
\ee
This action is purely algebraic. As we will see, such a property is the core of the harmonic analysis method
for solving differential equations.
\par
The Lie derivative associated to a Killing vector, acting on the vielbein, is
\be
l_{K_{\Lambda}}\cB^a\ll(y\rr)=W_{\Lambda}^i\ll(y\rr)C^{~~a}_{ib}\cB^b\ll(y\rr).
\ee
Then, if we define the {\it $H$--covariant Lie derivative}
\be
\cL_{K_{\Lambda}}\equiv l_{K_{\Lambda}}-W_{\Lambda}^i\ll(y\rr)T_i,
\ee
which satisfies all the properties of the Lie derivative, we have
\be
\cL_{K_{\Lambda}}\cB^a\ll(y\rr)=0.
\ee
On the coset representative the action of the $H$--covariant Lie derivative is
\be
\label{L=T}
\cL_{K_{\Lambda}}L\ll(y\rr)=T_{\Lambda}L\ll(y\rr).
\ee
\par
\subsubsection{Spin connection}
\par
Another useful structure we can build on our coset manifold is a Riemannian connection $\cB^a_{~b}$,
or {\it spin connection},
that defines a parallel transport. It is an $so\ll(n\rr)$--valued one--form defined by the vanishing torsion equation
\be
\cR^a\equiv d\cB^a-\cB^{ab}\wedge\cB_b=0.
\ee
The Riemannian curvature $\cR^a_{~b}$ is an $so\ll(n\rr)$--valued two--form defined by 
\be
\cR^{ab}\equiv d\cB^{ab}-\cB^{ac}\wedge\cB_c^{~b}=\cR^{ab}_{~~cd}\cB^c\wedge\cB^d.
\ee
Notice that in this way we define a parallel transport by $SO\ll(n\rr)$ transformations on the
vielbein; in other words, we select the $SO\ll(n\rr)$ group as the tangent group. The
vielbein is then in the vector $SO\ll(n\rr)$ representation, and all the fields on the
manifold are in $SO\ll(n\rr)$ representations. Being $H\subset SO\ll(n\rr)$, the
fields in irreducible representations of $SO\ll(n\rr)$ can be branched in fields 
in irreducible representations of $H$.
\par
Expanding the spin connection one finds
\footnote{in the following we call $T^H_i$ the generators of $\IH$ and $T_a^K$ the generators of $\IK$, for clarity of notations}
\be
\cB^a_{~b}=-C^{~~a}_{ib}\Omega^i+{1\over 2}C^{~~a}_{bc}\cB^c=-\ll(T_i^H\rr)^a_{~b}
\Omega^i+{1\over 2}C^{~~a}_{bc}\cB^c.
\label{spinexpansion}
\ee
The first term in this expression is valued in $\IH\subset so\ll(n\rr)$ (whose generators
are the $C^{~~a}_{bi}$), while the last term is valued in other $so\ll(n\rr)$ generators.
In other words, the spin connection contains the $H$--connection plus other $so\ll(n\rr)$--valued
terms.
\par
The spin connection naturally defines a $SO\ll(n\rr)$ covariant derivative
\be
\label{covder}
\cD^{SO\ll(n\rr)}=d-\cB^{ab}t^{SO\ll(n\rr)}_{ab}.
\ee
Substituting the (\ref{spinexpansion}), we find an expression of the form 
\beq
\cD^{SO\ll(n\rr)}&=&d+\ll(T_i^H\rr)^{ab}\Omega^it^{SO\ll(n\rr)}_{ab}+{1\over 2}C^{~~a}_{bc}\cB^c{t^{SO\ll(n\rr)}}_a^{~b}=\nn\\
&=&\cD^{H}+\IM_c\cB^c.
\label{defIM}
\eeq
A very useful property of the $SO\ll(n\rr)$ covariant derivative is that it commutes with the $H$--covariant Lie derivative:
\be
\ll[\cL_{K_{\Lambda}},\cD^{SO\ll(n\rr)}\rr]=0.
\label{LDDL}
\ee
It follows that, because of the Schur's lemma, $\cD^{SO\ll(n\rr)}$ acts irreducibly on $G$ representations, namely, it cannot
change a representation of $G$ in another one.
\par
\subsubsection{Rescalings}
\par
In general the metric (\ref{metricGH}) is not the only $G$--invariant metric on $G/H$;
it is unique only up to some particular {\it rescalings} of the vielbein
\be
\cB^a=r^a{\cB'}^a~~~\hbox{no sum on $a$}.
\ee
The (\ref{spinexpansion}) changes with rescalings by coefficients depending on the $r^a$'s, and the same happens
to the matrices $\IM$, but the properties (\ref{Hcovaction}), (\ref{LDDL})
remain satisfied. By means of these rescalings, it is sometimes
possible to obtain an Einstein metric, namely, a metric such that
\be
\cR^a_{~b}=\Lambda\delta^a_{b},
\ee
even if the non--rescaled metric is non--Einstein. Furthermore, by a global vielbein rescaling one can choose the value
of $\Lambda$. 
\par
In the following, to avoid confusion between the not rescaled vielbein and the rescaled one, we call $\Omega^a$ the
not rescaled vielbein, namely,
\be
\Omega\equiv L^{-1}dL=\Omega^iT^H_i+\Omega^aT^K_a
\label{omegadef}
\ee
and $\cB^a$ the rescaled vielbein
\be
\cB_a={1\over r_a}\Omega_a.
\label{rescaling}
\ee
So, for example, the (\ref{Hcovaction}) becomes
\be
\cD^HL^{-1}=-\Omega^aT_a^KL^{-1}=-r_a\cB^aT_a^KL^{-1},
\ee
or, expanding on the vielbein $\cD^H=\cB^a\cD^H_a$,
\be
\label{rescHcovaction}
\cD_a^HL^{-1}=-r_aT_aL^{-1}.
\ee
\par
\subsection{The Freund Rubin solution}
\par
As I said in chapter $1$, given a seven dimensional compact coset manifold $G/H$,
the (\ref{FreundRubin}) is a solution of eleven dimensional supergravity, with the
geometry of $AdS_4\times G/H$, and can be viewed as a four dimensional anti--de Sitter supergravity
with internal space $G/H$.
\par
Let us express this in the formalism of rheonomy (for a review on rheonomy, see \cite{castdauriafre}).
We use here the following conventions:
\beq
m,n&\hbox{flat indices on }AdS_4\nn\\
a,b&\hbox{flat indices on }G/H\nn\\
\mu\nu&\hbox{curved indices on }AdS_4\nn\\
\a,\b&\hbox{curved indices on }G/H\nn\\
\hat{a},\hat{b}&\hbox{eleven dimensional flat indices}\nn\\
M,N&SO\ll(\cN\rr)~\hbox{indices}\nn\\
x^{\mu}&\hbox{coordinates on }AdS_4\nn\\
y^{\a}&\hbox{coordinates on }G/H\nn\\
\theta&\hbox{fermionic coordinates in $AdS_4$ superspace}.
\eeq
We call $\tau_a$ the $SO\ll(7\rr)$ gamma matrices, which are $8\times 8$ and act on the
$G/H$ spinors (which are in the spinor representation of $SO\ll(7\rr)$):
\be
\ll\{\tau_a,\tau_b\rr\}=2\eta_{ab}=2{\rm ~diag}\ll(-,-,-,-,-,-,-\rr).
\ee
We suppose that it is possible to define an Einstein metric structure on $G/H$ such that
\be
\cR^{ac}_{~~bc}=12e^2\delta^a_b.
\ee
The $SO\ll(7\rr)$ generators in the spinor representation are
\be
t^{SO\ll(7\rr)}_{ab}={1\over 4}\tau_{ab}={1\over 8}\ll[\tau_a,\tau_b\rr].
\ee
\par
We call $\cN$ the number of independent $SO\ll(7\rr)$ real spinors $\eta_M\ll(y\rr)$ satisfying the equation
\be
\label{killspeqGH}
\cD^{SO\ll(7\rr)}\eta_M=\ll(d-{1\over 4}\cB^{ab}\tau_{ab}\rr)\eta_M=e\cB^a\tau_a\eta_M.
\ee
Notice that the $\eta_M$, which we call {\it Killing spinors on $G/H$}, are made of $\IC$--numbers, not
of grassmannian variables, and are then commuting.
\par
Let us consider now the $\cN$ extended $AdS_4$ supergravity. Its superspace is
\be
\cM_{4\cN\vert 4}={Osp\ll(\cN\vert 4\rr)\over SO\ll(1,3\rr)\times SO\ll(\cN\rr)}.
\ee
Let $\stackrel{o}{V}^m\!\ll(x,\theta\rr)$, $~\stackrel{o}{\omega}^{mn}\!\ll(x,\theta\rr)$, 
$~\stackrel{o}{A}^{MN}\!\ll(x,\theta\rr)$, $~\stackrel{o}{\psi}_M\!\ll(x,\theta\rr)$ be
the left--invariant one forms on $\cM_{4\cN\vert 4}$. They fulfill by definition the following Maurer Cartan
equations
\beq
d\stackrel{o}{V}^m-\stackrel{o}{\omega}^m_{~n}\wedge\stackrel{o}{V}^n-{1\over 2}\ii\stackrel{o}{\bar{\psi}}_M
\wedge\gamma^m\stackrel{o}{\psi}_M &=&0\nn\\
d\stackrel{o}{\omega}^{mn}-\stackrel{o}{\omega}^{mr}\wedge\stackrel{o}{\omega}_r^{~n}+16e^2
\stackrel{o}{V}^m\wedge\stackrel{o}{V}^n-2\ii e^2\stackrel{o}{\bar{\psi}}_M\wedge\gamma_5\gamma^{mn}
\stackrel{o}{\psi}_M &=&0\nn\\
d\stackrel{o}{A}^{MN}+e\stackrel{o}{A}^{MR}\wedge\stackrel{o}{A}_R^{~N}-4\ii\stackrel{o}{\bar{\psi}}_M\wedge
\gamma_5\stackrel{o}{\psi}_N &=&0\nn\\
d\stackrel{o}{\psi}_M-{1\over 4}\gamma^{mn}\stackrel{o}{\omega}_{mn}\stackrel{o}{\psi}_M-e\stackrel{o}{A}_M^{~N}
\wedge\stackrel{o}{\psi}_N-2e\gamma_5\gamma_m\stackrel{o}{V}^m\wedge\stackrel{o}{\psi}_M &=&0.
\eeq
\par
With all these objects we can build the Freund Rubin solution of eleven dimensional supergravity:
the following eleven dimensional forms 
\footnote{We leave implicit the spinor indices; remind that a four dimensional $AdS_4$ spinor has an index taking four values, a 
seven dimensional $G/H$ spinor has an index taking eight values, the eleven dimensional spinor has an index taking thirty-two values, 
and in fact the tensor product of an $AdS_4$ spinor and a $G/H$ spinor is an eleven dimensional spinor.}
\be
V^{\hat{a}}=\ll(V^m,B^a\rr),~\omega^{\hat{a}\hat{b}}=\ll(\omega^{mn},K^{ma},B^{ab}\rr)
\ee
\beq
V^m&=&\stackrel{o}{V}^m\ll(x,\theta\rr)\nn\\
\omega^{mn}&=&\stackrel{o}{\omega}^{mn}\ll(x,\theta\rr)\nn\\
\psi&=&\stackrel{o}{\psi}_M\ll(x,\theta\rr)\eta^M\ll(y\rr)\nn\\
B^a&=&\cB^a\ll(y\rr)+{1\over 8}\bar{\eta}_M\ll(y\rr)\tau^a\eta_N\ll(y\rr)\stackrel{o}{A}^{MN}\ll(x,\theta\rr)\nn\\
B^{ab}&=&\cB^{ab}\ll(y\rr)-{1\over 4}e\bar{\eta}_M\ll(y\rr)\tau^{ab}\eta_N\ll(y\rr)\stackrel{o}{A}^{MN}\ll(x,\theta\rr)\nn\\
K^{ma}&=&0
\eeq
and $A=\stackrel{o}{A}\ll(x,y,\theta\rr)$ three-form not globally defined (it is a section of a fiber bundle), such that
\beq
\label{fourform}
d\stackrel{o}{A}&=&e\epsilon_{mnrs}\stackrel{o}{V}^m\wedge\stackrel{o}{V}^n\wedge\stackrel{o}{V}^r\wedge\stackrel{o}{V}^s
+{1\over 2}\stackrel{o}{\bar{\psi}}_M\wedge\gamma^{mn}\stackrel{o}{\psi}^M\wedge\stackrel{o}{V}^m\wedge\stackrel{o}{V}^n+\nn\\
&&-\stackrel{o}{\bar{\psi}}_M\wedge\gamma_5\gamma_m\stackrel{o}{\psi}_N\wedge\stackrel{o}{V}^m\wedge\bar{\eta}_M\tau_a\eta_NB^a+
{1\over 2}\stackrel{o}{\bar{\psi}}_M\wedge\stackrel{o}{\psi}_N\bar{\eta}_M\tau_{ab}\eta_NB^a\wedge B^b,\nn\\
\eeq
satisfy the Maurer Cartan equations of eleven dimensional supergravity. 
\par
If we evaluate these superspace forms on a bosonic surface, i.e. at $\theta=0$, we get the fields of the $\cN$--extended
eleven dimensional Freund Rubin solution of supergravity:
\be
\label{FreundRubin2}
\begin{array}{ccc}
\begin{array}{ccc}
g_{\mu\nu}\ll(x,y\rr)&=&g^0_{\mu\nu}\ll(x\rr)\\
g_{\alpha\beta}\ll(x,y\rr)&=&g^0_{\alpha\beta}\ll(y\rr)\\
g_{\mu\alpha}&=&0\\
\end{array}&&
\begin{array}{ccc}
F_{\mu\nu\rho\sigma}&=&e\sqrt{g^0}\varepsilon_{\mu\nu\rho\sigma}\\
{\rm other~}F&=&0\\
\psi_{\mu}=\psi_{\alpha}&=&0\\
\end{array}\\
\end{array}
\ee
where $g^0_{\mu\nu}$ is the $AdS_4$ metric, $g^0_{\alpha\beta}$ is the $G$--invariant $G/H$ metric.
\par
This solution preserves $\cN$ supersymmetries. In fact, being $\psi_{\hat{a}}\ll(x,y\rr)=0$, the supersymmetry transformations of
the bosonic fields vanish. The supersymmetry transformations of the gravitino fields are:
\beq
\delta_{\epsilon}\psi_{\mu}\ll(x,y\rr)&=&\ll(\partial_{\mu}-{1\over 4}\omega_{\mu}^{mn}\gamma_{mn}+2e\gamma_5\gamma_{\mu}V_{\mu}^m\rr)
\epsilon\ll(x,y\rr)\nn\\
\delta_{\epsilon}\psi_{\a}\ll(x,y\rr)&=&\ll(\partial_{\a}-{1\over 4}\cB_{\a}^{ab}\tau_{ab}-e\tau_a\cB_{\a}^a\rr)
\epsilon\ll(x,y\rr).
\eeq
They vanish for 
\be
\epsilon\ll(x,y\rr)=\epsilon\ll(x\rr)\eta\ll(y\rr)
\ee
where $\epsilon\ll(x\rr)$ is an $AdS_4$ Killing spinor, satisfying 
\be
\label{killspeqAdS4}
\ll(\partial_m-{1\over 4}\omega_{m}^{rs}\gamma_{rs}+2e\gamma_5\gamma_{m}\rr)\epsilon\ll(x\rr)=0
\ee
and $\eta\ll(y\rr)$ is a $G/H$ Killing spinor, satisfying the (\ref{killspeqGH}). There are four independent solutions of the
(\ref{killspeqAdS4}), and, as I said, $\cN$ is the number of the independent solutions of the (\ref{killspeqGH}), then
there exist $4\cN$ independent one--component supersymmetry transformations leaving invariant the (\ref{FreundRubin2}). With the
conventions of four dimensional supergravity (where supercharges have four pseudo--real 
components), this means that the solution preserves $\cN$ supersymmetries.
\par
Notice that the Freund Rubin solution is a {\it spontaneous compactification}, in the sense that $AdS_4\times X_7$ is a 
solution of eleven dimensional supergravity, and then an allowed vacuum around which we can perform perturbation
theory; nothing has been added to the theory at hand. The key that allows this is the presence of a four--form 
field strength, to which we can give the expectation value of $\epsilon_{mnrs}$, the invariant tensor of $SO\ll(1,3\rr)$, breaking $11
\longrightarrow 4+7$.
\par
\subsubsection{The solution of the $G/H$ Killing spinor equation and holonomy}
\par
Given a coset manifold $G/H$ admitting an Einstein metric structure, we want to know if the corresponding Freund Rudin solution
is supersymmetric, and how much. As we have seen, this can be done by solving the $G/H$ Killing spinor equation (\ref{killspeqGH}).
\par
First of all we have to consider the integrability conditions of the (\ref{killspeqGH}). They are
\be
\label{integrcond}
\cC_{ab}\eta\equiv\ll(\cR^{cd}_{~~ab}-4e^2\delta^{cd}_{ab}\rr)\tau_{cd}\eta=\cC_{ab}^{~~cd}\tau_{cd}\eta.
\ee
We have to find the null eigenspinors of the $21$ $\cC_{ab}$ operators here defined, which are combination
of the $\tau_{ab}$ generators with coefficients $\cC_{ab}^{~~cd}$ (which are the components of the Weyl tensor) 
and then generate a subgroup of $SO\ll(7\rr)$. Being the $8$ dimensional spinor representation
of $SO\ll(7\rr)$ irreducible, the equation $\tau_{ab}\eta=0$ has no solutions, and then the (\ref{integrcond}) has
null eigenspinors only if the combinations $\cC_{ab}$ do not generate all $SO\ll(7\rr)$ but lie, with their commutators, in a subspace of
the $SO\ll(7\rr)$ algebra, under which the $8$ dimensional spinor representation of $SO\ll(7\rr)$ be reducible. This
algebra is called the {\it Weyl holonomy algebra} $\cG_{hol}$. It is slightly different from the usual holonomy algebra of riemannian geometry, namely, 
the algebra of transformations that can occur to a vector after parallel riemannian transport around a closed curve, which is the Riemann holonomy algebra;
the latter is generated by the Riemann tensor, not by the Weyl tensor (see \cite{vannieuw}); $\cG_{hol}$ is the holonomy algebra with respect to the parallel transport 
defined by the covariant derivative $\cD^{SO\ll(7\rr)}-e\cB^a\tau_a$. So, for example, the Riemannian holonomy algebra of $S^7$ is $SO\ll(7\rr)$, while the Weyl
holonomy algebra of $S^7$ is $\{0\}$. Notice that the generators in the (\ref{killspeqGH}), $\ll(\tau_{ab},\tau_a\rr)$, are the generators of $SO\ll(8\rr)$;
the Killing spinors, in fact, are covariantly constant under an $SO\ll(8\rr)\supset SO\ll(7\rr)$ group.
\par
So if $\cG_{hol}=SO\ll(7\rr)$, $\cN_{MAX}=0$. If $\cG_{hol}=G_2$, being  
\be
{\bf 8}\stackrel{{G_2\subset SO\ll(7\rr)}}{\longrightarrow}{\bf 7}\oplus{\bf 1},
\ee
$\cN_{MAX}=1$. If $\cG_{hol}=SU\ll(3\rr)$, being 
\be
{\bf 8}\stackrel{{SU\ll(3\rr)\subset SO\ll(7\rr)}}{\longrightarrow}{\bf 3}\oplus{\bf{\bar 3}}\oplus{\bf 1}\oplus{\bf 1},
\ee
$\cN_{MAX}=2$. If $\cG_{hol}=SU\ll(2\rr)$, being 
\be
{\bf 8}\stackrel{{SU\ll(2\rr)\subset SO\ll(7\rr)}}{\longrightarrow}{\bf 2}\oplus{\bf 2}\oplus{\bf 1}\oplus{\bf 1}\oplus{\bf 1}\oplus{\bf 1},
\ee
$\cN_{MAX}=4$. If $\cG_{hol}=\{0\}$, $\cN_{MAX}=8$. 
\par
Then, in order to find the solutions of the (\ref{killspeqGH}), one has to find the holonomy group, then to determine the 
null eigenspinors of the integrability condition $C_{ab}\ll(y\rr)\eta\ll(y\rr)=0$, and finally substitute these eigenspinors in
the (\ref{killspeqGH}) to check if they are actually solutions.
\par
\subsubsection{From Killing spinors to Killing vectors}
\par
There is an interesting property of $G/H$ Killing spinors. Given the $\cN$ Killing spinors $\eta_M\ll(y\rr)$, namely, the 
solutions of the (\ref{killspeqGH}), we can build the following $\cN\ll(\cN-1\rr)/2$ vectors on $G/H$
\be
k_{MN}^a\ll(y\rr)\equiv \bar{\eta}_{\ll[M\rr.}\tau^a\eta_{\ll.N\rr]}.
\label{vectorsfromspinors}
\ee
It can be shown that these are Killing vectors of $G/H$, generating an $SO\ll(\cN\rr)$ group which is then an isometry of
the coset manifold. This is the reason of the previously stressed property
\be
G=G'\times SO\ll(\cN\rr).
\ee
\par
\subsection{Four dimensional supergravity from Freund Rubin solution}
\par
Given a Freund Rubin solution of eleven dimensional supergravity, we can consider this classical
solution as a vacuum of the theory, and do perturbation theory taking as dynamical degrees of freedom 
the fluctuation around this vacuum (see \cite{bosonicmassformula}, \cite{castdauriafre}):
\beq
g_{\mu\nu}\ll(x,y\rr)&=&g^0_{\mu\nu}\ll(x\rr)+h_{\mu\nu}\ll(x,y\rr)\nn\\
g_{\alpha\beta}\ll(x,y\rr)&=&g^0_{\alpha\beta}\ll(x\rr)+h_{\alpha\beta}\ll(x,y\rr)\nn\\
g_{\mu\alpha}\ll(x,y\rr)&=&h_{\mu\alpha}\ll(x,y\rr)\nn\\
A_{\mu\nu\rho}\ll(x,y\rr)&=&A^0_{\mu\nu\rho}\ll(x\rr)+a_{\mu\nu\rho}\ll(x,y\rr)\nn\\
A_{\mu\nu\alpha}\ll(x,y\rr)&=&a_{\mu\nu\alpha}\ll(x,y\rr)\nn\\
A_{\mu\alpha\beta}\ll(x,y\rr)&=&a_{\mu\alpha\beta}\ll(x,y\rr)\nn\\
A_{\alpha\beta\gamma}\ll(x,y\rr)&=&a_{\alpha\beta\gamma}\ll(x,y\rr)\,.
\label{linearizedfields}
\eeq
The equations of eleven dimensional supergravity, linearized in these fluctuations, have
in general the form
\be
\label{differentialequation}
\left( \Box_x^{[E\,s]}+\xbox_y^{[\lambda_1\lambda_2\lambda_3]}\right)
\Phi_{[\lambda_1\lambda_2\lambda_3]}^{[E\,s]}(x,y)=0.
\ee
Here $\Phi_{[\lambda_1\lambda_2\lambda_3]}^{[E\,s]}(x,y)$ is a field transforming
in the irreducible representation $[E\,s]$ of $SO(3,2)$
and $[\lambda_1\lambda_2\lambda_3]$ of $SO(7)$
\footnote{$[\lambda_1\lambda_2\lambda_3]$ are the Dynkin labels of the $SO\ll(7\rr)$ UIR ($SO\ll(7\rr)$ has rank three).}, 
and depends both on the coordinates $x$ of anti--de Sitter space and on the coordinates $y$
of $G/H$. Notice that $\Phi$ has $SO\ll(7\rr)$ indices because, as I explained, a generic
field on $G/H$ is in an irreducible representation of $SO\ll(7\rr)$.
$\Box_x^{[E\,s]}$ is the kinetic operator for a field of energy and spin
$[E\,s]$ on $AdS_4$, and is well known from $AdS_4$ theory (see chapter $2$). $\xbox_y^{[\lambda_1\lambda_2\lambda_3]}$ is
the kinetic operator for a field of spin $[\lambda_1\lambda_2\lambda_3]$ on the
seven dimensional $G/H$. The operators $\xbox_y^{\ll[\lambda_1\lambda_2\lambda_3\rr]}$ 
are built with the $SO\ll(7\rr)$--covariant derivative $\cD^{SO\ll(7\rr)}$,
the Killing metric on $G/H$, and, for spinor fields, the gamma matrices $\tau_a$. They all have the property of the
$SO\ll(7\rr)$--covariant derivative to be {\it invariant operators}, namely, to commute with the $H$--invariant
Lie derivative:
\be
\ll[\xbox_y^{\ll[\lambda_1\lambda_2\lambda_3\rr]},\cL_{K_A}\rr]=0.
\label{invariantoperators}
\ee
As I explain in section \ref{harmonicanalysis}, we can expand the field $\Phi$ in a complete set of
eigenfunctions of $\xbox_y$, the $G/H$ harmonics:
\beq
\Phi\ll(x,y\rr)=\sum{\cal H}\ll(y\rr)\phi\ll(x\rr)\\
\xbox_y{\cal H}\ll(y\rr)=M{\cal H}\ll(y\rr).
\label{harmexp}
\eeq
The differential equation (\ref{differentialequation}) becomes
\be
\left(\Box_x+M\right)\phi\ll(x\rr)=0
\label{fourdimeq}
\ee
which is an equation for  a four dimensional supergravity field on $AdS_4$.
\par
Then the eleven dimensional supergravity linearized around the Freund Rubin solution looks like
the {\bf $\cN$--extended four dimensional supergravity on $AdS_4$}. 
\par
The explicit expression of the expansion (\ref{harmexp}) of the fields (\ref{linearizedfields}) is
\begin{eqnarray}
h_{mn}\left(x,y\right)
         &=& \Big( h_{mn}^I
\left(x\right)
                 - \frac{3}{M_{(0)^3}+32}{\cal D}_{(m}{\cal D}_{n)}
                 \left[(2+\sqrt{M_{(0)^3}+36}\,) S^I\left(x\right)\right.+ \nonumber\\ 
        & &  \left. +(2-\sqrt{M_{(0)^3}+36}\,)
                 \Sigma^I
\left(x\right)\right]
                 + \ft54 \delta_{mn}
                   \left[(6-\sqrt{M_{(0)^3}+36}\,)S^I\left(x\right)+\right.\nonumber\\
        & &    \left. +(6+\sqrt{M_{(0)^3}+36}\,)\Sigma^I
\left(x\right)\right]
            \Big) \, \cY^I \left(y\right)\,,
\nonumber \\
\nn\\
h_{ma}\left(x,y\right)
            &=& \big[
                (\sqrt{M_{(1)(0)^2}+16}-4)A_m^I\left(x\right) +
                (\sqrt{M_{(1)(0)^2}+16}+4)W_m^I
\left(x\right)\big] \, \cY^I_a\left(y\right) \,,
\nonumber \\
\nn\\
h_{ab}\left(x,y\right)
           &=& \phi^I\left(x\right) \cY^I_{(ab)}\left(y\right) - \delta_{ab}
           \left[(6-\sqrt{M_{(0)^3}+36}) S^I\left(x\right)+\right. \nonumber \\
 & & \left.  +(6+\sqrt{M_{(0)^3}+36})\Sigma^I
\left(x\right)\right] \, \cY^I\left(y\right) \,,
\nonumber \\
\nn\\
a_{mnr}\left(x,y\right)
             &=& 2 \, \varepsilon_{mnrp} \,{\cal D}_p (S^I\left(x\right)+\Sigma^I\left(x\right))
              \cY^I\left(y\right) \,,
\nonumber \\
\nn\\
a_{mna}\left(x,y\right)
             &=& \ft23 \, \varepsilon_{mnrs} \,
            ({\cal D}_r A_s^I\left(x\right) + {\cal D}_r W_s^I\left(x\right))\, \cY_a^I\left(y\right) \,,
\nonumber \\
\nn\\
a_{mab}\left(x,y\right)
             &=& Z_m^I\left(x\right) \cY^I_{[ab]}\left(y\right) \,,
\nonumber \\
\nn\\
a_{abc}\left(x,y\right)
            &=& \pi^I\left(x\right) \cY^I_{[abc]}\left(y\right) \,,
\nonumber \\
\nn\\
\psi_m\left(x,y\right)
            &=& \Big(
                \chi_m^I\left(x\right) +\frac{\ft47 M_{(1/2)^3}+8}{M_{(1/2)^3}+8}
                  \big[D_m \lambda_L^I\left(x\right) \big]_{3/2}
-\nonumber\\
 &&    +(6+\ft37 M_{(1/2)^3})\gamma_5\gamma_m \lambda_L^I
\left(x\right)
           \Big) \, \Xi^I\left(y\right) \,,
\nonumber \\
\nn\\
\psi_a &=& \lambda_T^I\left(x\right) \Xi_a^I\left(y\right) + \lambda_L^I\left(x\right)
                \big[ \nabla_a \Xi ^I\left(y\right) \big]_{3/2}
\,.\label{kkexpansion}
\end{eqnarray}
The conventions for the names of the harmonics $\cH^I$ and their eigenvalues are the 
following:
\begin{equation}
\begin{array}{|c|l|c|}
\hline
SO\ll(7\rr)~{\rm UIR}&{\rm Harmonic}~\cH&{\rm Eigenvalue~}M_{\ll[\lambda_1,\lambda_2,\lambda_3\rr]}\\\hline
\ll[0,0,0\rr]&{\cal Y} & M_{\left(0\right)^3} \\
\ll[1,0,0\rr]&{\cal Y}_{a},~~~~~\cD^a\cY_a=0 & M_{\left(1\right)\left(0\right)^2}\\
\ll[1,1,0\rr]&{\cal Y}_{\left[ab\right]},~~~\cD^a\cY_{\ll[ab\rr]}=0 & M_{\left(1\right)^2\left(0\right)}\\
\ll[1,1,1\rr]&{\cal Y}_{\left[abc\right]},~~~\cD^a\cY_{\ll[abc\rr]}=0 & M_{\left(1\right)^3}\\
\ll[2,0,0\rr]&{\cal Y}_{\left(ab\right)},~~~\eta^{ab}\cY_{\ll(ab\rr)}=\cD^a\cY_{\ll(ab\rr)}=0 & M_{\left(2\right)\left(0\right)^2}\\
\ll[{1\over 2},{1\over 2},{1\over 2}\rr]&\Xi & M_{\left(1\over 2\right)^3} \\
\ll[{3\over 2},{1\over 2},{1\over 2}\rr]&\Xi_a,~~~~~\tau^a\Xi_a=\cD^a\Xi_a=0 & M_{\left(3\over 2\right)\left(1\over 2\right)^2}\\
\hline
\end{array}
\end{equation}
I explain in section \ref{harmonicanalysis} how are defined the harmonics and why in the expansion (\ref{kkexpansion}) they have an index $I$, running in an UIR of $G$.
\par
To each of these $SO\ll(7\rr)$ UIRs does correspond an invariant operator on $G/H$ arising from linearization of the eleven dimensional supergravity equations.
They are the following (we call $\cD\equiv\cD^{SO\ll(7\rr)}$):
\begin{itemize}
\item $0$--form Hodge de Rahm operator (the Laplacian)
\be
\label{scalarop}
\xbox_y^{\ll[0,0,0\rr]}\cY\equiv\cD^a\cD_a\cY=M_{\left(0\right)^3}\cY.
\ee
\item $1$--form Hodge de Rahm operator
\be
\label{oneformop}
\xbox_y^{\ll[1,0,0\rr]}\cY^a=\ll(\cD^a\cD_a+24e^2\rr)\cY^a=M_{\ll(1\rr)\ll(0\rr)^2}\cY^a.
\ee
\item $2$--form Hodge de Rahm operator
\beq
\label{twoformop}
\xbox_y^{\ll[1,1,0\rr]}\cY^{\ll[ab\rr]}&=&\ll(\cD^a\cD_a+48e^2\rr)\cY^{\ll[ab\rr]}+\nn\\
&&-4\cR^{\ll[a~b\rr]}_{~\ll[c~d\rr]}\cY^{\ll[cd\rr]}=M_{\ll(1\rr)^2\ll(0\rr)}\cY^{\ll[ab\rr]}.
\eeq
\item $3$--form first order operator
\beq
\label{threeformop}
\xbox_y^{\ll[1,1,1\rr]}\cY^{\ll[abc\rr]}&=&{1\over 24}\epsilon^{abcd}_{~~~~efg}\cD_d\cY^{efg}=\nn\\
&=&M_{\ll(1\rr)^3}\cY^{\ll[abc\rr]}.
\eeq
\item Lichnerowicz operator
\beq
\xbox_y^{\ll[2,0,0\rr]}\cY^{\ll(ab\rr)}&=&\cD^c\cD_c\cY^{\ll(ab\rr)}+4\cR^{a~b}_{~c~d}\cY^{\ll(cd\rr)}+\nn\\
&&+2\cR^a_{~c}\cY^{\ll(bc\rr)}+2\cR^b_{~c}\cY^{\ll(ac\rr)}=M_{\ll(2\rr)\ll(0\rr)^2}\cY^{\ll(ab\rr)}.
\eeq
\item Dirac operator
\be
\label{diracop}
\xbox_y^{\ll[1/2,1/2,1/2\rr]}\Xi=\ll(\tau^a\cD_a-7e\rr)\Xi=M_{\ll(1/2\rr)^3}\Xi.
\ee
\item Rarita Schwinger operator
\be
\label{raritaschwingerop}
\xbox_y^{\ll[3/2,1/2,1/2\rr]}\Xi_a=\ll(\tau^a\cD_a-5e\rr)\Xi_a=M_{\ll(3/2\rr)\ll(1/2\rr)^2}\Xi.
\ee
\end{itemize}
\par
The $AdS_4$ fields appearing in the expansion (\ref{kkexpansion}) are the following:
\begin{itemize}
\item one spin $2$ field $h_{mn}\ll(x\rr)$, arising from the expansion of the eleven dimensional graviton along the $AdS_4$ directions;
\item two spin $1$ fields, $A_m\ll(x\rr),~W_m\ll(x\rr)$, arising from the expansions of the components $h_{ma}\ll(x,y\rr)$ of the eleven dimensional graviton, and from the
components $a_{mna}$ of the three form;
as the massless graviton gauges the symmetries in eleven dimensional supergravity, the massless vectors $A_m$ gauge the isometry $G$;
\item one spin $1$ field $Z_m\ll(x\rr)$, arising from the expansion of the components $a_{mab}$ of the eleven dimensional three form; in the (\ref{kkexpansion})
it is the coefficient of a two form $G/H$ harmonic $\cY_{\ll[ab\rr]}$; there is one massless $Z_m$ field for each harmonic two form $\cY_{\ll[ab\rr]}$ on $G/H$,
then the massless $Z_m$ are counted by the second Betti number $b_2$ of $G/H$;
\item two scalar fields $S\ll(x\rr),~\Sigma\ll(x\rr)$, arising from the expansion of the graviton and of the components $a_{mnr}$ of the three form;
\item one scalar field $\phi\ll(x\rr)$, arising from the expansion of the graviton along the $G/H$ directions;
\item one pseudo--scalar field $\pi\ll(x\rr)$ arising from the expansion of the components $a_{abc}$ of the three form;
\item two spinor fields $\lambda_L\ll(x\rr),~\lambda_T\ll(x\rr)$ arising from the expansion of the eleven dimensional gravitino;
\item one gravitino field $\chi_m$ arising from the expansion of the eleven dimensional gravitino along the $AdS_4$ directions.
\end{itemize}
\par
Substituting the harmonic expansion (\ref{kkexpansion}) of the eleven dimensional fields and the (\ref{scalarop}),$\dots$,(\ref{raritaschwingerop}) eigenvalue
equations into the linearized equation of supergravity (\ref{differentialequation}), one finds equations for the $AdS_4$ fields with masses given by 
the $G/H$ harmonic eigenvalues 
$M_{\ll(0\rr)^3},\dots,M_{\ll(3/2\rr)\ll(1/2\rr)^2}$. One finds \cite{bosonicmassformula}:
\begin{eqnarray}
m_h^2 & = & M_{\left(0\right)^3}\,, \nonumber \\
m_{\Sigma}^2&=& M_{\left(0\right)^3} +176+24\sqrt{ M_{\left(0\right)^3}+36}\,,\nonumber\\
m_S^2 &=& M_{\left(0\right)^3} +176-24\sqrt{ M_{\left(0\right)^3}+36}\,,\nonumber\\
m_{\phi}^2 &=& M_{\left(2\right)\left(0\right)^2} \,,\nonumber\\
m_{\pi}^2 &=& 16\left( M_{\left(1\right)^3}-2\right)\left( M_{\left(1\right)^3}-1\right)
\,,\nonumber\\
m_W^2 &=& M_{\left(1\right)\left(0\right)^2} + 48 + 12 
\sqrt{M_{\left(1\right)\left(0\right)^2}+16}\,,\nonumber\\
m_A^2 &=& M_{\left(1\right)\left(0\right)^2} + 48 - 12 
\sqrt{M_{\left(1\right)\left(0\right)^2}+16}\,,\nonumber\\
m_Z^2 & = &  M_{\left(1\right)^2\left(0\right)} \,,\nonumber\\
m_{\lambda_L} & = & -\left( M_{\left(1\over 2\right)^3} +16\right)\,,
\nonumber\\
m_{\lambda_T} & = & M_{\left(3\over 2\right)\left(1\over 2\right)^2}+8\,,\nonumber\\
m_{\chi} & = & M_{\left(1\over 2\right)^3}\,.\label{massform}
\end{eqnarray}
I remind that the masses of $AdS_4$ fields are related to their energies by the (\ref{massenergyAdS4}).
\par
Summarizing, if we want to find the mass spectrum of a four dimensional supergravity obtained 
by Freund Rubin compactification with a coset manifold $G/H$, we have to determine the spectrum of the invariant
operators (\ref{scalarop}),$\dots$,(\ref{raritaschwingerop}) on the coset manifold; from this, by the mass formula (\ref{massform}), we can find
all the masses of the $AdS_4$ fields in the supergravity. Looking at the expansion (\ref{kkexpansion}), we see that:
\begin{itemize}
\label{listfields}
\item for each eigenvalue of the zero--form harmonic $\cY\ll(y\rr)$ there are one graviton field $h_{mn}\ll(x\rr)$, one scalar field
$S\ll(x\rr)$ and one scalar field $\Sigma\ll(x\rr)$;
\item for each eigenvalue of the one--form harmonic $\cY^a\ll(y\rr)$ there are one vector field $A_m\ll(x\rr)$ and one vector field
$W_m\ll(x\rr)$;
\item for each eigenvalue of the two--form harmonic $\cY^{\ll[ab\rr]}\ll(y\rr)$ there is one vector field $Z_m\ll(x\rr)$;
\item for each eigenvalue of the three--form harmonic $\cY^{\ll[abc\rr]}\ll(y\rr)$ there is one pseudo--scalar field $\pi\ll(x\rr)$;
\item for each eigenvalue of the harmonic $\cY^{\ll(ab\rr)}\ll(y\rr)$ there is one scalar field $\phi\ll(x\rr)$;
\item for each eigenvalue of the spinor harmonic $\Xi\ll(y\rr)$ there is one spinor field $\lambda_L\ll(x\rr)$ (called 
the longitudinal spinor field) and one gravitino field $\chi_m\ll(x\rr)$;
\item for each eigenvalue of the spinor--vector harmonic $\Xi_a\ll(y\rr)$ there is one spinor field $\lambda_T\ll(x\rr)$ (called 
the transverse spinor field).
\end{itemize}
\par
\subsection{Supersymmetric mass relations}
\par
A useful tool for deriving the mass spectrum of a Freund Rubin supergravity are the {\it supersymmetric mass relations} 
\cite{univer}, \cite{castdauriafre}. The key point is that it is possible to build $G/H$ harmonics eigenfunctions
of invariant operators by means of other $G/H$ harmonics eigenfunctions of other invariant operators. The eigenvalues
of these harmonics are related, and then for each eigenvalue of the latter invariant operator there is one eigenvalue
of the former, given by relations which can be worked out. Then, using the (\ref{massform}), one
can translate these relations between $G/H$--harmonics eigenvalues into relations between $AdS_4$ fields masses.
\par
These relations can be understood from a different point of view. The different fields of $\cN$--extended $AdS_4$ supergravity
are related by supersymmetry transformations. While in Poincar\`e supersymmetry the fields in a same supermultiplet have the same mass,
in $AdS_4$ supersymmetry it is not so (see chapter $2$); however, the masses of the fields in a same supermultiplet are related. These
relations are precisely the ones which can be found with the method above explained.
\par
\begin{figure}
\centering
\begin{picture}(268,200)
\put (15,100){\circle{30}}
\put (12,97){\shortstack{\large{$2$}}}
\put (12,72){\shortstack{\large{$h$}}}
\put (60,100){\vector (-1,0){30}}
\put (75,100){\circle{30}}
\put (72,97){\shortstack{\large{$3\over 2$}}}
\put (72,72){\shortstack{\large{$\chi$}}}
\put (85.61,110.61){\vector (1,1){21.22}}
\put (85.61,89.39){\vector (1,-1){21.22}}
\put (117.43,142.43){\circle{30}}
\put (114.43,139.43){\shortstack{\large{$1^-$}}}
\put (114.43,114.43){\shortstack{\large{$Z$}}}
\put (117.43,57.57){\circle{30}}
\put (114.43,54.57){\shortstack{\large{$1^+$}}}
\put (104.43,29.57){\shortstack{\large{$A,W$}}}
\put (132.43,142.43){\vector (1,0){30}}
\put (132.43,57.57){\vector (1,0){30}}
\put (125.04,68.18){\vector (2,3){42}}
\put (177.43,142.43){\circle{30}}
\put (174.43,139.43){\shortstack{\large{$1\over 2$}}}
\put (174.43,114.43){\shortstack{\large{$\lambda_T$}}}
\put (177.43,57.57){\circle{30}}
\put (174.43,54.57){\shortstack{\large{$1\over 2$}}}
\put (174.43,29.57){\shortstack{\large{$\lambda_L$}}}
\put (192.43,142.43){\vector (1,0){30}}
\put (188.04,131.82){\vector (1,-1){32.5}}
\put (188.04,68.18){\vector (1,1){32.5}}
\put (192.43,57.57){\vector (1,0){30}}
\put (237.43,142.43){\circle{30}}
\put (234.43,139.43){\shortstack{\large{$0^+$}}}
\put (259.43,139.43){\shortstack{\large{$\phi$}}}
\put (237.43,100){\circle{30}}
\put (234.43,97){\shortstack{\large{$0^-$}}}
\put (259.43,97){\shortstack{\large{$\pi$}}}
\put (237.43,57.57){\circle{30}}
\put (234.43,54.57){\shortstack{\large{$0^+$}}}
\put (259.43,54.57){\shortstack{\large{$S,\Sigma$}}}
\end{picture}
\caption{Supersymmetry relations between the
Kaluza Klein fields: for every couple of fields linked
by an arrow there is a mass relation descending by
supersymmetry.\label{molecola}}
\end{figure}
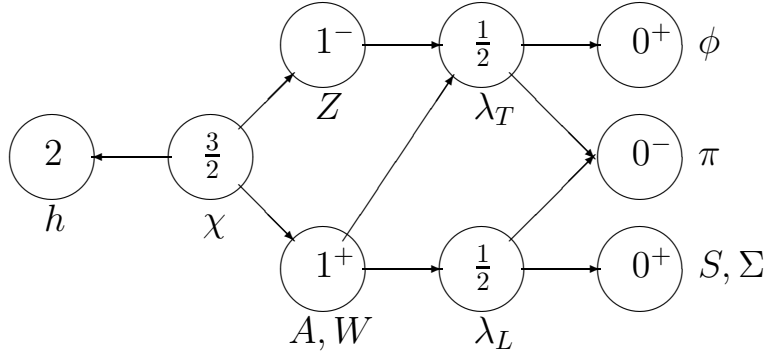

I do not review here the explicit calculations \cite{univer} which give the mass relations, I give only the result:
\begin{eqnarray}
m_h^2 =& m_\chi (m_\chi + 12)\,,
&
\nonumber \\
m_A^2 =&
m_\chi (m_{\chi} + 4) &{\rm  \hskip .5 cm    if  \hskip .5 cm  } m_{\chi} \geq -8
\,,
\nonumber \\
m_A^2 =&
m_\chi^2+2m_\chi+192 &{\rm  \hskip .5 cm    if  \hskip .5 cm  } m_{\chi} \leq -8
\,,
\nonumber \\
m_W^2 =&
m_\chi^2+2m_\chi+192 &{\rm  \hskip .5 cm    if  \hskip .5 cm  } m_{\chi} \geq -8
\,,
\nonumber \\
m_W^2 =&
m_\chi (m_{\chi} + 4) &{\rm  \hskip .5 cm    if  \hskip .5 cm  } m_{\chi} \leq -8
\,,
\nonumber \\
m_Z^2 =& (m_\chi + 8)(m_\chi + 4)
\,,\nonumber \\
\label{massrelationchi}
\end{eqnarray}
\begin{eqnarray}
m_{\pi}^2=& m_{\lambda_T}(m_{\lambda_T}+4)
 & \,,\nonumber \\
m_{\phi}^2=& m_{\lambda_T}(m_{\lambda_T}-4)
 & \,,\nonumber \\
m_A^2 =& m_{\lambda_T}^2-20\, m_{\lambda_T} + 96
 &{\rm  \hskip .5 cm    if  \hskip .5 cm  } m_{\lambda_T} \geq 4
\,,\nonumber \\
m_A^2 =& m_{\lambda_T}(m_{\lambda_T}+4)
 &{\rm  \hskip .5 cm    if  \hskip .5 cm  } m_{\lambda_T} < 4
\,,\nonumber \\
m_W^2 =& m_{\lambda_T}(m_{\lambda_T}+4)
 &{\rm  \hskip .5 cm    if  \hskip .5 cm  } m_{\lambda_T} \geq 4
\,,\nonumber \\
m_W^2 =& m_{\lambda_T}^2-20\, m_{\lambda_T} + 96
 &{\rm  \hskip .5 cm    if  \hskip .5 cm  } m_{\lambda_T} < 4
\,,\nonumber \\
m_Z^2 =& m_{\lambda_T}(m_{\lambda_T}-4)\,,
 &
\label{massrelationlambdaT}
\end{eqnarray}
\begin{eqnarray}
m_{\pi}^2=& m_{\lambda_L}(m_{\lambda_L}+4)
 & \,,\nonumber \\
m_S^2 =& \left(m_{\lambda_L}+24\right)\left(m_{\lambda_L}+20\right)
 &{\rm  \hskip .5 cm    if  \hskip .5 cm  } m_{\lambda_L} < -10
\,,\nonumber \\
m_S^2 =& m_{\lambda_L}(m_{\lambda_L}-4)
 &{\rm  \hskip .5 cm    if  \hskip .5 cm  } m_{\lambda_L} \geq -10
\,,\nonumber \\
m_{\Sigma}^2 =& m_{\lambda_L}(m_{\lambda_L}-4)
 &{\rm  \hskip .5 cm    if  \hskip .5 cm  } m_{\lambda_L} <-10
\,,\nonumber \\
m_{\Sigma}^2 =& \left(m_{\lambda_L}+24\right)\left(m_{\lambda_L}+20\right)
 &{\rm  \hskip .5 cm    if  \hskip .5 cm  } m_{\lambda_L} \geq -10
\,, \nonumber\\
m_A^2 =& m_{\lambda_L}^2-2\, m_{\lambda_L} + 192
 &{\rm  \hskip .5 cm    if  \hskip .5 cm  } m_{\lambda_L} <-8
\,,\nonumber \\
m_A^2 =& m_{\lambda_L}(m_{\lambda_L}+4)
 &{\rm  \hskip .5 cm    if  \hskip .5 cm  } m_{\lambda_L} \geq -8
\,,\nonumber \\
m_W^2 =& m_{\lambda_L}(m_{\lambda_L}+4)
 &{\rm  \hskip .5 cm    if  \hskip .5 cm  } m_{\lambda_L} < -8
\,,\nonumber \\
m_W^2 =& m_{\lambda_L}^2-2\, m_{\lambda_L} + 192
 &{\rm  \hskip .5 cm    if  \hskip .5 cm  } m_{\lambda_L} \geq -8\,.
\label{massrelationlambdaL}
\end{eqnarray}
These supersymmetry relations are pictorially represented
in Figure \ref{molecola}.
\par
\section{The $M^{111}$, $Q^{111}$ and $N^{010}$ spaces}
\par
Here and afterwards we set
\be
\kappa=1,~~e=1
\ee
which means $R_{AdS_4}=1/4$, in order to have dimensionless quantities.
\par
\subsection{$M^{111}$}
\par
\subsubsection{Definitions}
\par
The $M^{pqr}$ spaces are seven dimensional coset manifolds
\be
\label{321ov211}
M^{pqr}={G\over H}={SU\ll(3\rr)\times SU\ll(2\rr)\times U\ll(1\rr)\over SU\ll(2\rr)\times U\ll(1\rr)\times U\ll(1\rr)},
\ee
where the embedding of $H$ in $G$ is defined as I will explain in the following. They were introduced by E.Witten in the beginning
of the eighties \cite{kkwitten}, with the hope that the four dimensional theory arising from compactification on such a manifold, having
as symmetry group $SU\ll(3\rr)\times SU\ll(2\rr)\times U\ll(1\rr)$, could at the end describe standard model physics.
Then in \cite{cdfm111} the differential geometry of this manifold has been studied, and it has been shown that for every $M^{pqr}$ space 
an Einstein metric can be defined on it and then there exists a corresponding Freund Rubin solution of eleven dimensional supergravity, 
and that this solution preserves $\cN=2$ supersymmetry
if and only if $p=q$, namely, for the $M^{ppr}$ spaces. Other considerations on these manifolds have been 
given in \cite{castromwar} and, 
recently, in \cite{noi3}.
\par
Unfortunately, this was not the right way to obtain the standard model,
because chiral fermions cannot arise from these compactifications, and because the anti--de Sitter radius would be unphysical
\footnote{it is related to the coupling constant of the gauge symmetry $G$ by $e\sim l_p/R_{AdS}$, then if $G$ is the standard
model group $e\sim 1$ and $R_{AdS}\sim 10^{-33}$cm!}. However, these compactifications acquire a new meaning in the context of $AdS/CFT$ 
correspondence.
\par
The $M^{pqr}$ spaces can be defined as coset manifolds of the form (\ref{321ov211}) where 
$SU\ll(2\rr)\subset H$ is embedded in $SU\ll(3\rr)\subset G$, and this embedding is such that the 
fundamental representation of $SU\ll(3\rr)$ decomposes under $SU\ll(2\rr)\subset SU\ll(3\rr)$ as
\be
\label{32p1}
{\bf 3}\longrightarrow{\bf 2}\oplus{\bf 1}.
\ee
The (\ref{32p1}) defines univocally the embedding of $SU\ll(2\rr)$ in $SU\ll(3\rr)$ (modulo
isomorphisms); the embedding of the two $U\ll(1\rr)$ factors is encoded in the three
numbers $p,q,r$. To define exactly how these numbers determine the embedding of $U\ll(1\rr)\times U\ll(1\rr)$,
we give an explicit representation of the group $G$, by the following $6\times 6$ 
block--diagonal matrices:
\begin{equation}
G\ni g=
\begin{array}{c}
\left(\begin{array}{c|c|c}
SU\left(3\right)&0&0\\
\hline
0&SU\left(2\right)&0\\
\hline
0&0&U\left(1\right)
\end{array}
\right)\,,\\
\underbrace{\hspace{1.3 cm}}_{3}
\underbrace{\hspace{1.3 cm}}_{2}
\underbrace{\hspace{1.15 cm}}_{1}
\end{array}
\end{equation}
where the diagonal blocks contain the fundamental representations of
$SU(3)$, $SU(2)$ and $U(1)$ respectively.
The whole set of generators of $G$ is given by:
\begin{equation}
T_{\Lambda}\equiv\left(\ft{1}{2}\ii\l_1,\dots,\ft{1}{2}\ii\l_8,\ft{1}{2}\ii\s_1,\dots,\ft{1}{2}i\s_3,
iY\right),
\end{equation}
where $\l_i$ stands for the $i$-th Gell-Mann matrix (see appendix \ref{mpqrconventions})
 trivially extended
to a $6\times6$ matrix:
\begin{equation}
\l_i\longrightarrow\left(\begin{array}{ccc}
\l_i&0&0\\
0&0&0\\
0&0&0
\end{array}
\right)\,.
\end{equation}
Similarly $\s_m$ denotes the following extension of the Pauli matrices:
\begin{equation}
\s_m\longrightarrow\left(\begin{array}{ccc}
0&0&0\\
0&\s_i&0\\
0&0&0
\end{array}
\right)\,,
\end{equation}
and $Y$ is given by
\footnote{The normalizations of these generators are chosen to follow the literature \cite{cdfm111}, \cite{castdauriafre}, \cite{noi1}.
They are normalized so that
\be
{\rm Tr}\ll(T_{\Lambda}T_{\Lambda'}\rr)=-{1\over 2}\d_{\Lambda\Lambda'},
\ee
with the exception ${\rm Tr}\ll(YY\rr)=-1$. They are all orthogonal.}:
\begin{equation}
Y=\left(\begin{array}{ccc}
0&0&0\\
0&0&0\\
0&0&1
\end{array}
\right)\,.
\end{equation}
With these conventions, the $SU\ll(2\rr)\subset SU\ll(3\rr)$ satisfying
the (\ref{32p1}) is generated by $\lambda_1,\lambda_2,\lambda_3$. The remaining two $U\ll(1\rr)$ factors in $H$,
whose generators we call $Z',Z''$, are linear combinations of the three $U\ll(1\rr)$ factors in $G$ orthogonal
to $SU\ll(2\rr)$:
\be
\label{3u1}
\lambda_8,~\sigma_3,~Y.
\ee
What is relevant is the space generated by $Z',Z''$, not $Z',Z''$ themselves; this space is defined giving the 
combination of the three generators (\ref{3u1}) orthogonal to $Z',Z''$:
\be
\label{defZ}
Z\equiv p\ii{\sqrt{3}\over 2}\lambda_8+q\ii{1\over 2}\sigma_3+r\ii Y.
\ee
Then, a basis for the two abelian generators of $H$ is given by
\begin{eqnarray}
Z'=\sqrt{3}i\l_8+i\s_3-4iY\,,\label{z1}\\
Z''=-\ft{\sqrt{3}}{2}i\l_8+\ft{3}{2}i\s_3\,,
\label{z2}
\end{eqnarray}
which are orthogonal among themselves and with $Z$:
\footnote{But have different norms: for example, when $p=q=r=1$ 
${\rm Tr}\ll(ZZ\rr)=-3,~{\rm Tr}\ll(Z'Z'\rr)=-24,~{\rm Tr}\ll(Z''Z''\rr)=-6$.}
\begin{equation}
Tr(ZZ')=Tr(ZZ'')=Tr(Z'Z'')=0\,.
\end{equation}
\par
Summarizing, the orthogonal decomposition of the algebra $\IG$
\be
\IG=\IH\oplus \IK,
\ee
is given by:
\begin{eqnarray}
&G&\nn\\
SU\ll(3\rr)&:&\lambda_1,\dots,\lambda_8\nn\\
SU\ll(2\rr)&:&\sigma_1,\sigma_2,\sigma_3\nn\\
U\ll(1\rr)&:&Y\nn\\
&H&\nn\\
SU\ll(2\rr)&:&\lambda_{\dot m}~~~{\dot m}=1,2,3\nn\\
U\ll(1\rr)&:&Z'\nn\\
U\ll(1\rr)&:&Z''\nn\\
&K&\nn\\
\lambda_A&&A=4,5,6,7\nn\\
\sigma_m&&m=1,2\nn\\
Z&&\\
&{\rm where}&\nn
\end{eqnarray}
\be
Z\equiv p\ii{\sqrt{3}\over 2}\lambda_8+q\ii{1\over 2}\sigma_3+r\ii Y,
\ee
\be
Z',Z''\perp Z.
\ee
In this way, the embedding of $H$ in $G$ depends on the choice of the numbers $p,q,r$. 
\par
The generator $Z\in\IK$, with these conventions, is
\be
Z={1\over 2}\ii
\ll(\begin{array}{cccccc}
p&0&0&0&0&0\\
0&p&0&0&0&0\\
0&0&-2p&0&0&0\\
0&0&0&q&0&0\\
0&0&0&0&-q&0\\
0&0&0&0&0&2r\\
\end{array}\rr).
\ee
In order for $Z$ to be the generator of a compact $U\ll(1\rr)$, $p,q,r$ have to be rational; 
in fact only in this case the application 
\be
\phi\in I\subset\IR\longrightarrow e^{\ii Z\phi}
\ee
has a compact image. Since $Z$ is defined up to a multiplicative constant (equivalent to a rescaling of
$\phi$), we can take $p,q,r$ as integer numbers.
\par
\subsubsection{Differential geometry and supersymmetry}
\par
An explicit parametrization of the coset $G/H$ is given by the
seven coordinates \linebreak $(y^A,y^m,y^3)$:
\begin{equation}
L(y^A,y^m,y^3)=\exp(\ft{1}{2}i\l_Ay^A)\exp(\ft{1}{2}i\s_my^m)\exp(Zy^3)\ .
\end{equation}
Actually, it is not important that the parametrization is this one: the harmonic analysis
formalism does not depend on the coordinate choice.
\par
From the coset representative we can construct the
left-invariant one-forms on $G/H$ as:
\begin{equation}
\Omega(y)=L^{-1}(y){\rm d}L(y)=\Omega^{\Lambda}(y)T_{\Lambda}\,,
\end{equation}
which satisfies the Maurer-Cartan equations
\begin{equation}\label{Omega}
{\rm d}\Omega^{\Lambda}+\ft{1}{2}C^{\Lambda}_{\ \Sigma\Pi}
\Omega^{\Sigma}\wedge\Omega^{\Pi}=0
\end{equation}
with the structure constants of $G$:
\begin{equation}
\left[T_{\Sigma},T_{\Pi}\right]=C^{\Lambda}_{\ \Sigma\Pi}T_{\Lambda}\,.
\end{equation}
The one-forms $\Omega^{\Lambda}$ can be separated into a set
$\{\Omega^i\}$ corresponding to the generators of the subalgebra $\IH$
and a set $\{\Omega^{a}\}$ corresponding to the coset generators.
These latter can be identified with the $SU(3)\times SU(2)\times U(1)$
invariant seven-vielbein on $G/H$:
\begin{eqnarray}
\cB^{a} \equiv (\cB^A, \cB^m, \cB^3),\nonumber\\
\left\{\begin{array}{ccl}
\cB^A & = & {1\over a}\Omega^A,\\
\cB^m & = & {1\over b}\Omega^m,\\
\cB^3 & = & {1\over c}(\sqrt{3}\Omega^8+\Omega^3+
2\Omega^Y) = {12\over c}\Omega^Z,
\end{array}\right.
\end{eqnarray}
where the multiplicative coefficients define the more general rescaling preserving the 
$G$--isometry. The invariant forms $\Omega^i$ are:
\begin{equation}
\left\{\begin{array}{ccl}
\Omega^{\dot m}, &&\\
\Omega^{Z'} & = & \ft{1}{24}(\sqrt{3}\Omega^8+\Omega^3
 -4\Omega^Y),\\
\Omega^{Z''} & = & \ft{1}{12}(3\Omega^3-\sqrt{3}\Omega^8).
\end{array}\right.
\end{equation}
The spin-connection $\cB^{a}_{\ b}$ is determined from
the vielbein $\cB^{a}$ by imposing vanishing torsion:
\begin{equation}
{\rm d}\cB^{a}-\cB^{a}_{\ b}\wedge\cB^{b}=0,
\end{equation}
\begin{equation}
\left\{
\begin{array}{ccl}
\cB^{mn} &=& \e^{mn}\left(\Omega^3-{qb^2\over 2c}\cB^3\right),\\
\cB^{3m} &=& -{qb^2\over 2c}\e^{mn}\cB_n,\\
\cB^{mA} &=& 0,\\
\cB^{3A} &=& -{\sqrt{3}\over 2}{pa^2\over c}f^{8AB}\cB_B,\\
\cB^{AB} &=& f^{{\dot m}AB}\Omega_{\dot m}+f^{8AB}\Omega_8-
{\sqrt{3}\over 2}{pa^2\over c}f^{8AB}\cB^3.
\end{array}\right.
\end{equation}
\par
Working out the Ricci tensor one finds \cite{cdfm111} that for each value of the parameters $\ll(p,q,r\rr)$ there is one and only one
value of the rescalings $a,b,c$ such that
\be
\cR^a_{~b}=12\delta^a_b.
\ee
Working out the Weyl holonomy algebra of the $M^{pqr}$ manifolds so rescaled, one finds \cite{cdfm111} that if $p\neq q$, $\cG_{hol}=SO\ll(7\rr)$
and then $\cN=0$; if $p=q$, $\cG_{hol}=SU\ll(3\rr)$, then $\cN_{MAX}=2$, and substituting the null eigenspinors in the (\ref{killspeqGH})
one finds that actually $\cN=2$. So the only supersymmetric $M^{pqr}$ manifolds are the $M^{ppr}$, and have $\cN=2$ supersymmetry.
\par
Notice that, as we have seen, since the spaces $M^{ppr}$ preserve $\cN=2$ supersymmetries, their isometry group must have the form $G=G'\times SO\ll(2\rr)$.
In fact this is the case, being $G=SU\ll(3\rr)\times SU\ll(2\rr)\times U\ll(1\rr)$, $U\ll(1\rr)\simeq SO\ll(2\rr)$, so 
\be
G'=SU\ll(3\rr)\times SU\ll(2\rr).
\ee
\par
\subsubsection{$p,r$ in$~M^{ppr}$}
\par
Let us understand what implies the choice of $p,r$ in the $M^{ppr}$ manifolds. 
The simplest of the $M^{ppr}$ manifolds is $M^{110}$. In this case $Z''\perp Y$, so we can take $Z'\propto Y$, and
the $U\ll(1\rr)$ factor in $G$ decouple
\be
M^{110}={SU\ll(3\rr)\times SU\ll(2\rr)\over SU\ll(2\rr)\times U\ll(1\rr)}={{SU\ll(3\rr)\over SU\ll(2\rr)}\times SU\ll(2\rr)\over U\ll(1\rr)}.
\ee
It can be shown that $SU\ll(3\rr)/SU\ll(2\rr)=S^5$, and $SU\ll(2\rr)=S^3$ (locally), so
\be
M^{110}={S^5\times S^3\over U\ll(1\rr)}.
\ee
The $U\ll(1\rr)$ in the denominator is
\be
\label{zetasecond}
Z''=-{\ii\over 2}\ll(\sqrt{3}\lambda_8-3\sigma_3\rr)=-{
\ii\over 2}{\rm ~diag}\ll(1,1,-2,-3,3,0\rr),
\ee
so the ratio of the periods of the $U\ll(1\rr)$ actions on $SU\ll(3\rr)/SU\ll(2\rr)$ and on $SU\ll(2\rr)$ is $3/2$.
This manifold is simply connected (see chapter $4$).
\par
If $r\neq 0$, we can define $r'=r/p$, and the manifold is
\be
M^{ppr}=M^{11r'}={M^{110}\times U\ll(1\rr)\over U\ll(1\rr)}
\ee
where the $U\ll(1\rr)$ factor in the numerator is $Y$, and the $U\ll(1\rr)$ factor in the denominator is (throwing away the global $p^2$ 
multiplicative factor)
\be
Z'=r'\ii\ll(\sqrt{3}\lambda_8+\sigma_3\rr)-4\ii Y=r'\ii{\rm ~diag}\ll(1,1,-2,1,-1,-4/r'\rr).
\ee
Namely, the manifold $M^{ppr}$ is the product of $M^{110}$ with a new one dimensional manifold generated by $Y$, all
quotiented by an identification relation generated by $Z'$. So points of $M^{110}$ are identified, but are distinguished
by the new coordinate. This yields a manifold which sometimes coincide with $M^{110}$, but in general is
\be
M^{ppr}={M^{110}\over\ZZ_l}.
\ee
In fact points of $M^{110}$ which differ by integer powers of
\be
{\rm ~diag}\ll(e^{\ii{\pi\over 2}r'},e^{\ii{\pi\over 2}r'},e^{-\ii\pi r'},e^{\ii{\pi\over 2}r'},e^{-\ii{\pi\over 2}r'},1\rr)
\label{identified}
\ee
are identified. But these points, different in ${SU\ll(3\rr)\over SU\ll(2\rr)}
\times SU\ll(2\rr)$, could be the same point 
of $M^{110}$, namely, they could be already identified in $M^{110}$ by $Z''$
(\ref{zetasecond}). 
If we can find a value of $\phi$ such that ${\rm exp}\ll(\ii\phi Z''\rr)$ 
is equal to the (\ref{identified}), then $M^{ppr}=M^{110}$. A trivial calculation shows that it always happens if $r'$ is integer,
while if $r'=m/n$ (relative primes) there are $n$ points identified. In conclusion, taking $p$ and $r$ relative primes
\footnote{In \cite{castromwar}, \cite{castdauriafre} part of this discussion has been 
worked out, without taking into account the identification \cite{noi3} here discussed.},
\be
M^{ppr}={M^{110}\over\ZZ_p}.
\ee
\par
I will consider the simplest $M^{ppr}$ manifold, namely, the simply connected one; as we have shown, all the manifolds $M^{11r}$ with $r\ge 0$ integer
coincide; I call this manifold $M^{111}$. A more detailed geometrical and topological analysis of the $M^{111}$ space is done in the next chapter;
a result of this treatment useful in the interpretation of the harmonic analysis results is the following: the second Betti number of $M^{111}$
is $b_2=1$, namely, the manifold admits a family of homotopic non--trivial two--cycles (and a family of non--trivial two--forms).
\par
For $M^{111}$ with Einstein metric, taking $e=1$, the rescaled vielbein is
\footnote{here the relation $\Omega^Z={1\over 6}\ll(\sqrt{3}\Omega^8+\Omega^3+2\Omega^Y\rr)$ has been found in the following way: 
\beq
&\Omega=\Omega^3T_3+\Omega^8T_8+\Omega^YT_Y+\dots=\Omega^ZZ+\dots,\nn\\
&Z=\sqrt{3}T_8+T_3+T_Y,~~{\rm Tr}\ll(Z\Omega\rr)=-1/2\ll(\sqrt{3}\Omega^8+\Omega^3+2\Omega^Y\rr)=-3\Omega^Z.\nn
\eeq}
\begin{eqnarray}
\cB^{a} \equiv (\cB^A, \cB^m, \cB^3),\nonumber\\
\left\{\begin{array}{ccl}\label{B}
\cB^A & = & {\sqrt{3}\over 8}\Omega^A,\\
\cB^m & = & {\sqrt{2}\over 8}\Omega^m,\\
\cB^3 & = & {1\over 8}(\sqrt{3}\Omega^8+\Omega^3+
2\Omega^Y) = {3\over 4}\Omega^Z,
\end{array}\right.
\end{eqnarray}
and the spin connection is
\begin{equation}
\left\{
\begin{array}{ccl}
\cB^{mn} &=& \e^{mn}\left(\Omega^3-2\cB^3\right),\\
\cB^{3m} &=& -2\e^{mn}\cB_n,\\
\cB^{mA} &=& 0,\\
\cB^{3A} &=& -{4\over\sqrt{3}}f^{8AB}\cB_B,\\
\cB^{AB} &=& f^{{\dot m}AB}\Omega_{\dot m}+f^{8AB}\Omega_8-
{4\over\sqrt{3}}f^{8AB}\cB^3.
\end{array}\right.
\end{equation}
\par
\subsection{$Q^{111}$}
\par
The $Q^{pqr}$ spaces, found in the eighties \cite{dafrepvn}, are the following coset manifolds:
\be
Q^{pqr}={G\over H}={SU\ll(2\rr)\times SU\ll(2\rr)\times SU\ll(2\rr)\over U\ll(1\rr)\times U\ll(1\rr)}
\ee
where the embedding of the two $U\ll(1\rr)$ is the following; if we take
\be
\sigma^{\ll(1\rr)}_i,~\sigma^{\ll(2\rr)}_i,~\sigma^{\ll(3\rr)}_i
\ee
as the generators of the three $SU\ll(2\rr)$ factors in $G$, the maximal torus $U\ll(1\rr)\times U\ll(1\rr)\times U\ll(1\rr)\subset G$
is generated by
\be
\label{torusG}
\sigma^{\ll(1\rr)}_3,~\sigma^{\ll(2\rr)}_3,~\sigma^{\ll(3\rr)}_3;
\ee
the generators of $H=U\ll(1\rr)\times U\ll(1\rr)$, which we call $Z',Z''$, are the combinations of the generators (\ref{torusG}) 
orthogonal to
\be
Z\equiv {\ii\over 2}p\sigma^{\ll(1\rr)}_3+{\ii\over 2}q\sigma^{\ll(2\rr)}_3+{\ii\over 2}r\sigma^{\ll(3\rr)}_3.
\ee
So the embedding of $H$ in $G$ is completely defined by the three numbers $p,q,r$. In order for $Z$ to be the generator of a compact $U\ll(1\rr)$,
$p,q,r$ have to be rational numbers (as in the $M^{pqr}$ case), and we can take them integers and relative primes 
by rescaling $Z$ by a multiplicative constant.
\par
By studying the rescaling of the invariant vielbein, one finds that for each value of $p,q,r$ there is one and only one rescaling such that
\be
\label{einsteinQ}
\cR^a_{~b}=12\delta^a_b.
\ee
By studying the Weyl holonomy, one finds that if $\ll(p,q,r\rr)\neq\ll(p,p,p\rr)$ the Weyl holonomy is $SO\ll(7\rr)$ and so $\cN=0$. If
on the contrary $p=q=r$, $\cG_{hol}=SU\ll(3\rr)$, so $\cN_{MAX}=2$, and substituting in the (\ref{killspeqGH}) one finds that actually
in this case $\cN=2$. I will consider then the manifold $Q^{111}$, with the vielbein rescaled such that (\ref{einsteinQ}) is satisfied.
\par
In general, the isometry of a coset manifold $G/H$ is not $G$, but is 
\be
G\times \ll(N\ll(H\rr)\over H\rr)/U(1)^l
\ee
where $N\ll(H\rr)$ is the normalizer of $H$ in $G$, and $U(1)^l$ are the explicit
$U(1)$ factors common in $G$ and $N(H)/H$ . This because the generators of $N\ll(H\rr)$ not present in $H$ generate transformations
whose {\it right action} leave invariant the $G$--left invariant metric; the explicit
$U(1)$ factors commute with $G$, then their right action
coincide with their left action, and they are not new symmetries.
In the case of $M^{111}$, $\ll(N\ll(H\rr)/H\rr)/U(1)^l=\{0\}$, but in the case
of $Q^{111}$ it is $U\ll(1\rr)$. This is the reason for the apparent contradiction between the $\cN=2$ supersymmetry of $Q^{111}$ and the
lack of an explicit $SO\ll(2\rr)$ factor in $G=SU\ll(2\rr)\times SU\ll(2\rr)\times SU\ll(2\rr)$. However, it is possible to describe
the $Q^{111}$ manifold by taking into account the normalizer  from the start, in a form that exhibits explicitly the complete isometry in $G$:
\be
Q^{111}={SU\ll(2\rr)\times SU\ll(2\rr)\times SU\ll(2\rr)\times U\ll(1\rr)\over U\ll(1\rr)\times U\ll(1\rr)\times U\ll(1\rr)},
\ee
where the $U\ll(1\rr)^4\subset G$ is generated by
\be
\sigma^{\ll(1\rr)}_3,~\sigma^{\ll(2\rr)}_3,~\sigma^{\ll(3\rr)}_3,~Y,
\ee
and $H$ is generated by $Z',Z'',Z'''$ orthogonal to 
\be
Z=-{\ii\over 2\sqrt{3}}\ll(\sigma^{\ll(1\rr)}_3+\sigma^{\ll(2\rr)}_3+\sigma^{\ll(3\rr)}_3\rr)
+{\ii\over \sqrt{3}} Y.
\ee
This form, the one with the $SO\ll(2\rr)$ $R$--symmetry manifest, is the one which is convenient to use in harmonic analysis, because, as I explain in 
the next section, only in this way we get a spectrum of fields with well defined $R$--charge, ready to be organized in supermultiplets.
\par
\subsection{$N^{010}$}
\par
The $N^{pqr}$ spaces, found in the eighties \cite{castromwar} (see also \cite{newcast}), are the following coset manifolds:
\be
N^{pqr}={G\over H}={SU\ll(3\rr)\times U\ll(1\rr)\over U\ll(1\rr)\times U\ll(1\rr)}
\ee
where the embedding of the two $U\ll(1\rr)$ is the following; if we take the Gell Mann matrices (see appendix \ref{mpqrconventions})
as the generators of $SU\ll(3\rr)$ and
call $Y$ the generator of the additional $U\ll(1\rr)$ factor in $G$,
the generators of $H=U\ll(1\rr)\times U\ll(1\rr)$ are 
\begin{eqnarray}
  \label{eq:pqr}
  M &=& -\frac{\sqrt{2}}{RQ} 
    \left(
        \ft{i}{2} rp \sqrt{3} \lambda_8
      + \ft{i}{2} rq \lambda_3
      - \ft{i}{2} (3 p^2 + q^2) Y
    \right) \,,
    \nonumber \\
  N &=& -\frac{1}{Q}
    \left(
      - \ft{i}{2} q \lambda_8
      + \ft{i}{2} p \sqrt{3} \lambda_3
    \right) 
    \,,
\end{eqnarray}
with
\begin{eqnarray}
  \label{eq:RQTr}
  R=\sqrt{3 p^2 + q^2 + 2 r^2} \,, 
  \qquad
  Q=\sqrt{3 p^2 + q^2} \,.
\end{eqnarray}
$Z,M,N$ are orthonormalized to $-1/2$.
So the embedding of $H$ in $G$ is completely defined by the three numbers $p,q,r$. In order for $Z$ to be the generator of a compact $U\ll(1\rr)$,
$p,q,r$ have to be rational, and as usual we can take them integers relative primes.
As for the $M^{pqr}$ spaces, the local geometry depends only from the ratio $x=3p/q$, while its multiple connectivity depends on $r$.
\par
One can find that for each $p,q,r$ there are two different rescalings of the invariant vielbein such that
\be
\cR^a_{~b}=12\delta^a_b,
\ee
coincident only if $x=1$. We can call the corresponding Einstein manifolds $N_I^{pqr}$ and $N_{II}^{pqr}$. 
Studying the holonomy and the Killing spinor equation, one finds that:
\begin{itemize}
\item the Weyl holonomy of the $N_I^{pqr}$ spaces with $p\neq 0$ is $G_2$; they have then $\cN_{MAX}=1$, and actually they have $\cN=1$;
\item the Weyl holonomy of the $N_I^{0qr}$ spaces is $SU\ll(2\rr)$, so they
  have $\cN_{MAX}=4$; nevertheless, not all of the solutions of the 
integrability condition (\ref{integrcond}) are actually Killing spinors, namely, solutions of the (\ref{killspeqGH}): the space
$N_I^{010}$ admits $\cN=3$ Killing spinors, the other $N_I^{0qr}$ spaces admit $\cN=1$ Killing spinor;
\item the Weyl holonomy of the $N_{II}^{pqr}$ spaces is $G_2$, so $\cN_{MAX}=1$, and they have all $\cN=1$.
\end{itemize}
In the following, I will consider among these only the space $N_I^{010}$ 
(and omit the subscript $I$), which is the only Freund Rubin compactification admitting $\cN=3$ supersymmetries.
\par
In the manifold $N^{010}$ the generators are
\beq
Z&=&-{\ii\over 2}\lambda_3\nn\\
M&=&{\ii\over\sqrt{2}}Y\nn\\
N&=&{\ii\over 2}\lambda_8
\eeq
so the $U\ll(1\rr)$ decouples and we have
\be
N^{010}={SU\ll(3\rr)\over U\ll(1\rr)}
\ee
where the $U\ll(1\rr)$ generator is ${\ii\over 2}\lambda_8$. The normalizer of $\lambda_8$ in $SU\ll(3\rr)$ is $SU\ll(2\rr)$ (generated by
$\lambda_1,\lambda_2,\lambda_3\in\IK$), so the isometry of this manifold is
\be
SU\ll(3\rr)\times SU\ll(2\rr)
\ee
as foreseen by the fact that it has $\cN=3$ supersymmetry, and $SO\ll(3\rr)\simeq SU\ll(2\rr)$. The $G'$ group is then $SU\ll(3\rr)$.
\par
It has been shown in \cite{castromwar}
that this manifold can be realized in a way
that makes manifest the all $SU\ll(3\rr)\times SU\ll(2\rr)$ isometry, as
\be
N^{010}={SU\ll(3\rr)\times SU\ll(2\rr)\over SU\ll(2\rr)\times U\ll(1\rr)}
\ee
where the $U\ll(1\rr)\subset H$ is generated by
\be
T_8^H={\ii\over 2}\lambda_8,
\ee
and the $SU\ll(2\rr)\subset H$ is {\it diagonally embedded} into the two $SU\ll(2\rr)$ in $G$, namely, taking the Pauli matrices as generators
of the $SU\ll(2\rr)$ factor in $G$,
\be
T_i^H={\ii\over 2}\ll(\lambda_i+\sigma_i\rr)~~~i=1,2,3.
\ee
We call this $SU\ll(2\rr)\subset H$ $~~SU\ll(2\rr)^{diag}$. The generators of the subspace $\IK$ in the orthogonal decompositions $\IG=\IH\oplus\IK$ are
\be
T_a={\ii\over 2}\ll(\lambda_1-\sigma_1,\lambda_2-\sigma_2,\lambda_3-\sigma_3,\lambda_4,\lambda_5,\lambda_6,\lambda_7\rr).
\ee
This form, the one with the $SO\ll(\cN\rr)$ $R$--symmetry manifest, is the one appropriate for the harmonic analysis.
\par
\section{Harmonic analysis and mass spectra of Freund Rubin supergravities\label{harmonicanalysis}}
\par
Here I review the general theory of harmonic analysis on coset spaces, 
and its application for the derivation of mass spectra of Freund Rubin $G/H$ supergravities. For a more detailed 
treatment of this subject see \cite{salamstrat}, \cite{spectfer}, \cite{univer}, \cite{spec321}, \cite{castdauriafre}.
\par
\subsection{Harmonics on coset spaces}
\par
Let us consider as a first step a group manifold $G$. A complete functional basis on $G$ is
given by the matrix elements of the $G$ UIRs: any function 
\be
\Phi\ll(g\rr)~~~g\in G
\ee
can be expanded as
\be
\Phi\ll(g\rr)=\sum_{\ll(\mu \rr)}\sum_{m,n=1}^{{\rm dim}\ll(\mu\rr)}{c_{mn}^{\ll(\mu\rr)}
D_{mn}^{\ll(\mu\rr)}\ll(g\rr)}
\ee
where $\ll(\mu\rr)$ are the UIRs of $G$, $m,n$ run in these representations, and $D$ are
the elements of these representations. In fact, the $D_{mn}^{\ll(\mu\rr)}\ll(g\rr)$ satisfy
the orthogonality and completeness relations (see \cite{barut}, p.172)
\beq
\int_GdgD_{mn}^{\ll(\mu\rr)}\ll(g\rr)D_{sr}^{\ll(\nu\rr)}\ll(g^{-1}\rr)={{\rm vol}\ll(G\rr)\over{\rm vol}\ll(\mu\rr)}\delta_{mr}
\delta_{ns}\delta^{\ll(\mu\rr)\ll(\nu\rr)}\nn\\
\sum_{\ll(\mu\rr)}D_{mn}^{\ll(\mu\rr)}(g)D_{nm}^{\ll(\mu\rr)}({g'}^{-1}){\rm dim}\ll(\mu\rr)=
\delta\ll(g-g'\rr){\rm vol}\ll(G\rr).
\eeq
If $\Phi\ll(g\rr)$ transforms in an irreducible representation $\ll(\mu\rr)$ of $G$,
for example under left multiplication, namely
\be
\Phi^{\ll(\mu\rr)}_m\ll(g'g\rr)=D^{\ll(\mu\rr)}_{mn}\ll(g'\rr)\Phi^{\ll(\mu\rr)}_n\ll(g\rr),
\ee
then only a subset of the complete functional basis is present, the $D$'s that transform
in the same way, that is,
\be
D^{\ll(\mu\rr)}_{mn}~~~~~\mu,m~{\rm fixed}.
\ee
So in this case the expansion is shorter:
\be
\Phi^{\ll(\mu\rr)}_m\ll(g\rr)=\sum_nc^{\ll(\mu\rr)}_nD^{\ll(\mu\rr)}_{mn}\ll(g\rr).
\ee
\par
Let us now consider the functions $\Phi\ll(y\rr)$ on a coset manifold $G/H$.
The matrix elements
\[
D^{\ll(\mu\rr)}_{mn}\ll(L\ll(y\rr)\rr),
\]
with $L\ll(y\rr)$ coset representative of the coset space, $y$ coset coordinate, 
are a complete functional basis on $G/H$:
\be
\Phi\ll(L\ll(y\rr)\rr)=\sum_{\ll(\mu \rr)}\sum_{m,n=1}^{{\rm dim}\ll(\mu\rr)}{c_{mn}^{\ll(\mu\rr)}
D_{mn}^{\ll(\mu\rr)}\ll(L\ll(y\rr)\rr)}
\ee
satisfying
\beq
\int_{G/H}d\mu\!\ll(y\rr)D_{mn}^{\ll(\mu\rr)}\ll(g\rr)D_{sr}^{\ll(\nu\rr)}\ll(g^{-1}\rr)={{\rm vol}\ll(G/H\rr)\over{\rm vol}\ll(\mu\rr)}\delta_{mr}
\delta_{ns}\delta^{\ll(\mu\rr)\ll(\nu\rr)}\nn\\
\sum_{\ll(\mu\rr)}D_{mn}^{\ll(\mu\rr)}\ll(g\rr)D_{nm}^{\ll(\mu\rr)}({g'}^{-1}){\rm dim}\ll(\mu\rr)=
\delta\ll(g-g'\rr){\rm vol}\ll(G/H\rr)
\eeq
where $d\mu\!\ll(y\rr)$ is the invariant measure on $G/H$.
\par
We are interested on functions $\Phi\ll(L\ll(y\rr)\rr)$ on which an action of $H$ is defined,
that transform in an irreducible representation $\ll(\rho\rr)$ of $H$
\be
h\cdot\Phi^{\ll(\rho\rr)}_i\ll(L\ll(y\rr)\rr)\equiv D_{ij}\ll(h\rr)^{\ll(\rho\rr)}\Phi^{\ll(\rho\rr)}_j\ll(L\ll(y\rr)\rr)
\ee
where the index $i$ runs in $\ll(\rho\rr)$.
Which are the functions among the
\be
D_{mn}^{\ll(\mu\rr)}~~~m,n~{\rm running~in~}\ll(\mu\rr)~{\rm of~}G
\ee
that transform in this way? They are the functions
\be
\label{harmcondition}
D_{in}^{\ll(\mu\rr)}~~~i~{\rm running~in~}\ll(\rho\rr)~{\rm of~}H,~n~{\rm running~in~}\ll(\mu\rr)~{\rm of~}G
\ee
but $i$ has to run also in the $\ll(\mu\rr)$ of $G$, then not all the $G$ representations $\ll(\mu\rr)$ are 
appropriate, only the $\ll(\mu\rr)$ that satisfy the following condition:
the decomposition of $\ll(\mu\rr)$ with respect to $H\subset G$ must contain the $H$
irreducible representation $\ll(\rho\rr)$:
\be
\label{hdec}
\ll(\mu\rr)\stackrel{H}{\longrightarrow}\cdots\oplus\ll(\rho\rr)\oplus\cdots.
\ee
Only in this case $D_{mn}^{\ll(\mu\rr)}$ decomposes in $\ll(\dots,D_{in}^{\ll(\mu\rr)},D_{i'n}^{\ll(\mu\rr)},\dots\rr)$ and the $D_{in}^{\ll(\mu\rr)}$
actually exists.
The functions satisfying the (\ref{harmcondition}), (\ref{hdec}) are called {\bf $H-$harmonics} on $G/H$, and constitute a complete basis for the coset function 
$\Phi^{\ll(\rho\rr)}_i\ll(L\ll(y\rr)\rr)$. Its expansion is
\be
\label{salamexpansion}
\Phi^{\ll(\rho\rr)}_i\ll(L\ll(y\rr)\rr)={\sum_{\ll(\mu\rr)}}'\sum_n
c_n^{\ll(\mu\rr)}D_{in}^{\ll(\mu\rr)}\ll(L\ll(y\rr)\rr)
\ee
where $\sum'$ means a sum only on the representations
$\ll(\mu\rr)$ satisfying the property (\ref{hdec}).
Notice that the $H-$harmonics have both an index running in an irreducible representation 
of $G$ (on the right) and an index running in an irreducible representation of $H$ (on the left). The coefficients of the expansion
$c_n^{\ll(\mu\rr)}$ have an index of a representation of $G$ present in the expansion $\sum'$.
\par
\subsection{Differential operators on $H$ harmonics}
\par
The $H$--harmonics have a very powerful property: it is possible to express the action of differential operators
on them in an algebraic way. In fact, as we has seen in (\ref{rescHcovaction}), the action of the $H$--covariant derivative on the inverse coset representative is
\be
\cD_a^HL^{-1}=-r_aT_aL^{-1}~~~\hbox{no sum on $a$}
\ee
with $T_a$ generator of the subspace $\IK$ defined by the orthogonal decomposition $\IG=\IK\oplus\IH$, in the representation in which the inverse coset representative is
expressed, and $r_a$  rescaling of the vielbein. But the harmonics {\it are} the inverse coset representatives in the representation $\ll(\mu\rr)$:
\be
{{D^{\ll(\mu\rr)}}^i_{~n}}={L^{-1}}^i_{~n}.
\ee
More precisely, the harmonic in the $\ll(\rho,\mu^t\rr)$, ${{D^{\ll(\mu\rr)}}^i_{~n}}$, is obtained doing the decomposition (\ref{hdec}) of
the first index of ${{D^{\ll(\mu\rr)}}^m_{~n}}={L^{-1}}^m_{~n}$ and taking the $\ll(\rho\rr)$ term.
We consider the {\it inverse} coset representative because for simplicity of notation we want $H$ to act on their left, 
while it acts on the right of coset representatives.
\par
The action of $(T_a^{\ll(\mu\rr)})^m_{~n}$ on ${L^{-1}}^i_{~m}$ is 
\be
\cD^H_a\ll(D^i_{~n}\rr)=-r_a\ll(T_aD\rr)^i_{~n}=-r_a\ll(T_a\rr)^i_{m}D^m_{~n}
\label{actionTa}
\ee
where $\ll(T_a\rr)^i_{~m}$ is defined as the $\ll(\rho\rr)$ term in the decomposition (\ref{hdec}) of the index $n$ in $\ll(T_a\rr)^n_{~m}$, namely,
\be
\ll(T_a\rr)^n_{~m}=\ll\{\dots,\ll(T_a\rr)^i_{~m},\dots\rr\}.
\ee
\par
As we have seen, all the operators (\ref{scalarop}),$\dots$,(\ref{raritaschwingerop}) 
can be built with
the $SO\ll(7\rr)$ covariant derivative and the $G/H$ Killing metric. Furthermore, from the (\ref{defIM})
\be
\label{defIM2}
\cD^{SO\ll(7\rr)}=\cD^{H}+\IM_a\cB^a
\ee
we can write the $SO\ll(7\rr)$ covariant derivative in terms of the $H$ covariant derivative. Then, the action of all the operators 
(\ref{scalarop}),$\dots$,(\ref{raritaschwingerop}) on the harmonics can be expressed algebraically.
\par
\subsection{Harmonic expansion of supergravity fields}
\par
The fluctuations of eleven dimensional supergravity fields around the Freund-Rubin solution, defined in
(\ref{linearizedfields}), are fields on $AdS_4\times G/H$:
\be
\Phi_{[\lambda_1\lambda_2\lambda_3]{\hat a}}^{[E\,s]}(x,y)
\ee
where ${\hat a}$ is an index 
in the $[\lambda_1\lambda_2\lambda_3]$ representation of
$SO\ll(7\rr)$.  We leave implicit the spacetime index in the $[E\,s]$ of $SO\ll(3,2\rr)$
because we are not interested on it.
We know how to expand in harmonics a field lying in an $H$ representation, but 
$\Phi$ is in an $SO\ll(7\rr)$ representation. However $H\subset G$, embedding defined by the (\ref{HinSOn}):
\be
C^{~~b}_{ia}=-\ll(T^H_i\rr)_a^{~b}=\ll(T^H_i\rr)_{~a}^b,
\ee
then given the generators of $SO\ll(7\rr)$ in a generic representation, $t^{SO\ll(7\rr)}_{ab}$, the generators of $H$
in that representations are
\be
T^H_i=C_{ia}^{~~b}\ll(t^{SO\ll(7\rr)}\rr)^a_{~b}.
\ee
This defines the decomposition
of the $SO\ll(7\rr)$ irreducible representations $\ll[\lambda_1,\lambda_2,\lambda_3\rr]$ in $H$ irreducible representations.
In this way we can decompose the $\Phi$ field in fragments which are in irreducible representations of $H$
\be
\Phi_{[\lambda_1\lambda_2\lambda_3]{\hat a}}^{[E\,s]}(x,y)=\ll\{\Phi_{\ll(\rho_1\rr)i_1}^{[E\,s]}(x,y),\dots,
\Phi_{\ll(\rho_r\rr)i_r}^{[E\,s]}(x,y)\rr\}.
\ee
Each of these fragments can be expanded in $H$--harmonics 
\be
\label{littleexpansion}
\Phi_{\ll(\rho_{\xi}\rr)i_{\xi}}^{[E\,s]}(x,y)=\sum_{\ll(\mu\rr)}'\sum_{n=1}^{{\rm dim}
\ll(\mu\rr)}
~_{\xi}\phi_{\ll(\mu\rr)n}^{[E\,s]}\ll(x\rr)\cH_{\ll(\rho_{\xi}\rr)i_{\xi}}^{\ll(\mu\rr)n}\ll(L\ll(y\rr)\rr)~~~\xi=1,\dots,r
\ee
where we denote the harmonics $\cH_i^{~n}\equiv D_i^{~n}$.
Here each $\phi$ is one of the $AdS_4$ fields listed in section \ref{listfields}, and is in a representation of $G$.
So we have
\be
\Phi_{\ll[\lambda_1,\lambda_2,\lambda_3\rr]{\hat a}}^{\ll[E\,s\rr]}\ll(x,y\rr)=
\sum_{\ll(\mu\rr)}'\sum_n\ll\{_1\phi_{\ll(\mu\rr)n}^{[E\,s]}\ll(x\rr)\cH_{\ll(\rho_1\rr)i_1}^{\ll(\mu\rr)n}\ll(y\rr),
\dots,_r\phi_{\ll(\mu\rr)n}^{[E\,s]}\ll(x\rr)\cH_{\ll(\rho_1\rr)i_r}^{\ll(\mu\rr)n}\ll(y\rr)\rr\}.
\ee
This is the 
expansion given in the (\ref{kkexpansion})
\footnote{with little notation differences: in the (\ref{kkexpansion}) the name of the $G$ representation, $\ll(\mu\rr)$, 
is not explicit, and the index running in this representation is called $I$ instead of $n$},
where it was written in the simpler but less precise form $\Phi^{\ll[E\,s\rr]}_{\hat{a}}
\ll(x,y\rr)=\phi_n^{\ll[E\,s\rr]}\ll(x\rr)\cH_{\hat{a}}^n\ll(y\rr)$.
\par
\subsection{The mass spectrum from harmonic analysis}
\par
In order to find the masses of the $AdS_4$ supergravity fields we have to solve
the eigenvalue equations (\ref{scalarop}),$\dots$,(\ref{raritaschwingerop}), which all have the form
\be
\label{eigeneq}
\xbox_y^{\ll[\lambda_1,\lambda_2,\lambda_3\rr]}
\cH_{\ll[\lambda_1,\lambda_2,\lambda_3\rr]{\hat a}}^{\ll(\mu\rr)m}\ll(y\rr)=
M_{\ll[\lambda_1,\lambda_2,\lambda_3\rr]}\cH_{\ll[\lambda_1,\lambda_2,\lambda_3\rr]{\hat a}}^{\ll(\mu\rr)m}\ll(y\rr).
\ee
These eigevalues yield the masses of the fields by the formulae (\ref{massform}).
\par
{\bf Harmonic analysis allows us to find the eigenvalues of the (\ref{eigeneq}) 
with purely algebraic calculations, without solving any differential equation, and
without requiring an explicit coordinatization of the manifold.}
To do this, we have to write the (\ref{eigeneq}), which as we have seen is a shorthand
notation but not actually the exact expression, in a precise form.
First of all we have to consider the decomposition under $H$ of the given $SO\ll(7\rr)$ representation $\ll[\lambda_1,\lambda_2,\lambda_3\rr]$,
\be
\ll[\lambda_1,\lambda_2,\lambda_3\rr]\stackrel{H}{\longrightarrow}\ll(\rho_1\rr)\oplus\cdots\oplus\ll(\rho_r\rr)
\label{rdimvector}
\ee
where $r$ is the number of fragments in this decomposition.
We have to keep attention on the fragments which appear more than one time in the decomposition: $r$ is the number of $H$ UIRs
times their multiplicity.
\par
Then, we determine which $G$ UIRs satisfy the condition (\ref{hdec}), namely, do contain in their $H$ decompositions
the representations present in the (\ref{rdimvector}). The expansion contains only that $G$ representations $\ll(\mu\rr)$.
The eigenvalue equation has the form:
\be
\xbox_y^{\ll[\lambda_1,\lambda_2,\lambda_3\rr]}
\sum_{\ll(\mu\rr)}'\sum_n
\ll(\!\begin{array}{c}
_1\phi_{\ll(\mu\rr)n}^{[E\,s]}\ll(x\rr)\cH_{\ll(\rho_1\rr)i_1}^{\ll(\mu\rr)n}\ll(y\rr)\\
\vdots\\
_r\phi_{\ll(\mu\rr)n}^{[E\,s]}\ll(x\rr)\cH_{\ll(\rho_r\rr)i_r}^{\ll(\mu\rr)n}\ll(y\rr)\\
\end{array}\!\rr)
=M_{\ll[\lambda_1,\lambda_2,\lambda_3\rr]}
\sum_{\ll(\mu\rr)}'\sum_n
\ll(\!\begin{array}{c}
_1\phi_{\ll(\mu\rr)n}^{[E\,s]}\ll(x\rr)\cH_{\ll(\rho_1\rr)i_1}^{\ll(\mu\rr)n}\ll(y\rr)\\
\vdots\\
_r\phi_{\ll(\mu\rr)n}^{[E\,s]}\ll(x\rr)\cH_{\ll(\rho_r\rr)i_r}^{\ll(\mu\rr)n}\ll(y\rr)\\
\end{array}\!\rr).
\label{expandedeq}
\ee
Now, we determine the action of the $H$ covariant derivative on the fragments $\cH_{\ll(\rho_i\rr)i}^{\ll(\mu\rr)n}$ given by the (\ref{actionTa}), and 
by means of the (\ref{defIM2}), the action of the $SO\ll(7\rr)$ covariant derivative and then of the invariant operator
$\xbox_y^{\ll[\lambda_1,\lambda_2,\lambda_3\rr]}$. This action, in general, sends a fragment $\cH_{\ll(\rho_{\xi}\rr)i_{\xi}}^{\ll(\mu\rr)n}$ in
a fragment $\cH_{\ll(\rho_{\xi'}\rr)i_{\xi'}}^{\ll(\mu\rr)n}$ with $i_{\xi'}$ running in $\rho_{\xi'}$ which is 
another $H$ representation of the decomposition (\ref{hdec}). 
However, for an operator $\xbox_y^{\ll[\lambda_1,\lambda_2,\lambda_3\rr]}$
among the (\ref{scalarop}),$\dots$,(\ref{raritaschwingerop}), this new fragment has to
be present in the decomposition of $\ll[\lambda_1,\lambda_2,\lambda_3\rr]$.
So, if we consider the $r$ dimensional vector space with base vectors
\be
\label{defe}
e_{\xi}\equiv
\sum_{\ll(\mu\rr)}'\sum_n
\ll(\begin{array}{c}
0\\
\vdots\\
0\\
_{\xi}\phi_{\ll(\mu\rr)n}^{[E\,s]}\ll(x\rr)\cH_{\ll(\rho_{\xi}\rr)i_{\xi}}^{\ll(\mu\rr)n}\ll(y\rr)\\
0\\
\vdots\\
0\\
\end{array}\rr),
\ee
the invariant operator acts as an $r\times r$
numeric matrix on this vector space
\be
\label{truematrix}
\xbox_y^{\ll[\lambda_1,\lambda_2,\lambda_3\rr]}e_{\xi}=
\cM_{\xi}^{~\xi'}e_{\xi'}.
\label{eigeneqtrue}
\ee
All we have to do is to construct this matrix, whose entries depend on the labels of the $G$ UIR, and to find its eigenvalues
$M_{\ll[\lambda_1,\lambda_2,\lambda_3\rr]}$. The corresponding eigenvectors
are the supergravity fields. Substituting these eigenvalues in the (\ref{massform}) we find the masses of the
corresponding $AdS_4$ fields listed in section \ref{listfields}. In this way the complete spectrum of compactified supergravity
can be worked out.
\par
It is worth noting that the $AdS_4$ fields so found are in $G$ representations. I remind that $G=G'\times SO\ll(\cN\rr)$, 
and $SO\ll(\cN\rr)$ is the $R$--symmetry group of the theory. So in this way we find not only the masses of the fields with
various spins (namely, their $SO\ll(3,2\rr)$ UIRs), but also their $R$--symmetry labels. We can then organize them in
supermultiplets of $\cN$--extended supersymmetry. This is the reason we prefer coset spaces in a form which exhibits explicitly the
complete isometry. All the fields in a same supermultiplet have the same $G'$ labels, so each supermultiplet is in a $G'$ UIR.
The final result of the harmonic analysis, then, is the list of all the supermultiplets present in the theory and of their
$G'$ representations. 
\par
Furthermore, the mass relations (\ref{massrelationchi}), (\ref{massrelationlambdaT}), (\ref{massrelationlambdaL})
 give the mass of every field in a supermultiplet in terms of 
the mass of some other field in the same supermultiplet. This is a very strong check against errors, because 
from the analysis of an $SO\ll(7\rr)$ harmonics we know what we expect from other $SO\ll(7\rr)$ harmonics.
And this allows us to skip the more complicate operators: thanks of the constraint of supersymmetry - namely, 
the fields have to make supermultiplets with same $G'$ labels and masses related by the mass relations (\ref{massrelationchi}), 
(\ref{massrelationlambdaT}), (\ref{massrelationlambdaL}) - only the harmonic analysis of a part of the operators (\ref{scalarop}),$\dots$,
(\ref{raritaschwingerop}) is necessary. Finally, if when we start the harmonic analysis we do not know completely
the structure of the multiplets, we can use the harmonic analysis itself and the mass relations to fill the
blanks in our knowledge. This is what we have actually done: the derivation of the spectrum of $AdS_4\times M^{111}$
supergravity (see next section and \cite{noi1}) allowed us to complete the structure of $\cN=2$ supermultiplets, the derivation of
the spectrum of $AdS_4\times N^{010}$ supergravity \cite{piet}, \cite{noi4} allowed us to complete the structure of $\cN=3$ supermultiplets,
yielding the tables given in chapter $2$.
\par
\section{The mass spectrum of $AdS_4\times M^{111}$ supergravity}
\par
This section is based on the work done in the collaboration \cite{noi1}. However, part of the spectrum had been 
worked out previously \cite{spectfer}, \cite{spec321}.
I start giving the result, then I explain how it has been found. 
The conventions relative to $M^{111}$ space are given in appendix \ref{mpqrconventions}.
\par
\subsection{Representations of $G$ and $H$}
\par
\label{youngconventions}
Here we fix the conventions for
labelling the irreducible representations of 
\be
G'=SU\ll(3\rr)\times SU\ll(2\rr).
\ee
It has rank three, so that its irreducible representations are labeled
by three integer numbers. A representation of $SU\left(3\right)$ can be identified by a Young
diagram of the following type
\begin{eqnarray*}
\begin{array}{l}
\begin{array}{|c|c|c|c|c|c|}
\hline
             \hskip .3 cm & \cdots & \hskip .3 cm &
             \hskip .3 cm & \cdots & \hskip .3 cm \\
\hline
\end{array}\,\,,\\
\begin{array}{|c|c|c|}
             \hskip .3 cm & \cdots & \hskip .3 cm \\
\hline
\end{array}
\end{array}\\
\underbrace{\hskip 2.2 cm}_{M_2}
\underbrace{\hskip 2.2 cm}_{M_1}
\end{eqnarray*}
while an UIR of $SU\left(2\right)$ can be identified by a Young diagram
as follows
\begin{eqnarray*}
\begin{array}{|c|c|c|}
\hline
             \hskip .3 cm & \cdots & \hskip .3 cm \\
\hline
\end{array}\,\,.\\
\underbrace{\hskip 2.2 cm}_{2J}
\end{eqnarray*}
Hence we can take the  nonnegative integers $M_1,~M_2,~2J$, 
as the labels of a $G'$ irreducible representation.
\par
An UIR of $G=SU\ll(3\rr)\times SU\ll(2\rr)\times U\ll(1\rr)$ will then be denoted by 
\be
\ll[M_1.M_2,J,Y\rr]
\ee
where $Y$ is the charge of the $U\ll(1\rr)$ factor in $G$, namely, the hypercharge.
An UIR of $H=SU\ll(2\rr)\times U\ll(1\rr)\times U\ll(1\rr)$ will be denoted by 
\be
\ll[J^h,Z',Z''\rr]
\ee
where $Z',Z''$ are the charges of the $U\ll(1\rr)$'s generated by $Z'$ and $Z''$, and $J^h$
is the $SU\ll(2\rr)$ spin defined as above; the superscript $c$ distinguishes it
from the label $J$ of the $SU\ll(2\rr)\subset G$ representation.
\par
\subsection{Results}
Relying on the procedures explained in the following sections, we have found the  following results.
\par
Not every $G'$ representation  is actually present, but only
those representations that satisfy the following relations
\begin{equation}
M_2-M_1 \in 3 \, \ZZ \quad ; \quad J \in \IN.
\label{labcond}
\end{equation}
In the following pages,  for each type  of ${\cal N}=2$ multiplet
I list the  $G'$ representations through which it occurs in the
spectrum. I do this by writing bounds on the range of values for
the $M_1,~M_2,~2J~$ labels. The reader  should take into account
that, case by case, in addition to the specific bounds written, also the
general restriction (\ref{labcond}) has to be imposed.
\par
Furthermore for every multiplet, I give   the energy and hypercharge
values $E_0$ and $y_0$  of the Clifford vacuum. From the tables \ref{longgraviton},
\ref{longgravitino}, \ref{longvector}, \ref{shortgraviton},
\ref{shortgravitino}, \ref{shortvector}, \ref{hyper}, \ref{masslessgraviton},
\ref{masslessvector} it is straightforward to get the energies and the
hypercharges of all other  fields in each  multiplet.
\par
As a short--hand notation let us name $H_0$ the following quadratic form in the representation
labels:
\begin{equation}
\label{defH0}
H_0\equiv {64\over 3}\left(M_2+M_1+M_2M_1\right)+32J\left(J+1\right)+{32\over 9}
\left(M_2-M_1\right)^2.
\end{equation}
Up to multiplicative constants, the first two addenda $M_2 + M_1 + M_2\,
M_1$ and  $J(J+1)$   are the Casimirs of  $G^\prime = SU(3)
\times SU(2)$. The last addendum is contributed by the square of the hypercharge through
its relation with the $SU(3)$ representation implied by the geometry of the space.
\par
I remind that when $y_0=0$ the multiplet is real, while when $y_0\neq 0$ it is complex, 
and the number of the degrees of freedom is doubled; in the latter case, I write only the multiplet
with positive hypercharge (or, in few cases, negative); the one with negative
(positive) hypercharge, and conjugate flavour indices
$\ll(M_1,M_2,J\rr)\longrightarrow\ll(M_2,M_1,J\rr)$, is its complex conjugate.
\vskip 2cm
\centerline{LONG MULTIPLETS}
\begin{enumerate}
  \item {{\underline {\bf Long graviton multiplets}} 
\beq
\hbox{complex $(y_0\neq 0)$}&:&
\left(2\left(2\right),8\left(3\over 2\right),12\left(1\right),8\left(1\over 2\right),2\ll(0\rr)\right)\nn\\
\hbox{real $(y_0=0)$}&:&
\left(1\left(2\right),4\left(3\over 2\right),6\left(1\right),4\left(1\over 2\right),1\ll(0\rr)\right)\nn
\eeq
One long graviton multiplet (table \ref{longgraviton}) in each representation of the series
\begin{eqnarray}
&&\left\{M_2\ge M_1\ge 0,~J>{1\over 3}\left(M_2-M_1\right)\right\}\cup\nonumber\\
&&\left\{M_2\ge M_1>0,~J={1\over 3}\left(M_2-M_1\right)\right\}
\end{eqnarray}
with
\begin{equation}
h:~E_0={1\over 2}+{1\over 4}\sqrt{H_0+36},~y_0={2\over 3}\left(M_2-M_1\right)
\end{equation}}
  \item {{\underline {\bf Long gravitino multiplets}}
\beq
\hbox{complex $(y_0\neq 0)$}&:&
\left(2\left({3\over 2}\right),8\left(1\right),12\left(1\over 2\right),8\left(0\right)\right)\nn
\eeq
There are two different realizations of the long gravitino multiplet, $\chi^+$ with positive mass
and $\chi^-$ with negative mass.
\begin{itemize}
\item Four long gravitino multiplets (two $\chi^+$ and two $\chi^-$, table
\ref{longgravitino}) in each representation of the series
\[
\left\{M_2\ge M_1>0,~J>{1\over 3}\left(M_2-M_1\right)+1\right\}\cup
\]
\be
\left\{M_2\ge M_1>1,~J={1\over 3}\left(M_2-M_1\right)+1\right\}
\ee
with
\begin{eqnarray}
\chi^+:~&E_0=-{1\over 2}+{1\over 4}\sqrt{H_0+{32\over 3}\left(M_2-M_1\right)+16},
&y_0={2\over 3}\left(M_2-M_1\right)-1\nn\\
\chi^+:~&E_0=-{1\over 2}+{1\over 4}\sqrt{H_0-{32\over 3}\left(M_2-M_1\right)+16},
&y_0={2\over 3}\left(M_2-M_1\right)+1\nn\\
\chi^-:~&E_0={3\over 2}+{1\over 4}\sqrt{H_0+{32\over 3}\left(M_2-M_1\right)+16},
&y_0={2\over 3}\left(M_2-M_1\right)-1\nn\\
\chi^-:~&E_0={3\over 2}+{1\over 4}\sqrt{H_0-{32\over 3}\left(M_2-M_1\right)+16},
&y_0={2\over 3}\left(M_2-M_1\right)+1\,.\nn\\
\end{eqnarray}
\item Three long gravitino multiplets (one $\chi^+$ and two $\chi^-$, table
\ref{longgravitino}), in each representation of the series
\begin{equation}
\left\{M_2\ge M_1=1,~J={1\over 3}\left(M_2-M_1\right)+1\right\}
\end{equation}
with
\begin{eqnarray}
\chi^+:~&E_0=-{1\over 2}+{1\over 4}\sqrt{H_0+{32\over 3}\left(M_2-M_1\right)+16},
&y_0={2\over 3}\left(M_2-M_1\right)-1\nn\\
\chi^-:~&E_0={3\over 2}+{1\over 4}\sqrt{H_0-{32\over 3}\left(M_2-M_1\right)+16},
&y_0={2\over 3}\left(M_2-M_1\right)+1\nn\\
\chi^-:~&E_0={3\over 2}+{1\over 4}\sqrt{H_0+{32\over 3}\left(M_2-M_1\right)+16},
&y_0={2\over 3}\left(M_2-M_1\right)-1\,.\nn\\
\end{eqnarray}
\item Two long gravitino multiplets (one $\chi^+$ and one $\chi^-$, table
\ref{longgravitino}) in each representation of the series
\[
\left\{M_2>M_1> 0,~J={1\over 3}\left(M_2-M_1\right)
\right\}\cup
\]
\[
\left\{M_2>M_1> 0,~J={1\over 3}\left(M_2-M_1\right)-1
\right\}\cup
\]
\begin{equation}
\left\{M_2>M_1=0,~J\ge {1\over 3}\left(M_2-M_1\right)\right\}
\end{equation}
with
\begin{eqnarray}
\chi^+:~&E_0=-{1\over 2}+{1\over 4}\sqrt{H_0+{32\over 3}\left(M_2-M_1\right)+16},
&y_0={2\over 3}\left(M_2-M_1\right)-1\nn\\
\chi^-:~&E_0={3\over 2}+{1\over 4}\sqrt{H_0+{32\over 3}\left(M_2-M_1\right)+16},
&y_0={2\over 3}\left(M_2-M_1\right)-1\,.\nn\\
\end{eqnarray}
\item One long gravitino multiplet (a $\chi^-$, table \ref{longgravitino}),
in each representation of the series
\begin{equation}
\left\{M_2>M_1=0,~J={1\over 3}\left(M_2-M_1\right)-1\right\}
\end{equation}
with
\begin{equation}
\chi^-:~E_0={3\over 2}+{1\over 4}\sqrt{H_0+{32\over 3}\left(M_2-M_1\right)+16},
~y_0={2\over 3}\left(M_2-M_1\right)-1\,.
\end{equation}
\end{itemize}
}
\item {{\underline {\bf Long vector multiplets}}
\beq
\hbox{complex $(y_0\neq 0)$}&:&
\left(2\left(1\right),8\left(1\over 2\right),10\left(0\right)\right)\nn\\
\hbox{real $(y_0=0)$}&:&
\left(1\left(1\right),4\left(1\over 2\right),5\left(0\right)\right)\nn
\eeq
As already stressed there are different realizations of the long vector
 multiplet
  arising   from different fields of the $D=11$ theory. We have the $W$ vector
multiplets, the $A$ vector multiplets, and   the  $Z$ vector
multiplets.
\begin{itemize}
  \item One $W$ long vector multiplet (table \ref{longvector})
in each representation of the series
\begin{equation}
\left\{M_2\ge M_1\ge 0,~J\ge {1\over 3}\left(M_2-M_1\right)\right\}
\end{equation}
with
\begin{equation}
W:~E_0={5\over 2}+{1\over 4}\sqrt{H_0+36},~y_0={2\over 3}\left(M_2-M_1\right)\,.
\end{equation}
\item One $A$ long vector multiplet (table \ref{longvector})
in each representation of the series
\[
\left\{M_2\ge M_1=0,~J>{1\over 3}\left(M_2-M_1\right)+1\right\}\cup
\]
\[
\left\{M_2\ge M_1=1,~J>{1\over 3}\left(M_2-M_1\right)\right\}\cup
\]
\begin{equation}
\left\{M_2\ge M_1>1,~J\ge{1\over 3}\left(M_2-M_1\right)\right\}
\end{equation}
with
\begin{equation}
A:~E_0=-{3\over 2}+{1\over 4}\sqrt{H_0+36},~y_0={2\over 3}\left(M_2-M_1\right)\,.
\end{equation}
\item One $Z$ long vector multiplet (table \ref{longvector})
in each representation of the series
\begin{equation}
\left\{M_2\ge M_1>0,~J\ge {1\over 3}\left(M_2-M_1\right)\right\}
\end{equation}
with
\begin{equation}
Z:~E_0={1\over 2}+{1\over 4}\sqrt{H_0+4},~y_0={2\over 3}\left(M_2-M_1\right)\,.
\end{equation}
\item One $Z$ long vector multiplet (table \ref{longvector})
in each representation of the series
\[
\left\{M_2>M_1+3,~J\ge {1\over 3}\left(M_2-M_1\right)-2\right\}\cup
\]
\begin{equation}
\left\{M_1+3\ge M_2>1,~J>-{1\over 3}\left(M_2-M_1\right)+1\right\}
\end{equation}
with
\begin{equation}
Z:~E_0={1\over 2}+{1\over 4}\sqrt{H_0+{64\over 3}\left(M_2-M_1\right)-28},
~y_0={2\over 3}\left(M_2-M_1\right)-2\,.
\end{equation}
\end{itemize}
}
\end{enumerate}
\vskip 0.6cm
\centerline{SHORT MULTIPLETS}
\vskip 0.6cm
\centerline{They are always complex.}
\par
\begin{enumerate}
  \item {{\underline {\bf Short graviton multiplets}}
  $~~\left(2\left(2\right),6\left(3\over 2\right),6\left(1\right),2\ll({1\over 2}\rr)\right)$\par
\begin{itemize}
  \item One short graviton multiplet (table \ref{shortgraviton})
in each representation of the series
\begin{equation}
\cases{
M_2=3k\cr
M_1=0\cr
J=k\cr}~~k>0~{\rm integer}
\end{equation}
with
\begin{equation}
E_0=2k+2,~y_0=2k\,.
\end{equation}
\end{itemize} } 
  \item {{\underline {\bf Short gravitino multiplets}}
   $~~\left(2\left({3\over 2}\right),6\left(1\right),6\left(1\over 2\right),2\ll(0\rr)\right)$\par
  \begin{itemize}
\item One short gravitino multiplet ($\chi^+$, table \ref{shortgravitino}) in
each representation of the series
\begin{equation}
\cases{
M_2=3k+1\cr
M_1=1\cr
J=k+1\cr}~~
k\ge 0~{\rm integer}
\end{equation}
with
\begin{equation}
E_0=2k+{5\over 2},~y_0=2k+1\,.
\end{equation}
\item One short gravitino multiplet ($\chi^+$, table \ref{shortgravitino}) in
each representation of the series
\begin{equation}
\cases{
M_2=3k+3\cr
M_1=0\cr
J=k\cr}~~
k\ge 0~{\rm integer}
\end{equation}
with
\begin{equation}
E_0=2k+{5\over 2},~y_0=2k+1\,.
\end{equation}
\end{itemize}
}
  \item {{\underline {\bf  Short vector multiplets}}
 $~~\left(2\left(1\right),6\left(1\over 2\right),6\left(0\right)\right)$\par
  \begin{itemize}
\item One short vector multiplet ($A$, table \ref{shortvector}), in each
representation of the series
\begin{equation}
\cases{
M_2=3k+1\cr
M_1=1\cr
J=k\cr}~~k>0~{\rm integer}
\end{equation}
\begin{equation}
\cases{
M_2=3k\cr
M_1=0\cr
J=k+1\cr}~~k>0~{\rm integer}
\end{equation}
with
\begin{equation}
E_0=2k+1,~y_0=2k\,.
\end{equation}
\end{itemize}
}
\item {{\underline {\bf  Hypermultiplets}}
 $~~\left(2\left(1\over 2\right),4\left(0\right)\right)$\par
\begin{itemize}
\item One hypermultiplet (table \ref{hyper}) in each representation of the series
\begin{equation}
\label{hyp1}
\cases{
M_2=3k\cr
M_1=0\cr
J=k\cr}~~k>0~{\rm integer}
\end{equation}
\begin{equation}
E_0=\left|y_0\right|=2k\,.
\end{equation}
\end{itemize}
}
\end{enumerate}
\vskip 0.6cm
\centerline{MASSLESS MULTIPLETS}
\vskip 0.6cm
\centerline{They are always real.}
\par
\begin{enumerate}
\item {\underline{\bf The massless graviton multiplet}} (table \ref{masslessgraviton})
$~~\left(1\left(2\right),2\left(3\over 2\right),1\ll(1\rr)\right)$\\
in the singlet representation
\begin{equation}
M_2=M_1=J=0
\end{equation}
with
\begin{equation}
E_0=2,~y_0=0.
\end{equation}
In this multiplet the graviphoton is associated with the Killing vector of the $R$--symmetry group
$U\left(1\right)_R$.
\item {\underline{\bf The massless vector multiplet}} (table \ref{masslessvector})
$~~\left(1\left(1\right),2\left(1\over 2\right),2\left(0\right)\right)$\\
in  the {\em adjoint representation} of the $G^\prime$ group
\begin{eqnarray}
M_2=M_1=1,~J=0&&\label{rkilsu3}\\
M_2=M_1=0,~J=1&&\label{rkilsu2}
\end{eqnarray}
with
\begin{equation}
  E_0=1,~y_0=0.
\label{labellini}
\end{equation}
\item {\underline{\bf An additional massless vector multiplet}} in the singlet
representation of the gauge group
\begin{equation}
M_2=M_1=J=0\label{rbetti}
\end{equation}
with the same energy and hypercharges as in (\ref{labellini}) that
arises from the three--form $a_{mab}$ and is due to the existence
of one closed cohomology two--form on the $M^{111}$ manifold. This
multiplet is named the Betti multiplet.
\end{enumerate}
\par
Summarizing, the massless spectrum, besides the supergravity multiplet
contains  twelve vector multiplets: so the total number
of  massless gauge bosons is thirteen, one of them being the graviphoton.
In the low energy effective lagrangian we just couple to supergravity these twelve vector
multiplets. However we expect the gauging of
a thirteen--parameter group:
\begin{equation}
SU\left(3\right)\times SU\left(2\right)\times U(1)_R \times U\left(1\right)'
\end{equation}
the further $U\left(1\right)'$  being  associated with the Betti
multiplet. All Kaluza Klein states are neutral under
$U\left(1\right)'$ yet non perturbative states can carry
$U(1)^\prime$ charges. This actually happens, as I will show in chapter $4$.
\par
\subsection{Harmonic expansion on $M^{111}$}
\par
In terms of the structure constant of $G$, given in appendix \ref{mpqrconventions},
we find that the embedding of the algebra $\IH$ into the adjoint
representation of $SO(7)$ is
\begin{eqnarray}
(T_H)^{a}_{\ b}=C^{~~ a}_{Hb},\nonumber\\
(T_{Z'})^{a}_{\ b}=\left(
\begin{array}{ccc}
2\sqrt{3}f^{8AB}&0&0\\
0&2\e^{mn}&0\\
0&0&0
\end{array}\right),\label{TZp}\\
(T_{Z''})^{a}_{\ b}=\left(
\begin{array}{ccc}
-\sqrt{3}f^{8AB}&0&0\\
0&3\e^{mn}&0\\
0&0&0
\end{array}\right),\label{TZpp}\\
(T_{\dot m})^{a}_{\ b}=\left(
\begin{array}{ccc}
f^{{\dot m}AB}&0&0\\
0&0&0\\
0&0&0
\end{array}\right).\label{Tm}
\end{eqnarray}
This means that the $SO(7)$-indices of the various $n$-forms can be
split in the following subsets, each one transforming into an irreducible
representation of $H$:
\begin{equation}
\begin{array}{ccl}
{\cal Y}^{a} &=& \{{\cal Y}^A,{\cal Y}^m,{\cal Y}^3\}\\
{\cal Y}^{[ab]} &=& \{{\cal Y}^{AB},{\cal Y}^{Am},{\cal Y}^{mn},
{\cal Y}^{A3},{\cal Y}^{m3}\}\\
{\cal Y}^{[abc]} &=& \{{\cal Y}^{ABC},{\cal Y}^{ABm},{\cal Y}^{AB3},
{\cal Y}^{Amn},{\cal Y}^{Am3},{\cal Y}^{mn3}\}
\end{array}
\end{equation}
and the $SO(7)$ irreducible representations $[\l_1\l_2\l_3]$ break into the direct
sum of $H$ irreducible representations. The $[J^h,Z,Z']$ labels of every 
$H$--irreducible fragment can be read off from the action of $T_{\dot m}$, $T_Z$,
$T_{Z'}$ on that representation.
\par
The expansion (\ref{littleexpansion})
of a generic $SO(3,2)\!\!\times\!\!H$-irreducible field is
\begin{equation} \label{expansion}
\Phi_{[J^hZ'Z'']i_1\cdots i_{2J^h}}^{[E\,s]}(x,y) =
{\sum}'_{[M_1M_2J\,Y]}
\sum_{\zeta}\ \sum_m
\cH_{[J^h Z' Z'']i_1\cdots i_{2J^h}}^{[M_1M_2J\,Y]m\zeta}(y)\cdot
\varphi^{[E\,s]}_{[M_1M_2J\,Y]m\zeta}(x)\,.
\end{equation}
The coefficients $\varphi(x)$ of the expansion become
the space--time fields of the theory in $AdS_4$.
The first sum is over all the $G$ irreducible representations $[M_1M_2J\,Y]$ which
break into the given $H$-one.
 We call ${\sum}'$ the sum over this 
subset of the possible representations of $G$.
The subscripts $_{i_1,\cdots,i_{2J^h}}$ span the representation
space of $[J^hZ'Z'']$, while $m$ is a collective index which
spans the representation space of $[M_1M_2J\,Y]$.
Finally $\zeta$ accounts for the fact that the same $H$ irreducible representation
can be embedded in $G$ in different ways.
In fact, given an $SU(3)$ representation $[M_1,\,M_2]$, it can contain the 
$SU(2)\subset SU(3)$ representation $[J^h]$ in more than one way 
\be
\label{MMJ}
[M_1,\,M_2]\stackrel{SU(2)}{\longrightarrow}\cdots\oplus [J^h]\oplus\cdots
\oplus [J^h]\oplus\cdots\,.
\ee
The cases of interest for us are $J^h=0,\,1/2,\,1$. $J^h=0$ is contained only
in one way
\be
\label{mu0}
J^h=0\,:~~~
\begin{array}{l}
\begin{array}{|c|c|c|c|c|c|}
\hline
            1  & \cdots & 1 & 3 & \cdots & 3 \\
\hline
\end{array}\\
\begin{array}{|c|c|c|}
            2 & \cdots & 2 \\
\hline
\end{array}
\end{array}~,
\ee
while for $J^h=1/2,\,1$ we have:
\begin{equation}
\label{muab}
J^h=\frac{1}{2}:\ \left\{
\begin{array}{ccc}
\zeta=(a) & \to &
\begin{array}{l}
\begin{array}{|c|c|c|c|c|c|c|}
\hline
            1  & \cdots & 1 & 3 & \cdots & 3 & i\\
\hline
\end{array}\\
\begin{array}{|c|c|c|}
            2 & \cdots & 2 \\
\hline
\end{array}
\end{array}\\
 & & \\
\zeta=(b) & \to &
\begin{array}{l}
\begin{array}{|c|c|c|c|c|c|c|}
\hline
             \hskip .03 cm i \hskip .04 cm & 1
             & \cdots & 1 & 3 & \cdots & 3\\
\hline
\end{array}\\
\begin{array}{|c|c|c|c|c|}
            3 & 2 & \cdots & 2 \\
\hline
\end{array}
\end{array}
\end{array}\right.
\end{equation}
\par
\begin{equation}
\label{mucde}
J^h=1:\ \left\{
\begin{array}{ccc}
\zeta=(c) & \to & \frac{1}{2}\left(
\begin{array}{l}
\begin{array}{|c|c|c|c|c|c|c|c|}
\hline
            1 & \cdots & 1 & \hskip .03 cm i \hskip .03 cm
            & j & 3 & \cdots & 3\\
\hline
\end{array}\\
\begin{array}{|c|c|c|c|c|}
            2 & \cdots & 2 & 3\\
\hline
\end{array}
\end{array} + (i\leftrightarrow j)\right)\\
 && \\
\zeta=(d) & \to &
\begin{array}{l}
\begin{array}{|c|c|c|c|c|c|c|c|}
\hline
             \hskip .03 cm i \hskip .04 cm & j & 1
             & \cdots & 1 & 3 & \cdots & 3\\
\hline
\end{array}\\
\begin{array}{|c|c|c|c|c|c|}
            3 & 3 & 2 & \cdots & 2 \\
\hline
\end{array}
\end{array}\\
 && \\
\zeta=(e) & \to &
\begin{array}{l}
\begin{array}{|c|c|c|c|c|c|c|c|}
\hline
            1 & \cdots & 1 & 3 & \cdots & 3 &
            \hskip .03 cm i \hskip .03 cm & j\\
\hline
\end{array}\\
\begin{array}{|c|c|c|c|}
            2 & \cdots & 2 \\
\hline
\end{array}
\end{array}
\end{array}\right.
\end{equation}
%%%%%%%%%%%%%%%%%%%%%%%%%%%%%%%%%%%%%%%%%%%%%%%%%%%%%%%%%%%
%                                                         %
%              C O N S T R A I N T S                      %
%                                                         %
%%%%%%%%%%%%%%%%%%%%%%%%%%%%%%%%%%%%%%%%%%%%%%%%%%%%%%%%%%%
\subsection{The constraints on the irreducible representations}
\label{irrepconstraint}
As I said, the expansion of a
generic field contains only the harmonics whose $H$- and
$G$-quantum numbers are such that the $G$ representation, 
decomposed under $H$, contains the $H$ representation of 
the field. This fact poses some constraints on the 
$G$-quantum numbers.
\par
Depending on which constraints are satisfied by a certain $G$ representation,
only part of the harmonics is  present,
and only their corresponding four--dimensional fields 
appear in the spectrum. Then, in the $G$ representations in which
such field  disappear, there is  {\sl multiplet shortening}.
In the modern perspective of Kaluza Klein theory, the exact spectrum
of the short multiplets is crucial.
Hence the importance of analyzing this disappearance of harmonics with care.
\par
Every harmonic is defined by its $SU\left(2\right)\times U\left(1\right)'
\times U\left(1\right)''$ representation, identified by the labels
$[J^h~Z'~Z'']$. Substituting these values in equations (\ref{z1}), (\ref{z2}),
\beq
\ii\sqrt{3}\l_8+2\ii\ll(\frac{\ii}{2}\s_3\rr)-4\ii Y &=& Z'\nn\\
-\frac{\ii}{2}\sqrt{3}\l_8+3\ii\ll(\frac{\ii}{2}\s_3\rr)-4\ii Y &=& Z''~,
\label{constraintsequations}
\eeq
we can determine the constraints of the $G$ representations.
\par
The eigenvalue of $\sqrt{3}\l_8={\rm diag}(1,1,-2)$ depends on $\zeta$: it is 
$2(M_2-M_1)$ for the scalar, and
\beq
\zeta=(a)\,:&~~&\sqrt{3}\l_8=2(M_2-M_1)+3\nn\\
\zeta=(b)\,:&~~&\sqrt{3}\l_8=2(M_2-M_1)-3\nn\\
\zeta=(c)\,:&~~&\sqrt{3}\l_8=2(M_2-M_1)\nn\\
\zeta=(d)\,:&~~&\sqrt{3}\l_8=2(M_2-M_1)-6\nn\\
\zeta=(e)\,:&~~&\sqrt{3}\l_8=2(M_2-M_1)+6~.
\eeq
Simplifying $\frac{1}{2}\s_3$ one finds the first constraint, that is the
value of $Y$ in terms of $M_1$ and $M_2$. In the cases of interest for us
we have five possible expressions of $Y$, identifying five families of
$G$ representations which we denote with the superscripts
$^0$, $^+$, $^-$, $^{++}$ and $^{--}$:
\begin{eqnarray}
\left.
\begin{array}{rccl}
^0:&\quad Y & = & 2/3(M_2-M_1)\\
^{++}:&\quad Y & = & 2/3(M_2-M_1)-2\\
^{--}:&\quad Y & = & 2/3(M_2-M_1)+2
\end{array}
\right\}
&&\begin{array}{c}
{\rm for\ \ bosonic}\\
{\rm fields}
\end{array}\nonumber\\
\left.
\begin{array}{rccl}
^+:&\quad Y & = & 2/3(M_2-M_1)-1\\
^-:&\quad Y & = & 2/3(M_2-M_1)+1
\end{array}
\right\}
&&\begin{array}{c}
{\rm for\ \ fermionic}\\
{\rm fields~.}
\end{array}
\label{constraintY}
\end{eqnarray}
It is worth noting that the value of $Y$ identifies a $U\left(1\right)_R$ 
representation, so these five families of representations correspond
to the five possible representations of $U\left(1\right)_R$.
\par
The second constraint arising from (\ref{constraintsequations}) 
is the lower bound on the quantum number $J$,
since its third component $J_3=\frac{1}{2}\s_3$ 
is linked to $Y$.
We have three possibilities:
\begin{equation}
\label{constraintJ}
J \geq \left\{
\begin{array}{c}
 \left|Y/2\right|\\
\left|Y/2+1\right|\\
\left|Y/2-1\right|\,.
\end{array}\right.
\end{equation}
\par
The last kind of constraint refers to $M_1$ and $M_2$.
If $M_1,\,M_2$ are too small, the decomposition (\ref{MMJ}) 
can contain less $[J^h]$ representations, because not all of the 
Young tableaux (\ref{mu0}), (\ref{muab}), (\ref{mucde}) do exist. 
The conditions for the existence of the representations 
$[J^h]_{\zeta}$ in $[M_1,\,M_2]$ are:
\begin{equation}
\label{constraintM}
\begin{array}{|c||c|c|}\hline
{\rm constraints} & J^h & \zeta\\
\hline
\begin{array}{l}
M_1 \geq 0\\
M_2 \geq 0
\end{array}&0&-\\
\hline
\begin{array}{l}
M_1 \geq 1\\
M_2 \geq 0
\end{array}&\ft{1}{2}&(a)\\
\hline
\begin{array}{l}
M_1 \geq 0\\
M_2 \geq 1
\end{array}&\ft{1}{2}&(b)\\
\hline
\begin{array}{l}
M_1 \geq 1\\
M_2 \geq 1
\end{array}&1&(c)\\
\hline
\begin{array}{l}
M_1 \geq 0\\
M_2 \geq 2
\end{array}&1&(d)\\
\hline
\begin{array}{l}
M_1 \geq 2\\
M_2 \geq 0
\end{array}&1&(e)\\
\hline
\end{array}
\end{equation}
\par
We organize the series of the $G=G'\times U\left(1\right)_R$ 
representations in the following way.
The constraints  (\ref{constraintJ}) and (\ref{constraintM}),  with
the five  values of $Y$ in terms of $M_1,~M_2$ given by 
(\ref{constraintY}), define  the series of $G'$ representations that we list in
table \ref{defseries}. Every $G'$ representation, together with a 
superscript $^0,\ ^+,\ ^-,\ ^{++}$ or $^{--}$ that define the value of $Y$,
is  a $G$ representation. So the series of $G'$ representations
defined in table \ref{defseries} with such a superscript are 
series of  representations of the whole $G$ group.
\par
For each family of representations ($^0,\ ^+,\ ^-,\ ^{++}, $ and
$^{--}$) we call a series {\sl regular} if it contains the
maximum number of harmonics.
The regular series cover all the representations with $M_1$, $M_2$
and $J$ sufficiently high to satisfy all the inequality constraints.
When some of these inequalities are not satisfied instead, some of
the harmonics may be absent in the expansion.
The series $A_R,A_1,\dots,A_8$ are defined by means of the constraints arising in
the cases $^0,\,^+,\,^-$, while the series $B_R,B_1,\dots,B_{11}$ are defined
by means of the constraints arising in the cases $^{++},\,^{--}$.
\par
In tables \ref{0series}, \ref{+series}, \ref{-series}, \ref{++series}, \ref{--series},
we show which harmonics are present for the different series of $G$ representations.
The first column contains the name of each series.
The other columns contain the possible harmonics, each labeled
by its $H$-quantum numbers.
An asterisk denotes the presence of a given harmonic.
To obtain the constraints on the conjugate series 
it suffices to exchange $M_1$ and $M_2$, as explained in \cite{spectfer}, \cite{castdauriafre}.
\par
Tables \ref{defseries},$\dots$\ref{--series} are the results of the
analysis of equations (\ref{constraintsequations}) for the relevant
$[J^h,Z',Z'']$ values.
\begin{table}
\begin{footnotesize}
\[
\begin{array}{||c|c|c||}\hline
G'{\rm -name} & M_1,\ M_2 & J\ \ {\rm constraints}\\
\hline \hline
A_R & M_2>0,\ M_1>0 & J>\left|(M_2-M_1)/3\right|\\
\hline
A_1 & M_2>M_1>0 & J=(M_2-M_1)/3\\
\hline
A_2 & M_2>M_1>0 & J=(M_2-M_1)/3-1\\
\hline
A_3 & M_2>M_1=0 & J>(M_2-M_1)/3\\
\hline
A_4 & M_2>M_1=0 & J=(M_2-M_1)/3\\
\hline
A_5 & M_2>M_1=0 & J=(M_2-M_1)/3-1\\
\hline
A_6 & M_2=M_1>0 & J=0\\
\hline
A_7 & M_2=M_1=0 & J>0\\
\hline
A_8 & M_2=M_1=0 & J=0\\
\hline
B_R & M_2>1,\ M_1\geq 0 & J>\left|(M_2-M_1)/3-1\right|\\
\hline
B_1 & M_2>M_1+3 & J=(M_2-M_1)/3-1\\
\hline
B_2 & M_2>M_1+3 & J=(M_2-M_1)/3-2\\
\hline
B_3 & M_2=M_1+3 & J=(M_2-M_1)/3-1\\
\hline
B_4 & M_1\geq M_2>1 & J=-(M_2-M_1)/3+1\\
\hline
B_5 &M_1\geq M_2>1 & J=-(M_2-M_1)/3\\
\hline
B_6 & M_1\geq M_2=1 & J>-(M_2-M_1)/3+1\\
\hline
B_7 & M_1\geq M_2=1 & J=-(M_2-M_1)/3+1\\
\hline
B_8 & M_1\geq M_2=1 & J=-(M_2-M_1)/3\\
\hline
B_9 & M_1\geq M_2=0 & J>-(M_2-M_1)/3+1\\
\hline
B_{10} & M_1\geq M_2=0  & J=-(M_2-M_1)/3+1\\
\hline
B_{11} & M_1\geq M_2=0 & J=-(M_2-M_1)/3\\
\hline
\end{array}
\]
\end{footnotesize}
\caption{Series of $G'$ representations}
\label{defseries}
\end{table}
\begin{table}
\centering
\begin{footnotesize}
\[
\begin{array}{||c||c|c|c|c|c|c|c|c|c|c|c|c||} \hline
J^h&0&0&0&1/2&1/2&1/2&1/2&1/2&1/2&1&1&1\\
Z'&0&-2i&2i&3i&-3i&i&-i&5i&-5i&0&-2i&2i\\
Z''&0&-3i&3i&-3/2i&3/2i&-9/2i&9/2i&3/2i&-3/2i&0&-3i&3i\\
\mu& & & &\left({\rm a}\right)&\left({\rm b}\right)&\left({\rm a}\right)
&\left({\rm b}\right)&\left({\rm a}\right)&\left({\rm b}\right)
&\left({\rm c}\right)&\left({\rm c}\right)&\left({\rm c}\right)
\\\hline\hline
A^{\ 0}_R&*&*&*&*&*&*&*&*&*&*&*&*\\
\hline
A^{\ 0}_1&*&*& &*&*&*& & &*&*&*& \\
\hline
A^{*\ 0}_1&*& &*&*&*& &*&*& &*& &*\\
\hline
A^{\ 0}_2& &*& & & &*& & &*& &*& \\
\hline
A^{*\ 0}_2& & &*& & & &*&*& & & &*\\
\hline
A^{\ 0}_3&*&*&*& &*& &*& &*& & & \\
\hline
A^{*\ 0}_3&*&*&*&*& &*& &*& & & & \\
\hline
A^{\ 0}_4&*&*& & &*& & & &*& & & \\
\hline
A^{*\ 0}_4&*& &*&*& & & &*& & & & \\
\hline
A^{\ 0}_5& &*& & & & & & &*& & & \\
\hline
A^{*\ 0}_5& & &*& & & & &*& & & & \\
\hline
A^{\ 0}_6&*& & &*&*& & & & &*& & \\
\hline
A^{\ 0}_7&*&*&*& & & & & & & & & \\
\hline
A^{\ 0}_8&*& & & & & & & & & & & \\
\hline
\end{array}
\]
\end{footnotesize}
\caption{Harmonics content for the series of type $^0$}
\label{0series}
\end{table}
\begin{table}
\centering
\begin{footnotesize}
\[
\begin{array}{||c||c|c|c|c||}
\hline
J^h & 0 & 0 & 1/2 & 1/2\\
Z'  & 2i & 4i & -i & i\\
Z'' & -3i &0 & -3i/2 & 3i/2\\
\mu & & &\left({\rm b}\right)&\left({\rm b}\right)\\
\hline
\hline
A_R^{\ +} & * & * & * & *\\
\hline
A_1^{\ +} & * & * & * & * \\
\hline
A_1^{*+} & & * & & *\\
\hline
A_2^{\ +} & * & & * & \\
\hline
A_2^{*+} & & & & \\
\hline
A_3^{\ +} & * & * & * & *\\
\hline
A_3^{*+} & * & * & & \\
\hline
A_4^{\ +} & * & * & * & *\\
\hline
A_4^{*+} & & * & & \\
\hline
A_5^{\ +} & * & & * & \\
\hline
A_5^{*+} & & & & \\
\hline
A_6^{\ +} & & * & & * \\
\hline
A_7^{+} & * & * & & \\
\hline
A_8^{+} & & * & & \\
\hline
\end{array}
\]
\end{footnotesize}
\caption{Harmonics content for the series of type $^+$}
\label{+series}
\end{table}
\begin{table}
\centering
\begin{footnotesize}
\[
\begin{array}{||c||c|c|c|c||}
\hline
J^h & 0 & 0 & 1/2 & 1/2\\
Z' & -2i & -4i & i & -i\\
Z'' & 3i & 0 & 3i/2 & -3i/2\\
\mu & & &\left({\rm a}\right)&\left({\rm a}\right)\\
\hline
\hline
A_R^{\ -} & * & * & * & *\\
\hline
A_1^{\ -} & & * & & *\\
\hline
A_1^{*-} & * & * & * & *\\
\hline
A_2^{\ -} & & & &\\
\hline
A_2^{*-} & * & & * &\\
\hline
A_3^{\ -} & * & * & &\\
\hline
A_3^{*-} & * & * & * & *\\
\hline
A_4^{-} & & * & &\\
\hline
A_4^{*-} & * & * & * & *\\
\hline
A_5^{-} & & & &\\
\hline
A_5^{*-} & * & & * &\\
\hline
A_6^{-} & & * & & *\\
\hline
A_7^{-} & * & * & &\\
\hline
A_8^{-} & & * & &\\
\hline
\end{array}
\]
\end{footnotesize}
\caption{Harmonics content for the series of type $^-$}
\label{-series}
\end{table}
\begin{table}
\centering
\begin{footnotesize}
\[
\begin{array}{||c||c|c|c|c|c|c|c|c|c||} \hline
J^h&0&0&0&1/2&1/2&1/2&1&1&1\\
Z'&6i&8i&4i&3i&i&5i&0&-2i&2i\\
Z''&-3i&0&-6i&-3/2i&-9/2i&3/2i&0&-3i&3i\\
\mu & & & &\left({\rm b}\right)&\left({\rm b}\right)
&\left({\rm b}\right)&\left({\rm d}\right)
&\left({\rm d}\right)&\left({\rm d}\right)\\
\hline\hline
B^{\ ++}_R&*&*&*&*&*&*&*&*&*\\
\hline
B^{\ ++}_1&*& &*&*&*& &*&*& \\
\hline
B^{\ ++}_2& & &*& &*& & &*& \\
\hline
B^{\ ++}_3&*& & &*& & &*& & \\
\hline
B^{\ ++}_4&*&*& &*& &*&*& &*\\
\hline
B^{\ ++}_5& &*& & & &*& & &*\\
\hline
B^{\ ++}_6&*&*&*&*&*&*& & & \\
\hline
B^{\ ++}_7&*&*& &*& &*& & & \\
\hline
B^{\ ++}_8& &*& & & &*& & & \\
\hline
B^{\ ++}_9&*&*&*& & & & & & \\
\hline
B^{\ ++}_{10}&*&*& & & & & & & \\
\hline
B^{\ ++}_{11}& &*& & & & & & & \\
\hline
\end{array}
\]
\end{footnotesize}
\caption{Harmonics content for the
series of type $^{++}$}
\label{++series}
\end{table}
\begin{table}
\centering
\begin{footnotesize}
\[
\begin{array}{||c||c|c|c|c|c|c|c|c|c||} \hline
J^h&0&0&0&1/2&1/2&1/2&1&1&1\\
Z'&-6i&-8i&-4i&-3i&-i&-5i&0&2i&-2i\\
Z''&3i&0&6i&3/2i&9/2i&-3/2i&0&3i&-3i\\
\mu & & & &\left({\rm a}\right)&\left({\rm a}\right)
&\left({\rm a}\right)&\left({\rm e}\right)
&\left({\rm e}\right)&\left({\rm e}\right)\\
\hline\hline
B^{*\ --}_R&*&*&*&*&*&*&*&*&*\\
\hline
B^{*\ --}_1&*& &*&*&*& &*&*& \\
\hline
B^{*\ --}_2& & &*& &*& & &*& \\
\hline
B^{*\ --}_3&*& & &*& & &*& & \\
\hline
B^{*\ --}_4&*&*& &*& &*&*& &*\\
\hline
B^{*\ --}_5& &*& & & &*& & &*\\
\hline
B^{*\ --}_6&*&*&*&*&*&*& & & \\
\hline
B^{*\ --}_7&*&*& &*& &*& & & \\
\hline
B^{*\ --}_8& &*& & & &*& & & \\
\hline
B^{*\ --}_9&*&*&*& & & & & & \\
\hline
B^{*\ --}_{10}&*&*& & & & & & & \\
\hline
B^{*\ --}_{11}& &*& & & & & & & \\
\hline
\end{array}
\]
\end{footnotesize}
\caption{Harmonics content for the series
of type $^{--}$}
\label{--series}
\end{table}
%%%%%%%%%%%%%%%%%%%%%%%%%%%%%%%%%%%%%%%%%%%%%%%%%%%%%%%%%%%
%                                                         %
%     D I F F E R E N T I A L       C A L C U L U S       %
%                                                         %
%%%%%%%%%%%%%%%%%%%%%%%%%%%%%%%%%%%%%%%%%%%%%%%%%%%%%%%%%%%
\subsection{Differential calculus via harmonic analysis}
The Kaluza Klein kinetic operators $\xbox_y^{[\l_1\l_2\l_3]}$
act as finite dimensional matrices on
the harmonic subspaces of fixed $G$-quantum numbers:
\begin{equation}
\xbox_y^{[\l_1\l_2\l_3]} e^{[M_1M_2J\,Y]}_{\xi}(y)=
\cM\left([M_1M_2J\,Y]\right)_{\xi}^{\ \xi'}
e^{[M_1M_2J\,Y]}_{\xi'}(y)
\end{equation}
(see (\ref{defe}), (\ref{truematrix}) ).
\par
Let us now consider the explicit action of the covariant
derivative \eqn{covder} on the harmonics.
It is given by the (\ref{defIM})
\begin{equation}\label{MM}
\cD = d + \Omega^Ht_H + \cB^{a}\IM_{a} \equiv \cD^H + \cB^{a}\IM_{a}\,,
\end{equation}
where $t_H$ are the generators of $H$ and 
$\IM_{a}$ the part of the $SO(7)$-connection not belonging to $H$.
The zero-forms transform in the trivial representation of
the tangent space structure group $SO(7)$.
In other words, the $SO(7)$ generators in the scalar representation
vanish identically.
This means that the covariant derivatives equal the simple one:
$\cD = \cD^H = d$.
For the vector representation instead, we can easily compute the
matrices $(\IM_{c})^{a}_{\ b}$ from eq. \eqn{covder}:
\[
\begin{array}{l}
\IM_1=\left(\begin{array}{cc|c|cccc}
0&0&0&0&0&0&0\\
0&0&-2&0&0&0&0\\\hline
0&2&0&0&0&0&0\\\hline
0&0&0&0&0&0&0\\
0&0&0&0&0&0&0\\
0&0&0&0&0&0&0\\
0&0&0&0&0&0&0
\end{array}\right),
\qquad
\IM_2=\left(\begin{array}{cc|c|cccc}
0&0&2&0&0&0&0\\
0&0&0&0&0&0&0\\\hline
-2&0&0&0&0&0&0\\\hline
0&0&0&0&0&0&0\\
0&0&0&0&0&0&0\\
0&0&0&0&0&0&0\\
0&0&0&0&0&0&0
\end{array}\right),\\
~~~~~~~~~\begin{array}{c}
\underbrace{\hspace{1.2 cm}}_{m}
\underbrace{\hspace{0.7 cm}}_{3}
\underbrace{\hspace{2.2 cm}}_{A}\\
\end{array}\qquad
\end{array}
\]
\begin{eqnarray}
\IM_3=\frac{1}{3}\left(\begin{array}{cc|c|cccc}
0&2&0&0&0&0&0\\
-2&0&0&0&0&0&0\\\hline
0&0&0&0&0&0&0\\\hline
0&0&0&0&0&0&0\\
0&0&0&0&0&0&0\\
0&0&0&0&0&0&0\\
0&0&0&0&0&0&0
\end{array}\right),\nn\\
\IM_4=\left(\begin{array}{cc|c|cccc}
0&0&0&0&0&0&0\\
0&0&0&0&0&0&0\\\hline
0&0&0&0&-2&0&0\\\hline
0&0&0&0&0&0&0\\
0&0&2&0&0&0&0\\
0&0&0&0&0&0&0\\
0&0&0&0&0&0&0
\end{array}\right),\qquad
\IM_5=\left(\begin{array}{cc|c|cccc}
0&0&0&0&0&0&0\\
0&0&0&0&0&0&0\\\hline
0&0&0&2&0&0&0\\\hline
0&0&-2&0&0&0&0\\
0&0&0&0&0&0&0\\
0&0&0&0&0&0&0\\
0&0&0&0&0&0&0
\end{array}\right),\nn\\
\IM_6=\left(\begin{array}{cc|c|cccc}
0&0&0&0&0&0&0\\
0&0&0&0&0&0&0\\\hline
0&0&0&0&0&0&-2\\\hline
0&0&0&0&0&0&0\\
0&0&0&0&0&0&0\\
0&0&0&0&0&0&0\\
0&0&2&0&0&0&0
\end{array}\right),\qquad
\IM_7=\left(\begin{array}{cc|c|cccc}
0&0&0&0&0&0&0\\
0&0&0&0&0&0&0\\\hline
0&0&0&0&0&2&0\\\hline
0&0&0&0&0&0&0\\
0&0&0&0&0&0&0\\
0&0&-2&0&0&0&0\\
0&0&0&0&0&0&0
\end{array}\right).
\end{eqnarray}
We know that (\ref{actionTa})
\begin{equation}
\label{algebract}
\cD^H_a\ll(\cH^i_{~n}\rr)=-r_a\ll(T_a\cH\rr)^i_{~n}=-r_a\ll(T_a\rr)^i_{m}\cH^m_{~n}.
\end{equation}
By means of eq. \eqn{B} we can calculate the explicit components
of $\cD^H$, i.e. its projection along the vielbein:
\begin{eqnarray}
\cD^H = \cB^{a}\cD^H_{a} = -\Omega^{a}t_{a},\nonumber\\
\left\{\begin{array}{ccl}
\cD^H_A & = & -\ft{4}{\sqrt{3}}i\l_A,\\
\cD^H_m & = & -\ft{4}{\sqrt{2}}i\s_m,\\
\cD^H_3 & = & -\ft{4}{3}Z,
\end{array}\right.
\end{eqnarray}
where the coset generators $t_{a}$ act on the harmonics as follows.
$\l_A$ acts on the $SU(3)$ part of the $G$ representation of the harmonic.
The fundamental representation of $\lambda_A$ is given by the
Gell--Mann matrices (see Appendix \ref{convenzioni}). 
On a generic Young tableau $\lambda_A$ acts as the tensor
representation. 
\par
To give an example, let us consider 
the case $[M_1,M_2]=[2,2]$, $[J^h]=[1/2]$, $\zeta=(b)$. The index $i$ in the
(\ref{algebract}) (or, more precisely, its $SU(2)$ part) runs in the representation
\be
\begin{array}{l}
\begin{array}{|c|c|c|c|}
\hline
            1 & i\, & 3 & 3 \\
\hline
\end{array}\\
\begin{array}{|c|c|}
            2 & 3 \\
\hline
\end{array}
\end{array}
\,.
\ee
With this notation we write the $H$ representation as a fragment of the $G$ representation
(in the (\ref{algebract}) the index of the $G$ representation is $m$). Let us
consider the component $i=1$ of this representation,
\be
\begin{array}{l}
\begin{array}{|c|c|c|c|}
\hline
            1 & 1 & 3 & 3 \\
\hline
\end{array}\\
\begin{array}{|c|c|}
            2 & 3 \\
\hline
\end{array}
\end{array}
\,.
\ee
We can determine from (\ref{algebract}) the action of 
\be
\l_4=\ll(\begin{array}{ccc} 0 & 0 & 1 \\ 0 & 0 & 0 \\ 1 & 0 & 0 \\
\end{array}\rr)
\ee
on this component:
\begin{eqnarray}
\l_4\,
\begin{array}{l}
\begin{array}{|c|c|c|c|}
\hline
            1 & 1 & 3 & 3 \\
\hline
\end{array}\\
\begin{array}{|c|c|}
            2 & 3 \\
\hline
\end{array}
\end{array}=\hspace{10 cm}\nn\\
=\begin{array}{l}
\begin{array}{|c|c|c|c|}
\hline
            3 & 1 & 3 & 3 \\
\hline
\end{array}\\
\begin{array}{|c|c|}
            2 & 3 \\
\hline
\end{array}
\end{array}+
\begin{array}{l}
\begin{array}{|c|c|c|c|}
\hline
            1 & 3 & 3 & 3 \\
\hline
\end{array}\\
\begin{array}{|c|c|}
            2 & 3 \\
\hline
\end{array}
\end{array}+
\begin{array}{l}
\begin{array}{|c|c|c|c|}
\hline
            1 & 1 & 3 & 3 \\
\hline
\end{array}\\
\begin{array}{|c|c|}
            2 & 1 \\
\hline
\end{array}
\end{array}+2\,
\begin{array}{l}
\begin{array}{|c|c|c|c|}
\hline
            1 & 1 & 1 & 3 \\
\hline
\end{array}\\
\begin{array}{|c|c|}
            2 & 3 \\
\hline
\end{array}
\end{array}=\nn\\
=\begin{array}{l}
\begin{array}{|c|c|c|c|}
\hline
            3 & 1 & 3 & 3 \\
\hline
\end{array}\\
\begin{array}{|c|c|}
            2 & 3 \\
\hline
\end{array}
\end{array}+2\,
\begin{array}{l}
\begin{array}{|c|c|c|c|}
\hline
            1 & 1 & 1 & 3 \\
\hline
\end{array}\\
\begin{array}{|c|c|}
            2 & 3 \\
\hline
\end{array}
\end{array}.
\end{eqnarray}
Similarly, $\s^m\ (m=\{1,2\})$ acts as the $m$-th Pauli matrix
on the fundamental representation of $SU(2)$, and as its $n$-th
tensor power on the $n$-boxes $SU(2)$ Young tableau:
\be
\s^1\,
\begin{array}{|c|c|c|c|c|}
\hline
            1 & 1 & 1 & 2 & 2\\
\hline
\end{array}= 3\,
\begin{array}{|c|c|c|c|c|}
\hline
            1 & 1 & 2 & 2 & 2 \\
\hline
\end{array}+ 2\,
\begin{array}{|c|c|c|c|c|}
\hline
            1 & 1 & 1 & 1 & 2\\
\hline
\end{array}\,.
\ee
Finally, $Z$ acts trivially, multiplying the harmonic by its
$Z$-charge:
\be
Z=\frac{\ii}{2}\sqrt{3}\l_8+\frac{\ii}{2}\s_3+\ii Y
\ee
\begin{eqnarray}
Z\,\begin{array}{l}
\begin{array}{|c|c|c|c|}
\hline
            1 & 1 & 3 & 3 \\
\hline
\end{array}\\
\begin{array}{|c|c|}
            2 & 3 \\
\hline
\end{array}
\end{array}\otimes\,
\begin{array}{|c|c|c|c|c|}
\hline
            1 & 1 & 1 & 2 & 2 \\
\hline
\end{array}\ =\hspace{8cm}\nn\\
=\left(-\frac{3}{2}i-\frac{1}{2}i+iY\right)\,\begin{array}{l}
\begin{array}{|c|c|c|c|}
\hline
            1 & 1 & 3 & 3 \\
\hline
\end{array}\\
\begin{array}{|c|c|}
            2 & 3 \\
\hline
\end{array}
\end{array}\otimes\,
\begin{array}{|c|c|c|c|c|}
\hline
            1 & 1 & 1 & 2 & 2 \\
\hline
\end{array}\,.\nn\\
\end{eqnarray}
In the course of the calculations, one often encounters
the $H$-covariant Laplace-Beltrami operator on $G/H$:
\begin{equation}\label{HLaplace}
\d^{ab}\cD^H_{\,a}\cD^H_{\,b}=\frac{16}{3}\l_A\l_A+
\frac{16}{2}\s^m\s^m-\frac{16}{9}Z^2\,.
\end{equation}
The eigenvalues of the first operator, $\l_A\l_A$, are listed
in the following table:
$$
\begin{array}{|c|c||c|}\hline
J^h & \zeta & \l^A\l^A\quad{\rm eigenvalues}\\
\hline\hline
0 & - & 4(M_1+M_2+M_1M_2)\\
1/2 & (a) &2(4M_1+2M_1M_2-3)\\
1/2 & (b) &2(4M_2+2M_1M_2-3)\\
1 & (c) &4(M_1+M_2+M_1M_2-2)\\
1 & (d) &4(3M_2-M_1+M_1M_2-5)\\
1 & (e) &4(3M_1-M_2+M_1M_2-5)\\\hline
\end{array}
$$
while the eigenvalues of $\s^m\s^m$ depend on the
$SU\ll(2\rr)$ quantum numbers $J$ and $J_3$:
\begin{eqnarray}
\s^m\s^m &
\begin{array}{|c|c|c|c|c|c|}
\hline
             1 & \cdots & 1 & 2 & \cdots & 2 \\
\hline
\end{array} & = 4\left[J(J+1)-J_3^2\right]
\begin{array}{|c|c|c|c|c|c|}
\hline
             1 & \cdots & 1 & 2 & \cdots & 2 \\
\hline
\end{array}\,, \nonumber\\
& \underbrace{\hskip 2 cm}_{m_1}
\underbrace{\hskip 2 cm}_{m_2} & 
\end{eqnarray}
where $2J=m_1+m_2$ and $2J_3=m_1-m_2$.
The complete Kaluza Klein mass operator heavily depends
on the kind of field it acts on and will be analyzed in
detail in the next sections.
%%%%%%%%%%%%%%%%%%%%%%%%%%%%%%%%%%%%%%%%%%%%%%%%%%
%                                                 %
%           Z E R O     F O R M                   %
%                                                 %
%%%%%%%%%%%%%%%%%%%%%%%%%%%%%%%%%%%%%%%%%%%%%%%%%%%
\subsubsection{The zero-form}
The only representation into which the $[0,0,0]$ (i.e. the scalar)
of $SO(7)$ breaks under $H$, is obviously the $H$-scalar representation.
The question now is: which $G$-irreducible representations 
do contain the $H$-scalar?
From equations (\ref{constraintsequations}) we see that $Z'=Z''=0$ implies
\begin{equation}
2J_3=Y={2\over 3}\left(M_2-M_1\right).
\end{equation}
This means that
\begin{itemize}
\item $M_2-M_1\in 3\ZZ$
\item $J\in\IN$
\item $J\ge\left|{1\over 3}\left(M_2-M_1\right)\right|$
\item $Y={2\over 3}\left(M_2-M_1\right)$.
\end{itemize}
We will denote the scalar as
\begin{equation}
{\cal Y}(x,y) = [0|{\rm I}](x,y) \equiv {\sum}'_{\left[M_1M_2J\,Y\right]} \cH_{[000]}^{[M_1M_2 J Y]}(y)
\cdot S_{[M_1M_2 Y J]}(x).
\end{equation}
The Kaluza Klein mass operator for the zero-form ${\cal Y}$ is given by
\begin{equation}
\xbox^{[000]}{\cal Y} \equiv \cD_{b}\cD^{b}{\cal Y} = \cD^H_{b}\cD^{Hb}{\cal Y}.
\end{equation}
For the scalar, there are no $\IM$--connection terms.
So, by means of eq. \eqn{HLaplace}, the computation of its
eigenvalues, on the $G$--representations as listed above, is immediate:
\begin{eqnarray}
\xbox^{[000]}{\cal Y} \equiv M_{(0)^3}{\cal Y} &=& 
\left[\ft{64}{3}(M_1+M_2+M_1M_2)+ 32 J(J+1) +
\ft{32}{9}(M_2-M_1)^2\right]{\cal Y}=\nonumber\\
&=&H_0{\cal Y}
\end{eqnarray}
where $H_0$ is the same quantity defined in eq. (\ref{defH0}).
\par
As we see from the Kaluza Klein expansion  (\ref{kkexpansion}),
the eigenvalues of the zero--form harmonic allow us to determine
the masses of the $AdS_4$ graviton field $h$ and the scalar fields $S,\Sigma$.
%%%%%%%%%%%%%%%%%%%%%%%%%%%%%%%%%%%%%%%%%%%%%%%%%%%
%                                                 %
%              O N E    F O R M                   %
%                                                 %
%%%%%%%%%%%%%%%%%%%%%%%%%%%%%%%%%%%%%%%%%%%%%%%%%%%
\subsubsection{The one-form}
Let us decompose under $H$ the vector representation of $SO(7)$.
The generators of $H$ in this representation are given by
(\ref{TZp}), (\ref{TZpp}), (\ref{Tm}). We see that
\be
{\bf 7}\longrightarrow{\bf 4}\oplus{\bf 2}\oplus{\bf 1}\,.
\ee
It is convenient to move to a complex basis. The real four dimensional
representation of $SU(2)$ is a complex two dimensional representation, and
the real two dimensional representation is a complex one dimensional
representation. The change from the real to the complex basis can be performed
as follows (see also \cite{spec321}):
\begin{eqnarray}
{\cal Y}^A & = & \l^A_{3i}\langle 1 | {\rm I } \rangle_i+
\l^A_{i3}\langle 1 | {\rm I } \rangle_i^*\\
{\cal Y}^m & = & \s^m_{21}\langle 1 | {\rm I } \rangle_{.}+
\s^m_{12}\langle 1 | {\rm I } \rangle_{.}^*\\
{\cal Y}^3 & = & [ 1 , {\rm I} ]\,.
\end{eqnarray}
By applying (\ref{TZp}), (\ref{TZpp}), (\ref{Tm}) on these fragments one finds
the $Z'$, $Z''$ eigenvalues.
\par
As a result of this calculation, one finds that 
the decomposition under $H$ of the vector representation of 
$SO(7)$ is the following
\footnote{When we write a pair of complex conjugate 
representations we assume a conjugation relation between 
them. For example, by writing $[0,-2i,-3i]\oplus[0,2i,3i]$
we intend a complex representation of complex dimension one or
real dimension two.}:
\begin{equation}
[1,0,0] \to [0,0,0] \oplus [0,-2i,-3i] \oplus [0,2i,3i] \oplus
[\ft{1}{2},3i,-\ft{3}{2}i] \oplus [\ft{1}{2},-3i,\ft{3}{2}i].
\end{equation}
These $H$--irreducible fragments can be expanded as in (\ref{expansion})
\footnote{Using the same conventions as in 
\cite{spectfer}, \cite{univer}, \cite{castdauriafre}, \cite{spec321}, the reader might notice that
there appears a sign $(-1)^{J - J_3}$ upon taking the
complex conjugate of the fragments $\langle \dots | \dots \rangle_x$.
In order to reduce the notation we have absorbed this sign
in the $x$--space fields $\tilde W\langle \dots , \dots \rangle$.
This will be done for all the
complex conjugates henceforth.  } (summation over
the $G$-quantum numbers is intended):
\begin{eqnarray}
&& {\rm For\ type}\quad ^0:\nonumber\\
&&\nonumber\\
\langle 1 | {\rm I } \rangle_i &=& {\cal H}^{[1/2,3i,-{3i}/{2}](a)}_i \cdot
                    W\langle \ft12 , {\rm I} \rangle \,,\nonumber\\
\langle 1 | {\rm I } \rangle_i^* &=& \varepsilon^{ij}
{\cal H}^{[1/2,-3i,{3i}/{2}](b)}_j \cdot
                    \tilde{W}\langle \ft12 , {\rm I} \rangle \,,\nonumber\\
\langle 1 | {\rm I } \rangle_\cdot &=&
{\cal H}^{[0,-2i,-3i]} \cdot
                 W\langle 0 , {\rm I} \rangle \,,\nonumber\\
\langle 1 | {\rm I } \rangle_\cdot^* &=&
{\cal H}^{[0,2i,3i]} \cdot
                    \tilde{W}\langle 0 , {\rm I} \rangle\,,\nonumber\\
{}[1 | {\rm I } ]_\cdot &=&
{\cal H}^{[0,0,0]} \cdot W[ 0 , {\rm I} ] \,,\label{1f0serexpansion}
\end{eqnarray}
\begin{eqnarray*}
 &&{\rm For\ type}\quad ^{++} :\nonumber\\
 &&\nonumber\\
 \langle 1 | {\rm I } \rangle_i &=& {\cal H}^{[1/2,3i,-3/2i](b)}_i \cdot
W\langle \ft12 , {\rm II} \rangle\,,\\
\nonumber \\
&&{\rm For\ type}\quad ^{--}:\nonumber\\
 &&\nonumber\\
 \langle 1 | {\rm I } \rangle_i^* &=& -\varepsilon^{ij}{\cal H}^{[1/2,-3i,3/2i](a)}_j \cdot
\tilde{W}\langle \ft12 , {\rm II} \rangle\,.
\end{eqnarray*}
As we see, there are five different $AdS_4$ fields ($W,\tilde{W}$)
in the case of the $^0$ series, and one field in the case of the
$^{++}$ and $^{--}$ series.
So, for the regular $^0$ series the Laplace Beltrami operator acts 
on the $AdS_4$ fields as a $5\times 5$ matrix.
For the {\sl exceptional} series it acts as a matrix of lower dimension.
\par
The Laplace Beltrami operator for the transverse one-form field ${\cal Y}^{a}$,
is given by
\begin{equation}
\xbox^{[100]}{\cal Y}^{a} \equiv M_{(1)(0)^2}{\cal Y}^{a} = 2\cD_{b}\cD^{[b}{\cal Y}^{a]} =
(\cD^{b}\cD_{b}+24){\cal Y}^{a}\ ,
\end{equation}
where transversality of ${\cal Y}^{a}$ means that $\cD_{a}{\cal Y}^{a}=0$.
From the decomposition $\cD_{a}=\cD^{H}_{a}+\IM_{a}$ we obtain:
\begin{equation}
\xbox^{[100]}{\cal Y}^{a} =
(\cD^{Hb}\cD^{H}_{b}+24){\cal Y}^{a}+\eta^{gd}\left(2(\IM_{g})^{a}_{\ b}
\cD^H_{d}+(\IM_{g})^{a}_{\ e}(\IM_{d})^{e}_{\ b}\right) {\cal Y}^{b}\,.
\end{equation}
The matrix of this operator on the $AdS_4$ fields is given by
\begin{footnotesize}
\begin{eqnarray}\label{onematrix}
\begin{array}{|c||c|c|c|c|c|}
\hline
 M_{(1)(0)^2}
  & W\langle \ft12,{\rm I}  \rangle
  & \tilde {W}\langle\ft12,{\rm I}  \rangle
  & W\langle 0,{\rm I} \rangle
  & \tilde {W}\langle 0,{\rm I}  \rangle
  & W[0,{\rm I}]\\
\hline
\hline
 W\langle \ft12,{\rm I}\rangle & H_0\!-\!\frac{32(M_2-M_1)}{3} &
 0 & 0 & 0 & \frac{16M_1}{\sqrt{3}}\\
 \tilde{W}\langle\ft12 ,{\rm I}\rangle & 0 & H_0\!+\!\frac{32(M_2-M_1)}{3} &
 0 & 0 & \frac{16M_2}{\sqrt{3}}\\
 W\langle 0,{\rm I}\rangle & 0 & 0 & H_0\!+\!\frac{32(M_2-M_1)}{3} &
 0 & -\frac{8(2J+Y)}{\sqrt{2}}\\
 \tilde{W}\langle 0,{\rm I}\rangle & 0 & 0 & 0 & H_0\!-\!\frac{32(M_2-M_1)}{3} &
 \frac{8(2J-Y)}{\sqrt{2}}\\
 W[0,{\rm I}] & \frac{32(2+M_2)}{\sqrt{3}} & \frac{32(2+M_1)}{\sqrt{3}} &
 -\frac{16(2+2J-Y)}{\sqrt{2}} & \frac{16(2+2J+Y)}{\sqrt{2}} & H_0\!+\!48\\
\hline
\end{array}.\nn\\
\end{eqnarray}
\end{footnotesize}
Its eigenvalues are:
\begin{eqnarray}\label{eigenAoneform}
\lambda_1 &=& H_0 + \ft{32}{3}(M_2-M_1)\,,
\nonumber \\
\lambda_2 &=& H_0 - \ft{32}{3}(M_2-M_1)\,,
\nonumber \\
\lambda_3 &=& H_0\,,
\\
\lambda_4 &=& H_0 + 24 + 4 \sqrt{H_0 + 36}\,,
\nonumber \\
\lambda_5 &=& H_0 + 24 - 4 \sqrt{H_0 + 36}\,.
\nonumber
\end{eqnarray}
Actually, what we have just calculated are the eigenvalues of 
\begin{equation}\label{actualM100}
M_{(1)(0)^2}{\cal Y}^a + \cD^{a}\cD_{b}{\cal Y}^b\,.
\end{equation}
It coincides with $M_{(1)(0)^2}$ when acting on a transverse one-form.
But on a generic ${\cal Y}^{a}$, which possibly contains a longitudinal
term, the second part of \eqn{actualM100}, $\cD^{a}\cD_{b}{\cal Y}^{b}$,
is not inert.
Indeed, let us suppose
$$
{\cal Y}^{a}=\cD^{a}{\cal Y}
$$
for some scalar function ${\cal Y}$.
Then
\begin{equation}
\cD^{a}\cD_{b}{\cal Y}^{b}=\cD^{a}\cD_{b}\cD^{b}{\cal Y}=\cD^{a}M_{(0)^3}
{\cal Y}=M_{(0)^3}{\cal Y}^{a}.
\end{equation}
So, our actual operator \eqn{onematrix} contains the eigenvalues
of $M_{(0)^3}$, which are \emph{longitudinal} (hence \emph{non-physical})
for the one--form.
This fact is true also for the two-form.
\par
The eigenvalue $\lambda_3$ in \eqn{eigenAoneform} is the longitudinal one,
equal to the zero--form eigenvalue $H_0$.
The other four, instead, are transverse physical eigenvalues.
\par
The matrices corresponding to the exceptional series are easily
obtained from \eqn{onematrix} by removing the rows and the
columns of the fields that disappear in the expansions (\ref{1f0serexpansion}), 
as we read from table \ref{0series}.
We list the mass eigenvalues of each series:
\begin{footnotesize}
\begin{eqnarray}
\begin{array}{|c||c|}
\hline
\, A_R \, & \, \lambda_1, \lambda_2, \lambda_3, \lambda_4, \lambda_5 \, \cr
\hline
A_1 & \lambda_1,  \lambda_3, \lambda_4, \lambda_5 \cr
\hline
A_1^* & \lambda_2,  \lambda_3, \lambda_4, \lambda_5 \cr
\hline
A_2 & \lambda_1  \cr
\hline
A_2^* & \lambda_2  \cr
\hline
A_3 & \lambda_1,  \lambda_3, \lambda_4, \lambda_5 \cr
\hline
A_3^* & \lambda_2,  \lambda_3, \lambda_4, \lambda_5 \cr
\hline
A_4 & \lambda_1, \lambda_3, \lambda_4  \cr
\hline
A_4^* & \lambda_2, \lambda_3, \lambda_4  \cr
\hline
A_5 & \lambda_1 \cr
\hline
A_5^* & \lambda_2 \cr
\hline
A_6 & \lambda_3, \lambda_4, \lambda_5 \cr
\hline
A_7 & \lambda_3, \lambda_4, \lambda_5 \cr
\hline
A_8 & \lambda_4 \cr
\hline
\end{array}
\label{eigenvaluesAoneform}
\end{eqnarray}
\end{footnotesize}
For the series of type $^{++}$ the operator $M_{(1)(0)^2}$ 
acts as a $1\times 1$ matrix on the $AdS_4$ fields and has eigenvalue:
\begin{equation}
H_0+\ft{32}{3}(M_2-M_1)
\label{eigenBoneform}
\end{equation}
for the series $B_R, B_1, B_3, B_4, B_6$ and $B_7$.
For the type $^{--}$-series the eigenvalue is the conjugate one
($M_2 \leftrightarrow M_1$) in the conjugate series.
\par
We can use the eigenvalues of the one--form harmonic to determine (see 
next section and \cite{univer}) 
the masses of the $AdS_4$vector field $A,W$.
%%%%%%%%%%%%%%%%%%%%%%%%%%%%%%%%%%%%%%%%%%%%%%%%%%%
%                                                 %
%              T W O    F O R M                   %
%                                                 %
%%%%%%%%%%%%%%%%%%%%%%%%%%%%%%%%%%%%%%%%%%%%%%%%%%%
\subsubsection{The two-form}
Under the action of $H=SU(2)\times U(1)'\times U(1)''$
the 21 components of the $SO(7)$ two-form transform into the
completely reducible representation:
\begin{eqnarray}
[1,1,0] & \to & [1,0,0] \oplus [0,0,0] \oplus [0,0,0]
\oplus [0,6i,-3i] \oplus [0,-6i,3i] \oplus\nonumber\\
 && \oplus [1/2,i,-9/2i] \oplus [1/2,-i,9/2i] \oplus [1/2,5i,3/2i]
\oplus [1/2,-5i,-3/2i] \nonumber\\
 && \oplus [1/2,3i,-3/2i] \oplus [1/2,-3i,3/2i] \oplus [0,-2i,-3i]
\oplus [0,2i,3i]\,.
\end{eqnarray}
The  decomposition of the two--form in $H$--irreducible fragments is as follows:
\begin{eqnarray*}
{\cal Y}^{AB} & = & -i\l^{[A}_{\ i3}\l^{B]}_{\ 3i}\ [2|{\rm I}]_{.}-
i\l^{[A}_{\ i3}\,\l^{B]}_{\ 3j}\,\varepsilon^{ik}[2|{\rm I}]_{jk}+\\
 && \l^{[A}_{\ i3}\l^{B]}_{\ 3j}\,\varepsilon^{ik}\big<2|{\rm I}\big>_{jk}+
\l^{[A}_{\ 3i}\l^{B]}_{\ j3}\,\varepsilon^{ik}\big<2|{\rm I}\big>^*_{jk}+\\
 && \l^{[A}_{\ 3i}\l^{B]}_{\ 3j}\,\varepsilon^{ij}\big<2|{\rm I}\big>_{.}+
\l^{[A}_{\ i3}\l^{B]}_{\ j3}\,\varepsilon^{ij}\big<2|{\rm I}\big>^*_{.}\\
{\cal Y}^{Am} & = & \l^A_{\ 3i}\,\s^m_{\ 21}\big<2|{\rm II}\big>_i
+\l^A_{\ i3}\,\s^m_{\ 12}\big<2|{\rm II}\big>^*_i+\\
 && \l^A_{\ 3i}\s^m_{\ 12}\big<2|{\rm III}\big>_i+
\l^A_{\ i3}\s^m_{\ 21}\big<2|{\rm III}\big>^*_i\\
{\cal Y}^{mn} & = & \varepsilon^{mn}[2|{\rm II}]_{.}\\
{\cal Y}^{m3} & = & \s^m_{\ 21}\big<2|{\rm II}\big>_{.}+\s^m_{\ 12}\big<2|{\rm II}\big>^*_{.}\\
{\cal Y}^{A3} & = & \l^A_{\ 3i}\big<2|{\rm I}\big>_i+\l^A_{\ i3}\big<2|{\rm I}\big>^*_i\,,
\end{eqnarray*}
where:
\begin{eqnarray*}
[2|{\rm I}]_{.} & = & \cH^{[0,0,0]}\ Z [0,{\rm I}|\r]\\
\,[2|{\rm II}]_{.} & = & \cH^{[0,0,0]}\ Z [0,{\rm II}|\r]\\
\big<2|{\rm I}\big>_{.} & = & \cH^{[0,6i,-3i]}\ Z\big<0,{\rm I}|\r\big>\\
\big<2|{\rm I}\big>^*_{.} & = & \cH^{[0,-6i,3i]}\ \tilde{Z}
\big<0,{\rm I}|\r\big>\\
\big<2|{\rm II}\big>_{.} & = & \cH^{[0,-2i,-3i]}\ Z
\big<0,{\rm II}|\r\big>\\
\big<2|{\rm II}\big>^*_{.} & = & \cH^{[0,2i,3i]}\ \tilde{Z}
\big<0,{\rm II}|\r\big>\\
\big<2|{\rm I}\big>_i & = & \cH^{[1/2,3i,-3/2i](a)}_i\ Z
\big<1/2,{\rm I}|\r\big>\\
\big<2|{\rm I}\big>^*_i & = & \varepsilon^{ij}\cH^{[1/2,-3i,3/2i](b)}_j\ \tilde{Z}
\big<1/2,{\rm I}|\r\big>\\
\big<2|{\rm II}\big>_i & = & \cH^{[1/2,i,-9/2i](a)}_i\ Z
\big<1/2,{\rm II}|\r\big>\\
\big<2|{\rm II}\big>^*_i & = & \varepsilon^{ij}\cH^{[1/2,-i,9/2i](b)}_j\ \tilde{Z}
\big<1/2,{\rm II}|\r\big>\\
\big<2|{\rm III}\big>_i & = & \cH^{[1/2,5i,3/2i](a)}_i\ Z
\big<1/2,{\rm III}|\r\big>\\
\big<2|{\rm III}\big>^*_i & = & \varepsilon^{ij}\cH^{[1/2,-5i,-3/2i](b)}_j\ \tilde{Z}
\big<1/2,{\rm III}|\r\big>\\
\,[2|{\rm I}]_{ij} & = & \cH^{[1,0,0](c)}_{ij}\ Z[1,{\rm I}|\r]\\
\big<2|{\rm I}\big>_{ij} & = & \cH^{[1,0,0](d)}_{ij}\ Z
\big<1,{\rm I}|\r\big>\\
\big<2|{\rm I}\big>^*_{ij} & = & \varepsilon^{ik}\e_{jl}\cH^{[1,0,0](e)}_{kl}\ \tilde{Z}
\big<1,{\rm I}|\r\big>\,.
\end{eqnarray*}
%%%%%%%%%%%%%%%%%%%%%%%%%%%%%%%%%%%%%%%%%%%%%%%%%%%
%                                                 %
%      T W O   F O R M    O P E R A T O R         %
%                                                 %
%%%%%%%%%%%%%%%%%%%%%%%%%%%%%%%%%%%%%%%%%%%%%%%%%%%
%\subsection{The two-form mass operator}
The Laplace Beltrami operator for the transverse two-form field ${\cal Y}^{ab}$,
is given by
\begin{equation}
\xbox^{[110]}{\cal Y}^{[ab]} \equiv M_{(1)^2(0)}{\cal Y}^{[ab]} =
3\cD_{g}\cD^{[g}{\cal Y}^{ab]} = (\cD^{g}\cD_{g}+48){\cal Y}^{[ab]}-
4\cR^{[a\ b]}_{\ [g\ d]}{\cal Y}^{[gd]}
\end{equation}
From the decomposition $\cD_{a}{\cal Y}^{bg}=\cD^H_{a}{\cal Y}^{bg}
+(\IM_{a})^{b}_{\ d}{\cal Y}^{dg}+(\IM_{a})^{g}_{\ d}{\cal Y}^{bd}$
we obtain:
\begin{eqnarray}
\xbox^{[110]}{\cal Y}^{[ab]} = \left\{48\,\d^{[a\ b]}_{\ [g\ d]}-
4\cR^{[a\ b]}_{\ [g\ d]} + \right. & \nonumber\\
 & \nonumber\\
+2\eta^{mn}(\IM_{m})^{[a}_{\ \,[g}(\IM_{n})^{b]}_{\ \,d]}+
2\eta^{mn}(\IM_{m}\IM_{n})_{\ [g}^{[a}d^{b]}_{\ d]}\, + & \!\!\!\left.
4\eta^{mn}(\IM_{m})_{\ [g}^{[a}d^{b]}_{\ d]}\cD^H_{n}
\right\} {\cal Y}^{[gd]}.
\end{eqnarray}
For the regular $G$ representations of type $^0$ this operators acts on 
$AdS_4$ fields as the following $11\times 11$ matrix:\\
Columns one to three:
\begin{footnotesize}
\begin{eqnarray*}
\begin{array}{|c||c|c|c|}
\hline
  M_{(1)^2(0)}
    & Z[0,{\rm I}]
    & Z[0,{\rm II}]
    & Z[1,{\rm I}]
\cr
\hline
\hline
% row 1
   Z[0,{\rm I}] &
   H_0\!+\!32 & \!-\!16 &
   \ft{16}{\sqrt{3}}i(M_2\!+\!2)
   \cr
% row 2
   Z[0,{\rm II}] &
   \!-32 & H_0\!+\!16 & 0
   \cr
% row 3
   Z[1,{\rm I}] &
   0 & 0 & H_0
   \cr
% row 4
   Z \langle \ft12, {\rm I} \rangle &
   \!-\ft{16}{\sqrt{3}}iM_1 & 0 & \ft{16}{\sqrt{3}}i(M_1\!+\!2)
   \cr
% row 5
   \tilde{Z} \langle \ft12, {\rm I} \rangle &
   \ft{16}{\sqrt{3}}iM_2 & 0 & \!-\ft{16}{\sqrt{3}}i(M_2\!\!+\!\!2)
   \cr
% row 6
   Z \langle 0, {\rm II} \rangle &
   0 & \ft{16}{3\sqrt{2}}i(M_2\!-\!M_1\!+\!3J) & 0
   \cr
% row 7
   \tilde{Z} \langle 0, {\rm II}\rangle &
   0 & \!\!-\ft{16}{3\sqrt{2}}i(M_2\!\!-\!\!M_1\!\!-\!\!3J) & 0
   \cr
% row 8
   Z \langle 0, {\rm I} \rangle &
   0 & 0 & 0
   \cr
% row 9
   \tilde{Z} \langle 0, {\rm II}  \rangle &
   0 & 0 & 0
   \cr
% row 10
   Z \langle \ft{1}{2}, {\rm III}  \rangle &
   0 & 0 & 0
   \cr
% row 11
   \tilde{Z} \langle \ft{1}{2}, {\rm III}  \rangle &
   0 & 0 & 0
   \cr
\hline
\end{array}
\end{eqnarray*}
\end{footnotesize}
Columns four to seven:
\begin{tiny}
\begin{eqnarray*}
\begin{array}{|c||c|c|c|c|}
\hline
  M_{(1)^2(0)}
    & Z \langle \ft12, {\rm I} \rangle
    & \tilde{Z} \langle \ft12, {\rm I} \rangle
    & Z \langle 0, {\rm II} \rangle
    & \tilde{Z} \langle 0 , {\rm II} \rangle
\cr
\hline
\hline
% row 1
   Z[0,{\rm I}] & \!-\ft{16}{\sqrt{3}}i(M_1\!+\!2)& 0
   & 0 & 0
   \cr
% row 2
   Z[0,{\rm II}]  & 0 & 0
   & \ft{32}{2\sqrt{2}}i(M_2\!\!-\!\!M_1\!\!-\!\!3J)
   & -\ft{32}{2\sqrt{2}}i(M_2\!\!-\!\!M_1\!\!-\!\!3J)
   \cr
% row 3
   Z[1,{\rm I}]  & \!-\ft{16}{\sqrt{3}}iM_2 & \ft{16}{\sqrt{3}}iM_1
   & 0 & 0
   \cr
% row 4
   Z \langle \ft12, {\rm I} \rangle &
   \!H_0\!\!+\!\!32\!\!-\!\!\ft{32}{3}(M_2\!\!-\!\!M_1)  & 0
   & 0 & 0
   \cr
% row 5
   \tilde{Z} \langle \ft12, {\rm I} \rangle  &
   0 & \!H_0\!\!+\!\!32\!\!+\!\!\ft{32}{3}(M_2\!\!-\!\!M_1)
   & 0 & 0
   \cr
% row 6
   Z \langle 0, {\rm II} \rangle & 0 & 0
   & \!H_0\!\!+\!\!32\!\!+\!\!\ft{32}{3}(M_2\!\!-\!\!M_1)
   & 0
   \cr
% row 7
   \tilde{Z} \langle 0, {\rm II}\rangle & 0 & 0
   & 0 &\! H_0\!\!+\!\!32\!\!-\!\!\ft{32}{3}(M_2\!\!-\!\!M_1)
   \cr
% row 8
   Z \langle 0, {\rm I} \rangle & \!-\ft{8}{3\sqrt{2}}(M_2\!-\!M_1\!+\!3J) & 0
   & -\ft{16}{\sqrt{3}}M_1 & 0
   \cr
% row 9
   \tilde{Z} \langle 0, {\rm II}  \rangle
    & 0 & \!-\ft{16}{3\sqrt{2}}(M_2\!-\!M_1\!-\!3J)
    & 0 & -\ft{16}{\sqrt{3}} M_2
   \cr
% row 10
   Z \langle \ft{1}{2}, {\rm III}  \rangle+
   & \!-\ft{8}{3\sqrt{2}}(M_2\!-\!M_1\!-\!3J)& 0
   & 0 & -\ft{16}{\sqrt{3}} M_1
   \cr
% row 11
   \tilde{Z} \langle \ft{1}{2}, {\rm III}  \rangle
   & 0 & \!-\ft{16}{3\sqrt{2}}(M_2\!-\!M_1+\!3J)
   & -\ft{16}{\sqrt{3}}M_2
   & 0
   \cr
\hline
\end{array}
\end{eqnarray*}
\end{tiny}
\par
Columns eight to eleven:
\begin{footnotesize}
\begin{eqnarray*}
\begin{array}{|c||c|c|c|c|}
\hline
    & Z\langle \ft12,{\rm II}\rangle
    & \tilde{Z} \langle \ft12, {\rm II} \rangle
    & Z\langle \ft12, {\rm III} \rangle
    & \tilde{Z} \langle \ft12, {\rm III} \rangle
\cr
\hline
\hline
% row 1
   Z[0,{\rm I}] &
   0 & 0 & 0 & 0
   \cr
% row 2
   Z[0,{\rm II}]  & 0 & 0 & 0 & 0\cr
% row 3
   Z[1,{\rm I}] & 0 & 0 & 0 & 0
   \cr
% row 4
   Z \langle \ft12, {\rm I} \rangle
    & \ft{32}{3\sqrt{2}}(M_2\!\!-\!\!M_1\!\!-\!\!3J) & 0 &
   \ft{32}{3\sqrt{2}}(M_2\!\!-\!\!M_1\!\!+\!\!3J) & 0
   \cr
% row 5
   \tilde{Z} \langle \ft12, {\rm I} \rangle
   & 0 & \ft{32}{3\sqrt{2}}(M_2\!\!-\!\!M_1\!\!+\!\!3J) & 0 &
   \ft{32}{3\sqrt{2}}(M_2\!\!-\!\!M_1\!\!-\!\!3J)
   \cr
% row 6
   Z\langle 0,{\rm II}\rangle &
   \!\!-\ft{32}{\sqrt{3}}(M_2\!\!+\!\!2) & 0 & 0 &
   \!\!-\ft{32}{\sqrt{3}}(M_1\!\!+\!\!2)
   \cr
% row 7
   \tilde{Z}\langle 0,{\rm II}\rangle & 0 &
   \!\!-\ft{32}{\sqrt{3}}(M_1\!\!+\!\!2) &
   \!\!-\ft{32}{\sqrt{3}}(M_2\!\!+\!\!2) & 0
   \cr
% row 8
   Z\langle \ft12,{\rm II}\rangle & H_0 & 0 & 0 & 0
   \cr
% row 9
   \tilde{Z} \langle \ft12, {\rm II} \rangle & 0 & H_0 & 0 & 0
   \cr
% row 10
   Z\langle \ft12, {\rm III} \rangle & 0 & 0 &
   H_0\!\!-\ft{64}{3}(M_2\!\!-\!\!M_1) & 0
   \cr
% row 11
   \tilde{Z} \langle \ft12, {\rm III} \rangle & 0 & 0 & 0 &
   H_0\!\!+\!\!\ft{64}{3}(M_2\!\!-\!\!M_1)
   \cr
\hline
\end{array}
\end{eqnarray*}
\end{footnotesize}
This matrix has the following eigenvalues:
\begin{eqnarray}
\lambda_1 &=& H_0 + \ft{32}{3}(M_2-M_1)\,,
\nonumber \\
\lambda_2 &=& H_0 - \ft{32}{3}(M_2-M_1)\,,
\nonumber \\
\lambda_3 &=& H_0\,,
\nonumber \\
\lambda_4 &=& H_0 + 24 + 4 \sqrt{H_0 + 36}\,,
\nonumber \\
\lambda_5 &=& H_0 + 24 - 4 \sqrt{H_0 + 36}\,,
\nonumber \\
\lambda_6 &=& H_0 + \ft{32}{3}(M_2-M_1) + 16 + 4 \sqrt{H_0 + \ft{32}{3}(M_2-M_1) + 16}
\,, \nonumber \\
\lambda_7 &=& H_0 + \ft{32}{3}(M_2-M_1) + 16 - 4 \sqrt{H_0 + \ft{32}{3}(M_2-M_1) + 16}
\,, \nonumber \\
\lambda_8 &=& H_0 - \ft{32}{3}(M_2-M_1) + 16 + 4 \sqrt{H_0 - \ft{32}{3}(M_2-M_1) + 16}
\,, \nonumber \\
\lambda_9 &=& H_0 - \ft{32}{3}(M_2-M_1) + 16 - 4 \sqrt{H_0 - \ft{32}{3}(M_2-M_1) + 16}
\,, \nonumber \\
\lambda_{10} &=& \lambda_{11} = H_0 + 32 \,.
\nonumber \\
\label{eigenAtwoform}
\end{eqnarray}
The eigenvalues $\lambda_1,\lambda_2,\lambda_4,\lambda_5$, equal
to the one--form physical ones, are the longitudinal eigenvalues.
The other seven are the physical two-form eigenvalues.
\par
As in the case of the one--form, by removing rows and columns we find the matrix
of each exceptional series, and the corresponding eigenvalues:
\begin{footnotesize}
\begin{eqnarray}
\begin{array}{|c||c|}
\hline
\, A_R \, & \, \lambda_1, \lambda_2, \lambda_3, \lambda_4, \lambda_5 , \lambda_6,
\lambda_7, \lambda_8, \lambda_9, \lambda_{10}, \lambda_{11} \,\cr
\hline
A_1 & \, \lambda_1, \lambda_3, \lambda_4, \lambda_5 , \lambda_6,
\lambda_7, \lambda_{10}, \lambda_{11} \, \cr
\hline
A_1^* &  \, \lambda_2, \lambda_3, \lambda_4, \lambda_5 , \lambda_8,
\lambda_9, \lambda_{10}, \lambda_{11} \,  \cr
\hline
A_2 & \, \lambda_1,  \lambda_6,
\lambda_7 \,  \cr
\hline
A_2^* & \, \lambda_2,  \lambda_8,
\lambda_9 \, \cr
\hline
A_3 & \, \lambda_1, \lambda_4, \lambda_5 , \lambda_6,
\lambda_7, \lambda_{10}, \lambda_{11} \, \cr
\hline
A_3^* & \, \lambda_2, \lambda_4, \lambda_5 , \lambda_8,
\lambda_9, \lambda_{10}, \lambda_{11} \, \cr
\hline
A_4 &  \, \lambda_1, \lambda_4, \lambda_6,
\lambda_7, \lambda_{10} \, \cr
\hline
A_4^* & \, \lambda_2, \lambda_4, \lambda_8,
\lambda_9, \lambda_{10} \, \cr
\hline
A_5 & \, \lambda_1, \lambda_6 \,\cr
\hline
A_5^* & \, \lambda_2, \lambda_8 \, \cr
\hline
A_6 & \, \lambda_3, \lambda_4, \lambda_5, \lambda_{10}, \lambda_{11}\,  \cr
\hline
A_7 &  \, \lambda_4, \lambda_5, \lambda_{10}, \lambda_{11} \, \cr
\hline
A_8 & \, \lambda_3, \lambda_4 \, \cr
\hline
\end{array}
\label{eigenvaluesAtwoform}
\end{eqnarray}
\end{footnotesize}
The two--form operator matrix in the representations $^{++}$ is the following
$5\times 5$ matrix:\\
Columns one to two:
\begin{footnotesize}
\begin{eqnarray*}
\begin{array}{|c||c|c|}
\hline
  M_{(1)^2(0)}
    & Z\langle0,{\rm I}\rangle
    & Z\langle{1\over 2},{\rm I}\rangle
\cr
\hline
\hline
% row 1
   Z\langle 0,{\rm I}\rangle &
   H_0-{32\over 3}\left(M_2-M_1\right) & 
   {16\over\sqrt{3}}\left(M_1+2\right) 
   \cr
% row 2
   Z\langle{1\over 2},{\rm I}\rangle &
   \ft{32}{\sqrt{3}}M_2 & H_0+32
   \cr
% row 3
   Z\langle{1\over 2},{\rm II}\rangle &
   0 &  -\ft{16}{3\sqrt{2}}\left(M_2-M_1+3J-3\right)
   \cr
% row 4
   Z\langle{1\over 2},{\rm III}\rangle &
   0 & -\ft{16}{3\sqrt{2}}\left(M_2-M_1-3J-3\right) 
   \cr
% row 5
  Z\langle1,{\rm I}\rangle &
    0 & \ft{32}{\sqrt{3}}\left(M_2-1\right) 
   \cr
\hline
\end{array}
\end{eqnarray*}
\end{footnotesize}
Columns three to five:
\begin{footnotesize}
\begin{eqnarray*}
\begin{array}{|c||c|c|c|}
\hline
  M_{(1)^2(0)}
    & Z\langle{1\over 2},{\rm II}\rangle
    & Z\langle{1\over 2},{\rm III}\rangle
    & Z\langle1,{\rm I}\rangle
\cr
\hline
\hline
% row 1
   Z\langle0,{\rm I}\rangle &
   0&0&0
   \cr
% row 2
   Z\langle{1\over 2},{\rm I}\rangle &
   \ft{32}{3\sqrt{2}}\left(M_2-M_1-3J-6\right) &
   \ft{32}{3\sqrt{2}}\left(M_2-M_1+3J\right) &
   {16\over\sqrt{3}}\left(M_1+3\right)
   \cr
% row 3
   Z\langle{1\over 2},{\rm II}\rangle &
   H_0+{32\over 3}\left(M_2-M_1\right)-32 & 0 & 0
   \cr
% row 4
   Z\langle{1\over 2},{\rm III}\rangle &
   0 & H_0-{32\over 3}\left(M_2-M_1\right)+32 & 0
   \cr
% row 5
  Z\langle1,{\rm I}\rangle &
    0 & 0 & H_0+{32\over 3}\left(M_2-M_1\right)-32
   \cr
\hline
\end{array}
\end{eqnarray*}
\end{footnotesize}
It has eigenvalues
\begin{eqnarray}
\lambda_1 &=& H_0 + \ft{32}{3}(M_2 - M_1)
\,, \nonumber \\
\lambda_2 &=& H_0 + \ft{32}{3}(M_2 - M_1) + 16 + 4  \sqrt{H_0+ \ft{32}{3}(M_2-M_1)+16}
\,, \nonumber \\
\lambda_3 &=& H_0 + \ft{32}{3}(M_2 - M_1) + 16 - 4 \sqrt{H_0+ \ft{32}{3}(M_2-M_1)+16}
\,, \nonumber \\
\lambda_4 &=& H_0 + 32
\,, \nonumber \\
\lambda_5 &=& H_0 + \ft{64}{3}(M_2-M_1) -32
\,.
\label{eigenBtwoform}
\end{eqnarray}
The eigenvalue $\lambda_1$, equal to the physical eigenvalue of the
($^{++}$) one--form, is longitudinal.
The other four are the physical eigenvalues.
\begin{footnotesize}
\begin{eqnarray}
\begin{array}{|c||c|}
\hline
\, B_R \, & \, \lambda_1, \lambda_2, \lambda_3, \lambda_4, \lambda_5 \, \cr
\hline
\, B_1 \, & \, \lambda_1, \lambda_2,  \lambda_3, \lambda_5 \, \cr
\hline
\, B_2 \, & \, \lambda_5 \, \cr
\hline
\, B_3 \, & \, \lambda_1, \lambda_2, \lambda_3 \, \cr
\hline
\, B_4 \, & \, \lambda_1, \lambda_2, \lambda_3, \lambda_4 \,  \cr
\hline
\, B_5 \, & \,  \lambda_4 \, \cr
\hline
\, B_6 \, & \,  \lambda_1, \lambda_2, \lambda_3, \lambda_4 \, \cr
\hline
\, B_7 \, & \, \lambda_1, \lambda_2, \lambda_4 \, \cr
\hline
\, B_8 \, & \, \lambda_4 \, \cr
\hline
\, B_9 \, & \, \lambda_4 \, \cr
\hline
\, B_{10} \, & \, \lambda_4 \, \cr
\hline
\end{array}
\label{eigenvaluesBtwoform}
\end{eqnarray}
\end{footnotesize}
For the $^{--}$ representations, the eigenvalues are the conjugates
($M_2 \leftrightarrow M_1$) of the ones in (\ref{eigenvaluesBtwoform}).
\par
We can use the eigenvalues of the two--form harmonic to determine 
the masses of the $AdS_4$ vector field $Z$.
%%%%%%%%%%%%%%%%%%%%%%%%%%%%%%%%%%%%%%%%%%%%%%%%%%%
%                                                 %
%          T H R E E    F O R M                   %
%                                                 %
%%%%%%%%%%%%%%%%%%%%%%%%%%%%%%%%%%%%%%%%%%%%%%%%%%%
\subsubsection{The three-form}
The $H$ decomposition of the three--form in $H$--irreducible fragments has been done
in \cite{spec321}: 
\begin{eqnarray*}
{\cal Y}^{ABC} & = & \varepsilon^{ABCD}\,\{\l^D_{3i}\langle 3| {\rm I}
\rangle_i+\l^D_{i3}\langle 3| {\rm I}\rangle_i^* \}\,\,,\\
{\cal Y}^{ABm} & = & \l^A_{i3}\l^B_{j3}\varepsilon^{ij}\{\s^m_{21}
\langle 3|{\rm II}\rangle_{.}+\s^m_{12}\langle 3|{\rm III}\rangle_{.}\}+
 \l^A_{3i}\l^B_{3j}\varepsilon^{ij}\{\s^m_{12}\langle 3|{\rm II}\rangle_{.}^*+
\s^m_{21}\langle 3|{\rm III}\rangle_{.}^*\}+\\
 && +i\l^{[A}_{i3}\l^{B]}_{3i}\{\s^m_{21}\langle 3|{\rm IV}\rangle_{.}+
\s^m_{12}\langle 3|{\rm IV}\rangle_{.}^*\}+\l^{[A}_{i3}\l^{B]}_{3j}
\{\s^m_{21}\varepsilon^{ik}\langle 3|{\rm I}\rangle_{kj}-
\s^m_{12}\varepsilon^{jk}\langle 3|{\rm I}\rangle_{ik}^*\}\,\,,\\
{\cal Y}^{AB3} & = & \l^A_{3i}\l^B_{3j}\varepsilon^{ij}\langle 3| {\rm I}\rangle_{.}+
i\l^A_{i3}\l^B_{j3}\varepsilon^{ij}\langle 3| {\rm I}\rangle_{.}^*+
i\l^{[A}_{i3}\l^{B]}_{3i}[3|{\rm I}]_{.}
 +\l^{[A}_{i3}\l^{B]}_{3j}\varepsilon^{ik}[3|{\rm I}]_{kj}\,\,,\\
{\cal Y}^{Amn} & = & \varepsilon^{mn}\{ \l^A_{3i}\langle 3| {\rm II}
\rangle_i+\l^A_{i3}\langle 3| {\rm II}\rangle_i^* \}\,\,,\\
{\cal Y}^{Am3} & = & \l^A_{3i}\{\s^m_{12}\langle 3| {\rm IV}\rangle_i+
\s^m_{21}\langle 3| {\rm III}\rangle_i \}+
\l^A_{i3}\{ \s^m_{21}\langle 3| {\rm IV}\rangle_i^*+
\s^m_{12}\langle 3| {\rm III}\rangle_i^* \}\,,\\
{\cal Y}^{mn3} & = & \varepsilon^{mn}[3| {\rm II}]_{.}\,\,,
\end{eqnarray*}
where the fragments of type $^0$ are:
\begin{eqnarray*}
\langle 3 | {\rm I } \rangle_{ij} &=& {\cal H}^{[1,-2i,-{3i}](c)}_{ij} \cdot
                    \pi\langle 1 , {\rm I} \rangle \,,
\\
\langle 3 | {\rm I } \rangle_{ij}^* &=& -\varepsilon^{ik} \varepsilon^{jl}
            {\cal H}^{[1,2i,3i](c)}_{kl} \cdot
                    \tilde{\pi}\langle 1 , {\rm I} \rangle \,,
\\
{}[ 3 | {\rm I } ]_{ij} &=& {\cal H}^{[1,0,0](c)}_{ij} \cdot
                    \pi[ 1 , {\rm I} ] \,,
\\
\langle 3 | {\rm I } \rangle_i &=& {\cal H}^{[1/2,3i,-{3i}/{2}](a)}_i \cdot
                    \pi\langle \ft12 , {\rm I} \rangle \,,
\\
\langle 3 | {\rm I } \rangle_i^* &=& \varepsilon^{ij}
          {\cal H}^{[1/2,-3i,{3i}/{2}](b)}_j \cdot
                    \tilde {\pi}\langle \ft12 , {\rm I} \rangle \,,
\\
\langle 3 | {\rm II } \rangle_i &=& {\cal H}^{[1/2,3i,-{3i}/{2}](a)}_i \cdot
                    \pi\langle \ft12 , {\rm II} \rangle \,,
\\
\langle 3 | {\rm II } \rangle_i^* &=& \varepsilon^{ij}
          {\cal H}^{[1/2,-3i,{3i}/{2}](b)}_j \cdot
                    \tilde {\pi}\langle \ft12 , {\rm II} \rangle \,,
\\
\langle 3 | {\rm III } \rangle_i &=& {\cal H}^{[1/2,i,-{9i}/{2}](a)}_i \cdot
                    \pi\langle \ft12 , {\rm III} \rangle \,,
\\
\langle 3 | {\rm III } \rangle_i^* &=& \varepsilon^{ij}
          {\cal H}^{[1/2,-i,{9i}/{2}](b)}_j \cdot
                    \tilde {\pi}\langle \ft12 , {\rm III} \rangle \,,
\\
\langle 3 | {\rm IV } \rangle_i &=& {\cal H}^{[1/2,5i,{3i}/{2}](a)}_i \cdot
                    \pi\langle \ft12 , {\rm IV} \rangle \,,
\\
\langle 3 | {\rm IV } \rangle_i^* &=& \varepsilon^{ij}
          {\cal H}^{[1/2,-5i,-{3i}/{2}](b)}_j \cdot
                    \tilde {\pi}\langle \ft12 , {\rm IV} \rangle \,,
\\
\langle 3 | {\rm IV } \rangle_\cdot &=&
{\cal H}^{[0,-2i,-3i]} \cdot
                 \pi\langle 0 , {\rm IV} \rangle \,,
\\
\langle 3 | {\rm IV } \rangle_\cdot^* &=&
{\cal H}^{[0,2i,3i]} \cdot
                    \tilde{\pi}\langle 0 , {\rm IV} \rangle \,,
\\
{}[3 | {\rm I } ]_\cdot &=&
{\cal H}^{[0,0,0]} \cdot
                    \pi[ 0 , {\rm I} ] \,,
\\
{}[3 | {\rm II } ]_\cdot &=&
{\cal H}^{[0,0,0]} \cdot
                    \pi[ 0 , {\rm II} ] \,,
\end{eqnarray*}
while the fragments of type $^{++}$ are:
\begin{eqnarray*}
\langle 3 | {\rm I } \rangle_{ij} &=& {\cal H}^{[1,-2i,-{3i}](d)}_{ij} \cdot
                    \pi\langle 1 , {\rm I} \rangle \,,
\\
\langle 3 | {\rm I } \rangle_{ij}^* &=& \varepsilon^{ik} \varepsilon^{jl}
            {\cal H}^{[1,2i,3i](d)}_{kl} \cdot
                    \tilde{\pi}\langle 1 , {\rm I} \rangle \,,
\\
{}[ 3 | {\rm I } ]_{ij} &=& {\cal H}^{[1,0,0](d)}_{ij} \cdot
                    \pi[ 1 , {\rm I} ] \,,
\\
\langle 3 | {\rm I } \rangle_i &=& {\cal H}^{[1/2,3i,-{3i}/{2}](b)}_i \cdot
                    \pi\langle \ft12 , {\rm I} \rangle \,,
\\
\langle 3 | {\rm II } \rangle_i &=& {\cal H}^{[1/2,3i,-{3i}/{2}](b)}_i \cdot
                    \pi\langle \ft12 , {\rm II} \rangle \,,
\\
\langle 3 | {\rm III } \rangle_i &=& {\cal H}^{[1/2,i,-{9i}/{2}](b)}_i \cdot
                    \pi\langle \ft12 , {\rm III} \rangle \,,
\\
\langle 3 | {\rm IV } \rangle_i &=& {\cal H}^{[1/2,5i,{3i}/{2}](b)}_i \cdot
                    \pi\langle \ft12 , {\rm IV} \rangle \,,
\\
\langle 3 | {\rm I } \rangle_\cdot &=&
{\cal H}^{[0,6i,-3i]} \cdot
                 \pi\langle 0 , {\rm I} \rangle \,,
\\
\langle 3 | {\rm II } \rangle_\cdot^* &=&
{\cal H}^{[0,8i,0]} \cdot
                    \tilde{\pi}\langle 0 , {\rm II} \rangle \,,
\\
\langle 3 | {\rm III } \rangle_\cdot^* &=&
{\cal H}^{[0,4i,-6i]} \cdot
                    \tilde{\pi}\langle 0 , {\rm III} \rangle \,.
\end{eqnarray*}
The fragments that are present in the type $^{--}$ series are the complex
conjugates of the fragments above.
%%%%%%%%%%%%%%%%%%%%%%%%%%%%%%%%%%%%%%%%%%%%%%%%%%%
%                                                 %
%   T H R E E    F O R M    O P E R A T O R       %
%                                                 %
%%%%%%%%%%%%%%%%%%%%%%%%%%%%%%%%%%%%%%%%%%%%%%%%%%%
%\subsection{The three-form mass operator}
The Laplace Beltrami operator for the transverse three-form
${\cal Y}^{[abc]}$, is a first-order differential operator, given by
\begin{eqnarray}
\xbox^{[111]}{\cal Y}^{[abc]} \equiv M_{(1)^3}{\cal Y}^{[abc]} =
\ft{1}{24}\e^{abcd}_{\ \ \ \ mnr}
\cD_{d}{\cal Y}^{mnr} =\hspace{6 cm}\nonumber\\
=\ft{1}{24}\e^{abgd}_{\ \ \ \ \ mnr}\left[
\cD^H_{d}{\cal Y}^{mnr}+(\IM_{d})^{m}_{\ s}{\cal Y}^{snr}+
(\IM_{d})^{n}_{\ s}{\cal Y}^{msr}+
(\IM_{d})^{r}_{\ s}{\cal Y}^{mns}\right].\nonumber\\
\end{eqnarray}
For  the regular series of type $^0$  this operator acts on the $AdS_4$ fields as a
$15\times 15$ matrix:\\
Columns one to five:
\begin{footnotesize}
\begin{eqnarray}
\begin{array}{|c||c|c|c|c|c|}
\hline
   M_{(1)^3} &
        \pi\langle 1,{\rm I} \rangle &
        \tilde{\pi}\langle 1,{\rm I} \rangle &
        \pi [1,{\rm I}] &
        \pi \langle \ft12, {\rm I} \rangle &
        \tilde{\pi}\langle \ft12, {\rm I}  \rangle \cr
\hline
\hline
    \pi\langle 1,{\rm I} \rangle         &
    {Y} & 0 &
  {\frac{-\left( 2\,J + {Y} \right) }{2\,{\sqrt{2}}}} & 0 & 0
   \cr
     \tilde{\pi}\langle 1,{\rm I} \rangle      &
 0 & -{Y} & {\frac{-2\,J + {Y}}{2\,{\sqrt{2}}}}
   & 0 & 0 \cr
     \pi [1,{\rm I}]   &
     {\frac{-2 - 2\,J + {Y}}{{\sqrt{2}}}} &
  -{\frac{2 + 2\,J + {Y}}{{\sqrt{2}}}} & 1 & 0 & 0 \cr
     \pi \langle \ft12, {\rm I} \rangle   &
 0 & 0
   & 0 & 0 & 0 \cr
         \tilde{\pi}\langle \ft12, {\rm I}  \rangle     &
 0 & 0 & 0 & 0 & 0 \cr
     \pi \langle \ft12, {\rm II} \rangle          &
 0 & 0 &
  {\frac{{\frac{i}{2}}\,\left( 2 + {M_1} \right) }{{\sqrt{3}}}} &
  -i\,{Y} & 0 \cr
         \tilde{\pi}\langle \ft12, {\rm II}  \rangle      &
 0 & 0 &
  {\frac{{\frac{i}{2}}\,\left( 2 + {M_2} \right) }{{\sqrt{3}}}} & 0
   & -i\,{Y} \cr
     \pi \langle \ft12, {\rm III} \rangle                &
 {\frac{2 + {M_1}}{2\,{\sqrt{3}}}} & 0
   & 0 & {\frac{-\left( 2\,J + {Y} \right) }{2\,{\sqrt{2}}}} & 0
   \cr
        \tilde{\pi}\langle \ft12, {\rm III}  \rangle              &
 0 & {\frac{2 + {M_2}}{2\,{\sqrt{3}}}} & 0 & 0 &
  {\frac{2\,J - {Y}}{2\,{\sqrt{2}}}} \cr
     \pi \langle \ft12, {\rm IV} \rangle                   &
 0 &
  {\frac{2 + {M_1}}{2\,{\sqrt{3}}}} & 0 &
  {\frac{2\,J - {Y}}{2\,{\sqrt{2}}}} & 0 \cr
        \tilde{\pi}\langle \ft12, {\rm IV}  \rangle            &
  {\frac{2 + {M_2}}{2\,{\sqrt{3}}}} & 0 & 0 & 0 &
  {\frac{-\left( 2\,J + {Y} \right) }{2\,{\sqrt{2}}}} \cr
     \pi \langle 0 , {\rm IV} \rangle                    &
 0 & 0
   & 0 & 0 & 0 \cr
        \tilde{\pi}\langle 0 , {\rm IV}  \rangle              &
 0 & 0 & 0 & 0 & 0 \cr
 \pi[0,{\rm I}|\rho]           &
 0 & 0 & 0 & 0 & 0 \cr
\pi[0,{\rm II}|\rho]               &
 0
   & 0 & 0 & {\frac{2\,i\,\left( 2 + {M_2} \right) }{{\sqrt{3}}}}
   & {\frac{-2\,i\,\left( 2 + {M_1} \right) }{{\sqrt{3}}}} \cr
\hline
\end{array}
\end{eqnarray}
\end{footnotesize}
Columns six to ten:
\begin{footnotesize}
\begin{eqnarray}
\begin{array}{|c||c|c|c|c|c|}
\hline
        &  \pi\langle \ft12, {\rm II}  \rangle
        & \tilde{\pi}\langle \ft12, {\rm II} \rangle
        & \pi\langle \ft12, {\rm III}  \rangle
        & \tilde{\pi}\langle \ft12, {\rm III} \rangle
        & \pi\langle \ft12, {\rm IV} \rangle
 \cr
\hline
\hline
% row 1
    \pi\langle 1,{\rm I} \rangle            &
   0 & 0 & {\frac{2\,{M_2}}{{\sqrt{3}}}} & 0 & 0 \cr
% row 2
    \tilde{\pi} \langle 1,{\rm I} \rangle          &
   0 & 0
   & 0 & {\frac{2\,{M_1}}{{\sqrt{3}}}} &
  {\frac{2\,{M_2}}{{\sqrt{3}}}} \cr
% row 3
    \pi [1,{\rm I}]            &
  {\frac{-2\,i\,{M_2}}{{\sqrt{3}}}} &
  {\frac{-2\,i\,{M_1}}{{\sqrt{3}}}} & 0 & 0 & 0 \cr
% row 4
\pi\langle \ft12, {\rm I} \rangle                &
  i\,{Y} & 0 & {\frac{-2 - 2\,J + {Y}}{{\sqrt{2}}}}
   & 0 & {\frac{2 + 2\,J + {Y}}{{\sqrt{2}}}} \cr
% row 5
\tilde{\pi}\langle \ft12, {\rm I} \rangle                 &
  0 & i\,{Y} & 0 & {\frac{2 + 2\,J + {Y}}{{\sqrt{2}}}} &
  0 \cr
% row 6
\pi\langle \ft12, {\rm II}  \rangle                &
   0 & 0 & 0 & 0 & 0 \cr
% row 7
 \tilde{\pi}\langle \ft12, {\rm II} \rangle               &
   0 & 0 & 0 & 0 & 0 \cr
% row 8
 \pi\langle \ft12, {\rm III}  \rangle                &
    0 & 0 & 1 & 0
   & 0 \cr
% row 9
\tilde{\pi}\langle \ft12, {\rm III} \rangle                &
    0 & 0 & 0 & 1 & 0 \cr
% row 10
  \pi\langle \ft12, {\rm IV} \rangle              &
    0 & 0 & 0 & 0 & -1 \cr
% row 11
  \tilde{\pi}\langle \ft12, {\rm IV} \rangle                &
   0 & 0 & 0 & 0 & 0 \cr
% row 12
    \pi \langle 0, {\rm IV} \rangle            &
    0 & 0 & {\frac{-i\,\left( 2 + {M_2} \right) }
    {{\sqrt{3}}}} & 0 & 0 \cr
% row 13
     \tilde{\pi} \langle 0, {\rm IV} \rangle           &
   0 & 0 & 0 &
  {\frac{i\,\left( 2 + {M_1} \right) }{{\sqrt{3}}}} &
  {\frac{i\,\left( 2 + {M_2} \right) }{{\sqrt{3}}}} \cr
% row 14
    \pi[0,{\rm I}]             &
  -{\frac{2 + {M_2}}{{\sqrt{3}}}} &
  -{\frac{2 + {M_1}}{{\sqrt{3}}}} & 0 & 0 & 0 \cr
% row 15
   \pi[0,{\rm II}]             &
   0 & 0 & 0 & 0
   & 0 \cr
\hline
  \end{array}
\end{eqnarray}
\end{footnotesize}
Columns eleven to fifteen:
\begin{footnotesize}
\begin{eqnarray}
\begin{array}{|c||c|c|c|c|c|}
\hline
     & \tilde{\pi}\langle \ft12, {\rm IV} \rangle
     & \pi \langle 0, {\rm IV} \rangle
     & \tilde{\pi} \langle 0, {\rm IV} \rangle
     & \pi[0,{\rm I}]
     & \pi[0,{\rm II}]
\cr
\hline
\hline
% row 1
 \pi\langle 1,{\rm I} \rangle         &
   {\frac{2\,{M_1}}{{\sqrt{3}}}} & 0 & 0 & 0 & 0 \cr
% row 2
  \tilde{\pi}\langle 1,{\rm I} \rangle        &
   0 & 0
   & 0 & 0 & 0 \cr
% row 3
\pi [1,{\rm I}]          &
   0 & 0 & 0 & 0 & 0 \cr
% row 4
 \pi\langle \ft12, {\rm I} \rangle         &
  0 & 0 & 0 & 0 &
  {\frac{-i\,{M_1}}{{\sqrt{3}}}} \cr
% row 5
 \tilde{\pi} \langle \ft12, {\rm I} \rangle         &
  {\frac{-2 - 2\,J + {Y}}{{\sqrt{2}}}} & 0 & 0 & 0 &
  {\frac{i\,{M_2}}{{\sqrt{3}}}} \cr
% row 6
 \pi\langle \ft12, {\rm II}  \rangle         &
  0 & 0 & 0 &
  -{\frac{{M_1}}{{\sqrt{3}}}} & 0 \cr
% row 7
   \tilde{\pi} \langle \ft12, {\rm II} \rangle       &
  0 & 0 & 0 &
  -{\frac{{M_2}}{{\sqrt{3}}}} & 0 \cr
% row 8
  \pi\langle \ft12, {\rm III}  \rangle        &
  0 &
  {\frac{i\,{M_1}}{{\sqrt{3}}}} & 0 & 0 & 0 \cr
% row 9
     \tilde{\pi}\langle \ft12, {\rm III} \rangle     &
  0 & 0 &
  {\frac{-i\,{M_2}}{{\sqrt{3}}}} & 0 & 0 \cr
% row 10
  \pi\langle \ft12, {\rm IV} \rangle        &
  0 & 0 &
  {\frac{-i\,{M_1}}{{\sqrt{3}}}} & 0 & 0 \cr
% row 11
 \tilde{\pi}\langle \ft12, {\rm IV} \rangle          &
   -1 &
  {\frac{i\,{M_2}}{{\sqrt{3}}}} & 0 & 0 & 0 \cr
% row 12
     \pi \langle 0, {\rm IV} \rangle     &
  {\frac{-i\,\left( 2 + {M_1} \right) }{{\sqrt{3}}}} &
  -{Y} & 0 & {\frac{2\,J + {Y}}{2\,{\sqrt{2}}}} & 0
   \cr
% row 13
    \tilde{\pi} \langle 0, {\rm IV} \rangle      &
   0 & 0 & {Y} &
  {\frac{-2\,J + {Y}}{2\,{\sqrt{2}}}} & 0 \cr
% row 14
  \pi[0,{\rm I}]        &
   0 &
  {\frac{2 + 2\,J - {Y}}{{\sqrt{2}}}} &
  -{\frac{2 + 2\,J + {Y}}{{\sqrt{2}}}} & -1 & -1 \cr
% row 15
  \pi[0,{\rm II}]        &
   0 & 0 & 0
   & -2 & 0 \cr
\hline
\end{array}
\end{eqnarray}
\end{footnotesize}
This matrix has the following eigenvalues:
\begin{eqnarray}
\lambda_1 &=& \ft14 \sqrt{H_0 + \ft{32}{3}(M_2-M_1)+16} \,,
\nonumber \\
\lambda_2 &=& \ft14 \sqrt{H_0 - \ft{32}{3}(M_2-M_1)+16} \,,
\nonumber \\
\lambda_3 &=& - \ft14 \sqrt{H_0 + \ft{32}{3}(M_2-M_1)+16} \,,
\nonumber \\
\lambda_4 &=& - \ft14 \sqrt{H_0 - \ft{32}{3}(M_2-M_1)+16} \,,
\nonumber
\\
\lambda_5 &=& \ft14 \sqrt{H_0 + 36} - \ft12 \,,
\nonumber \\
\lambda_6 &=& -\ft14 \sqrt{H_0 + 36} - \ft12 \,,
\nonumber \\
\lambda_7 &=& -\ft14 \sqrt{H_0 + 4} + \ft12 \,,
\nonumber \\
\lambda_8 &=& \ft14 \sqrt{H_0 + 4} + \ft12 \,,
\nonumber \\
\lambda_9 &=& \dots = \lambda_{15} = 0 \,.
\label{eigenAthreeform}
\end{eqnarray}
We note that seven eigenvalues are $0$.
They correspond to the longitudinal three-forms
(${\cal Y}^{(3)}=\cD\wedge {\cal Y}^{(2)}$), which are annihilated by
$\xbox^{[111]}$ ($=\,^{*}\cD\wedge$).
\par
As in the cases of the one--form and of the two--form, by removing
rows and columns we find the matrix for each exceptional series, and
the corresponding eigenvalues:
\begin{footnotesize}
\begin{eqnarray}
\begin{array}{|c||c|}
\hline
\, A_R \, & \, \lambda_1, \lambda_2, \lambda_3, \lambda_4, \lambda_5, \lambda_6,
               \lambda_7, \lambda_8  \, \cr
\hline
\, A_1 \, & \, \lambda_1,  \lambda_3,  \lambda_5, \lambda_6,
               \lambda_7, \lambda_8  \cr
\hline
\, A_1^* \, & \,  \lambda_2, \lambda_4, \lambda_5, \lambda_6,
               \lambda_7, \lambda_8  \cr
\hline
\, A_2 \, & \, \lambda_1,  \lambda_3 \cr
\hline
\, A_2^* \, & \, \lambda_2,  \lambda_4 \cr
\hline
\, A_3 \, & \, \lambda_1,  \lambda_3, \lambda_5, \lambda_6
               \cr
\hline
\, A_3^* \, & \, \lambda_2,  \lambda_4, \lambda_5, \lambda_6
               \cr
\hline
\, A_4 \, & \, \lambda_1, \lambda_3,  \lambda_6
               \, \cr
\hline
\, A_4^* \, & \, \lambda_2, \lambda_4,  \lambda_6
               \, \cr
\hline
\, A_5 \, & \, \lambda_1, \lambda_3 \cr
\hline
\, A_5^* \, & \, \lambda_2, \lambda_4 \cr
\hline
\, A_6 \, & \,  \lambda_5, \lambda_6,
               \lambda_7, \lambda_8  \cr
\hline
\, A_7 \, & \, \lambda_5, \lambda_6
               \, \cr
\hline
\, A_8 \, & \, \lambda_6,
               \lambda_8  \cr
\hline
\end{array}
\label{eigenvaluesAthreeform}
\end{eqnarray}
\end{footnotesize}
%%%%%%%%%%%%%%%%%%%%%%%%%%%%%%%%%%%%%%%%%%%%%%%%%%%%%%%%%%%%%%%%
%                            + +                               %
%%%%%%%%%%%%%%%%%%%%%%%%%%%%%%%%%%%%%%%%%%%%%%%%%%%%%%%%%%%%%%%%
The two--form operator matrix for the regular series of type $^{++}$ is
the following $10\times 10$ matrix:\\
Columns one to five:
\begin{footnotesize}
\begin{eqnarray}
\begin{array}{|c||c|c|c|c|c|}
\hline
    & \pi \langle 1,{\rm I} \rangle
    & \tilde {\pi} \langle 1,{\rm I} \rangle
    & \pi[1,{\rm I}]
    & \pi \langle \ft12, {\rm I} \rangle
    & \pi \langle \ft12, {\rm II} \rangle
\cr
\hline
\hline
% row 1
  \pi \langle 1,{\rm I} \rangle                 &
  {Y} & 0 &
  {\frac{-\left( 2\,J + {Y} \right) }{2\,{\sqrt{2}}}} & 0 & 0
   \cr
% row 2
    \tilde {\pi} \langle 1,{\rm I} \rangle               &
 0 & -{Y} & {\frac{-\left( -2\,J + {Y} \right) }
    {2\,{\sqrt{2}}}} & 0 & 0 \cr
% row 3
  \pi[1,{\rm I}]                 &
  {\frac{-2 - 2\,J + {Y}}{{\sqrt{2}}}} &
  {\frac{2 + 2\,J + {Y}}{{\sqrt{2}}}} & 1 & 0 &
  {\frac{-2\,i\,\left( -1 + {M_2} \right) }{{\sqrt{3}}}} \cr
% row 4
   \pi \langle \ft12, {\rm I}   \rangle                 &
   0 & 0
   & 0 & 0 & i\,{Y} \cr
% row 5
         \pi \langle \ft12, {\rm II}  \rangle          &
  0 & 0 &
  {\frac{i\,\left( 3 + {M_1} \right) }{{\sqrt{3}}}} &
  -i\,{Y} & 0 \cr
% row 6
   \pi \langle \ft12, {\rm III}   \rangle                 &
 {\frac{3 + {M_1}}{{\sqrt{3}}}} & 0 &
  0 & {\frac{-\left( 2\,J + {Y} \right) }{2\,{\sqrt{2}}}} & 0
   \cr
% row 7
      \pi \langle \ft12, {\rm IV}    \rangle              &
 0 & -{\frac{3 + {M_1}}{{\sqrt{3}}}} & 0 &
  {\frac{2\,J - {Y}}{2\,{\sqrt{2}}}} & 0 \cr
% row 8
     \pi \langle 0, {\rm I} \rangle              &
  0 & 0 & 0 & 0 &
  {\frac{i\,\left( 2 + {M_1} \right) }{{\sqrt{3}}}} \cr
% row 9
   \tilde{\pi} \langle 0, {\rm II}  \rangle                &
 0 & 0 & 0
   & 0 & 0 \cr
% row 10
       \tilde{\pi} \langle 0, {\rm III}  \rangle             &
 0 & 0 & 0 & 0 & 0 \cr
\hline
\end{array}
\end{eqnarray}
\end{footnotesize}
                %%%%%%%%%%%%%%%%%%%%%%%%%%%%%%%%%%%%%%%%
Columns six to ten:
\begin{footnotesize}
\begin{eqnarray}
\begin{array}{|c||c|c|c|c|c|}
\hline
    &\pi \langle \ft12, {\rm III}  \rangle
    &\pi \langle \ft12, {\rm IV} \rangle
    &\pi \langle 0, {\rm I} \rangle
    & \tilde{\pi} \langle 0, {\rm II}  \rangle
    &\tilde{\pi} \langle 0, {\rm III}  \rangle
\cr
\hline
\hline
% row 1
   \pi \langle 1,{\rm I} \rangle             &
   {\frac{2\,\left( -1 + {M_2} \right) }{{\sqrt{3}}}} & 0 & 0
   & 0 & 0 \cr
% row 2
         \tilde {\pi \langle 1,{\rm I} \rangle}       &
   0 & {\frac{-2\,\left( -1 + {M_2} \right) }
    {{\sqrt{3}}}} & 0 & 0 & 0 \cr
% row 3
   \pi[1,{\rm I}]             &
   0 & 0 & 0 & 0 & 0 \cr
% row 4
     \pi \langle \ft12, {\rm I}  \rangle            &
  {\frac{-2 - 2\,J + {Y}}{{\sqrt{2}}}} &
  {\frac{2 + 2\,J + {Y}}{{\sqrt{2}}}} & 0 & 0 & 0 \cr
% row 5
       \pi \langle \ft12, {\rm II}  \rangle         &
 0 & 0 &
  {\frac{-2\,i\,{M_2}}{{\sqrt{3}}}} & 0 & 0 \cr
% row 6
   \pi \langle \ft12, {\rm III}  \rangle             &
 1 & 0 & 0 & 0
   & {\frac{-2\,{M_2}}{{\sqrt{3}}}} \cr
% row 7
   \pi \langle \ft12, {\rm IV} \rangle             &
 0 & -1 & 0 &
  {\frac{2\,{M_2}}{{\sqrt{3}}}} & 0 \cr
% row 8
    \pi \langle 0, {\rm I} \rangle            &
 0 & 0 & -1 &
  -{\frac{2 + 2\,J + {Y}}{{\sqrt{2}}}} &
  {\frac{2 + 2\,J - {Y}}{{\sqrt{2}}}} \cr
% row 9
   \tilde{\pi} \langle 0, {\rm II}  \rangle             &
 0 &
  {\frac{2 + {M_1}}{{\sqrt{3}}}} &
  {\frac{-2\,J + {Y}}{2\,{\sqrt{2}}}} & {Y} & 0 \cr
% row 10
       \tilde{\pi} \langle 0, {\rm III}  \rangle          &
  -{\frac{2 + {M_1}}{{\sqrt{3}}}} & 0 &
  {\frac{2\,J + {Y}}{2\,{\sqrt{2}}}} & 0 & -{Y} \cr
\hline
\end{array}
\end{eqnarray}
\end{footnotesize}
It has eigenvalues:
\begin{eqnarray}
\lambda_1 &=& \ft14 \sqrt{H_0 + \ft{32}{3}(M_2 - M_1) +16} \,,
\nonumber \\
\lambda_2 &=& - \ft14 \sqrt{H_0 + \ft{32}{3}(M_2 - M_1) +16} \,,
\nonumber \\
\lambda_3 &=& \ft14 \sqrt{H_0 + 36} - \ft12 \,,
\nonumber \\
\lambda_4 &=& - \ft14 \sqrt{H_0 + 36} - \ft12 \,,
\nonumber \\
\lambda_5 &=& - \ft14 \sqrt{H_0 + \ft{64}{3}(M_2 - M_1) - 28} + \ft12 \,,
\nonumber \\
\lambda_6 &=&  \ft14 \sqrt{H_0 + \ft{64}{3}(M_2 - M_1) - 28} + \ft12 \,,
\nonumber \\
\lambda_7 &=& \dots = \lambda_{10} = 0 \,.
\label{eigenBthreeform}
\end{eqnarray}
The complete table of eigenvalues for the type $^{++}$ series is:
\begin{footnotesize}
\begin{eqnarray}
\begin{array}{|c||c|}
\hline
\, B_R \, & \, \lambda_1, \lambda_2, \lambda_3, \lambda_4, \lambda_5, \lambda_6 \cr
\hline
\, B_1 \, & \, \lambda_1, \lambda_2,  \lambda_5, \lambda_6 \cr
\hline
\, B_2 \, & \, \lambda_5, \lambda_6 \cr
\hline
\, B_3 \, & \, \lambda_1, \lambda_2 \cr
\hline
\, B_4 \, & \, \lambda_1, \lambda_2, \lambda_3, \lambda_4 \cr
\hline
\, B_5 \, & \, \lambda_3, \lambda_4 \cr
\hline
\, B_6 \, & \, \lambda_1, \lambda_2, \lambda_3, \lambda_4 \cr
\hline
\, B_7 \, & \, \lambda_2, \lambda_3, \lambda_4 \cr
\hline
\, B_8 \, & \,  \lambda_4 \cr
\hline
\, B_9 \, & \, \lambda_3, \lambda_4 \cr
\hline
\, B_{10} \, & \,  \lambda_4 \cr
\hline
\, B_{11} \, & \,  \lambda_4 \cr
\hline
\end{array}
\label{eigenvaluesBthreeform}
\end{eqnarray}
\end{footnotesize}
For the representations of the $^{--}$ series, the eigenvalues are the 
conjugates of the ones in \eqn{eigenvaluesBthreeform}.
%%%%%%%%%%%%%%%%%%%%%%%%%%%%%%%%%%%%%%%%%%%%%%%%%%%%%%%%%%%
%                                                         %
%                     S P I N O R                         %
%                                                         %
%%%%%%%%%%%%%%%%%%%%%%%%%%%%%%%%%%%%%%%%%%%%%%%%%%%%%%%%%%%
\subsubsection{The spinor}
The harmonic analysis of the eight--component Majorana spinor
has been completely worked out in \cite{spectfer}. We reformulate these results 
in our framework, in order to facilitate the matching of the spectrum with the ${\cal N}=2$ multiplets.
\par
The decomposition of the spinor
in its $H$--irreducible   components is
\begin{eqnarray}
\eta=
\left(
\matrix{
\langle \ft12 | {\rm I} \rangle_i \cr
\langle \ft12 | {\rm I} \rangle_\cdot \cr
\langle \ft12 | {\rm II} \rangle_\cdot \cr
-i\sigma_2 \langle \ft12 | {\rm I} \rangle_i^* \cr
\langle \ft12 | {\rm II} \rangle^*_\cdot \cr
-\langle \ft12 | {\rm I} \rangle^*_\cdot
}
\right)
\end{eqnarray}
where
\begin{eqnarray}
\langle \ft12 | {\rm I} \rangle_i
&=& {\cal H}_i^{[1/2, -i, -3i/2]\xi} \cdot \chi\langle \ft12, {\rm I} \rangle
\,, \nonumber \\
\langle \ft12 | {\rm I} \rangle_\cdot
 &=& {\cal H}^{[0, 2i, -3i]} \cdot \chi\langle 0, {\rm I} \rangle
\,, \nonumber \\
\langle \ft12 | {\rm II} \rangle_\cdot
 &=& {\cal H}^{[0, -4i, 0]} \cdot \chi\langle 0, {\rm II} \rangle
\,, \nonumber \\
\langle \ft12 | {\rm I} \rangle_i^*
 &=& \pm \varepsilon^{ij} {\cal H}_j^{[1/2, i, 3i/2]\xi} \cdot
\tilde \chi\langle \ft12, {\rm I} \rangle
\,, \nonumber \\
\langle \ft12 | {\rm I} \rangle^*_\cdot
 &=& {\cal H}^{[0, -2i, 3i]} \cdot \tilde \chi\langle 0, {\rm I} \rangle
\,, \nonumber \\
\langle \ft12 | {\rm II} \rangle^*_\cdot &=& {\cal H}^{[0, -4i, 0]} \cdot
\tilde \chi\langle 0, {\rm II} \rangle.
\end{eqnarray}
The fragments of type $^+$ are
\begin{eqnarray}
\langle \ft12 | {\rm I} \rangle_i &=& {\cal H}_i^{[1/2, -i, -3i/2](b)} \cdot \chi\langle \ft12, {\rm I} \rangle
\,, \nonumber \\
\langle \ft12 | {\rm I} \rangle_\cdot &=& {\cal H}^{[0, 2i, -3i]} \cdot \chi\langle 0, {\rm I} \rangle
\,, \nonumber \\
\langle \ft12 | {\rm I} \rangle_i^* &=&   \varepsilon^{ij} {\cal H}_j^{[1/2, i, 3i/2](b)} \cdot
\tilde \chi\langle \ft12, {\rm I} \rangle
\,, \nonumber \\
\langle \ft12 | {\rm II} \rangle_\cdot^* &=&
{\cal H}^{[0, 4i, 0]} \cdot \tilde \chi\langle 0, {\rm II} \rangle
\,.\nn\\
\end{eqnarray}
\par
For the regular series $^+$ the spinor operator acts on the $AdS_4$ fields
as a $4\times 4$ matrix, whose eigenvalues are:
\begin{eqnarray}
\lambda_1 &=& - 6 + \sqrt{H_0 + 36} \,,
\nonumber \\
\lambda_2 &=& - 6 - \sqrt{H_0 + 36} \,,
\nonumber \\
\lambda_3 &=& - 8 + \sqrt{H_0 + 16 + \ft{32}{3}(M_2 - M_1)} \,,
\nonumber \\
\lambda_4 &=& - 8 - \sqrt{H_0 + 16 + \ft{32}{3}(M_2 - M_1)} \,.
\nonumber \\
\label{eigenA+spinor}
\end{eqnarray}
The eigenvalues for each exceptional series are
\begin{footnotesize}
\begin{eqnarray}
\begin{array}{|c||c|}
\hline
A_R^+, A_1^+, A_3^+, A_4^+ & \, \lambda_1, \lambda_2, \lambda_3, \lambda_4 \, \cr
\hline
A_2^+, A_5^+ & \, \lambda_3, \lambda_4 \, \cr
\hline
A_1^{+*}, A_6^+ & \, \lambda_1, \lambda_2 \, \cr
\hline
A_3^{+*}, A_7^+ & \, \lambda_1, \lambda_2 \, \cr
\hline
A_4^{+*}, A_8^+ & \, \lambda_1  \, \cr
\hline
\end{array}
\label{eigenvaluesA+spinor}
\end{eqnarray}
\end{footnotesize}
The fragments of type $^-$ are
\begin{eqnarray}
\langle \ft12 | {\rm I} \rangle_i &=& {\cal H}_i^{[1/2, -i, -3i/2](a)} \cdot \chi\langle \ft12, {\rm I} \rangle
\,, \nonumber \\
\langle \ft12 | {\rm II} \rangle_\cdot &=& {\cal H}^{[0, -4i, 0]} \cdot \chi\langle 0, {\rm II} \rangle
\,, \nonumber \\
\langle \ft12 | {\rm I} \rangle_i^* &=&  \varepsilon^{ij} {\cal H}_j^{[1/2, i, 3i/2](a)} \cdot
\tilde{ \chi\langle \ft12, {\rm I} \rangle }
\,, \nonumber \\
\langle \ft12 | {\rm I} \rangle_\cdot^* &=& {\cal H}^{[0, -2i, 3i]} \cdot \tilde{ \chi\langle 0, {\rm II} \rangle }
\,.
\end{eqnarray}
\par
For the regular series $^-$ the spinor operator acts on the $AdS_4$ fields
as a $4\times 4$ matrix, whose eigenvalues are:
\begin{eqnarray}
\lambda_1 &=& - 6 + \sqrt{H_0 + 36} \,,
\nonumber \\
\lambda_2 &=& - 6 - \sqrt{H_0 + 36} \,,
\nonumber \\
\lambda_3 &=& - 8 + \sqrt{H_0 + 16 - \ft{32}{3}(M_2 - M_1)} \,,
\nonumber \\
\lambda_4 &=& - 8 - \sqrt{H_0 + 16 - \ft{32}{3}(M_2 - M_1)} \,.
\nonumber \\
\label{eigenA-spinor}
\end{eqnarray}
The eigenvalues for each exceptional series are:
\begin{footnotesize}
\begin{eqnarray}
\begin{array}{|c||c|}
\hline
A_R^-, A_1^{-*}, A_3^{-*}, A_4^{-*}
& \, \lambda_1, \lambda_2, \lambda_3, \lambda_4 \, \cr
\hline
A_2^{-*}, A_5^{-*} & \, \lambda_3, \lambda_4 \, \cr
\hline
A_1^-, A_6^- & \, \lambda_1, \lambda_2 \, \cr
\hline
A_3^-, A_7^- & \, \lambda_1, \lambda_2 \, \cr
\hline
A_4^-, A_8^- & \, \lambda_1  \, \cr
\hline
\end{array}
\label{eigenvaluesA-spinor}
\end{eqnarray}
\end{footnotesize}
%%%%%%%%%%%%%%%%%%%%%%%%%%%%%%%%%%%%%%%%%%%%%%%%%%%%%%%%%%%
%                                                         %
%                 M A T C H I N G                         %
%                                                         % 
%%%%%%%%%%%%%%%%%%%%%%%%%%%%%%%%%%%%%%%%%%%%%%%%%%%%%%%%%%%
\subsection{Matching the spectrum with the $Osp(2|4)$ multiplets}
As already mentioned, the structures of the long multiplets that arise
from ${\cal N}=2$ compactifications of eleven--dimensional supergravity
in $AdS_4$ has been found in \cite{multanna}.
The structure and the $G'$ representations of the long graviton, the long gravitino
and the massless multiplets are known since the eighties \cite{multanna}. 
The structure of the long vector multiplet can be very easily derived, as shown in
chapter $2$. However this is not the case for the the short multiplets: the method of norms 
become very cumbersome after the $B_2$ sector.
So we have joined our knowledge on long multiplets and on the part of
short multiplets arising from the simpler norm calculation, which are
shown in chapter $2$, and information arising from harmonic analysis,
in order to get two results at the same time:
\begin{enumerate}
\item filling the blanks in the structure of $\cN=2$ short multiplets, 
verifying which fields disappear, and if there are new shortening conditions;
\item finding the complete spectrum of this supergravity, even the part of
the spectrum which has not been directly found from harmonic analysis.
\end{enumerate}
In this section I  rewrite the tables of the $\cN=2$ supermultiplets, 
already given in chapter two, because I have to assign to each field its name following the definitions
given in section \ref{listfields}. This is necessary in order to follow the reasoning of filling the 
multiplets with these fields. In each of these tables, the fields whose presence in the 
corresponding multiplet can be established by means of the norm evaluation discussed in
chapter $2$ are denoted by an asterisk, in order to distinguish them from the fields whose 
presence is established by the discussion below, utilizing
the harmonic analysis results and the $\cN=2\longrightarrow\cN=1$ decomposition. 
The fields in the long and massless multiplets have all the asterisk because as I said 
the structure of these multiplets was known before performing harmonic analysis.
\par
We use a procedure of {\it exhaustion},
i.e. one starts with one of the
four different types of multiplets for which
all the masses of a certain field component 
are most easily retrieved
(this is for instance the case for the graviton field of the graviton multiplet)
and using the mass relations (\ref{massrelationchi}), (\ref{massrelationlambdaL}), (\ref{massrelationlambdaT}),
one calculates all the masses
of the other types of fields present in the multiplet.
One uses also the information that all the fields in  a 
multiplet are in the same irreducible 
$G'=SU\left(3\right)\times SU\left(2\right)$ representation
and that their hypercharges are related according to the group
theoretical structure of the multiplets shown in tables
\ref{longgraviton},$\dots$,\ref{masslessvector}.
So one knows in which $G$ representation  to find
the other fields of the multiplet, whose masses have been determined. 
Then, upon using the relations (\ref{massform}), these masses are 
compared with the eigenvalues
of the invariant operators on the spinor, the one--form,
the two--form or the three--form depending on the type of field one is 
considering. The upshot of this is
that some of these eigenvalues yield all the masses
obtained from the mass relations.
However, the remaining eigenvalues signal the
existence of some extra masses which then pertain to other
fields that are to be found in other multiplets.
In this way one establishes the existence of
new unknown multiplets and determines their structure
by filling out their field content.
After repeatedly applying this procedure one
will have filled out all the existing multiplets
in the spectrum.
\par
I should remark here that we did not calculate the eigenvalues 
of the Lichnerowicz  and Rarita--Schwinger operators 
$M_{(2)(0)^2}$ and $M_{(3/2)(1/2)^2}$.
However we succeeded in finding the complete
multiplet structure without making use of this.
The $AdS_4$ fields whose spectrum is determined by  
$M_{(2)(0)^2}$ and $M_{(3/2)(1/2)^2}$ are the scalar field $\phi$
and the transverse spinor field $\lambda_T$ (see (\ref{kkexpansion}) ).
We can fill the multiplets without knowing the spectrum of 
these two fields with the
help of the ${\cal N}=2\rightarrow {\cal N}=1$ decompositions
(\ref{N21longgraviton}),$\dots$,(\ref{N21hyper}).
If we know every field of a multiplet
except for $\phi$ and $\lambda_T$, we can deduce which $\phi$ and $\lambda_T$
are present by trying to organize the ${\cal N}=2$ multiplet  in ${\cal N}=1$ multiplets. 
There is no ambiguity, because no ${\cal N}=1$ multiplet is built using $\phi$ and 
$\lambda_T$ fields alone.
In particular, a Wess Zumino multiplet with one $\lambda_T$ and two $\phi$ 's is
not allowed, since it has to contain both a scalar and a pseudoscalar.
\par
In practice one starts with the graviton multiplet
since the masses of the graviton field in the different
representations are immediate to derive, being the eigenvalues
of the scalar operator $M_{(0)^3}$.
By means of the above procedure,
one {\it exhausts} all the spin--$\ft32$ fields
 in the graviton multiplet comparing
the masses of the spin--$\ft32$ fields in the graviton multiplet
with the eigenvalues of the operator $M_{(1/2)^3}$.
The spin--$\ft32$ fields that provide the remaining
eigenvalues of the operator $M_{(1/2)^3}$,
can only be the highest--spin component gravitino fields
of the gravitino multiplet and hence we know all the masses of the gravitinos
in the gravitino multiplet.  At this stage we can repeat the same procedure.
We use the eigenvalues of the one--form operator
$M_{(1)(0)^2}$ to identify the vector fields $A$ and $W$
and we use the eigenvalues of the two--form operator
$M_{(1)^2 (0)}$ to identify the vector fields $Z$ in the graviton
and the gravitino multiplet. The remaining vector fields constitute the
highest--component vector fields of the vector multiplet.
Then we determine the masses of the longitudinal spinors, provided by the
eigenvalues of the operator $M_{(1/2)^3}$, and we find the longitudinal
spinors of the gravitino and vector multiplet.
The remaining longitudinal spinors  belong to hypermultiplets.
At the end we determine the masses of the scalars $S,\Sigma$, that are
provided by the eigenvalues of $M_{(0)^3}$, and of the pseudoscalar $\pi$,
provided by the eigenvalues of the three--form operator $M_{(1)^3}$.
At this point, the  matching of the spectrum with the multiplets will be complete.
\par
Since we are in particular interested in multiplet shortening, it is of
utmost importance
to pay attention to what happens with the eigenvalues in the exceptional series.
As it is clear from tables (\ref{eigenvaluesAoneform}), (\ref{eigenvaluesAthreeform}),
(\ref{eigenvaluesBthreeform}) of the eigenvalues, there are always less eigenvalues
present when the operators act on the harmonics in the exceptional
series. This is reflected into the fact that certain field components are
not present in the multiplets, thus multiplet shortening.
\par
In the next sections I give a detailed discussion of the matching
of the multiplets. Doing so I 
show that the information
collected about the invariant operators on the
zero form, the one--form, the two--form, the three--form and the spinor
is in perfect agreement with the group theoretical information
given in \cite{multanna} and in chapter $2$ of this thesis.
%%%%%%%%%%%%%%%%%%%%%%%%%%%%%%%%%%%%%%%%%%%%%%%%%%%%%%%%%%%%%%%%%%%%%%%%%%%%%%%%%%%%%%%%
\begin{table}
\centering
\begin{tabular}{||c|c|c|c|c|c||}
\hline
& Spin &
Energy &
Hypercharge &
Mass ($^2$)&
Name \\ 
\hline
\hline
$*$& $2$      & $E_0+1    $  & $y_0$     & $16(E_0+1)(E_0-2)$  & $h$ \\
$*$&$\ft32$  & $E_0+\ft32$  & $y_0-1$   & $-4E_0-4$  & $\chi^-$ \\
$*$&$\ft32$  & $E_0+\ft32$  & $y_0+1$   & $-4E_0-4$  & $\chi^-$ \\
$*$&$\ft32$  & $E_0+\ft12$  & $y_0-1$   & $4E_0-8 $  & $\chi^+$ \\
$*$&$\ft32$  & $E_0+\ft12$  & $y_0+1$   & $4E_0-8 $  & $\chi^+$ \\
$*$&$1$      & $E_0+2    $  & $y_0$     & $16E_0(E_0+1)$    & $W$ \\
$*$&$1$      & $E_0+1    $  & $y_0-2$   & $16E_0(E_0-1)$    & $Z$ \\
$*$&$1$      & $E_0+1    $  & $y_0+2$   & $16E_0(E_0-1)$    & $Z$ \\
$*$&$1$      & $E_0+1    $  & $y_0$     & $16E_0(E_0-1)$    & $Z$ \\
$*$&$1$      & $E_0+1    $  & $y_0$     & $16E_0(E_0-1)$    & $Z$ \\
$*$&$1$      & $E_0      $  & $y_0$     & $16(E_0-1)(E_0-2)$& $A$  \\
$*$&$\ft12$  & $E_0+\ft32$  & $y_0-1$   & $4E_0$     &  $\lambda_T$  \\
$*$&$\ft12$  & $E_0+\ft32$  & $y_0+1$   & $4E_0$     &  $\lambda_T$ \\
$*$&$\ft12$  & $E_0+\ft12$  & $y_0-1$   & $-4E_0+4$  &  $\lambda_T$ \\
$*$&$\ft12$  & $E_0+\ft12$  & $y_0+1$   & $-4E_0+4$  &  $\lambda_T$ \\
$*$&$0$      & $E_0+1    $  & $y_0$     & $16E_0(E_0-1)  $  & $\phi$ \\
\hline
\end{tabular}\\[.13in]
\caption{$M^{111}$ Kaluza Klein fields in the ${\cal N}=2$ long graviton multiplet with $y_0\ge 0$}
\label{longgraviton}
\end{table}
%%%%%%%%%%%%%%%%%%%%%%%%%%%%%%%%%%%%%%%%%%%%%%%%%%%%%%%%%%%%%%%%%%%%%%%%%%%%%%%%%%%%%
\begin{table}
\centering
\begin{small}
\begin{tabular}{||c|c|c|c|c|c|c|c||}
\hline
&$\!$Spin$\!$ &
$\!$Energy$\!$ &
$\!$Hypercharge$\!$ &
$\!$Mass ($^2$)$\!$&
$\!$Name$\!$ &
$\!$Mass ($^2$)$\!$&
$\!$Name $\!$\\
\hline
\hline
$*$&$\ft32$  & $E_0+1$      & $y_0$   & $4E_0-6$  & $\chi^+$
                & $-4E_0-2$ & $\chi^-$  \\
$*$&$1$      & $E_0+\ft32$  & $y_0-1$ & $16(E_0-\ft12)(E_0+\ft12)$ & $Z$
                & $16(E_0-\ft12)(E_0+\ft12)$ & $W$ \\
$*$&$1$      & $E_0+\ft32$  & $y_0+1$ & $16(E_0-\ft12)(E_0+\ft12)$ & $Z$
                & $16(E_0-\ft12)(E_0+\ft12)$ & $W$ \\
$*$&$1$      & $E_0+\ft12$  & $y_0-1$ & $16(E_0-\ft32)(E_0-\ft12)$ & $A$
                & $16(E_0-\ft32)(E_0-\ft12)$ & $Z$ \\
$*$&$1$      & $E_0+\ft12$  & $y_0+1$ & $16(E_0-\ft32)(E_0-\ft12)$ & $A$
                & $16(E_0-\ft32)(E_0-\ft12)$ & $Z$ \\
$*$&$\ft12$  & $E_0+2$      & $y_0$   & $4E_0+2$  &  $\lambda_T$
                & $-4E_0-2$ &  $\lambda_L$ \\
$*$&$\ft12$  & $E_0+1$      & $y_0-2$ & $-4E_0+2$ &  $\lambda_T$
                & $-4E_0-2$ &  $\lambda_T$ \\
$*$&$\ft12$  & $E_0+1$      & $y_0$   & $-4E_0+2$ &  $\lambda_T$
                & $4E_0-2$  &  $\lambda_T$  \\
$*$&$\ft12$  & $E_0+1$      & $y_0+2$ & $-4E_0+2$ &  $\lambda_T$
                & $4E_0-2$  &  $\lambda_T$ \\
$*$&$\ft12$  & $E_0+1$      & $y_0$   & $-4E_0+2$ &  $\lambda_T$
                & $4E_0-2$  &  $\lambda_T$  \\
$*$&$\ft12$  & $E_0$        & $y_0$   & $4E_0-6$  &  $\lambda_L$
                & $-4E_0+6$ &  $\lambda_T$ \\
$*$&$0$      & $E_0+\ft32$  & $y_0-1$ & $16(E_0-\ft12)(E_0+\ft12)$  & $\phi$
                & $16(E_0-\ft12)(E_0+\ft12)$  & $\pi$ \\
$*$&$0$      & $E_0+\ft32$  & $y_0+1$ & $16(E_0-\ft12)(E_0+\ft12)$  & $\phi$
                & $16(E_0-\ft12)(E_0+\ft12)$  & $\pi$ \\
$*$&$0$      & $E_0+\ft12$  & $y_0-1$ & $16(E_0-\ft32)(E_0-\ft12)$  & $\pi$
                & $16(E_0-\ft32)(E_0-\ft12)$  & $\phi$ \\
$*$&$0$      & $E_0+\ft12$  & $y_0+1$ & $16(E_0-\ft32)(E_0-\ft12)$  & $\pi$
                & $16(E_0-\ft32)(E_0-\ft12)$  & $\phi$ \\
\hline
\end{tabular}\\[.13in]
\end{small}
\caption{$M^{111}$ Kaluza Klein fields in the ${\cal N}=2$ long gravitino multiplets $\chi^+$ and $\chi^-$ with $y_0\ge 0$}
\label{longgravitino}
\end{table}
%%%%%%%%%%%%%%%%%%%%%%%%%%%%%%%%%%%%%%%%%%%%%%%%%%%%%%%%%%%%%%%%%%%%%%%%%%%%%%%%%%%
\begin{table}
\centering
\begin{footnotesize}
\begin{tabular}{||c|c|c|c|c|c|c|c|c||}
\hline
&$\!$Spin$\!$ &
$\!$Energy$\!$ &
$\!$Hypercharge$\!$ &
$\!$Mass ($^2$)$\!$&
$\!$Name$\!$ &
$\!$Name $\!$&
$\!$Mass ($^2$)$\!$&
$\!$Name $\!$\\
\hline
\hline
$*$&$1$      & $E_0+1$      & $y_0$   & $16E_0(E_0-1)$  & $A$
                 & $W$     & $16E_0(E_0-1)$  & $Z$      \\
$*$&$\ft12$  & $E_0+\ft32$  & $y_0-1$ & $-4E_0$  &   $\lambda_T$
                 &   $\lambda_L$  & $4E_0$   &   $\lambda_T$   \\
$*$&$\ft12$  & $E_0+\ft32$  & $y_0+1$ & $-4E_0$  &   $\lambda_T$
                  &   $\lambda_L$  & $4E_0$   &   $\lambda_T$   \\
$*$&$\ft12$  & $E_0+\ft12$  & $y_0-1$ & $4E_0-4$ &   $\lambda_L$
                 &   $\lambda_T$   & $-4E_0+4$&   $\lambda_T$ \\
$*$&$\ft12$  & $E_0+\ft12$  & $y_0+1$ & $4E_0-4$ &   $\lambda_L$
                 &   $\lambda_T$   & $-4E_0+4$&   $\lambda_T$  \\
$*$&$0$      & $E_0+2$      & $y_0$   & $16E_0(E_0+1)$  & $\phi$
                 & $\Sigma$  & $16E_0(E_0+1)$  & $\pi$  \\
$*$&$0$      & $E_0+1$      & $y_0-2$ & $16E_0(E_0-1)$  & $\pi$
                  & $\pi$   & $16E_0(E_0-1)$  & $\phi$   \\
$*$&$0$      & $E_0+1$      & $y_0+2$ & $16E_0(E_0-1)$  & $\pi$
                  & $\pi$   & $16E_0(E_0-1)$  & $\phi$   \\
$*$&$0$      & $E_0+1$      & $y_0$   & $16E_0(E_0-1)$  & $\pi$
                  & $\pi$    & $16E_0(E_0-1)$  & $\phi$   \\
$*$&$0$      & $E_0$        & $y_0$   & $\!16(E_0\!-\!2)(E_0\!-\!1)$  & $S$
                  & $\phi$  & $\!16(E_0\!-\!2)(E_0\!-\!1)$  & $\pi$  \\
\hline
\end{tabular}\\[.13in]
\end{footnotesize}
\caption{$M^{111}$ Kaluza Klein fields in the ${\cal N}=2$ long vector multiplets $A$,$W$ and $Z$ with $y_0\ge 0$}
\label{longvector}
\end{table}
%%%%%%%%%%%%%%%%%%%%%%%%%%%%%%%%%%%%%%%%%%%%%%%%%%%%%%%%%%%%%%%%%%%%%%%%%%%%%%%%%%%%%%
\begin{table}
\centering
\begin{footnotesize}
\begin{tabular}{||c|c|c|c|c|c||}
\hline
&Spin &
Energy &
Hypercharge &
Mass ($^2$)&
Name \\
\hline
\hline
&$2$      & $y_0+3$        & $y_0$     & $16y_0(y_0+3)$  & $h$        \\
&$\ft32$  & $y_0+\ft72$    & $y_0-1$& $-4y_0-12$  & $\chi^-$      \\
$*$&$\ft32$  & $y_0+\ft52$    & $y_0+1$   & $4y_0 $  & $\chi^+$             \\
$*$&$\ft32$  & $y_0+\ft52$    & $y_0-1$   & $4y_0 $  & $\chi^+$             \\
$*$&$1$      & $y_0+3    $    & $y_0-2$& $16(y_0+2)(y_0+1)$    & $Z$        \\
&$1$      & $y_0+3    $    & $y_0$     & $16(y_0+2)(y_0+1)$    & $Z$     \\
$*$&$1$      & $y_0+2      $  & $y_0$     & $16y_0(y_0+1)$& $A$         \\
$*$&$\ft12$  & $y_0+\ft52$  & $y_0-1$& $-4y_0-4$  &  $\lambda_T$    \\
\hline
\end{tabular}\\[.13in]
\end{footnotesize}
\caption{$M^{111}$ Kaluza Klein fields in the ${\cal N}=2$ short graviton multiplet with $y_0>0$}
\label{shortgraviton}
\end{table}
%%%%%%%%%%%%%%%%%%%%%%%%%%%%%%%%%%%%%%%%%%%%%%%%%%%%%%%%%%%%%%%%%%%%%%%%%%%%%%%%%%%%%%%
\begin{table}
\centering
\begin{footnotesize}
\begin{tabular}{||c|c|c|c|c|c||}
\hline
&Spin &
Energy &
Hypercharge &
Mass ($^2$)&
Name \\
\hline
\hline
&$\ft32$  & $y_0+\ft52$    & $y_0$         & $4y_0$                  & $\chi^+$  \\
&$1$      & $y_0+3$        & $y_0-1$  & $16(y_0+1)(y_0+2)$      & $Z$    \\
$*$&$1$      & $y_0+2$        & $y_0+1$       & $16y_0(y_0+1)$       & $A$  \\
$*$&$1$      & $y_0+2$        & $y_0-1$       & $16y_0(y_0+1)$       & $A$  \\
&$\ft12$   & $y_0+\ft52$   & $y_0$         & $-4y_0-4$            &  $\lambda_T$ \\
$*$&$\ft12$  & $y_0+\ft52$    & $y_0-2$         & $-4y_0-4$            &  $\lambda_T$   \\
$*$&$\ft12$  & $y_0+\ft32$    & $y_0$         & $4y_0$                  &  $\lambda_L$  \\
$*$&$0$       &  $y_0+3$   & $y_0\pm 1$     & $16(y_0+1)(y_0+2)$ &  $\phi$  \\
\hline
\end{tabular}\\[.13in]
\end{footnotesize}
\caption{$M^{111}$ Kaluza Klein fields in the ${\cal N}=2$ short gravitino multiplet $\chi^+$ with $y_0 >0$}
\label{shortgravitino}
\end{table}
%%%%%%%%%%%%%%%%%%%%%%%%%%%%%%%%%%%%%%%%%%%%%%%%%%%%%%%%%%%%%%%%%%%%%%%%%%%%%%%%%%%%%%%%%%%%%
\begin{table}
\centering
\begin{footnotesize}
\begin{tabular}{||c|c|c|c|c|c||}
\hline
&Spin &
Energy &
Hypercharge &
Mass ($^2$)&
Name  \\
\hline
\hline
&$1$      & $y_0+2$      & $y_0$      & $16y_0(y_0+1)$   & $A$       \\
&$\ft12$  & $y_0+\ft52$  & $y_0\pm 1$ & $-4y_0-4$    & $\lambda_T$   \\
$*$&$\ft12$  & $y_0+\ft32$  & $y_0+1$    & $4y_0$       & $\lambda_L$   \\
$*$&$\ft12$  & $y_0+\ft32$  & $y_0-1$    & $4y_0$       & $\lambda_L$   \\
$*$&$0$      & $y_0+2$      & $y_0-2$ & $16y_0(y_0+1)$  & $\pi$     \\
&$0$      & $y_0+2$      & $y_0$      & $16y_0(y_0+1)$   & $\pi$         \\
$*$&$0$      & $y_0+1$      & $y_0$      & $16y_0(y_0-1)$   & $S$       \\
\hline
\end{tabular}\\[.13in]
\end{footnotesize}
\caption{$M^{111}$ Kaluza Klein fields in the ${\cal N}=2$ short vector multiplet $A$ with $y_0 >0$}
\label{shortvector}
\end{table}
%%%%%%%%%%%%%%%%%%%%%%%%%%%%%%%%%%%%%%%%%%%%%%%%%%%%%%%%%%%%%%%%%%%%%%%%%%%%%%%%%%%%%%%%%%%%
\begin{table}
\centering
\begin{footnotesize}
\begin{tabular}{||c|c|c|c|c|c||}
\hline
&Spin &
Energy &
Hypercharge &
Mass ($^2$)&
Name  \\
\hline
\hline
$*$&$\ft12$  & $y_0+\ft12$  & $y_0-1$    & $4y_0-4$       & $\lambda_L$   \\
$*$&$0$      & $y_0+1$      & $y_0-2$    & $16y_0(y_0-1)$   & $\pi$     \\
$*$&$0$      & $y_0$        & $y_0$      & $16(y_0-2)(y_0-1)$   & $S$       \\
\hline
\end{tabular}\\[.13in]
\end{footnotesize}
\caption{$M^{111}$ Kaluza Klein fields in the ${\cal N}=2$ hypermultiplet with $y_0>0$}
\label{hyper}
\end{table}
%%%%%%%%%%%%%%%%%%%%%%%%%%%%%%%%%%%%%%%%%%%%%%%%%%%%%%%%%%%%%%%%%%%%%%%%%%%%%%%%%%%%%%%%%%%%
\begin{table}
\centering
\begin{footnotesize}
\begin{tabular}{||c|c|c|c|c|c||}
\hline
&Spin &
Energy &
Hypercharge &
Mass ($^2$)&
Name \\
\hline
\hline
$*$&$2$      & $3$        & $0$     & $0$  & $h$     \\
$*$&$\ft32$  & $\ft52$    & $-1$   & $0$  & $\chi^+$\\
$*$&$\ft32$  & $\ft52$    & $+1$   & $0$  & $\chi^+$\\
$*$&$1$      & $2$        & $0$     & $0$  & $A$     \\
\hline
\end{tabular}\\[.13in]
\end{footnotesize}
\caption{$M^{111}$ Kaluza Klein fields in the ${\cal N}=2$ massless graviton multiplet}
\label{masslessgraviton}
\end{table}
%%%%%%%%%%%%%%%%%%%%%%%%%%%%%%%%%%%%%%%%%%%%%%%%%%%%%%%%%%%%%%%%%%%%%%%%%%%%%%%%%%%%%%%%%%
\begin{table}
\centering
\begin{footnotesize}
\begin{tabular}{||c|c|c|c|c|c|c|c||}
\hline
&&&&&&&\\
&Spin &
Energy &
Hypercharge &
Mass ($^2$)&
Name &
Mass ($^2$)&
Name \\
\hline
\hline
$*$&$1$      & $2$      & $0$      & $0$  & $A$
                       & $0$  & $Z$         \\
$*$&$\ft12$  & $\ft32$  & $-1$    & $0$  & $\lambda_L$
                   & $0$  & $\lambda_T$ \\
$*$&$\ft12$  & $\ft32$  & $+1$    & $0$  & $\lambda_L$
                   & $0$  & $\lambda_T$ \\
$*$&$0$      & $2$      & $0$ & $0$  & $\pi$
                   & $0$ & $\phi$      \\
$*$&$0$      & $1$        & $0$      & $0$  & $S$
                   & $0$  & $\pi$   \\
\hline
\end{tabular}\\[.13in]
\end{footnotesize} 
\caption{$M^{111}$ Kaluza Klein fields in the ${\cal N}=2$ massless vector multiplets $A$ and $Z$}
\label{masslessvector}
\end{table}
%%%%%%%%%%%%%%%%%%%%%%%%%%%%%%%%%%%%%%%%%%%%%%%%%%%%%%%%%%%%%%%%%%%%%%%%%%%%%%%%%%%%%%%%%%%%%%
\par
\subsubsection{The graviton multiplet}
As pointed out above, the graviton multiplet is the appropriate multiplet
to start with. In particular
we look at the spin--two graviton field.
The mass of the graviton is given by the eigenvalue of the scalar operator
(see eq.s (\ref{massform}) ):
\begin{equation}
m_h^2 = M_{(0)^3} \equiv H_0 \,.
\end{equation}
Using table \ref{0series} we find that its
harmonics can sit in
 all the $G$ representations of the series
\begin{eqnarray}
A_R^0, A_1^0, A_1^{*0}, A_3^0, A_3^{*0}, A_4^0, A_4^{*0}, A_6^0, A_7^0, A_8^0 \,.
\label{gravitonseries}
\end{eqnarray}
Remember that the superscripts $^0$ mean that the hypercharge is 
$Y={2\over 3}\left(M_2-M_1\right)$.
\par
Using the group--theoretical
information of the long graviton multiplet (see table \ref{longgraviton})
we find the energy and hypercharge $\left(E_0,y_0\right)$ 
of the graviton multiplet
\footnote{Remember that  $E_0, y_0$ denote the energy and
hypercharge of the Clifford vacuum of the multiplet}
\begin{eqnarray}
E_0 &=& \ft14 \sqrt{H_0+36}+\ft12\,\nonumber\\
y_0 &=&  {2\over 3}\left(M_2-M_1\right)\,,
\end{eqnarray}
and using table \ref{longgraviton} we find the energies and
hypercharges of all the fields in the multiplet. In particular, we see that the
gravitinos are in $U(1)_R$ representations  $^+,~^-$, the $A,W$ vectors in
$U(1)_R$ representations $^0$, the $Z$  vectors in $U(1)_R$ representations $^0$, 
$^{++}$, $^{--}$.
From the mass of the graviton we deduce, using the mass relations
(\ref{massrelationchi}),
the masses of the gravitinos and 
vectors present in the graviton multiplet,
\begin{eqnarray}
m_{\chi^\pm} &=& - 6 \pm \sqrt{H_0 + 36} \,,
\label{massgravitini} \\
m_A^2 &=& H_0 + 48 - 8 \sqrt{H_0 + 36} \,,
\nonumber \\
m_W^2 &=& H_0 + 48 + 8 \sqrt{H_0 + 36} \,,
\nonumber \\
m_Z^2 &=& H_0 + 32 \,.
\label{massesbosonsgraviton}
\end{eqnarray}
From equations (\ref{massform}), we predict the presence
of  the eigenvalues
$M_{(1/2)^3}=m_{\chi^\pm}$ for the spinor. Indeed, looking at
(\ref{eigenA+spinor}), we see that the two eigenvalues
$\lambda_1$ and $\lambda_2$
come from spin--$\ft32$ fields that belong to the graviton multiplet.
To find out whether there are some short graviton multiplets
present in the spectrum, we
now use table \ref{eigenvaluesA+spinor}.
The absence of these eigenvalues $\lambda_1$ or
$\lambda_2$ in some of the exceptional series
implies the existence of a short graviton multiplet in that
particular $G'$ series.
Let us look at it more closely. For instance, for $A_2^+$ and $A_5^+$,
there is none of the eigenvalues $\lambda_1$ or $\lambda_2$.
This would imply a graviton multiplet without gravitino fields. But fortunately,
the series
$A_2$ and $A_5$ do not contain representations of $G^\prime$ in which there is a
graviton field, see (\ref{gravitonseries}). Considering the rest of
table \ref{eigenvaluesA+spinor} and also table \ref{eigenvaluesA-spinor},
we find three types of graviton multiplets:
a long graviton multiplet and two types of short graviton multiplets.
The long graviton multiplet
contains four spinors $\chi$: $\chi^{+}$ with hypercharge
$y_0 \pm 1$ and $\chi^{-}$ with hypercharge $y_0 \pm 1$.
They  are found in the $G^\prime$ representations of
$A_R, A_1, A_1^*, A_3, A_3^*, A_6, A_7$.
Then there is a short graviton multiplet in the
series $A_4$ and $A_4^*$. From
tables \ref{eigenvaluesA+spinor} and \ref{eigenvaluesA-spinor},
one sees that they contain the two $\chi^{+}$
with hypercharge $y_0 \pm 1$, but only one $\chi^{-}$,
i.e. for $A_4$ we have one $\chi^{-}$ with $y_0-1$, and for $A_4^*$ we have
one $\chi^{-}$ with $y_0+1$. We also find the massless multiplet
in $A_8$ for which none of the spin--$\ft32$ fields $\chi^{-}$ are present.
\par
At this stage, we know that the spin--$\ft32$ fields that
correspond to the eigenvalues $\lambda_1$ and $\lambda_2$ in
(\ref{eigenA+spinor}) and (\ref{eigenA-spinor}) sit
in the graviton multiplets. However, there are also
spin--$\ft32$ fields that yield the eigenvalues
$\lambda_3$ and $\lambda_4$ in
(\ref{eigenA+spinor}) and (\ref{eigenA-spinor}).
They can only be gravitinos of the gravitino multiplets
in the spectrum.
So now we know the highest components of gravitino multiplets,
their energies, hypercharges and $G'$ representations.
But before we continue with the gravitino multiplet, let us look at the
vectors of the graviton multiplet.
\par
Let us consider $A$ and $W$ first.
We know that, if present, they should be in the series (\ref{gravitonseries}).
Using equations (\ref{massform}) we see that their $M_{(1)(0)^2}$
eigenvalues would then be
\begin{eqnarray}
M_{(1)(0)^2}^A &=& H_0 + 24 + 4 \sqrt{H_0 + 36} \,,
\nonumber \\
M_{(1)(0)^2}^W &=& H_0 + 24 - 4 \sqrt{H_0 + 36} \,.
\end{eqnarray}
Indeed, these eigenvalues are present, namely for
$A$ we find $\lambda_4$ and for
$W$ we find $\lambda_5$
of eq. (\ref{eigenAoneform}).
To determine whether, in the exceptional series, the
vector $A$ or the vector $W$ is present we use table
\ref{eigenvaluesAoneform}. The absence of one of the vectors
will imply shortening of the graviton multiplet.
Studying the spin $3/2$ fields, we have found that there are
long graviton multiplets in the series $A_R, A_1, A_1^*, A_3, A_3^*, A_6, A_7$ 
and short graviton multiplets in the series $A_4, A_4^*$ . This is
confirmed here: in the former 
series both the $A$ and $W$ fields are present, 
in the latter only the field $A$ is present.
For the massless multiplet
of $A_8$ we also see that only the vector $A$ is present.
\par
Let us look at the vector $Z$ in the graviton multiplet.
We know that the $Z$ vectors should be in the same $G'$ representations
of the graviton:
\begin{equation}
A_R, A_1, A_1^*, A_3, A_3^*, A_4, A_4^*, A_6, A_7, A_8
\end{equation}
and that two $Z$ vectors should be in the series $^0$, one in the series $^{++}$ 
and one in the series $^{--}$.
For the operator $M_{(1)^2(0)}$ on the two--form
we predict, using eq.s (\ref{massform}), the presence of the eigenvalue
\begin{eqnarray}
M_{(1)^2(0)}^Z &=& H_0 +32
\,.
\end{eqnarray}
Indeed, it corresponds to $\lambda_{10}$ and $\lambda_{11}$ in (\ref{eigenAtwoform})
for the series $^0$, and $\lambda_4$ in (\ref{eigenBtwoform})
for the series $^{++}$ (and $^{--}$, which are the series of
the conjugate representations of $^{++}$ ($M_2 \leftrightarrow M_1$)).
So we see that for the long graviton multiplets all the vectors $Z$ are present.
Using the fact that
\begin{eqnarray}
B_R \cup B_4 \cup B_5 \cup B_6 \cup B_7 \cup B_8 \cup B_9 \cup B_{10}
&=&
A_R \cup A_1 \cup A_1^* \cup A_3 \cup A_3^* \cup A_4 \cup A_6 \cup A_7,
\nonumber \\
B_R^* \cup B_4^* \cup B_5^* \cup B_6^* \cup B_7^* \cup B_8^* \cup B_9^* \cup B_{10}^*
&=&
A_R \cup A_1 \cup A_1^* \cup A_3 \cup A_3^* \cup A_4^* \cup A_6 \cup A_7,
\label{overlapgraviton}
\nonumber \\
\end{eqnarray}
and tables \ref{eigenvaluesAtwoform} and \ref{eigenvaluesBtwoform}
we find that
for the short graviton multiplets of $A_4$ we have two $Z$'s, one with
hypercharge $y$ and one with hypercharge $y-2$; for the short graviton multiplets
of $A_4^*$ we have two $Z$'s, one with hypercharge $y$ and one with hypercharge $y+2$;
for the massless graviton multiplet we have no vectors $Z$.
\par
To determine which $\lambda_T$ fields and  scalar fields
$\phi$ are present, we use the ${\cal N}=2\rightarrow {\cal N}=1$
decomposition of the multiplets (\ref{N21longgraviton}),$\dots$,(\ref{N21hyper}). We already know where 
$\lambda_T$ and $\phi$ are located in the long graviton multiplet
from table \ref{longgraviton} \cite{multanna}.
From the decomposition of the long $\cN=2$ graviton multiplet (\ref{N21longgraviton}) we see that it
is made of four ${\cal N}=1$ massive multiplets: one graviton, two gravitino
and a vector multiplet. Harmonic analysis teaches us that in the short
graviton multiplet there are three gravitino fields and three vector fields.
The only possible structure of the short graviton multiplet is then the one
displayed in chapter $2$ and in table \ref{shortgraviton}.
\par
The  multiplet that we have found in the representation of
series $A_8$ is in fact the massless graviton multiplet.
In this case the field $A$ becomes the graviphoton.
The final structure of the short graviton multiplet
and the massless graviton multiplet
is displayed in tables \ref{shortgraviton} and 
\ref{masslessgraviton} respectively.
\subsubsection{The gravitino multiplet}
\par
As already previously explained, 
we know the $M_{(1/ 2)^3}$ eigenvalues 
and the $G$ representations of the spin--$\ft32$ in the
gravitino multiplet from the matching of the graviton multiplet. Their masses
are given by equations (\ref{massform}),
\begin{eqnarray}
m_{\chi^{+}}&=& - 8 + \sqrt{H_0 + 16 + \ft{32}{3}(M_2 - M_1)} = \lambda_3
\nonumber \\
m_{\chi^{-}}&=& - 8 - \sqrt{H_0 + 16 + \ft{32}{3}(M_2 - M_1)} = \lambda_4
\label{masseschiC+}
\end{eqnarray}
for series of type $^+$ and
\begin{eqnarray}
m_{\chi^{+}}&=& - 8 + \sqrt{H_0 + 16 - \ft{32}{3}(M_2 - M_1)} = \lambda_3
\nonumber \\
m_{\chi^{-}}&=& - 8 - \sqrt{H_0 + 16 - \ft{32}{3}(M_2 - M_1)} = \lambda_4
\label{gravitinisA-}
\end{eqnarray}
for series of type $^-$. Each of the above four different eigenvalues
gives rise to gravitino multiplets of different types
and/or in different $G'$ representations. Now we look at tables \ref{eigenvaluesA+spinor}
and \ref{eigenvaluesA-spinor} and see that we have gravitino multiplets
for the series $A^\pm_R$ and $A^\pm_1$. We  consider the
gravitino multiplets in the series of type $^+$ only. The
gravitino multiplets in the series of type $^-$ coming from (\ref{gravitinisA-})
can be obtained be taking the conjugates of the gravitino multiplets
in the series of type $^+$.
\par
We start with $\chi^+$ in the series of type $^+$.
The energy and hypercharge $(E_0,~y_0)$ of the gravitino multiplets are given by,
\begin{eqnarray}
E_0 &=& \ft14 \sqrt{H_0 +16 +\ft{32}{3}(M_2 - M_1)} - \ft12 \nonumber\\
y_0  &=& {2\over 3}\left(M_2-M_1\right)-1\,.
\end{eqnarray}
Let us look at the vectors in the gravitino multiplets.
As we know from group theory (see table \ref{longgravitino}) we
should find a vector
with hypercharge $y_0+1$ and energy $E_0+\ft12$, in the series $^0$.
However
group theory does not tell us whether it is the vector
$A$ or the vector $W$. But since we 
know that in series of type $^+$ we have
$m_{\chi^+} \geq -8$, we can use
the mass relations (\ref{massrelationchi}) to derive
\begin{eqnarray}
m_A^2 &=& H_0 +\ft{32}{3}(M_2-M_1) + 48 - 12 \sqrt{H_0 + \ft{32}{3}(M_2-M_1)+ 16}
\end{eqnarray}
or
\begin{eqnarray}
m_W^2 &=& m_\chi^2 + 2 m_\chi + 192 \,.
\end{eqnarray}
We see from table \ref{longgravitino} that it is the $A$ vector 
which is present in the $\chi^+$ gravitino multiplet and not $W$.
Hence, comparing with the formula (\ref{massform})
in order to find $A$, we expect the following eigenvalue
\begin{equation}
M_{(1)(0)^2}^A = H_0 + \ft{32}{3}(M_2 - M_1) \,
\label{lambdaone}
\end{equation}
for the $M_{(1)(0)^2}$ operator. Looking at table \ref{eigenAoneform}we see that it is indeed present: $\lambda_1$.
Looking at table \ref{eigenvaluesAoneform} we see that it appears
in the series $A_R^0, A_1^0, A_2^0, A_3^0, A_4^0, A_5^0$.
We also find a vector $A$ with hypercharge $y_0-1$ in series $^{++}$. Indeed,
using
\begin{equation}
B_R \cup B_1 \cup B_3 \cup B_4 \cup B_6 \cup B_7
=
A_R \cup A_1 \cup A_2 \cup A_3 \cup A_4 \cup A_5 \,,
\end{equation}
we see that (\ref{lambdaone}) is an eigenvalue of the one--form operator
$M_{(1)(0)^2}$ in series $^{++}$ as given in (\ref{eigenBoneform}).
Both the spin--$1$ fields $A$ with $y_0-1$ and $y_0+1$
of the gravitino multiplet for
$\chi^+$ are present and there are no other left with eigenvalue
(\ref{lambdaone}).
For the vector $Z$ sector, we expect the presence of two states
with mass
\begin{equation}
m_Z^2 = H_0 + 16 + \ft{32}{3} (M_2 - M_1)
         -4 \sqrt{H_0 +16 +\ft{32}{3}(M_2 - M_1)},
\label{massz32}
\end{equation}
one in the $G$ representations of type $^0$, the other in the 
representations $^{++}$ or $^{--}$ (depending on the $G$ representation
of the gravitino).
The mass (\ref{massz32}) 
corresponds to $\lambda_7$ in (\ref{eigenAtwoform}) and $\lambda_3$
in (\ref{eigenBtwoform}). From this we see that $Z$ is present except for
series $A_5$, and series $B_7$. The series $A_5$ and $B_7$ have no overlap.
So we conclude that we have long gravitino multiplets except if the
multiplet sits in a representation of $A_5$ or $B_7$.
For the gravitino multiplet with $\chi^+$ in the series $^+$, we now look
at the mass of the scalar $\pi$,
\begin{equation}
m_\pi^2 = 16 \, (\ft14\sqrt{H_0+16+\ft{32}{3}}-1) (\ft14\sqrt{H_0+16+\ft{32}{3}}-2)\,.
\end{equation}
From eq.s (\ref{massform}) we predict the eigenvalue
\begin{eqnarray}
M_{(1)^3}^\pi = \ft14\sqrt{H_0+16+\ft{32}{3}(M_2-M_1)}
\end{eqnarray}
which we do find as $\lambda_1$ in (\ref{eigenAthreeform})
in series $A_R^0, A_1^0, A_2^0, A_3^0, A_4^0$
(see (\ref{eigenvaluesAthreeform}))
and  as  $\lambda_1$ in (\ref{eigenBthreeform}) in the series
$B_R^{++}, B_1^{++}, $ $ B_3^{++}, B_4^{++}, B_6^{++}$ (see
(\ref{eigenvaluesBthreeform})).
So none of the fields $\pi$ with $y_0-1$ and $y_0+1$ is
present in the short gravitino multiplets with $\chi^+$ in the series of type
$^+$ .
Let us now consider the spin--$\ft12$ field $\lambda_L^+$.
Looking at the expansion (\ref{kkexpansion}),
we see that $\lambda_L$ appears in the expansion of the spinor.
So we can check whether it is present in the
gravitino multiplet with $\chi^+$ in the series $^+$.
Its mass is (\ref{massform})
\begin{eqnarray}
m_{\lambda_L^+}=-8+\sqrt{H_0+16+\ft{32}{3}(M_2-M_1)}\,,
\end{eqnarray}
so, from eq.s (\ref{massform}) we expect the
eigenvalue
\begin{eqnarray}
M_{(1/2)^3}^{\lambda_L^+}=-8-\sqrt{H_0+16+\ft{32}{3}(M_2-M_1)}
\end{eqnarray}
which we do find  as $\lambda_4$ in (\ref{eigenA+spinor})
in $A_R^+, A_1^+, A_2^+, A_3^+, A_4^+,  A_5^+$ (see
(\ref{eigenvaluesA+spinor})).
So the field $\lambda_L^+$ is present in both long and short
gravitino multiplets with hypercharge $y_0$. In fact
it has to be there since it provides the Clifford vacuum
of the representation.
For the short gravitino multiplets
we have found which of  the fields $\phi$ and $\lambda_T$ are 
present by using the ${\cal N}=2\rightarrow {\cal N}=1$
decomposition (\ref{N21shortgravitino}) and 
by calculating the norms of the states (see chapter $2$).
The result is displayed in table \ref{shortgravitino}.
\par
Let us consider $\chi^-$ for the series of type $^+$.
It has mass $m_{\chi^-}$ from (\ref{masseschiC+}).
The energy and hypercharge $(E_0,~y_0)$ of the multiplet are
\begin{eqnarray}
E_0 &=& \ft14\sqrt{H_0+16+\ft{32}{3}(M_2-M_1)}+\ft32\,.\nonumber\\
y_0 &=& {2\over 3}\left(M_2-M_1\right)-1\,.
\end{eqnarray}
We now have $m_{\chi}\leq -8$.
So, using the mass relations for $W$ we find
\begin{eqnarray}
m_W^2 &=& H_0 +\ft{32}{3}(M_2-M_1) + 48 + 12 \sqrt{H_0 + \ft{32}{3}(M_2-M_1)+ 16}\,.
\end{eqnarray}
Thus in this case it is $W$ that is present and not $A$.
We find the same eigenvalue (\ref{lambdaone}), so we conclude that $W$
is present in all types of gravitino multiplets with $\chi^-$ in series of type $^+$.
For $Z$ we have
\begin{eqnarray}
m_Z^2 = H_0 +16 +\ft{32}{3}(M_2-M_1)+4\sqrt{H_0+16+\ft{32}{3}(M_2-M_1)}
\end{eqnarray}
which, according to eq.s (\ref{massform}),
 has to be an eigenvalue of the two--form mass operator. Indeed,
for series of type $^0$ it corresponds to $\lambda_6$, which is present in series
$A_R, A_1, A_2, A_3, A_4, A_5$ (see (\ref{eigenvaluesAtwoform})).
Notice that these  are the same series of representations as the ones
in which we found $\chi^+$. For the series $^{++}$ we find $\lambda_2$, which is
present in the series $B_R, B_1, B_3, B_4, B_6, B_7$
(see (\ref{eigenvaluesBtwoform})), which are again the same series
of representations
as for $\chi^+$. The fields $\pi$ present have mass,
\begin{equation}
m_\pi^2 = 16 \, (-\ft14\sqrt{H_0+16+\ft{32}{3}(M_2-M_1)}-1)
(-\ft14\sqrt{H_0+16+\ft{32}{3}(M_2-M_1)}-2)\,.
\end{equation}
So we predict the eigenvalue
\begin{equation}
M_{(1)^3}^{\pi}= - \ft14 \sqrt{H_0+16+\ft{32}{3}(M_2-M_1)}
\end{equation}
Indeed it is $\lambda_3$ in (\ref{eigenAthreeform}),
present in the series
$A_R, A_1, A_2, A_3, A_4, A_5$
(\ref{eigenvaluesAthreeform}) and
$\lambda_2$ in (\ref{eigenvaluesBthreeform}), present
in the series $B_R, B_1, B_3, B_4,$ $ B_6, B_7$
(\ref{eigenBthreeform}).
We conclude that all the gravitino multiplets with $\chi^-$
are long gravitino multiplets.
\subsubsection{The vector multiplet}
\par
What are the vector field we have been left with?
They have to be the highest components of the vector multiplets.
Well, we have a multiplet with highest component vector $A$ with
eigenvalue $\lambda_5$ in (\ref{eigenAoneform}). We have
a vector multiplet with highest vector component $W$ with
eigenvalue $\lambda_4$ in (\ref{eigenAoneform}). We have
some vector multiplets with highest vector component $Z$
with eigenvalues $\lambda_3$ in (\ref{eigenAtwoform}),
$\lambda_5$ in (\ref{eigenBtwoform}) and
$\lambda_5^*$ in the series $^{--}$. All  these eigenvalues give
rise to the existence of different types of
vector multiplets in different representations of $G^\prime$.
\par
Let us start with $A$. We call this the $A$--vector multiplet.
It has eigenvalue $\lambda_5$ in (\ref{eigenAoneform}).
Its energy and hypercharge are
\begin{eqnarray}
E_0 &=& \ft14 \sqrt{H_0 + 36} - \ft32 \,\nonumber\\
y_0 &= & {2\over 3}\left(M_2-M_1\right)
\end{eqnarray}
and the mass of the field component $A$ is
\begin{equation}
m_A^2 = H_0 + 96 - 16\,\sqrt{H_0 + 36} \,.
\label{massAvector}
\end{equation}
This eigenvalue is present in  the series
$A_R^0, A_1^0, A_1^{*0}, A_3^0, A_3^{*0}, A_6^0, A_7^0$. We now figure
out for which of these there is shortening. From the table \ref{longvector} we see that
$\pi$ has the same mass as $A$ (\ref{massAvector}), and using
eq.s (\ref{massform}) we conclude that we should find the eigenvalues
\begin{eqnarray}
M_{(1)^3}^\pi = \ft14 \sqrt{H_0 + 36} -\ft12 \,,
\end{eqnarray}
which is present: $\lambda_5$ in
$A_R^0, A_1^0, A_1^{0*}, A_3^0, A_3^{0*}, A_6^0, A_7^0$
(\ref{eigenAthreeform}) (\ref{eigenvaluesAthreeform}).
It is also present as $\lambda_3$ in
$B_R^{++}, B_4^{++}, B_5^{++}, B_6^{++}, B_7^{++}, B_9^{++}$
(\ref{eigenBthreeform}) (\ref{eigenvaluesBthreeform}). Considering
(\ref{overlapgraviton}) this seems strange at
first sight. However, what happens is that here we discover a
scalar $\pi$ in the series $A_4$ of a
hypermultiplet. We can see this as follows.
Suppose the eigenvalue were also present in series $B_8$
and series $B_{10}$. Then the eigenvalue $\lambda_3$ would
appear in the representations of $B$ that are on the right--hand side
of (\ref{overlapgraviton}). So we would find the field $\pi$ in the
$G^\prime$ representations
$A_R, A_1, A_1^*, A_3, A_3^*, A_6, A_7$ and in $A_4$, with $Y=\ft23(M_2-M_1)-2$.
The series $A_4$ and $B_8$ and $B_{10}$ have no overlap.
Consequently,
the $\pi$ in $A_4$ can not belong to the $A$--vector multiplet and
thus has to be a scalar of a hypermultiplet. Similarly, we find $\pi$
in $B_R^{--*}, B_4^{--*} B_5^{--*}, B_6^{--*},$ $B_7^{--*}, 
B_8^{--*}, B_9^{--*}, B_{10}^{--*}$. With the
same reasoning, we conclude that $\pi$ in $A_4^*$ with $Y=\ft23(M_2-M_1)+2$
has to be a scalar of some hypermultiplet.
However, $\lambda_3$ does not sit in the series $B_8, B_8^*, B_{10}, B_{10}^*$.
So we conclude that we get shortening in these series. Now we
get different types of short vector multiplets. This is due to fact the
$B_8$ and $B_8^*$ have overlap, namely if $M_1=M_2=1, J=0$ and
that also $B_{10}$ and $B_{10}^*$ have overlap, namely
for the representation $M_1=M_2=0, J=1$.
For the representations in the series $B_8$ and $B_{10}$
with $M_1 > M_2 = 1$,
we find that the field $\pi$ with hypercharge $y-2$
in the long vector multiplet decouples. The representations
\begin{eqnarray}
&M_1=M_2, &J=1\nonumber\\
&M_1=M_2=1, &J=0\label{shvecreps}
\end{eqnarray}
yield massless  vector multiplets.
They contain the vectors that gauge $SU(2)$ and $SU(3)$
respectively.
\par
Let us now figure out whether we can learn something about the
presence of $\phi$, $S$ and $\Sigma$ in the $A$--vector multiplet. The table \ref{hyper} 
gives the mass,
\begin{eqnarray}
m_{\phi,S/\Sigma}^2 = 16 \, E_0 (E_0 + 1) = H_0 + 48 - 4 \sqrt{H_0 + 36} \,.
\end{eqnarray}
Looking at eq.s (\ref{massform}), we see that the entry in the table can not be
$S$ or $\Sigma$, but has to be $\phi$. If we look at the other
$\phi, S/\Sigma$ in the table with  mass
\begin{eqnarray}
m_{\phi,S/\Sigma}^2 = 16 \, (E_0 - 2)(E_0 - 1)
                    = H_0 + 176 - 24 \sqrt{H_0 + 36} \,,
\end{eqnarray}
we see that it is the mass for the field $S$.
So at this place in the table we
find the field $S$. The field $S$ is found in the series
$A_R^0, A_1^0, A_1^{0*}, A_3^0, A_3^{0*}, 
A_4^0, A_4^{0*}, A_6^0, A_7^0, A_8^0$. So
it is always present in the $A$--vector multiplets. Besides, we
get some extra $S$--fields that are to be put in the hypermultiplets
in the series $A_4, A_4^*, A_8$.
\par
To conclude the discussion of the $A$ vector multiplet, there is
shortening of $A$--vector multiplets in series
$B_8, B_{8}^*$ and $B_{10}, B_{10}^*$.
In the representation (\ref{shvecreps}) there are massless vector
multiplets, in the other $B_8,B_8^*,B_{10},B_{10}^*$ representations
there are short vector multiplets.
The $\phi$ and $\lambda_T$ contents of the short vector multiplets
can be determined by using the ${\cal N}=2\rightarrow {\cal N}=1$
decomposition (\ref{N21shortvector}).
The structure of the long vector multiplet and
the short vector multiplet
is displayed in table \ref{longvector} and \ref{shortvector}
respectively.
\par
Let us now consider the vector multiplet with highest vector component $W$.
We will call this the $W$-- vector multiplet.
We expect eigenvalue $\lambda_5$
in (\ref{eigenAoneform}) and (\ref{eigenvaluesAoneform}),
which we find in series
$A_R^0, A_1^0, A_1^{0*}, A_3^0, A_3^{0*},
A_4^0, A_4^{0*}, A_6^0, A_7^0, A_8^0$.
This multiplet has energy and hypercharge,
\begin{eqnarray}
E_0 &=& \ft14 \sqrt{H_0 + 36} +\ft52 \,,\nonumber \\
y_0 &=& {2\over 3}\left(M_2-M_1\right),
\end{eqnarray}
the $W$ field has mass
\begin{equation}
m_W^2 = H_0 + 96 + 16 \sqrt{H_0 +36} \,.
\end{equation}
For the fields $\pi$, we expect to find the eigenvalues $\lambda_6$ in
series $A_R^0$, $A_1^0$, $A_1^{*0}$, $A_3^0$, $A_3^{*0}$, $A_4^0$, 
$A_4^{*0}$, $A_6^0$, $A_7^0$, $A_8^0$ 
(\ref{eigenAthreeform}), (\ref{eigenvaluesAthreeform}), and
$\lambda_4$ in series
$B_R^{++}$, $B_4^{++}$, $B_5^{++}$, $B_6^{++}$, $B_7^{++}$, $B_8^{++}$, 
$B_9^{++}$, $B_{10}^{++}$, $B_{11}^{++}$ 
(\ref{eigenBthreeform}), (\ref{eigenvaluesBthreeform}),
and $\lambda_4^*$ in series
$B_R^{--}$, $B_4^{--}$, $B_5^{--}$, $B_6^{--}$, $B_7^{--}$, $B_8^{--}$, 
$B_9^{--}$, $B_{10}^{--}$, $B_{11}^{--}$. 
Using
\begin{eqnarray}
B_{11} &=& A_4 \cup A_8 \,, \nonumber \\
B_{11}^* &=& A_4^* \cup A_8 \,,
\end{eqnarray}
and (\ref{overlapgraviton}), we see that all these $^0,~^{++},$ and $^{--}$ series coincide.
Thus all the
fields $\pi$ in the table of \cite{multanna} are
always present and we find no fields $\pi$ that have to be put in
other multiplets. So the $W$--vector multiplet is always long.
Which of the fields $\phi, S/\Sigma$ are present? Let us look at
$\phi, S/\Sigma$ with mass
\begin{eqnarray}
m_{\phi, S/\Sigma}^2 = 16 \, E_0 (E_0 + 1) = H_0 + 176 + 24 \sqrt{H_0 + 36}
\,.
\end{eqnarray}
From eq.s (\ref{massform}) we see
that it is the field $\Sigma$ that is  present in the series
$A_R^0, A_1^0, A_1^{0*},$ $ A_3^0, A_3^{0*}, A_4^0,$ $ A_4^{0*},
A_6^0,$ $ A_7^0, A_8^0$.
So this confirms that there is no shortening and we do not
find any extra fields $\Sigma$ that are to be put in the
hypermultiplets.
Let us look at
$\phi, S/\Sigma$ with mass
\begin{eqnarray}
m_{\phi, S/\Sigma}^2 = 16 \, (E_0 - 2) (E_0 - 1) = H_0 + 48 + 8 \sqrt{H_0 + 36}
\,.
\end{eqnarray}
This can only be the field $\phi$.
So we conclude that the $W$--vector multiplets are always long vector multiplets.
And there are no scalar left that have to be put in hypermultiplets.
Its structure is displayed in table \ref{longvector}.
\par
Let us now look at the $Z$--vector multiplet with eigenvalue $\lambda_3$
in series $A_R$, $A_1$, $A_1^*$, $A_6$, $A_8$ (\ref{eigenAtwoform})
(\ref{eigenvaluesAtwoform}). 
The multiplet has energy and hypercharge
\begin{eqnarray}
E_0 &=& \ft14\sqrt{H_0 + 4} + \ft12 \,,\nonumber \\
y_0 &=& {2\over 3}\left(M_2-M_1\right),
\end{eqnarray}
the field $Z$ has mass
\begin{equation}
m_Z^2 = H_0 \,.
\end{equation}
What about the two fields $\pi$? Let us look at $\pi$ with mass
\begin{eqnarray}
m_\pi^2 = 16\, E_0 (E_0 + 1)
        = H_0 + 16 +  \sqrt{H_0 + 4}\,.
\end{eqnarray}
From eq.s (\ref{massform}) we expect there to be
$\lambda_7$ in (\ref{eigenAthreeform}). Indeed, it is present in series
$A_R^0, A_1^0, A_1^{*0}, A_6^0$. So we get shortening in the singlet representation 
$A_8$. For $\pi$ with mass
\begin{eqnarray}
m_\pi^2 = 16\, (E_0 -2)(E_0 - 1) \,,
\end{eqnarray}
we find $\lambda_8$ in series $A_R,^0 A_1^0, A_1^{0*}, A_6^0, A_8^0$.
So finally, we conclude that for this type of $Z$--vector multiplet
(with $\lambda_3$ in (\ref{eigenAtwoform})) there is
shortening in series $A_8$, which yields the
massless Betti multiplet.
The structure of the long $Z$--vector multiplet and the
massless Betti multiplet is displayed in tables \ref{longvector}
and \ref{masslessvector} respectively.
\par
Let us now look at the $Z$--vector multiplet with $\lambda_5$ in
(\ref{eigenBtwoform}). It appears in series $B_R, B_1, B_2$
(\ref{eigenvaluesBtwoform}). 
The multiplet has energy and hypercharge
\begin{eqnarray}
E_0 &=& \ft14 \sqrt{H_0 + \ft{64}{3}(M_2- M_1) -28} +\ft12\, \nonumber\\
y_0 &=& {2\over 3}\left(M_2-M_1\right)-2,
\end{eqnarray}
the field $Z$ has mass
\begin{equation}
m_Z^2 = H_0 + \ft{64}{3}(M_2-M_1) -32 \,.
\end{equation}
What about the presence of the fields $\pi$?
For $\pi$ with mass
\begin{eqnarray}
m_\pi^2 = 16 \, (E_0-2)(E_0-1) \,,
\end{eqnarray}
we expect the eigenvalue $\lambda_5$ in (\ref{eigenBthreeform}), which
is found in the series $B_R^{++}, B_1^{++}, B_2^{++}$ (\ref{eigenvaluesBthreeform}).
For $\pi$ with mass
\begin{eqnarray}
m_\pi^2 = 16 \, E_0(E_0+1) \,,
\end{eqnarray}
we expect $\lambda_6$
in (\ref{eigenBthreeform}), which
is found in the series $B_R^{++}, B_1^{++}, B_2^{++}$ (\ref{eigenvaluesBthreeform}).
So we conclude that for the $Z$--vector multiplet (with vector $Z$ with
eigenvalue $\lambda_5$ in (\ref{eigenBtwoform})), there is never
shortening. We do not find extra scalars that are to be put
in hypermultiplets either.
The structure of this long $Z$ vector multiplet
is displayed in table \ref{longvector}.
\par
For the $Z$--vector multiplet with $\lambda_5^*$ in series
$B_R^*, B_1^*, B_2^*$, one just takes the conjugate of the previous results.
\subsubsection{The hypermultiplet}
After having put the scalars $\pi$ in the right places in the
graviton, the gravitino and the vector multiplet, we are only left with scalars
$\pi$ in series $A_4^0$ and $A_4^{0*}$ and $S$ in series $A_4,^0 A_4^{0*}, A_8^0$.
\par
So for each representation of $A_4$ we find a hypermultiplet
with energy
\begin{eqnarray}
E_0 = \ft14 \sqrt{H_0 + 36} -\ft32
\end{eqnarray}
containing
the field $\pi$ with hypercharge $Y=\ft23(M_2-M_1)-2$ and mass
\begin{eqnarray}
m_\pi^2= H_0 + 96 - 16\,\sqrt{H_0+36}
\end{eqnarray}
and the field $S$ with $Y=\ft23(M_2 - M_1)$ and mass
\begin{eqnarray}
m_S^2 = H_0 +176 - 24\, \sqrt{H_0+36}\,.
\end{eqnarray}
The scalars
of this hypermultiplet are complete if we add the scalars $\pi$ and $S$
of $A_4^*$, which are in fact the complex conjugates of the scalars in
$A_4$. From the eigenvalues of the operator $M_{(1/ 2)^3}$
we find the $\lambda_L$ necessary to fill all the hypermultiplets.
The structure of the hypermultiplets is displayed in
the table \ref{hyper}. 
\par
In order to correctly match the fields with the multiplets, it is
important to note that in the singlet $G$ representation $M_1=M_2=J=Y=0$
the scalar $S$ is absent. This is due to the fact that, from the Kaluza Klein expansion 
(\ref{kkexpansion}) of
the eleven-dimensional field $h_{mn}\left(x,y\right)$, the scalar $S$ 
appears in the expressions $(6-\sqrt{M_{(0)^3}+36})S^I\left(x\right)$ and
${\cal D}_{(m}{\cal D}_{n)}(2+\sqrt{M_{(0)^3}+36})S^I\left(x\right)$. 
The coefficient of the former, $6-\sqrt{M_{(0)^3}+36}$, disappears
in the singlet representation. The latter become a pure gauge term,
due to the freedom of coordinate reparametrization, being the graviton
in the singlet $G$ representation the massless graviton.
\vskip 1cm
At this point we have done the complete matching of the multiplets 
with the spectrum of Laplace Beltrami operators. It is reassuring that all the fields
we have found have been organized in ${\cal N}=2~AdS_4$ multiplets.
An important result is that we have established the existence of short multiplets.
From the expressions of the energies and hypercharges
$(E_0,y_0)$ we have found, we can easily derive that what we expect on
unitarity bounds and shortening conditions is confirmed:
\begin{itemize}
\item for all the long multiplets
\[
E_0>\left|y_0\right|+s_0+1
\]
\item for all the short graviton, gravitino and vector multiplets
\[
E_0=\left|y_0\right|+s_0+1
\]
\item for all the hypermultiplets
\[
E_0=\left|y_0\right|\ge {1\over 2}
\]
\item for all the massless multiplets
\[
E_0=s_0+1~~y_0=0.
\]
\end{itemize}
\par
\section{The mass spectra of $AdS_4\times N^{010}$ and $AdS_4\times Q^{111}$ supergravities}
\par
\subsection{$N^{010}$}
\par
Here I do not review the harmonic analysis on $N^{010}$, worked out in
\cite{piet} (see also, for the geometry, \cite{newcast}).
I simply give the result, namely, the spectrum of $Osp\ll(3\vert 4\rr)$ multiplets \cite{noi4}.
In this case we have
\be
G=SU\ll(3\rr)\times SU\ll(2\rr)=G'\times SU\ll(2\rr)
\ee
where $SU\ll(2\rr)$ is the $R$--symmetry. The $\cN=3$ supermultiplets are then organized in $SU\ll(3\rr)$ UIRs,
which I denote as usual with the Young labels $M_1,M_2$, while $J$ denotes the isospin (see chapter $2$)
of a field in a supermultiplet.
\par
\subsubsection{Long multiplets} 
There are long multiplets for the following $SU\left(3\right)$ representations: 
\begin{equation} 
\cases{M_1=k~~~~~~~~k\ge 0\cr M_2=k+3j~~j\ge 0\cr} 
\end{equation} 
$k,j$ integers. 
\begin{itemize} 
\item For every $SU\left(3\right)$ representation with $k\ge 0,~j\ge 2$ there is only 
one of the following multiplets, that are long: 
\begin{equation} 
\begin{array}{|c|c|c|} 
\hline 
{\rm multiplet} & {isospin} & {\rm energy} \\ 
\hline 
SD\left(E_0,2,J_0\right) & j\le J_0\le k+j & E_0={1\over 4}\sqrt{H_0+36} \\ 
\hline 
SD\left(E_0,3/2,J_0\right) & j\le J_0\le k+j & E_0={1\over 4}\sqrt{H_0+36}-{3\over 2} \\ 
\hline 
SD\left(E_0,3/2,J_0\right) & j\le J_0\le k+j & E_0={1\over 4}\sqrt{H_0+36}+{3\over 2} \\ 
\hline 
\end{array} 
\end{equation} 
\item For every $SU\left(3\right)$ representation with $k\ge 0,~j=1$ there is only 
one of the following multiplets, that are long: 
\begin{equation} 
\begin{array}{|c|c|c|} 
\hline 
{\rm multiplet} & {isospin} & {\rm energy} \\ 
\hline 
SD\left(E_0,2,J_0\right) & 1\le J_0\le k+1 & E_0={1\over 4}\sqrt{H_0+36} \\ 
\hline 
SD\left(E_0,3/2,J_0\right) & 1\le J_0< k+1 & E_0={1\over 4}\sqrt{H_0+36}-{3\over 2} \\ 
\hline 
SD\left(E_0,3/2,J_0\right) &1\le J_0\le k+1 & E_0={1\over 4}\sqrt{H_0+36}+{3\over 2} \\ 
\hline 
\end{array} 
\end{equation} 
\item For every $SU\left(3\right)$ representation with $k\ge 0,~j=0$ there is only 
one of the following multiplets, that are long: 
\begin{equation} 
\begin{array}{|c|c|c|} 
\hline 
{\rm multiplet} & {isospin} & {\rm energy} \\ 
\hline 
SD\left(E_0,2,J_0\right) & 0\le J_0< k & E_0={1\over 4}\sqrt{H_0+36} \\ 
\hline 
SD\left(E_0,3/2,J_0\right) & 0\le J_0< k & E_0={1\over 4}\sqrt{H_0+36}-{3\over 2} \\ 
\hline 
SD\left(E_0,3/2,J_0\right) &0\le J_0\le k & E_0={1\over 4}\sqrt{H_0+36}+{3\over 2} \\ 
\hline 
\end{array} 
\end{equation} 
\end{itemize} 
\subsubsection{Short multiplets} 
There are the following short multiplets in the following $SU\left(3\right)$ 
 representations: 
\begin{itemize} 
\item There is only one massive short graviton multiplet $SD\left(J_0+3/2,2,J_0\right)$ 
in each of the representations: 
\begin{equation} 
M_1=k,~~M_2=k,~~k\ge 1. 
\end{equation} 
It has 
\begin{equation} 
E_0=k+3/2,~J_0=k. 
\end{equation} 
\item There is only one massive short gravitino multiplet $SD\left(J_0+1,3/2,J_0\right)$ 
in each of the representations: 
\begin{equation} 
M_1=k,~~M_2=k+3,~~k\ge 0. 
\end{equation} 
It has 
\begin{equation} 
E_0=k+2,~J_0=k+1. 
\end{equation} 
\item There is only one massive short vector multiplet $SD\left(J_0,1,J_0\right)$ 
in each of the representations: 
\begin{equation} 
M_1=k,~~M_2=k,~~k\ge 2. 
\end{equation} 
It has 
\begin{equation} 
E_0=k,~J_0=k. 
\end{equation} 
\end{itemize} 
\subsubsection{Massless multiplets} 
The massless sector of the theory is composed by the following multiplets.
\begin{itemize} 
\item There is  one massless graviton multiplet in the representation: 
\begin{equation} 
M_1=M_2=0\,.
\end{equation} 
It has 
\begin{equation} 
E_0=3/2,~J_0=0. 
\end{equation} 
This multiplet  has the standard field content expected for the 
$\mathcal{N}=3$ supergravity multiplet in four--dimensions, 
namely  one massless graviton, three massless gravitinos 
that gauge    ${\cal N}=3$ supersymmetry,  three 
massless vector fields (organized in a $J_0=1$ adjoint representation of $SO(3)_R$) 
that gauge the  $R$-symmetry and one spin one--half field. 
\item There is one massless vector multiplet in each of the representations: 
\begin{eqnarray} 
M_1=M_2=1\label{killing}\\ 
M_1=M_2=0\label{betti}\,.
\end{eqnarray} 
They have: 
\begin{equation} 
E_0=1,~J_0=1. 
\end{equation} 
The multiplet (\ref{killing}) contains the gauge vectors of the 
$SU\left(3\right)$ isometry. 
The multiplet (\ref{betti}) is the Betti multiplet \cite{spec321}, 
related to the non--trivial cohomology of $N^{010}$ in degree two. 
\end{itemize} 
It is worth noting that before the harmonic analysis on $N^{010}$ was performed, the complete
structure of short $\cN=3$ supermultiplets was not known (only the short vector multiplet structure
had been derived \cite{frenico}). As in the $M^{111}$ case, the
partial knowledge of $Osp\ll(3\vert 4\rr)$ UIRs joined with harmonic analysis of part of the
operators (\ref{scalarop}),$\dots$,(\ref{raritaschwingerop}) yielded both the complete spectrum
of $AdS_4\times N^{010}$ supergravity given above and the complete structure of $\cN=3$ UIRs given
in chapter $2$.
\par
\subsection{$Q^{111}$}
\par
The harmonic analysis of 
\beq
Q^{111}&=&{SU\ll(2\rr)\times SU\ll(2\rr)\times SU\ll(2\rr)\over U\ll(1\rr)\times U\ll(1\rr)}\nn\\
&=&{SU\ll(2\rr)\times SU\ll(2\rr)\times SU\ll(2\rr)\times U\ll(1\rr)\over U\ll(1\rr)\times U\ll(1\rr)\times U\ll(1\rr)}
\eeq
has not been carried out yet (at the moment it is work in progress \cite{N010Q111sp}).
So the complete spectrum of supergravity on $AdS_4\times Q^{111}$ is not known at the moment. Nevertheless the
spectrum of the scalar operator (\ref{scalarop}), namely the laplacian $\cD^a\cD_a$, has been found long ago by C. Pope
\cite{popelast}, not by harmonic analysis but by means of explicit resolution of the differential equations.
\par
First of all an explicit coordinatization of the manifold $Q^{111}$ has been found, by noting that $Q^{111}$ is an $U\ll(1\rr)$ fiber 
bundle on $S^2\times S^2\times S^2$. In the next chapter I will go into detail of this coordinate description, for $Q^{111}$ and $M^{111}$. 
\par
Then, in terms of this coordinate system, the eigenvalues of
\be
\cD^a\cD_a\cH\ll(y\rr)=M_{\ll(0\rr)^3}\cH\ll(y\rr)
\ee
have been found. Here I only give the result of this calculation, which will be an useful hint for the construction of the dual
conformal theory in the next chapter.
\par
I  remind that the labels of the $G'=\ll(SU\ll(2\rr)\rr)^3$ UIRs are given by the three $SU\ll(2\rr)$-spins
\be
\ll[J^{\ll(1\rr)},J^{\ll(2\rr)},J^{\ll(3\rr)}\rr].
\label{3spins}
\ee
For each value of the three labels (\ref{3spins}), integer of half integer (differently from the case of $M^{111}$, where only
integer $SU\ll(2\rr)$ spins were allowed), there is an eigenvalue of the scalar operator, which is
\be
M_{\ll(0\rr)^3}=32\ll(J^{\ll(1\rr)}\ll(J^{\ll(1\rr)}+1\rr)+J^{\ll(2\rr)}\ll(J^{\ll(2\rr)}+1\rr)+J^{\ll(3\rr)}\ll(J^{\ll(3\rr)}+1\rr)\rr).
\ee
Then, looking at the mass formula (\ref{massform}), we find that for every $G'$ representation $\ll[J^{\ll(1\rr)},J^{\ll(2\rr)},J^{\ll(3\rr)}\rr]$
there are the following $AdS_4$ fields:
\begin{itemize}
\item one graviton field $h_{mn}\ll(x\rr)$ with mass squared $m_h^2=M_{\ll(0\rr)^3}$;
\item one scalar field $S\ll(x\rr)$ with mass squared $m_{\Sigma}^2=M_{\left(0\right)^3} +176+24\sqrt{ M_{\left(0\right)^3}+36}$;
\item one scalar field $\Sigma\ll(x\rr)$ with mass squared 
$m_S^2=M_{\left(0\right)^3} +176-24\sqrt{ M_{\left(0\right)^3}+36}$;
\end{itemize}
but we do not know anything about the fields $\phi\ll(x\rr)$, 
$\pi\ll(x\rr)$, $W\ll(x\rr)$, $A\ll(x\rr)$, $Z\ll(x\rr)$, 
$\lambda_L\ll(x\rr)$, $\lambda_T\ll(x\rr)$, $\chi\ll(x\rr)$.
\par
Notice, however, that from the table \ref{hyper}, found by studying the $M^{111}$ spectrum but having a more general validity,
we see that every hypermultiplet (and then chiral superfield) has a field $S$ as lowest energy field; it is then reasonable
that the chiral superfields (which, as we will see in next chapter, are the fundamental degrees of freedom of the conformal
theory on the boundary) are in the flavour representations 
\be
J^{\ll(1\rr)}=J^{\ll(2\rr)}=J^{\ll(3\rr)}=k/2~~~k\in\ZZ.
\ee
\newpage
$~$
\newpage
\par
\chapter{Superconformal field theories dual to $AdS_4\times\left(G\over H\right)_7$ supergravities}
\par
The purpose of this chapter is to determine the conformal theory on
a collection of M2-branes sitting at the singular point
of the cone (\ref{cone}) ${\cal C}(X_7)$ (here named {\it conifold}), 
where $X_7=Q^{111}$ or $X_7=M^{111}$
\footnote{The construction of the SCFT for $X_7=N^{010}$ is in preparation \cite{N010Q111sp}.}. Such a theory 
is dual, by $AdS/CFT$ correspondence, to the supergravities on
$AdS_4\times Q^{111}$ and $AdS_4\times M^{111}$, which have been studied in chapter $3$.
If we find such a theory, this would be a strong check to $AdS/CFT$ correspondence.
\par
While for branes sitting at orbifold singularities there is a straightforward
method for identifying the conformal theory living on the world-volume
\cite{douglasmoore}, \cite{maldasonn}, for
conifold singularities much less is known \cite{uranga}, \cite{morpless}. The strategy
of describing the conifold as a deformation of an orbifold singularity
used in \cite{witkleb}, \cite{morpless} and identifying the superconformal theory
as the IR limit of the deformed orbifold theory, seems more difficult to be applied in
three dimensions \footnote{See however \cite{tatar} where a similar approach
for $Q^{111}$ was attempted without, however, providing a match with Kaluza
Klein spectra. Another attempt in this direction was also given in
\cite{dallagat}.}. We will then use the intuition
from geometry in order to identify the fundamental degrees of freedom of the
superconformal theory and to compare them with the results of the KK expansion.
\par
We expect to find the superconformal fixed points dual to
$AdS$-compactifications as the IR limits of three-dimensional gauge theories.
In the maximally supersymmetric case $AdS_4\times S^7$, for example,
the superconformal theory is the IR limit of the ${\cal N}=8$
supersymmetric gauge theory \cite{Maldyads}. In three dimensions, the
gauge coupling constant is dimensionful and a gauge theory is certainly not
conformal. However, the theory becomes conformal in the IR, where the
coupling constant blows up. In this simple case, the identification of
the superconformal theory living on the world-volume of the M2-branes
follows from considering M-theory on a circle. The M2-branes
become D2-branes in type IIA, whose world-volume supports the  ${\cal N}=8$
gauge theory with a dimensionful coupling constant related to the radius
of the circle. The near horizon geometry of D2-branes is not anymore
AdS \cite{maldasonn}, since the theory is not conformal. The AdS
background and conformal invariance is recovered by sending the radius to
infinity; this corresponds to sending the gauge theory coupling to infinity
and probing the IR of the gauge theory.
\par
We expect a similar behaviour for other three dimensional gauge theories.
As a difference with four--dimensional CFT's corresponding to $AdS_5$
backgrounds, which always have exact marginal directions labeled
by the coupling constants (the type IIB dilaton is a free parameter of
the supergravity solution), these three dimensional fixed points
may also be isolated. The only universal parameter in  M-theory
compactifications is $ l_p$, which is related to the number of colours
$N$, that is also the number of M2-branes (see chapter $1$). The
$1/N$ expansion in the gauge theory corresponds to the $R_{AdS}/ l_p$
expansion of M-theory through the relation $R_{AdS}/ l_p\sim N^{1/6}$
(\ref{Rlp}), \cite{Maldyads}.
For large $N$, the M-theory solution is weakly coupled and supergravity
can be used for studying the gauge theory.
\par
The relevant degrees of freedom at the superconformal fixed points
are in general different from the elementary fields of the
supersymmetric gauge theory. For example, vector multiplets are not
conformal in three dimensions and they should be replaced by
some other multiplets of the superconformal group by dualizing
the vector field to a scalar. Let us again consider the simple example
of ${\cal N}=8$. The degrees of freedom at the superconformal point
 are contained in a supermultiplet with eight real
scalars and eight fermions, transforming in representations of the
global R-symmetry $SO(8)$. This is the same content of the ${\cal N}=8$
vector multiplet, when the vector field is dualized into a scalar.
The change of variable from a vector to a scalar, which is well-defined
in an abelian theory, is obviously a non-trivial and not even well-defined
operation in a non-abelian theory. The scalars degrees of freedom at the
superconformal point parametrize the flat space transverse to the M2-branes. In this case,
the moduli space of vacua of the abelian ${\cal N}=8$ gauge theory,
corresponding to a single M2-brane, is isomorphic to the transverse space.
The case with $N$ M2-branes is obtained by promoting the theory to a
non-abelian one. We want to follow a similar procedure for the conifold cases.
\par
For branes at the conifold singularity of ${\cal C}(X_7)$ there is no obvious way
of reducing the system to a simple configuration of D2-branes in type IIA
and read the field content by using standard brane techniques \footnote{
This possibility exists for orbifold singularities and was exploited
in \cite{pz}, \cite{fkpz}, \cite{gomis} for ${\cal N}=4$ and in \cite{ahnC3} for
${\cal N}=2$.}. We can nevertheless use the intuition from geometry
for identifying the relevant degrees of freedom at the superconformal
point. We need an abelian gauge theory whose moduli space of vacua
is isomorphic to ${\cal C}(X_7)$. The moduli space of vacua of
${\cal N}=2$ theories have two different branches touching at a point, the
Coulomb branch parametrized by the {\it vev} of the scalars in the vector
multiplets and the Higgs branch parametrized by the {\it vev} of the
scalars in the chiral multiplets.
The Higgs branch is the one we are interested in.
Each of the two branches excludes the other, so we
can consistently set the scalars in the vector multiplets to zero
(see section \ref{scalapot} for a discussion of the scalar potential).
We can find what we need in toric geometry.
Indeed, this latter describes certain complex manifolds as K\"ahler
quotients \cite{kalquot} associated to symplectic actions of a product
of $U(1)$'s on some ${\IC }^p$.
This is completely equivalent to imposing the
D-term equations for an abelian ${\cal N}=2,D=3$ gauge theory and dividing
by the gauge group or, in other words, to finding the moduli space
of vacua of the theory. Fortunately, both the cone over $Q^{111}$ and
that over $M^{111}$ have a toric geometry description. This description
 was already 
used for studying these spaces in \cite{tatar}, \cite{dallagat}. Here,
we will consider
a different point of view. 
We can then easily find abelian gauge theories whose moduli space of vacua
(the Higgs branch component) is isomorphic to these two particular conifolds.
These abelian gauge theories will be then promoted to non abelian ones,
whose IR fixed point will be our candidates as $AdS/CFT$--duals to the
supergravities developed in chapter $3$. We will find strong arguments
that these theories are actually dual, giving in this way a non--trivial
check of $AdS/CFT$ correspondence.
\par
A comment on the nomenclature. Most authors call the fundamental superfields of the
gauge theory {\it supersingletons}. Actually this denomination is misleading, because
they do not belong to the supersingleton representation of $Osp\ll(\cN\vert 4\rr)$
(see  tables \ref{N2supersingleton}, \ref{N3shortvector} for the $\cN=2$, $\cN=3$ cases),
but, in the case of $\cN=2$,  are chiral supermultiplets (see table \ref{N2hyper}).
However, they are non unitary representations of the supergroup: in our
case as we will show they have $E_0=\ll|y_0\rr|<1/2$; this is not a problem, because these
superfields are degrees of freedom of the gauge theory, which does not have the $Osp\ll(2\vert 4\rr)$
isometry, while the fundamental fields of the conformal theory are composite superfields, sitting
in $Osp\ll(2\vert 4\rr)$ UIRs. There  are two reasons by which most authors call supersingletons the
fundamental fields of the gauge theory; the first is that in the case $X_7=S^7$ this is true; the second 
is an analogy: as the supersingleton superfields, they cannot
be degrees of freedom of the conformal theory, while the composite superfields made by them can.
\par 
In section $1$ I build, using rheonomy formalism (for a review on rheonomy
see \cite{castdauriafre}), the generic three dimensional $\cN=2$ supersymmetric
gauge theory, writing the lagrangian and fixing the basis for the discussion of
the next sections; furthermore, I write the $\cN=4$ and $\cN=8$ theories as $\cN=2$ 
theories with constraints on the field content and the representations. In 
section $2$ I discuss the geometry of the two manifolds $Q^{111}$, $M^{111}$ as fiber bundles and as
toric manifolds,
and show how to find the abelian gauge theories associated with these
toric descriptions. 
In section $3$ I generalize these abelian gauge theories to  
non abelian ones, which are our candidates to be the $AdS/CFT$--duals
to supergravity on $AdS_4\times Q^{111}$ and $AdS_4\times M^{111}$. 
I show that these theories perfectly reproduce the complete spectrum of
shortened supergravity multiplets found in chapter $3$. In section $4$ I
address the issue of the so--called baryonic operators, and show that they
correspond to non--perturbative states of supergravity. They allow us
to find the conformal weights of the fundamental fields of the gauge theory. 
In section $5$ I draw the conclusions.
\par
The content of the present chapter refers to results obtained within the collaborations
\cite{noi3}, \cite{noi2}.
\par
\section{${\cal N}=2$ three dimensional gauge theories
and their rheonomic construction}
\label{n2d3gauge}
\par
As a first step, we
construct a generic ${\cal N}=2$ gauge theory with an arbitrary
gauge group and an arbitrary number of chiral multiplets in generic
interaction. We are mostly interested in the final formulae for the
scalar potential, which will be used in section \ref{scalapot},
but we provide a complete construction of the lagrangian and of the
supersymmetry transformation rules. To this effect we utilize the
method of rheonomy \cite{castdauriafre} that yields the
result for the lagrangian and the supersymmetry rules in
component form avoiding the too much implicit notation of superfield formulation.
Furthermore, we study the restrictions that guarantee
an enlargement of supersymmetry to ${\cal N}=4$ or ${\cal N}=8$; in fact,
even if in the case of $M^{111}$ and $Q^{111}$ the conformal theories
do not seem to arise from deformation of more supersymmetric theories
(as in \cite{witkleb}), in other cases this phenomenom could occur.
\par
The first step in the rheonomic construction of a rigid supersymmetric theory
 involves writing the structural equations of rigid superspace. Then, 
we have to solve them in terms of the rheonomic expansion of the curvatures.
Finally, we will write the superspace lagrangian in term of these curvatures, and,
projecting this lagrangian on the bosonic three--dimensional surface $\cM_3$ we
find the space--time lagrangian. This is the gauge theory lagrangian; we can introduce
the YM coupling constant by scaling some or all of the scalar fields (actually in our
theory we rescale all of them) by
\be
z^i\longrightarrow g_{YM}z^i,
\ee
and multiplying the entire lagrangian by $1/g_{YM}^2$. The conformal IR fixed point is retrieved
sending 
\be
g_{YM}\longrightarrow 0.
\ee
\par
\subsection{${\cal N}=2, \, d=3$ rigid superspace}
The $d\!\!=\!\!3,~{\cal N}$--extended superspace is
viewed as the supercoset space:
\begin{equation}
{\cal M}_{3\vert\cN}=\frac{ISO(1,2|{\cal N})}
{SO(1,2) }\, \equiv \, \frac{Z\left[ ISO(1,2|{\cal N})\right ]
}{SO(1,2) \times \IR^{{\cal N}({\cal N}-1)/2}}
\end{equation}
where $ISO(1,2\vert{\cal N})$ (see section \ref{superspaces}) 
is the ${\cal N}$--extended Poincar\'e superalgebra
in three dimensions. It is the subalgebra of $Osp({\cal N}\vert 4)$
(see eq. (\ref{ospD})) spanned by the generators $J_m$, $P_m$, $q^i$.
The central extension $Z\left[ ISO(1,2|{\cal N})\right ]$ which is
not contained in $Osp({\cal N}\vert 4)$ is obtained by adjoining to
$ISO(1,2|{\cal N})$ the central charges that generate the subalgebra
$\IR^{{\cal N}({\cal N}-1)/2}$.
Specializing our analysis to the case ${\cal N}\!\!=\!\!2$, we can
define the new generators:
\begin{equation}
\left\{\begin{array}{ccl}
Q&=&q^+=\ft{1}{\sqrt 2}(q^1-iq^2)\\
Q^c&=&iq^-=\ft{1}{\sqrt 2}(iq^1-q^2)\\
Z&=&Z^{12}
\end{array}\right.\,.
\end{equation}
\par
Before going on, I have to clarify the notations.
In doing this computation, the conventions for  two--component
spinors are slightly modified with respect to the ones of chapter $2$, in order 
to simplify the notations and avoid the explicit writing of spinor indices.
The Grassman coordinates of  ${\cal N}\!\!=\!\!2$ three-dimensional
superspace introduced in equation (\ref{complexthet}) , $\theta^\pm_\a$, are
renamed $\theta$ and $\theta^c$.
The reason for the superscript ``$\,^c\,$'' is that, in three
dimensions the upper and lower components of the four--dimensional
$4$--component spinor are charge conjugate. In fact, the charge conjugation is
defined by:
\begin{equation}\label{conjugations}
\theta^c\equiv C_{[3]}\overline\theta^T\,,\qquad
\overline\theta\equiv\theta^\dagger\g^0\,,
\end{equation}
where $C_{[3]}$ is the $d=3$ charge conjugation matrix:
\begin{equation}
\left\{\begin{array}{ccc}
C_{[3]}\g^m C_{[3]}^{-1}&=&-(\g^m)^T\\
\g^0\g^m(\g^0)^{-1}&=&(\g^m)^\dagger\,.
\end{array}\right.
\end{equation}
The lower case gamma matrices are $2\!\times\!2$
and provide a realization of the $d\!=\!2\!+\!1$ Clifford algebra:
\begin{equation}
  \{\gamma^m \, , \, \gamma^n \} = \eta^{mn}
\label{so21cli}
\end{equation}
Utilizing  the following explicit basis:
\begin{equation}\label{gammamatrices}
\left\{\begin{array}{ccl}
\g^0&=&\sigma^2\\
\g^1&=&-i\sigma^3\\
\g^2&=&-i\sigma^1
\end{array}\right.\qquad C_{[3]}=-i\sigma^2\,,
\end{equation}
both $\g^0$ and $C_{[3]}$ become proportional to $\varepsilon_{\a\b}$.
This implies that in equation (\ref{conjugations}) the role of the
matrices $C_{[3]}$ and $\g^0$ is just to convert  upper  into lower
$SL(2,\IC)$ indices and viceversa.
\par
The relation between the two notations for the spinors is summarized
in the following table:
\begin{equation}
\begin{array}{|c|c|}
\hline
(\theta^+)^\a&\theta\\
(\theta^+)_\a&\overline\theta^c\\
(\theta^-)^\a&-i\theta^c\\
(\theta^-)_\a&-i\overline\theta\\
\hline
\end{array}
\end{equation}
With the second set of conventions the spinor indices can
be ignored since the contractions are always made between
barred (on the left) and unbarred (on the right) spinors.
\par
The left invariant one--form $\Omega$ on ${\cal M}_{3\vert \cal N}$ is:
\begin{equation}
\Omega=V^mP_m-\ft{1}{2}i\o^{mn}J_{mn}+\overline{\psi^c}Q+\overline{\psi}Q^c
+i{A}Z\,.
\end{equation}
The superalgebra (\ref{ospD}) defines  all the structure constants
apart from those relative to the central charge that are trivially
determined. Hence we can write:
\begin{eqnarray}
d\Omega-\Omega\wedge\Omega&=&\left(dV^m-\o^m_{\ n}\wedge V^n
+i\overline{\psi}\wedge\g^m\psi
+i\overline{\psi}^c\wedge\g^m\psi^c\right)P_m+\nonumber\\
&&-\ft{1}{2}i\left(d\o^{mn}-\o^m_{\ p}\wedge\o^{pn}\right)J_{mn}+\nonumber\\
&&+\left(d\overline{\psi}^c
+\ft{1}{2}i\o^{mn}\wedge\overline{\psi}^c\g_{mn}\right)Q+\nonumber\\
&&+\left(d\overline{\psi}
-\ft{1}{2}i\o^{mn}\wedge\overline{\psi}\g_{mn}\right)Q^c+\nonumber\\
&&+i\left(d{A}+i\overline{\psi}^c\wedge\psi^c-i\overline{\psi}\wedge\psi
\right)Z\,.
\end{eqnarray}
Imposing the Maurer-Cartan equation $d\Omega-\Omega\wedge\Omega=0$
is equivalent to imposing flatness in superspace, i.e. global
supersymmetry. So we have
\begin{equation}
\left\{\begin{array}{ccl}
dV^m-\omega^m_{~n}\wedge V^n&=&-i\overline{\psi}^c\wedge\g^m\psi^c
-i\overline{\psi}\wedge\g^m\psi\\
d\omega^{mn}&=&\omega^m_{~p}\wedge\omega^{pn}\\
d\overline{\psi}&=&\frac{1}{2}\omega^{mn}\wedge\overline{\psi}\g_{mn}\\
d\overline{\psi}^c&=&-\frac{1}{2}\omega^{mn}\wedge\overline{\psi}^c\g_{mn}\\
d{A}&=&-i\overline{\psi}^c\wedge\psi^c
-i\overline{\psi}\wedge\psi\,.
\end{array}\right.
\end{equation}
The simplest solution for the supervielbein and connection is:
\begin{equation}
\left\{\begin{array}{ccl}
V^m&=&dx^m-i\overline{\theta}^c\g^md\theta^c
-i\overline{\theta}\g^m d\theta\\
\omega^{mn}&=&0\\
\psi&=&d\theta\\
\psi^c&=&d\theta^c\\
{A}&=&-i\overline{\theta}^c\,d\theta^c
-i\overline{\theta}\,d\theta\,.
\end{array}\right.
\end{equation}
The superderivatives discussed in section \ref{superfields} (compare with eq.(\ref{supdervcf}) ),
\begin{equation}
\left\{\begin{array}{ccl}
D_m&=&\partial_m\\
D&=&{\partial\over\partial\overline{\theta}}-i\g^m\theta\partial_m\\
D^c&=&{\partial\over\partial\overline{\theta}^c}-i\g^m\theta^c\partial_m\\
\end{array}\right.,
\end{equation}
are the vectors dual to these one--forms.
%%%%%%%%%%%%%%%%%%%%%%%%%%%%%%%%%%%%%%%%%%%%%%%%%%%%%%%%%%%%%%%%
%
%             N = 2         D = 3
%
%%%%%%%%%%%%%%%%%%%%%%%%%%%%%%%%%%%%%%%%%%%%%%%%%%%%%%%%%%%%%%%%
\subsection{Rheonomic construction of the ${\cal N}=2,~d=3,$ lagrangian}
As stated we are interested in the generic form of ${\cal N}=2,~d=3$
super Yang Mills theory coupled to $n$ chiral multiplets arranged into a generic
representation $\cal R$ of the gauge group $\cal G$.
\par
In ${\cal N}=2,~d=3$ supersymmetric theories, two formulations
are allowed: the on--shell and the off--shell one.
In the on--shell formulation which contains only
the physical fields, the supersymmetry transformations rules
 close the supersymmetry algebra only
upon use of  the field equations.
On the other hand the off--shell formulation contains further auxiliary, non
dynamical fields that make it possible for
 the supersymmetry transformations rules to close the
supersymmetry algebra identically.
By solving the field equations of the auxiliary fields
these latter can be eliminated   and the on--shell formulation can be retrieved.
We  adopt the off--shell formulation.
%%%%%%%%%%%%%%%%%%%%%%%%%%%%%%%%%%%%%%%%%%%%%%%%%%%%%%%%%%%%%%%%
%
%                  V E C T O R       M U L T I P L E T
%
%%%%%%%%%%%%%%%%%%%%%%%%%%%%%%%%%%%%%%%%%%%%%%%%%%%%%%%%%%%%%%%%
\subsubsection{The gauge multiplet}
The three--dimensional ${\cal N}=2$ vector multiplet contains the following
Lie-algebra valued fields:
\begin{equation}\label{vectorm}
\left({\cal A},\l,\l^c,M,P\right)\,,
\end{equation}
where ${\cal A}={\cal A}^It_I$ is the real gauge connection one--form,
$\l$ and $\l^c$ are two complex Dirac spinors (the \emph{gauginos}),
$M$ and $P$ are real scalars; $P$ is an auxiliary field.
\par
The field strength is:
\begin{equation}\label{defF}
F=d{\cal A}+i{\cal A}\wedge{\cal A}\,.
\end{equation}
The covariant derivative on the other fields of the gauge multiplets
is defined as:
\begin{equation}\label{defnablagauge}
\nabla X=dX+i\left[{\cal A},X\right]\,.
\end{equation}
From (\ref{defF}) and (\ref{defnablagauge}) we obtain the Bianchi identity:
\begin{equation}
\nabla^2 X=i\left[F,X\right]\,.
\end{equation}
The rheonomic parametrization of the \emph{curvatures} is given by:
\begin{equation}\label{vectorB}
\left\{\begin{array}{ccl}
F&=&F_{mn}V^mV^n-i\overline{\psi}^c\g_m\l V^m-i\overline{\psi}\g_m\l^cV^m
+iM\left(\overline{\psi}\psi-\overline{\psi}^c\psi^c\right)\\
\nabla\l&=&V^m\nabla_m\l+\nslash M\psi^c-F_{mn}\g^{mn}\psi^c
+iP\psi^c\\
\nabla\l^c&=&V^m\nabla_m\l^c-\nslash M\psi-F_{mn}\g^{mn}\psi
-iP\psi\\
\nabla M&=&V^m\nabla_mM+i\overline{\psi}\l^c-i\overline{\psi}^c\l\\
\nabla P&=&V^m\nabla_mP+\overline{\psi}\nslash\l^c-\overline{\psi}^c\nslash\l
-i\overline{\psi}\left[\l^c,M\right]-i\overline{\psi}^c\left[\l,M\right]
\end{array}\right.
\end{equation}
and we also have:
\begin{equation}
\left\{\begin{array}{ccl}
\nabla F_{mn}&=&V^p\nabla_pF_{mn}+i\overline{\psi}^c\g_{[m}\nabla_{n]}\l
+i\overline{\psi}\g_{[m}\nabla_{n]}\l^c\\
\nabla\nabla_mM&=&V^n\nabla_n\nabla_mM+i\overline{\psi}\nabla_m\l^c
-i\overline{\psi}^c\nabla_m\l+\overline{\psi}^c\g_m\left[\l,M\right]
+\overline{\psi}\g_m\left[\l^c,M\right]\\
\nabla\nabla_m\l&=&V^n\nabla_n\nabla_m\l+\nabla_m\nabla_nM\g^n\psi^c
-\nabla_mF_{np}\g^{np}\psi^c+\\
&&+i\nabla_mP\psi^c+\overline{\psi}\g_m\left[\l^c,\l\right]\\
\nabla_{[p}F_{mn]}&=&0\\
\nabla_{[m}\nabla_{n]}M&=&i\left[F_{mn},M\right]\\
\nabla_{[m}\nabla_{n]}\l&=&i\left[F_{mn},\l\right]\,.
\end{array}\right.
\end{equation}
The off--shell formulation of the theory contains an arbitrariness
in the choice of the functional dependence of the auxiliary fields
on the physical fields.
Consistency with the Bianchi identities forces the generic expression of $P$ as a function
of $M$ to be:
\begin{equation}\label{defP}
P^I=2\a M^I+\zeta^{\tilde I}{\cal C}_{\tilde I}^{\ I}\,,
\end{equation}
where $\a,~\zeta^{\tilde I}$ are arbitrary real parameters and
${\cal C}_{\tilde I}^{\ I}$ is the projector on the center 
$Z[\cal G]$ of the gauge Lie algebra.
The terms in the lagrangian proportional to $\a$ and $\zeta$ are separately
supersymmetric.
In the bosonic lagrangian, the part proportional to $\a$ is a Chern Simons
term, while the part proportional to $\zeta$ constitutes the Fayet Iliopoulos
term.
Note that the Fayet Iliopoulos terms are associated only with a
central abelian subalgebra of the gauge algebra $\cal G$.
\par
Enforcing (\ref{defP}) we get the following equations of motion for the spinors:
\begin{equation}
\left\{\begin{array}{c}
\nslash\l=2i\a\l-i\left[\l,M\right]\\
\\
\nslash\l^c=2i\a\l^c+i\left[\l^c,M\right]\,.
\end{array}\right.
\end{equation}
Taking the covariant derivatives of these, we obtain the equations of motion
for the bosonic fields:
\begin{equation}
\left\{\begin{array}{c}
\nabla_m\nabla^m M=-4\a^2M-2\a\b-2\left[\overline{\l},\l\right]\\
\nabla^n F_{mn}=-\a\e_{mnp}F^{np}-\ft{i}{2}\left[\nabla_mM,M\right]\,.
\end{array}\right.
\end{equation}
Using the rheonomic approach we find the following superspace lagrangian
for the gauge multiplet:
\begin{equation}\label{gaugeLag}
{\cal L}_{gauge}={\cal L}_{gauge}^{Maxwell}
+{\cal L}_{gauge}^{Chern-Simons}
+{\cal L}_{gauge}^{Fayet-Iliopoulos}\,,
\end{equation}
where
\begin{eqnarray}
{\cal L}_{gauge}^{Maxwell}&=&Tr\left\{-2F^{mn}\left[F
+i\overline{\psi}^c\g_m\l V^m+i\overline{\psi}\g_m\l^cV^m
-2iM\overline{\psi}\psi\right]V^p\epsilon_{mnp}\right.+\nonumber\\
&+&\ft{1}{3}F_{qr}F^{qr}V^m V^n V^p\epsilon_{mnp}
-\ft{1}{2}i\epsilon_{mnp}\left[\nabla\overline{\l}\g^m\l
+\nabla\overline{\l}^c\g^m\l^c\right]
V^n V^p+\nonumber\\
&+&\epsilon_{mnp}{\cal M}^m\left[\nabla M-i\overline{\psi}\l^c
+i\overline{\psi}^c\l\right] V^nV^p
-\ft{1}{6}{\cal M}^d{\cal M}_d\epsilon_{mnp}V^mV^nV^p+\nonumber\\
&+&2\nabla M \overline{\psi}^c\g_c\l V^p
-2\nabla M \overline{\psi}\g_p\l^cV^p+\nonumber\\
&+&2F\overline{\psi}^c\l+2F\overline{\psi}\l^c
+i\overline{\l}^c\l\overline{\psi}^c\g_m\psi V^m+
i\overline{\l}\l^c\overline{\psi}\g_m\psi^cV^m+\nonumber\\
&+&\left.\ft{1}{6}{\cal P}^2V^mV^nV^p\epsilon_{mnp}
-4i(\overline{\psi}\psi)M\left[\overline{\psi}^c\l
+\overline{\psi}\l^c\right]\right\}\,,
\end{eqnarray}
\begin{eqnarray}
{\cal L}_{gauge}^{Chern-Simons}&=&\a Tr\left\{
-2\left(\cA\wedge F-i\cA\wedge \cA\wedge \c A\right)
-\ft{2}{3}MP\epsilon_{mnp}V^m V^n V^p\right.+\nonumber\\
&+&\ft{2}{3}\overline{\l}\l\epsilon_{mnp}V^m V^n V^p
+2M\epsilon_{mnp}\left[\overline{\psi}^c\g^m\l
+\overline{\psi}\g^m\l^c\right]V^n V^p+\nonumber\\
&&\left.-4iM^2\overline{\psi}\g_m\psi V^m\right\}
\end{eqnarray}
\begin{eqnarray}
{\cal L}_{gauge}^{Fayet-Iliopoulos}&=&Tr\left\{\zeta{\cal C}\left[
-\ft{1}{3}P\e_{mnp}V^mV^nV^p
\e_{mnp}\left(\overline{\psi}^c\g^m\l
-\overline{\psi}\g^m\l^c\right)V^nV^p+
\right.\right.\nonumber\\
&&\left.\left.-4iM\overline{\psi}\g_m\psi V^m
-4i{\cal A}\overline{\psi}\psi\right]\right\}\,.
\end{eqnarray}
%%%%%%%%%%%%%%%%%%%%%%%%%%%%%%%%%%%%%%%%%%%%%%%%%%%%%%%%%%%%%%%%
%
%             C H I R A L          M U L T I P L E T
%
%%%%%%%%%%%%%%%%%%%%%%%%%%%%%%%%%%%%%%%%%%%%%%%%%%%%%%%%%%%%%%%%
\subsubsection{Chiral multiplet}
The chiral multiplet contains the following fields:
\begin{equation}\label{chiralm}
\left(z^i,\chi^i,H^i\right)
\end{equation}
where $z^i$ are complex scalar fields which parametrize a  K\"ahler
manifold. Since we are interested in microscopic theories
with canonical kinetic terms we take this K\"ahler manifold to be
flat and we choose its metric to be the constant
$\eta_{ij^*}\equiv \mbox{diag}(+,+,\dots,+)$. The other fields in
the chiral multiplet are $\chi^i$ which is a two components Dirac spinor
and $H^i$ which is a complex scalar auxiliary field.
The index $i$ runs in the representation $\cal R$ of $\cal G$.
\par
The covariant derivative of the fields $X^i$ in the chiral multiplet is:
\begin{equation}
\nabla X^i=dX^i+i\eta^{ii^*}{\cal A}^I(T_I)_{i^*j}X^j\,,
\end{equation}
where $(T_I)_{i^*j}$ are the hermitian generators of $\cal G$
in the representation $\cal R$.
The covariant derivative of the complex conjugate fields
$\overline X^{i^*}$ is:
\begin{equation}
\nabla\overline X^{i^*}=d\overline X^{i^*}
-i\eta^{i^*i}{\cal A}^I(\overline T_I)_{ij^*}\overline X^{j^*}\,,
\end{equation}
where
\begin{equation}
(\overline T_I)_{ij^*}\equiv\overline{(T_I)_{i^*j}}=(T_I)_{j^*i}\,.
\end{equation}
The rheonomic parametrization of the curvatures is given by:
\begin{equation}\label{chiralB}
\left\{\begin{array}{ccl}
\nabla z^i&=&V^m\nabla_m z^i+2\overline{\psi}^c\chi^i\\
\nabla\chi^i&=&V^m\nabla_m\chi^i-i\nslash z^i\psi^c+H^i\psi
-M^I(T_I)^i_{\,j}z^j\psi^c\\
\nabla H^i&=&V^m\nabla_m H^i-2i\overline{\psi}\nslash\chi^i
-2i\overline{\psi}\l^I(T_I)^i_{\,j} z^j
+2M^I(T_I)^i_{\,j}\overline{\psi}\chi^j
\end{array}\right.\,.
\end{equation}
We can choose the auxiliary fields $H^i$ to be the derivatives of an
arbitrary antiholomorphic superpotential $\overline W(\overline z)$:
\begin{equation}\label{defW^*}
H^i=\eta^{ij^*}\frac{\partial\overline W(\overline z)}{\partial z^{j^*}}
=\eta^{ij^*}\partial_{j^*}\overline W\,.
\end{equation}
Enforcing eq. (\ref{defW^*}) we get the following equations of motion
for the spinors:
\begin{equation}\label{chimotion}
\left\{\begin{array}{c}
\nslash\chi^i=i\eta^{ij^*}\partial_{j^*}\partial_{k^*}
\overline W\chi^{ck^*}-\l^I(T_I)^i_{\,j} z^j-iM^I(T_I)^i_{\,j}\chi^j\\
\\
\nslash\chi^{ci^*}=i\eta^{i^*j}\partial_j\partial_k W\chi^k
+\l^{cI}(\overline T_I)^{i^*}_{\,j^*}\overline z^{j^*}
-iM^I(\overline T_I)^{i^*}_{\,j^*}\chi^{cj^*}
\end{array}\right.\,.
\end{equation}
Taking the differential of (\ref{chimotion}) one obtains the equation of
motion for $z$:
\begin{eqnarray}
\Box z^i&=&\eta^{ii^*}\partial_{i^*}\partial_{j^*}\partial{k^*}
\overline W\left(\overline{\chi}^{j^*}\chi^{ck^*}\right)
-\eta^{ij^*}\partial_{j^*}\partial_{k^*}
\overline W(\overline z)\partial_i W+\nonumber\\
&&+P^I(T_I)^i_{\,j}z^j-M^IM^J(T_IT_J)^i_{\,j}z^j
-2i\overline{\l}^I(T_I)^i_{\,j}\chi^j\,.
\end{eqnarray}
%%%%%%%%%%%%%%%%%%%%%%%%%%%%%%%%%%%%%%%%%%%%%%%%%%%%%%%%%%%%%%%%
%
%           N = 2      L A G R A N G I A N
%
%%%%%%%%%%%%%%%%%%%%%%%%%%%%%%%%%%%%%%%%%%%%%%%%%%%%%%%%%%%%%%%%
The first order Lagrangian for the chiral multiplet (\ref{chiralm}) is:
\begin{equation}\label{chiralLag}
{\cal L}_{chiral}={\cal L}_{chiral}^{Wess-Zumino}
+{\cal L}_{chiral}^{superpotential}\,,
\end{equation}
where
\begin{eqnarray}
{\cal L}_{chiral}^{Wess-Zumino}&=&\e_{mnp}\overline{\Pi}^{m\,i^*}
\eta_{i^*j}\left[\nabla z^j
-2\overline{\psi}^c\chi^j\right]V^nV^p+\nonumber\\
&+&\e_{mnp}\Pi^{m\,i}\eta_{ij^*}\left[\nabla \overline z^{j^*}
-2\overline{\chi}\psi^{c\,j^*}\right]V^nV^p+\nonumber\\
&-&\ft{1}{3}\e_{mnp}\eta_{ij^*}\Pi_q^{\,i}
\overline{\Pi}^{q\,j^*}V^mV^nV^p+\nonumber\\
&+&i\e_{mnp}\eta_{ij^*}\left[\overline{\chi}^{j^*}\g^m\nabla\chi^i
+\overline{\chi}^{c\,i}\g^m\nabla\chi^{c\,j^*}
\right]V^nV^p+\nonumber\\
&+&4i\eta_{ij^*}\left[\nabla z^i\overline{\psi}\g_m\chi^{c\,j^*}
-\nabla \overline z^{j^*}\overline{\chi}^{c\,i}\g_m\psi
\right]V^m+\nonumber\\
&-&4i\eta_{ij^*}\left(\overline{\chi}^{j^*}\g_m\chi^i\right)
\left(\overline{\psi}^c\psi^c\right)V^m
-4i\eta_{ij^*}\left(\overline{\chi}^{j^*}\chi^i\right)
\left(\overline{\psi}^c\g_m\psi^c\right)V^m+\nonumber\\
&+&\ft{1}{3}\eta_{ij^*}H^i\overline H^{j^*}\e_{mnp}V^mV^nV^p
+2\left(\overline{\psi}\psi\right)
\eta_{ij^*}\left[\overline z^{j^*}\nabla z^i
-z^i\nabla \overline z^{j^*}\right]+\nonumber\\
&+&2i\e_{mnp}z^iM^I(T_I)_{ij^*}\overline{\chi}^{j^*}\g^m\psi^c
V^nV^p+\nonumber\\
&+&2i\e_{mnp}\overline z^{j^*}M^I(T_I)_{j^*i}
\overline{\chi}^{c\,i}\g^m\psi V^nV^p+\nonumber\\
&-&\ft{2}{3}M^I(T_I)_{ij^*}\overline{\chi}^{j^*}\chi^i
\e_{mnp}V^mV^nV^p+\nonumber\\
&+&\ft{2}{3}i\left[\overline{\chi}^{j^*}\l^I(T_I)_{j^*i} z^i
-\overline{\chi}^{c\,i}\l^{c\,I}(T_I)_{ij^*}\overline z^{j^*}\right]
\e_{mnp}V^mV^nV^p+\nonumber\\
&+&\ft{1}{3}z^iP^I(T_I)_{ij^*}\overline z^{j^*}\e_{mnp}V^mV^nV^p+\nonumber\\
&-&\left(\overline{\psi}^c\g^m\l^I(T_I)_{ij^*}\right)z^i\overline z^{j^*}
\e_{mnp}V^nV^p+\nonumber\\
&+&\left(\overline{\psi}\g^m\l^{c\,I}(T_I)_{ij^*}\right)z^i\overline z^{j^*}
\e_{mnp}V^nV^p+\nonumber\\
&-&\ft{1}{3}M^IM^J\,z^i(T_IT_J)_{ij^*}\overline z^{j^*}
\e_{mnp}V^mV^nV^p+\nonumber\\
&+&4iM^I(T_I)_{ij^*}z^i\overline z^{j^*}\overline{\psi}\g_m\psi V^m\,,
\end{eqnarray}
and
\begin{eqnarray}
{\cal L}_{chiral}^{superpotential}&=&
-2i\e_{mnp}\left[\overline{\chi}^{j^*}\g^m\partial_{j^*}\overline W
(\overline z)\psi+\overline{\chi}^{c\,j}\g^m\partial_j\overline W(z)
\psi^c\right]V^nV^p+\nonumber\\
&+&\ft{1}{3}\left[\partial_i\partial_jW(z)\overline{\chi}^{c\,i}\chi^j
+\partial_{i^*}\partial_{j^*}\overline W(\overline z)
\overline{\chi}^{i^*}\chi^{c\,j^*}\right]\e_{mnp}V^mV^nV^p+\nonumber\\
&-&\ft{1}{3}\left[H^i\partial_iW(z)
+\overline H^{j^*}\partial_{j^*}\overline W(\overline z)
-\eta_{ij^*}H^i\overline H^{j^*}\right]\e_{mnp}V^mV^nV^p+\nonumber\\
&-&4i\left[W(z)+\overline W(\overline z)\right]
\overline{\psi}\g_m\psi^cV^m\,.
\end{eqnarray}
%%%%%%%%%%%%%%%%%%%%%%%%%%%%%%%%%%%%%%%%%%%%%%%%%%%%%%%%%%%%%%%%
%
%      N = 2    S P A C E T I M E     L A G R A N G I A N
%
%%%%%%%%%%%%%%%%%%%%%%%%%%%%%%%%%%%%%%%%%%%%%%%%%%%%%%%%%%%%%%%%
\subsubsection{The space--time Lagrangian}
In the rheonomic approach (\cite{castdauriafre}), the total
three--dimensional ${\cal N}\!\!=\!\!2$ lagrangian:
\begin{equation}
{\cal L}^{{\cal N}=2}={\cal L}_{gauge}+{\cal L}_{chiral}
\end{equation}
is a closed ($d{\cal L}^{{\cal N}=2}=0$) three--form defined in
superspace.
The action is given by the integral of ${\cal L}^{{\cal N}=2}$ on
a generic \emph{bosonic} three--dimensional surface ${\cal M}_3$
in superspace:
\begin{equation}
S=\int_{{\cal M}_3}{\cal L}^{{\cal N}=2}\,.
\end{equation}
Supersymmetry transformations can be viewed as global translations in
superspace which move ${\cal M}_3$.
Then, being ${\cal L}^{{\cal N}=2}$ closed, the action is invariant under
global supersymmetry transformations.
\par
We choose as bosonic surface the one defined by:
\begin{equation}
\theta=d\theta=0\,.
\end{equation}
Then the space--time lagrangian, i.e. the pull--back of
${\cal L}^{{\cal N}=2}$ on ${\cal M}_3$, is:
\begin{equation}\label{N=2stLag}
{\cal L}^{{\cal N}=2}_{st}={\cal L}^{kinetic}_{st}
+{\cal L}^{fermion~mass}_{st}+{\cal L}^{potential}_{st}\,,
\end{equation}
where
\begin{eqnarray}\label{N=2chiralst}
\LL_{st}^{kinetic}&=&\left\{
\eta_{ij^*}\nabla_m z^i\nabla^m\overline z^{j^*}
+i\eta_{ij^*}\left(\overline{\chi}^{j^*}\nslash\chi^i
+\overline{\chi}^{c\,i}\nslash\chi^{c\,j^*}\right)
+\right.\nonumber\\
&&-g_{IJ}F^I_{mn}F^{J\,mn}
+\ft{1}{2}g_{IJ}\nabla_m M^I\nabla^m M^J+\nonumber\\
&&+\left.\ft{1}{2}ig_{IJ}\left(\overline{\l}^I\nslash\l^J
+\overline{\l}^{c\,I}\nslash\l^{c\,J}\right)\right\}d^3x\\
&&\nonumber\\
\LL_{st}^{fermion~mass}&=&
\left\{i\left(\overline{\chi}^{c\,i}\partial_i\partial_jW(z)\chi^j
+\overline{\chi}^{i^*}\partial_{i^*}\partial_{j^*}\overline W(\overline z)
\chi^{c\,j^*}\right)+\right.\nonumber\\
&&-f_{IJK}M^I\overline{\l}^J\l^K
-2\overline{\chi}^{i^*}M^I(T_I)_{ij^*}\chi^{j^*}+\nonumber\\
&&+2i\left(\overline{\chi}^{i^*}\l^I(T_I)_{i^*j}z^j
-\overline{\chi}^{c\,i}\l^I(T_I)_{ij^*}\overline z^{j^*}\right)+\nonumber\\
&&\left.+2\a g_{IJ}\overline{\l}^I\l^J\right\}d^3x\\
\LL_{st}^{potential}&=&-U(z,\overline z,H,\overline H,M,P)d^3x\,,
\end{eqnarray}
and
\begin{eqnarray}\label{defU}
U(z,\overline z,H,\overline H,M,P)&=&
H^i\partial_iW(z)
+\overline H^{j^*}\partial_{j^*}\overline W(\overline z)
-\eta_{ij^*}H^i\overline H^{j^*}+\nonumber\\
&&-\ft{1}{2}g_{IJ}P^IP^J-z^iP^I(T_I)_{ij^*}\overline z^{j^*}+\nonumber\\
&&+z^iM^I(T_I)_{ij^*}\eta^{j^*k}M^J(T_J)_{kl^*}
\overline z^{l^*}+\nonumber\\
&&+2\a g_{IJ}M^IP^J+\zeta^{\tilde I}{\cal C}_{\tilde I}^{\ I}g_{IJ}P^J\,.
\end{eqnarray}
From the variation of the lagrangian with respect to the auxiliary
fields $H^i$ and $P^I$ we find:
\begin{eqnarray}
H^i&=&\eta^{ij^*}\partial_{j^*}\overline W(\overline z)\,,\\
P^I&=&D^I(z,\overline z)+2\a M^I+\zeta^{\tilde I}{\cal C}_{\tilde I}^{\ I}\,,
\end{eqnarray}
where
\begin{equation}
D^I(z,\overline z)=-\overline z^{i^*}(T_I)_{i^*j}z^j\,.
\end{equation}
Substituting this expression in the potential (\ref{defU}) we obtain:
\begin{eqnarray}
U(z,\overline z,M)&=&
\partial_i W(z)\eta^{ij^*}\partial_{j^*}\overline W(\overline z)+\nonumber\\
&&+\ft{1}{2}g^{IJ}\left(\overline z^{i^*}(T_I)_{i^*j}z^j\right)
\left(\overline z^{k^*}(T_J)_{k^*l}z^l\right)+\nonumber\\
&&+\overline z^{i^*}M^I(T_I)_{i^*j}\eta^{jk^*}M^J(T_J)_{k^*l}
z^l+\nonumber\\
&&+2\a^2g_{IJ}M^IM^J+2\a\zeta^{\tilde I}{\cal C}_{\tilde I}^{\ I}g_{IJ}M^J
+\ft{1}{2}\zeta^{\tilde I}{\cal C}_{\tilde I}^{\ I}g_{IJ}
\zeta^{\tilde J}{\cal C}_{\tilde J}^{\ J}+\nonumber\\
&&-2\a M^I\left(\overline z^{i^*}(T_I)_{i^*j}z^j\right)
-\zeta^{\tilde I}{\cal C}_{\tilde I}^{\ I}
\left(\overline z^{i^*}(T_I)_{i^*j}z^j\right)\,.\label{truedefU}
\end{eqnarray}
%%%%%%%%%%%%%%%%%%%%%%%%%%%%%%%%%%%%%%%%%%%%%%%%%%%%%%%%%%%%%%%%
%
%          N = 4     L A G R A N G I A N
%
%%%%%%%%%%%%%%%%%%%%%%%%%%%%%%%%%%%%%%%%%%%%%%%%%%%%%%%%%%%%%%%%
\subsection{A particular ${\cal N}=2$ theory: ${\cal N}=4$}
A  general   lagrangian for
 matter coupled rigid ${\cal N}=4, d=3 $ super Yang Mills theory  is easily
obtained from the dimensional reduction of the
${\cal N}=2,d=4$ gauge theory (see \cite{BertFre}).
The bosonic sector of this latter lagrangian   is the following:
\begin{eqnarray}\label{N=4Lag}
\LL_{bosonic}^{{\cal N}=4}&=&-{1\over g_{YM}^2}g_{IJ}F^I_{mn}F^{J\,mn}
+\ft{1}{2 g_{YM}^2}g_{IJ}\nabla_m M^I\nabla^m M^J+\nonumber\\
&&+{2\over g_{YM}^2}g_{IJ}\nabla_m\overline Y^I\nabla^mY^J
+\ft{1}{2}Tr\left(\nabla_m\overline{\bf Q}\nabla^m{\bf Q}\right)+\nonumber\\
&&-{1\over g_{YM}^2}g_{IN}f^I_{JK}f^N_{LM}\,M^J\overline Y^K\,M^L Y^M
-M^IM^JTr\left(\overline{\bf Q}(T_I\,T_J){\bf Q}\right)+\nonumber\\
&&-{2\over g_{YM}^2}g_{IN}f^I_{JK}f^N_{LM}\,\overline Y^JY^K\,\overline Y^LY^M
-\overline Y^IY^J\,Tr\left(\overline{\bf Q}\left\{T_I,T_J\right\}
{\bf Q}\right)+
\nonumber\\
&&-\ft{1}{4}g^2_{_{YM}}g_{IJ}Tr\left(\overline{\bf Q}(T^I){\bf Q}\,
\overline{\bf Q}(T^J){\bf Q}\right)\,.
\end{eqnarray}
The bosonic matter field content is given by two kinds of fields.
First we have a complex field $Y^I$ in the adjoint representation
of the gauge group, which belongs to a chiral multiplet.
Secondly, we have an $n$-uplet of quaternions
${\bf Q}$, which parametrize a (flat)\footnote{ Once again we choose
the  HyperK\"ahler manifold to be flat since we are interested in microscopic
theories with canonical kinetic terms} HyperK\"ahler manifold:
\begin{equation}
{\bf Q}=\left(\begin{array}{ccl}
Q^1&=&q^{1|0}\unity-iq^{1|x}\s_x\\
Q^2&=&q^{2|0}\unity-iq^{2|x}\s_x\\
&\cdots&\\
Q^A&=&q^{A|0}\unity-iq^{A|x}\s_x\\
&\cdots&\\
Q^n&=&q^{n|0}\unity-iq^{n|x}\s_x
\end{array}\right)\,.
\qquad\begin{array}{l}
q^{A|0},q^{A|x}\in\IR\\
\\
A\in\{1,\ldots,n\}\\
\\
x\in\{1,2,3\}
\end{array}
\end{equation}
The quaternionic conjugation is defined by:
\begin{equation}
\overline Q^A=q^{A|0}\unity+iq^{A|x}\s_x\,.
\end{equation}
In this realization, the quaternions are represented by
matrices of the form:
\begin{equation}
Q^A=\left(\begin{array}{cc}
u^A&i\overline v_{A^*}\\
iv_A&\overline u^{A^*}
\end{array}\right)\qquad
\overline Q^A=\left(\begin{array}{cc}
\overline u^{A^*}&-i\overline v_{A^*}\\
-iv_A&u^A
\end{array}\right)\qquad\begin{array}{l}
u^A=q^{A|0}-iq^{A|3}\\
v^A=-q^{A|1}-iq^{A|2}\,.
\end{array}
\end{equation}
The generators of the gauge group ${\cal G}$ have a triholomorphic
action on the flat HyperK\"ahler manifold, namely they respect the
three complex structures \cite{kalquot}. Explicitly this triholomorphic action
 on {\bf Q} is the following:
\begin{eqnarray}
\d^I{\bf Q}&=&i\hat T^I{\bf Q}\nonumber\\
&&\nonumber\\
\d^I\left(\begin{array}{cc}
u^A&i\overline v_{A^*}\\
iv_A&\overline u^{A^*}
\end{array}\right)&=&
i\left(\begin{array}{cc}
T^I_{A^*B}&\\
&-\overline T^I_{AB^*}
\end{array}\right)\left(\begin{array}{cc}
u^B&i\overline v_{B^*}\\
iv_B&\overline u^{B^*}
\end{array}\right)
\end{eqnarray}
where the $T^I_{A^*B}$ realize a representation
of ${\cal G}$ in terms of $n\times n$ hermitian matrices.
We define $\overline T_{AB^*}\equiv\left(T_{A^*B}\right)^*$, so,
being the generators hermitian ($T^*=T^T$), we can write:
\begin{equation}
T_{A^*B}=\overline T_{BA^*}.
\end{equation}
We can rewrite eq. (\ref{N=4Lag}) in the form:
\begin{eqnarray}\label{N=4bosonicLag}
\LL_{bosonic}^{{\cal N}=4}&=&-\frac{1}{g^2_{_{YM}}}g_{IJ}F^I_{mn}F^{J\,mn}
+\frac{1}{2g^2_{_{YM}}}g_{IJ}\nabla_m M^I\nabla^m M^J+\nonumber\\
&&+\frac{2}{g^2_{_{YM}}}\g_{IJ}\nabla_m\overline Y^I\nabla^m Y^J
+\nabla_m\overline u\nabla^m u+\nabla_m\overline v\nabla^mv+\nonumber\\
&&-\frac{2}{g^2_{_{YM}}}M^I M^J\overline Y^R f_{RIL}f^L_{\,JS}Y^S
-M^I M^J\left(\overline u T_IT_J u
+\overline v\overline T_I\overline T_J v\right)+\nonumber\\
&&-\frac{2}{g^2_{_{YM}}}g_{IJ}\left[\overline Y,Y\right]^I
\left[\overline Y,Y\right]^J
-2\overline Y^IY^J\left(\overline u\{T_I,T_J\} u
+\overline v\{\overline T_I,\overline T_J\} v\right)+\nonumber\\
&&-2g^2_{_{YM}}g_{IJ}\left(vT^Iu\right)\left(\overline v\overline T^J
\overline u\right)
-\ft{1}{2}g^2_{_{YM}}g_{IJ}\left[\left(\overline u T^I u\right)
\left(\overline uT^Ju\right)+\right.\nonumber\\
&&\left.+(\overline v\overline T^I v)(\overline v
\overline T^J v)-2(\overline u T^I u)(
\overline v\overline T^J v)\right]\,.
\end{eqnarray}
By comparing the bosonic part of (\ref{N=2stLag}) with (\ref{N=4bosonicLag}),
we see that in order for a ${\cal N}\!\!=\!\!2$ lagrangian to be also
${\cal N}\!\!=\!\!4$ supersymmetric, the matter content of the theory
and the form of the superpotential are constrained.
The chiral multiplets have to be in an adjoint plus a
generic quaternionic representation of $\cal G$.
So the fields $z^i$ and the gauge generators are
\begin{equation}
z^i=\left\{\begin{array}{l}
\sqrt{2}Y^I\\
g_{YM}u^A\\
g_{YM}v_A
\end{array}\right.\qquad
T^I_{i^*j}=\left\{\begin{array}{l}
f^I_{\,JK}\\
(T^I)_{A^*B}\\
-(\overline T^I)_{AB^*}
\end{array}\right.\,.
\end{equation}
Moreover, the holomorphic superpotential $W(z)$ has to be of the form:
\begin{equation}
W\left(Y,u,v\right)=2g^4_{_{YM}}\delta^{AA^*}Y^I\,v_A(T_I)_{A^*B}u^B\,.
\end{equation}
Substituting these choices in the supersymmetric lagrangian
(\ref{N=2stLag}) we obtain the general ${\cal N}\!\!=\!\!4$ lagrangian
expressed in ${\cal N}\!\!=\!\!2$ language.
\par
Since the action of the gauge group is triholomorphic there is a
triholomorphic momentum map associated with each gauge group
generator (see \cite{ALE}, \cite{damia}, \cite{BertFre}).
\par
The momentum map is given by:
\begin{equation}
{\cal P}=\ft{1}{2}\left(\overline{\bf Q}\,\hat T\,{\bf Q}\right)=
\left(\begin{array}{cc}
{\cal P}_3&{\cal P}_+\\
{\cal P}_-&-{\cal P}_3
\end{array}\right)\,,
\end{equation}
where
\begin{eqnarray}
{\cal P}_3^I&=&i\left(\overline uT^Iu-\overline v\overline T^Iv\right)=
-iD^I\nonumber\\
{\cal P}_+^I&=&2i\overline v\overline T^I\overline u=
ig^{-4}_{_{YM}}\ \partial\overline W/\partial \overline Y_I\nonumber\\
{\cal P}_-^I&=&-2ivT^Iu=
-ig^{-4}_{_{YM}}\ \partial W/\partial Y_I\,.
\end{eqnarray}
So the superpotential can be written as:
\begin{equation}
W=ig^4_{_{YM}}Y_I{\cal P}_-^I\,.
\end{equation}
%%%%%%%%%%%%%%%%%%%%%%%%%%%%%%%%%%%%%%%%%%%%%%%%%%%%%%%%%%%%%%%%
%
%          N = 8     L A G R A N G I A N
%
%%%%%%%%%%%%%%%%%%%%%%%%%%%%%%%%%%%%%%%%%%%%%%%%%%%%%%%%%%%%%%%%
\subsection{A particular ${\cal N}=4$ theory: ${\cal N}=8$}
In this section we discuss the further conditions under which the
${\cal N}\!\!=\!\!4$ three dimensional lagrangian previously
derived acquires an ${\cal N}\!\!=\!\!8$ supersymmetry.
To do that we will compare the four dimensional ${\cal N}\!\!=\!\!2$
lagrangian of \cite{BertFre} with the four dimensional ${\cal N}\!\!=\!\!4$
lagrangian of \cite{N8DF} (rescaled by a factor ${4\over g_{_{YM}}^2}$),
whose bosonic part is:
\begin{eqnarray}\label{N=8piphiLag}
{\cal L}^{{\cal N}=4~D=4}_{bosonic}&=&{1\over g_{_{YM}}^2}
\left\{-F^{\underline m\underline n}F_{\underline m\underline n}+
{1\over 4}\nabla^{\underline m}\phi^{AB}\nabla_{\underline m}\phi^{AB}+
{1\over 4}\nabla^{\underline m}\pi^{AB}\nabla_{\underline m}\pi^{AB}
+\right.\nonumber\\
&+&{1\over 64}\left(
\left[\phi^{AB},\phi^{CD}\right]\left[\phi^{AB},\phi^{CD}\right]+
\left[\pi^{AB},\pi^{CD}\right]\left[\pi^{AB},\pi^{CD}\right]
+\right.\nonumber\\
&+&\left.\left.2\left[\phi^{AB},\pi^{CD}\right]
\left[\phi^{AB},\pi^{CD}\right]\right)\right\}\,.
\end{eqnarray}
The fields $\pi^{AB}$ and $\phi^{AB}$ are Lie-algebra valued:
\begin{equation}
\left\{\begin{array}{ccl}
\pi^{AB}&=&\pi^{AB}_It^I\\
\phi^{AB}&=&\phi^{AB}_It^I
\end{array}\right.\,,
\end{equation}
where $t^I$ are the generators of the gauge group $\cal G$.
They are the real and imaginary parts
of the complex field $\rho$:
\begin{equation}
\left\{\begin{array}{ccl}
\rho^{AB}&=&\ft{1}{\sqrt{2}}\left(\pi^{AB}+i\phi^{AB}\right)\\
\overline{\rho}_{AB}&=&\ft{1}{\sqrt{2}}\left(\pi^{AB}
-i\phi^{AB}\right)
\end{array}\right.\,.
\end{equation}
$\rho^{AB}$ transforms in the representation $\bf 6$ of a
global $SU(4)$-symmetry of the theory.
Moreover, it satisfies the following pseudo-reality condition:
\begin{equation}
\rho^{AB}=-\ft{1}{2}i\e^{ABCD}\overline{\rho}_{CD}\,.
\end{equation}
In terms of $\rho$ the lagrangian (\ref{N=8piphiLag}) can be
rewritten as:
\begin{equation}\label{N=8rhoLag}
{\cal L}^{{\cal N}=8}_{bosonic}={1\over 2g_{_{YM}}^2}\left\{
-F^{\underline m\underline n}F_{\underline m\underline n}
+\nabla_{\underline m}\overline{\rho}_{AB}\nabla^{\underline m}\rho^{AB}
+{1\over 16}\left[\overline{\rho}_{AB}\rho^{CD}\right]\left[\rho^{AB},
\overline{\rho}_{CD}\right]\right\}\,.
\end{equation}
The $SU(2)$ global symmetry of the ${\cal N}\!\!=\!\!2,~D\!\!=\!\!4$
theory can be diagonally embedded into the $SU(4)$ of
the ${\cal N}\!\!=\!\!4,~D\!\!=\!\!4$ theory:
\begin{equation}
{\cal U}=\left(\begin{array}{cc}
U&0\\
0&\overline U
\end{array}\right)\ \in SU(2)\subset SU(4)\,.
\end{equation}
By means of this embedding, the $\bf 6$ of SU(4) decomposes as
${\bf 6}\longrightarrow{\bf 4+1+1}$.
Correspondingly, the pseudo-real field $\rho$ can be splitted into:
\begin{eqnarray}
\rho^{AB}&=&\left(\begin{array}{cccc}
0&\sqrt 2Y&g_{_{YM}} u&ig_{_{YM}}\overline v\\
-\sqrt 2Y&0&ig_{_{YM}} v&g_{_{YM}}\overline u\\
-g_{_{YM}} u&-ig_{_{YM}} v&0&-\sqrt 2\overline Y\\
-ig_{_{YM}}\overline v&-g_{_{YM}}\overline u&\sqrt 2\overline Y&0
\end{array}\right)=\nonumber\\
&&\nonumber\\
&=&\left(\begin{array}{cc}
i\sqrt 2\s^2\otimes Y&g_{_{YM}}Q\\
&\\
-g_{_{YM}}Q^T&-i\sqrt 2\s^2\otimes\overline Y
\end{array}\right)\,,
\end{eqnarray}
where $Y$ and $Q$ are Lie-algebra valued.
The global $SU(2)$ transformations act as:
\begin{equation}
\rho\longrightarrow{\cal U}\rho{\cal U}^T=\left(\begin{array}{cc}
i\sqrt 2\s^2\otimes Y&g_{_{YM}}UQU^{\dagger}\\
&\\
-g_{_{YM}}\left(UQU^{\dagger}\right)^T&-i\sqrt 2\s^2\otimes\overline Y
\end{array}\right)\,.
\end{equation}
Substituting this expression for $\rho$ into (\ref{N=8rhoLag})
and dimensionally reducing to three dimensions, we obtain the
lagrangian (\ref{N=4Lag}).
In other words the ${\cal N}\!\!=\!\!4,~D\!\!=\!\!3$ theory is
enhanced to ${\cal N}\!\!=\!\!8$ provided the hypermultiplets
are in the adjoint representation of $\cal G$.
\par
\section{Geometry of $Q^{111}$, $M^{111}$ and abelian gauge theories}
\par
In this section I perform a geometrical analysis of the $Q^{111}$, $M^{111}$ manifolds deeper 
than that given in chapter $3$. In several points this is only sketched, without proofs.
More details and proofs (and more mathematical rigor) can be found in \cite{noi3}.
\par
\subsection{$Q^{111}$ and $M^{111}$ as fiber bundles and toric manifolds}
\par
\subsubsection{$Q^{111}$ and $M^{111}$ as fiber bundles}
\par
As a premise, I remind the well known result that 
the complex projective spaces $\IP_1,~\IP_2$ are isomorphic to the following coset manifolds:
\beq
\IP_1&=&{SU\ll(2\rr)\over U\ll(1\rr)}\nn\\
\IP_2&=&{SU\ll(3\rr)\over SU\ll(2\rr)\times U\ll(1\rr)}.
\eeq
Now, let us start our geometric analysis.
\par
The manifold 
\be
Q^{111}={SU\ll(2\rr)\times SU\ll(2\rr)\times SU\ll(2\rr)\over U\ll(1\rr)\times U\ll(1\rr)}
\ee
is a fiber bundle with base space
\be
{SU\ll(2\rr)\times SU\ll(2\rr)\times SU\ll(2\rr)\over U\ll(1\rr)\times U\ll(1\rr)\times U\ll(1\rr)}=\IP_1\times \IP_1\times \IP_1
\ee
and fiber $U\ll(1\rr)$, namely,
\be
Q^{111}=E\ll(\IP_1\times\IP_1\times\IP_1,U\ll(1\rr)\rr).
\ee
The description of the fibration encodes the same information as the numbers $\ll(p,q,r\rr)=\ll(1,1,1\rr)$
in the description of chapter $3$, as I will show afterwards with the help of toric geometry.
\par
If we extend the fibration from $U\ll(1\rr)$ to
\be
U\ll(1\rr)\times\IR^+ = \IC^*,
\ee
we find the conifold (\ref{cone}) on $Q^{111}$, with $\IR^+$ parametrized by the coordinate $r$; so we have
\be
\cC\ll(Q^{111}\rr)=E\ll(\IP_1\times\IP_1\times\IP_1,\IC^*\rr).
\ee
\par
The manifold 
\be
M^{111}={SU\ll(3\rr)\times SU\ll(2\rr)\times U\ll(1\rr)\over SU\ll(2\rr)\times U\ll(1\rr)\times U\ll(1\rr)}
\ee
is a fiber bundle with base space
\be
{SU\ll(3\rr)\times SU\ll(2\rr)\times U\ll(1\rr)\over SU\ll(2\rr)\times U\ll(1\rr)\times U\ll(1\rr)\times U\ll(1\rr)}=\IP_2\times \IP_1\
\ee
and fiber $U\ll(1\rr)$, namely,
\be
M^{111}=E\ll(\IP_2\times\IP_1,U\ll(1\rr)\rr).
\ee
The description of the fibration encodes the same information as the numbers $\ll(p,q,r\rr)=\ll(1,1,1\rr)$
in the description of chapter $3$, as I will show afterwards with the help of toric geometry.
\par
If we extend the fibration from $U\ll(1\rr)$ to $U\ll(1\rr)\times\IR^+$,
we find the conifold (\ref{cone}) on $M^{111}$:
\be
\cC\ll(M^{111}\rr)=E\ll(\IP_2\times\IP_1,\IC^*\rr).
\ee
\par
\subsubsection{Toric manifolds}
\par
I do not review here the theory of toric manifolds (for a complete treatment of
toric geometry see \cite{fulton}),
I use only few concepts of that theory which are useful in our derivation. 
In general, a toric manifold can be seen as a manifold
\be
\label{toricdef}
{\IC^n/F\over\ll(\IC^*\rr)^k},
\ee
where $F\subset\IC^n$ is a null measure set. For simplicity, in the following
we will not consider $F$, even if in a rigorous treatment it should be taken into 
account.
\par
The toric manifold (\ref{toricdef}) can be parametrized by $n$ complex coordinates
\be
\ll(X_1,X_2,\dots,X_n\rr)
\label{torcoords}
\ee
on which $k$ equivalence relations are defined, describing the action of $\ll(\IC^*\rr)^k$ on $\IC^n$:
\beq
\label{toreqrels}
&(X_1,X_2,\dots,X_n)\sim \ll((\lambda_1^{p^1_1}\lambda_2^{p^1_2}\cdots
\lambda_k^{p^1_k})X_1,~
(\lambda_1^{p^2_1}\lambda_2^{p^2_2}\cdots\lambda_k^{p^2_k})X_2,\dots,
(\lambda_1^{p^n_1}\lambda_2^{p^n_2}\cdots\lambda_k^{p^n_k})X_n\rr)\nn\\
&\ll(\lambda_1,\dots,\lambda_k\rr)\in\ll(\IC^*\rr)^k.
\eeq
The matrix
\be
\ll(\begin{array}{cccc}
p^1_1 & p^1_2 & \cdots & p^1_k \\
p^2_1 & p^2_2 & \cdots & p^2_k \\
\dots & \dots & \dots & \dots \\
p^n_1 & p^n_2 & \dots & p^n_k \\
\end{array}\rr)
\label{matricetta1}
\ee
codifies the embedding of $\ll(\IC^*\rr)^k$ in $\IC^n$. 
For example, for $k=1$ the matrix with one 
column whose all the entries are equal to $1$ represents the projective space 
\be
\IP^{n-1}={\IC^n\over\IC^*}.
\ee
The other toric manifolds can be seen as generalizations of the projective spaces.
\par
\subsubsection{$\cC\ll(Q^{111}\rr)$ as a toric manifold}
\par
The base space of $\cC\ll(Q^{111}\rr)$ is a toric manifold
\be
\IP_1\times\IP_1\times\IP_1={\IC^2\over\IC^*}\times{\IC^2\over\IC^*}\times{\IC^2\over\IC^*}=
{\IC^6\over\ll(\IC^*\rr)^3}\,,
\ee
which can be described by six homogeneous coordinates 
\beq
&\ll(A_i,B_i,C_i\rr)~~~i=1,2\nn\\
&\ll(A_i,B_i,C_i\rr)\sim\ll(\lambda_1 A_i,\lambda_2 B_i, \lambda_3 C_i\rr),
\eeq
where each couple of coordinates describes one of the three $\IP_1$ factors.
This is a toric manifold described by the matrix
\be
\ll(\begin{array}{ccc}
1&0&0\\ 0&1&0\\ 0&0&1\\
\end{array}\rr)
\ee
(to be precise, it has six rows and columns, not three, but they are equal in pairs).
$\cC\ll(Q^{111}\rr)$, being a fiber bundle with base space $\IP_1\times\IP_1\times\IP_1$ and fiber $\IC^*$,
is a toric manifold with one $\IC^*$ less in the denominator:
\be
\label{toricCQ}
\cC\ll(Q^{111}\rr)=E\ll({\IC^6\over\ll(\IC^*\rr)^3},\IC^*\rr)={\IC^6\over\ll(\IC^*\rr)^2}.
\ee
It is simple to see in this context how the fiber bundle structure can implement the information of the
embedding of $H=\ll(U\ll(1\rr)\rr)^3$ in $G=\ll(SU\ll(2\rr)\rr)^3\times U\ll(1\rr)$ yielding $Q^{111}$, namely, the
choice $p=q=r=1$. 
To describe the fibration, we have to add a further coordinate on the matrix representing the toric manifold,
the coordinate $y$. On the coordinates $\ll(A_i,B_i,C_i,y\rr)$ there is an action of a
$\ll(\IC^*\rr)^3$ group, whose compact part is the action of the $\ll(U\ll(1\rr)\rr)^3$ group in $Q^{111}$.
This latter action is generated by (see chapter $3$)
\beq
Z'&=&-{\ii\over 2\sqrt{3}}\ll(\sigma^{\ll(1\rr)}_3-\sigma^{\ll(3\rr)}_3\rr)\nn\\
Z''&=&-{\ii\over 2\sqrt{3}}\ll(-\sigma^{\ll(1\rr)}_3+\sigma^{\ll(2\rr)}_3\rr)\nn\\
Z'''&=&-{\ii\over 2\sqrt{3}}\ll(\sigma^{\ll(1\rr)}_3+\sigma^{\ll(2\rr)}_3+\sigma^{\ll(3\rr)}_3
\rr)-\ii{\sqrt{3}\over 2} Y.
\eeq
Then the toric manifold (\ref{toricCQ}) is described by
\be
\ll(\begin{array}{ccc|c}
1&0&-1&0\\ -1&1&0&0\\ 1&1&1&-3\\
\end{array}\rr)\,.
\ee
We can eliminate the coordinate $y$ by fixing $\lambda_3=-1/3y$, getting a matrix with a row and a column less:
\be
\ll(\begin{array}{ccc}
1&0&-1\\ -1&1&0\\ 
\end{array}\rr)\,.
\ee
This matrix, describing the toric form of $\cC\ll(Q^{111}\rr)$, 
codifies the choice of the embedding $H\subset G$ for the $Q^{111}$ space.
\par
To retain the $\ZZ_3$ symmetry which exchange the three $\IP_1$ factors, we prefer to maintain the three
equivalence relations, making them dependent. So the matrix representing $\cC\ll(Q^{111}\rr)$ becomes
\be
\label{matrixQ}
\ll(\begin{array}{ccc}
1&0&-1\\ -1&1&0 \\ 0& -1 & 1\\
\end{array}\rr),
\ee
which means that the equivalence relations are
\be
\label{eqrelCQ}
\ll(A_i,B_i,C_i\rr)\sim\ll(\lambda_1\lambda_3^{-1} A_i,\lambda_1^{-1}\lambda_2 B_i, \lambda_2^{-1}\lambda_3 C_i\rr),
~~~\lambda_1,\lambda_2,\lambda_3\in\IC^*.
\ee
The matrix (\ref{matrixQ}) has rank $2$, and the group in the denominator of the coset (\ref{toricCQ}) can be written as
\be
\ll(\IC^*\rr)^2={\ll(\IC^*\rr)^3\over\IC^*_{diag}}.
\ee
We can fix the $\ll(\IR^+\rr)^2\subset\ll(\IC^*\rr)^2$ gauge in the manifold (\ref{toricCQ}) by imposing the further condition
on the coordinates
\footnote{If furthermore we impose
\be
\ll|A_1\rr|^2+\ll|A_2\rr|^2=1
\ee
we fix another $\IR^+$ gauge, and we get the $Q^{111}$ space itself, but we will not do that, being mainly
interested in the cone that is the space transverse to the stack of $M2$--branes.}
\beq
\ll|A_1\rr|^2+\ll|A_2\rr|^2&=&\ll|B_1\rr|^2+\ll|B_2\rr|^2\nn\\
\ll|B_1\rr|^2+\ll|B_2\rr|^2&=&\ll|C_1\rr|^2+\ll|C_2\rr|^2.
\label{DtermQ}
\eeq
If we interpret the toric description of the cone as a K\"ahler quotient, the (\ref{DtermQ}) equations have the interpretation of $D$--terms.
\par
Summarizing, the cone on $Q^{111}$ can be described by the six complex coordinates $(A_i,B_i,C_i)$ with the constraints (\ref{DtermQ}) and
the equivalence relations
\be
\label{eqrelQ}
\ll(A_i,B_i,C_i\rr)\sim\ll(e^{\ii\a}e^{-\ii\g} A_i,e^{-\ii\a}e^{\ii\b} B_i, e^{-\ii\b}e^{\ii\g} C_i\rr),~~~\a,\b,\g\in\IR.
\ee
Under the actions of these three $U\ll(1\rr)$'s
the coordinates of $Q^{111}$ have the following charges:
\beq
A_i&:&~~\ll(1,-1,0\rr)\nn\\
B_i&:&~~\ll(0,1,-1\rr)\nn\\
C_i&:&~~\ll(-1,0,1\rr).\label{ABCabeliancharges}
\eeq
I stress that one of these $U\ll(1\rr)$'s is decoupled and has no role in our discussion. The group acting on $\IC^6/\ll(\IR^+\rr)^2$ is 
\be
U\ll(1\rr)^2={U\ll(1\rr)^3\over U\ll(1\rr)_{diag}}.
\ee
\par
\subsubsection{$\cC\ll(M^{111}\rr)$ as a toric manifold}
\par
The base space of $\cC\ll(M^{111}\rr)$ is a toric manifold
\be
\IP_2\times\IP_1={\IC^5\over\ll(\IC^*\rr)^2},
\ee
which can be described by five homogeneous coordinates 
\beq
&\ll(U_i,V_A\rr)~~~i=1,2,3;~~A=1,2\nn\\
&\ll(U_i,V_A\rr)\sim\ll(\lambda_1 U_i,\lambda_2 V_A\rr).
\eeq
where $U_i$ and $V_A$ describe the $\IP_2$ and $\IP_1$ factors respectively.
This is a toric manifold described by the matrix
\be
\ll(\begin{array}{cc}
1&0\\ 0&1\\ 
\end{array}\rr)
\ee
(which actually has five rows and columns).
$\cC\ll(M^{111}\rr)$, being a fiber bundle with base space $\IP_2\times\IP_1$ and fiber $\IC^*$,
is a toric manifold with one $\IC^*$ less in the denominator:
\be
\label{toricCM}
\cC\ll(M^{111}\rr)=E\ll({\IC^5\over\ll(\IC^*\rr)^2},\IC^*\rr)={\IC^5\over\IC^*}.
\ee
The matrix describing the toric form of $\cC\ll(M^{111}\rr)$
codifies the choice of the embedding $H\subset G$ for the $M^{111}$ space.
\par
As in the $Q^{111}$ case, we can derive the toric description of this manifold, finding  
how the fiber bundle structure can implement the information of the
embedding of $H=U\ll(1\rr)\times U\ll(1\rr)$ in $G=SU\ll(3\rr)\times SU\ll(2\rr)\times U\ll(1\rr)$ in $M^{111}$, namely, the
choice $p=q=r=1$. 
To describe the fibration, we have to add a further coordinate on the matrix representing the toric manifold,
the coordinate $y$. On the coordinates $\ll(U_i,V_A,y\rr)$ there is an action of a
$\ll(\IC^*\rr)^2$ group, whose compact part is the action of the $\ll(U\ll(1\rr)\rr)^2$ group in $M^{111}$.
This latter action is generated by (see chapter $3$)
\beq
Z'&=&\ii\sqrt{3}\lambda_8+\ii\sigma_3-4\ii Y\nn\\
Z''&=&-\ii{\sqrt{3}\over 2}\lambda_8+{3\over 2}\ii\sigma_3.
\eeq
Here we have to keep attention to the normalizations. The explicit forms of these generators are
\beq
Z'&=&\ii~{\rm diag}\ll(1,1,-2,1,-1,-4\rr)\nn\\
Z''&=&{\ii\over 2}~{\rm diag}\ll(-1,-1,2,3,-3,0\rr).
\eeq
Then the equivalence relations of these generators on the coordinates $\ll(U_i,V_A,y\rr)$ are (in toric language)
\be
\ll(\begin{array}{cc|c}
2&1&-4\\ -2&3&0\\ 
\end{array}\rr)\,.
\ee
We can eliminate the coordinate $y$ by fixing $\lambda_1=-1/{4y}$, getting a matrix with a row and a column less:
\be
\ll(\begin{array}{cc}
-2&3\\
\end{array}\rr).
\ee
In analogy with the $Q^{111}$ case (and in order to get reasonable results in the non--abelian extension) we prefer to maintain the two
equivalence relations, making them dependent. So the matrix representing $\cC\ll(M^{111}\rr)$ becomes
\be
\label{matrixM}
\ll(\begin{array}{ccc}
2&-3\\ -2&3 \\
\end{array}\rr),
\ee
which means that the equivalence relations are
\be
\ll(U_i,V_A\rr)\sim\ll(\ll(\lambda_1\rr)^2\ll(\lambda_2\rr)^{-2} U_i,\ll(\lambda_1\rr)^{-3}\ll(\lambda_2\rr)^3 V_A\rr)
\ee
(that is, defining $\rho=\lambda_1/\lambda_2$, $\ll(U_i,V_A\rr)\sim\ll(\rho^2 U_i,\rho^{-3} V_A\rr)$).
The matrix (\ref{matrixM}) has rank $1$, and the group in the denominator of the coset (\ref{toricCM}) is
\be
\IC^*={\ll(\IC^*\rr)^2\over\IC^*_{diag}}.
\ee
We can fix the $\IR^+\subset\IC^*$ gauge in the manifold (\ref{toricCM}) by imposing the further condition
on the coordinates
\footnote{If furthermore we impose
\be
\ll|V_1\rr|^2+\ll|V_2\rr|^2=1
\ee
we fix another $\IR^+$ gauge, and we get the $M^{111}$ space itself.}
\be
\ll|U_1\rr|^2+\ll|U_2\rr|^2+\ll|U_3\rr|^2=\ll|V_1\rr|^2+\ll|V_2\rr|^2.
\label{DtermM}
\ee
If we interpret the toric description of the cone as a K\"ahler quotient, the
(\ref{DtermM}) equation has the interpretation of $D$--term.
\par
Summarizing, the cone on $M^{111}$ can be described by the five complex coordinates $(U_i,V_A)$ with the constraint (\ref{DtermM}) and
the equivalence relations
\be
\label{eqrelM}
\ll(U_i,V_A\rr)\sim\ll(e^{2\ii\a}e^{-2\ii\b} U_i,e^{-3\ii\a}e^{3\ii\b} V_A,\rr),~~~\a,\b\in\IR.
\ee
Under the actions of these two dependent $U\ll(1\rr)$'s
the coordinates of $M^{111}$ have the following charges:
\beq
U_i&:&~~\ll(2,-2\rr)\nn\\
V_A&:&~~\ll(-3,3\rr).\label{UVabeliancharges}
\eeq
I stress that one of these $U\ll(1\rr)$'s is decoupled and has no role in our discussion. The group acting on $\IC^5/\ll(\IR^+\rr)^2$ is 
\be
U\ll(1\rr)={U\ll(1\rr)^2\over U\ll(1\rr)_{diag}}.
\ee
\par
\subsection{The abelian theories}
\par
\label{abeliantheories}
\subsubsection{The abelian theory for $Q^{111}$}
\par
Given the toric description of $\cC\ll(Q^{111}\rr)$, the identification of an abelian ${\cal N}=2$
gauge theory whose Higgs branch reproduces the conifold is straightforward.
The fields appearing in the toric description should represent the
fundamental degrees of freedom of the gauge theory. They have definite
transformation properties under the gauge group. Out of
them we can also build some gauge invariant combinations, which should
represent the operators of the conformal theory and which should
be matched with the KK spectrum. Geometrically, this
corresponds to describing the cone as an affine submanifold of some
${\IC}^p$.
This is a standard procedure, which converts the definition of
a toric manifold in terms of D-terms to an equivalent one in terms of
binomial equations in ${\IC}^p$.
In this case, we have an embedding in ${\IC}^8$. We first construct
all the $U(1)$ invariants (in this case there are $8=2\times 2 \times 2$ of them)
\begin{equation}
X^{ijk}=A^i B^j C^k, \qquad\qquad i,j,k=1,2.
\label{embedding}
\end{equation}
They satisfy a set of binomial equations which cut out the image of
our conifold ${\cal C}(Q^{111})$ in ${\IC}^8$.
These equations are actually the $9$ quadrics
\begin{eqnarray}
0&=& \left( \epsilon \sigma^A \right)_{ij} \, X^{i \ell p} \, X^{j m q} \,
\epsilon_{\ell m} \, \epsilon_{pq}\,, \nonumber\\
0&=& \left( \epsilon \sigma^A\right)_{\ell m} \, X^{i \ell p} \, X^{j m q} \,
\epsilon_{ij} \, \epsilon_{pq}\,, \nonumber\\
0&=& \left( \epsilon \sigma^A\right)_{pq} \, X^{i \ell p} \, X^{j m q} \,
\epsilon_{ij} \, \epsilon_{\ell m}\,.
\label{sigXX}
\end{eqnarray}
Indeed, there is a general method to obtain the embedding equations of the
cones over algebraic homogeneous varieties based on representation theory.
\footnote{The $9$ equations were
already given in \cite{tatar} although their representation
theory interpretation was not given there.}  
If we want to summarize this general method (see \cite{noi3}) in few words, we can say the
following. Through eq. (\ref{embedding}) we see that the coordinates
$X^{ijk}$ of ${\IC}^8$ are assigned to a certain representation
$\cal R$ of the isometry group $SU(2)^3$. In our case such a
representation is ${\cal R}=(J^{\ll(1\rr)}=\ft{1}{2},J^{\ll(2\rr)}=\ft{1}{2},J^{\ll(3\rr)}=\ft{1}{2})$.
The products $X^{i_1j_1k_1}X^{i_2j_2k_2}$ belong to the symmetric product
$Sym^2({\cal R})$, which in general branches into various
representations, one of highest weight plus several subleading ones.
On the cone, however, only the highest weight representation survives
while all the subleading ones vanish.
Imposing that such subleading representations
are zero corresponds to writing the embedding equations.
This has far reaching consequences in the conformal field theory, since
provides the definition of the chiral ring. In principle all
the representations appearing in the $k$-th symmetric tensor power of ${\cal R}$
could correspond to primary conformal operators. Yet the attention should be restricted
to those that do not vanish modulo the equations of the cone, namely
modulo the ideal generated by the representations of subleading weights.
In other words, only the highest weight representation contained in
the $Sym^k({\cal R})$ gives a true chiral operator. 
This is what matches the Kaluza Klein spectra found through
harmonic analysis.  Two points should be stressed. In general the number of
embedding equations is larger than the codimension of the algebraic
locus. For instance $8-4 < 9$, i.e. the cone is not a complete 
intersection.
The $9$ equations (\ref{sigXX}) define the
ideal $I$ of ${\IC}[X]:={\IC}[X^{111},\ldots,X^{222}]$
cutting the cone ${\cal C}(Q^{111})$.
The second point to stress is the double interpretation of the embedding
equations.
The fact that $Q^{111}$ leads to ${\cal N}=2$
supersymmetry means that it is Sasakian, i.e. it is a circle bundle
over a suitable complex three--fold.
If considered in ${\IC}^8$ the ideal $I$ cuts out the conifold
${\cal C}(Q^{111})$. Being homogeneous, it can also be
regarded as cutting out an algebraic variety in ${\IP}^7$.
This is ${\IP}^1 \times {\IP}^1 \times {\IP}^1$,
namely the base of the $U(1)$ fibre-bundle $Q^{111}$.
\par
It follows from this discussion that the invariant operators
$X^{ijk}$ of eq. (\ref{embedding}) can be naturally associated with the building blocks
of the gauge invariant composite operators of our CFT.
Holomorphic combinations of the $X^{ijk}$ should span the set of
chiral operators of the theory.
As stated above, the set
of embedding equations (\ref{sigXX}) imposes  restrictions on the allowed
representations of $SU(2)^3$ and hence on the existing operators.
If we put the definition of $X^{ijk}$ in terms of the fundamental
fields $A,B,C$ into the equations (\ref{sigXX}), we see that they are
automatically satisfied when the theory is abelian. Since we want
eventually to promote $A,B,C$ to non-abelian fields, these equations
become non-trivial because the fields do not commute
anymore. They essentially assert that the chiral operators
we may construct out of the $X^{ijk}$ are totally symmetric in the
exchange of the various $A,B,C$, that is they belong to
the highest weight representations we mentioned above.
\par
It is clear that the two different geometric descriptions of the conifold,
the first in terms of the variables $A,B,C$ and the second in terms of
the $X$, correspond to the two possible parametrization of the moduli
space of vacua of an ${\cal N}=2$ theory, one in terms of {\it vevs} of
the fundamental fields and the second in terms of gauge invariant
chiral operators.
\par
We notice that this discussion closely parallels the analogous one
in \cite{witkleb}, \cite{nekr}. $Q^{111}$ is indeed a close relative of $T^{11}$.
\par
\subsubsection{The abelian theory for $M^{111}$}
\par
Given the toric description of $\cC\ll(M^{111}\rr)$, 
we can  identify the corresponding  abelian ${\cal N}=2$ gauge theory.
The fields $U,V$ should represent the
fundamental degrees of freedom of the gauge theory. As before, we can find a second
representation of our manifold in terms of an embedding in some ${\IC}^p$
with coordinates representing the chiral operators of our CFT.
In this case, we have an embedding in ${\IC}^{30}$. We again construct
all the $U(1)$ invariants (in this case there are 30 of them) and we
find that they are assigned to the $({\bf 10},{\bf 3})$ of
$SU(3)\times SU(2)$. The embedding equations of the conifold into ${\IC}^{30}$
correspond to the statement that in the Clebsch--Gordon expansion of
the symmetric product $({\bf 10},{\bf 3})\otimes_s ({\bf 10},{\bf3})$
all representations different from the highest
weight one should vanish. This yields $325$ equations grouped into
$5$ irreducible representations (see \cite{noi3}).
\par
As in the $Q^{111}$ case, the $X^{ij\ell\vert AB}$ can be associated
with the building blocks of the gauge invariant composite operators of our CFT
and the ideal generated by the embedding equations (see \cite{noi3}) imposes many restrictions
on the existing conformal operators. Actually, as we try to make clear
in the explicit comparison with Kaluza Klein data (see section
\ref{supertesso}), the entire spectrum is fully determined by the
structure of the ideal above. Indeed,
as it should be clear from the previous group theoretical description of
the embedding equations, the result of the constraints is to select
chiral operators which are totally symmetrized in the $SU(3)$ and $SU(2)$
indices.
\par
\section{The non-abelian theory and the comparison with KK spectrum}
\par
In the previous section, we explicitly constructed an abelian theory
whose moduli space of vacua
reproduces the cone over the two manifolds $Q^{111}$ and $M^{111}$.
These can be easily promoted to non-abelian ones.
Once this is done, we can compare the expected spectrum
of short operators in the CFT with the KK spectrum. 
\par
\subsection{The case of $Q^{111}$}
\begin{figure}[ht]
\begin{center}
\epsfxsize = 8cm
\epsffile{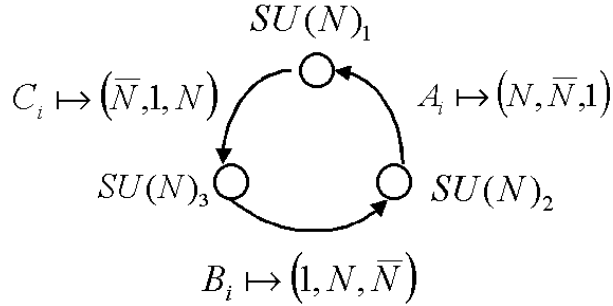}
\vskip  0.2cm
\hskip 2cm
\unitlength=1.1mm
\end{center}
\caption{Gauge group $SU(N)_1 \times SU(N)_2 \times SU(N)_3$ and colour
representation assignments of the fundamental  fields
$A_i, \, B_j,\, C_\ell$ in the $Q^{111}$ world volume gauge theory.
}
\label{ABCcolour}
\end{figure}

The theory for $Q^{111}$ becomes $SU(N)\times SU(N)\times SU(N)$ with
three series of chiral fields in the following representations of
the gauge group
\begin{equation}
A_i:\qquad ({\bf N},\bar {\bf N},{\bf 1}),\qquad\qquad
B_l:\qquad ({\bf 1,N},\bar {\bf N}),\qquad\qquad
C_p:\qquad (\bar {\bf N},{\bf 1,N})\, .\label{repps}
\end{equation}
The representations of the fundamental fields have been chosen in such a
way that they reduce to the abelian theory discussed in the previous section (eq. (\ref{ABCabeliancharges}) ).
The field content can be conveniently encoded in a quiver diagram,
where nodes represent the gauge groups and links matter fields in the
bi-fundamental representation of the groups they are connecting.
The quiver diagram for $Q^{111}$ is pictured in figure \ref{ABCcolour}.
%\iffigs
The global symmetry of the gauge theory is $SU(2)^3$,
where each of the doublets of chiral fields transforms in the fundamental
representation of one of the $SU(2)$'s.
\par
Notice that we are considering $SU(N)$ gauge group and not the naively expected
$U(N)$. The reason is that there is compelling evidence
\cite{Wittenads}, \cite{wittenaharon}, \cite{bariowit} that the $U(1)$ factors
are washed out in the near horizon limit. Since in three dimensions
$U(1)$ theories may give rise to CFT's in the IR, it is an important point
to check whether $U(1)$ factors are described by the $AdS$-solution or not.
A first piece of evidence that the supergravity solutions are dual to
$SU(N)$ theories, and not $U(N)$, comes from the absence in the KK spectrum
(even in the maximal supersymmetric case) of KK modes corresponding
to colour trace of single fundamental fields of the CFT, which are non
zero only for $U(N)$ gauge groups. A second evidence is the existence
of states dual to baryonic operators in the non-perturbative spectrum of
these Type II or M-theory compactifications; baryons
exist only for $SU(N)$ groups. We will find baryons in the spectrum of both
$Q^{111}$ and $M^{111}$: this implies that, for the compactifications
discussed in this paper, the  gauge group of the CFT is $SU(N)$.
\par
In the non-abelian case, we expect that the
generic point of the moduli space corresponds to N separated branes.
Therefore,  the space of vacua of the theory should
reduce to the symmetrization of N copies
of $Q^{111}$. To get rid of unwanted light non-abelian degrees of freedom,
we would like to introduce, following \cite{witkleb}, a superpotential
for our theory. Unfortunately, the obvious candidate for this job
\begin{equation}
\epsilon^{ij}\epsilon^{mn}\epsilon^{pq}{\rm Tr}(A_iB_mC_pA_jB_nC_q)
\label{supoQ}
\end{equation}
is identically zero. Here the close analogy with $T^{11}$ and
reference \cite{witkleb} ends.
\par
We consider now the spectrum of KK excitations of $Q^{111}$.
As we have seen in chapter $3$, there is a
chiral multiplet in the 
\be
\label{donalfonso}
J^{\ll(1\rr)}={k\over 2}~;~~J^{\ll(2\rr)}={k\over 2}~;~~J^{\ll(3\rr)}={k\over 2}
\ee
representation of $SU(2)^3$
for each integer value of k, with dimension $E_0=k$. We naturally associate
these multiplets with the series of composite operators
\begin{equation}
{\rm Tr}(ABC)^k,
\label{chiralQ}
\end{equation}
where the $SU(2)$'s indices are totally symmetrized. A first
important result, following from the existence of these hypermultiplets
in the KK spectrum, is  that the dimension of the
combination $ABC$ at the superconformal point must be 1.
\par
We see that the predictions from the KK spectrum are in perfect
agreement with the geometric discussion in the previous section.
Operators which are not totally symmetric in the flavour indices do not
appear in the spectrum. The agreement with the proposed CFT, however,
is only partial. The chiral operators predicted by supergravity certainly
exist in the gauge theory. However, we can construct many more chiral
operators which are not symmetric in flavour indices.
They do not have any counterpart in the KK spectrum.
The superpotential in the case of $T^{11}$ \cite{witkleb} had the double
purpose of getting rid of the unwanted non-abelian degrees of freedom
and of imposing, via the equations of motion, the total symmetrization
for chiral and short operators which is predicted both by geometry and by
supergravity. Here, we are not so lucky, since there is no superpotential.
We can not consider superpotentials
of dimension bigger than that considered before
(for example, cubic or quartic in $ABC$) because
the superpotential ~(\ref{supoQ}) is
the only one which has dimension compatible with the supergravity
predictions. \footnote{For a three dimensional theory to be conformal
the dimension of the superpotential must be 2.}
We need to suppose that all the non symmetric operators
are not conformal primary. Since the relation between
R-charge and dimension is only valid for conformal chiral operators,
such operators are not protected  and
therefore may have enormous anomalous dimension, disappearing
from the spectrum.
Simple examples of chiral but not conformal operators are those
obtained by derivatives of the superpotential.
Since we do not have a superpotential here, we have to suppose that
both the elimination of the unwanted coloured massless states as well
as the disappearing of the non-symmetric chiral operators
emerges as a non-perturbative IR effect.
\par
\subsection{The case of $M^{111}$}
Let us now consider $M^{111}$.
The non-abelian theory is now $SU(N)\times SU(N)$ with
chiral matter in the following representations of
the gauge group
\begin{equation}
U^i\in Sym^2({\IC}^N)\otimes Sym^2({\IC}^{N*}),\qquad\qquad
V^A\in Sym^3({\IC}^{N*})\otimes Sym^3({\IC}^N).
\label{reppsM}
\end{equation}
The representations of the fundamental fields have been chosen in such a
way that they reduce to the abelian theory discussed in the previous
section (eq. (\ref{UVabeliancharges}) ), match with the KK spectrum and imply the existence of baryons
predicted by supergravity.
Comparison with supergravity, which will be made soon,
justifies, in particular,
the choice of colour symmetric representations.
\par
\begin{figure}[ht]
\begin{center}
\epsfxsize = 8cm
\epsffile{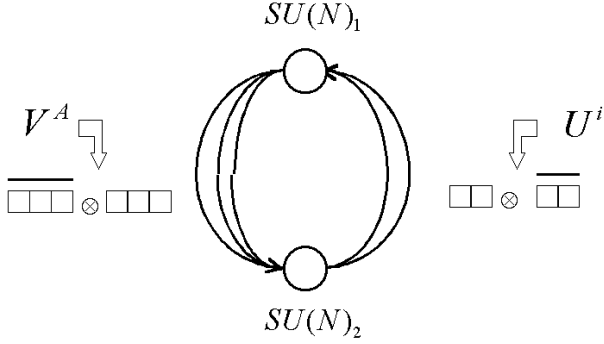}
\vskip  0.2cm
\hskip 2cm
\unitlength=1.1mm
\end{center}
\caption{Gauge group $U(N)_1 \times U(N)_2 $ and colour
representation assignments of the fundamental  fields
$V^A$ and $U^i$ in the $M^{111}$ world volume gauge theory.
}
\label{UVcolour}
\end{figure}

The field content can be conveniently encoded in the  quiver diagram
 in figure \ref{UVcolour}.
\par
%\iffigs
The global symmetry of the gauge theory is $SU(3)\times SU(2)$,
with the chiral fields $U$ and $V$ transforming in the fundamental
representation of $SU(3)$ and $SU(2)$, respectively.
\par
We next compare the expectations from gauge theory
with the KK spectrum \cite{noi1}.
Let us start with the hypermultiplet spectrum.
There is exactly one hypermultiplet in the  symmetric representation of
$SU(3)$ with $3k$ indices and the symmetric representation of $SU(2)$ with
$2k$ indices, namely,
\be
\label{donM111}
\ll[M_1,M_2,2J\rr]=\ll[3k,0,2k\rr]
\ee
for each integer $k\ge 1$.
The dimension of the operator is $E_0=2k$.
We naturally identify these states with the totally symmetrized
chiral operators
\begin{equation}
{\rm Tr} (U^3V^2)^k.
\label{chiralM}
\end{equation}
One immediate consequence of the supergravity analysis is that
the combination $U^3V^2$ has dimension 2 at the superconformal fixed point.
\par
Once again, we are not able to write any superpotential of dimension 2.
The natural candidate is the dimension two flavour singlet
\begin{equation}
\epsilon_{ijk}\epsilon_{AB}\left( U^i U^j U^k V^A V^B\right) _{\mbox{
colour singlet}}
\label{supoM}
\end{equation}
which however vanishes identically.
There is no superpotential that might help in the elimination
of unwanted light coloured degrees of freedom and that might
eliminate all the non symmetric chiral operators that we can construct out
of the fundamental fields. Once again, we have to suppose that, at the
superconformal fixed point in the IR, all the non totally symmetric
operators are not conformal primaries.
\par
\subsection{The scalar potential}
\par
\label{scalapot}
Let us now consider more closely the scalar potential of the ${\cal N}=2$ 
world--volume gauge theories we have conjectured to be associated with the
$Q^{111}$ and $M^{111}$ compactifications.
\par
In complete generality, the scalar potential of a three dimensional
${\cal N}=2$
gauge theory with an arbitrary gauge group and an arbitrary number of
chiral multiplets in generic representations of the gauge--group has the 
form (\ref{truedefU})
\begin{eqnarray}
U(z,\overline z,M)&=&
\partial_i W(z)\eta^{ij^*}\partial_{j^*}\overline W(\overline z)+\nonumber\\
&&+\ft{1}{2}g^{IJ}\left(\overline z^{i^*}(T_I)_{i^*j}z^j\right)
\left(\overline z^{k^*}(T_J)_{k^*l}z^l\right)+\nonumber\\
&&+\overline z^{i^*}M^I(T_I)_{i^*j}\eta^{jk^*}M^J(T_J)_{k^*l}
z^l+\nonumber\\
&&+2\a^2g_{IJ}M^IM^J+2\a\zeta^{\tilde I}{\cal C}_{\tilde I}^{\ I}g_{IJ}M^J
+\ft{1}{2}\zeta^{\tilde I}{\cal C}_{\tilde I}^{\ I}g_{IJ}
\zeta^{\tilde J}{\cal C}_{\tilde J}^{\ J}+\nonumber\\
&&-2\a M^I\left(\overline z^{i^*}(T_I)_{i^*j}z^j\right)
-\zeta^{\tilde I}{\cal C}_{\tilde I}^{\ I}
\left(\overline z^{i^*}(T_I)_{i^*j}z^j\right)\,.
\end{eqnarray}
If we put the Chern Simons and the Fayet Iliopoulos terms to zero
$\alpha =  \zeta^{\tilde J} =0$, the scalar potential
becomes the sum of three quadratic forms:
\begin{equation}
 U(z,\overline z,M)= |\partial W (z) |^2 + \ft{1}{2}g^{IJ}\, D_I(z,\overline z
 \,)\, D_J(z,\overline z) + M^I \, M^J \, K_{IJ}(z,\overline z),
\label{trequadra}
\end{equation}
where the real functions
\begin{equation}
D^I(z,\overline z)=-\overline z^{i^*}(T_I)_{i^*j}z^j
\end{equation}
are the $D$--terms, namely the on--shell values of the vector multiplet auxiliary
fields, while by definition we have put
\begin{equation}
   K_{IJ}(z,\overline z) \stackrel{\mbox{\scriptsize def}}{=}
   \overline z^{i^*}(T_I)_{i^*j}\eta^{jk^*}(T_J)_{k^*l}z^l.
\label{KIJ}
\end{equation}
If the quadratic form $M_I \, M_J \, K_{IJ}(z , \bar z)  $ is
positive definite, then the vacua of the gauge theory are singled out
by the three conditions
\begin{eqnarray}
\frac{\partial W}{\partial z^i} & = & 0, \label{suppoteq}\\
D^I(z,\bar z) & = & 0, \label{Dtermeq} \\
M_I \, M_J \, K_{IJ}(z , \bar z) & = & 0. \label{MMKeq}
\end{eqnarray}
The basic relation between the candidate superconformal gauge theory
$CFT_3$ and the compactifying $7$--manifold $X_7$ that we have used in
eq.s (\ref{DtermQ}, \ref{DtermM}) is that, in the Higgs branch ($ \langle M_I \rangle =0$),
the space of vacua of $CFT_3$, described by eq.s (\ref{suppoteq}, \ref{Dtermeq}, \ref{MMKeq}), should be equal to
the product of
$N$ copies of $X_7$:
\begin{equation}
  \mbox{vacua of gauge theory} = \underbrace{ X_7 \, \times \, \dots\,\times \, X_7}_{N}/\Sigma_N\,.
\label{NcopieM7}
\end{equation}
Indeed, if there are $N$ M2--branes in the game, each of them can be
placed somewhere in $X_7$ and the vacuum is described by giving all
such locations. In order for this to make sense it is necessary
that
\begin{itemize}
  \item The Higgs branch should be distinct from the Coulomb branch
  \item The vanishing of the D--terms should indeed be a geometric
  description of (\ref{NcopieM7}).
\end{itemize}
Let us apply our general formula to the two cases under consideration and see that
these conditions are indeed verified.
\subsubsection{The scalar potential in the $Q^{111}$ case}
Here the gauge group is
\begin{equation}
 {\cal G}=SU(N)_1 \times SU(N)_2 \times SU(N)_{3}
\label{Unkvolte}
\end{equation}
in the non--abelian case $N>1$ and
\begin{equation}
 {\cal G}=U(1)_1 \times U(1)_2 \times U(1)_{3}
\label{U1kvolte}
\end{equation}
in the abelian case $N=1$. The chiral fields $A_i, B_j , C_\ell$
are in the $SU(2)^3$ flavour
representations $({\bf 2},{\bf 1},{\bf 1})$, $({\bf 1},{\bf 2},{\bf 1})$, 
$({\bf 1},{\bf 1},{\bf 2})$  and in
the colour $SU(N)^3$ representations $({\bf N},\bar {\bf N},{\bf 1})$,
$({\bf 1,N},\bar {\bf N})$,
$(\bar {\bf N},{\bf 1,N})$, respectively (see fig.\ref{ABCcolour}).
We can arrange the chiral fields into a column vector:
\begin{equation}
  {\vec z}  \, = \, \left( \matrix{A_i\cr B_j \cr C_\ell \cr } \right).
\label{zabc}
\end{equation}
Naming $(t_I)_{\phantom{\Lambda}\Sigma}^{\Lambda}$ the $N \times N$ hermitian
matrices such that $\mbox{i}\, t_I$ span
the $SU(N)$ Lie algebra ($I=1,\dots,N^2-1$), the generators of the gauge
group acting on the chiral fields can be written as follows:
\begin{eqnarray}
  T^{[1]}_I &=& \left( \begin{array}{ccc}
    t_I \otimes {\bf 1} & 0 & 0 \\
    0 & 0 & 0 \\
    0 & 0 & -{\bf 1} \otimes t_I \
  \end{array}\right) , \quad T^{[2]}_I = \left( \begin{array}{ccc}
    -{\bf 1} \otimes t_I & 0 & 0 \\
    0 & t_I \otimes {\bf 1}& 0 \\
    0 & 0 & 0 \
  \end{array}\right), \nonumber\\
  \quad  T^{[3]}_I &=& \left( \begin{array}{ccc}
    0 & 0 & 0 \\
    0 & -{\bf 1} \otimes t_I & 0 \\
    0 & 0 & t_I \otimes {\bf 1} \
  \end{array}\right).
\label{Un3gene}
\end{eqnarray}
Then the $D^2$--terms appearing in the scalar potential take the
following form:
\begin{eqnarray}
 \mbox{$D^2$-terms}&=& \ft{1}{2}\,\Bigl [ \sum_{I=1}^{N^2-1} \, \left(
 {\bar A}^i \,\left(  t_I \otimes {\bf 1}\right)  \,
 A_i - {\bar C}^i \, \left( {\bf 1} \otimes t_I \right) \, C_i \right)
 ^2+\nonumber\\
 &&
 +\sum_{I=1}^{N^2-1} \, \left( {\bar B}^i \,\left(  t_I\otimes {\bf 1}\right)  \,
 B_i - {\bar A}^i \, \left( {\bf 1} \otimes t_I \right) \, A_i \right) ^2+\nonumber\\
 &&+\sum_{I=1}^{N^2-1} \, \left( {\bar C}^i \, \left( t_I \otimes {\bf 1}\right)
 C_i - {\bar B}^i \, \left( {\bf 1} \otimes t_I \right) \, B_i \right) ^2 \Bigr ].
\label{Dterm}
\end{eqnarray}
The part of the scalar potential involving the gauge multiplet
scalars is instead given by:
\begin{eqnarray}
\mbox{$M^2$--terms} & = & M_1^I \, M_1^J \, \left(
 {\bar A}^i \,\left(  t_I t_J \otimes {\bf 1}\right)  \,
 A_i + {\bar C}^i \, \left( {\bf 1} \otimes t_I t_J \right) \, C_i
 \right)+\nonumber\\
 &&+   M_2^I \, M_2^J \,\left( {\bar B}^i \,\left(  t_I t_J\otimes {\bf 1}\right)  \,
 B_i + {\bar A}^i \, \left( {\bf 1} \otimes t_I t_J \right) \, A_i \right) +\nonumber\\
&& + M_3^I \, M_3^J \,  \left( {\bar C}^i \, \left( t_I t_J \otimes {\bf 1}\right)
 C_i + {\bar B}^i \, \left( {\bf 1} \otimes t_I t_J \right) \, B_i
 \right)+\nonumber\\
 && - \, 2 \,M_1^I \, M_2^J \, {\bar A}^i \, \left( t_I  \otimes t_J \right)
 A_i - \, 2 \, M_2^I \, M_3^J \, {\bar B}^i \, \left( t_I  \otimes t_J \right)
 B_i +\nonumber\\
 && - \, 2 \,M_3^I \, M_1^J \, {\bar C}^i \, \left( t_I  \otimes t_J \right)
 C_i.\nn\\
\label{mmterm}
\end{eqnarray}
In the abelian case we simply get:
\begin{eqnarray}
 \mbox{$D^2$-terms}&=& \ft{1}{2}\,\Bigl [ \left (
\vert A_1 \vert^2 + \vert A_2 \vert^2- \vert  C_1 \vert^2 - \vert  C_2 \vert^2
\right)^2
 +\nonumber\\
 &&
 +\left (
\vert B_1 \vert^2 + \vert B_2 \vert^2- \vert  A_1 \vert^2 - \vert  A_2 \vert^2
\right)^2+\nonumber\\
 &&+\left (
\vert C_1 \vert^2 + \vert C_2 \vert^2- \vert  B_1 \vert^2 - \vert  B_1 \vert^2
\right)^2 \Bigr ],
\label{Dtermab}
\end{eqnarray}
\begin{eqnarray}
 \mbox{$M^2$-terms}&=&  \Bigl [ \left (
\vert A_1 \vert^2 + \vert A_2 \vert^2\right)  (M_1 - M_2)^2
+\nonumber\\
&&+\left (
\vert B_1 \vert^2 + \vert B_2 \vert^2\right)  (M_2 - M_3)^2
+\nonumber\\
&&+ \left( \vert  C_1 \vert^2 + \vert  C_2 \vert^2\right)  (M_3 -M_1)^2
 \Bigr ].
\label{mmtermab}
\end{eqnarray}
Eq.s (\ref{Dtermab}) and (\ref{mmtermab}) are what we have used in our
toric description of  $Q^{111}$ as the manifold of gauge--theory
vacua in the Higgs branch. Indeed it is evident from
eq. (\ref{mmtermab}) that if  we give non vanishing {\it vev} to the
chiral fields, then we are forced to put \linebreak $<M_1>=<M_2>=<M_3>=m$.
Alternatively, if we give non trivial {\it vevs} to the vector
multiplet scalars $M_i$, then we are forced to put
$<A_i>=<B_j>=<C_\ell>=0$ which confirms that the Coulomb branch is
separated from the Higgs branch.
\par
Finally, from eq.s (\ref{Dterm}, \ref{mmterm}) we can retrieve
the vacua describing  $N$ separated branes. Each chiral field has
two colour indices and is actually a matrix. Setting
\begin{eqnarray}
 < A_{i\vert \Sigma}^{\phantom{i\vert
  \Sigma}\Lambda}>&=&\delta^{\Lambda}_\Sigma \, a_i^\Lambda,
  \nonumber\\
 < B_{i\vert \Sigma}^{\phantom{i\vert
  \Sigma}\Lambda}>&=&\delta^{\Lambda}_\Sigma \, b_i^\Lambda,
  \nonumber\\
  <C_{i\vert \Sigma}^{\phantom{i\vert
  \Sigma}\Lambda}>&=&\delta^{\Lambda}_\Sigma \, c_i^\Lambda,
\label{NbraneQ}
\end{eqnarray}
a little work shows that the potential (\ref{Dterm}) vanishes if each of the
$N$--triplets
$a_i^\Lambda, b_j^\Lambda, c_\ell^\Lambda$ separately satisfies the
$D$--term equations, yielding the toric description of a $Q^{111}$
manifold (\ref{DtermQ}). Similarly, for each abelian generator
belonging to the Cartan subalgebra of $U_i(N)$ and having a non trivial action
on $a_i^\Lambda, b_j^\Lambda, c_\ell^\Lambda$  we have
$<M_1^\Lambda>=<M_2^\Lambda>=<M_3^\Lambda>=m^\Lambda$.
\subsubsection{The scalar potential in the $M^{111}$ case}
Here the gauge group is
\begin{equation}
 {\cal G}=SU(N)_1 \times SU(N)_2
\label{SUn2volte}
\end{equation}
in the non--abelian case $N>1$ and
\begin{equation}
 {\cal G}=U(1)_1 \times U(1)_2
\label{U12volte}
\end{equation}
in the abelian case $N=1$. The chiral fields $U_i, V_A$
are in the $SU(3) \times SU(2)$ flavour
representations $({\bf 3,1})$, $({\bf 1,2})$ respectively. As for colour,
they are in the $SU(N)^2$ representations
$Sym^2({\IC}^N)\otimes Sym^2({\IC}^{N*})$,
$Sym^3({\IC}^{N*})\otimes Sym^3({\IC}^N)$
respectively (see fig. \ref{UVcolour}).
As before, we can arrange the chiral fields into a column vector:
\begin{equation}
  {\vec z}  \, = \, \left( \matrix{U_i\cr V_A \cr } \right).
\label{zuuuvv}
\end{equation}
Naming $(t^{[3]}_I)_{\phantom{\Lambda\Sigma\Gamma}
\Xi\Delta\Theta}^{\Lambda\Sigma\Gamma}$ the   hermitian
matrices generating $SU(N)$ in the three--times symmetric
representation  and  $(t^{[2]}_I)_{\phantom{\Lambda\Sigma}
\Xi\Delta}^{\Lambda\Sigma}$ the same generators
in the two--times symmetric representation, the generators of the gauge
group acting on the chiral fields can be written as follows:
\begin{eqnarray}
  T^{[1]}_I &=& \left( \begin{array}{cc}
    t^{[2]}_I \otimes {\bf 1} & 0  \\
 0 & -{\bf 1} \otimes t^{[3]}_I \
  \end{array}\right) ,
  \quad T^{[2]}_I = \left( \begin{array}{cc}
    -{\bf 1} \otimes  t^{[2]}_I & 0  \\
    0 & t^{[3]}_I \otimes {\bf 1} \\
\end{array}\right).
\label{SUn2gene}
\end{eqnarray}
Then the $D^2$--terms appearing in the scalar potential take the
following form:
\begin{eqnarray}
 \mbox{$D^2$-terms}&=& \ft{1}{2}\,\Bigl [ \sum_{I=1}^{N^2-1} \, \left(
 {\bar U}^i \,\left(  t^{[2]}_I \otimes {\bf 1}\right)  \,
 U_i - {\bar V}^A \, \left( {\bf 1} \otimes t^{[3]}_I \right) \, V_A \right)^2+\nonumber\\
 &&
 +\sum_{I=1}^{N^2-1} \, \left(
 {\bar U}^i \,\left( {\bf 1} \otimes t^{[2]}_I  \right)  \,
 U_i - {\bar V}^A \, \left(  t^{[3]}_I \otimes {\bf 1}  \right) \, V_A \right)^2 \Bigr ],
\label{DtermMb}
\end{eqnarray}
while the part of the scalar potential involving the gauge multiplet
scalars is given by
\begin{eqnarray}
\mbox{$M^2$--terms} & = & M_1^I \, M_1^J \, \left(
 {\bar U}^i \,\left(  t^{[2]}_I t^{[2]}_J \otimes {\bf 1}\right)  \,
 U_i + {\bar V}^A \, \left( {\bf 1} \otimes  t^{[3]}_I t^{[3]}_J \right) \,
 V_A
 \right)+\nonumber\\
 &&+   M_2^I \, M_2^J \, \left(
 {\bar U}^i \,\left( {\bf 1} \otimes t^{[2]}_I t^{[2]}_J  \right)  \,
 U_i + {\bar V}^A \, \left( t^{[3]}_I t^{[3]}_J \otimes {\bf 1}    \right) \,
 V_A \right) +\nonumber\\
 && - \, 2 \,M_1^I \, M_2^J \, {\bar U}^i \, \left( t^{[2]}_I  \otimes t^{[2]}_J \right)
 U_i - \, 2 \, M_2^I \, M_1^J \, {\bar V}^A \, \left( t^{[3]}_I  \otimes t^{[3]}_J \right)
 V_A.\nonumber\\
\label{mmtermM}
\end{eqnarray}
In the abelian case we simply get
\begin{eqnarray}
 \mbox{$D^2$-terms}&=& \ft{1}{2}\,\Bigl \{ \left [ 2
\left( \vert U_1 \vert^2 + \vert U_2 \vert^2+ \vert  U_3 \vert^2\right)  -
3 \left( \vert  V_1 \vert^2 + \vert  V_1 \vert^2\right)
\right]^2
 +\nonumber\\
 &&
 +\left [ 2
\left( \vert U_1 \vert^2 + \vert U_2 \vert^2+ \vert  U_3 \vert^2\right)  -
3 \left( \vert  V_1 \vert^2 + \vert  V_2 \vert^2\right)
\right]^2 \Bigr \},
\label{DtermabM}
\end{eqnarray}
\begin{eqnarray}
 \mbox{$M^2$-terms}&=&   \left [
4 \left( \vert U_1 \vert^2 + \vert U_2 \vert^2 +\vert U_3 \vert^2\right)
+9 \left( \vert V_1 \vert^2 + \vert V_2 \vert^2 \right) \right] (M_1 - M_2)^2.
\label{mmtermabM}\nonumber\\
\end{eqnarray}
Once again from eq.s (\ref{DtermabM}) and (\ref{mmtermabM}) we see
that the Higgs and Coulomb branches are separated.
Furthermore, in eq. (\ref{DtermabM}) we recognize  the
toric description of  $M^{111}$ as the manifold of gauge--theory
vacua in the Higgs branch (see eq. (\ref{DtermMb})).
\par
As before, from eq.s (\ref{DtermabM}, \ref{mmtermabM}) we can retrieve
the vacua describing  $N$ separated branes. In this case the colour
index structure is more involved and we must set
\begin{eqnarray}
 < U_{i\vert \Lambda\Lambda}^{\phantom{i\vert
  \Lambda\Lambda}\Lambda\Lambda}>&=& \, u_i^\Lambda,
  \nonumber\\
 < V_{A\vert \Lambda\Lambda\Lambda}^{\phantom{i\vert
  \Lambda\Lambda\Lambda}\Lambda\Lambda\Lambda}>&=& \, v_A^\Lambda.
\label{NbraneM}
\end{eqnarray}
A little work shows that the potential (\ref{Dterm}) vanishes if each of the
$N$--doublets
$u_i^\Lambda, v_A^\Lambda $ separately satisfies the
$D$--term equations yielding the toric description of a $M^{111}$
manifold (\ref{DtermMb}). Similarly, for each abelian generator
belonging to the Cartan subalgebra of $U_i(N)$ and having a non trivial action
on $u_i^\Lambda, v_A^\Lambda $  we have
$<M_1^\Lambda>=<M_2^\Lambda>=m^\Lambda$.
\par
\subsection{Conformal superfields and comparison with the KK spectrum}
\par
\label{supertesso}
Starting from the choice of the fundamental fields of the gauge theory and of the
chiral ring (inherited from the geometry of the
compact manifold), we can build all sort of candidate conformal superfields
for both theories $M^{111}$ and $Q^{111}$. In the first case,
where the full spectrum of $Osp(2|4)\times SU(3)\times SU(2)$
supermultiplets has already been determined through harmonic analysis
(see chapter $3$, \cite{noi1}), relying on the conversion vocabulary between $AdS_4$
bulk supermultiplets and boundary superfields established in section \ref{vocabulary} \cite{noi2}, we can make a detailed comparison
of the Kaluza Klein predictions with the candidate conformal
superfields available in the gauge theory. In particular we find the
gauge theory interpretation of the entire spectrum of short
multiplets. The corresponding short superfields are in the right
$SU(3)\times SU(2)$ representations and have the right conformal
dimensions. Applying the same scheme to the case of $Q^{111}$, we
can use the gauge theory to make predictions about the spectrum of
short multiplets one should find in Kaluza Klein harmonic expansions.
The partial results already known from harmonic analysis on
$Q^{111}$ are in agreement with these predictions.
\par
In addition, looking at the $M^{111}$ spectrum, one finds
that there is a rich collection of long multiplets whose conformal
dimensions are rational and seem to be protected from acquiring
quantum corrections. This is in full analogy with results
obtained in the four--dimensional theory associated with the
$T^{11}$ manifold \cite{gubser}, \cite{sergiotorino}. 
Actually, we find an even larger class of
such {\sl rational} long multiplets. For a subclass of them the gauge
theory interpretation is clear while for others it is not immediate.
Their presence, which seems universal in all coset models, indicates
some general protection mechanism that has still to be clarified.
\par
The fundamental superfields of the $M^{111}$ theory are the
following ones:
\begin{eqnarray}
U^{i\vert\Lambda\Sigma}_{\phantom{i\vert\Lambda\Sigma} \underline{\Gamma\Delta}}(x,\theta)
&=&u^{i\vert\Lambda\Sigma}_{\phantom{i\vert\Lambda\Sigma} \underline{\Gamma\Delta}}(x)+
\left(\lambda_u^{\alpha}\right)^{i\vert\Lambda\Sigma}_{\phantom{i\vert\Lambda\Sigma}
\underline{\Gamma\Delta}}(x)~\theta^+_{\alpha}\nonumber,\\
V^{A\vert\underline{\Gamma\Delta\Theta}}_{\phantom{A\vert\underline{\Gamma\Delta\Theta}}
\Lambda\Sigma\Pi}(x,\theta)
&=&v^{A\vert\underline{\Gamma\Delta\Theta}}_{\phantom{A\vert\underline{\Gamma\Delta\Theta}}
\Lambda\Sigma\Pi}(x)+
\left(\lambda_v^{\alpha}\right)^{A\vert
\underline{\Gamma\Delta\Theta}}_{\phantom{A\vert\underline{\Gamma\Delta\Theta}}
\Lambda\Sigma\Pi}(x)~\theta^+_{\alpha},
\label{supsingM}
\end{eqnarray}
where $(i,A)$ are $SU(3)\times SU(2)$ {\sl flavour} indices, $(\Lambda,
\underline{\Lambda})$ are
$SU(N)\times SU(N)$ {\sl colour} indices while  $\alpha$ is a world volume
spinorial index of $SO(1,2)$.
The fundamental superfields are chiral superfields, so they satisfy $E_0=|y_0|$.
\par
$U^i$ is in the fundamental representation ${\bf 3}$  of
$SU(3)_{\rm flavour}$ and in the $\left( \Box\!\Box,{\Box\!\Box}^\star\right) $ of
$(SU(N)\times SU(N))_{\rm colour}$.
$V^A$ is in the fundamental representation ${\bf 2}$  of
$SU(2)_{\rm flavour}$ and in the $\left(
{\Box\!\Box\!\Box}^\star,
\Box\!\Box\!\Box\right)$
of $\left(SU(N)\times SU(N)\right)_{\rm colour}$. In eq.s (\ref{supsingM})
we have followed the conventions that lower $SU(N)$ indices transform
in the fundamental representation, while upper $SU(N)$ indices
transform in the complex conjugate of the fundamental representation.
\par
In the next section, studying the non perturbative baryon state,
we will unambiguously establish the conformal weights of the
fundamental superfields $U,V$ (or, more precisely, the conformal weights of the
Clifford vacua $u,v$) that are:
\begin{equation}
E_0(u)=y_0(u)={4\over 9},~~~E_0(v)=y_0(v)={1\over 3}.
\label{hsupsingM}
\end{equation}
For the $Q^{111}$ theory the fundamental superfields are instead the
following ones:
\begin{eqnarray}
A^{\phantom{\Lambda_1}\Gamma_2}_{i_1\vert\Lambda_1}(x,\theta)
&=&a^{\phantom{\Lambda_1}\Gamma_2}_{i_1\vert\Lambda_1}(x)+
\left(\lambda_a^{\alpha}\right)^{\phantom{\Lambda_1}\Gamma_2}_{i_1\vert\Lambda_1}
(x)~\theta^+_{\alpha}\nonumber,\\
B^{\phantom{\Lambda_2}\Gamma_3}_{i_2\vert\Lambda_2}(x,\theta)
&=&b^{\phantom{\Lambda_2}\Gamma_3}_{i_2\vert\Lambda_2}(x)+
\left(\lambda_b^{\alpha}\right)^{\phantom{\Lambda_2}\Gamma_3}_{i_2\vert\Lambda_2}
(x)~\theta^+_{\alpha}\nonumber,\\
C^{\phantom{\Lambda_3}\Gamma_1}_{i_3\vert\Lambda_3}(x,\theta)
&=&c^{\phantom{\Lambda_3}\Gamma_1}_{i_3\vert\Lambda_3}(x)+
\left(\lambda_c^{\alpha}\right)^{\phantom{\Lambda_3}\Gamma_1}_{i_3\vert\Lambda_3}
(x)~\theta^+_{\alpha},
\label{supsingQ}
\end{eqnarray}
where $i_\ell$ $(\ell=1,2,3)$ are flavour indices of $SU(2)_1 \times
SU(2)_2 \times SU(2)_3$, while $\Lambda_\ell$  $(\ell=1,2,3)$ are
colour indices of $SU(N)_1 \times SU(N)_2 \times SU(N)_3$. Also in
this case we know (see next section) their conformal dimensions
through the calculation of the conformal dimension of the
baryon operators. We have:
\begin{equation}
E_0(a)=E_0(b)=E_0(c)=y_0(a) =y_0(b)=y_0(c) ={1\over 3}.
\label{hsupsingQ}
\end{equation}
\subsubsection{Chiral operators}
\par
When the gauge group is $U(1)^N$, there is a simple interpretation for
the ring of the chiral superfields: they  describe the oscillations
of the $M2-$branes in the $7$ compact transverse directions, so
they should have the form of a parametric description of the manifold.
As we have seen,
$M^{111}$ embedded in ${\IP}^{29}$, can be parametrized
by
\begin{equation}
X^{ijl\vert AB}=U^iU^jU^kV^AV^B.
\end{equation}
Furthermore, the embedding equations can be reformulated in the following
way. In a product
\begin{equation}
X^{i_1j_1l_1\vert A_1B_1}\, X^{i_2j_2l_2\vert A_2B_2}\dots X^{i_kj_kl_k\vert A_kB_k}
\label{prodchir}
\end{equation}
only the highest weight representation of $SU(3)\times SU(2)$,
that is the completely symmetric in the $SU(3)$ indices
and completely symmetric in the $SU(2)$ indices, survives.
So the ring of the chiral superfields should be composed by superfields of the form
\begin{equation}
\label{hyperr}
\Phi^{\left(i_1j_1l_1\dots i_kj_kl_k\right)\left(A_1B_1\dots A_kB_k\right)}=
\underbrace{U^{i_1}U^{j_1}U^{l_1}V^{A_1}V^{B_1}
\dots U^{i_k}U^{j_k}U^{l_k}V^{A_k}V^{B_k}}_{k}.
\end{equation}
First of all, we note that a product of chiral superfields is always
a chiral superfield, that is, a field satisfying the equation
(see section \ref{superfields})
\begin{equation}
\label{eqchiral}
{\cal D}^+_{\a}\Phi=0,
\end{equation}
whose general solution has the form
\begin{equation}
\Phi(x,\theta)=S(x)+\lambda^{\a}(x)\theta^+_{\a}+\pi(x)\theta^{+\a}\theta^+_{\a}.
\label{phisupfi}
\end{equation}
%%%%%%%%%%%%%%%%%%%%%%%%%%%%%%%%%%%%%%%%%%%%%%%%%%%
% Young tableaux %%%%%%%%%%%%%%%%%%%%%%%%%%%%%%%%%%
%%%%%%%%%%%%%%%%%%%%%%%%%%%%%%%%%%%%%%%%%%%%%%%%%%%
Following the notation of section \ref{youngconventions},
we identify the flavour representations with three nonnegative integers $M_1,~M_2,~2J$.
The superfields (\ref{hyperr}) are in the same  $Osp(2|4)\times SU(3)\times SU(2)$
representations as the bulk hypermultiplets that were determined
through harmonic analysis:
\begin{equation}
\label{reprhyper}
\cases{
M_1=3k\cr
M_2=0\cr
J=k\cr
E_0=y_0=2k\cr}
~~~k>0\,.
\end{equation}
In particular, it is worth noticing that every block $UUUVV$ is in the
$(\Box\!\Box\!\Box,\Box\!\Box)_{\rm flavour}$ and has conformal weight
\begin{equation}
\label{3492132}
3\cdot\left(4\over 9\right)+2\cdot\left(1\over 3\right)=2,
\end{equation}
as in the Kaluza Klein spectrum.
As a matter of fact, the conformal weight of a product of chiral fields
equals the sum of the weights of the single components, as in a free
field theory.
This is due to the relation $E_0=|y_0|$ satisfied by the chiral superfields and
to the additivity of the hypercharge.
\par
When the gauge group is promoted to $SU(N)\times SU(N)$, the coordinates
become tensors (see (\ref{supsingM})).
Our conclusion about the composite operators is that the only primary chiral
superfields are those which preserve the structure (\ref{hyperr}).
So, for example, the lowest lying operator is:
\begin{equation}
U^{\Lambda\Sigma}_{\phantom{\Lambda\Sigma}i\vert(\underline{\Lambda\Sigma}}
U^{\Gamma\Delta}_{\phantom{\Gamma\Delta}j\vert\underline{\Gamma\Delta}}
U^{\Theta\Xi}_{\phantom{\Theta\Xi}\ell\vert\underline{\Theta\Xi})}
V^{\underline{\Lambda\Sigma\Gamma}}_{\phantom{\underline{\Lambda\Sigma\Gamma}}A\vert
(\Lambda\Sigma\Gamma}
V^{\underline{\Delta\Theta\Xi}}_{\phantom{\underline{\Delta\Theta\Xi}}B\vert
\Delta\Theta\Xi)},
\end{equation}
where the colour indices of every $SU(N)$ are symmetrized.
The generic primary chiral superfield has the form (\ref{hyperr}),
with all the colour indices symmetrized before being contracted.
The choice of symmetrizing the colour indices is not arbitrary:
if we impose symmetrization on the
flavour indices, it necessarily follows that also the colour
indices are symmetrized (see \cite{noi3} for a proof
of this fact). Clearly, the $Osp(2|4)
\times SU(3)\times SU(2)$ representations (\ref{reprhyper}) of these
fields are the same as in the abelian case, namely those
predicted by the $AdS/CFT$ correspondence.
\par
It should be noted that in the $4$--dimensional analogue of these
theories, namely in the $T^{11}$ case \cite{witkleb} \cite{sergiotorino},
the restriction of the primary conformal fields to the
geometrical chiral ring occurs through the derivatives of the quartic
superpotential. As we already noted, in the $D=3$ theories there is no
superpotential of dimension $2$ which can be introduced and,
accordingly, the embedding equations defining the vanishing
ideal cannot be given as derivatives of a single holomorphic
"function". It follows that there is some other non perturbative
and so far unclarified mechanism that suppresses the chiral
superfields not belonging to the highest weight representations.
\par
Let us know consider the case of the $Q^{111}$ theory. Here, as
already pointed out, the complete Kaluza Klein spectrum is still under
construction \cite{N010Q111sp}. Yet the information available in the
literature, given at the end of chapter $3$, is sufficient to make a comparison between the
Kaluza Klein predictions and the gauge theory at the level of the
chiral multiplets (and also of the graviton multiplets as I show below).
Looking at table \ref{hyper}, we learn that in the
$AdS_4 \times M^{111}$ compactification, each hypermultiplet contains a
scalar state $S$ of energy label $E_0=|y_0|$, which is actually the
Clifford vacuum of the representation and corresponds to the world volume
field $S$ of eq.(\ref{phisupfi}). It is reasonable to guess that the same
happens in the $AdS_4 \times Q^{111}$ compactification.
From the general bosonic mass--formulae 
(\ref{massform}), we know that $S$ is related to traceless
deformations of the internal metric and its mass is determined by the
spectrum of the scalar laplacian on $X_7$. 
In (\ref{massform}) we have
\begin{equation}
  m^2_S=H_0 +176 -24 \, \sqrt{H_0+36}
\label{massadiS}
\end{equation}
which, combined with the general $AdS_4$ relation between scalar
masses and energy labels $16(E_0 -2)(E_0-1)=m^2$ (\ref{massenergyAdS4}), yields the formula
\begin{equation}
  E_0=\ft{3}{2} + \ft{1}{4}\sqrt{180+H_0 -24\sqrt{36+H_0}}
\label{eoformul}
\end{equation}
for the conformal weight of candidate hypermultiplets in terms of the
scalar laplacian eigenvalues. These are already known for $Q^{111}$
(see chapter $3$):
\begin{equation}
  H_0 =32 \, \left( J^{\ll(1\rr)}\ll(J^{\ll(1\rr)}+1\rr) +
J^{\ll(2\rr)}\ll(J^{\ll(2\rr)}+1\rr) + J^{\ll(3\rr)}\ll(J^{\ll(3\rr)}+1\rr)
 -\ft{1}{4}  Y^2 \right ),
\label{H0q111}
\end{equation}
where $(J^{\ll(1\rr)},J^{\ll(2\rr)},J^{\ll(3\rr)} )$ denotes the $SU(2)^3$ flavour representation and
$y$ the $R$--symmetry $U(1)$ charge. From our knowledge of the geometrical chiral ring
of $Q^{111}$ and from our
calculation of the conformal weights of the fundamental superfields, on the
gauge theory side we expect the following chiral operators:
\begin{equation}
  \Phi_{i_1 j_1 \ell_1,\dots\, i_k j_k \ell_k} = \mbox{Tr}\, \left(
  A_{i_1} B_{j_1} C_{\ell_1} \, \dots \, A_{i_k} B_{j_k} C_{\ell_k} \right)
\label{phiq111}
\end{equation}
in the following $Osp(2\vert 4) \times SU(2) \times SU(2) \times
SU(2)$ representation:
\begin{eqnarray}
 Osp(2\vert 4) &:& \mbox{hypermultiplet with} \cases{ \matrix{E_0 & = & k\cr
 y_0 &=& k \cr}}\label{osprep}\\
 SU(2)\times SU(2) \times SU(2) &:& J^{\ll(1\rr)}=J^{\ll(2\rr)}=J^{\ll(3\rr)} =\ft{1}{2} \, k\\
 &&k\ge 1.\nonumber
\label{hypq111}
\end{eqnarray}
Inserting the representation (\ref{hypq111}) into eq. (\ref{H0q111})
we obtain $H_0=16 k^2 + 48 k$ and, using this value in
eq. (\ref{eoformul}), we retrieve the conformal field theory prediction
$E_0 = k$. This shows that the hypermultiplet spectrum found in
Kaluza Klein harmonic expansions on $Q^{111}$ agrees with the chiral superfields
predicted by the conformal gauge theory.
\subsubsection{Conserved currents of the world volume gauge theory}
The supergravity mass--spectrum on $AdS_4 \times X_7$, where $X_7$ is
an Einstein space admitting $\cN=2$ Killing spinors, contains a number of {\it ultrashort} or {\it massless}
$Osp(2\vert 4)$ multiplets that correspond to the unbroken local gauge
symmetries of the vacuum. These are:
\begin{enumerate}
  \item The massless ${\cal N}=2$ graviton multiplet
  \item The massless ${\cal N}=2$ vector multiplets of the flavour group
  $G'$
  \item The massless ${\cal N}=2$ vector multiplets associated with
  the non--trivial harmonic 2--forms of $X_7$ (the Betti multiplets).
\end{enumerate}
Each of these massless multiplets must have a suitable gauge theory
interpretation. Indeed, also on the gauge theory side, the ultra--short
multiplets are associated with the symmetries of the theory (global
in this case) and are given by the corresponding conserved Noether
currents.
\par
We begin with the  stress--energy  superfield
$T_{\alpha\beta}$ which has a pair of symmetric $SO(1,2)$ spinor
indices and satisfies the conservation equation
\begin{equation}
\label{eqemtenssor}
{\cal D}^+_{\a}T^{\alpha\beta}={\cal D}^-_{\a}T^{\alpha\beta}=0.
\end{equation}
In components, the $\theta$--expansion of this superfield yields  the
stress energy tensor $T_{\mu\nu}(x)$, the ${\cal N}=2$ supercurrents
$j_\mu^{A\alpha}(x)$ ($A=1,2$) and the $U(1)$ R--symmetry current
$J_\mu^R(x)$. Obviously $T^{\alpha\beta}$ is a singlet with respect to the flavour
group $G'$ and it has
\begin{equation}
E_0=2,~~y_0=0,~~s_0=1.
\label{stressE0}
\end{equation}
This corresponds to the massless graviton multiplet of the bulk and explains
the first entry in the above enumeration.
\par
To each generator of the flavour symmetry group there corresponds, via Noether
theorem, a conserved vector supercurrent. This is a scalar superfield
$J^I(x,\theta)$ transforming in the adjoint representation of the flavour group $G'$
and satisfying the conservation equations
\begin{equation}
\label{eqcurrents}
{\cal D}^{+\alpha}{\cal D}^+_{\alpha}J^I={\cal D}^{-\alpha}{\cal D}^-_{\alpha}J^I=0.
\end{equation}
These superfields have
\begin{equation}
E_0=1,~~y_0=0,~~s_0=0
\end{equation}
and correspond to the ${\cal N}=2$ massless vector multiplets of
$G'$ that propagate on the bulk. This explains the second
item of the above enumeration.
\par
In the specific theories under consideration, we can easily construct
the flavour currents  in terms of the fundamental superfields:
\begin{equation}
\begin{array}{rl}
  M^{111} & \cases{ \begin{array}{ccc}
     J^{\phantom{SU(3)\vert j}i}_{SU(3)\vert j} &
    = & U^{i\vert\Lambda\Sigma}_{\phantom{i\vert\Lambda\Sigma}
    \underline{\Lambda\Sigma}} \, \bar{U}_{j\vert\Lambda\Sigma}^{\phantom{i\vert\Lambda\Sigma}
    \underline{\Lambda\Sigma}}\, - \,
    {1\over 3}\delta^i_j \, U^{\ell\vert\Lambda\Sigma}_{\phantom{\ell\vert\Lambda\Sigma}
    \underline{\Lambda\Sigma}} \,
    \bar{U}_{\ell\vert\Lambda\Sigma}^{\phantom{\ell\vert\Lambda\Sigma}
    \underline{\Lambda\Sigma}} \\
    \null & \null & \\
   J^{\phantom{SU(2)\vert B}A}_{SU(2)\vert B} &
    = & V^{A\vert\underline{\Lambda\Sigma\Gamma}}_{
    \phantom{A\vert\underline{\Lambda\Sigma\Gamma}}
     \Lambda\Sigma\Gamma} \, \bar{V}_{B\vert\underline{\Lambda\Sigma\Gamma}}^{
    \phantom{B\vert\underline{\Lambda\Sigma\Gamma}}
     \Lambda\Sigma\Gamma} \, - \,
    {1\over 2}\delta^A_B \, V^{C\vert\underline{\Lambda\Sigma\Gamma}}_{
    \phantom{A\vert\underline{\Lambda\Sigma\Gamma}}
     \Lambda\Sigma\Gamma} \, \bar{V}_{C\vert\underline{\Lambda\Sigma\Gamma}}^{
    \phantom{C\vert\underline{\Lambda\Sigma\Gamma}}
     \Lambda\Sigma\Gamma} \, \\
  \end{array}} \\
  \null & \null \\
  Q^{111} & \cases{\begin{array}{ccc}
    J^{\phantom{SU(2)_1\vert j_1}i_1}_{SU(2)_1\vert j_1} & = &
    A^{i_1\vert\Gamma_1}_{
    \phantom{i_1\vert\Gamma_1}\Lambda_2}\, \bar{A}_{j_1\vert\Gamma_1}^{
    \phantom{j_1\vert\Gamma_1}\Lambda_2} \, - \, \ft{1}{2} \,
    \delta^{i_1}_{j_1} \, A^{\ell_1\vert\Gamma_1}_{
    \phantom{i_1\vert\Gamma_1}\Lambda_2}\, \bar{A}_{\ell_1\vert\Gamma_1}^{
    \phantom{j_1\vert\Gamma_1}\Lambda_2}
      \\
      \null & \null & \null \\
   J^{\phantom{SU(2)_2\vert j_2}i_2}_{SU(2)_2\vert j_2} & = &
    B^{i_2\vert\Gamma_2}_{
    \phantom{i_2\vert\Gamma_2}\Lambda_3}\, \bar{B}_{j_2\vert\Gamma_2}^{
    \phantom{j_1\vert\Gamma_2}\Lambda_3} \, - \, \ft{1}{2} \,
    \delta^{i_2}_{j_2} \, B^{\ell_2\vert\Gamma_2}_{
    \phantom{i_1\vert\Gamma_2}\Lambda_3}\, \bar{B}_{\ell_2\vert\Gamma_2}^{
    \phantom{j_1\vert\Gamma_2}\Lambda_3}
      \\
     \null & \null & \null \\
   J^{\phantom{SU(2)_3\vert j_3}i_3}_{SU(2)_3\vert j_3} & = &
    C^{i_3\vert\Gamma_3}_{
    \phantom{i_3\vert\Gamma_2}\Lambda_1}\, \bar{C}_{j_3\vert\Gamma_3}^{
    \phantom{j_1\vert\Gamma_3}\Lambda_1} \, - \, \ft{1}{2} \,
    \delta^{i_3}_{j_3} \, C^{\ell_3\vert\Gamma_3}_{
    \phantom{i_1\vert\Gamma_3}\Lambda_1}\, \bar{C}_{\ell_3\vert\Gamma_3}^{
    \phantom{j_1\vert\Gamma_3}\Lambda_1}~.
      \\
  \end{array}
  }
\end{array}
\label{flaviocurro}
\end{equation}
These currents satisfy eq.(\ref{eqcurrents}) and are in the right
representations of $SU(3)\times SU(2)$. Their hypercharge is $y_0=0$.
The conformal weight is not the one obtained by a naive sum,
being the theory interacting.
As we have seen in chapter $2$, the conserved currents satisfy
$E_0=|y_0|+1$, hence $E_0=1$.
\par
Let us finally identify the gauge theory superfields associated with
the Betti multiplets. As we have stressed, the non
abelian gauge theory has $SU(N)^p$ rather than $U(N)^p$ as gauge
group. The abelian gauge symmetries that were used to obtain the
toric description of the manifold $M^{111}$ and $Q^{111}$ in the one--brane case
$N=1$ are not promoted to gauge symmetries in the many brane regime
$N\to \infty$. Yet, they survive as exact global symmetries of the
gauge theory. The associated conserved currents provide the
superfields corresponding to the massless Betti multiplets
found in the Kaluza Klein spectrum of the bulk. As the reader can
notice, the $b_2$ Betti number of each manifold always agrees with the
number of independent $U(1)$ groups needed to give a toric
description of the same manifold. It is therefore fairly easy to
identify the Betti currents of our gauge theories. For instance
for the $M^{111}$ case the Betti current is
\begin{equation}
 J_{\rm Betti} \, = \, 2\,  U^{\ell\vert\Lambda\Sigma}_{\phantom{\ell\vert\Lambda\Sigma}
    \underline{\Lambda\Sigma}} \,
    \bar{U}_{\ell\vert\Lambda\Sigma}^{\phantom{\ell\vert\Lambda\Sigma}
    \underline{\Lambda\Sigma}} \, - \, 3 \, V^{C\vert\underline{\Lambda\Sigma\Gamma}}_{
    \phantom{A\vert\underline{\Lambda\Sigma\Gamma}}
     \Lambda\Sigma\Gamma} \, \bar{V}_{C\vert\underline{\Lambda\Sigma\Gamma}}^{
    \phantom{C\vert\underline{\Lambda\Sigma\Gamma}}
     \Lambda\Sigma\Gamma} \,.
\label{betcurro}
\end{equation}
The two Betti currents of $Q^{111}$ are similarly
written down from the toric description. Since the Betti
currents are conserved, according to what shown in
chapter $2$  they satisfy
$E_0=|y_0|+1$. Since the hypercharge is zero, we have $E_0=1$ and the
Betti currents provide the gauge theory interpretation of the
massless Betti multiplets.
\subsubsection{Gauge theory interpretation of the short multiplets}
\label{protcorto}
Using the massless currents above reviewed and the
chiral superfields, one has all the building blocks necessary to
construct the constrained superfields that correspond to all the
short multiplets found in the Kaluza Klein spectrum.
\par
As shown in chapter $2$, short $Osp(2\vert 4)$ multiplets
correspond to shortened superfields defined
imposing a suitable differential constraint, invariant with respect
to Poincar\'e supersymmetry \cite{noi2}. 
Using chiral superfields and conserved currents as
building blocks, we can construct candidate short superfields that
satisfy the appropriate differential constraint and the unitarity bounds  (\ref{N2bounds}).
Then we can compare their flavour representations with
those of the short multiplets obtained in Kaluza Klein expansions.
In the case of
the $M^{111}$ theory, where the Kaluza Klein spectrum is known, we
find complete agreement and hence we explicitly verify the $AdS/CFT$ correspondence.
For the
$Q^{111}$ manifold we make instead a prediction in the reverse
direction: the gauge theory realization predicts the outcome of
harmonic analysis. While we wait for the construction of the complete
spectrum \cite{N010Q111sp}, we can partially verify the correspondence
using the information available at the moment, namely the spectrum of
the scalar laplacian.
\subsubsection{Superfields corresponding to the short graviton multiplets}
The gauge theory interpretation of these multiplets is quite simple.
Consider the superfield
\begin{equation}
  \Phi_{\a\b}(x,\theta)=T_{\alpha\beta}(x,\theta) \, \Phi_{\rm chiral}(x,\theta),
\label{Phicorto}
\end{equation}
where $T_{\alpha\beta}$ is the stress energy tensor
(\ref{eqemtenssor}) and $\Phi_{\rm chiral}(x,\theta)$ is a chiral
superfield. By construction, the superfield (\ref{Phicorto}), at least in 
the abelian case, satisfies the equation
\begin{equation}
\label{eqshortgraviton}
{\cal D}^+_{\a}\Phi^{\a\b}=0
\end{equation}
and then, as shown in chapter $2$, it corresponds to a short graviton
multiplet on the bulk. It is natural to extend this identification to the
non-abelian case.
\par
Given the chiral multiplet spectrum (\ref{reprhyper})
and the dimension of the stress energy current (\ref{stressE0}), we
immediately get the spectrum of superfields (\ref{Phicorto}) for the case
$M^{111}$:
\begin{equation}
\label{reprshortgraviton}
\cases{
M_1=3k\cr
M_2=0\cr
J=k\cr
E_0=2k+2,~~y_0=2k\cr}
~~~k>0\,.
\end{equation}
This exactly coincides with the spectrum of short graviton multiplets
found in Kaluza Klein theory through harmonic analysis.
\par
For the $Q^{111}$ case  the same analysis gives the
following prediction for the short graviton multiplets:
\begin{equation}
\label{reprshortgravq111}
\cases{
J^{\ll(1\rr)}=J^{\ll(2\rr)}=J^{\ll(3\rr)}=\ft{1}{2}k\cr
E_0=k+2,~~y_0=k\cr}
~~~k>0\,.
\end{equation}
We can make a consistency check on this prediction just relying on the
spectrum of the laplacian (\ref{H0q111}). Indeed, looking at table \ref{shortgraviton},
we see that in a short graviton multiplet the
mass of the spin two particle is
\begin{equation}
  m_h^2 = 16 y_0 (y_0+3).
\label{tab4predi}
\end{equation}
Looking instead at equation (\ref{massform}), we see that such a mass
is equal to the eigenvalue of the scalar laplacian $m^2_h= H_0$.
Therefore, for consistency of the prediction (\ref{reprshortgravq111}),
we should have  $H_0=16k(k+3)$ for the representation $J^{\ll(1\rr)}=J^{\ll(2\rr)}=J^{\ll(3\rr)}=k/2;
Y=k$. This is indeed the value provided by eq. (\ref{H0q111}).
\par
It should be noted that when we write the operator (\ref{Phicorto}),
it is understood that {\it all colour indices are symmetrized before taking
the contraction.}
\subsubsection{Superfields corresponding to the short vector multiplets}
\par
Consider next the superfields of the following type:
\begin{equation}
\label{shortvectora}
\Phi (x,\theta)=J(x,\theta) \, \Phi_{\rm chiral}(x,\theta),
\end{equation}
where $J$ is a conserved vector current of the type analyzed in
eq. (\ref{flaviocurro}) and $\Phi_{\rm chiral}$ is a chiral
superfield. By construction, the superfield (\ref{shortvectora}), at least in
the abelian case,
satisfies the constraint
\begin{equation}
\label{eqshortvector}
{\cal D}^{+\a}{\cal D}^+_{\a}\Phi =0
\end{equation}
and then, according to the analysis of section \ref{vocabulary}, it can describe
a  short vector multiplet propagating into the  bulk.
\par
In principle, the flavour irreducible representations occurring in the superfield
(\ref{shortvectora}) are those originating from the tensor product decomposition
\begin{equation}
  ad  \otimes  {\cal R}_{\rho_k} = {\cal R}_{\chi_{max}} \oplus
  \sum_{\chi < \chi_{max}} {\cal R}_{\chi},
\label{chicchimax}
\end{equation}
where $ad$ is the adjoint representation, $\rho_k$ is the flavour weight of the
chiral field at level $k$, $\chi_{max}$ is the highest weight occurring in
the product $ad\otimes {\cal R}_{\rho_k}$ and $\chi < \chi_{max}$ are the lower
weights occurring in the same decomposition.
\par
Let us assume that the quantum mechanism that  suppresses
all the candidate chiral superfields of subleading weight does the
same suppression also on the short vector superfields
(\ref{shortvectora}). Then in the sum appearing on the l.h.s of eq. (\ref{chicchimax})
we keep only the first term and, as we show in a moment, we reproduce
the Kaluza Klein spectrum of short vector multiplets.
As we see, there is just a universal rule that presides at the
selection of the flavour representations in all sectors of the
spectrum. It is the restriction to the maximal weight. This is the
group theoretical implementation of the ideal that defines
the conifold as an algebraic locus in ${\IC}^p$. We already
pointed out that, differently from the $D=4$ analogue of these conformal gauge
theories, the ideal cannot be implemented through a superpotential.
An equivalent way of imposing the result is to assume that the colour indices have
to be completely symmetrized: such a symmetrization automatically
selects the highest weight flavour representations.
\par
Let us now explicitly verify the matching with Kaluza Klein spectra.
We begin with the $M^{111}$ case. Here the highest weight representations
occurring in the tensor product of the adjoint $(M_1=M_2=1,J=0)\oplus
(M_1=M_2=0,J=1)$ with the chiral spectrum (\ref{reprhyper}) are
$M_1=3k+1,M_2=1,J=k$ and $M_1=k,M_2=0,J=k+1$. Hence the spectrum of
vector fields (\ref{shortvectora}) limited to highest weights is
given by the following list of $Osp(2|4)\times SU(2)
\times SU(3)$ UIRs:
\begin{equation}
\label{reprshortvectorsu3}
\cases{
M_1=3k+1\cr
M_2=1\cr
J=k\cr
E_0 =2k+1,~~y_0=2k \cr}
~~~k>0
\end{equation}
and
\begin{equation}
\label{reprshortvectorsu2}
\cases{
M_1=3k\cr
M_2=0\cr
J=k+1\cr
E_0=2k+1,~~y_0=2k\cr}
~~~k>0\,.
\end{equation}
This is precisely the result found in chapter $3$.
\par
For the $Q^{111}$ case our gauge theory realization predicts the following short
vector multiplets:
\begin{equation}
\label{reprshortvecq111}
\cases{
J^{\ll(1\rr)}=\ft{1}{2}k+1\cr
J^{\ll(2\rr)}=\ft{1}{2}k\cr
J^{\ll(3\rr)}=\ft{1}{2}k\cr
E_0=k+1,~~y_0=k\, \cr}
~~~k>0
\end{equation}
and all the other are obtained from
(\ref{reprshortvecq111}) by permuting the role of the three $SU(2)$
groups. Looking at table \ref{hyper}, we see that in the 
${\cal N}=2$ short multiplet emerging from M--theory compactification on
$AdS_4 \times M^{111}$ the lowest energy state is a scalar $S$, and
we guess that the same happens in the $X_7=Q^{111}$ case.
It has squared mass
\begin{equation}
  m_S^2=16 y_0 (y_0 -1).
\label{ceccovect}
\end{equation}
Hence, recalling eq. (\ref{massadiS}) and combining it with
(\ref{ceccovect}), we see that for consistency of our predictions we
must have
\begin{equation}
  H_0 +176 -24 \sqrt{H_0+36} = 16 k (k-1)
\label{consishort}
\end{equation}
for the representations (\ref{reprshortvecq111}). The quadratic
equation (\ref{consishort}) implies $H_0=16 k^2 + 80 k +64$ which is
precisely the result  obtained by inserting the values (\ref{reprshortgravq111})
into Pope's formula (\ref{H0q111}) for the laplacian eigenvalues.
Hence, also the short vector multiplets seems to follow a general pattern
identical in all $\cN=2$ compactifications.
\par
We can finally wonder why there are no short vector multiplets
obtained by multiplying the Betti currents with chiral superfields.
The answer might be the following. From the flavour view point these
would not be highest weight representations occurring in the tensor
product of  the  constituent fundamental superfields. Hence they are suppressed from
the spectrum.
\subsubsection{Superfields corresponding to the short gravitino multiplets}
\par
The spectrum of $M^{111}$ derived in chapter $3$ contains
various series of short gravitino multiplets. We can provide their
gauge theory interpretation through the following superfields.
Consider:
\begin{eqnarray}
\label{shortgravitino1}
&
{\Phi'}_{\a jB}^{\left(ii_1j_1\ell_1\dots i_kj_k\ell_k\right)
\left(AC_1D_1\dots C_kD_k\right)}=
&\nonumber\\
&=\left(U\bar{U}\left({\cal D}^+_{\a}V\bar{V}\right)+
V\bar{V}\left({\cal D}^+_{\a}U\bar{U}\right)\right)^{i~A}_{~j~~B}
\underbrace{U^{i_1}U^{j_1}U^{\ell_1}V^{C_1}V^{D_1}
\dots U^{i_k}U^{j_k}U^{\ell_k}V^{C_k}V^{D_k}}_{k}&\nonumber\\
\end{eqnarray}
and
\begin{eqnarray}
\label{shortgravitino2}
&{\Phi''}_{\a}^{\left(ij\ell i_1j_1\ell_1\dots i_kj_k\ell_k\right)
\left(C_1D_1\dots C_kD_k\right)}=&\nonumber\\
&
=\left(U^iU^jU^\ell V^A{\cal D}^-_{\a}V^B\epsilon_{AB}\right)
\underbrace{U^{i_1}U^{j_1}U^{\ell_1}V^{C_1}V^{D_1}
\dots U^{i_k}U^{j_k}U^{\ell_k}V^{C_k}V^{D_k}}_{k}\,,&\nonumber\\
\end{eqnarray}
where all the colour indices are symmetrized before being contracted.
By construction the superfields (\ref{shortgravitino1},\ref{shortgravitino2}),
at least in the abelian case,
satisfy  the equation
\begin{equation}
\label{eqshortgravitino}
{\cal D}^+_{\a}\Phi^{\a}=0
\end{equation}
and then, as explained in section \ref{superfields}, they correspond to short gravitino
multiplets propagating on the bulk.
We can immediately check that their highest weight flavour
representations yield the spectrum of $Osp(2|4)\times SU(2)
\times SU(3)$ short gravitino multiplets. Indeed for (\ref{shortgravitino1}),(\ref{shortgravitino2})
we respectively have:
\begin{equation}
\label{reprshortgravitino1}
\cases{
M_1=3k+1\cr
M_2=1\cr
J=k+1\cr
E_0=2k+{5\over 2},~~y_0=2k+1\cr}
~~~k\ge 0\,,
\end{equation}
and
\begin{equation}
\label{reprshortvectorgravitino2}
\cases{
M_1=3k+3\cr
M_2=0\cr
J=k\cr
E_0=2k+{5\over 2},~~y_0=2k+1\cr}
~~~k\ge 0\,.
\end{equation}
We postpone the analysis of short gravitino multiplets on $Q^{111}$
to \cite{N010Q111sp} since this requires a more extended knowledge of
the spectrum.
\subsubsection{Long multiplets with rational protected dimensions}
\label{protlungo}
Let us now observe that, in complete analogy to what happens for the
$T^{11}$ conformal spectrum one dimension above \cite{gubser}, \cite{sergiotorino},
also in the case of $M^{111}$ there is a large class of long
multiplets with rational conformal dimensions. Actually this seems to
be a general phenomenon in all Kaluza Klein compactifications on
homogeneous spaces $G/H$. Indeed, although the $Q^{111}$ spectrum is
not yet completed \cite{N010Q111sp}, we can already see from its laplacian spectrum
(\ref{H0q111}) that a similar phenomenon occurs also there. More
precisely, while the short multiplets saturate the unitarity bound
and have a conformal weight related to the hypercharge and maximal
spin by equations  (\ref{N2bounds}), the {\it rational long multiplets}
satisfy a quantization condition of the conformal dimension of the
following form
\begin{equation}
E_0=|y_0|+s_0+1+\lambda,~~~\lambda \, \in \, {\IN}.
\label{quantcondo}
\end{equation}
\par
Inspecting the  $M^{111}$ spectrum, we find the following long rational multiplets:
\begin{itemize}
\item {\it Long rational graviton multiplets}
\par
In the series
\begin{equation}
\cases{
M_1=0,~M_2=3k,~J=k+1\cr
M_1=1,~M_2=3k+1,~J=k\cr}
\end{equation}
and conjugate ones we have
\begin{equation}
y_0=2k,~~E_0=2k+3=|y_0|+3
\end{equation}
corresponding to
\begin{equation}
\lambda=1.
\end{equation}
\item {\it Long rational gravitino multiplets}
\par
In the series of representations
\begin{equation}
M_1=1,~M_2=3k+1,~J=k+1
\end{equation}
(and conjugate ones) for the gravitino  multiplets of type $\chi^-$
 we have
\begin{equation}
y_0=2k+1,~~E_0=2k+{9\over 2}=|y_0|+{7\over 2},
\end{equation}
while in the series
\begin{equation}
M_1=0,~M_2=3k+3,~J=k
\end{equation}
(and conjugate ones)
for the same type of gravitinos we get
\begin{equation}
y_0=2k+1,~~E_0=2k+{9\over 2}=|y_0|+{7\over 2}.
\end{equation}
Both series fit into the quantization rule (\ref{quantcondo}) with:
\begin{equation}
\lambda=2.
\end{equation}
\item {\it Long rational vector multiplets}
\par
In the series
\begin{equation}
M_1=0,~M_2=3k,~J=k
\end{equation}
(and conjugate ones) for the  vector multiplets of type $W$ we have
\begin{equation}
y_0=2k,~~E_0=2k+4=|y_0|+4,
\end{equation}
that fulfills the quantization condition (\ref{quantcondo}) with
\begin{equation}
\lambda=3.
\end{equation}
For the same vector multiplets of type $W$, in the series
\begin{equation}
\cases{
M_1=0,~M_2=3k,~J=k+1\cr
M_1=1,~M_2=3k+1,~J=k\cr}
\end{equation}
(and conjugate ones) we have
\begin{equation}
y_0=2k,~~E_0=2k+10=|y_0|+10,
\end{equation}
that satisfies the quantization condition (\ref{quantcondo}) with
\begin{equation}
\lambda=9.
\end{equation}
\end{itemize}
The generalized presence of these rational long multiplets hints at
various still unexplored quantum mechanisms that, in the conformal
field theory, protect certain operators from acquiring anomalous
dimensions. At least for the long graviton multiplets, characterized by
$\lambda=1$, the corresponding protected superfields can be guessed, in
analogy whith \cite{T11spectrum}. 
If we take the superfield of a short vector multiplet
$J(x,\theta) \, \Phi_{\rm chiral}(x,\theta)$ and we multiply it by the
stress--energy superfield $T_{\alpha\beta}(x,\theta)$, namely if we consider a
superfield of the form
\begin{equation}
  \Phi \sim \mbox{conserved vector current} \, \times \, \mbox{stress energy
  tensor} \, \times \, \mbox{chiral operator},
\label{guessa}
\end{equation}
we reproduce the right $Osp(2\vert 4)\times SU(3) \times SU(2)$
representations of the long rational graviton multiplets of
$M^{111}$. The soundness of such an interpretation can be checked
by looking at the graviton multiplet spectrum on $Q^{111}$. This is
already available since it is once again determined by the laplacian
spectrum. Applying formula eq. (\ref{guessa}) to the $Q^{111}$ gauge theory
leads to predict the following spectrum of long rational multiplets:
\begin{equation}
\label{ratlongq111}
\cases{
J^{\ll(1\rr)}=\ft{1}{2}k+1\cr
J^{\ll(2\rr)}=\ft{1}{2}k\cr
J^{\ll(3\rr)}=\ft{1}{2}k\cr
E_0=k+1,~~y_0=k\, \cr}
~~~k>0
\end{equation}
and all the other are obtained from
(\ref{ratlongq111}) by permuting the role of the three $SU(2)$
groups. Looking at table \ref{longgraviton}, we see that in a
graviton multiplet the spin two particle has mass
\begin{equation}
  m_h^2=16(E_0+1)(E_0-2),
\label{emmemgrave}
\end{equation}
which for the candidate multiplets (\ref{emmemgrave}) yields
\begin{equation}
  m_h^2=16(k+4)(k+1).
\label{emmecandid}
\end{equation}
On the other hand, looking at equation (\ref{massform}) we see
that the squared mass of the graviton is just the eigenvalue of the scalar
laplacian $m_h^2=H_0$. Applying formula (\ref{H0q111}) to the
representations of (\ref{ratlongq111}) we indeed find
\begin{equation}
  H_0 = 16 k^2 +80 k +64 = 16(k+4)(k+1).
\label{bingo!}
\end{equation}
It appears, therefore, that the generation of rational long graviton multiplets
is based on the universal mechanism codified by the ansatz
(\ref{guessa}), proposed in \cite{T11spectrum} and 
applicable to all compactifications. Why these
superfields have protected conformal dimensions is still to be
clarified within the framework of the superconformal gauge theory.
The superfields  leading to rational long multiplets with much
higher values of $\lambda$, like the cases $\lambda=3$ and
$\lambda=9$ that we have found, are more difficult to guess. Yet
their appearance seems to be a general phenomenon and this, as we
have already stressed, hints at general protection mechanisms that
have still to be investigated.
\par
\section{The baryons}
\par
\label{bariobetti}
There is one important property that $M^{111}$, $Q^{111}$ and
$T^{11}$ share. These manifolds have non-zero Betti numbers ($b_2=b_5=2$
for $Q^{111}$, $b_2=b_5=1$ for $M^{111}$ and $b_2=b_3=1$ for $T^{11}$).
This implies the existence of non-perturbative states in the supergravity
spectrum associated with branes wrapped on non-trivial cycles. They
can be interpreted as baryons in the CFT \cite{bariowit} \cite{gubserkleb}.
\par
The existence of non-zero Betti numbers implies the existence of new
global $U(1)$ symmetries which do not come from the geometrical symmetries
of the coset manifold, as was pointed out long time ago.
The massless vector multiplets associated with these symmetries
were discovered in \cite{univer}, \cite{spec321}.
They  have the property that the entire KK spectrum is neutral
and only non-perturbative states can be charged.
The massless vectors, dual to the conserved currents, arise from the
reduction of the 11-dimensional 3-form
along the non-trivial 2-cycles. This definition implies that
non-perturbative objects made with M2 and M5 branes are charged
under these $U(1)$ symmetries.
\par
We can identify the Betti multiplets 
with baryonic symmetries. This was first pointed out in \cite{witkleb2}, \cite{sergiotorino}
for the case of $T^{11}$ and discussed for orbifold models in
\cite{morpless}. The existence of baryons in the proposed CFT's
is due to the choice of $SU(N)$ (as opposed to $U(N)$) as gauge group.
In the $SU(N)$ case, we can form the
gauge invariant operators $\mbox{det}\, (A)$, $\mbox{det}\, (B)$ and $\mbox{det}\, (C)$ for
$Q^{111}$ and $\mbox{det}\, (U)$ and $\mbox{det}\, (V)$ for $M^{111}$ (defined below).
The baryon symmetries act on fields in the same way as the $U(1)$
factors that we used for defining our abelian
theories in section \ref{abeliantheories}.
They disappeared in the non-abelian
theory associated to the conifolds, but the very same fact that they
can be consistently incorporated in the theory means that they must exist as
global symmetries. It is easy to check that no
operator corresponding to KK states is charged under these $U(1)$'s.
The reason is that the KK spectrum is made out with the combinations
$X=ABC$ or $X=U^3V^2$ defined in section \ref{abeliantheories} 
which, by definition, are $U(1)$ invariant variables.
The only objects that are charged under the $U(1)$ symmetries
are the baryons.
\par
Baryons have dimensions which diverge with $N$ and
can not appear in the KK spectrum. They are indeed non-perturbative
objects associated with wrapped branes \cite{bariowit}, \cite{gubserkleb}.  We see that
the baryonic symmetries have the right properties to be associated
with the Betti multiplets: the only charged objects
are non-perturbative states. This identification can be strengthened
by noticing that the only non-perturbative branes in M-theory have
an electric or magnetic coupling to the eleven dimensional
three-form.
Since for our
manifolds, both $b_2$
and $b_5$ are greater than 0, we have the choice of wrapping both
M2 and M5-branes.
M2 branes wrapped around a non-trivial two-cycle are certainly charged
under the massless vector in the Betti multiplet which is obtained by
reducing the three-form on the same cycle. Since a non-trivial 5-cycle
is dual to a 2-cycle, a similar remark applies also for M5-branes.
We identify M5-branes as baryons because they have a mass
(and therefore a conformal dimension) which, as we will show, goes like $N$.
\par
What follows from the previous discussion and is probably quite general,
is that there is a close relation between the $U(1)$'s entering
the brane construction of the gauge theory, the baryonic symmetries
and the Betti multiplets. The previous remarks apply as well to
CFT associated with orbifolds of $AdS_4\times S^7$. In the case of
$T^{11}$,$Q^{111}$ and $M^{111}$, the baryonic symmetries are also
directly related to the $U(1)$'s entering the toric description
of the manifold.
\par
\subsection{Dimension of the fundamental superfields and the baryon operators}
\label{barioformul}
A crucial check of
our conjectured conformal gauge theories comes from a direct
computation of the conformal weight of the fundamental superfields
\begin{equation}
  \mbox{fundamental superfields}=\cases{\begin{array}{cccl}
    U^i & V^A & \null &\mbox{in the $M^{111}$ theory}  \\
    \null & \null & \null \\
    A_i & B_j & C_\ell &\mbox{in the $Q^{111}$ theory}  \
  \end{array}}
\label{supsingleMQ}
\end{equation}
whose colour index structure and $\theta$-expansion are explicitly
given in the formulae (\ref{supsingM}), (\ref{supsingQ}). If the
non--abelian gauge theory has the $SU(N) \times \dots \times SU(N)$
gauge groups illustrated by the quiver diagrams of fig.s \ref{ABCcolour}
and \ref{UVcolour}, then we can consider the following chiral
operators:
\begin{eqnarray}
 \mbox{det}U&\equiv & U_{i_1\vert \Lambda^1_1\Sigma^1_1}^{\Lambda^2_1\Sigma^2_1} \,
 \dots \,
  U_{i_N\vert  \Lambda^1_N\Sigma^1_N}^{\Lambda^2_N\Sigma^2_N }\,
  \epsilon^{\Lambda^1_1 \dots \Lambda^1_N} \,\epsilon^{\Sigma^1_1 \dots \Sigma^1_N}
  \epsilon_{\Lambda^2_1 \dots \Lambda^2_N} \,\epsilon_{\Sigma^2_1 \dots \Sigma^2_N}
\label{operUUU}\\
  \mbox{det}V&\equiv & V_{A_1\vert \Lambda^1_1\Sigma^1_1\Gamma^1_1}^{\Lambda^2_1
  \Sigma^2_1\Gamma^2_1} \,
 \dots \,
  V_{A_N\vert  \Lambda^1_N\Sigma^1_N\Gamma^1_N}^{\Lambda^2_N\Sigma^2_N \Gamma^2_N}\,
  \epsilon^{\Lambda^1_1 \dots \Lambda^1_N} \,\epsilon^{\Sigma^1_1 \dots \Sigma^1_N}
  \epsilon^{\Gamma^1_1 \dots \Gamma^1_N}
  \epsilon_{\Lambda^2_1 \dots \Lambda^2_N} \,\epsilon_{\Sigma^2_1 \dots \Sigma^2_N}
  \epsilon_{\Gamma^2_1 \dots \Gamma^2_N}\nonumber\\
 & & \label{operVV} \\
\mbox{det}A&\equiv & A_{i_1\vert \Lambda^1_1}^{\Lambda^2_1} \, \dots \,
  A_{i_N\vert \Lambda^1_N}^{\Lambda^2_N }\, \epsilon^{\Lambda^1_1 \dots \Lambda^1_N} \,
  \epsilon_{\Lambda^2_1\dots \Lambda^2_N}
\label{q111baryopA}\\
\mbox{det}B&\equiv & B_{i_1\vert \Lambda^2_1}^{\Lambda^3_1} \, \dots \,
  B_{i_N\vert \Lambda^2_N}^{\Lambda^3_N }\, \epsilon^{\Lambda^2_1 \dots \Lambda^2_N} \,
  \epsilon_{\Lambda^3_1\dots \Lambda^3_N}
\label{q111baryopB}\\
  \mbox{det}C&\equiv & C_{i_1\vert \Lambda^3_1}^{\Lambda^1_1} \, \dots \,
  C_{i_N\vert \Lambda^3_N}^{\Lambda^1_N }\, \epsilon^{\Lambda^3_1 \dots \Lambda^3_N} \,
  \epsilon_{\Lambda^1_1\dots \Lambda^1_N}.
\label{q111baryopC}
\end{eqnarray}
If these operators are truly chiral primary fields, then their
conformal dimensions are obviously given by
\begin{equation}
\begin{array}{ccccccccccc}
h[\mbox{det} \, U] & = & h[U] \, \times \, N  & ; &
h[\mbox{det} \, V] & = & h[V] \, \times \, N  & \null & \null & \null & \null \\
h[\mbox{det} \, A] & = & h[A] \, \times \, N & ; &
h[\mbox{det} \, B] & = & h[B] \, \times \, N & ; &
h[\mbox{det} \, C] & = & h[C] \, \times \, N \\
\end{array}
\label{trulychir}
\end{equation}
and their flavour representations are:
\begin{eqnarray}
\mbox{det} \, U & \Rightarrow & (M_1=N, M_2=0, J=0) \label{repdetU},\\
\mbox{det} \, V & \Rightarrow & (M_1=0, M_2=0, J=N/2)  \label{repdetV},\\
\mbox{det} \, A & \Rightarrow & (J^{\ll(1\rr)}=N/2, J^{\ll(2\rr)}=0, J^{\ll(3\rr)}=0) \label{repdetA},\\
\mbox{det} \, B & \Rightarrow & (J^{\ll(1\rr)}=0, J^{\ll(2\rr)}=N/2, J^{\ll(3\rr)}=0)  \label{repdetB},\\
\mbox{det} \, C & \Rightarrow & (J^{\ll(1\rr)}=0, J^{\ll(2\rr)}=0, J^{\ll(3\rr)}=N/2)  \label{repdetC}\,.
\end{eqnarray}
\par
The interesting fact is that the conformal operators
(\ref{operUUU},...,\ref{q111baryopC}) can be reinterpreted as
solitonic supergravity states obtained by wrapping a $5$--brane on a
non--trivial supersymmetric $5$--cycle. This gives the possibility of calculating
directly the mass of such states and, as a byproduct, the conformal
dimension of the individual fundamental superfields. All what is involved is
a geometrical information, namely the ratio of the volume of the
$5$--cycles to the volume of the entire compact $7$--manifold. In
addition, studying the stability subgroup of the supersymmetric
$5$--cycles, we can also verify that the gauge--theory predictions
(\ref{repdetU},...,\ref{repdetC}) for the flavour representations are
the same one obtains in supergravity looking at the state as a
wrapped solitonic $5$--brane.
\par
To establish these results we need to  derive a general
mass--formula for baryonic states corresponding to wrapped
$5$--branes. This formula is obtained by considering various
relative normalizations.
\subsubsection{The M2 brane solution and normalizations of the seven
manifold metric and volume}
\label{branenormalizations}
\par
Let us write the curvatures of the Freund Rubin solution (\ref{FreundRubin})
\begin{equation}
\begin{array}{ccccccc}
  R^{mn} & = &  -16 e^2 \, E^m \wedge E^n & \Rightarrow & R^{ m  r}_{ n r} & = &
  - 24 \, e^2 \, \delta^m_n \\
  \cR^{ a   b} & = & \cR^{ a  b}_{c  d} \, \cB^{
c} \, \cB^{ d} & \mbox{with} & \cR^{ a  b}_{ c 
b} & = & 12 \, e^2 \, \delta^{ a}_{ c} \\
  F^{[4]} & = &  e \, \varepsilon_{mnrs} \, E^m \, \wedge\, E^{n}\, \wedge\,
E^r \, \wedge\, E^s\,, & \null & \null & \null  & \null
\end{array}
\label{freurub}
\end{equation}
where $E^m$ ($m=0,1,2,3$) is the vielbein of anti--de Sitter space
$AdS_4$, $R^{mn}$ is the corresponding curvature $2$--form, $\cB^{
a}$ ($a = 4,\dots, 10$) is the vielbein of $X_7$ and $\cR^{ a  b}$ is the
corresponding curvature. In these normalizations, both the internal and space--time vielbeins
do not have their physical dimension of a length $[E^m]_{phys}=[\cB^{ a}]_{phys}=\ell$,
since one has reabsorbed the Planck length $ l_p$ into their definition
by working in natural units where the $D=11$ gravitational constant $G_{11}$ has been
set equal to $1\over 8\pi$. Physical units are reinstalled through the following rescaling:
\begin{eqnarray}
  E^m &=& \frac{1}{\kappa^{2/9}}\, \hat{E}^m,\nonumber\\
  \cB^a &=& \frac{1}{\kappa^{2/9}}\, \hat{\cB}^a,\nonumber\\
  F^{[4]}_{mnrs} &=& \kappa^{11/9}\, \hat{F}^{[4]}_{mnrs},\nonumber\\
  \kappa^2 & = & 8 \pi G_{11}  \sim  l_p^9.
\label{riscal}
\end{eqnarray}
After such a rescaling, the relations between the Freund Rubin parameter and
the curvature scales for both $AdS_4$ and $X_7$ become
\begin{eqnarray}
\mbox{Ricci}_{\mu\nu}^{AdS} & = & - 2 \, \Lambda  \, g_{\mu\nu} \label{ricciout} \\
\mbox{Ricci}_{\a \b} \ &=& \Lambda  g_{\a \b}
\label{ricciin} \\
\Lambda & \stackrel{\mbox{\scriptsize def} }{=} & 24 \, \frac{e^2}{\kappa^{4/9}}.
\label{riccio}
\end{eqnarray}
Note that in eq. (\ref{riccio}) we have used the
normalization of the Ricci tensor which is standard in the general relativity
literature and is twice the normalization of the Ricci tensor $R^{ab}_{cb}$ appearing in
eq. (\ref{freurub}) and in chapter $3$. Furthermore eq.s (\ref{freurub}) were written in
flat indices while eq.s (\ref{ricciout}, \ref{ricciin}) are written in
curved indices.
\par
In the solvable coordinates \cite{torinos7}, \cite{Maldyads} 
defined in chapters $1,2$, the anti--de Sitter metric is:
\begin{eqnarray}
  ds^2_{AdS_4} &=& R_{AdS}^2 \left[ \rho^2 \left( -dt^2 + dx_1^2 +
  dx_2^2 \right) + \frac{d\rho^2}{\rho^2} \right], \nonumber\\
  \mbox{Ricci}_{\mu\nu}^{AdS}&=& - \frac{3}{R^2_{AdS}}\, g_{\mu\nu},\nonumber\\
\label{adsolv}
\end{eqnarray}
which yields the relation anticipated in chapter $1$:
\begin{equation}
  R_{AdS}= \frac{\kappa^{2/9}}{4 \, e}=\frac{1}{2}\,
  \sqrt{\frac{6}{\Lambda}}.
\label{adsrad}
\end{equation}
\par
As I said, we can consider the exact M2--brane
solution of $D=11$ supergravity that has the cone ${\cal C}(X_7)$ over $X_7$
as transverse space. The $D=11$ bosonic action can be written as
\begin{equation}
I_{11}=\int d^{11}x \sqrt{-g} ~({R\over\kappa^2}-3\, \hat{F}^2_{[4]} ) + {288} \sigma\int
\hat{F}_{[4]} \wedge
\hat{F}_{[4]} \wedge \hat{A}_{[3]}
\label{elevenaction}
\end{equation}
(where the coupling constant for the last term is $\sigma=\kappa$) 
and the exact $M2$--brane solution is as follows:
\begin{eqnarray}
ds^2_{M2} & = & \left( 1+\frac{R^6}{r^6}\right) ^{-2/3}(-dt^2 + dx_1^2 +dx_2^2)+
\left( 1+\frac{R^6}{r^6}\right) ^{1/3} \, ds^2_{cone},  \nonumber\\
ds^2_{cone} & = & dr^2 +r^2\frac{\Lambda}{6} ds^2_{X_7}, \nonumber\\
A^{[3]} & = & dt\wedge dx_1 \wedge dx_2 \,\left( 1+\frac{R^6}{r^6}\right) ^{-1},
\label{ghm2bra}
\end{eqnarray}
where $ds^2_{X_7}$ is the Einstein metric on $X_7$, with Ricci tensor
as in eq. (\ref{riccio}), and $ds^2_{cone}$ is the corresponding Ricci
flat metric on the associated cone. When we go near the horizon, $r\to
0$, the metric (\ref{ghm2bra}) is approximated by
\begin{equation}
  ds^2_{M2} \approx {r^{4}\over R^4}(-dt^2 + dx_1^2 +dx_2^2) \, +
  R^2 \, \frac{dr^2}{r^2} + R^2 \, \frac{\Lambda}{6} \,
  ds_{X_7}^2.
\label{approxi}
\end{equation}
The Freund Rubin solution $AdS_4 \times X_7$ is obtained by setting
\begin{equation}
  \rho= {2\over R^3} \, r^2
\label{rhotor}
\end{equation}
and by identifying
\begin{equation}
  R_{AdS}= \frac{R}{2} \quad \Leftrightarrow \quad \Lambda = {6\over R^2}\,.
\label{agnisco}
\end{equation}
\subsubsection{The dimension of the baryon operators}
\label{barioquila}
Having fixed the normalizations, we can now compute the mass of a M5-brane
wrapped around a non-trivial supersymmetric cycle of $X_7$ and the conformal
 dimension of the associated baryon operator.
\par
The parameter $R^6$ appearing in the M2-solution is obviously proportional to
the number $N$ of membranes generating the $AdS$-background and, by dimensional
analysis, to $l_p^6$. The exact relation for the maximally supersymmetric case
$AdS_4\times S^7$ is (see chapter $1$)
\begin{equation}
  R_{AdS}= {l_p\over 2}\left (2^5\pi^2 N\right )^{1/6} \quad \Leftrightarrow \quad R^6=2^5 \pi^2 N l_p^6.
\label{rel}
\end{equation}
\par
We can easily adapt this formula to the case of $AdS_4\times X_7$ by noticing
that, by definition,  the number of M2-branes $N$ is determined by the
flux of the RR
three-form through $X_7$, $\int_{X_7} *F^{[4]}$. As a consequence, $N$ and
the volume of $X_7$ will appear in all the relevant formulae in the
combination $N/{\rm Vol}(X_7)$. We therefore obtain the general formula
\begin{equation}
\sqrt{{\Lambda\over 6}}={1\over R}=\left ( {{\rm Vol}(X_7)\over {\rm Vol}(S^7)}\right )^{1/6}
{1\over l_p(2^5\pi^2 N)^{1/6}}.
\label{general}
\end{equation}
\par
We can now consider the solitonic particles in $AdS_4$ obtained by
wrapping M2- and
M5-branes on the
non-trivial 2- and 5-cycles of $X_7$, respectively.
They are associated with boundary operators with conformal
dimensions that diverge in the
large $N$ limit. The exact dependence on $N$ can be easily
estimated.
The mass of a p-brane wrapped on a p-cycle is given by
$T_p\times {\rm Vol(p-cycle)}\sim l_p^{-\left(p+1\right)}\Lambda^{-{p\over 2}}\sim 
l_p^{-\left(p+1\right)}$. Once the
mass of the non-perturbative states is known, the dimension $E_0$ of the
associated
boundary operator is given by the relation 
\footnote{In general $m^2\simeq E_0^2/R^2_{AdS}$; with the conventions of
chapter 3, $R_{AdS}=1/4$ and $m^2\simeq E_0^2/16$; whith the convention
(\ref{general}), $m^2\simeq 2\Lambda/3 E_0^2$.}
\be
\label{mE0}
m^2={2\Lambda\over 3}(E_0-1)(E_0-2)
\simeq{2\Lambda\over 3}E_0^2\,.
\ee
From equation ~(\ref{general}) we learn that
$l_p\sim N^{-1/6}$.
We see that M2-branes correspond to operators with dimension $\sqrt{N}$
while M5-branes to operators with dimension of order $N$. The natural
candidates for the baryonic operators we are looking for are therefore
the wrapped five-branes.
\par
We can easily write a more precise formula for the dimension of the baryonic
operator associated with a wrapped  M5-brane, following the analogous 
computation in \cite{gubserkleb}. For this, we need the exact
expression for the M5 tension which can be found, for example, in
\cite{dealwis}. We find
\begin{equation}
m={1\over (2\pi )^5l_p^6}{\rm Vol(5-cycle)}.
\label{int}
\end{equation}
\par
Using equations (\ref{general}), (\ref{mE0}),
and substituting $V(S^7)=\pi^4R^7/3$, we obtain the
formula for the dimension of a baryon,
\begin{equation}
E_0={\pi N\over \Lambda}{{\rm Vol(5-cycle)}\over {\rm Vol}(X_7)},
\label{baryondim}
\end{equation}
where the volume is evaluated with the internal metric normalized so that
(\ref{ricciin}) is true.
\par
As a check, we can compute the dimension of a Pfaffian operator in the
${\cal N} = 8$ theory with gauge group $SO(2N)$. The theory contains
adjoint scalars which can be represented as antisymmetric matrices
$\phi_{ij}$ and we can form the gauge invariant baryonic operator
$\epsilon_{i_1,...,i_{2N}}\phi_{i_1i_2}....\phi_{i_{2N-1}i_{2N}}$
with dimension $N/2$. The internal manifold
is ${\IR\IP}^7$ \cite{bariowit}, \cite{ahn}, a supersymmetric preserving ${\ZZ}_2$
projection of original $AdS_4\times S^7$ case, corresponding to the $SU(N)$
gauge group. We obtain the Pfaffian by wrapping an M5-brane on a ${\IR\IP}^5$
submanifold. Equation ~(\ref{baryondim}) gives
\begin{equation}
E_0= {\pi N\over \Lambda}{{\rm Vol}{\IR\IP}^5\over {\rm Vol}{\IR\IP}^7}=
{\pi N\over \Lambda}{{\rm Vol}S^5\over {\rm Vol}S^7}= N/2,
\end{equation}
as expected.
\par
\subsection{The case of $M^{111}$}
\par
\subsubsection{Cohomology of $M^{111}$}
Let us now compute the cohomology
of $M^{111}$. The
first Chern class of $L$ is $c_1 = 2 \omega_1 + 3 \omega_2$, where
$\omega_1$ (resp. $\omega_2$) is the generator of the second cohomology
group of ${\IP}^1$
(resp. ${\IP}^{2}$).
In this case the Gysin sequence \cite{BT} gives:
\begin{eqnarray}
&H^0( M^{111} ) = H^7( M^{111} )= \ZZ,&\nonumber\\ 
& 0 \longrightarrow H^1 ( M^{111} ) \longrightarrow \ZZ \stackrel{c_1}{\longrightarrow} \ZZ \oplus \ZZ
\longrightarrow H^2( M^{111} ) \longrightarrow 0,
&\nonumber\\
& 0 \longrightarrow H^3 ( M^{111} ) \longrightarrow \ZZ \oplus \ZZ \stackrel{c_1}{\longrightarrow} \ZZ
\oplus \ZZ \longrightarrow H^4( M^{111} ) \longrightarrow 0,
&\nonumber\\
& 0 \longrightarrow H^5 ( M^{111} ) \longrightarrow \ZZ\oplus \ZZ \stackrel{c_1}{\longrightarrow} \ZZ
 \longrightarrow H^6( M^{111} ) \longrightarrow 0.
&
\label{m111gys}
\end{eqnarray}
The first $c_1$ sends $1 \in H^0(M_a)$ to $c_1 \in H^2 (M_a)$. Its
kernel is zero, and its image is $\ZZ$. Accordingly, $H^2 ( M^{111} ) =
\ZZ \cdot \pi^*(\omega_1 + \omega_2) $.
The second $c_1$ sends $(\omega_1 ,\omega_2) \in \ZZ \oplus \ZZ =
H^2(M_a)$ to $(3 \omega_1 \omega_2, 2\omega_1 \omega_2 +3 \omega_2^2) \in
\ZZ \oplus \ZZ = H^4 (M_a)$. Its kernel vanishes and therefore $H^3( M^{111} )= 0$.
Its cokernel is $\ZZ_9 = H^4( M^{111} )$ generated by $\pi^*(\omega_1 \omega_2
+ \omega_2^2)$.
Finally, the last $c_1$ sends $\omega_1 \omega_2$ and $ \omega_2^2
\in H^4 (M_a)= \ZZ \oplus \ZZ $ respectively to $ 3 \omega_1 \omega_2^2 $
and $ 2 \omega_1 \omega_2^2 \in H^6(M_a)$. This map is surjective,
so $H^6( M^{111} )=0$ and its kernel is generated by $ \beta = -2 \omega_1
\omega_2 + 3\omega_2^2$. Hence $H^5( M^{111} ) = \ZZ \cdot \alpha$, with $\pi_*
\alpha = \beta$.
%%%%%%%%%%%%%%%%%%%%%%%%%%%%%%%%%%%%%%%%%%%%%
% Parte di Davide su M111%%%%%%%%%%%%%%%%%%%
%%%%%%%%%%%%%%%%%%%%%%%%%%%%%%%%%%%%%%%%%%%%
\subsubsection{Explicit description of the $U\ll(1\rr)$ fibration for $M^{111}$}
We proceed next to an explicit description
of the fibration structure of  $M^{111}$  as
 a $U(1)$-bundle over ${\IP}^{2}\times{\IP}^1$.
 We construct an atlas of local trivializations and we give the
 appropriate transition functions. This is important for our
  discussion of the supersymmetric cycles leading to the
 baryon states.
 \par
We take $\tau\in[0,4\pi)$ as a local coordinate on the fibre and
$(\tilde\theta,\tilde\phi)$ as local coordinates on ${\IP}^1\simeq S^2$.
To describe ${\IP}^{2}$ we have to be a little bit careful.
${\IP}^{2}$ can be covered by the three patches $W_\alpha\simeq{\IC}^2$ in
which one of the three homogeneous coordinates, $U_\alpha$, does not vanish.
The set not covered by one of these $W_\alpha$ is homeomorphic to $S^2$.
We choose to parametrize $W_3$ as in \cite{gibbonspope}:
\begin{equation}
\label{C2coord}
\left\{\begin{array}{c}
\zeta^1=U_1/U_3=\tan\mu\,\cos(\theta/2)\,e^{i(\psi+\phi)/2}\\
\zeta^2=U_2/U_3=\tan\mu\,\sin(\theta/2)\,e^{i(\psi-\phi)/2}
\end{array}\right.,
\end{equation}
where
\begin{equation}
\left\{\begin{array}{l}
\mu\in(0,\pi/2)\\
\theta\in(0,\pi)\\
0\leq(\psi+\phi)\leq 4\pi\\
0\leq(\psi-\phi)\leq 4\pi
\end{array}\right.\,.
\label{psiphi}
\end{equation}
These coordinates cover the whole $W_3\simeq{\IC}^2$ except for the
trivial coordinate singularities $\mu=0$ and $\theta=0,\pi$.
Furthermore $\theta$ and $\phi$ can be extended to the complement of $W_3$.
Indeed, the ratio
\begin{equation}
z=\zeta^1/\zeta^2=\tan^{-1}(\theta/2)\,e^{i\phi}
\label{fubstud}
\end{equation}
is well defined in the limit $\mu\to\pi/2$ and it constitutes
the usual stereographic map of $S^2$ onto the complex plane
(see the next discussion of $Q^{111}$ and in particular figure
\ref{S2patches}).
\par
\begin{figure}[ht]
\begin{center}
\epsfxsize = 8cm
\epsffile{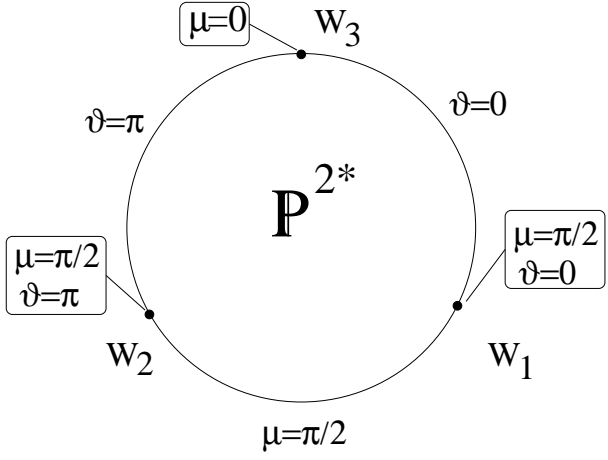}
\vskip  0.2cm
\hskip 2cm
\unitlength=1.1mm
\end{center}
\caption{Schematic representation of the atlas on ${\IP}^{2}$.
The three patches $W_\alpha$ cover the open ball and part of the
boundary circle, which constitutes the set of coordinate singularities.
This latter is made of three $S^2$'s: $\{\theta=0\}$,
$\{\theta=\pi\}$ and $\{\mu=\pi/2\}$, which touch each other
at the three points marked with a dot.
Each $W_\alpha$ covers the whole ${\IP}^{2}$ except for one of the spheres
(for example, $W_3$ does not cover $\{\mu=\pi/2\}$).
The three \emph{most singular} points are covered by only one
patch (for example, $\{\mu=0\}$ is covered by the only $W_3$).
}
\label{CP2patches}
\end{figure}

We must be careful in treating
some one-forms near the coordinate singularities.
In particular, $d\psi$ and $d\phi$ are not well defined on the
three $S^2$ which are not covered by one of the patches $W_\alpha$:
$\{\mu=\pi/2\}$, $\{\theta=0\}$ and $\{\theta=\pi/2\}$
(see figure  \ref{CP2patches}.)
%%%%%%%%%%%%%%%%%%%%%%%%%%%%%%%%%%%%%%%%%%%%%%%%%
%        CP^2   PATCHES   FIGURE
%%%%%%%%%%%%%%%%%%%%%%%%%%%%%%%%%%%%%%%%%%%%%%%%%
%\iffigs
%\fi
Actually, except for the three points of these spheres that
are covered by only one patch ($\{\mu=0\}\in W_3$, $\{\mu=\pi/2,
\theta=0\}\in W_1$, $\{\mu=\pi/2,\theta=\pi\}\in W_2$), one particular
combination of $d\psi$ and $d\phi$ survives, as it is illustrated in
table (\ref{1forms}).
%%%%%%%%%%%%%%%%%%%%%%%%%%%%%%%%%%%%%%%%%%%%%%%%%%%%%%%%%%%%%%%
%          SINGULARITIES       TABLE
%%%%%%%%%%%%%%%%%%%%%%%%%%%%%%%%%%%%%%%%%%%%%%%%%%%%%%%%%%%%%%%
\begin{equation}
\label{1forms}
\begin{array}{|c|c|c|}
\hline
{\rm coordinate}&{\rm regular}&{\rm singular}\\
{\rm singularity}&{\rm one-form}&{\rm one-forms}\\
\hline\hline
\theta=0&d\psi+d\phi&\alpha d\psi+\beta d\phi~(\alpha\neq\beta)\\
\theta=\pi&d\psi-d\phi&\alpha d\psi-\beta d\phi~(\alpha\neq\beta)\\
\mu=\pi/2&d\phi&\alpha d\psi\\
\hline
\end{array}
\end{equation}
The singular one-forms become well defined if we multiply them
by a function having a double zero at the coordinate
singularities.
\par
We come now to the description of the fibre bundle $M^{111}$.
We cover the base ${\IP}^{2}\times{\IP}^1$ with six open
charts ${\cal U}_{\alpha\pm}=W_\alpha\times H_\pm$ ($\alpha=1,2,3$) on which we
can define a local fibre coordinate $\tau_{\alpha\pm}\in[0,4\pi)$.
The transition functions are given by:
\begin{equation}\label{Mtrans}
\left\{\begin{array}{l}
\tau_{1\beta}=\tau_{3\gamma}-3(\psi+\phi)+2(\beta-\gamma)\tilde\phi\,,\qquad
(\beta,\gamma=\pm1)\\
\tau_{1\beta}=\tau_{2\gamma}-6\phi+2(\beta-\gamma)\tilde\phi\,.
\end{array}\right.
\end{equation}
On this principal fibre bundle we can easily introduce a $U(1)$ Lie
algebra valued connection which, on the various patches
of the base space, is described by the following one--forms:
\begin{eqnarray}
\left\{\begin{array}{l}
{\cal A}_{1\pm}=-\frac{3}{2}(\cos2\mu\!+\!1)(d\psi+d\phi)
-\frac{3}{2}(\cos2\mu\!-\!1)(\cos\theta\!-\!1)d\phi
+2(\pm 1\!-\!\cos\tilde\theta)d\tilde\phi\,,\\
{\cal A}_{2\pm}=-\frac{3}{2}(\cos2\mu\!+\!1)(d\psi-d\phi)
-\frac{3}{2}(\cos2\mu\!-\!1)(\cos\theta\!+\!1)d\phi
+2(\pm 1\!-\!\cos\tilde\theta)d\tilde\phi\,,\\
{\cal A}_{3\pm}=-\frac{3}{2}(\cos2\mu-1)(d\psi+\cos\theta d\phi)
+2(\pm 1-\cos\tilde\theta)d\tilde\phi\,.\\
\end{array}\right.\nn\\
\label{m111connec}
\end{eqnarray}
Due to (\ref{Mtrans}), the one-form $(d\tau\!-\!{\cal A})$ is a global angular form
\cite{BT}.
It can then be taken as the $7$-th vielbein of the following
$SU(3)\times SU(2)\times U(1)$
invariant metric on $M^{111}$:
\begin{equation}
ds^2_{M^{111}}=c^2(d\tau\!-\!{\cal A})^2+ds^2_{\,{\IP}^{2}}+
ds^2_{\,{\IP}^1}.
\label{fibrametra}
\end{equation}
The one-form ${\cal A}$ is the connection of the Hodge-K\"ahler bundle on 
${\IP}^{2}\times{\IP}^1$.
\vskip 0.3cm
\leftline{ {\underline {\it Einstein Metric}}}
\vskip 0.3cm
%%%%%%%%%%%%%%%%%%%%%%%%%%%%%%%%%%%%%%%%%%%%%%%%%%%%%%%%%
% From timpo.tex %%%%%%%%%%%%%%%%%%%%%%%%%%%%%%%%%%%%%%%%
%%%%%%%%%%%%%%%%%%%%%%%%%%%%%%%%%%%%%%%%%%%%%%%%%%%%%%%%%
The Einstein metric on the homogeneous space $M^{111}$ can be written in terms
of the vielbein given in chapter $3$ and found in \cite{cdfm111}. However,
the same metric can be expressed (see \cite{pagepopeM}) in  the
coordinate frame we have just utilized to describe the fibration
structure and which is convenient for our discussion of the
supersymmetric $5$--cycles. In this frame it is
\begin{eqnarray}
\label{metrica}
ds^2_{M^{111}}&=&\frac{3}{32 \Lambda}\Bigl [
d\tau-3 \sin^2 \mu \,\left(d\psi
+\cos{\theta}d\phi\right)+2\cos{\tilde{\theta}}
d\tilde{\phi} \Bigr ]^2 +\nonumber\\
&+& \frac{9}{2 \, \Lambda}\left[d\mu^2+{1\over 4}\sin^2{\mu}\cos^2{\mu}^2\left(d\psi
+\cos{\theta}d\phi\right)^2+\right.\nonumber\\
&+&\left.{1\over 4}\sin^2{\mu}\left(d\theta^2
+\sin^2{\theta}d\phi^2\right)\right]
+\frac{3}{4 \Lambda
}\left( d\tilde{\theta}^2+\sin^2{\tilde{\theta}}d\tilde{\phi}^2\right)\,.
\end{eqnarray}
The second and the third
addenda are the ${\IP}^{2}$ and $S^2$ metric on the base manifold of the
$U(1)$ fibration, while the first term is the fibre metric. In other
words, one recognizes the structure of the metric anticipated in
(\ref{fibrametra}). The parameter $\Lambda$ appearing in the metric
(\ref{metrica}) is the internal cosmological constant defined by
eq. (\ref{ricciin}).
%%%%%%%%%%%%%%%%%%%%%%%%%%%%%%%%%%%%%%%%%%%%%%%%%%%%%%%%%%%%
%%%%%%%%%%%%%%%%%%%%%%%%%%%%%%%%%%%%%%%%%%%%%%%%%%%%%%%%%%%%%%
% The supersymmetric 5-cycles %%%%%%%%%%%%%%%%%%%%%%%%%%%%%%%%
%%%%%%%%%%%%%%%%%%%%%%%%%%%%%%%%%%%%%%%%%%%%%%%%%%%%%%%%%%%%%%
%%%%%%%%%%%%%%%%%%%%%%%%%%%%%%%%%%%%%%%%%%%%%%%%%%%%%%%%%%%%%%
% The supersymmetric 5-cycles %%%%%%%%%%%%%%%%%%%%%%%%%%%%%%%%
%%%%%%%%%%%%%%%%%%%%%%%%%%%%%%%%%%%%%%%%%%%%%%%%%%%%%%%%%%%%%%
\subsubsection{The baryonic $5$--cycles of $M^{111}$ and their volume}
\label{bryn}
As we saw above, the relevant homology group of $M^{111}$ for the
calculation of the baryonic masses is
\begin{equation}
H_5(M^{111},{\IR})={\IR}\,.
\end{equation}
Let us consider the following two five-cycles, belonging to the same homology class:
\begin{eqnarray}
{\cal C}^1:\left\{\begin{array}{c}
\tilde\theta={\tilde \theta}_0 =const\\
\tilde\phi={\tilde \phi}_0=const
\end{array}\right.\,, \label{cycca1}\\
{\cal C}^2:\left\{\begin{array}{c}
\theta=\theta_0 =const\\
\phi=\phi_0 = const
\end{array}\right.\,.
\label{cycca2}
\end{eqnarray}
The two representatives
(\ref{cycca1}, \ref{cycca2}) are distinguished by their
different stability subgroups which we calculate in the next subsection.
\vskip 0.3cm
\leftline{ {\underline {\it Volume of the $5$--cycles}}}
\vskip 0.3cm
The volume of the cycles (\ref{cycca1}, \ref{cycca2}) is easily computed by pulling back
the metric (\ref{metrica}) on ${\cal C}^1$ and ${\cal C}^2$, that
have the topology of a $U(1)$-bundle over ${\IP}^{2}$ and
${\IP}^1\times{\IP}^1$ respectively:
\begin{eqnarray}
{\rm Vol}({\cal C}^1)=\oint_{{\cal C}^1}\sqrt{g_1}=
9\left(8\Lambda/3\right)^{-5/2}\int\sin^3\mu\cos\mu\sin\theta\,
d\tau d\mu d\psi d\theta d\phi
=\frac{9\pi^3}{2}\left(\frac{3}{2\Lambda}\right)^{5/2}\nn\\
\label{volcyc1}
\end{eqnarray}
\begin{eqnarray}
{\rm Vol}({\cal C}^2)=\oint_{{\cal C}^2}\sqrt{g_2}=
6\left(8\Lambda/3\right)^{-5/2}\int\sin\mu\cos\mu\sin\tilde\theta\,
d\tau d\mu d\psi d\tilde\theta d\tilde\phi
=6\pi^3\left(\frac{3}{2\Lambda}\right)^{5/2}\,.\nn\\
\label{volcyc2}
\end{eqnarray}
The volume of $M^{111}$ is instead given by
\begin{eqnarray}
{\rm Vol}(M^{111})=\oint_{M^{111}}\sqrt{g}=
18\left(8\Lambda/3\right)^{-7/2}\int\sin^3\mu\cos\mu\sin\theta
\sin\tilde\theta\, d\tau d\mu d\psi d\theta d\phi d\tilde\theta d\tilde\phi+
\nonumber\\
=\frac{27\pi^4}{2\Lambda}\left(\frac{3}{2\Lambda}\right)^{5/2}\,.\nn\\
\label{volm111}
\end{eqnarray}
The results (\ref{volcyc1}, \ref{volcyc2}, \ref{volm111}) can be
inserted into the general formula (\ref{baryondim}) to calculate the
conformal weights (or energy labels) of five-branes wrapped on the
cycles ${\cal C}^1$ and ${\cal C}^2$. We obtain:
\begin{equation}
\left\{\begin{array}{c}
E_0({\cal C}^1)=N/3\\
\\
E_0({\cal C}^2)=4N/9
\end{array}\right.\,.
\label{risultone}
\end{equation}
As stated above, the result (\ref{risultone}) is essential in proving
that the conformal weight of the elementary world--volume fields
$V^A$, $U^i$ are
\begin{equation}
  h\left[ V^4 \right]=1/3 \quad , \quad h\left[ U^i \right]=4/9
\label{hUV}
\end{equation}
respectively. To reach such a conclusion we need to
identify the states obtained by wrapping the five--brane on ${\cal
C}^1, {\cal C}^2$ with operators in the flavour representations
$M_1=0,M_2=0,J=N/2$ and $M_1=N,M_2,J=0$, respectively. This
conclusion is reached by
studying the stability subgroups of the supersymmetric $5$--cycles.
\par
This matches with the previous result (\ref{donM111}) on the
spectrum of chiral operators, which are predicted of the form
\begin{equation}
{\rm Tr}\ll(U^3V^2\rr)^k
\label{chiropm111}
\end{equation}
and should have conformal weight $E=2k$. Indeed, we have 
\be
3\times {4\over 9}+2\times{1\over 3}=2\,!!!
\label{mastercheck}
\ee
\par
\subsubsection{The flavour representations of the baryons}
\label{stab5M}
\par
To find the flavour representations of these non--perturbative
states we follow an argument
introduced by Witten \cite{bariowit}, where they are found by studying
the stability subgroups of the five--cycles $H\ll(\cC^i\rr)$.
As shown in \cite{bariowit}, the collective degrees of
freedom $c$ of the wrapped $5$--brane soliton live on the coset manifold
$G/H({\cal C}^i)$, where $G$ is the isometry group of $X_7$. The
wave--function $\Psi(c)$ of the soliton must be expanded in harmonics on
$G/H({\cal C}^i)$ characterized by having  charge $N$ under the
baryon number $U(1)_B \subset H({\cal C}^i)$. Minimizing the energy
operator (the laplacian) on such harmonics one obtains the
corresponding $G$ representation and hence the flavour assignment of
the baryon. 
\par
Let us now consider the stability subgroups
\begin{equation}
  H({\cal C}_i) \subset G = SU(3) \times SU(2) \times U(1)
\label{stabalg}
\end{equation}
of the two cycles (\ref{cycca1}, \ref{cycca2}). Let us begin with the
first cycle defined by (\ref{cycca1}). As we have previously said, this is the
restriction of the $U(1)$-fibration to ${\IP}^{2}\times\{p\}$, $p$ being a
point of ${\IP}^1$.
Hence, the stability subgroup of the cycle ${\cal C}^1$ is:
\begin{equation}
  H\left({\cal C}^1\right) = SU(3) \times U(1)_R \times U(1)_{B,1}
\label{hcyc1}
\end{equation}
where $U(1)_R$ is the R--symmetry $U(1)$ appearing as a factor in
$SU(3) \times SU(2) \times U(1)_R$ while $U(1)_{B,1} \subset SU(2)$  is
a maximal torus.
\par
Turning to the case of the second cycle (\ref{cycca2}), which is the restriction of the
$U(1)$-bundle to the product of a hyperplane of ${\IP}^{2}$ and ${\IP}^1$,
its stabilizer is
\begin{equation}
  H\left({\cal C}^2\right) = SU(2)\times U(1)_{B,2} \times SU(2)\times U(1)_R,
\label{hcyc2}
\end{equation}
where $SU(2) \times U(1)_R$ is the group appearing as a factor in
$SU(3) \times SU(2) \times U(1)_R$, $U(1)_{B,2} \subset SU(3)$ is the
subgroup generated by $h_1=\diag(1,-1,0)$ and $SU(2)\times U(1)_{B,2}\subset SU(3)$ is the
stabilizer of the first basis vector of ${\IC}^3$.
\par
Following the procedure introduced by Witten in \cite{bariowit} we
should now quantize the {\it collective coordinates} of the
non--perturbative baryon state obtained by {\it wrapping} the
five--brane on the $5$--cycles we have been discussing. As explained
in Witten's paper this leads to  quantum mechanics on the
homogeneous manifold $G/H(\cal C)$. In our case the collective
coordinates of the baryon live on the following spaces:
\begin{equation}
  \mbox{space of collective coordinates} \,\quad \rightarrow \quad \frac{G}{H(\cal C)}=
   \cases{\begin{array}{cc}
    \frac{SU(2)}{U(1)_{B,1}} \simeq {\IP}^1 & \mbox{for ${\cal C}^1$} \\
    \null & \null \\
    \frac{SU(3)}{SU(2)\times U(1)_{B,2}} \simeq {\IP}^2 & \mbox{for ${\cal C}^2$} \\
  \end{array}\cr}.
\label{collecti}
\end{equation}
The wave function $\Psi\left( \mbox{collec. coord.}\right) $ is in
Witten's phrasing a section of a line bundle of degree $N$. This
happens because the baryon has {\it baryon number} $N$, namely it has
charge $N$ under the additional massless vector multiplet that is associated with
a harmonic $2$--form and appears
in the Kaluza Klein spectrum since $\mbox{dim} H_2(M^{111}) = 1 \ne
0$. These are the Betti multiplets mentioned in Section \ref{bariobetti}.
Following Witten's reasoning there is a morphism
\begin{equation}
 \mu^i: \quad  U(1)_{Baryon} \hookrightarrow H({\cal C}^i) \quad
 i=1,2
\label{morfismo}
\end{equation}
of the non perturbative baryon number group  into the stability subgroup
of the $5$--cycle. Clearly the image of such a morphism must be a
$U(1)$--factor in $H({\cal C})$ that has a non trivial action on the
collective coordinates of the baryons. Clearly in the case of our two
baryons we have:
\begin{equation}
  \mbox{Im} \, \mu^i =\, U(1)_{B,i} \quad  i=1,2\,.
\label{immorf}
\end{equation}
The name given to these groups anticipated the conclusions of such an
argument.
\par
Translated into the language of harmonic analysis, Witten's statement
that the baryon wave function should be a section of a line bundle
with degree $N$ means that we are supposed to consider harmonics on
$G/H({\cal C})$ which, rather than being scalars of $H({\cal C})$, are
in the $1$--dimensional representation of $U(1)_B$ with charge $N$.
%(for a review on representation theory studied with complex fiber bundles,
%see \cite{reina})
According to the general rules of harmonic analysis (see chapter $3$) 
we are supposed to collect
all the representations of $G$ whose reduction with respect to $H({\cal
C})$ contains the prescribed representation of $H({\cal C})$. In the
case of the first cycle, in view of eq. (\ref{hcyc1}) we want all
representations of $SU(2)$ that contain the state $J^{\ll(3\rr)}=N$. Indeed
the generator of $U(1)_{B,1}$ can always be regarded as
the third component of angular momentum by means of a change of
basis. The representations with this property are those characterized
by:
\begin{equation}
  2 J = N+2k,\qquad k \ge 0.
\label{boundo}
\end{equation}
Since the laplacian on $G/H({\cal C})$ has eigenvalues proportional to
the Casimir
\begin{equation}
  \Box_{SU(2)/U(1)} = \mbox{const} \, \times \, J(J+1),
\label{laplacio}
\end{equation}
the harmonic satisfying the constraint (\ref{boundo}) and with
minimal energy is just that with
\begin{equation}
  2J=N.
\label{JisN}
\end{equation}
This shows that under the flavour group the baryon associated with the
first cycle is neutral with respect to $SU(3)$ and transforms in the
$N$--times symmetric representation of $SU(2)$. This perfectly
matches, on the superconformal field theory side, with our candidate
operator (\ref{operVV}).
\par
Equivalently the choice of the representation $2 J= N$ corresponds
with the identification of the baryon wave--function with a {\it
holomorphic section (=zero mode)} of the $U(1)$--bundle under
consideration, i.e. with a section of the corresponding line bundle.
Indeed such a line bundle is, by definition, constructed over
${\IP}^1$ and declared to be of degree $N$, hence it is ${\cal
O}_{{\IP}^1}(N)$. Representation-wise a section of ${\cal
O}_{{\IP}^1}(N)$ is just an element of the $J=N/2$
representation, namely it is the $N$ times symmetric of $SU(2)$.
\par
Let us now consider the case of the second cycle. Here the same
reasoning instructs us to consider all representations of $SU(3)$
which, reduced with respect to $U(1)_{B,2}$, contain a state of charge
$N$. Moreover, directly aiming at zero mode,
we can assign the baryon wave--function
to a holomorphic section of a line bundle on ${\IP}^2$, which
must correspond to characters of the parabolic subgroup $SU(2)\times U(1)_{B,2}$.
As before the degree $N$ of this line bundle uniquely characterizes it as
${\cal O}(N)$.
In the language of Young tableaux, the corresponding $SU(3)$ representation is
\begin{equation}
  M_1 = 0 \, ; \, M_2 =N,\label{m1m2N}
\end{equation}
i.e. the representation of this baryon state is the $N$--time
symmetric of the dual of $SU(3)$ and this perfectly matches with the
complex conjugate of the candidate
conformal operator \ref{operUUU}. In other words we have constructed
the antichiral baryon state. The chiral one obviously has the same
conformal dimension.
\subsubsection{These $5$--cycles are supersymmetric}
\label{supsym5M}
The $5$--cycles we have been considering in the above subsections
have to be supersymmetric in order for the conclusions we have been
drawing to be correct. Indeed all our arguments have been based on
the assumption that the $5$--brane wrapped on such cycles is a
$BPS$--state. This is true if the $5$--brane action localized on the
cycle is $\kappa$--supersymmetric.
\par
The $\kappa$-symmetry projection operator for a five-brane is
\begin{equation}\label{kprojector}
P_\pm=\frac{1}{2}\left(\unity\pm {\rm i}\,\frac{\rm 1}{5!\,\sqrt{g}}\epsilon^
{\alpha\beta\gamma\delta\varepsilon}
\partial_\alpha X^M\partial_\beta X^N\partial_\gamma X^P
\partial_\delta X^Q\partial_\varepsilon X^R\,\Gamma_{MNPQR}\right)\,,
\end{equation}
where the functions $X^M(\sigma^\alpha)$ define the embedding of
the five-brane into the eleven dimensional spacetime, and $\sqrt{g}$
is the square root of the determinant of the induced metric on the
brane.
The gamma matrices $\Gamma_{MNPQR}$, defining the spacetime
spinorial structure, are the pullback through the vielbein
of the constant gamma matrices $\Gamma_{ABCDE}$ satisfying the
standard Clifford algebra:
\begin{equation}
\Gamma_{MNPQR}=e^A_{\,M}e^B_{\,N}e^C_{\,P}e^D_{\,Q}e^E_{\,R}
\Gamma_{ABCDE}\,.
\end{equation}
A possible choice of vielbein for ${\cal C}(M^{111})\times{\cM}_3$, namely
the product of  the
metric cone over $M^{111}$  times three dimensional Minkowski space is the following
one:
\begin{equation}\label{Mvielbein}
\left\{\begin{array}{ccl}
e^1&=&\frac{1}{2\sqrt2}\,r\,d\tilde\theta\\
e^2&=&\frac{1}{2\sqrt2}\,r\sin\tilde\theta d\tilde\phi\\
e^3&=&\frac{1}{8}\,r\left(d\tau+3\sin^2\mu(d\psi+\cos\theta d\phi)
+2\cos\tilde\theta d\tilde\phi\right)\\
e^4&=&\frac{\sqrt3}{2}\,r\,d\mu\\
e^5&=&\frac{\sqrt3}{4}\,r\sin\mu\cos\mu
\left(d\psi+\cos\theta d\phi\right)\\
e^6&=&\frac{\sqrt3}{4}\,r\sin\mu
\left(\sin\psi d\theta-\cos\psi\sin\theta d\phi\right)\\
e^7&=&\frac{\sqrt3}{4}\,r\sin\mu
\left(\cos\psi d\theta+\sin\psi\sin\theta d\phi\right)\\
e^{8}&=&dr\\
e^9&=&dx^1\\
e^{10}&=&dx^2\\
e^0&=&dt\\
\end{array}\right.\,.
\end{equation}
In these coordinates the embedding equations of the two cycles
(\ref{cycca1}), (\ref{cycca2}) are very simple, so we have
\begin{equation}
\frac{1}{5!}\epsilon^{\alpha\beta\gamma\delta\varepsilon}
\partial_\alpha X^M\partial_\beta X^N\partial_\gamma X^P
\partial_\delta X^Q\partial_\varepsilon X^R\,\Gamma_{MNPQR}=
\left\{\begin{array}{c}
\Gamma_{\tau\mu\theta\psi\phi}\\
\Gamma_{\tau\mu\tilde\theta\psi\tilde\phi}
\end{array}\right.\,,
\end{equation}
for ${\cal C}^1$ and ${\cal C}^2$ respectively.
By means of the vielbein (\ref{Mvielbein}) these gamma matrices
are immediately computed:
\begin{equation}
\left\{\begin{array}{c}
\Gamma_{\tau\mu\theta\psi\phi}=\left(\frac{3}{32}\right)^2
r^5\sin^3\mu\cos\mu\sin\theta\,
\Gamma_{34567}\\
\Gamma_{\tau\mu\tilde\theta\psi\tilde\phi}=\frac{3}{512}r^5
\sin\mu\cos\mu\sin\tilde\theta\,\Gamma_{31245}
\end{array}\right.\,,
\end{equation}
while the square root of the determinant of the metric on the
two cycles is easily seen to be
\begin{equation}
\left\{\begin{array}{c}
\sqrt{g_1}=\left(\frac{3}{32}\right)^2r^5\sin^3\mu\cos\mu\sin\theta\\
\sqrt{g_1}=\frac{3}{512}r^5\sin\mu\cos\mu\sin\tilde\theta
\end{array}\right.\,.
\end{equation}
So, for both cycles, the $\kappa$-symmetry projector
(\ref{kprojector})
reduces to the projector of a five dimensional hyperplane embedded
in flat spacetime:
\begin{equation}
P_\pm=\left\{\begin{array}{c}
\frac{1}{2}\left(\unity\pm{\rm i}\,\Gamma_{34567}\right)\\
\frac{1}{2}\left(\unity\pm{\rm i}\,\Gamma_{31245}\right)
\end{array}\right.\,.
\label{project}
\end{equation}
The important thing to check is that the projectors (\ref{project})
are non--zero on the two Killing spinors of the space ${\cal C}(M^{111})\times 
\cM_3$. Indeed, this latter has not $32$ preserved supersymmetries, rather it
has only $8$ of them. In order to avoid long and useless calculations
we just argue as follows. Using the gamma--matrix basis of \cite{cdfm111}, the
Killing spinors are already known. We have:
\begin{equation}
\begin{array}{rclcrcl}
  \Gamma_0&=&\gamma_0 \,\otimes \,{\bf 1}_{8 \times 8} & ; &
  \Gamma_8&=&\gamma_1 \,\otimes \,{\bf 1}_{8 \times 8} \\
  \Gamma_9&=&\gamma_2 \,\otimes \,{\bf 1}_{8 \times 8} & ; &
  \Gamma_{10}&=&\gamma_3 \,\otimes \,{\bf 1}_{8 \times 8}\\
  \Gamma_i&=&\gamma_5 \,\otimes \,\tau_i & (i=1,\dots,7) &\null&\null&\null  \\
\end{array}
\label{gammole}
\end{equation}
where $\gamma_{0,1,2,3}$ are the usual $4 \times 4$ gamma matrices in
four--dimensional space--time, while $\tau_i$ are the $8\times 8$
gamma--matrices satisfying the $SO(7)$ Clifford algebra in the form:
$\{ \tau_i \, , \, \tau_j \}= -\delta_{ij}$. For these matrices we
take the representation given in the Appendix of \cite{cdfm111}, which
is well adapted to the intrinsic description of the $M^{111}$
metric through Maurer--Cartan forms. In this
basis the Killing spinors were calculated in \cite{cdfm111} and have
the following form:
\begin{eqnarray}
  \mbox{Killing spinors} &=& \epsilon (x)\,\otimes \,\eta \quad ;
  \quad \eta = \left(\begin{array}{c}
   {\bf 0} \\
\hline
    {\bf u} \\
    \hline
   {\bf 0} \\
    \hline
    \epsilon {\bf u}^\star \\
\end{array}\right)\,,
\label{kilspino}
\end{eqnarray}
where
\begin{eqnarray}
 {\bf u} &=& \left( \begin{array}{c}
   a+{\rm i} b \\
   0 \
 \end{array}\right) \, \quad ; \quad \,\epsilon {\bf u}^\star= \left( \begin{array}{cc}
   0 & 1 \\
   -1 & 0 \
 \end{array}\right) \,{\bf u}^\star=\left( \begin{array}{c}
   0 \\
   -a+{\rm i} b \
 \end{array}\right)
\label{uspino}
\end{eqnarray}
and where the $8$--component spinor was written in $2$--component blocks.
\par
In the same basis, using notations of \cite{cdfm111}, we have:
\begin{equation}
\begin{array}{rcccl}
\Gamma_{34567} & = & \gamma_5 \, \otimes \, U_8 \, U_4 \, U_5 \, U_6 \, U_7
\otimes \sigma_3 &=&
{\rm i} \, \gamma_5 \, \otimes \, \left( \begin{array}{c|c|c|c}
 - {\bf 1}_{2\times 2} & 0 & 0 & 0 \\
  \hline
  0 &  {\bf 1}_{2\times 2} & 0 & 0 \\
  \hline
  0 & 0 &  {\bf 1}_{2\times 2} & 0 \\
  \hline
  0 & 0 & 0 &  - {\bf 1}_{2\times 2}
\end{array}\right)\,,\\
\null & \null & \null & \null & \null \\
\Gamma_{31245} & = & \gamma_5 \, \otimes \,{\rm i} U_8 \, U_4 \, U_5 \,
\otimes \, {\bf 1} &=&{\rm i} \, \gamma_5 \, \otimes \, \left( \begin{array}{c|c|c|c}
 \sigma_3 & 0 & 0 & 0 \\
  \hline
  0 &  \sigma_3 & 0 & 0 \\
  \hline
  0 & 0 &  \sigma_3 & 0 \\
  \hline
  0 & 0 & 0 &  \sigma_3
\end{array}\right)\,.\\
\end{array}
\label{expligamma}
\end{equation}
As we see, by comparing eq. (\ref{project}) with eq. (\ref{kilspino})
and (\ref{expligamma}), the $\kappa$--supersymmetry projector reduces
for both cycles to a chirality projector on the $4$--component
space--time part $\epsilon(x)$. As such, the $\kappa$--supersymmetry
projector always admits non vanishing eigenstates implying that the
cycle is supersymmetric. The only flaw in the above argument is
that the Killing spinor (\ref{kilspino}) was determined in
\cite{cdfm111} using as vielbein basis the suitably rescaled Maurer--Cartan forms
$\cB^3$, $\cB^{m}$,  $(m=1,2)$ and $\cB^A$, $(A=4,5,6,7)$ (see chapter $3$).
Our choice (\ref{Mvielbein}) does not correspond to the same vielbein basis.
However, a little inspection shows that it differs only by some
$SO(4)$ rotation in the space of ${\IP}^{2}$ vielbein $4,5,6,7$.
Hence we can turn matters around and ask what happens to the Killing spinor
(\ref{kilspino}) if we apply an $SO(4)$ rotation in the directions
$4,5,6,7$. It suffices to check the form of the gamma--matrices
$[\tau_A \, , \, \tau_B]$ which are the generators of such rotations.
Using again the Appendix of \cite{cdfm111} we see that such $SO(4)$
generators are of the form
\begin{equation}
 {\rm i} \,  \left( \begin{array}{c|c|c|c}
 \sigma_i & 0 & 0 & 0 \\
  \hline
  0 &  \sigma_i & 0 & 0 \\
  \hline
  0 & 0 &  \sigma_i & 0 \\
  \hline
  0 & 0 & 0 &  \sigma_i
\end{array}\right) \quad  \mbox{or} \quad  {\rm i} \,  \left( \begin{array}{c|c|c|c}
 \sigma_i & 0 & 0 & 0 \\
  \hline
  0 &  -\sigma_i & 0 & 0 \\
  \hline
  0 & 0 &  \sigma_i & 0 \\
  \hline
  0 & 0 & 0 &  -\sigma_i
\end{array}\right),
\end{equation}
so that  the $SO(4)$ rotated Killing spinor is of the same form as in
eq.(\ref{kilspino}) with, however, ${\bf u}$ replaced by ${\bf u}^\prime = A {\bf u}$
where $A\in SU(2)$. It is obvious that such an $SU(2)$ transformation does
not alter our conclusions. We can always decompose ${\bf u}^\prime$
into $\sigma_3$ eigenstates and associate the $\sigma_3$--eigenvalue
with the chirality eigenvalue, so as to satisfy the
$\kappa$--supersymmetry projection. Hence, our $5$--cycles are indeed
supersymmetric.
\par
\subsection{The case of $Q^{111}$}
\par
\subsubsection{Cohomology of $Q^{111}$}
As for the cohomology \cite{BT}, the
first Chern class of $L$ is $c_1 = \omega_1 + \omega_2 + \omega_3$, where
$\omega_i$ are the generators of the second cohomology
group of the ${\IP}^1$'s.
Reasoning as for $ M^{111} $, one gets
\begin{eqnarray}
& H^1(Q^{111}, \ZZ) =
H^3(Q^{111}, \ZZ) =
H^6(Q^{111}, \ZZ) = 0,&
\nonumber\\ & H^2(Q^{111}, \ZZ) = \ZZ \cdot \omega_1 \oplus \ZZ \cdot \omega_2,
&
\nonumber\\ &
H^4(Q^{111}, \ZZ) = \ZZ_2 \cdot (\omega_1 \omega_2 +\omega_1 \omega_3
+ \omega_2 \omega_3),
&
\nonumber\\ &
H^5(Q^{111}, \ZZ) = \ZZ \cdot \alpha \oplus \ZZ \cdot \beta,&
\label{q111gys}
\end{eqnarray}
where $\pi_* \alpha = \omega_1 \omega_2 - \omega_1 \omega_3$,
$\pi_* \beta = \omega_1 \omega_2 - \omega_2 \omega_3$ and the pullbacks
are left implicit.
\subsubsection{Explicit description of the $U\ll(1\rr)$ fibration for $Q^{111}$}
The coset space $Q^{111}$ is a $U(1)$-fibre bundle over
${\IP}^1\times{\IP}^1\times{\IP}^1\simeq S^2\times S^2\times S^2$.
We can parametrize the base manifold with polar coordinates
$(\theta_i,\phi_i)$, $i=1,2,3$.
We cover the base with eight coordinate patches,
$H_{\alpha\beta\gamma}$ $(\alpha,\beta,\gamma=\pm 1)$ and choose local coordinates
for the fibre, $\psi_{\alpha\beta\gamma}\in[0,4\pi)$.
Every patch is the product of three open sets, $H^i_{\pm}$,
each one describing a coordinate patch for a single two-sphere,
as indicated in fig. \ref{S2patches}:
\begin{equation}
H_{\alpha\beta\gamma}=H^1_\alpha\times H^2_\beta\times H^3_\gamma.
\end{equation}
%%%%%%%%%%%%%%%%%%%%%%%%%%%%%%%%%%%%%%%%%%%%%%%%%
%          S^2    PATCH  FIGURE
%%%%%%%%%%%%%%%%%%%%%%%%%%%%%%%%%%%%%%%%%%%%%%%%%
\iffigs
\begin{figure}[ht]
\begin{center}
\epsfxsize = 5cm
\epsffile{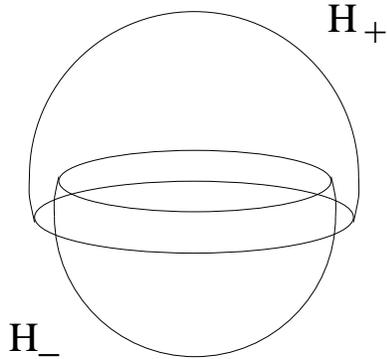}
\vskip  0.2cm
\hskip 2cm
\unitlength=1.1mm
\end{center}
\caption{Two coordinate patches for the sphere.
They constitute the base for a local trivialization of a fibre
bundle on $S^2$.
Each patch covers only one of the poles, where the coordinates
$(\theta,\phi)$ are singular.
}
\label{S2patches}
\end{figure}
\fi
To describe the total space we have to specify the transition maps
for $\psi$ on the intersections of the patches.
These maps for the generic $Q^{pqr}$ space are
\begin{equation}\label{maps}
\psi_{\alpha_1\beta_1\gamma_1}=\psi_{\alpha_2\beta_2\gamma_2}+p(\alpha_1-\alpha_2)\phi_1+q(\beta_1-\beta_2)\phi_2
+r(\gamma_1-\gamma_2)\phi_3\,.
\end{equation}
For example, in the case of interest, $Q^{111}$, we have
\begin{equation}
\psi_{+-+}=\psi_{++-}-2\phi_2+2\phi_3\,.
\end{equation}
We note that these maps are well defined, being all
the $\psi$'s and $\phi$'s defined modulo $4\pi$ and $2\pi$ respectively.
\par
It is important to note that $\theta$ and $\phi$ are clearly not good
coordinates for the whole $S^2$.
The most important consequence of this fact is that the one-form
$d\phi$ is not extensible to the poles.
To extend it to one of the poles, $d\phi$ has to be multiplied
by a function which has a double zero on that pole, such as
$\sin^2\frac{\theta}{2}\,d\phi$.
\par
We can define a $U(1)$-connection $\cal A$ on the base
$S^2\times S^2\times S^2$ by specifying it on each patch $H_{\alpha\beta\gamma}$
\footnote{It is worth noting that the connection $\cal A$ is
chosen to be well defined on the coordinate singularities of each patch,
i.e. on the product of the three $S^2$ poles covered by the patch.}:
\begin{eqnarray}
{\cal A}_{\alpha\beta\gamma}=(\alpha-\cos\theta_1)d\phi_1+(\beta-\cos\theta_2)d\phi_2+
(\gamma-\cos\theta_3)d\phi_3\,.
\end{eqnarray}
Because of the fibre-coordinate transition maps (\ref{maps}), the
one-form $(d\psi-{\cal A})$ is globally well defined on $Q^{111}$.
In other words the different one-forms $(d\psi_{\alpha\beta\gamma}-
{\cal A}_{\alpha\beta\gamma})$
defined on the corresponding $H_{\alpha\beta\gamma}$, coincide on the intersections
of the patches.
We can therefore define an $SU(2)^3\times U(1)$-invariant metric on the
total space by:
\begin{equation}
ds^2_{Q^{111}}=c^2(d\psi-{\cal A})^2+a^2ds^2_{S^2\times S^2\times S^2}\,.
\end{equation}
The Einstein metric of this family is given by
\begin{equation}\label{Qmetric}
ds^2_{Q^{111}}=\frac{3}{8\Lambda}(d\psi-{\cal A})^2
+\frac{3}{4\Lambda}\sum_{i=1}^3
\left(d\theta_i^2+\sin^2\theta_i\,d\phi_i^2\right)\,,
\label{metq111}
\end{equation}
where $\Lambda$ is the compact space cosmological constant defined in
eq.(\ref{ricciin}). The Einstein metric (\ref{metq111}) was
originally found in \cite{dafrepvn}, 
using the intrinsic geometry of coset manifolds and using Maurer--Cartan
forms. An explicit form was also given using stereographic
coordinates on the three $S^2$. In the coordinate form of
eq. (\ref{metq111}) the Einstein metric of $Q^{111}$ was later given in \cite{pagepopeQ}.
%%%%%%%%%%%%%%%%%%%%%%%%%%%%%%%%%%%%%%%%%%%%%%%%%%%%%%%%%%%%%%%%%
%             Q 111    H O M O L O G Y
%%%%%%%%%%%%%%%%%%%%%%%%%%%%%%%%%%%%%%%%%%%%%%%%%%%%%%%%%%%%%%%%%
\subsubsection{The baryonic $5$--cycles of $Q^{111}$ and their volume}
\label{brynQ}
The relevant homology group of $Q^{111}$ for the
calculation of the baryonic masses is
\begin{equation}
H_5(Q^{111},{\IR})={\IR}^2\,.
\end{equation}
Three (dependent) five-cycles spanning $H_5(Q^{111})$ are the restrictions
of the $U(1)$-fibration to the product of two of the three ${\IP}^1$'s.
Using the above metric (\ref{metq111}) one easily computes the volume of these
cycles. For instance
\begin{equation}
{\rm Vol(cycle)}=\oint_{\pi^{-1}({\IP}^1_1\times{\IP}^1_2)}
\left(\frac{3}{8\Lambda}\right)^{5/2}
4\sin\theta_1\,\sin\theta_2\ d\theta_1\,d\theta_2\,d\phi_1\,d\phi_2\,d\psi=
\frac{\pi^3}{4}\left(\frac{6}{\Lambda}\right)^{5/2}\,.
\label{volcycq}
\end{equation}
The volume of the whole space $Q^{111}$ is
\begin{equation}
{\rm Vol(Q^{111})}=\oint_{Q^{111}}\left(\frac{3}{8\Lambda}\right)^{7/2}
8\,\prod_{i=1}^{3}\sin\theta_i\,d\theta_i d\phi_i\,d\psi
=\frac{\pi^4}{8}\left(\frac{6}{\Lambda}\right)^{7/2}\,.
\label{volq111}
\end{equation}
Just as in the $M^{111}$ case, inserting the above results (\ref{volcycq}, \ref{volq111})
into the general formula (\ref{baryondim}) we obtain the conformal weight of
the  baryon operator corresponding to the five-brane wrapped on this cycle:
\begin{equation}
E_0=\frac{N}{3}\,.
\label{Ccweight}
\end{equation}
\par
The other two cycles can be obtained from this by permuting the role of the three
${\IP}^1$'s and their volume is the same.
This fact agrees with the symmetry which exchanges the
fundamental fields $A$, $B$ and $C$ of the conformal theory, or the
three gauge groups $SU(N)$. Indeed, naming $SU(2)_i$ ($i=1,2,3$) the
three $SU(2)$ factors appearing in the isometry group of $Q^{111}$,
the stability subgroup of the first of the cycles described above is
\begin{eqnarray}
  H({\cal C}^1) &=& SU(2)_1 \times SU(2)_2 \times U(1)_{B,3}\nonumber\\
  U(1)_{B,3} &\subset & SU(2)_3
\label{hcycq111}
\end{eqnarray}
so that the collective coordinates of the baryon state live on
${\IP}^1 \simeq SU(2)_3/U(1)_{B,3}$. This result is obtained by
an argument completely analogous to that used in the analysis of
$M^{111}$ $5$--cycles and leads to a completely analogous conclusion.
The baryon state is in the $J^{\ll(1\rr)}=0, \, J^{\ll(2\rr)} = 0, \, J^{\ll(3\rr)}=N/2$ flavour
representation. In the conformal field theory the corresponding
baryon operator is the chiral field (\ref{q111baryopC})
and the result (\ref{Ccweight}) implies that the conformal weight of
the $C_i$ elementary world--volume field is
\begin{equation}
  h[C_i]=\frac{1}{3}.
\label{hCfield}
\end{equation}
The stability subgroup of the permuted
cycles is obtained permuting the indices $1,2,3$ in eq. (\ref{hcycq111})
and we reach the obvious conclusion
\begin{equation}
  h[A_i]=h[B_j]=h[C_\ell]=\frac{1}{3}.
\label{hABfield}
\end{equation}
This matches with the previous result (\ref{donalfonso}) on the
spectrum of chiral operators, which are predicted of the form
\begin{equation}
  \mbox{chiral operators} = \mbox{Tr} \left(  A_{i_1} \, B_{j_1} \, C_{\ell_1} \,
  \dots \,A_{i_k} \, B_{j_k} \, C_{\ell_k}\right)
\label{chiropq111}
\end{equation}
and should have conformal weight $E=k$. Indeed, we have $k \times (
\frac{1}{3}+\frac{1}{3}+\frac{1}{3}) =k$ !
\par
\section{Conclusions}
We saw, using geometrical intuition,
 that there is  a set of fundamental fields which are likely to be
the fundamental degrees of freedom of the CFT's corresponding to
$Q^{111}$ or $M^{111}$. The entire KK spectrum and
the existence of baryons of given quantum numbers
can be explained in terms of them. This fact (expecially the formula
(\ref{mastercheck}) ) constitutes in my opinion a strong non--trivial 
check of the $AdS/CFT$ correspondence.
\par
Candidate three-dimensional gauge theories which should
flow in the IR to the superconformal fixed points dual to the $AdS_4$
compactifications have also been discussed in this thesis.
The fundamental fields are the elementary chiral
multiplets of these gauge theories.
\par
The main problem which has not been solved is the existence of chiral operators 
in the gauge theory that have no counterpart in the KK spectrum. These are
the non completely flavour symmetric chiral operators. Their existence
 is due to
the fact that, differently from the case of $T^{11}$, we are not able
to write any superpotential of dimension two.  
If the proposed gauge theories are correct, the dynamical
 mechanism responsible for
the disappearing of the non symmetric operators in the IR
has still to be clarified. It is probably of non--perturbative nature.
\par
It would be quite helpful to have a description of the conifold
as  a deformation of an orbifold singularity \cite{witkleb}, \cite{morpless}.
It would provide an holographic description of the RG flow between two
different CFT theories and it would also help in checking whether the
proposed gauge theories are correct or require to be slightly
modified by the introduction of new fields.
Another direction of possible improvement of our theory consists in
considering the Chern Simons coupling, which we have set to zero.
%%%%%%%%%%%%%%%%%%%%%%%%%%%%%%%%%%%%%%%%%%%%%%%%%%%%%%%%
%%%%%%%% APPENDIX %%%%%%%%%%%%%%%%%%%%%%%%%%%%%%%%%%%%%% 
%%%%%%%%%%%%%%%%%%%%%%%%%%%%%%%%%%%%%%%%%%%%%%%%%%%%%%%%
\appendix
\chapter{Conventions for the $M^{111}$ space
\label{mpqrconventions}}
\par
\label{convenzioni}
The
 Gell--Mann matrices are:
\begin{eqnarray}
    \lambda_1    =
    \left(\begin{array}{ccc}
          0   &  1  & 0 \\
          1   &  0   & 0 \\
          0   &  0  & 0 \\
          \end{array}
    \right)\,, \quad
    \lambda_2    =
    \left(\begin{array}{ccc}
          0   & -i   & 0 \\
          i   &  0  & 0  \\
          0   &  0  & 0  \\
          \end{array}
    \right)\,, \quad
    \lambda_3    =
    \left(\begin{array}{ccc}
          1   & 0   & 0 \\
          0   & -1  & 0 \\
          0   & 0   & 0 \\
          \end{array}
    \right)\,,
\nonumber \\
    \lambda_4    =
    \left(\begin{array}{ccc}
           0  & 0   & 1 \\
           0  & 0   & 0 \\
           1  &  0  & 0 \\
          \end{array}
    \right)\,, \quad
    \lambda_5    =
    \left(\begin{array}{ccc}
           0  & 0   & -i \\
           0  & 0   &  0\\
           i  & 0   &  0\\
          \end{array}
    \right)\,,\quad
    \lambda_6    =
    \left(\begin{array}{ccc}
           0  & 0   & 0 \\
           0  & 0   & 1 \\
            0 & 1   & 0 \\
          \end{array}
    \right)\,, \quad
\nonumber \\
    \lambda_7    =
    \left(\begin{array}{ccc}
           0  & 0   & 0 \\
           0  & 0   & -i \\
            0 & i   & 0 \\
          \end{array}
    \right)\,,
    \lambda_8    = \ft{1}{\sqrt{3}}
    \left(\begin{array}{ccc}
           1  &  0  & 0 \\
           0  &  1  & 0 \\
           0  &  0  & -2 \\
          \end{array}
    \right)\,.
\nonumber \\
\end{eqnarray}
The Pauli matrices are:
\begin{equation}
\sigma_1 =
    \left(\begin{array}{cc}
           0  &  1  \\
           1  &  0  \\
          \end{array}
    \right) \,,
\sigma_2 =
    \left(\begin{array}{cc}
           0  &  -i  \\
           i  &  0  \\
          \end{array}
    \right) \,,
\sigma_3 =
    \left(\begin{array}{cc}
           1  &  0  \\
           0  &  -1  \\
          \end{array}
    \right) \,.
\end{equation}
The structure constants of $SU(3)$ are given by $f_{ijk}=f_{[ijk]}$,
$[\lambda_i, \lambda_j]=2i f_{ijk} \lambda_k$
\begin{eqnarray}
f_{123} &=& 1 \,, 
\nonumber \\
f_{147}&=& \ft12\,, \quad f_{156}=-\ft12\,, \quad f_{246} = \ft12\,,
f_{257} = \ft12 \,, \quad f_{345} = \ft12 \,, \quad f_{367} = -\ft12 \,,
\nonumber \\ 
f_{458} &=& \ft{\sqrt{3}}{2} \,, \quad f_{678} = \ft{\sqrt{3}}{2}\,.
\end{eqnarray}
The generators of $G=SU\left(3\right)\times SU\left(2\right)\times U\left(1\right)$ are:
\begin{eqnarray}
SU\left(3\right) : &~~~&{i\over 2}\lambda_1,\dots,{i\over 2}\lambda_8\nonumber\\
SU\left(2\right) : &~~~&{i\over 2}\sigma_1,\dots,{i\over 2},\sigma_3\nonumber\\
U\left(1\right):         &~~~&i Y\,.
 \nonumber
\end{eqnarray}
The orthogonal decomposition gives
\begin{equation}
\IG=\IH\oplus\IK
\end{equation}
where $\IH$ is a subalgebra of $\IG$, and $\IK$ is a representation of $\IH$.
\par
The generators of $H=SU\left(2\right)\times U\left(1\right)'\times U\left(1\right)''
$ are:
\begin{eqnarray}
SU\left(2\right) : &~~~&{i\over 2}\lambda_{\dot{m}}={i\over 2}\lambda_1,\dots,{i\over 2}
\lambda_3\nonumber\\
U\left(1\right)':         &~~~& Z'=\sqrt{3}i\lambda_8+i\sigma_{3}-4i
Y\nonumber\\
U\left(1\right)'':         &~~~& Z''=-{\sqrt{3}\over 2}i\lambda_8+{3\over 2}i\sigma_{3}\nonumber
\end{eqnarray}
so the generators of the orthogonal space $\IK$ are
\begin{eqnarray}
{i\over 2}\lambda_A&=&{i\over 2}\lambda_4\dots{i\over 2}\lambda_7,\nonumber\\
\sigma_{m}&=&{i\over 2}\sigma_1,{i\over 2}\sigma_2\nonumber\\
Z&=&{\sqrt{3}\over 2}i\lambda_8+{1\over 2}i\sigma_{3}+iY\,.
\end{eqnarray}
\par
Due to this decomposition we divide the indices into six groups:
\begin{eqnarray}
&\dot{m},\dot{n}=1,2,3,\nonumber\\
&Y,\nonumber\\
&m,n=1,2,\nonumber\\
&3,\\
&A,B,C=4,5,6,7,\nonumber\\
&8\,.\nonumber
\end{eqnarray}
\par
Other indices used in this context are:
\begin{eqnarray}
\Sigma,\Lambda: &~~~&{\rm indices~of~the~adjoint~representation~of~}G\nonumber\\
a,b:&~~~&{\rm indices~of~the~vector~representation~of~}SO\left(7\right)\nonumber\\
i,j:&~~~&{\rm indices~of~the~vector~representation~of~}SU\left(2\right)\,.
\end{eqnarray}
\par
Our conventions for the $\varepsilon$ tensors are the following:
\begin{equation}
\begin{array}{lll} 
SU\left(2\right)\subset G: & \varepsilon^{mn} & \varepsilon^{12}=-1 \\
SU\left(3\right)\subset G:  & \varepsilon^{\dot{m}\dot{n}\dot{r}} &
 \varepsilon^{\dot{1}\dot{2}\dot{3}}=1 \\
SU\left(2\right)\subset H:  & \varepsilon^{\dot{m}\dot{n}} &
\varepsilon^{\dot{1}\dot{2}}=1 \\
SO\left(7\right)^c:  & \varepsilon^{abcdefg} &
\varepsilon^{1234567}=-1\,.
\end{array}
\end{equation}
%%%%%%%%%%%%%%%%%%%%%%%%%%%%%%%%%%%%%%%%%%%%%%%%%%%%%%%%%%%%%%%%%%%%%%%%%
%%%%%%%%%%%%%%%%%%%%%%%%%%%%%%%%%%%%%%%%%%%%%%%%%%%%%%%%%%%%%%%%%%%%%%%%%
%     BIBLIOGRAFIA
%%%%%%%%%%%%%%%%%%%%%%%%%%%%%%%%%%%%%%%%%%%%%%%%%%%%%%%%%%%%%%%%%%%%%%%%%
%%%%%%%%%%%%%%%%%%%%%%%%%%%%%%%%%%%%%%%%%%%%%%%%%%%%%%%%%%%%%%%%%%%%%%%%%

\end{document}